\documentclass[12pt, a4paper, twoside]{book}

\ifx\HCode\UnDef\else\hypersetup{tex4ht}\fi

\usepackage{amssymb}
\usepackage{graphicx,epsfig}
\usepackage{color}
\usepackage{simplewick}
\usepackage{slashed}
\usepackage[retainorgcmds]{IEEEtrantools}
\usepackage{verbatim}
\usepackage{setspace}
\usepackage{dsfont} 
\usepackage{fancyhdr}
\usepackage{appendix}
\usepackage{tocvsec2}
\usepackage[pdftex,colorlinks=true,linkcolor=blue,citecolor=red,pdftitle={The Yang-Mills Vacuum Wave Functional in 2+1 Dimensions},pdfauthor={Sebastian Krug}]{hyperref}

\fancypagestyle{plain}{
\fancyhf{}

\fancyfoot[LE,RO]{\thepage}
}

\makeatletter
\def\cleardoublepage{\clearpage\if@twoside \ifodd\c@page\else
    \hbox{}
    \thispagestyle{plain}
    \newpage
    \if@twocolumn\hbox{}\newpage\fi\fi\fi}
\makeatother \clearpage{\pagestyle{plain}\cleardoublepage}

\def\tr{\mathrm{Tr}}

\def\p{\partial}
\def\e{\epsilon}
\def\de{\delta}
\def\th{\theta}
\def\al{\alpha}
\def\sd{\slashed{\delta}}
\def\T{\mathcal{T}}
\def\H{\mathcal{H}}
\def\V{\mathcal{V}}
\def\A{\mathcal{A}}
\def\D{\mathcal{D}}

\def\O{\mathcal{O}}
\def\m{{e^2C_A\over2\pi}}
\def\half{{1\over2}}
\def\SE{Schr\"odinger equation}

\def\vz {\vec{z}}

\def\vw {\vec{w}}

\def\d{\delta}

\def\bdel{\bar{\partial}}

\newcommand{\nn}{\nonumber}
\newcommand{\be}{\begin{equation}}
\newcommand{\ee}{\end{equation}}
\newcommand{\bea}{\begin{eqnarray}}
\newcommand{\eea}{\end{eqnarray}}
\newcommand{\bE}{\begin{IEEEeqnarray}}
\newcommand{\eE}{\end{IEEEeqnarray}}

\def \beq  {\begin{equation}}
\def \eeq  {\end{equation}}
\def \beqar {\begin{eqnarray}}
\def \eeqar {\end{eqnarray}}

\begin{document}

\begin{titlepage}
    \begin{center}
        \vspace*{1cm}
        
        \Huge
        \textbf{The Yang-Mills Vacuum Wave Functional in 2+1 Dimensions}
        
        \vspace{1cm}
        \LARGE        
        \textbf{Sebastian Krug}
        
        \vspace{0.75cm}
        \large
         A thesis presented for the degree of\\
         \LARGE
        Doctor of Philosophy\\

          \vspace{1.5cm}
          \large
 	Supervisor:\\
 	\LARGE
 	Dr.~Antonio Pineda
        
        \vfill

        
        \includegraphics[width=0.5\textwidth]{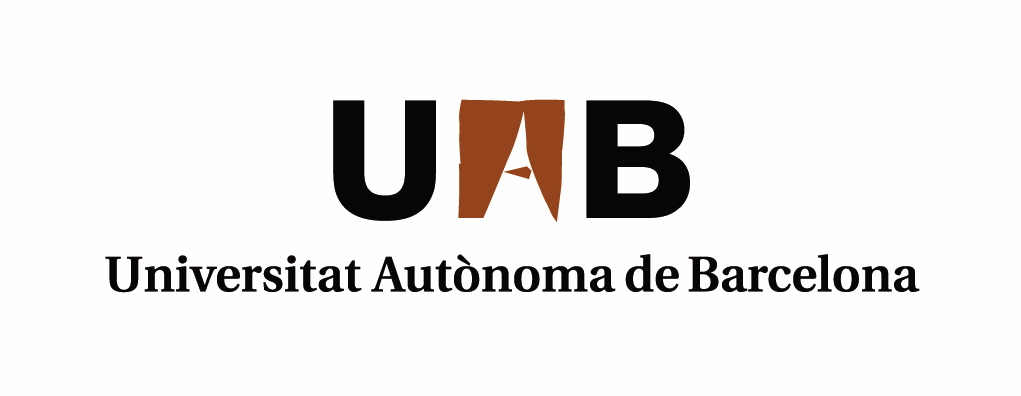}
        
        \large
        Departament de F\'isica; Unitat de F\'isica Te\`orica\\
        Doctorat en F\'isica
 
        \vspace{0.5cm}       
        \Large        
        January 2014
        
    \end{center}
\end{titlepage}

\newpage
\thispagestyle{empty}
\mbox{}

\frontmatter

\section*{Acknowledgments -- Agradecimientos -- Danksagungen}

It is with great pleasure that I thank my supervisor Antonio Pineda for the support he gave me throughout these four years. He was willing to help and to share his vast knowledge at any time. It has been illuminating to work with him and I very much appreciate having had this opportunity.

\vspace{0.5cm}

I also want to express my gratitude to Matthias Jamin, \v{S}tefan Olejn{\'\i}k, Joan Soto, Cristina Manuel, and Oriol Pujol\`as for agreeing to form the evaluation committee for this thesis.

\vspace{0.5cm}

Furthermore, I especially want to thank V.~Parameswaran Nair and Dimitra Karabali for valuable scientific discussions, as well as for the opportunity to visit them at CUNY.

\vspace{0.5cm}

During my time at IFAE I met a lot of amazing people, and I am happy to thank them for a great time: mis hermanos Alvise and Clara, as well as Alberto, Diogo, Javi, Joan Antoni, Juanjo, Linda, Marc, Mariona, Mat(t)eo, Max, Nikos, Oriol, Pablo, Sandeepan, Sergi, Simone, Volker and many more.

\vspace{0.5cm}

Much\'isimas gracias a Bere y a toda la gente de Barcelona por convertir mi tiempo aqu\'i en una experiencia excepcional: Ben, N\'uria, Xavi, Milos, Camille, Violant, Kevin, Kai, Ana, Ali, Juancar, Irati, Cisco, Moli, Rebecca, Albert, Elsa\ldots\ por mencionar solo algunos.

\vspace{0.5cm}

Gracias tambi\'en al Ministerio de Educaci\'on, Cultura y Deporte por patrocinar este trabajo con una beca FPU (AP2009-1492).

\vspace{0.5cm}

A very special thanks, once again, goes to Felix Pahl for sharing his impressive knowledge of both physics and English grammar. I am of course responsible for all remaining mistakes.

\vspace{0.5cm}

Ich bedanke mich ganz herzlich bei meiner Familie f\"ur die Unterst\"utzung w\"ahrend dieser Zeit (und anderen Zeiten).

\vspace{0.5cm}

Vor allem, und ganz besonders, gilt mein Dank Franziska Rakel, einerseits f\"ur die Unter-st\"utzung \"uber die letzten Jahre, aber haupts\"achlich daf\"ur, dass sie ein wunderbarer \linebreak Mensch ist. Diese Arbeit ist ihr gewidmet. Danke! 

\newpage
\thispagestyle{empty}
\mbox{}

\newpage
\section*{Abstract}
We investigate Yang-Mills theory in 2+1 dimensions in the Schr\"odinger representation. Three dimensional Yang-Mills theory is relevant on the one hand, because it is the lowest dimensional Yang-Mills theory with propagating degrees of freedom, 
and on the other hand, because it provides the high temperature limit of four dimensional QCD. The Schr\"odinger picture is interesting because it is well suited to explore properties of the vacuum state in the non-perturbative regime. Yet, not much analytical work has been done on this subject, and even the topic of perturbation theory in the Schr\"odinger representation is not well developed, especially in the case of gauge theories. In a paper by Hatfield [Phys.\ Lett.\  B {\bf 147}, 435 (1984)] the vacuum wave functional for SU(2) theory was computed to $\O(e)$. In the non-perturbative regime, the most sophisticated analytical approach has been developed by Karabali et al.~in a series of papers (see [Nucl.\ Phys.\  B {\bf 824}, 387 (2010)] and references therein). This thesis aims to put perturbation theory in the Schr\"odinger representation on more solid ground by computing the vacuum wave functional for a general gauge group SU$(N_c)$ up to $\O(e^2)$, utilizing modifications of these two methods. This is important since it provides us with a tool for testing non-perturbative approaches, which should reproduce the perturbative result in an appropriate limit.  

Furthermore, regularization and renormalization are also not well understood in the Schr\"odinger picture. The regularization method proposed by Karabali et al.~leads to conflicting results when applied to the computation of the vacuum wave functional with the two different methods mentioned above. We aim to clarify how regularization should be implemented and develop a new regularization approach, which brings these two expressions into agreement, providing a strong check of the regularization employed. We argue that this regularization procedure is not specific to the cases studied here. It should be applied in the same way to any quantum field theory in any dimension in the Schr\"odinger picture. This is the main result of the thesis.

We then go on to illustrate how physical observables can be computed in the non-perturbative regime, using the trial wave functional proposed in [Nucl.\ Phys.\  B {\bf 824}, 387 (2010)].
Among other observables, we compute the static potential at long distances, for which we find corrections not compatible with a linear potential. 

Finally, we also discuss the possibility of extending this approach to 3+1 dimensions.


\newpage
\section*{Notation and Conventions}
Throughout this thesis we use the acronyms QCD and QED for Quantum chromodynamics and Quantum electrodynamics, respectively. We also use LO and NLO for leading order and next-to-leading order, respectively, and VEV for vacuum expectation value. We use the abbreviations Ref., Chap., Sec., App.~and Eq.~for reference, chapter, section, appendix and equation, respectively, as well as the plural forms Refs., Chaps., Secs., Apps.~and Eqs. 
The expressions ``Wilson line" and ``string" are used synonymously.\\
Furthermore we employ the following conventions:

\begin{itemize}
\item We use units such that $\hbar=c=1$.
\item The metric tensor in 2+1 dimensions is $\eta_{\mu\nu}=\mathrm{diag}(-1,+1,+1)$.
\item Greek indices $\mu,\nu,\al,\ldots$ label the components of vectors and tensors in 2+1 space-time dimensions and take the values $0,1,2$, while Latin indices $i,j,k,\ldots$ label their spatial components only, taking the values $1,2$. Spatial vectors are indicated by arrows, e.g.~$\vec x=(x_1,x_2)$.
\item Color indices in the adjoint representation are $a,b,c,\ldots$ and take the values $1,\ldots,N_c^2-\nolinebreak1$.
\item If not noted otherwise, the Einstein summation convention over repeated indices (space-time as well as color) is employed.
\item The SU$(N_c)$ generators are $T^a$, with $(T^a)_{bc}=-if^{abc}$ in the adjoint representation, and $[T^a,T^b]=if^{abc}T^c$. The quadratic Casimir operators are $C_A=N_c$ in the adjoint and $C_F=\frac{N_c^2-1}{2N_c}$ in the fundamental representation.
\item Color carrying fields are $A_{\mu}=-iT^aA_{\mu}^a$, $B=-iT^aB^a$, $J=J^aT^a$ (sic) and $\theta=-i\theta^aT^a$.
\item Integration in position space is written as $\int_x \equiv \int d^dx$, and in momentum space as  $\int_\slashed{k} \equiv \int \frac{d^dk}{(2\pi)^d}$. Delta functions in momentum space are written as $\slashed{\delta}(\vec{k})\equiv (2\pi)^d\delta^{(d)}(\vec{k})$. Typically $d=2$, except for a small portion of Chap.~\ref{chap:TonV}, where $d=3$.
\item The convention for the Fourier transformation for all fields is
\[
\phi(\vec x)=\int_\slashed{k} e^{i{\vec k}\cdot{\vec x}} \phi(\vec k)
\,,
\qquad
\frac{\delta}{\delta\phi(\vec{x})}
=\int_\slashed{k} e^{-i{\vec k}\cdot{\vec x}}
\frac{\delta}{\delta \phi(\vec{k})}
\,. \nn
\]
\end{itemize}

\tableofcontents

\newpage
\thispagestyle{empty}
\mbox{}
\newpage
\thispagestyle{empty}
\mbox{}

\mainmatter

\chapter{Introduction}

Yang-Mills theories, gauge theories based on an SU$(N_c)$ gauge group, are crucial to our understanding of the physics of the fundamental forces that govern our world. They form the basis of our description of the strong force (based on SU(3)), as well as of the unified electroweak interaction (based on SU(2) $\times$ U(1)). Quantum chromodynamics (QCD) describes hadrons through elementary fermions (quarks and antiquarks) that carry an SU(3) charge, called color, interacting via the interchange of gauge bosons, called gluons. In contrast to photons (the gauge fields of Quantum electrodynamics (QED), which is based on the abelian group U(1)), the gauge bosons of a non-abelian theory also carry the charge of the interaction, which implies that they are self-interacting. 

In the case of the electroweak force the Higgs mechanism splits the gauge sector into three massive self-interacting bosons and the massless photon. The latter does not interact with itself, while the masses of the former tame the infrared behavior of the non-abelian gauge theory. This makes it possible to compute observables in general, and the vacuum state in particular, using weak coupling techniques. No such mechanism exists for the strong interaction, however, where the gluons remain self-interacting and massless, and the strength of the coupling increases towards lower energies, which is why the QCD vacuum is non-trivial, and yet to be understood quantitatively. 

This has several important consequences. One is color confinement, the fact that only states that transform as a singlet under color transformations appear in experiments. In particular, no free quarks, which are color triplets, or free gluons, which are color octets, are observed. Qualitatively this can be explained by the fact that the potential energy between static color sources, unlike the gravitational or electromagnetic potential, increases linearly with distance, due to the self-interaction of the gluons. In reality, quark-antiquark pairs are created as the separation increases, which then hadronize, meaning that they form color neutral (``white") bound states. A quantitative description of this phenomenon, however, is still lacking, making it one of the longest standing and most important problems in particle physics. 

Another consequence of self-interacting gauge bosons is the prediction of purely gluonic bound states, which as of now remains to be experimentally confirmed. As color confinement only allows color neutral states, there can be no free single gluon states, only singlets made up of two or more gluons. In contrast to QED, where single photon states with a continuous energy spectrum are possible, these bound states of gluons, called glueballs, have to have a finite mass, but the precise mechanism for the generation of this mass remains unknown.

In the quest for a better understanding of QCD, an approach which considers the case of a large number of colors $N_c$ has been studied (presented in Ref.~\cite{tHooft1}). In the limit $N_c\to\nolinebreak\infty$, gluons and quarks decouple; thus a good grasp of pure gluodynamics is crucial for a successful application of this method. For all of these reasons, it is important (yet difficult) to thoroughly investigate Yang-Mills theories, which describe the dynamics of the gauge bosons. As up to now it has been impossible to solve them in the physical case of 3+1 dimensions, one has to devise sensible simplifications.

A common one is to consider the theory at weak coupling and to calculate observables in perturbation theory. This approach has led to several major successes in the description of electroweak and high-energy QCD events, but it does not provide an understanding of low-energy QCD phenomena, in particular confinement. Since the strong coupling constant is not small at low energies, perturbation theory breaks down in this limit, because all orders in the perturbation series are important, and higher orders cannot be neglected. Nevertheless, most of the time we will consider the weak coupling limit in this thesis. 
It is important because perturbation theory provides us with a controllable tool for testing non-perturbative approaches, which should reproduce the perturbative result in an appropriate limit. Furthermore it allows us to address conceptual questions about the computational method that we use, which are independent of the magnitude of the coupling constant.

Another, independent way to achieve simplification is to reduce the number of space-time dimensions considered, and to try to draw information from these simpler cases on how to approach the physical case of 3+1 dimensions. Yang-Mills theory in 1+1 dimensions is exactly solvable (see Ref.~\cite{'tHooft:1974hx}), but, since it has no dynamical degrees of freedom, it is of limited informational value. In 2+1 dimensions the theory is more interesting, as it does contain propagating degrees of freedom, while still being easier to handle, in particular because it is super-renormalizable. An introduction to this topic is given in Ref.~\cite{Feynman:1981ss}. Furthermore, 2+1 dimensional Yang-Mills theory is amenable to a non-perturbative analysis devised by Karabali, Nair and collaborators in Refs.~\cite{Karabali:1995ps,Karabali:1996je,Karabali:1996iu,Karabali:1997wk,Karabali:1998yq,Karabali:2009rg} that makes extensive use of two-dimensional conformal field theory (which is very different from conformal field theory in any other dimension), thus making it an ideal testing ground for this approach. While we hope to gain information about the 3+1 dimensional case by studying the lower dimensional theory, Yang-Mills theory in 2+1 dimensions also has an important physical application: High temperature QCD in 3+1 dimensions, which is needed for the description of processes in the early universe and which can be tested with heavy ion collision experiments that are performed at the RHIC and the LHC, can be approximated by Yang-Mills theory in 3 euclidean dimensions (\cite{Gross:1980br,Appelquist:1981vg}). Relevant observables in this regime, like the magnetic screening mass, can thus be computed by way of analytic continuation from 2+1 dimensional Yang-Mills theory. Most of the time in this thesis, we will work on Yang-Mills theory in 2+1 dimensions, but in Chap.~\ref{chap:TonV} we will also give a brief glimpse of a possible extension to 3+1 dimensions of the methods applied here.


\bigskip
There are three equivalent representations of quantum field theory (QFT): operator, path integral, and Schr\"odinger representation. While the first two are well known, the Schr\"odinger representation, which 
makes use of wave functionals and functional differential equations, is less so. Nevertheless, all three approaches are equivalent, and they can benefit from each other. For example, the quantum effective action can be obtained from the vacuum wave functional (see \cite{Nair:2011ey}). In practice, specific problems are often solved most conveniently in one particular framework. In this thesis we will focus on the Schr\"odinger representation, which is very well suited to obtain information about the Yang-Mills vacuum, in particular because it allows for 
a straightforward way to go beyond perturbation theory, hence allowing for computations outside of the weak coupling regime. Yet, not much analytical work has been done on this subject, and even the topic of perturbation theory in the Schr\"odinger representation is not well developed, especially in the case of gauge theories. In a paper by Hatfield (Ref.~\cite{Hatfield:1984dv}) the vacuum wave functional for SU(2) theory was computed to $\O(e)$. In the non-perturbative regime, the most sophisticated analytical approach is the one developed in Refs.~\cite{Karabali:1995ps,Karabali:1996je,Karabali:1996iu,Karabali:1997wk,Karabali:1998yq,Karabali:2009rg}. This thesis aims to put perturbation theory in the Schr\"odinger representation on more solid ground,
utilizing modifications of these two methods.   

Furthermore, regularization and renormalization are also not well understood in the Schr\"odinger picture.
Here it proves advantageous to work in 2+1 dimensions: Since 2+1 dimensional Yang-Mills theory is super-renormalizable we do not need to worry about renormalization of the parameters of the theory. Regularization, however, must be addressed, and doing this is one of the main parts of this thesis. In this work, we aim to clarify how regularization in the Schr\"odinger representation should be implemented.

\bigskip

Because the Schr\"odinger representation is less well known we give a short introduction to this topic in Chap.~\ref{chap:Schrödinger} in order to make this thesis self-contained. As all representations of QFT are equivalent, the determination of the ground-state (or vacuum) wave functional of Yang-Mills theory, $\Psi[{\vec A}]$, is tantamount to solving it, because any observable (for instance the static potential or the spectrum of the theory) can then be obtained by the computation of the expectation value of the corresponding operator, as we will see in Chap.~\ref{chap:Schrödinger}. Even if the exact solution is not known, properly chosen trial functions may give valuable information on the vacuum via variational methods (see for instance \cite{Kovner:2004rg}).

\bigskip

We are still far from obtaining the exact ground-state wave functionals of non-abelian Yang-Mills theories. Even obtaining approximate expressions is very complicated. This is also true in the weak coupling limit. One reason is due to the requirement that the wave functional, in addition to satisfying the Schr\"odinger equation, has to be gauge invariant. This constraint is imposed by the Gauss law. Therefore, one cannot use standard quantum-mechanical perturbation theory in a straightforward manner. A procedure to overcome this problem was devised in the case of SU(2), for 3+1 dimensions, and was applied to ${\cal O}(e)$ in the weak coupling expansion, in Ref.~\cite{Hatfield:1984dv}. 
This method (which we shall call method (A)) can also be applied to the 2+1 dimensional case and a general group SU$(N_c)$ without major modifications, and it can be used to compute the terms at higher orders. We do so in Chap.~\ref{chap:Comparison} and obtain the ${\cal O}(e^2)$ expression for a general group SU$(N_c)$ in three dimensions. 

A different approach (method (B)) which reformulates the Schr\"odinger equation in terms of gauge invariant variables was worked out in Refs.~\cite{Karabali:1995ps,Karabali:1996je,Karabali:1996iu,Karabali:1997wk,Karabali:1998yq,Karabali:2009rg} in order to understand the strong coupling limit and confinement in three dimensions. It can, however, be easily reformulated to be used in a weak coupling expansion. This is done in Chap.~\ref{chap:Comparison} in order to obtain the vacuum wave functional at $\O(e^2)$. 

Both approaches have their benefits and drawbacks, so considering both is in some sense complementary. The wave functionals found in the two ways should, however, be identical. Due to the complexity of the expressions, comparing the two results is not an easy task, and we have to develop a systematic scheme to accomplish this. We do this in Chap.~\ref{chap:Comparison} and find that up to $\O(e)$ they are identical, while at $\O(e^2)$ they agree to a large extent but not completely. The discrepancy is due to regularization issues, which we address in Chap.~\ref{chap:Regularization}.

\bigskip

The regularization of the Schr\"odinger equation and the vacuum wave functional in QFT is a complicated subject. Whereas some formal aspects have been studied quite a while ago in Refs.~\cite{Symanzik:1981wd,Luscher:1985iu}, there have not been many quantitative studies of the regularization of the Yang-Mills vacuum wave functional. In three dimensions, the most detailed analyses have been carried out using method (B) (see, for instance, the discussions in Refs.~\cite{Karabali:1997wk,Agarwal:2007ns}, in particular in the appendix of the last reference). While it might seem that in method (B) regularization has already been completely taken into account, we find in Chap.~\ref{chap:Regularization} that the regularization procedure has to be modified to obtain the correct Yang-Mills vacuum wave functional.

The result of Chap.~\ref{chap:Comparison} using method (A) was obtained without any regularization of the functional Schr\"odinger equation at all. In Chap.~\ref{chap:Regularization} we carefully regularize the computation, finding that also for this method a new contribution has to be added to the result.
We then compare these new, modified results of both methods and find that they agree to ${\cal O}(e^2)$. This is a strong check of our computations and of the regularization method used.

Since this regularization method is independent of the specific theory, in Chap.~\ref{chap:Regularization} we actually give the general prescription for the implementation of regularization in the Schr\"odinger representation for a general QFT. In brief, we find that the regulator of the Hamiltonian in the Schr\"odinger representation has to be included throughout the determination of the vacuum wave functional, since removing it too early may lead to the loss of contributing terms. The insight gained here can be generalized to other QFTs and also to the four dimensional case.
 
In addition, the vacuum wave functional obtained in this way allows us to give an estimate for the magnetic screening mass.



\bigskip

In Chap.~\ref{chap:Potential} we move away from the perturbative regime. The true power of the Schr\"odinger representation lies in its ability to easily incorporate resummation schemes and \linebreak non-perturbative terms, so it does not necessarily depend on a weak coupling expansion. In Ref.~\cite{Karabali:1998yq} a strong coupling expansion for the vacuum wave functional was developed, which relies on the fact that the potential term $\V$ of the Yang-Mills Hamiltonian, viewed as a functional, is an eigenfunction of the kinetic operator $\T$. This is apparent in terms of the variables used in method (B), but it seems to be wrong in terms of the original gauge fields (method (A)) -- as long as only unregularized operators are considered. 
Once both the kinetic and potential operators are regularized, we find in a perturbative expansion that also in terms of gluon fields, $\V$ is an eigenfunction of $\T$. 
Nevertheless, we find that the corresponding eigenvalue depends on the regulator. This suggests that there may be a problem with using this strong coupling expansion to obtain the vacuum wave functional.


Another expansion scheme was developed in Ref.~\cite{Karabali:2009rg}, leading to a new proposal for the vacuum wave functional, which is claimed to interpolate between the weak coupling and the strong coupling regime, and to be a good approximation for all scales. It is given as an expansion in $e^2/m$ (where $m$ is a mass scale that appears in the computation), which corresponds to a resummation of a perturbative series. This expansion parameter is of $\O(1)$, so its use can only be justified a posteriori. The vacuum wave functional derived from this more general approach can be used to compute observables in all coupling regimes.
In Chap.~\ref{chap:Potential} we give estimates of the gluon condensate and of the correlator of the chromomagnetic field. In Ref.~\cite{Karabali:2009rg} this wave functional has been applied with great success to the computation of the static potential between a quark and an antiquark, predicting a linearly increasing potential at long distances from first principles. While this is an impressive result, there are some issues with it (in particular in light of the results of Chap.~\ref{chap:Regularization}, which demand a modification of the weak coupling limit of this vacuum wave functional), which we investigate in Chap.~\ref{chap:Potential}. In order to have more control over the computation we reformulate the wave functional of Ref.~\cite{Karabali:2009rg} in terms of the gauge fields. Computing the static potential with this trial functional, however, we find terms at next-to-leading order in $e^2/m$ which are cubic in the separation. This suggests either that $e^2/m$ is not a good expansion parameter for the computation of the static potential, or that the vacuum wave functional proposed in Ref.~\cite{Karabali:2009rg} should be modified along the lines of the findings of Chap.~\ref{chap:Regularization}. 

\bigskip

In Chap.~\ref{chap:TonV} we investigate the possibility of extending the gauge invariant approach to four dimensions. In Ref.~\cite{Freidel:2006qz}, in analogy with method (B), a third formulation of the Hamiltonian approach was devised, which we shall call method (C). Like method (B), it employs a reformulation in terms of gauge invariant variables, albeit different ones. In particular, these new variables are real, thus avoiding the problem of laborious checks for reality of the wave functional like the one we employ in Chap.~\ref{chap:Comparison}. The main advantage of this method is, however, that it may also be applied to 3+1 dimensional Yang-Mills theory. In Chap.~\ref{chap:TonV} we will first introduce it in 2+1 dimensions and then extend it to the 3+1 dimensional case. We propose a Hamiltonian which differs from the one of Ref.~\cite{Freidel:2006qz}, where a different regularization scheme was employed and some terms were dropped because they were argued to be subleading. Nevertheless, taking the results of Chap.~\ref{chap:Regularization} into account, it seems erroneous to neglect these terms. Here, we hence give an example of how the 2+1 dimensional theory can inform the theory in 3+1 dimensions, and why it is worthwhile to study it. The understanding of how QFTs in the Schr\"odinger representation should be regularized that we gain in this thesis is independent of both the specific QFT and the dimensionality and can thus be generalized.

\bigskip

As most calculations in this subject are very lengthy, this thesis is equipped with an extensive set of appendices, in order to keep the chapters as clear as possible.

\chapter{The Schr\"odinger Representation of Quantum Field Theory}
\label{chap:Schrödinger}
In order to make this thesis self-contained (and to establish our conventions and notation) we will give an introduction to the Schr\"odinger representation of QFT in this chapter, following mainly the presentation in  
\cite{Hatfield:1992rz}. Another helpful introduction to the topic can be found in \cite{Jackiw:1995be}. The Schr\"odinger picture is well known from, and widely used in, ordinary quantum mechanics: The time-dependence of observables is encoded in the states, while the operators are time-independent. Canonical quantization is implemented by demanding commutation relations for conjugated operators. In the coordinate representation, position operators are represented by their eigenvalues, momentum operators by differential operators, and states by wave functions. The \SE $ $ thus becomes a differential equation whose solutions represent the spectrum of the theory. This same picture can be applied to field theory: By using field operators instead of position operators, whose eigenvalues are functions instead of numbers, the states are represented by wave functionals and the \SE $ $ becomes a functional differential equation. In the first section we will explain the formalism with the help of the simplest example of a QFT:  the case of a real scalar field without interaction. In Sec.~\ref{sec:abelian} we will deal with the complications that arise when working with gauge theories. We start with the abelian (non-interacting) example, U(1) gauge theory, which has physical relevance in describing the photon field. Non-abelian (interacting) gauge theories will then be the topic of the remainder of the thesis. Since this thesis is mostly concerned with 2+1 dimensional field theory, all of our examples will be in this framework, but the derivations in this chapter hold for general space-time dimensions $D=d+1$.

\section{Free scalar field theory}
Using the metric
\be
\eta_{\mu\nu}=\mathrm{diag}(-1,+1,+1)\,,
\ee
the Lagrangian density of a free scalar field in 2+1 dimensions is given by
\be
{\cal L}=-\half(\p_\mu\phi\p^\mu\phi+m^2\phi^2)\,,
\ee
and the conjugated momentum of the field is
\be
\Pi(x)=\frac{\p\cal L}{\p(\p_0\phi(x))}=\p_0\phi(x) \,.
\ee
In order to quantize the field we promote the field to operator status and impose the equal time commutator
\be
[\hat\phi(t_0, \vec x),\hat\Pi(t_0, \vec y)]=i\de^{(2)}(\vec x-\vec y)\,, \label{phi-commutator}
\ee
while all other commutators vanish. 

In the Schr\"odinger picture we now choose the states to be time-dependent and the operators to be time-independent. In the coordinate representation a basis of the Fock space is chosen such that the (now time-independent) field operator $\hat\phi(\vec x)$ is diagonal. Analogously, one can have a momentum representation, which we ignore here for reasons that will become apparent later (see footnote \ref{footnote1} of Chap.~\ref{chap:Comparison}).

Let $|\phi\rangle$ be an eigenstate of $\hat\phi(\vec x)$ with eigenvalue $\phi(\vec x)$:
\be
\hat\phi(\vec x)|\phi\rangle=\phi(\vec x)|\phi\rangle\,,
\ee
then the coordinate representation of the (now time-dependent) states $|\Psi\rangle$ is given by the wave functional
\be
\Psi[\phi(\vec x),t]=\langle\phi|\Psi\rangle\,,
\ee
a functional of the ordinary function $\phi(\vec x)$, and we have the completeness relation
\be
\mathds{1}=\int\D\phi\,|\phi\rangle\langle\phi| := \int \prod_{\vec x} d\phi_{\vec x} |\phi_{\vec x}\rangle\langle\phi_{\vec x}|\,.
\ee

The commutator, Eq.~(\ref{phi-commutator}), is realized by\footnote{in terms of the coordinate basis: $\langle\tilde\phi|\hat\Pi(\vec x)|\phi\rangle=-i\frac{\de}{\de\tilde\phi(\vec x)} \prod_{\vec y} \de[\tilde\phi(\vec y)-\phi(\vec y)]$.}
\be
\hat\Pi(\vec x)=-i\frac{\de}{\de\phi(\vec x)}\,,
\ee
and so the Hamilton operator\footnote{Here and in the following, we use the notation ($d=2$): $\int_x \equiv \int d^dx$, $\int_\slashed{k} \equiv \int \frac{d^dk}{(2\pi)^d}$, $\slashed{\delta}(\vec{k})\equiv (2\pi)^d\delta^{(d)}(\vec{k})$, and so on.}
\be
\hat\H=\half\int_x (\hat\Pi(\vec x)^2+|\vec\nabla\hat\phi(\vec x)|^2+m^2\hat\phi^2(\vec x))
\ee
turns into a functional differential operator
\be
\H=\half\int_x \left(-\frac{\de^2}{\de\phi^2(\vec x)}+|\vec\nabla\phi(\vec x)|^2+m^2\phi^2(\vec x)\right)\,,
\ee
and the \SE $ $ $ i\frac{\p}{\p t}|\Psi\rangle = \hat\H|\Psi\rangle$ turns into a functional differential equation. Since $\H$ is time-independent, the time-dependence of the wave functionals can be separated out
\be
\Psi[\phi,t]=e^{-iEt}\Psi[\phi],
\ee
leading to the time-independent functional \SE:
\be
\half\int_x \left(-\frac{\de^2}{\de\phi^2(\vec x)}+|\vec\nabla\phi(\vec x)|^2+m^2\phi^2(\vec x)\right)\Psi[\phi]=E\Psi[\phi]\,. \label{SE-phi}
\ee
Note that for the ground state the energy can be normalized to zero by moving it to the left-hand side of the equation and absorbing it in the $\phi^2$ term as a counterterm.

Once the functional \SE $ $ is solved, the vacuum expectation value (VEV) of a general operator $\hat{\cal O}$ can be computed by functionally integrating over all possible field configurations, weighted by the ground-state functional:
\be
\langle\hat{\cal O}\rangle= \frac{\langle\Psi_0|\hat{\cal O}|\Psi_0\rangle}{\langle\Psi_0|\Psi_0\rangle} = \frac{ \int\D\phi\D\tilde\phi\, \langle\Psi_0|\phi\rangle\langle\phi|\hat{\cal O}|\tilde\phi\rangle\langle\tilde\phi|\Psi_0\rangle}{\int\D\phi\, \langle\Psi_0|\phi\rangle\langle\phi|\Psi_0\rangle}  = \frac{ \int\D\phi\, \Psi_0^*[\phi]{\cal O}\, \Psi_0[\phi]}{ \int\D\phi\, \Psi_0^*[\phi]\Psi_0[\phi]}  \,. \label{VEV}
\ee

Since in the Schr\"odinger representation the dynamics are in the states, in the case of interacting theories, S-Matrix elements can be obtained by projecting the interacting initial and final states onto each other:
\be
S_{\al\beta}=\langle\Psi_\al|\Psi_\beta\rangle  = \int\D\phi\,\Psi_\al^*[\phi] \Psi_\beta[\phi] \,.
\ee

Let us now solve the \SE $ $ for the ground state, i.e.~Eq.~(\ref{SE-phi}) with vanishing right-hand side. Since we are talking of the ground state, we expect the wave functional to be real and to have zero nodes. 
Therefore, it can be written as the exponential of a well behaved functional $F[\phi]$, meaning that it does not diverge for finite $\phi$:
\be
\Psi[\phi]=e^{-F[\phi]}\,.
\ee
Inserting this in Eq.~(\ref{SE-phi}) yields
\be
\int_x \left(\frac{\de^2 F[\phi]}{\de\phi^2(\vec x)} -\left(\frac{\de F[\phi]}{\de\phi(\vec x)}\right)^2\right) = \int_x \phi(\vec x)\left(-\vec\nabla^2+m^2\right) \phi(\vec x)\,. \label{SEF-phi}
\ee
It is easiest to solve functional differential equations in momentum space, so we take the Fourier transforms
\be
\label{FT}
\phi(\vec x)=\int_\slashed{k} e^{i{\vec k}\cdot{\vec x}} \phi(\vec k)
\,,
\qquad
\frac{\delta}{\delta\phi(\vec{x})}
=\int_\slashed{k} e^{-i{\vec k}\cdot{\vec x}}
\frac{\delta}{\delta \phi(\vec{k})}
\,.
\ee
We will use this same convention for all fields and functional derivatives throughout this thesis.

If we take $F$ to be quadratic in $\phi$, the square of the first derivative is so, too, and it can then be matched to the right-hand side of Eq.~(\ref{SEF-phi}). The second derivative is then a pure number that can be absorbed in the ground-state energy, or, equivalently, in a counterterm. Making the Ansatz
\be
F[\phi]=\int_\slashed{k}\phi(\vec k)\phi(-\vec k)\tilde g(\vec k)
\ee
and plugging it into the Fourier transformation of Eq.~(\ref{SEF-phi}), leads to the algebraic equation
\be
4\tilde g^2(\vec k)=\vec k^2+m^2\,,
\ee
and thus to
\be
\Psi[\phi]=\exp\left[-\half\int_\slashed{k}\phi(\vec k)\sqrt{\vec k^2+m^2}\phi(-\vec k)\right]\,.
\ee
We use the positive square root because it leads to a normalizable wave functional. From here it is possible to move on to the wave functionals of excited states, but we will not consider them in this thesis (see Sec.~10.1 of \cite{Hatfield:1992rz} for details on this topic), instead we will now look at the vacuum wave functional of another, more physical theory.

\section{Abelian gauge theory}
\label{sec:abelian}

Pure (non-interacting) photon field theory, i.e.~QED without fermions is described by a U(1) gauge theory. Its Lagrangian density is 
\be
{\cal L}=-\frac{1}{4}F^{\mu\nu}F_{\mu\nu} =\half(\vec E^2-B^2) \label{U1Lagrangian}
\,,
\ee
where 
$eF_{\mu\nu}=[D_{\mu},D_{\nu}]$, $D_{\mu}=\partial_{\mu} +eA_{\mu}$. The magnetic field 
\be
B=\frac{1}{2}\epsilon_{jk}F_{jk} =\frac{1}{2}\epsilon_{jk}(\partial_jA_k-\partial_kA_j) =: {\vec \nabla}\times\vec{A}
\ee
(where ${\vec A}\times {\vec B} \equiv \epsilon_{ij}A_iB_j$ and $\vec \nabla_i \equiv \partial_i=\partial/\partial x^i$) is a scalar field in 2+1 dimensions (recall that due to our convention for the metric there is no sign difference between upper and lower spatial indices). $F_{\mu\nu}$ and hence the Lagrangian are invariant under gauge transformations of the photon:
\be
A_\mu(x)\to A_\mu^g(x)=A_\mu(x)+\frac{1}{e}\p_\mu g(x) \,. \label{AbelGauge}
\ee

It is a well known problem of canonical quantization that while the conjugate momenta of the gauge fields $A_i$ are
\be
\Pi_i=\frac{\p\cal L}{\p(\p_0A_i)}=\p_0A_i - \p_iA_0=-E_i \,,
\ee
the conjugate momentum of $A_0$ vanishes identically:
\be
\Pi_0=\frac{\p\cal L}{\p(\p_0A_0)}\equiv 0 \,,
\ee
which impedes the direct employment of the canonical commutation relations, Eq.~(\ref{phi-commutator}). This is solved by choosing a particular gauge in which to quantize the electro-magnetic field. One possibility is to use Coulomb gauge, which requires a modification of the commutators and thus leads to directional functional derivatives, which are difficult to handle. In addition this gauge breaks explicit Lorentz invariance. Another option is to work in Lorentz gauge, in which both Lorentz invariance and the canonical commutation relations can be maintained. This choice, however, has the disadvantage that the action has to be modified, leading to a more complicated Hamiltonian. Furthermore one needs to carry along unphysical (scalar and longitudinal) photons and the quantization requires a constraint. Here we opt for a compromise and choose the temporal gauge
\be
A_0=0\,.
\ee  
This has the advantage that we can keep a simple Hamiltonian and the canonical commutation relations, but, as it is only a partial gauge condition, we have to deal with longitudinal photons and a constraint to keep the residual gauge freedom under control. Also in this case we lose explicit Lorentz invariance.

In temporal gauge we work with the spatial components only, ${\vec A}=(A_1, A_2)$. We have the equal time commutators
\be
\left[\hat E_i(t_0, \vec{x}),\hat A_j(t_0, \vec{y})\right] = i\de_{ij}\de^{(2)}(\vec{x}-\vec{y})\,, \label{A-commutators}
\ee
and the Hamiltonian
\be
\hat\H=\hat\T+\hat\V={1\over2}\int_x \left(\hat{E}_i^2(\vec x)+\hat B^2(\vec x)\right)\,,
\ee
where we introduced the kinetic operator $\hat\T$ and the potential $\hat\V$. In the coordinate representation we once again choose a basis of the Fock space in which $\hat A_i(\vec x)$ is diagonal. We then represent $\hat A_i(\vec x)$ by its eigenfunction $A_i(\vec x)$, and
\be
\hat E_i(\vec x) =i\frac{\de}{\de A_i(\vec x)}
\ee
is a differential representation of the commutators Eq.~(\ref{A-commutators}). The Hamiltonian again becomes a functional differential operator and the \SE $ $ in momentum space\footnote{The conventions for the Fourier transformation are the same as in Eq.~(\ref{FT})} reads
\be
\frac{1}{2}\int_\slashed{k} \left(-\frac{\delta}{\delta \vec{A}(\vec{k})} \cdot \frac{\delta}{\delta \vec{A}(-\vec{k})} + (\vec{k}\times\vec{A}(\vec{k}))(\vec{k}\times\vec{A}(-\vec{k})) \right) \Psi[{\vec A}] = E \Psi[{\vec A}]
\,. \label{SEA}
\ee
The temporal gauge is only a partial gauge, since gauge transformations with $\p_0g(x)=0$ leave $A_0=0$ unaffected, leaving us with a residual gauge freedom of the form of Eq.~(\ref{AbelGauge}) with time-independent $g(\vec x)$. Therefore, additionally to the \SE $ $ we now have to solve the so-called Gauss law constraint, which means that the generator of the residual gauge transformations (called the Gauss law operator $I$) has to vanish on physical states:
\be
I\,|\Psi\rangle = \vec\nabla\cdot\vec E\,|\Psi\rangle =0 \Longleftrightarrow \vec\nabla\cdot\frac{\de}{\de \vec A}\Psi[{\vec A}] =0\,. \label{GLA}
\ee
This is equivalent to the request to only consider gauge invariant wave functionals.

When solving the \SE, Eq.~(\ref{SEA}), for the vacuum wave functional, the same arguments for a Gaussian functional as in the previous section apply, so we make the Ansatz
\be
\Psi[{\vec A}]=\exp\left[-G[\vec A]\right]=\exp\left[-\int_\slashed{k}{A}_i(\vec{k}){A}_j(-\vec{k})g_{ij}({\vec k})\right] \,.
\ee
The tensor structure of $g_{ij}({\vec k})$ can be fixed by the Gauss law, Eq.~(\ref{GLA}), which for a free field theory in momentum space reads
\bea
\label{GL_free}
\vec{k}\cdot \frac{\delta G[{\vec A}]}{\delta \vec{A}^a(\vec{k})} = 0
\,. 
\eea
It implies that $g_{ij}({\vec k})$ can only depend on the transverse component of the momentum. Therefore
\be
g_{ij}({\vec k})=g({\vec k}){\cal P}_{ij}({\hat k})
\,,
\ee 
where ${\cal P}_{ij}=\delta_{ij}-k_ik_j/{\vec k}^2$ is the projector to the transverse component. We can now solve Eq.~(\ref{SEA}) and determine $g({\vec k})$. As the equation is quadratic there are again two solutions, of which we take the one that leads to a normalizable wave functional, which is
\be 
\Psi[{\vec A}]=\exp\left[- \frac{1}{2}\int_\slashed{k}\frac{1}{E_k} (\vec{k}\times\vec{A}(\vec{k})) (\vec{k}\times\vec{A}(-\vec{k})) \right]
\,,\label{GA}
\ee
where $E_k \equiv |\vec k|$. One can see that, even in the free-field case, the implementation of the Gauss law is not trivial. 

\bigskip

One way from here towards interacting theories would be to include fermions. These can be introduced in terms of Grassmann-valued fields. We will, however, follow a different route, and study interacting (non-abelian) gauge theories. In contrast to the cases of non-interacting field theories considered in this chapter, we are still far from obtaining the exact vacuum wave functionals of (non-trivial) interacting theories. Hence, we have to devise sensible approximation schemes. We investigate two approaches in the following chapter. In Sec.~\ref{sec:Comp:Hatfield} we will extend the approach considered in this section to non-abelian gauge theories. We are then forced to rely on perturbation theory and solve the \SE $ $ order by order. Also, the implementation of the Gauss law becomes tedious at higher orders.

Reformulating the Hamiltonian in terms of gauge invariant field variables is an elegant way to bypass the need for the Gauss law constraint. We will study such an approach in Sec.~\ref{sec:KNY}.

\chapter{Analysis of the Yang-Mills Vacuum Wave Functional at $\O\left(e^2\right)$}
\label{chap:Comparison}

The content of this chapter was published in Ref.~\cite{Krug:2013yq}. 

\section{Introduction}

We compute the ground-state (or vacuum) wave functional of Yang-Mills theory in 2+1 dimensions in a weak coupling expansion up to ${\cal O}(e^2)$. We use two different methods: (A) One extends to ${\cal O}(e^2)$ and to a general gauge group the computation performed in Ref.~\cite{Hatfield:1984dv} to ${\cal O}(e)$ for SU(2) (An alternative procedure has also been considered in Ref.~\cite{Chan:1986bv} and worked out to ${\cal O}(e)$); (B) The other method is based on the weak coupling limit of the reformulation of the Schr\"odinger equation in terms of gauge invariant variables \cite{Karabali:1995ps,Karabali:1996je,Karabali:1996iu, Karabali:1997wk,Karabali:1998yq}, and on the approximated expression obtained in Ref.~\cite{Karabali:2009rg} for the wave functional. 
   
Method (A), outlined by Hatfield~\cite{Hatfield:1984dv} was developed for four dimensions and SU(2), but it can also be applied to the three dimensional case and a general group SU$(N_c)$ without major modifications. 
The ${\cal O}(e)$ result agrees with the expression obtained by transforming the four dimensional result of Ref.~\cite{Hatfield:1984dv} to the expected three dimensional counterpart. The solutions obtained with this method satisfy the Schr\"odinger equation by construction but not necessarily the Gauss law, though it can be explicitly shown that it does at ${\cal O}(e)$. We then compute the ${\cal O}(e^2)$ wave functional in what is a completely new result. Again, this result satisfies the Schr\"odinger equation by construction, but at this order it is not possible to explicitly check the Gauss law, due to the complexity of the resulting expressions. The resulting wave functional  is explicitly real (as expected for the ground-state functional) and we name it $\Psi_{GL}[{\vec A}]$, where $GL$ stands for the explicit use of the Gauss law.

The fact that gauge invariance can not be guaranteed in general is one important drawback of the previous method. The reason is that the Gauss law is only implemented partially for some terms in some intermediate expressions. Moreover, even this partial implementation of the Gauss law is difficult to automatize, as at each order it has to be tailored somewhat. 

A possible solution to the previous problem is the reformulation of the Schr\"odinger equation in terms of gauge invariant variables. One such formulation was originally worked out in Refs.~\cite{Karabali:1995ps,Karabali:1996je,Karabali:1996iu,Karabali:1997wk,Karabali:1998yq} 
(for some introductory notes see \cite{Schulz:2000td}) and, more recently, in Ref.~\cite{Karabali:2009rg}, where a modified approximation scheme was devised. The authors use a change of field variables, which become complex, to simplify the problem. Even though the original motivation of those works was to understand the strong coupling limit (the opposite limit we are considering in this chapter), it is not difficult to see that the approximation scheme worked out in Ref.~\cite{Karabali:2009rg} could be easily reformulated to provide with a systematic expansion of the weak coupling limit. 
We use this reformulation to compute the ground-state wave functional to ${\cal O}(e^2)$. The vacuum wave functional is a function of the gauge invariant variables $J^a$, which we then transform to the original gauge variables ${\vec A}^a$. The resulting expression is gauge invariant by construction and also satisfies the Schr\"odinger equation by construction. We name it $\Psi_{GI}[{\vec A}]\equiv \Psi_{GI}[J({\vec A})]$, where $GI$ stands for the use of the gauge invariant degree of freedom. However, the explicit expression has the very unpleasant feature of having a non-trivial imaginary term. 

We have then obtained two different expressions for the vacuum wave functional: $\Psi_{GL}[{\vec A}]$ and $\Psi_{GI}[{\vec A}]$, which actually look completely different. 
At ${\cal O}(e)$ it is possible to show, after several manipulations and using the symmetries of the integrals, that they are equal (hence, both of them are real and gauge invariant at this order). Such brute force approach happens to be unfeasible at ${\cal O}(e^2)$ due to the complexity of the expressions. We need an organizing principle for the comparison. The approach we follow is to rewrite $\Psi_{GL}[{\vec A}]$ in terms of the gauge invariant variable $J$ and a gauge dependent field $\theta$. All $\theta$ dependent terms should vanish if $\Psi_{GL}[{\vec A}]$ is going to satisfy the Gauss law, and we explicitly show that this happens. This means that both $\Psi_{GL}[{\vec A}]$ and $\Psi_{GI}[{\vec A}]$ are gauge invariant. We would then say that they should be equal, since both satisfy the Schr\"odinger equation. We actually find (after a rather lengthy computation) that they are almost but not completely equal. The difference is proportional to a bilinear real term. This is puzzling but there is a reason behind it: $\Psi_{GL}[{\vec A}]$ and $\Psi_{GI}[{\vec A}]$ satisfy ``different" Schr\"odinger equations. $\Psi_{GL}[{\vec A}]$ 
was obtained using the unregularized Schr\"odinger equation, whereas $\Psi_{GI}[{\vec A}]$ was obtained after the 
Schr\"odinger equation in terms of $J^a$ variables was regularized. In this last case, regularization produces an extra term in the Schr\"odinger equation, producing in turn an extra term in the wave functional. We will follow up on this issue in Chap.~\ref{chap:Regularization}.

  Irrespectively of the above, this comparison allows to rewrite $\Psi_{GI}[{\vec A}]$  in an explicitly real form. This is by far non-trivial, as the initial $\Psi_{GI}[J]$ was explicitly complex and dependent on complex variables. In particular there is a delicate cancellation between terms such that, after transforming this expression back to real variables, the wave function becomes real (actually in our comparison we work the other way around and transform $\Psi_{GL}[{\vec A}]$, which is real, in terms of the complex variables). This is an important test of several parts of the computation done in Ref. \cite{Karabali:2009rg}. 

We believe that the weak coupling reformulation of the approach followed in Ref.~\cite{Karabali:2009rg} can be helpful to understand the meaning of the partial resummations performed in the approximation scheme used in this reference, though we do not explore this issue here. Our ${\cal O}(e)$ or ${\cal O}(e^2)$ wave functional can also be used to test different trial functionals in the literature that claim to have the proper weak and strong coupling limit. Typically, they reproduce the leading order weak coupling expansion but not the ${\cal O}(e)$ corrections. This is certainly the case with covariantization approaches where the exponent of the wave functional is approximated by a bilinear term in the $B$ fields (see for instance \cite{Greensite:2007ij,Greensite:2011pj}). Therefore, our results can hint to how those trial functions could be improved to correctly incorporate corrections in the weak coupling limit. 

The organization of this chapter is as follows: In Sec.~\ref{sec:Comp:Hatfield} we apply method (A) and obtain $\Psi_{GL}[{\vec A}]$ up to ${\cal O}(e^2)$. Method (B) is applied in Sec.~\ref{sec:KNY} where we compute $\Psi_{GI}[{\vec A}]$ up to ${\cal O}(e^2)$. We develop a comparison principle in Sec.~\ref{sec:comp} and use it to compare the two wave functionals obtained in the two previous sections. In Sec.~\ref{sec:Summary:Comparison} we summarize the results of this chapter. In order to keep the presentation clear we relegate lengthy calculations to Apps.~\ref{F1comp} and \ref{F2comp}.
\section{Determination of $\Psi_{GL}[{\vec A}]$}
\label{sec:Comp:Hatfield}

In Yang-Mills theory the gauge fields are matrix-valued (in particular they are SU$(N_c)$ matrices) and the Lagrangian is a generalization of the U(1) Lagrangian, Eq.~(\ref{U1Lagrangian}). It reads
\be
{\cal L}=-\frac{1}{4}G^{\mu\nu,a}G_{\mu\nu}^a
\,,
\ee

where 
\be
G_{\mu\nu}^a=\partial_{\mu}A_{\nu}^a-\partial_{\nu}A_{\mu}^a+ef^{abc}A^b_{\mu}A^c_{\nu}
\,,
\ee
$eG_{\mu\nu}=[D_{\mu},D_{\nu}]$, $D_{\mu}=\partial_{\mu} +eA_{\mu}$, $A_{\mu}=-iT^aA_{\mu}^a$, 
$G^{\mu\nu}=-iT^aG^{\mu\nu}_a$, $T^a$ are the SU$(N_c)$ generators  (with $(T^a)_{bc}=-if^{abc}$ in the adjoint representation), and $[T^a,T^b]=if^{abc}T^c$. The quadratic Casimir operators are $C_A=N_c$ in the adjoint and $C_F=\frac{N_c^2-1}{2N_c}$ in the fundamental representation.

As in Chap.~\ref{chap:Schrödinger} we will work in the Hamiltonian formalism and partially fix the gauge to $A_0=0$. Under (residual) gauge transformations with a time-independent matrix-valued function $g(\vec x)$ the fields transform as
\be
A_i\to A_i^g=gA_ig^{-1}+\frac{1}{e}g\p_ig^{-1} \label{GaugeTrafo} \,.
\ee
The chromomagnetic field is
\be
B^a=\frac{1}{2}\epsilon_{jk}(\partial_jA_k-\partial_kA_j+e[A_j,A_k])^a
 = {\vec \nabla}\times\vec{A}^a +\frac{e}{2}f^{abc}\vec{A}^b\times\vec{A}^c\,,
\ee
with $B=-iT^aB^a$ (recall that ${\vec A}\times {\vec B} \equiv \epsilon_{ij}A_iB_j$ is a scalar, and that we use the metric $\eta_{\mu\nu}=\mathrm{diag}(-1,+1,+1)$). 

In Ref. \cite{Hatfield:1984dv} the wave functional was computed to ${\cal O}(e)$ at weak coupling. It is possible to generalize the method used in this reference. We do so here and compute the ground-state wave functional to ${\cal O}(e^2)$. The ground-state wave functional has to satisfy the Schr\"odinger equation\footnote{\label{footnote1} At this point it becomes clear why we work in the coordinate representation instead of the momentum ($E_i^a$) representation, even though $E_i^a$ are gauge invariant fields: The potential contains terms of $\O(A^4)$, we would thus have to solve a fourth order functional differential equation.}:
\be
\H\Psi_{GL}[{\vec A}] = \frac{1}{2}\int_x\left(-\frac{\delta}{\delta \vec{A}^a(\vec{x})} \cdot \frac{\delta}{\delta \vec{A}^a(\vec{x})} + B^a(\vec{x}) B^a(\vec{x}) \right) 
\Psi_{GL}[{\vec A}] = E \Psi_{GL}[{\vec A}]
\,, \label{eq:SE}
\ee
which is the generalization of Eq.~(\ref{SEA}), and the Gauss law constraint, which in the non-abelian case reads
\be
I^a\Psi_{GL}[{\vec A}] = (\vec{D}\cdot\vec{E})^a\Psi_{GL}[{\vec A}] =
i\left(
\vec \nabla \cdot \frac{\delta }{\delta {\vec A}_a}+ef^{abc}{\vec A}_b \cdot \frac{\delta }{\delta {\vec A}_c}
\right)\Psi_{GL}[{\vec A}]=0 \label{NonAbGL}
\,.
\ee

Again, because we are talking of the ground state, we expect the wave functional to be real and to have zero nodes (see \cite{Feynman:1981ss} for a thorough discussion). Therefore, it can be written as the exponential of a functional $F[\vec A]$ that does not diverge for finite ${\vec A}$:
\be
\Psi_{GL}[{\vec A}]=e^{-F_{GL}[{\vec A}]}=e^{-F_{GL}^{(0)}[{\vec A}]-eF_{GL}^{(1)}[{\vec A}]-e^2F_{GL}^{(2)}[{\vec A}]+{\cal O}(e^3)}\,,
\ee
and satisfies the Gauss law
\be
\left(
\vec \nabla \cdot \frac{\delta }{\delta {\vec A}_a}+ef^{abc}{\vec A}_b \cdot \frac{\delta }{\delta {\vec A}_c}
\right)F_{GL}[{\vec A}]=0
\,.
\ee

In order to compute $F$, we will do a perturbative expansion in the coupling constant $e$, assuming that it is smaller than any other scale that appears.

\subsection{Order $e^0$}
At lowest order the \SE $ $ is
\be 
\label{Sch_Free}
\int_\slashed{k} \frac{\delta F_{GL}^{(0)}[{\vec A}]}{\delta \vec{A}^a(\vec{k})} \cdot \frac{\delta F_{GL}^{(0)}[{\vec A}] }{\delta \vec{A}^a(-\vec{k})} = \int_\slashed{k} (\vec{k}\times\vec{A}^a(\vec{k})) (\vec{k}\times\vec{A}^a(-\vec{k}))
\,,
\ee
which has to be solved together with the lowest order Gauss law:
\bea
\label{GL_free}
\vec{k}\cdot \frac{\delta F_{GL}^{(0)}[{\vec A}]}{\delta \vec{A}^a(\vec{k})} = 0
\,. 
\eea
$F_{GL}^{(0)}$ can be obtained in several ways. It is equivalent to solving the Schr\"odinger equation of the free theory with the free Gauss law, in other words, $N_c^2-1$ replicas of photon field theory, Eq.~(\ref{GA}):
\be 
\label{F2}
F_{GL}^{(0)}[{\vec A}] = \frac{1}{2}\int_\slashed{k}\frac{1}{|\vec k|} (\vec{k}\times\vec{A}^a(\vec{k})) (\vec{k}\times\vec{A}^a(-\vec{k}))
\,,\label{FGL0}
\ee

\subsection{Order $e$}

At ${\cal O}(e)$ the Schr\"odinger equation splits into two equations (organized by powers of ${\vec A}$):
\be 
\label{OeF1}
\int_\slashed{k} \frac{\delta F_{GL}^{(0)}[{\vec A}]}{\delta \vec{A}^a(-\vec{k})} \cdot \frac{\delta F_{GL}^{(1)}[{\vec A}] }{\delta \vec{A}^a(\vec{k})} = \frac{i}{2}f^{abc}\int_{\slashed{k_1},\slashed{k_2},\slashed{k_3}} \slashed{\delta}\left(\sum_{i=1}^3 \vec{k}_i\right) (\vec{k}_1\times\vec{A}^a(\vec{k}_1)) (\vec{A}^b(\vec{k}_2)\times\vec{A}^c(\vec{k}_3))
\,,
\ee
\be 
\int_\slashed{k} \frac{\delta^2 F_{GL}^{(1)}[{\vec A}]}{\delta \vec{A}^a(-\vec{k})\delta \vec{A}^a(\vec{k})} = 
0
\,,
\ee
and the Gauss law constraint reads\footnote{Note that in $d=2$:
$
\vec{A}^c(-\vec{k}-\vec{p})\cdot \left(\vec{p}\times\left(\vec{p}\times\vec{A}^b(\vec{p})\right)\right)= -\left(\vec{p}\times\vec{A}^b(\vec{p})\right) (\vec{p}\times\vec{A}^c(-\vec{k}-\vec{p}))
$.\\ Other useful relations are  $({\vec k}\cdot {\vec A})({\vec k}\times {\vec B})-({\vec k}\times {\vec A})({\vec k}\cdot {\vec B})={\vec k}^2
({\vec A}\times {\vec B})$ and  $\epsilon_{ij}\epsilon_{kl}=\delta_{ik}\delta_{jl}-\delta_{il}\delta_{jk}$.}
\bea 
\vec{k}\cdot \frac{\delta F_{GL}^{(1)}[{\vec A}]}{\delta \vec{A}^a(\vec{k})} 
&=&
-i f^{abc}\int_{\slashed{p_1},\slashed{p_2}}\vec{A}^b(\vec{p}_1) \cdot \frac{\delta F_{GL}^{(0)}[{\vec A}]}{\delta \vec{A}^c(\vec{p}_2)} 
\slashed{\delta}({\vec p}_1-{\vec p}_2+{\vec k})
\nn\\
&=& 
-i f^{abc}\int_{\slashed{p}}\frac{1}{|\vec{p}|} (\vec{p}\times\vec{A}^b(-\vec{k}-\vec{p}))  \left(\vec{p}\times\vec{A}^c(\vec{p})\right) 
\,.
\eea

Using Eq.(\ref{FGL0}) the left-hand side of Eq.~(\ref{OeF1}) can be rewritten as follows:
\bea
&&\int_{\slashed{p}}\frac{1}{|\vec{p}|} (\vec{p}\times\vec{A}^a(\vec{p}))  \left(\vec{p}\times \frac{\delta F_{GL}^{(1)}[{\vec A}]}{\delta \vec{A}^a(\vec{p})}\right) \nn\\
&&\qquad =
\int_{\slashed{p}}\frac{1}{|\vec{p}|} 
\left\{
{\vec p}^2\left(\vec{A}^a({\vec p})\cdot \frac{\delta F_{GL}^{(1)}[{\vec A}]}{\delta \vec{A}^a(\vec{p})}\right) 
-
\left({\vec p}\cdot {\vec A}^a({\vec p})\right)
\left(\vec{p}\cdot \frac{\delta F_{GL}^{(1)}[{\vec A}]}{\delta \vec{A}^a(\vec{p})}\right) 
\right\}
\,,
\eea
where the second term of the right-hand side is known because of the Gauss law.

We are now in the position to obtain $F^{(1)}_{GL}$. We profit from the fact that the kernel can be taken to be completely symmetric\footnote{Any term antisymmetric in any of the two indices will vanish when multiplied by the gauge fields. This means that the kernel is not completely determined, as such terms can always be added.} under the interchange of any two fields $A_{i,a_i,x_i}$,  $A_{j,a_j,x_j}$. Therefore, the density of
$
\int_{\slashed{p}} |{\vec p}| \left(\vec{A}^a({\vec p})\cdot \frac{\delta F_{GL}^{(1)}[{\vec A}]}{\delta \vec{A}^a(\vec{p})}\right) 
$
can be related with the density of $ F_{GL}^{(1)}[{\vec A}]$. More specifically, if for a functional $F\left[\vec{A}^{a_1}({\vec k}_1),\ldots,\vec{A}^{a_n}({\vec k}_n)\right]$ of $n$ fields we have
\be
\int_{\slashed{p}}|{\vec p}| \left(\vec{A}^a({\vec p})\cdot \frac{\delta F[{\vec A}]}{\delta \vec{A}^a(\vec{p})}\right) = \int_{\slashed{k_1},\ldots,\slashed{k_n}}D\left[\vec{A}^{a_1}({\vec k}_1),\ldots,\vec{A}^{a_n}({\vec k}_n)\right]\,, \label{density1}
\ee
then
\be
 F[{\vec A}] = \int_{\slashed{k_1},\ldots,\slashed{k_n}}\frac{1}{|{\vec k}_1|+\ldots+|{\vec k}_n|}D\left[\vec{A}^{a_1}({\vec k}_1),\ldots,\vec{A}^{a_n}({\vec k}_n)\right]\,. \label{density2}
\ee
With this we finally obtain
\begin{IEEEeqnarray}{rCl}
\label{F1H}
F_{GL}^{(1)}[{\vec A}] &=&  i f^{abc} \int_{\slashed{k_1},\slashed{k_2},\slashed{k_3}}\slashed{\delta}\left(\sum_{i=1}^3 \vec{k}_i\right) \Bigg\{ \frac{1}{2(\sum_i^3|\vec{k}_i|)} (\vec{k}_1\times\vec{A}^a(\vec{k}_1)) (\vec{A}^b(\vec{k}_2)\times\vec{A}^c(\vec{k}_3))\nn\\
&& -\frac{1}{(\sum_i^3|\vec{k}_i|)|\vec{k}_1||\vec{k}_3|} (\vec{k}_1\cdot\vec{A}^a(\vec{k}_1)) (\vec{k}_3\times\vec{A}^b(\vec{k}_2)) (\vec{k}_3\times\vec{A}^c(\vec{k}_3)) \Bigg\} 
\,,
\end{IEEEeqnarray}
which is the three dimensional version of Hatfield's result (except for a different sign convention for $e$).

\subsection{Order $e^2$}

At ${\cal O}(e^2)$ the Schr\"odinger equation leads to the following equality
\be
\label{Sch_e2}
\frac{1}{2}\int_x
\left(
\frac{\delta^2 F_{GL}^{(2)}}{(\delta A_i^a)^2}-\frac{\delta F_{GL}^{(1)}}{\delta A_i^a}\frac{\delta F_{GL}^{(1)}}{\delta A_i^a}
-2\frac{\delta F_{GL}^{(0)}}{\delta A_i^a}\frac{\delta F_{GL}^{(2)}}{\delta A_i^a}
+\frac{1}{4}f^{abc}f^{ade}(\vec{A}^b\times\vec{A}^c)(\vec{A}^d\times\vec{A}^e)
\right)
=0
\,.
\ee
At this order $F^{(2)}_{GL}$ can have contributions with four, two and zero fields (there are no contributions with three or one field): 
$F^{(2)}_{GL}=F_{GL}^{(2,4)}+F_{GL}^{(2,2)}+F_{GL}^{(2,0)}$. There is no need to compute $F_{GL}^{(2,0)}$, as it just changes the normalization of the state, which we do not fix, or alternatively can be absorbed in a redefinition of the ground-state energy. Then, 
Eq.~(\ref{Sch_e2}) can be split into two terms with two and four fields respectively:
\be
\label{Sch_e2_4}
\frac{1}{2}\int_x
\left(
-\frac{\delta F_{GL}^{(1)}}{\delta A_i^a}\frac{\delta F_{GL}^{(1)}}{\delta A_i^a}
-2\frac{\delta F_{GL}^{(0)}}{\delta A_i^a}\frac{\delta F_{GL}^{(2,4)}}{\delta A_i^a}
+\frac{1}{4}f^{abc}f^{ade}(\vec{A}^b\times\vec{A}^c)(\vec{A}^d\times\vec{A}^e)
\right)
=0
\,,
\ee
and
\be
\label{Sch_e2_2}
\frac{1}{2}\int_x
\left(
\frac{\delta^2 F_{GL}^{(2,4)}}{(\delta A_i^a)^2}
-2\frac{\delta F_{GL}^{(0)}}{\delta A_i^a}\frac{\delta F_{GL}^{(2,2)}}{\delta A_i^a}
\right)
=0
\,.
\ee
$F^{(0)}_{GL}$ and $F^{(1)}_{GL}$ have already been determined (see Eqs.~(\ref{FGL0}) and (\ref{F1H})) and can be inserted into Eqs.~(\ref{Sch_e2_4}) and (\ref{Sch_e2_2}), but we still have to implement the Gauss law, which at this order reads
\be
{\vec k} \cdot \frac{\delta F_{GL}^{(2,4)}}{\delta {\vec A}^a({\vec k})}=-if^{abc}\int_{\slashed{p_1},\slashed{p_2}}{\vec A}^b({\vec p}_1)
\cdot
\frac{\delta F_{GL}^{(1)}}{\delta A_i^a({\vec p}_2)}
\slashed{\delta}({\vec p}_1-{\vec p}_2+{\vec k})
\,,
\ee 
\be
{\vec k} \cdot \frac{\delta F_{GL}^{(2,2)}}{\delta {\vec A}^a({\vec k})}=0
\,.
\ee 
One first solves Eq.~(\ref{Sch_e2_4}) and determines $F_{GL}^{(2,4)}$. Afterwards $F_{GL}^{(2,2)}$ is fixed by Eq.~(\ref{Sch_e2_2}). The procedure to obtain $F_{GL}^{(2,4)}$  is similar to the one used for $F_{GL}^{(1)}$. The dependence on $F_{GL}^{(2,4)}$ is encoded in the 2nd term of Eq.~(\ref{Sch_e2_4}), which we rewrite in the following way
\bea
&&
\int_{\slashed{p}}\frac{1}{|{\vec p}|} (\vec{p}\times\vec{A}^a(\vec{p}))  \left(\vec{p}\times \frac{\delta F_{GL}^{(2,4)}[{\vec A}]}{\delta \vec{A}^a(\vec{p})}\right) \nn\\
&&\qquad =
\int_{\slashed{p}}\frac{1}{|{\vec p}|} 
\left\{
{\vec p}^2\left(\vec{A}^a(\vec{p})\cdot \frac{\delta F_{GL}^{(2,4)}[{\vec A}]}{\delta \vec{A}^a(\vec{p})}\right) 
-
\left({\vec p}\cdot {\vec A}^a({\vec p})\right)
\left(\vec{p}\cdot \frac{\delta F_{GL}^{(2,4)}[{\vec A}]}{\delta \vec{A}^a(\vec{p})}\right) 
\right\}
\,.
\eea
Once again the second term on the right-hand side is given by the Gauss law, which allows us to isolate $F_{GL}^{(2,4)}$. As above we use the fact that the kernel can be taken to be completely symmetric under the interchange of fields $A_{i,a_i,x_i}$, which lets us relate the density of
$
\int_{\slashed{p}}|{\vec p}| \left(\vec{A}^a({\vec p})\cdot \frac{\delta F_{GL}^{(2,4)}[{\vec A}]}{\delta \vec{A}^a(\vec{p})}\right) 
$
with the density of $F_{GL}^{(2,4)}[{\vec A}]$ and we finally obtain
\begin{IEEEeqnarray}{l}
\label{F4fromF3}
F_{GL}^{(2,4)} = -\frac{1}{2}\int_{\slashed{p},\slashed{k_1},\slashed{k_2},\slashed{q_1},\slashed{q_2}}\frac{1}{\sum^2_i(|\vec{k}_i|+|\vec{q}_i|)}\left(\frac{\delta F_{GL}^{(1)}}{\delta A^a_i(\vec{p})}\right)[\vec{k}_1,\vec{k}_2]\left(\frac{\delta F_{GL}^{(1)}}{\delta A^a_i(-\vec{p})}\right)[\vec{q}_1,\vec{q}_2] \nn\\
\qquad -if^{b_1b_2c}\int_{\slashed{p},\slashed{k_1},\slashed{k_2},\slashed{q_1},\slashed{q_2}}\frac{\slashed{\delta}(\vec{q}_1+\vec{q}_2-\vec{p})}{\sum_i(|\vec{k}_i|+|\vec{q}_i|)|\vec{q}_1|}(\vec{q}_1\cdot\vec{A}^{b_1}(\vec{q}_1))\left(\vec{A}^{b_2}(\vec{q}_2)\cdot\frac{\delta F_{GL}^{(1)}}{\delta \vec{A}^c(\vec{p})}[\vec{k}_1,\vec{k}_2]\right) \nn\\
\qquad +\frac{1}{8}f^{a_1a_2c}f^{b_1b_2c}\int_{\slashed{k_1},\slashed{k_2},\slashed{q_1},\slashed{q_2}}\frac{\slashed{\delta}(\sum_i(\vec{k}_i+\vec{q}_i))}{\sum_i(|\vec{k}_i|+|\vec{q}_i|)}(\vec{A}^{a_1}(\vec{k}_1)\times\vec{A}^{a_2}(\vec{k}_2))(\vec{A}^{b_1}(\vec{q}_1)\times\vec{A}^{b_2}(\vec{q}_2))
\,,
\end{IEEEeqnarray}
which explicitly reads
\begin{IEEEeqnarray}{l}
\label{F24GL}
F_{GL}^{(2,4)}=f^{abc}f^{cde}\int_{\slashed{k_1},\slashed{k_2},\slashed{q_1},\slashed{q_2}}
\slashed{\delta}\left(\sum_i (\vec{k}_i+\vec{q}_i)\right) \frac{1}{|\vec{k}_1|+|\vec{k}_2|+|\vec{q}_1|+|\vec{q}_2|} \Bigg\{  \nn\\
\quad \frac{1}{2(|\vec{k}_1|+|\vec{k}_2|+|\vec{k}_1+\vec{k}_2|)(|\vec{q}_1|+|\vec{q}_2|+|\vec{q}_1+\vec{q}_2|)} \Bigg\{\left(\vec{A}^d(\vec{q}_1)\times\vec{A}^e(\vec{q}_2)\right)  \nn\\
\qquad \times \Bigg[ -\frac{1}{4}|\vec{k}_1+\vec{k}_2|^2 \vec{A}^a(\vec{k}_1)\times\vec{A}^b(\vec{k}_2) \nn\\
\qquad\qquad +\frac{|\vec{k}_1+\vec{k}_2|}{|\vec{k}_2|}(\vec{k}_1+\vec{k}_2)\times\vec{A}^a(\vec{k}_1) (\vec{k}_2\cdot\vec{A}^b(\vec{k}_2)) +\frac{(\vec{k}_1+\vec{k}_2)\cdot\vec{k}_2}{|\vec{k}_1||\vec{k}_2|} (\vec{k}_1\cdot\vec{A}^a(\vec{k}_1)) (\vec{k}_2\times\vec{A}^b(\vec{k}_2)) \nn\\
\qquad\qquad +(\vec{k}_1\times\vec{A}^a(\vec{k}_1)) (\vec{k}_1+\vec{k}_2)\cdot\vec{A}^b(\vec{k}_2) \Bigg] \nn\\
\qquad +  (\vec{k}_1\times\vec{A}^a(\vec{k}_1)) (\vec{q}_1\times\vec{A}^d(\vec{q}_1)) \left(\vec{A}^b(\vec{k}_2)\cdot\vec{A}^e(\vec{q}_2)\right)\nn\\
\qquad + \frac{1}{|\vec{k}_1||\vec{k}_2|}\left[2\vec{k}_2\cdot\vec{A}^e(\vec{q}_2)-\frac{\vec{q}_1\cdot\vec{k}_2}{|\vec{q}_1||\vec{k}_2|}\vec{q}_2\cdot\vec{A}^e(\vec{q}_2)\right] (\vec{k}_1\cdot\vec{A}^a(\vec{k}_1)) (\vec{k}_2\times\vec{A}^b(\vec{k}_2)) (\vec{q}_1\times\vec{A}^d(\vec{q}_1)) \nn\\
\qquad + \frac{1}{|\vec{k}_1|} (\vec{k}_1\cdot\vec{A}^a(\vec{k}_1)) (\vec{k}_1+\vec{k}_2) \times\vec{A}^b(\vec{k}_2) \Bigg[  \frac{1}{|\vec{q}_2|} (\vec{q}_1+\vec{q}_2)\times\vec{A}^d(\vec{q}_1) (\vec{q}_2 \cdot\vec{A}^e(\vec{q}_2)) \nn\\
\qquad\qquad + \frac{2}{|\vec{q}_1+\vec{q}_2|} (\vec{q}_1\times\vec{A}^d(\vec{q}_1)) (\vec{q}_1+\vec{q}_2) \cdot\vec{A}^e(\vec{q}_2) \Bigg] \nn\\
\qquad -\frac{2(\vec{q}_1+\vec{q}_2)\cdot\vec{q}_1}{|\vec{k}_1+\vec{k}_2||\vec{k}_1| |\vec{q}_1||\vec{q}_2|}  (\vec{k}_1\cdot\vec{A}^a(\vec{k}_1)) (\vec{k}_1+\vec{k}_2) \times\vec{A}^b(\vec{k}_2) (\vec{q}_1\times\vec{A}^d(\vec{q}_1)) (\vec{q}_2 \cdot\vec{A}^e(\vec{q}_2)) \nn\\
\qquad +\frac{2\vec{k}_1\times\vec{k}_2}{|\vec{k}_1||\vec{k}_2| |\vec{q}_1+\vec{q}_2| |\vec{q}_2|}  (\vec{k}_1\cdot\vec{A}^a(\vec{k}_1)) (\vec{k}_2 \times\vec{A}^b(\vec{k}_2)) (\vec{q}_2\times\vec{A}^d(\vec{q}_1)) (\vec{q}_2 \times\vec{A}^e(\vec{q}_2)) \nn\\
\qquad +\frac{2}{|\vec{q}_1+\vec{q}_2||\vec{q}_2|}  (\vec{k}_1\times\vec{A}^a(\vec{k}_1)) (\vec{k}_1+\vec{k}_2) \times\vec{A}^b(\vec{k}_2) (\vec{q}_2\times\vec{A}^d(\vec{q}_1)) (\vec{q}_2 \times\vec{A}^e(\vec{q}_2)) \nn\\
\qquad -\frac{1}{|\vec{k}_2||\vec{q}_2|}  (\vec{k}_2\times\vec{A}^a(\vec{k}_1)) (\vec{k}_2 \times\vec{A}^b(\vec{k}_2)) (\vec{q}_2\times\vec{A}^d(\vec{q}_1)) (\vec{q}_2 \times\vec{A}^e(\vec{q}_2))
\Bigg\} \nn\\
\quad +\frac{1}{8} \left(\vec{A}^a(\vec{k}_1)\times\vec{A}^b(\vec{k}_2)\right) \left(\vec{A}^d(\vec{q}_1)\times\vec{A}^e(\vec{q}_2)\right) \nn\\
\quad + \frac{1}{|\vec{k}_1|(|\vec{q}_1|+|\vec{q}_2|+|\vec{q}_1+\vec{q}_2|)}  (\vec{k}_1\cdot\vec{A}^a(\vec{k}_1)) \Bigg\{ \frac{1}{2}  (\vec{k}_1+\vec{k}_2) \times\vec{A}^b(\vec{k}_2)  \left(\vec{A}^d(\vec{q}_1)\times\vec{A}^e(\vec{q}_2)\right) \nn\\
\quad\quad 
-  (\vec{q}_1 \times\vec{A}^d(\vec{q}_1) ) \left(\vec{A}^b(\vec{k}_2)\times\vec{A}^e(\vec{q}_2)\right) \nn\\
\quad \quad-  \frac{1}{|\vec{q}_1+\vec{q}_2||\vec{q}_2|} (\vec{k}_1+\vec{k}_2) \times\vec{A}^b(\vec{k}_2) (\vec{q}_1+\vec{q}_2)\times\vec{A}^d(\vec{q}_1) (\vec{q}_2 \cdot\vec{A}^e(\vec{q}_2)) \nn\\
\quad \quad +  \frac{1}{|\vec{q}_1||\vec{q}_2|} (\vec{q}_2 \times\vec{A}^b(\vec{k}_2)) (\vec{q}_1\cdot\vec{A}^d(\vec{q}_1)) (\vec{q}_2 \times\vec{A}^e(\vec{q}_2)) \nn\\
\quad \quad  -  \frac{1}{|\vec{q}_1+\vec{q}_2||\vec{q}_2|} (\vec{k}_1+\vec{k}_2) \cdot\vec{A}^b(\vec{k}_2) (\vec{q}_2\times\vec{A}^d(\vec{q}_1)) (\vec{q}_2 \times\vec{A}^e(\vec{q}_2)) 
\Bigg\}
\Bigg\}
\,. 
\end{IEEEeqnarray}

Proceeding analogously for $F_{GL}^{(2,2)}$ we obtain
\be
\label{F22GLa}
F_{GL}^{(2,2)}=\frac{1}{2}\int_{\slashed{p},\slashed{k_1},\slashed{k_2}} \frac{1}{\sum_{i}^2|{\vec k}_i|}\slashed{\delta}(\vec{p}+\vec{k_1}+\vec{k_2})
\left(\frac{\delta^2 F_{GL}^{(2,4)}}{\delta A^a_i(\vec{p})\delta A^a_i(-\vec{p})}\right)[\vec{k}_1,\vec{k}_2]
\,.
\ee
A direct computation of this object turns out to be extremely cumbersome. We will need to wait until Sec.~\ref{sec:comp}, where we will be able to relate $F^{(2,2)}_{GL}$ with a known term of $F^{(2,2)}_{GI}$. Its explicit expression in terms of the ${\vec A}$ fields can be found in Eq.~(\ref{F2GLb}).

We have thus obtained the wave functional to ${\cal O}(e^2)$ by extending the method first devised in Ref.~\cite{Hatfield:1984dv} to the next order. The different contributions to $\Psi_{GL}[{\vec A}]$ are summarized in Eqs.~(\ref{F2}), (\ref{F1H}), (\ref{F24GL}) and (\ref{F2GLb}).
This result satisfies the Schr\"odinger equation by construction. It is also explicitly real. On the other hand, we can not claim (a priori) that the Gauss law is satisfied, as it has only been used in some intermediate computations. At ${\cal O}(e)$ it is possible to directly check that the Gauss law is satisfied. A direct check at ${\cal O}(e^2)$ turns out to be extremely difficult to obtain, due to the complexity of the expressions involved. In Sec.~\ref{sec:comp} we will devise a method to test the gauge invariance of the 
expression obtained in this section. Finally we want to stress that the computation we have performed in this section has been carried 
out without any regularization. The final result happens to be finite but formal manipulations have been performed on potentially divergent expressions. We will come back to this issue in Sec.~\ref{sec:comp} and in more detail in Sec.~\ref{sec:Reg:Hatfield}, where we find that the implementation of regularization does not change $F^{(0)}_{GL}$, $F^{(1)}_{GL}$ and $F^{(2,4)}_{GL}$ computed in this section. $F^{(2,2)}_{GL}$, however, will have to be modified.

\section{Determination of $\Psi_{GI}[\vec{A}]$}
\label{sec:KNY}

In the previous section we have been able to compute the ground-state wave functional at weak coupling at ${\cal O}(e^2)$. 
However, it is difficult to automatize the method. First, regularization issues have been completely skipped in the previous computation and, second, the Gauss law is implemented in a partial, and somewhat ad hoc, manner. This last problem could be overcome by reformulating the Schr\"odinger equation in terms of gauge invariant variables. One such formulation was originally worked out in Refs.~\cite{Karabali:1995ps,Karabali:1996je,Karabali:1996iu,Karabali:1997wk,Karabali:1998yq}\footnote{While in those references the regularization of the Schr\"odinger equation was also addressed, we will find in Chap.~\ref{chap:Regularization} that a different regularization method is needed to achieve agreement between $\Psi_{GL}$ and $\Psi_{GI}$.}.  Here we mainly follow Ref.~\cite{Karabali:2009rg}, where a modified approximation scheme was devised.  Even though the original motivation of those works was to understand the strong coupling limit, it is not difficult to see that the approximation scheme worked out in Ref.~\cite{Karabali:2009rg} could be reformulated to provide with a systematic expansion of the weak coupling limit. We do so here and compute the ground-state wave functional to ${\cal O}(e^2)$. In order to arrive at the gauge invariant fields, called $J$, a series of field variable transformations has to be used. First one defines the holomorphic and anti-holomorphic gauge fields
\be
\label{AbarA}
A:=\frac{1}{2}\left(A_1+iA_2\right) \,,\qquad \bar A:=\frac{1}{2}\left(A_1-iA_2\right)\,,
\ee
which makes it convenient to also change the space and momentum components to complex variables in the following way (note that $k$ and $z$ are defined with different signs):
\begin{IEEEeqnarray}{rClrCl}
z &=& x_1-ix_2, \qquad &\bar z &=& x_1+ix_2, \nn\\
k &=& \frac{1}{2}(k_1+ik_2),  &\bar k &=& \frac{1}{2}(k_1-ik_2), \quad {\vec k}\cdot {\vec x}=\bar k \bar z+kz, \\
\partial &=& \frac{1}{2}\left(\partial_1+i\partial_2\right), 
&\bar \partial &=& \frac{1}{2}\left(\partial_1-i\partial_2\right),  \quad \partial\bar\partial=\frac{1}{4}\vec\nabla^2\,.  \nn
\end{IEEEeqnarray}
$A$ and $\bar A$ are still gauge-dependent degrees of freedom, so we define SL($N$,$\mathbb{C}$) matrices $M$ and $M^\dagger$ by
\be
A=-\frac{1}{e}(\partial M) M^{-1} \quad \mathrm{and}\quad \bar A=\frac{1}{e}M^{\dagger-1} (\bar\partial M^{\dagger}) \label{AofM}
\,,
\ee
which transform as
\be
M\to gM \quad \mathrm{and}\quad M^\dagger\to M^\dagger g^\dagger
\ee
under gauge transformations, Eq.~(\ref{GaugeTrafo}). This allows us to define
the gauge invariant field
\be
H=M^\dagger M \label{H}
\,,
\ee
and the gauge invariant current\footnote{The anti-holomorphic current $\bar J=\frac{2}{e}H^{-1}\bar\p H$ is related to $J$ via a reality condition
\be
\p\bar J = H^{-1}(\bar\p J)H\,, \label{RealityCon}
\ee
which implies that there is only one gauge invariant degree of freedom in 2+1 dimensions.}
\be
J=\frac{2}{e}\p H H^{-1}\ =J^aT^a\,. \label{J}
\ee
We will then use the following change of variables: $(A_1,A_2) \rightarrow (A,\bar A) \rightarrow (J(A,\bar A), \bar A(A,\bar A))$, where the relation between the variables is the following:
\bea
\bar A^a&=&\bar A^a \\
\nn
J^a&=&2i\left(M^{\dagger}\right)^{ac}A^c+\frac{2}{e}\left((\partial M^{\dagger})M^{\dagger-1}\right)^a=-\frac{1}{\bar\partial}{\vec \nabla}\times\vec{A}^a +{\cal O}(e)\,.
\eea 
Inverting Eqs.~(\ref{AofM}) yields (for a more compact expression see Eq.~(5) of \cite{Karabali:1996iu})
\bE{rCl}
\label{Mexp}
M(\vec x)&=&1-e\frac{4}{\vec \nabla^2}(\bar \partial A)+e^2\frac{4}{\vec \nabla^2}\bar \partial A\frac{4}{\vec \nabla^2}\bar \partial A
+O(e^3) \\
&=&1-e\int_y G(\bar x;\bar y) A(\vec y) + e^2\int_{y,z}G(\bar x;\bar z)A(\vec z)G(\bar z;\bar y)A(\vec y)+ O(e^3) \label{MofA} \,, \\
M^{\dagger}(\vec x) &=& 1+e\frac{4}{\vec \nabla^2}( \partial \bar A)+e^2\frac{4}{\vec \nabla^2} \partial \left(\frac{4}{\vec \nabla^2} \partial\bar A\right) \bar A + O(e^3) \\
&=&1+e\int_y \bar{G}(x;y) \bar A(\vec y) + e^2\int_{y,z}\bar G(x;z) \bar G(z;y)\bar A(\vec y) \bar A(\vec z)+ O(e^3)  \label{MdaggerofAbar}
\,,
\eE
with the Green's functions:
\bea
\bar G(x;y) &\equiv& \bar G(x-y) = \frac{1}{\bar \partial_x}\delta^{(2)}(\vec x -\vec y)=-i\int \frac{d^2k}{(2\pi)^2}e^{i\vec k \cdot(\vec x -\vec y)}\frac{1}{\bar k}=\frac{1}{\pi}\frac{(\bar x-\bar y)}{(x-y)(\bar x-\bar y)+\epsilon^2}
\,,\nn\\ \label{Gbar}
\\
 G(\bar x; \bar y) &\equiv& G(\bar x-\bar y) = \frac{1}{\partial_x}\delta^{(2)}(\vec x -\vec y)=-i\int \frac{d^2k}{(2\pi)^2}e^{i\vec k \cdot(\vec x -\vec y)}\frac{1}{k}=\frac{1}{\pi}\frac{(x-y)}{(x-y)(\bar x-\bar y)+\epsilon^2}
\,.\nn\\ \label{G}
\eea
Also, a useful relation reads 
%
\be
\frac{1}{\bar \partial} \left(\left(\frac{1}{\bar \partial} \bar A^a\right) \bar A^b\right)
=
-\frac{1}{\bar \partial} \left(\bar A^a \frac{1}{\bar \partial} \bar A^b\right)
+
\left(\frac{1}{\bar \partial} \bar A^a\right)
\left(\frac{1}{\bar \partial} \bar A^b\right)
\,,
\ee
which can easily be checked in momentum space.
We also need ($T_F=1/2$)
\bea
\left(M^{\dagger}\right)^{ac}=\frac{1}{T_F}Tr[T^aM^{\dagger}T^cM^{\dagger-1}]
\,. \label{Mab}
\eea
and the analogue for $M^{ac}$ (note that $M^{-1}_{ac}=M_{ca}$). With this definition one can easily check the following identity
\be
M^{\dagger}_{cg}f^{gbh}M^{\dagger-1}_{hd} = -f^{cdf}M^{\dagger-1}_{bf} \,. \label{Mf}
\ee 

More useful relations are:
\bea
D&=&\p+eA=M\p M^{-1}\,,\qquad \bar{D}=\bar\p+e\bar A=M^{\dagger-1} \bar\partial M^{\dagger} \,,\\
\left(\frac{1}{\bar D}\right)^{de}_{yx} \label{DbarInv}
&=&
\bar G(y-x)\left[M^{\dagger-1}(\vec y)M^{\dagger}(\vec x)\right]_{de}\,, \\
\frac{\delta M^{\dagger}_{cd}(\vec y)}{\delta \bar A^b(\vec x)}&=&
e\left(\frac{1}{\bar D}\right)^{de}_{yx}(-f_{ebh})M_{hc}^{\dagger-1}(\vec x)=
e\left(\frac{1}{\bar D}\right)^{eb}_{yx} f_{edh} M_{hc}^{\dagger-1}(\vec y)
\,. \label{MdaggerOverAbar}
\eea
\bea
\frac{\delta J^c(\vec y)}{\delta A^b(\vec x)}&=&2i M^{\dagger}_{cb}(\vec y)\delta(\vec y-\vec x)
\,, \label{JOverA}
\\
\frac{\delta J^c(\vec y)}{\delta \bar A^b(\vec x)}&=&2\left[i\frac{\delta M^{\dagger}_{cd}(\vec y)}{\delta \bar A^b(\vec x)}A_d(\vec y) 
+\frac{1}{e}\frac{\delta }{\delta \bar A^b(\vec x)}\left((\partial M^{\dagger}(\vec y))M^{\dagger-1}(\vec y)\right)_c
\right]
\,.
\eea

With Eqs.~(\ref{H}), (\ref{J}), (\ref{DbarInv}) and (\ref{MdaggerOverAbar}) we find
\bE{rCl}
M^\dagger_{dh}(\vec z) D_z^{he}\left(\bar{D}^{-1}\right)^{ea}_{zx} &=& M^\dagger_{dh}(\vec z) \left(M(\vec z)\p_z M^{-1}(\vec z)\right)^{he} \left(M^{\dagger-1}(\vec z)\bar{G}(z-x) M^\dagger(\vec x)\right)^{ea} \\
&=& \left(H(\vec z)\p_z H^{-1}(\vec z)\bar{G}(z-x) M^\dagger(\vec x)\right)^{da} \\
&=& \left(\p_z \bar{G}(z-x)\right)M^\dagger_{da}(\vec x) -\frac{e}{2}\bar{G}(z-x) \left(J(\vec z)M^\dagger(\vec x)\right)^{da} \\
&=& \left(\p_z \bar{G}(z-x)\right)M^\dagger_{da}(\vec x) + \frac{ie}{2}\bar{G}(z-x) f^{edf} J^e(\vec z)M^{\dagger}_{fa}(\vec x)\,,
\eE
or, more compact and for further reference:
\bE{rCl}
\label{JOverAbar}
\frac{\delta J^d(\vec z)}{\delta \bar{A}^a(\vec x)} 
&=& -2i\left(\mathcal{D}_z\bar{G}(z-x)M^{\dagger}(\vec x)\right)^{da}\,,\\
\mathcal{D}^{mn} &=& \p_z \delta^{mn} + \frac{ie}{2} f^{mnc} J^c(\vec z)\,.
\eE

The Gauss law operator can be written in a compact form in terms of $\bar A$ and $J$:
\be
I^a(\vec x)=({\vec D}\cdot {\vec E})^a(\vec x)
=
i\int_y\left(
D_x^{ab}\frac{\delta J^c(\vec y)}{\delta A^b(\vec x)}+{\bar D}_x^{ab}\frac{\delta J^c(\vec y)}{\delta \bar A^b(\vec x)}
\right)\frac{\delta }{\delta J^c(\vec y)}
+i{\bar D}_x^{ab}\frac{\delta }{\delta \bar A^b(\vec x)}
\,.
\ee
Not surprisingly the dependence on $J$ drops out, since it is possible to prove, using Eqs.~(\ref{JOverA}) and (\ref{JOverAbar}) that 
\be
 D_x^{ab}\frac{\delta J^c(\vec y)}{\delta A_i^b(\vec x)}+{\bar D}_x^{ab}\frac{\delta J^c(\vec y)}{\delta \bar A_i^b(\vec x)}=0
\,. \label{noJinGausslaw}
\ee 
Therefore we obtain
\be
I^a(\vec x)=i{\bar D}_x^{ab}\frac{\delta }{\delta \bar A^b(\vec x)}
\ee
for the Gauss law operator.

In Refs.~\cite{Karabali:1998yq,Karabali:2009rg} the Hamiltonian was written as a pure function of $J$ up to terms proportional to the Gauss law, which vanish when applied to physical (gauge-invariant) states. If we drop those terms the Hamiltonian reads\footnote{Note that in Ref.~\cite{Karabali:1998yq} the normalization of $J$ is different.}
\bea
\label{HamiltonianNair}
\H &=&  {2\over \pi} \int _{w,z} 
 {1\over (z-w)^2} {\d \over {\d J_a (\vw)}} {\d \over {\d
J_a (\vz)}} +{1\over 2} \int_z : \bdel J^a(\vec z) \bdel J^a(\vec z) :
\\
&&
+
 i e \int_{w,z} f_{abc} {J^c(w) \over \pi (z-w)} {\d \over {\d J_a (\vw)}} {\d \over {\d
J_b (\vz)}} 
+\frac{e^2C_A}{2\pi}  \int J_a (\vz) {\d \over {\d J_a (\vz)}} 
\,,
\nonumber
\eea
which we split into $\H=\H^{(0)}+\H_I$, where $\H^{(0)}$ is the first line 
and $\H_I$ the second. It is important to note that the last term in Eq.~(\ref{HamiltonianNair}) only appears after regularization of a divergent integral. We will give a thorough derivation of this Hamiltonian in Sec.~\ref{sec:RegulatingKNY}.

We can now obtain the vacuum wave functional in powers of $e$. We write 
\be
\label{WFJ}
\Psi_{GI}[J] = \exp (-F_{GI}[J])\,,
\ee 
where (following the notation of \cite{Karabali:2009rg})
\bea
-2F_{GI}[J] &=& \int f^{(2)}_{a_1 a_2}(\vec x_1, \vec x_2)\ J^{a_1}(\vec x_1) J^{a_2}(\vec x_2) ~+~
\frac{e}{2}\ f^{(3)}_{a_1 a_2 a_3}(\vec x_1,\vec  x_2,\vec  x_3)\ J^{a_1}(\vec x_1) J^{a_2}(\vec x_2) J^{a_3}(\vec x_3)
\nn\\
&&\hskip .2in~+~
\frac{e^2}{4}\ f^{(4)}_{a_1 a_2 a_3 a_4}(\vec x_1,\vec  x_2,\vec  x_3,\vec  x_4)\ J^{a_1}(\vec x_1) J^{a_2}(\vec x_2) J^{a_3}(\vec x_3)
J^{a_4}(\vec x_4)~+~\ldots\label{rec2}
\eea
and the kernels $f^{(2)}_{a_1 a_2}(\vec x_1, \vec x_2)$, 
$f^{(3)}_{a_1 a_2 a_3}(\vec x_1, \vec x_2, \vec x_3)$, {\it etc}., have the expansions
\bea
f^{(2)}_{a_1 a_2}(\vec x_1, \vec x_2) &=& f^{(2)}_{0~a_1 a_2}(\vec x_1, \vec x_2) +
e^2 f^{(2)}_{2~a_1 a_2}(\vec x_1, \vec x_2) +\ldots\nonumber\\
f^{(3)}_{a_1 a_2 a_3}(\vec x_1, \vec x_2, \vec x_3)&=& f^{(3)}_{0~a_1 a_2 a_3}(\vec x_1, \vec x_2, \vec x_3) + e^2 f^{(3)}_{2~a_1 a_2 a_3}(\vec x_1, \vec x_2, \vec x_3)
+\ldots\label{rec3}\\
f^{(4)}_{a_1 a_2 a_3 a_4}(\vec x_1, \vec x_2, \vec x_3, \vec x_4) &=& f^{(4)}_{0~a_1 a_2 a_3 a_4}(\vec x_1, \vec x_2, \vec x_3, \vec x_4) +\ldots
\,.
\nn 
\eea

Acting with the Hamiltonian of Eq.~(\ref{HamiltonianNair}) onto this expansion of the wave functional and equating terms of equal numbers of $J$'s we obtain recursion relations for the kernels. These read
\beqar
 \label{rec4}
&& 2\frac{e^2C_A}{2\pi}~ f^{(2)}_{a_1 a_2}(\vec x_1, \vec x_2) + 4 \int_{x,y}  f^{(2)}_{a_1 a}(\vec x_1, \vec x) (\bar{\Omega}^0)_{ab}(\vec x,\vec y) f^{(2)}_{b a_2}(\vec y, \vec x_2) +V_{ab}  
\\
&&+e^2 \left[ 6 \int_{x,y} \!\! f^{(4)}_{a_1 a_2 a b }(\vec x_1, \vec x_2, \vec x,\vec y) (\bar{\Omega}^0)_{ab}(\vec x,\vec y) + 3 \int_{x,y} \!\! f^{(3)}_{a_1 a b  }(\vec x_1, \vec x,\vec y) (\bar{\Omega}^1)_{ab a_2}(\vec x,\vec y, \vec x_2)\right] 
= 0\nn
\eeqar
for the term with 2 $J$'s, while for the terms with $p \ge 3$ $J$'s the recursion relation is
\beqar
&&\frac{e^2C_A}{2\pi} p f^{(p)}_{a_1\cdots a_p} + \sum_{n=2}^{p} n(p+2-n) f^{(n)}_{a_1\cdots a_{n-1} a}(\bar{\Omega}^0)_{ab} f^{(p-n+2)}_{b a_n\cdots a_p} \nonumber \\
&&+ \sum_{n=2}^{p-1} n(p+1-n) f^{(n)}_{a_1\cdots a_{n-1}a} (\bar{\Omega}^1)_{ab a_p} f^{(p-n+1)}_{b a_n\cdots a_{p-1}} \nonumber\\
&&+ e^2 \left[ \frac{(p+1)(p+2)}{2}\ f^{(p+2)}_{a_1\cdots a_p a b}(\bar{\Omega}^0)_{ab} +\frac{p(p+1)}{2}\ f^{(p+1)}_{a_1\cdots a_{p-1} a b} (\bar{\Omega}^1)_{ab a_p}\right] =0 \label{rec5}
\,.
\eeqar
In these equations, we have used the abbreviations (following \cite{Karabali:2009rg})
\beqar
(\bar{\Omega}^0)_{ab}(\vec x,\vec y) &=& \delta_{ab} \partial_y \bar{G}(\vec x,\vec y) \nonumber 
\,,
\\
(\bar{\Omega}^1)_{abc}(\vec x,\vec y,\vec z) &=& -\frac{i}{2}\ f^{abc} \left[ \delta(\vec z-\vec y) + \delta(\vec z-\vec x)\right] \bar{G}(\vec x,\vec y) 
\,,
\nonumber \\
V_{ab}(\vec x,\vec y) &=& \delta_{ab} \int_z \bar{\partial}_z \delta(\vec z-\vec x) ~\bar{\partial}_z \delta(\vec z-\vec y) \label{rec6}
\,.
\eeqar
These equations are the same as the ones in Ref.~\cite{Karabali:2009rg} (which we have checked explicitly). Note that the splitting into $H^{(0)}$ and $H_I$ was different there, since the last term in Eq.~(\ref{HamiltonianNair}) was included in $H^{(0)}$.

If one were able to solve the set of Eqs.~(\ref{rec4}-\ref{rec5}) exactly, one would obtain the exact vacuum functional, without any truncation. Therefore, those equations are a good playground on which to try different resummation schemes (as it was done in Ref.~\cite{Karabali:2009rg}). Here we focus on the weak coupling expansion and solve those equations iteratively. There is a caveat, though: In Chap.~\ref{chap:Regularization} we find that a different regularization method should be employed, leading to the kinetic term given in Eq.~(\ref{TregKKNallorders}), and therefore to different recursion relations (see Eq.~(\ref{recreg})). In order to test the proposal of Ref.~\cite{Karabali:2009rg} we will, however, continue to work with Eqs.~(\ref{rec4}-\ref{rec5}) in this section.

At the lowest (zeroth) order in $e$, we have to solve Eq.~(\ref{rec4}) for 
$f^{(2)}_{0\ a_1 a_2}(\vec x_1, \vec x_2) $ with $e=0$. Note that this equation is quadratic in $f^{(2)}$, thus it has two solutions. We take the normalizable one, compatible with perturbation theory:

\be
\label{rec7}
f^{(2)}_{0\ a_1 a_2}(\vec x_1, \vec x_2)  = \delta_{a_1 a_2}
\frac{\bar \partial_{x_1}^2}{\sqrt{-\vec \nabla_{x_1}^2}}
\delta^{(2)}(\vec{x}_1- \vec{x}_2)
\Longleftrightarrow f^{(2)}_{0\ a_1 a_2}(\vec k) = -\frac{\bar{k}^2}{E_k} \delta_{a_1 a_2} \,,
\ee
where $E_k=|\vec k|$. 

At higher orders it is better to work in momentum space. We define
\bea
f^{(3)}_{a_1 a_2 a_3}(\vec x_1, \vec x_2, \vec x_3)&=& \int_{\slashed{k_1} \cdots \slashed{k_3}} \exp\left( i \sum_i^3 \vec k_i\cdot \vec x_i\right) \ f^{(3)}_{a_1 a_2 a_3}(\vec k_1, \vec k_2, \vec k_3)
\,, \\
f^{(4)}_{a_1 a_2 a_3 a_4}(\vec x_1, \vec x_2, \vec x_3, \vec x_4) &=& \int_{\slashed{k_1} \cdots \slashed{k_4}} \exp\left( i \sum_i^4 \vec k_i\cdot \vec x_i\right)\ f^{(4)}_{a_1 a_2 a_3 a_4}(\vec k_1, \vec k_2, \vec k_3, \vec k_4).
\label{rec8}
\eea
The recursive solution of Eqs.~(\ref{rec4}-\ref{rec5}) to order $e^2$ gives the following lowest order expressions for the cubic and quartic kernels:
\be
f^{(3)}_{0\ a_1 a_2 a_3}(\vec k_1, \vec k_2, \vec k_3) = -\frac{f^{a_1 a_2 a_3}}{24}\ (2\pi)^2 \delta (\vec k_1+\vec k_2+\vec k_3)\  g^{(3)}(\vec k_1,\vec k_2,\vec k_3)\label{rec50}
\,,
\ee
\be
f^{(4)}_{0\ a_1 a_2; b_1 b_2}(\vec k_1, \vec k_2; \vec q_1, \vec q_2) = \frac{f^{a_1 a_2 c} f^{b_1 b_2 c}}{64}\ (2\pi)^2 \delta (\vec k_1+\vec k_2+\vec q_1+\vec q_2)\ g^{(4)}(\vec k_1, \vec k_2; \vec q_1, \vec q_2) \label{rec51}
\,,
\ee
where
\beq \label{rec52}
g^{(3)}(\vec k_1,\vec k_2,\vec k_3) = \frac{16}{E_{k_1}\! + E_{k_2}\! + E_{k_3}}\left \{ \frac{\bar k_1 \bar k_2 (\bar k_1 - \bar k_2)}{E_{k_1} E_{k_2}} + {cycl.\ perm.} \right \}
\,,
\eeq
\begin{equation}\label{rec53}
\begin{array}{cl}
g^{(4)}(\vec k_1, \vec k_2; \vec q_1, \vec q_2)& =\ \vspace{.2in} \displaystyle \frac{1}{E_{k_1}\! + E_{k_2}\! + E_{q_1}\! + E_{q_2}} \\
\vspace{.2in}
&\!\!\!\!\displaystyle \left \{ g^{(3)}(\vec k_1, \vec k_2, -\vec k_1-\vec k_2)\ \frac{k_1 + k_2}{\bar k_1 +\bar k_2}\ g^{(3)}(\vec q_1, \vec q_2, -\vec q_1-\vec q_2) \right . \\
\vspace{.2in}
&\displaystyle -  \left [ \frac{(2\bar k_1 + \bar k_2)\,\bar k_1}{E_{k_1}} - \frac{(2\bar k_2 + \bar k_1)\,\bar k_2}{ E_{k_2}}\right ]\frac{4}{\bar k_1+\bar k_2}\  g^{(3)}(\vec q_1, \vec q_2, -\vec q_1-\vec q_2) \\
&\displaystyle -  \left .   g^{(3)}(\vec k_1, \vec k_2, -\vec k_1-\vec k_2)\ \frac{4}{\bar q_1+\bar q_2}\left [ \frac{(2\bar q_1 + \bar q_2)\,\bar q_1}{ E_{q_1}} - \frac{(2\bar q_2 + \bar q_1)\,\bar q_2}{ E_{q_2}}\right ] \right\} \,.
\end{array}
\end{equation}

Note that the various $f^{(n)}$ are not fixed completely, since they are multiplied by local sources. Therefore, only the completely symmetric combination is determined, any antisymmetric term would vanish when multiplied by the sources, as they form a completely symmetric function.
 
Using the expressions for
 $f^{(3)}_0$, $f^{(4)}_0$ in Eq.~(\ref{rec4}), the order $e^2$-term in $f^{(2)}$ is given by
\be
f_{2\ a_1 a_2}^{(2)}(\vec k) = \delta_{a_1 a_2}\frac{C_A}{2\pi}
{{\bar k}^2 \over E^2_k}
\left[ 
1 +
N
\right]
\label{f22Weak}
\,,
\ee
where
\be
N=\frac{E_k}{{\bar k}^2}\left(\int \frac{d^2 p}{32\pi}\ \frac{1}{\bar p}\ g^{(3)}(\vec k,\vec p,-\vec p-\vec k)\ +\ \int \frac{d^2 p}{64\pi}\ \frac{p}{\bar p}\ g^{(4)}(\vec k,\vec p;-\vec k,-\vec p)\right) \label{N}
\,.
\ee

It is possible to perform this integration, albeit numerically. 
The potentially divergent terms vanish after doing the integration over the phase of the complex number. We obtain
\be
N= 0.025999\,(8\pi) 
\,.
\ee
Note that it is real. This is not trivial to predict a priori since $g^{(3)/(4)}$ are complex functions. As we will see this is a strong check of the computation. The kernels $f^{(n)}$, $n\geq 5$, become non-trivial only beyond ${\cal O}(e^2)$.

Note that the results above are nothing but Taylor expansions of the analogous set of equations in Ref.~\cite{Karabali:2009rg} to the appropriate order. In practice this means setting $m=0$ in their computation and adding the first term in Eq.~(\ref{f22Weak}). This 
last term will play a very important role in the comparison with the results of the previous section. 

Once we have an (approximated) expression for $\Psi_{GI}[J]$ we can transform it back to the 
original ${\vec A}$ variables: $\Psi_{GI}[J({\vec A})]\equiv \Psi_{GI}[{\vec A}]$. In principle, as it is a gauge invariant quantity, it should be possible to write it in terms of the gauge covariant quantities $\vec B$ and $\vec D$. However, since we work order by order in $e$, we do not need this. On the other hand, rotational $O(2)$ symmetry is preserved explicitly.\\
We will use the following relation to transform $J$ fields into $\vec{A}$ fields (where the derivatives are in the adjoint representation: $DB=\partial B+e[A,B]$; and we have defined $J=J^aT^a$):
\be
\bar \partial^nJ=-iM^{\dagger}(\bar D^{n-1}B)M^{\dagger-1}\,, \label{JofB}
\ee
as well as Eqs.~(\ref{MofA}) and (\ref{MdaggerofAbar}).

\subsection{Order $e^0$}

In this way at ${\cal O}(e^0)$ we obtain
\begin{IEEEeqnarray}{rCl}
 -2F^{(0)}_{GI}[{\vec A}]
 &=& -\int_\slashed{k}\frac{1}{E_k} (\vec{k}\times\vec{A}^a(\vec{k})) (\vec{k}\times\vec{A}^a(-\vec{k}))
 \,, \label{F0GI3}
 \end{IEEEeqnarray}
which is the expected free-field expression.

\subsection{Order $e$}
 
At ${\cal O}(e)$ we obtain
\begin{IEEEeqnarray}{rCl}
\label{F1Nair}
F^{(1)}_{GI}[{\vec A}]
&=& i f^{abc} \int_{\slashed{k_1},\slashed{k_2},\slashed{k_3}}\slashed{\delta}\left(\sum_{i=1}^3 \vec{k}_i\right) \Bigg\{ \frac{1}{2|\vec{k}_1|} (\vec{k}_1\times\vec{A}^a(\vec{k}_1)) (\vec{A}^b(\vec{k}_2)\times\vec{A}^c(\vec{k}_3))\nn\\
&&- \frac{1}{|\vec{k}_3|\vec{k}_1^2} \left(  \frac{\vec{k}_1\times\vec{k}_2+i\vec{k}_1\cdot\vec{k}_2}{(|\vec{k}_1|+|\vec{k}_2|+|\vec{k}_3|)|\vec{k}_2|} +i  \right) (\vec{k}_1\times\vec{A}^a(\vec{k}_1))(\vec{k}_2\times\vec{A}^b(\vec{k}_2)) (\vec{k}_3\times\vec{A}^c(\vec{k}_3))\nn\\
&& + \frac{1}{|\vec{k}_3|\vec{k}_1^2} (\vec{k}_1\cdot\vec{A}^a(\vec{k}_1)) (\vec{k}_2\times\vec{A}^b(\vec{k}_2)) (\vec{k}_3\times\vec{A}^c(\vec{k}_3)) \Bigg\}
\,.
\end{IEEEeqnarray}
This term stems from a combination of $f^{(3)}$ and $f^{(2)}$ terms, as we have to remember that $J$ has an expansion in $e$ itself. 
Using the invariance of the integrals under interchange of integration variables and the fact that the delta function allows to write one momentum in terms of the other two, it is possible, however far from obvious,
to show that the imaginary term of Eq.~(\ref{F1Nair}) vanishes and that the real part is equal to Eq.~(\ref{F1H}). The details are given in App.~\ref{F1comp}.

\subsection{Order $e^2$}

At ${\cal O}(e^2)$ we obtain
\begin{IEEEeqnarray}{rCl}
\label{F22GI}
 -2F^{(2,2)}_{GI}
 &=& \frac{C_A}{2\pi}\int_\slashed{k}\frac{1}{|\vec{k}|^2} (\vec{k}\times\vec{A}^a(\vec{k})) (\vec{k}\times\vec{A}^a(-\vec{k}))[1+N] 
\,.
\end{IEEEeqnarray}
This term is associated with the $f^{(2)}_2$ term.

 For the term with four gauge fields we obtain
\begin{IEEEeqnarray}{rCl}
&&
  -2\mathrm{Re}F^{(2,4)}_{GI}= 
  \\
  &&
  \frac{1}{4}f^{a_1a_2c}f^{b_1b_2c}\int_{\slashed{k_1},\slashed{k_2},\slashed{q_1},\slashed{q_2}} \slashed{\delta}(\vec{k_1}+\vec{k_2}+\vec{q_1}+\vec{q_2})\frac{1}{|\vec{k_1}+\vec{k_2}|} \left(\vec{A}^{a_1}(\vec{k_1})\times\vec{A}^{a_2}(\vec{k_2})\right) \left(\vec{A}^{b_1}(\vec{q_1})\times\vec{A}^{b_2}(\vec{q_2})\right)
\nn  \\
&& 
+ f^{a_1a_2c}f^{b_1b_2c}\int_{\slashed{k_1},\slashed{k_2},\slashed{q_1},\slashed{q_2}} \slashed{\delta}(\vec{k_1}+\vec{k_2}+\vec{q_1}+\vec{q_2})\frac{1}{\vec{k_2}^2} \left(\frac{1}{|\vec{k}_1+\vec{k_2}|} -\frac{1}{|\vec{k}_1|}\right) \nn\\
&&\qquad\qquad (\vec{k}_1\times\vec{A}^{a_1}(\vec{k}_1)(\vec{k}_2\cdot\vec{A}^{a_2}(\vec{k}_2))  (\vec{A}^{b_1}(\vec{q}_1)\times\vec{A}^{b_2}(\vec{q}_2)) 
\nn\\&& 
-f^{a_1a_2c}f^{b_1b_2c} \int_{\slashed{k_1},\slashed{k_2},\slashed{k_3},\slashed{k_4}}\slashed{\delta}\left(\sum_{i=1}^4\vec{k}_1\right) \Bigg\{\left(\frac{1}{|\vec{k}_1+\vec{k}_2|}-\frac{1}{|\vec{k}_3|}\right)\frac{1}{\vec{k_2}^2\vec{k_4}^2}(\vec{k}_1\times\vec{A}^{a_1}(\vec{k}_1))(\vec{k}_3\times\vec{A}^{b_1}(\vec{k}_3))\nn\\ 
  &&\quad\Bigg((\vec{k_2}\cdot\vec{A}^{a_2}(\vec{k_2}))(\vec{k}_4\cdot\vec{A}^{b_2}(\vec{k}_4)) - (\vec{k_2}\times\vec{A}^{a_2}(\vec{k_2})) (\vec{k}_4\times\vec{A}^{b_2}(\vec{k}_4)) \Bigg)\nn\\
  &&\quad + \frac{1}{|\vec{k}_2|(\vec{k}_3+\vec{k}_4)^2\vec{k}_3^2}(\vec{k}_1\times\vec{A}^{a_1}(\vec{k}_1)) (\vec{k}_2\times\vec{A}^{a_2}(\vec{k}_2)) \nn\\
  &&\quad \Bigg((\vec{k}_3\cdot\vec{A}^{b_1}(\vec{k}_3)) (\vec{k}_3+\vec{k}_4)\cdot\vec{A}^{b_2}(\vec{k}_4) - (\vec{k}_3\times\vec{A}^{b_1} (\vec{k}_3)) (\vec{k}_3+\vec{k}_4)\times\vec{A}^{b_2}(\vec{k}_4)\Bigg)\Bigg\}
\nn\\
&& 
+f^{a_1a_2c}f^{b_1b_2c}\int_{\slashed{k_1},\slashed{k_2},\slashed{k_3},\slashed{k_4}} \slashed{\delta}\left(\sum_{i=1}^4\vec{k}_i\right) \frac{\vec{k}_1\times\vec{k}_2}{(|\vec{k}_1|+|\vec{k}_2|+|\vec{k}_3+\vec{k}_4|)|\vec{k}_1||\vec{k}_2|} \nn\\
&&\qquad \Big(\frac{2}{|\vec{k}_3+\vec{k}_4||\vec{k}_1|} + \frac{1}{(\vec{k}_3+\vec{k}_4)^2} \Big) \nn\\
&&\qquad (\vec{k}_1\times\vec{A}^{a_1}(\vec{k}_1)) (\vec{k}_2\times\vec{A}^{a_2}(\vec{k}_2)) (\vec{A}^{b_1}(\vec{k}_3) \times \vec{A}^{b_2}(\vec{k}_4)) 
\nn\\
&& 
-2 f^{a_1a_2c}f^{b_1b_2c} \int_{\slashed{k_1},\slashed{k_2},\slashed{k_3},\slashed{k_4}} \slashed{\delta}\left(\sum_{i=1}^4\vec{k}_i\right)  (\vec{k}_1\times\vec{A}^{a_1}(\vec{k}_1) )(\vec{k}_2\times\vec{A}^{a_2}(\vec{k}_2) )(\vec{k}_3\times\vec{A}^{b_1}(\vec{k}_3))\nn\\
&&\quad \frac{1}{(|\vec{k}_1|+|\vec{k}_2|+|\vec{k}_3+\vec{k}_4|)|\vec{k}_1|}\Bigg( \frac{1}{|\vec{k}_3+\vec{k}_4|\vec{k}_2^2} \vec{k}_2\times \vec{A}^{b_2} (\vec{k}_4) \nn\\
&&\qquad  +\frac{1}{|\vec{k}_3+\vec{k}_4|\vec{k}_2^2 \vec{k}_4^2}\Big( \vec{k}_2 \times (\vec{k}_3-\vec{k}_1) (\vec{k}_4\cdot \vec{A}^{b_2} (\vec{k}_4)) + \vec{k}_2 \cdot (\vec{k}_3-\vec{k}_1) (\vec{k}_4\times \vec{A}^{b_2} (\vec{k}_4)) \Big)\nn\\
&&\qquad  -\frac{1}{|\vec{k}_2|\vec{k}_3^2 \vec{k}_4^2}  \Big(  \vec{k}_1 \times \vec{k}_3 (\vec{k}_4\cdot \vec{A}^{b_2} (\vec{k}_4)) - \vec{k}_1 \cdot \vec{k}_3 (\vec{k}_4\times \vec{A}^{b_2} (\vec{k}_4)) \Big)\nn\\
&&\qquad  + \frac{1}{|\vec{k}_2||\vec{k}_3+\vec{k}_4|^2 \vec{k}_3^2} \Big(\vec{k}_1 \times \vec{k}_3 (\vec{k}_3 + \vec{k}_4)\cdot \vec{A}^{b_2} (\vec{k}_4) - \vec{k}_1 \cdot \vec{k}_3 (\vec{k}_3 + \vec{k}_4)\times \vec{A}^{b_2} (\vec{k}_4)\Big)\Bigg)  
\nn\\&&
- f^{a_1a_2c}f^{b_1b_2c} \int_{\slashed{k_1},\slashed{k_2},\slashed{k_3},\slashed{k_4}} \slashed{\delta}\left(\sum_{i=1}^4\vec{k}_i\right) (\vec{k}_1\times\vec{A}^{a_1}(\vec{k}_1) )(\vec{k}_2\times\vec{A}^{a_2}(\vec{k}_2) )(\vec{k}_3\times\vec{A}^{b_1}(\vec{k}_3))(\vec{k}_4\times\vec{A}^{b_2}(\vec{k}_4)) \nn\\
&&
\quad \frac{1}{(\sum_i |\vec{k}_i|) (|\vec{k}_1|+|\vec{k}_2|+|\vec{k}_3+\vec{k}_4|)(|\vec{k}_3|+|\vec{k}_4|+|\vec{k}_1+\vec{k}_2|)|\vec{k}_1||\vec{k}_3|}  \nn\\
&&
\quad\Bigg\{  \frac{\vec{k}_1^2 \vec{q}_1^2-(\vec{k}_1\times\vec{k}_2)(\vec{q}_1\times\vec{q}_2)}{|\vec{k}_2||\vec{k}_4|(\vec{k}_1+\vec{k}_2)^2} 
\nn\\&&
\quad - \frac{|\vec{k}_2|}{|\vec{k}_1+\vec{k}_2|} \left(2\left(2\frac{\vec{q}_1\cdot\vec{q}_2}{\vec{q}_2^2} +1\right) +4\frac{(\vec{k}_1\times\vec{k}_2)(\vec{q}_1\times\vec{q}_2)}{\vec{k}_2^2\vec{q}_2^2}\right) \left(1 - \frac{|\vec{k}_3|+|\vec{k}_4|+|\vec{k}_1+\vec{k}_2|}{|\vec{k}_1+\vec{k}_2|} \right) \nn\\\nn
&&
\quad +  \left(\left(2\frac{\vec{k}_1\cdot\vec{k}_2}{\vec{k}_2^2} +1\right) \left(2\frac{\vec{q}_1\cdot\vec{q}_2}{\vec{q}_2^2} +1\right) -4\frac{(\vec{k}_1\times\vec{k}_2)(\vec{q}_1\times\vec{q}_2)}{\vec{k}_2^2\vec{q}_2^2}  \right) \left(1 - 2\frac{|\vec{k}_3|+|\vec{k}_4|+|\vec{k}_1+\vec{k}_2|}{|\vec{k}_1+\vec{k}_2|} \right) \Bigg\}   
\,,
\end{IEEEeqnarray}

\begin{IEEEeqnarray}{l}
 -2i\mathrm{Im}F^{(2,4)}_{GI} =\nn\\
i f^{a_1a_2c}f^{b_1b_2c} \int_{\slashed{k_1},\slashed{k_2},\slashed{k_3},\slashed{k_4}} \slashed{\delta}\left(\sum_{i=1}^4\vec{k}_i\right) (\vec{k}_1\times\vec{A}^{a_1}(\vec{k}_1)) (\vec{k}_2\times\vec{A}^{a_2}(\vec{k}_2)) (\vec{A}^{b_1}(\vec{k}_3) \times \vec{A}^{b_2}(\vec{k}_4))\nn\\
\quad \Bigg\{\frac{1}{(|\vec{k}_1|+|\vec{k}_2|+|\vec{k}_1+\vec{k}_2|)|\vec{k}_1||\vec{k}_2|} \left(\frac{\vec{k}_1^2+2\vec{k}_1\cdot\vec{k}_2}{|\vec{k}_1+\vec{k}_2||\vec{k}_1|} - \frac{\vec{k}_1^2+\vec{k}_1\cdot\vec{k}_2}{|\vec{k}_1+\vec{k}_2|^2} \right) - \frac{1}{\vec{k_2}^2} \left(\frac{1}{|\vec{k}_1+\vec{k_2}|} -\frac{1}{|\vec{k}_1|}\right) \Bigg\} \nn\\
+ if^{a_1a_2c}f^{b_1b_2c} \int_{\slashed{k_1},\slashed{k_2},\slashed{k_3},\slashed{k_4}} \slashed{\delta}\left(\sum_{i=1}^4\vec{k}_i\right) (\vec{k}_1\times\vec{A}^{a_1}(\vec{k}_1))(\vec{k_2}\times\vec{A}^{a_2}(\vec{k_2})) (\vec{k}_3\times\vec{A}^{b_1}(\vec{k}_3))(\vec{k}_4\cdot\vec{A}^{b_2}(\vec{k}_4)) \nn\\ 
\quad \Bigg\{\frac{1}{\vec{k}_1^2\vec{k}_2^2\vec{k}_3^2\vec{k}_4^2(\vec{k}_3+\vec{k}_4)^2} \Big( 2\vec{k}_1^2\vec{k}_3^2|\vec{k}_1+\vec{k}_2| -|\vec{k}_1|\vec{k}_3^2(\vec{k}_1+\vec{k}_2)^2 -\vec{k}_1^2|\vec{k}_3|(\vec{k}_1+\vec{k}_2)^2 + \vec{k}_1^2|\vec{k}_2|(\vec{k}_1+\vec{k}_2)^2 \nn\\ 
\qquad\quad + \vec{k}_1^2|\vec{k}_2|\vec{k}_3\cdot(\vec{k}_1+\vec{k}_2) \Big)    \nn\\
\qquad +2 \frac{1}{(|\vec{k}_1|+|\vec{k}_2|+|\vec{k}_3+\vec{k}_4|)|\vec{k}_1|} \Bigg\{\frac{\vec{k}_2 \cdot (2\vec{k}_1+\vec{k}_2)}{|\vec{k}_3+\vec{k}_4|\vec{k}_2^2 \vec{k}_4^2} -\frac{\vec{k}_1 \cdot \vec{k}_3}{|\vec{k}_2|\vec{k}_3^2 \vec{k}_4^2} + \frac{\vec{k}_1 \cdot \vec{k}_3}{|\vec{k}_2||\vec{k}_3+\vec{k}_4|^2 \vec{k}_3^2}\Bigg\} \Bigg\} \nn\\
+ if^{a_1a_2c}f^{b_1b_2c} \int_{\slashed{k_1},\slashed{k_2},\slashed{k_3},\slashed{k_4}} \slashed{\delta}\left(\sum_{i=1}^4\vec{k}_i\right) (\vec{k}_1\times\vec{A}^{a_1}(\vec{k}_1))(\vec{k_2}\times\vec{A}^{a_2}(\vec{k_2})) (\vec{k}_3\times\vec{A}^{b_1}(\vec{k}_3))(\vec{k}_3\cdot\vec{A}^{b_2}(\vec{k}_4)) \nn\\ 
\quad \Bigg\{ \frac{1}{|\vec{k}_2|(\vec{k}_3+\vec{k}_4)^2\vec{k}_3^2} -2 \frac{1}{(|\vec{k}_1|+|\vec{k}_2|+|\vec{k}_3+\vec{k}_4|)|\vec{k}_1|} \frac{1}{|\vec{k}_2||\vec{k}_3+\vec{k}_4|^2 \vec{k}_3^2} (\vec{k}_1 \cdot \vec{k}_3)\Bigg\} \nn\\
+ if^{a_1a_2c}f^{b_1b_2c} \int_{\slashed{k_1},\slashed{k_2},\slashed{k_3},\slashed{k_4}} \slashed{\delta}\left(\sum_{i=1}^4\vec{k}_i\right) (\vec{k}_1\times\vec{A}^{a_1}(\vec{k}_1))(\vec{k_2}\times\vec{A}^{a_2}(\vec{k_2})) (\vec{k}_3\cdot\vec{A}^{b_1}(\vec{k}_3)) (\vec{k}_3\times\vec{A}^{b_2}(\vec{k}_4)) \nn\\   
\qquad  \frac{1}{|\vec{k}_2|(\vec{k}_3+\vec{k}_4)^2\vec{k}_3^2} \nn\\
+ 2i f^{a_1a_2c}f^{b_1b_2c} \int_{\slashed{k_1},\slashed{k_2},\slashed{k_3},\slashed{k_4}} \slashed{\delta}\left(\sum_{i=1}^4\vec{k}_i\right)  (\vec{k}_1\times\vec{A}^{a_1}(\vec{k}_1) )(\vec{k}_2\times\vec{A}^{a_2}(\vec{k}_2) )(\vec{k}_3\times\vec{A}^{b_1}(\vec{k}_3)) (\vec{k}_3 \times \vec{A}^{b_2} (\vec{k}_4))\nn\\
\quad \frac{1}{(|\vec{k}_1|+|\vec{k}_2|+|\vec{k}_3+\vec{k}_4|)|\vec{k}_1|}\frac{1}{|\vec{k}_2||\vec{k}_3+\vec{k}_4|^2 \vec{k}_3^2} (\vec{k}_1 \times \vec{k}_3) \nn\\
+2i f^{a_1a_2c}f^{b_1b_2c} \int_{\slashed{k_1},\slashed{k_2},\slashed{k_3},\slashed{k_4}} \slashed{\delta}\left(\sum_{i=1}^4\vec{k}_i\right) (\vec{k}_1\times\vec{A}^{a_1}(\vec{k}_1) )(\vec{k}_2\times\vec{A}^{a_2}(\vec{k}_2) )(\vec{k}_3\times\vec{A}^{b_1}(\vec{k}_3))(\vec{k}_4\times\vec{A}^{b_2}(\vec{k}_4)) \nn\\
\quad \Bigg[\frac{2(\vec{k}_1\times\vec{k}_2)\Bigg\{|\vec{k}_2||\vec{k}_3| (\vec{k}_3\cdot\vec{k}_4)   - |\vec{k}_1| |\vec{k}_4|^3 +(\vec{k}_1+\vec{k}_2)^2 \left(2\vec{k}_3\cdot\vec{k}_4 +\vec{k}_4^2\right)   \Bigg\} }{(\sum_i |\vec{k}_i|) (|\vec{k}_1|+|\vec{k}_2|+|\vec{k}_3+\vec{k}_4|)(|\vec{k}_3|+|\vec{k}_4|+|\vec{k}_1+\vec{k}_2|)|\vec{k}_1|\vec{k}_2^2|\vec{k}_3|\vec{k}_4^2(\vec{k}_1+\vec{k}_2)^2}  
  \nn\\
\qquad +\frac{2(\vec{k}_1\times\vec{k}_2) \left(2\vec{k}_3\cdot\vec{k}_4 +\vec{k}_4^2\right) }{(|\vec{k}_1|+|\vec{k}_2|+|\vec{k}_3+\vec{k}_4|)(|\vec{k}_3|+|\vec{k}_4|+|\vec{k}_1+\vec{k}_2|)|\vec{k}_1|\vec{k}_2^2|\vec{k}_3|\vec{k}_4^2|\vec{k}_1+\vec{k}_2|} \nn\\
\qquad +\frac{1}{(|\vec{k}_1|+|\vec{k}_2|+|\vec{k}_3+\vec{k}_4|)|\vec{k}_1|} \Bigg\{\frac{-2 \vec{k}_2 \times \vec{k}_1}{|\vec{k}_3+\vec{k}_4|\vec{k}_2^2 \vec{k}_4^2} - \frac{\vec{k}_1 \times \vec{k}_3}{|\vec{k}_2|\vec{k}_3^2 \vec{k}_4^2} + \frac{ \vec{k}_1 \times \vec{k}_3}{|\vec{k}_2||\vec{k}_3+\vec{k}_4|^2 \vec{k}_3^2}  \Bigg\}   \Bigg]
\,.
\end{IEEEeqnarray}

The last two equations can be rewritten in several ways, yet, without an organizing principle, their sizes remain more or less the same. 

The resulting expression for the ground-state wave functional seems to have a non-vanishing imaginary term. This is at odds with expectations, and with the result of Sec.~\ref{sec:Comp:Hatfield}. The real part does not look at all like the result obtained in that section either. We discuss this puzzling situation in the next section. 

\section{Comparison of the two approaches}
\label{sec:comp}

If we compare the expressions we have found for the ground-state wave functional in Secs.~\ref{sec:Comp:Hatfield} and \ref{sec:KNY} we see that they look completely different. Even more so, whereas $\Psi_{GL}$ is explicitly real, $\Psi_{GI}$ has, a priori, a non-vanishing imaginary term. Only the ${\cal O}(e^0)$ expressions are trivially equal. 
Starting at ${\cal O}(e)$ we can get agreement between both expressions after quite lengthy and non-trivial rearrangements. 

At ${\cal O}(e^2)$ a direct comparison by brute force turns out to be completely impossible. In order to compare expressions we need an organizing principle to split the comparison into pieces. The procedure we follow is to rewrite $\Psi_{GL}$ in terms of $J$ and $\bar A$ (actually we will use the variable $\theta$ defined below\footnote{The field $\theta$ could be interpreted as a kind of generator of complex $SL(N,\mathbb{C})$ gauge transformations, see Ref.~\cite{Karabali:1998yq}.}). If $\Psi_{GL}$ and $\Psi_{GI}$ are going to be equal, all terms proportional to $\bar A$ (or $\theta$) should vanish. Moreover, to a given order in $e$ the polynomial in $\bar A$ is finite so only a finite number of terms need to be compared. 

In order to perform this comparison to ${\cal O}(e^2)$ we need the following relations:
\begin{IEEEeqnarray}{rCl}
M^\dagger &\equiv& e^{e\theta}=1 + e\theta +\frac{e^2}{2}\theta^2 +{\cal O}(e^3) 
\,,\\
M^{\dagger -1} &=& 1 - e\theta +\frac{e^2}{2}\theta^2 +{\cal O}(e^3) 
\,,\\
A &=& -\frac{1}{2}M^{\dagger-1}J M^{\dagger} +\frac{1}{e}M^{\dagger-1}\partial M^{\dagger} \nn\\
&=& -\frac{1}{2}\left[J-e[\theta,J]+\frac{e^2}{4}\left[\theta,[\theta,J]\right]\right]
+\partial\theta   - \frac{e}{2}[\theta,\partial\theta] +\frac{e^2}{3!}\left[\theta,[\theta,\partial\theta]\right]+{\cal O}(e^3) 
\,,\\
\bar{A} &=& \frac{1}{e}M^{\dagger-1} \bar{\partial} M^{\dagger} 
= \bar{\partial}\theta - \frac{e}{2}[\theta,\bar \partial \theta] +\frac{e^2}{3!}\left[\theta,[\theta,\bar \partial\theta]\right]+{\cal O}(e^3) 
\,,\\
A^a(\vec{k}) &=& -\frac{i}{2}J^a(\vec{k}) +ik\theta^a(\vec{k})
+\frac{ie}{2}f^{abc}\int_\slashed{q}\theta^b(\vec{k}-\vec{q})J^c(\vec{q}) 
-\frac{ie}{2}f^{abc}\int_\slashed{q}q\,\theta^b(\vec{k}-\vec{q})\theta^c(\vec{q})
\nn\\
&& + \frac{ie^2}{4}f^{bcd}f^{dea}\int_\slashed{q}\int_\slashed{p} \theta^b(\vec{k}-\vec{q}-\vec{p}) J^c(\vec{q}) \theta^e(\vec{p})  \nn\\
&&
- \frac{ie^2}{3!}f^{bcd}f^{dea}\int_\slashed{q}\int_\slashed{p} \theta^b(\vec{k}-\vec{q}-\vec{p}) q\theta^c(\vec{q}) \theta^e(\vec{p}) 
+{\cal O}(e^3) 
\,, \label{AinJ} \\
\bar{A}^a(\vec{k}) &=& i\bar{k}\theta^a(\vec{k}) 
-\frac{ie}{2}f^{abc}\int_\slashed{q}\bar{q}\,\theta^b(\vec{k}-\vec{q})\theta^c(\vec{q}) \nn\\
&&
-\frac{ie^2}{3!}f^{bcd}f^{dea}\int_{\slashed{q},\slashed{p}}[k\bar{q}-\bar{k}q]\,\theta^b(\vec{k}-\vec{q}-\vec{p})\theta^c(\vec{q})\theta^e(\vec{p})
+{\cal O}(e^3)
\,, \label{AbarinJ}
\end{IEEEeqnarray}
where $\theta = -i \theta^a T^a$, and we define the Fourier transform of $\theta$ and $J$ following the same conventions as in Eq.~(\ref{FT}).

For the ${\cal O}(e^0)$ and the ${\cal O}(e)$ contributions of $F_{GL}$ it is possible to show that the $\theta$ terms vanish and the rest agrees with $F_{GI}$ in a direct fashion by just inserting the relations (\ref{AinJ}) and (\ref{AbarinJ}) into $F_{GL}^{(0)}$ and $F_{GL}^{(1)}$ and summing coefficients of terms with equal numbers of $J$'s and $\theta$'s. This is, of course, not surprising, since we already showed in Eq.~(\ref{F1equal}), that $F_{GI}^{(1)}[\vec A]=F_{GL}^{(1)}[\vec A]$. However, for the ${\cal O}(e^2)$ contributions, even after these simplifications, a brute force attack on the problem leads to expressions too large and complicated to directly show the equality of both expressions.\\
At this respect it is better to use some intermediate expressions of the $\Psi_{GL}$ computation that better agree with the structure of the $\Psi_{GI}$ result in terms of $J$. Particularly relevant for us is Eq.~(\ref{F4fromF3}), which relates $F^{(2,4)}_{GL}$ with $(\delta F^{(1)}_{GL})/(\delta \vec{A})$. We can write $F^{(1)}_{GL}[J,\theta]\equiv F^{(1)}_{GL}[\vec{A}(J,\theta)]$ in terms of $g^{(3)}$. Using 
\begin{IEEEeqnarray}{rCl}
\label{dFdA}
\frac{\delta}{\delta A^a_i(\vec{p})} &=& \int_q\frac{\delta A^b(\vec{q})}{\delta A^a_i(\vec{p})}\frac{\delta}{\delta A^b(\vec{q})} 
+ \int_q\frac{\delta \bar{A}^b(\vec{q})}{\delta A^a_i(\vec{p})}\frac{\delta}{\delta \bar{A}^b(\vec{q})} \nn\\
&=& \int_{q_1,q_2}\frac{\delta A^b(\vec{q}_1)}{\delta A^a_i(\vec{p})}\frac{\delta J^c(\vec{q}_2)}{\delta A^b(\vec{q}_1)}\frac{\delta}{\delta J^c(\vec{q}_2)} \nn\\
&&\quad + \int_{q_1,q_2}\frac{\delta \bar{A}^b(\vec{q}_1)}{\delta A^a_i(\vec{p})}\left(\frac{\delta J^c(\vec{q}_2)}{\delta \bar{A}^b(\vec{q}_1)}\frac{\delta}{\delta J^c(\vec{q}_2)}+\delta(\vec{q}_1-\vec{q}_2)\frac{\delta}{\delta \bar{A}^b(\vec{q}_2)}\right) \nn\\
&=&\frac{1}{2}\left(\delta_{1i}+i\delta_{2i}\right)(2i)\frac{\delta}{\delta J^a(\vec{p})}   + \frac{1}{2}\left(\delta_{1i}-i\delta_{2i}\right) \left(-2i\frac{p}{\bar{p}}\frac{\delta}{\delta J^a(\vec{p})}+\frac{\delta}{\delta \bar{A}^a(\vec{p})}\right) \nn\\
&& +{\cal O}(e)  \label{funderiv}
\,,
\end{IEEEeqnarray}
we have
\begin{IEEEeqnarray}{rCl}
&&
\frac{\delta F^{(1)}_{GL}}{\delta A^a_i(\vec{p})}= -if^{aa_1a_2} \int_{\slashed{k_1},\slashed{k_2}}\slashed{\delta}\left(\vec{k}_1+\vec{k}_2+\vec{p}\right) \nn\\
&&
\quad
\Bigg\{\Bigg\{ (\delta_{1i}+i\delta_{2i})\frac{g^{(3)}(\vec{k}_1,\vec{k}_2,\vec{p})}{32} + (\delta_{1i}-i\delta_{2i})\left(-\frac{p}{\bar{p}}\frac{g^{(3)}(\vec{k}_1,\vec{k}_2,\vec{p})}{32}+\frac{\bar{k}_2^2}{2\bar{p}|\vec{k}_2|}\right)\Bigg\}J^{a_1}(\vec{k}_1)J^{a_2}(\vec{k}_2) \nn\\
&&
\quad
+\Bigg\{(\delta_{1i}+i\delta_{2i})\left( \frac{\bar{p}^2}{|\vec{p}|}-\frac{\bar{k}_1^2}{|\vec{k}_1|}\right)  \nn\\
&&\qquad\quad  - (\delta_{1i}-i\delta_{2i}) \left(\frac{1}{4}|\vec{p}|+\frac{\bar{k}_1}{|\vec{k}_1|}\left(-\frac{p}{\bar{p}}(\bar{k}_1+\bar{k}_2)+k_2\right) \right)\Bigg\}J^{a_1}(\vec{k}_1)\theta^{a_2}(\vec{k}_2) \nn\\
&&
\quad
+\Bigg\{(\delta_{1i}+i\delta_{2i})2\frac{\bar{p}k_1\bar{k}_2}{|\vec{p}|} -(\delta_{1i}-i\delta_{2i}) 2\frac{p k_1\bar{k}_2}{|\vec{p}|}\Bigg\} \theta^{a_1}(\vec{k}_1)\theta^{a_2}(\vec{k}_2) \Bigg\}+  O(e)\,.
\end{IEEEeqnarray}
With this we can write $F^{(2,4)}_{GL}[J,\theta]$ as a second order polynomial in $g^{(3)}$. This gives us the guiding principle to try to reconstruct $g^{(4)}$, which is also a second order polynomial in $g^{(3)}$. This term should be proportional to $J^4$ and we find that indeed it is.\\
In Eq.~(\ref{F4fromF3}) one can see that all terms in $F^{(2,4)}_{GL}[J,\theta]$ have a prefactor of $\frac{1}{|\vec{k}_1|+|\vec{k}_2|+|\vec{q}_1|+|\vec{q}_2|}$. As we need the gauge ($\theta$) dependent terms to cancel with the corresponding terms from $F_{GL}^{(0)}$ and $F_{GL}^{(1)}$, that don't have this prefactor, we find a second guiding principle, which is to rewrite the $\theta$ dependent terms of $F^{(2,4)}_{GL}[J,\theta]$ in such a way, that this prefactor drops out and then try to find a form similar to the gauge dependent contributions of $F_{GL}^{(0)}$ and $F_{GL}^{(1)}$. To do so we extensively use the Jacobi identity and the invariance of the integrals under interchange of integration variables, as well as the delta function. We also use the fact that the integration kernels can be taken to be completely symmetric under the interchange of the variables of two equal fields (for instance $J^{a_1}(\vec{k}_1)J^{a_2}(\vec{k}_2)$). Still the computation is very tedious and highly non-trivial, therefore we give the details in App.~\ref{F2comp}. In the end we obtain
\begin{IEEEeqnarray}{rCl}
\label{F0GLGI}
F^{(0)}_{GL} &=& \frac{1}{2}\int_\slashed{k}\frac{\bar{k}^2}{|\vec{k}|} J^a(\vec{k})J^a(-\vec{k}) 
+e\int_{\slashed{k_1},\slashed{k_2},\slashed{k_3}}\slashed{\delta}\left(\sum_{i=1}^3 \vec{k}_i\right) \frac{\bar k_3^2}{|\vec k_3|} f^{abc} J^a(\vec k_1 )\theta^b(\vec k_2 )J^c(\vec k_3 )\nn\\
&& -ef^{abc}\int_{\slashed{k_1},\slashed{k_2},\slashed{k_3}}\slashed{\delta}\left(\sum_{i=1}^3 \vec{k}_i\right) \frac{\bar{k}_1 (k_1\bar{k}_3-\bar{k}_1k_3)}{|\vec{k}_1|} J^a(\vec{k}_1)\theta^b(\vec{k}_2)\theta^c(\vec{k}_3) \nn\\
&& + \frac{e^2}{2}f^{a_1a_2c}f^{b_1b_2e}\int_{\slashed{k_1},\slashed{k_2},\slashed{q_1},\slashed{q_2}} 
\slashed{\delta}\left(\sum^2_i(\vec{k}_i+\vec{q}_i)\right)  J^{a_1}(\vec{k}_1) \theta^{a_2}(\vec{k}_2)J^{b_1}(\vec{q}_1)\theta^{b_2}(\vec{q}_2)   \nn\\
&&\qquad\qquad \times \left( \frac{(\bar{k}_1+\bar{k}_2)^2}{|\vec{k}_1+\vec{k}_2|} - \frac{\bar{k}_1^2}{|\vec{k}_1|} \right)
\nn\\
&&+ e^2f^{a_1a_2c}f^{b_1b_2c}\int_{\slashed{k}_1,\slashed{k}_2,\slashed{q}_1,\slashed{q}_2}\slashed{\delta}\left(\sum_i(\vec{k}_i+\vec{q}_i)\right)  J^{a_1}(\vec{k}_1)\theta^{a_2}(\vec{k}_2) \theta^{b_1}(\vec{q}_1) \theta^{b_2}(\vec{q}_2)    \nn\\
&&\qquad\qquad \times \Bigg( \frac{1}{3} \frac{1}{|\vec{k}_1|} \bar{k}_1(k_1\bar{q}_2-\bar{k}_1q_2)  +\frac{1}{|\vec{q}_1+\vec{q}_2|}(\bar{q}_1+\bar{q}_2) (q_2\bar{q}_1-\bar{q}_2q_1)\, \Bigg)
 \nn\\
&& -2e^2f^{a_1a_2c}f^{b_1b_2c}\int_{\slashed{k_1},\slashed{k_2},\slashed{q_1},\slashed{q_2}} 
\slashed{\delta}\left(\sum_i(\vec{k}_i+\vec{q}_i)\right)  \theta^{a_1}(\vec{k}_1)\theta^{a_2}(\vec{k}_2)\theta^{b_1}(\vec{q}_1)  \theta^{b_2}(\vec{q}_2)    \frac{\bar k_2 k_1\bar q_2q_1}{|\vec{k}_1+\vec{k}_2|} \nn\\
&&+O(e^3), 
\end{IEEEeqnarray}

\begin{IEEEeqnarray}{rCl}
\label{F1GLGI}
F^{(1)}_{GL} &=&  -f^{abc} \int_{\slashed{k_1},\slashed{k_2},\slashed{k_3}}\slashed{\delta}\left(\sum_{i=1}^3 \vec{k}_i\right) 
\frac{g^{(3)}(\vec{k}_1,\vec{k}_2,\vec{k}_3)}{96} 
 J^a(\vec{k}_1)J^b(\vec{k}_2)J^c(\vec{k}_3) 
 \nn
\\
&&
 -f^{abc} \int_{\slashed{k_1},\slashed{k_2},\slashed{k_3}}\slashed{\delta}\left(\sum_{i=1}^3 \vec{k}_i\right) 
\frac{\bar{k}_3^2}{|\vec{k}_3|} J^a(\vec{k}_1) \theta^b(\vec{k}_2)   J^c(\vec{k}_3) 
 \nn\\
&& -2f^{abc}\int_{\slashed{k_1},\slashed{k_2},\slashed{k_3}}\slashed{\delta}\left(\sum_{i=1}^3 \vec{k}_i\right)
\frac{\bar{k}_1k_2\bar{k}_3}{|\vec{k}_1|}J^a(\vec{k}_1)\theta^b(\vec{k}_2)\theta^c(\vec{k}_3) 
\nn\\
&& 
-ef^{a_1a_2c}f^{b_1b_2c}\int_{\slashed{k_1},\slashed{k_2},\slashed{q_1},\slashed{q_2}}\slashed{\delta}\left(\sum(\vec{k}_i+\vec{q}_i)\right) J^{a_1}(\vec{k}_1)J^{a_2}(\vec{k}_2) J^{b_1}(\vec{q}_1) \theta^{b_2}(\vec{q}_2) \nn\\
&&\qquad\qquad \times
\frac{g^{(3)}(\vec{k}_1,\vec{k}_2,-\vec{k}_1-\vec{k}_2)}{32}  \nn\\
&&
+ef^{a_1a_2c}f^{b_1b_2c} \int_{\slashed{k_1},\slashed{k_2},\slashed{q_1},\slashed{q_2}}  \slashed{\delta}\left(\sum_{i=1}^2 (\vec{k}_i+\vec
{q}_i)\right)   J^{a_1}(\vec{k_1})  J^{a_2}(\vec{k}_2) \theta^{b_1}(\vec{q}_1) \theta^{b_2}(\vec{q}_2)\nn\\
&&
\qquad\qquad  \times\Bigg(\frac{\bar{q}_2q_1}{\bar{q}_1+\bar{q}_2} \frac{g^{(3)}(\vec{k}_1,\vec{k}_2, -\vec{k}_1-\vec{k}_2)}{16} - \frac{\bar{q}_2}{(\bar{q}_1+\bar{q}_2)} \frac{\bar{k}_2^2}{2|\vec{k}_2|}\Bigg)  \nn\\
&&
 -ef^{a_1a_2c}f^{b_1b_2c} \int_{\slashed{k_1},\slashed{k_2},\slashed{q_1},\slashed{q_2}} 
\slashed{\delta}\left(\sum_{i=1}^2 (\vec{k}_i+\vec{q}_i)\right) J^{a_1}(\vec{k}_1) \theta^{a_2}(\vec{k}_2) J^{b_1}(\vec{q}_1)\theta^{b_2}(\vec{q}_2) \nn\\
&&\qquad\qquad \times
\left( \frac{(\bar{k}_1+\bar{k}_2)^2}{|\vec{k}_1+\vec{k}_2|} -\frac{ \bar{k}_1^2}{|\vec{k}_1|} \right) 
 \nn\\
&& 
-ef^{a_1a_2c}f^{b_1b_2c}\int_{\slashed{k_1},\slashed{k_2},\slashed{k_3},\slashed{q}}\slashed{\delta}\left(\sum_{i=1}(\vec{k}_i+\vec{q}_i)\right)    J^{a_1}(\vec{k}_1)\theta^{a_2}(\vec{k}_2) \theta^{b_1}(\vec{q}_1)\theta^{b_2}(\vec{q}_2)\nn\\
&&
\qquad\qquad \times\Bigg( \frac{\bar k_1}{|\vec{k}_1|} (k_1\bar{q}_2-\bar{k}_1q_2) +4\frac{(\bar k_1 +\bar k_2)}{|\vec{k}_1+\vec{k}_2|}q_1\bar q_2\Bigg) \nn \\
&&
+4ef^{a_1a_2c}f^{b_1b_2c}\int_{\slashed{k_1},\slashed{k_2},\slashed{q_1},\slashed{q_2}} \slashed{\delta}\left(\sum^2_i(\vec{k}_i+\vec{q}_i)\right)  \theta^{a_1}(\vec{k}_1)\theta^{a_2}(\vec{k}_2)\theta^{b_1}(\vec{q}_1)  \theta^{b_2}(\vec{q}_2)  \frac{k_1\bar k_2 q_1\bar q_2}{|\vec{q}_1+\vec{q}_2|} \nn\\ 
&&+O(e^2),
\end{IEEEeqnarray}

\begin{IEEEeqnarray}{rCl}
\label{F24GLGI}
F^{(2,4)}_{GL} &=& -\frac{f^{a_1a_2c}f^{b_1b_2c}}{512} \int_{\slashed{k_1},\slashed{k_2},\slashed{q_1},\slashed{q_2}}
\slashed{\delta}\left(\sum^2_i(\vec{k}_i+\vec{q}_i)\right) g^{(4)}(\vec k_1,\vec k_2;\vec q_1,\vec q_2) \nn\\
&&
\qquad\qquad \times J^{a_1}(\vec{k}_1)J^{a_2}(\vec{k}_2)J^{b_1}(\vec{q}_1)J^{b_2}(\vec{q}_2) \nn\\
&& +\frac{f^{a_1a_2c}f^{b_1b_2c}}{32} \int_{\slashed{k_1},\slashed{k_2},\slashed{q_1},\slashed{q_2}} 
\slashed{\delta}\left(\sum^2_i(\vec{k}_i+\vec{q}_i)\right) g^{(3)}(\vec{k}_1,\vec{k}_2,-\vec{k}_1-\vec{k}_2) \nn\\
&&
\qquad\qquad \times J^{a_1}(\vec{k}_1)J^{a_2}(\vec{k}_2)J^{b_1}(\vec{q}_1)\theta^{b_2}(\vec{q}_2)\nn\\
&& + \frac{1}{2} f^{a_1a_2c}f^{b_1b_2c} \int_{\slashed{k_1},\slashed{k_2},\slashed{q_1},\slashed{q_2}} 
\slashed{\delta}\left(\sum^2_i(\vec{k}_i+\vec{q}_i)\right)  J^{a_1}(\vec{k}_1)\theta^{a_2}(\vec{k}_2)J^{b_1}(\vec{q}_1)\theta^{b_2}(\vec{q}_2) \nn\\
&& \qquad\qquad \times  \left( \frac{(\bar{k}_1+\bar{k}_2)^2}{|\vec{k}_1+\vec{k}_2|} -  \frac{\bar{k}_1^2}{|\vec{k}_1|} \right) \nn\\
&& - f^{a_1a_2c}f^{b_1b_2c} \int_{\slashed{k_1},\slashed{k_2},\slashed{q_1},\slashed{q_2}} \slashed{\delta}
\left(\sum^2_i(\vec{k}_i+\vec{q}_i)\right)    J^{a_1}(\vec{k}_1)J^{a_2}(\vec{k}_2)\theta^{b_1}(\vec{q}_1)\theta^{b_2}(\vec{q}_2) \nn\\
&& \qquad\qquad \times \left(\frac{q_1\bar{q}_2}{\bar{q}_1+\bar{q}_2} \frac{g^{(3)}(\vec{k}_1,\vec{k}_2, -\vec{k}_1-\vec{k}_2)}{16} -   \frac{\bar{q}_2}{\bar{q}_1+\bar{q}_2} \frac{\bar{k}_2^2}{2|\vec{k}_2|}\right) \nn\\
&& + 2 f^{a_1a_2c}f^{b_1b_2c}\int_{\slashed{p},\slashed{k_1},\slashed{k_2},\slashed{q_1},\slashed{q_2}} 
\slashed{\delta}\left(\sum_i(\vec{k}_i+\vec{q}_i)\right) J^{a_1}(\vec{k}_1)\theta^{a_2}(\vec{k}_2) \theta^{b_1}(\vec{q}_1)\theta^{b_2}(\vec{q}_2)  \nn\\
&& \qquad\qquad \times q_1\bar{q}_2 \left(\frac{\bar{k}_1+\bar{k}_2}{|\vec{k}_1+\vec{k}_2|} - \frac{\bar{k}_1}{|\vec{k}_1|} \right) 
\nn\\
&&  -2 f^{a_1a_2c}f^{b_1b_2c}\int_{\slashed{k_1},\slashed{k_2},\slashed{q_1},\slashed{q_2}} \slashed{\delta}\left(\sum_i(\vec{k}_i+\vec{q}_i)\right) \frac{ k_1\bar{k}_2q_1\bar{q}_2}{|\vec{k}_1+\vec{k}_2|} \theta^{a_1}(\vec{k}_1)\theta^{a_2}(\vec{k}_2)\theta^{b_1}(\vec{q}_1)  \theta^{b_2}(\vec{q}_2) \nn\\
&&+O(e)\,.
\end{IEEEeqnarray}

We now move to $F^{(2,2)}_{GL}$, which is associated to a one-loop computation. We have already mentioned in Sec.~\ref{sec:Comp:Hatfield} that its direct determination in terms of ${\vec A}$ fields is not feasible. Again, we follow the strategy of rewriting $F^{(2,2)}_{GL}$ in terms of $J$ and $\theta$. For this we use Eq.~(\ref{F24GLGI}), which we plug into Eq.~(\ref{F22GLa}) after having rewritten the functional derivatives in terms of $J$ and $\bar A$ using Eq.~(\ref{funderiv}). The calculation simplifies a lot and we find
\begin{IEEEeqnarray}{rCl}
F^{(2,2)}_{GL}  &=& - \frac{C_A}{32}\int_{\slashed{p},\slashed{k}}\frac{1}{|\vec{k}|}\left( \frac{1}{\bar{p}}  g^{(3)}(\vec k,\vec p,-\vec k-\vec p) + \frac{1}{2} \frac{p}{\bar{p}}  g^{(4)}(\vec p,\vec k;-\vec p,-\vec k) \right) J^{a}(\vec{k})J^{a}(-\vec{k}) +{\cal O}(e)\,. \label{FGL22}
\nn\\
\end{IEEEeqnarray}
This result allows us to write $F^{(2,2)}_{GL}$ in terms of the gauge fields. It reads
\be
\label{F2GLb}
F^{(2,2)}_{GL}  =-N\frac{C_A}{4\pi}\int_\slashed{k}\frac{1}{|\vec{k}|^2} (\vec{k}\times\vec{A}^a(\vec{k})) (\vec{k}\times\vec{A}^a(-\vec{k})) 
\,,
\ee
where $N$ has been defined in Eq.~(\ref{N}).

We can now combine all the different contributions (in an, again, not completely trivial computation). We obtain the following equalities
\be
F_{GL}[\vec A(J,\theta)]=F_{GI}[J]+\frac{C_Ae^2}{4\pi}\int_\slashed{k}\frac{\bar{k}^2}{|\vec{k}|^2} J^a(\vec{k})J^a(-\vec{k}) +{\cal O}(e^3)
\,,
\ee
or in terms of the gauge fields
\be
F_{GI}[J({\vec A})]=F_{GL}[\vec A]-\frac{C_Ae^2}{4\pi}\int_\slashed{k}\frac{1}{|\vec{k}|^2} (\vec{k}\times\vec{A}^a(\vec{k})) (\vec{k}\times\vec{A}^a(-\vec{k})) +{\cal O}(e^3)
\,.
\ee
The first equality implies that $F_{GL}[\vec A]$ is gauge invariant to ${\cal O}(e^2)$, the second that $F_{GI}[J]$ is real to ${\cal O}(e^2)$.
We stress that $F^{(0)}_{GL}$, $F^{(1)}_{GL}$, and $F^{(2,4)}_{GL}$ are real, which is not evident at all as written in Eqs.~(\ref{F0GLGI}), (\ref{F1GLGI}) and (\ref{F24GLGI}).

Overall we get complete agreement except for one bilinear real extra term in $F_{GI}$. Its origin can be traced back to the appearance of the last term of the Schr\"odinger equation in Eq.~(\ref{HamiltonianNair}). In turn this term appears from an anomaly-like computation only after the kinetic operator has been regularized. Note that $F_{GL}$ was obtained without regularizing the theory, working with formal expressions. The existence of very lengthy and complicated expressions in the intermediate steps impedes in practice the identification of the divergences. We expect these divergences to particularly affect $F_{GL}^{(2,2)}$, since we have functional derivatives acting on the wave functional density (see Eq.~(\ref{F22GLa})) that effectively produce contractions of fields and internal integrals over momenta. Therefore, even if the final result was finite, one could have missed contributions of this kind. For the other terms of $F$ we have got a double check, which gives us strong confidence in our result. 

\section{Conclusions}
\label{sec:Summary:Comparison}

We have computed the Yang-Mills vacuum wave functional in three dimensions at weak coupling with ${\cal O}(e^2)$ precision. We have used two different methods to solve the Schr\"odinger functional equation: 
(A) One of them generalizes to ${\cal O}(e^2)$ the method followed by Hatfield at ${\cal O}(e)$~\cite{Hatfield:1984dv}. We have named the 
result $\Psi_{GL}[{\vec A}]$. 
(B) The other uses the weak coupling version of the gauge invariant formulation of the Schr\"odinger equation and the ground-state wave functional followed by Karabali, Nair, and Yelnikov \cite{Karabali:2009rg}. We have named the result $\Psi_{GI}[J]$. Each method has its own strengths and weaknesses, and they are to some extent complementary.

The computations performed with method (A) are relatively simple and the results are explicitly real. The generalization to four 
dimensions of the ${\cal O}(e^2)$ computation does not present major conceptual problems. Note that this is the order at which we expect to start to see the running of the coupling constant in $D=4$. On the other hand, such a computation has two major drawbacks. First, the implementation of the Gauss law is not done in a systematic way, only partially in some intermediate steps. Therefore, we cannot guarantee a priori that the final result is gauge invariant. Since the results grow rapidly in size and complexity, a direct check turns out to be unfeasible. Actually we were only able to check the Gauss law with the help of method (B). The main drawback, however, is that the computation has been performed with an unregularized kinetic operator. Whereas all computations can formally be carried out obtaining a finite result, some terms may be missed in this way. 

The computations with method (B) are somewhat more involved. Rather lengthy expressions appear when we rewrite the wave functional in terms of the gauge fields ${\vec A}$, which, moreover, look complex. Trying to prove by brute force that the result is real turns out to be impossible. Actually, we only manage to prove it after a careful comparison with the result of method (A). 
Moreover, a possible generalization to four dimensions does not look trivial. 
On the other hand, method (B) is particularly appealing, as it directly works with gauge-invariant degrees of freedom. Therefore, the 
Gauss law is automatically satisfied and it is not necessary to explicitly impose this constraint. Note also that the set of Eqs.~(\ref{rec4}) and (\ref{rec5}) can be solved recursively. Therefore, it could be possible to automatize the computation and obtain the wave functionals at higher orders with a combination of algebraic/numeric programing. Finally, and most importantly, the kinetic operator had been regularized. This produced non-trivial 
contributions. 

We have compared both results. It is impossible to show that they are equal in a direct way. The strategy we follow helps a lot, yet it continues to be extremely complicated to prove the equality of the two expressions.  As we have already mentioned, this comparison has allowed us on the one hand to prove that $\Psi_{GL}$ is indeed gauge invariant and on the other hand that $\Psi_{GI}$ is real. Most interestingly, the agreement between both results is almost complete except for one extra term that appears with method (B). This term shows up from an anomaly-like computation once the theory is regularized. Such a contribution does not show up in method (A). As we will show in the next chapter, this is due to the fact that no regularization was used in this computation. This result is potentially very interesting because it is precisely this term that produces the mass gap 
and a linearly rising potential in the strong coupling limit in Ref.~\cite{Karabali:1998yq}. Therefore, it is important to understand how such a term can be generated in a regularized version of the Schr\"odinger formalism in terms of the gauge fields, as this contribution has not been checked with an independent method so far. However, since regularization in the Schr\"odinger formalism with gauge variables is, to a large extent, uncharted territory, this requires a dedicated study. We address this issue in the following chapter and also revisit the regularization with method (B), with the aim of resolving the discrepancy between the two wave functionals. In that analysis we find new contributions for both methods which bring them into agreement. 

In this context, it may be worth mentioning that supersymmetric extensions of Yang-Mills theory with ${\cal N} \geq 2$ do not have this term~\cite{Agarwal:2012bn}. This is not completely unexpected, as the introduction of supersymmetry improves the ultraviolet behavior of the theory. This may lead to convergent integrals and the disappearance of the extra term. 

Finally, we expect that the inclusion of matter fields in the theory will not produce major changes to the general procedure.

\chapter{Regularization of the Yang-Mills Vacuum Wave Functional at $\O\left(e^2\right)$}
\label{chap:Regularization}

The content of this chapter was published in Ref.~\cite{Krugz}.

\section{Introduction}
 In the previous chapter we computed the Yang-Mills vacuum wave functional in three dimensions at weak coupling to ${\cal O}(e^2)$, using two different methods: (A) One extends the computation performed in Ref.~\cite{Hatfield:1984dv}
; (B) The other uses the weak coupling limit of the reformulation of the Schr\"odinger equation developed in  \cite{Karabali:1995ps,Karabali:1996je,Karabali:1996iu, Karabali:1997wk,Karabali:1998yq,Karabali:2009rg}. 

In the comparison between both results we obtained almost complete agreement, except for one term. We concluded that this discrepancy could be due to regularization issues, which had not been systematically addressed. In this chapter we fill this gap and provide with the complete expression of the Yang-Mills vacuum wave functional in three dimensions with ${\cal O}(e^2)$ precision for the first time.

The regularization of the Schr\"odinger equation and the vacuum wave functional in quantum field theories is a complicated subject. Whereas some formal aspects have been studied quite a while ago in Refs.~\cite{Symanzik:1981wd,Luscher:1985iu}, there have not been many quantitative studies of the regularization of the Yang-Mills vacuum wave functional. In three dimensions, the most detailed analyses have been carried out using method (B) (see, for instance, the discussions in Refs.~\cite{Karabali:1997wk,Agarwal:2007ns}, in particular in the appendix of the last reference). It is claimed in those references that the regularization has been completely taken into account. According to this, the result obtained in the previous chapter using method (B) (which corresponds to the weak coupling limit of the approximated expression obtained in Ref.~\cite{Karabali:2009rg} for the wave functional) should be the correct one. We will actually see that this is not so and that the regularization procedure has to be modified to obtain the correct Yang-Mills vacuum wave functional in three dimensions at weak coupling. This produces a new contribution that has to be added to the result obtained in Sec.~\ref{sec:KNY}.

The result given in the previous chapter using method (A) was obtained without regularizing the functional Schr\"odinger equation. 
It directly works with the gauge variables ${\vec A}$, but it has the complication that the Gauss law constraint has to be implemented by hand. 
In the intermediate steps potentially divergent expressions were found, which, nevertheless could be handled formally (assuming that the symmetries of the classical theory survive) obtaining a finite result. In this chapter we carefully regularize the computation using method (A). Out of this analysis a 
new contribution has to be added to the result obtained in Sec.~\ref{sec:Comp:Hatfield}.

The new results obtained for the Yang-Mills vacuum wave functional in three dimensions at weak coupling to ${\cal O}(e^2)$ with the methods (A) and (B) agree with each other. This is a strong check of our computations and of the regularization methods used here. On the other 
hand our results imply that the weak coupling limit of the expression obtained in Ref.~\cite{Karabali:2009rg} for the wave functional is not 
correct with ${\cal O}(e^2)$ precision (though it is at ${\cal O}(e)$).


The outline of this chapter is the following: In Sec.~\ref{sec:Reg} we regularize the Schr\"odinger equation. In Sec.~\ref{sec:Reg:Hatfield} we compute the wave functional using the method (A) with ${\cal O}(e^2)$ precision. In Sec.~\ref{sec:KKN} we rewrite the regularized version of the Schr\"odinger equation obtained in Sec.~\ref{sec:Reg} in terms of the gauge invariant variables, and compute the wave functional using the method (B) with ${\cal O}(e^2)$ precision. We also discuss the reason why the Schr\"odinger equation used in Ref.~\cite{Karabali:2009rg} is not sufficient to obtain the complete expression for the vacuum wave functional to ${\cal O}(e^2)$. Sec.~\ref{sec:Conclusions:Regularization} summarizes the results of this chapter.

\section{The regularized  \SE}
\label{sec:Reg} 

In Chap.~\ref{chap:Comparison} we used the unregularized \SE $ $, Eq.~(\ref{eq:SE}), which reads
\be
\frac{1}{2}\int_x\left(-\frac{\delta}{\delta \vec{A}^a(\vec{x})} \cdot \frac{\delta}{\delta \vec{A}^a(\vec{x})} + B^a(\vec{x}) B^a(\vec{x}) \right) 
\Psi = 0
\,. \label{GLH}
\ee

In order to regularize the kinetic operator we separate the points at which the differential operators act. As we want to preserve gauge invariance, we do this by introducing a Wilson line and a regularized delta function
\be
\delta_\mu(\vec x, \vec v)={\mu^2 \over \pi} e^{-(\vec{x}-\vec{v})^2\mu^2}\,,
\ee
such that after the removal of the regulator $\mu\to\infty$, one recovers the original expression:
\be
\mathcal{T}=-{1\over2}\int_x\frac{\delta}{\delta A_i^a(\vec x)}\frac{\delta}{\delta A_i^a(\vec x)} 
\longrightarrow 
\mathcal{T}_{reg}=-{1\over2}\int_{x,v} \delta_\mu(\vec x, \vec v)\frac{\delta}{\delta A_i^a(\vec x)} \Phi_{ab}(\vec x,\vec v) \frac{\delta}{\delta A_i^b(\vec v)}\,.
\label{Treg_middleString}
\ee
The first functional derivative also acts on the Wilson line, which ensures that the regularized kinetic operator is still hermitian.

The Wilson line is the path-ordered exponential of the gauge fields along a curve $\mathcal{C}$:
\be
\Phi(\mathcal{C};\vec x,\vec v)=\mathcal{P}e^{-e\int^{\vec x}_{\vec v}dz^iA_i(\vec z)} = \mathcal{P}e^{-e\int_0^1 ds\, \dot{z}^i(s)A_i(\vec z(s))}\,,
\ee
where $\vec{z}(s)$ is the parametrization of $\mathcal{C}$.
The Wilson line transforms as
\be
\Phi(\mathcal{C};\vec x,\vec v) \to \left(g(\vec x)\Phi(\mathcal{C};\vec x,\vec v)g^\dagger(\vec v)\right)_{ab}
\ee
under gauge transformations Eq.~(\ref{GaugeTrafo}).

\begin{figure}
\centerline{
\includegraphics[width=.49\textwidth,clip]{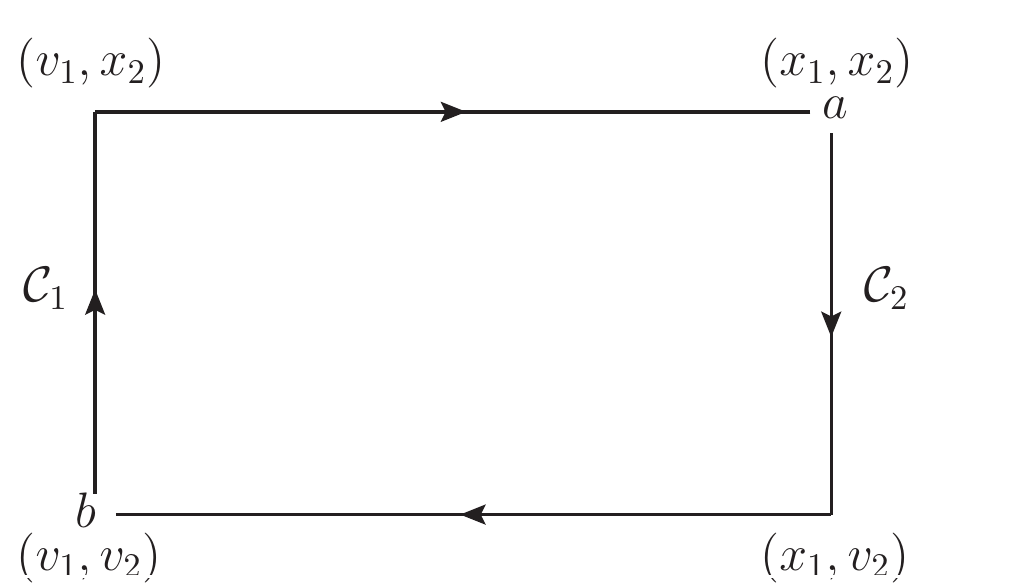}}
\caption{\it Curves $\mathcal{C}_1$ and $\mathcal{C}_2$ used to define $\Phi_{ab}(\vec x,\vec v)$ in Eqs.~(\ref{Phiabe2}) and (\ref{Phi}). \label{fig:curve}}
\end{figure}

The physical results should be independent of the curve $\mathcal{C}$. Nevertheless, for convenience, 
we choose the Wilson line to be symmetric under the combined interchange of color indices and endpoints:
\be
\Phi_{ab}(\mathcal{C};\vec x,\vec v) = \Phi_{ba}(\mathcal{C};\vec v,\vec x)\,.
\ee
For the computations in perturbation theory we need an explicit realization of the Wilson line. We choose the symmetric combination of two paths that go in straight lines (see Fig.~\ref{fig:curve}), so that up to ${\cal O}(e^2)$ the Wilson line reads:
\bE{rCl}
\label{Phiabe2}
\Phi_{ab}(\vec x,\vec v)  &\equiv& \half\left(\Phi_{ab}(\mathcal{C}_1;\vec x,\vec v) +\Phi_{ba}(\mathcal{C}_2;\vec v,\vec x) \right) \\
&=& \de_{ab}-\frac{e}{2}\left(\int_{v_2}^{x_2}ds_2 A_2(v_1,s_2) + \int_{v_1}^{x_1}ds_1 A_1(s_1,x_2)\right)_{ab} \nn\\
&&\qquad -\frac{e}{2}\left(\int_{x_2}^{v_2}ds_2 A_2(x_1,s_2) + \int_{x_1}^{v_1}ds_1 A_1(s_1,v_2)\right)_{ba} \nn\\
&&\qquad +\frac{(-e)^2}{2}\left(\int_{v_2}^{x_2}ds_2 A_2(v_1,s_2) \int_{v_2}^{s_2}ds'_2 A_2(v_1,s'_2) \right. \nn\\
&&\qquad\qquad\qquad\qquad \left.  + \int_{v_1}^{x_1}ds_1 A_1(s_1,x_2) \int_{v_1}^{s_1}ds'_1 A_1(s'_1,x_2) \right. \nn\\
&&\qquad\qquad\qquad\qquad \left. + \int_{v_1}^{x_1}ds_1 A_1(s_1,x_2) \int_{v_2}^{x_2}ds_2 A_2(v_1,s_2) \right)_{ab} \nn\\
&&\qquad +\frac{(-e)^2}{2}\left(\int_{x_2}^{v_2}ds_2 A_2(x_1,s_2) \int_{x_2}^{s_2}ds'_2 A_2(x_1,s'_2) \right. \nn\\
&&\qquad\qquad\qquad\qquad \left. + \int_{x_1}^{v_1}ds_1 A_1(s_1,v_2) \int_{x_1}^{s_1}ds'_1 A_1(s'_1,v_2) \right. \nn\\
&&\qquad\qquad\qquad\qquad \left. + \int_{x_1}^{v_1}ds_1 A_1(s_1,v_2) \int_{x_2}^{v_2}ds_2 A_2(x_1,s_2) \right)_{ba} + {\cal O}(e^3)
\nn
\eE
Note that $A_i^{ab}=-f^{abc}A_i^c$ and $(A_iA_j)^{ab}=f^{adc}f^{dbe}A_i^cA_j^e$. 

It is possible to write $\Phi_{ab}(\vec x,\vec v) $ in a more compact way using the Bars variables \cite{Bars:1978xy}:
\bE{rCl}
 \Phi_{ab}(\vec x,\vec v) &=&{1\over 2}\left((M_1(\vec x)M_1^{-1}(v_1,x_2)M_2(v_1,x_2)M_2^{-1}(\vec v))^{ab} \right. \nn\\
 &&\qquad\qquad \left.+(M_2(\vec x)M_2^{-1}(x_1,v_2)M_1(x_1,v_2)M_1^{-1}(\vec v))^{ab}\right)\,, \label{Phi}
\eE
where  (no sum over repeated spatial indices in Eqs.~(\ref{M_iPathordered}-\ref{MjOverAi})) 
\be
M_i(\vec x)=\mathcal{P}e^{-e\int^{\vec x}_{\infty}dz_iA_i(\vec z)} \label{M_iPathordered}
\ee
represents the Wilson line for a straight spatial curve $\mathcal{C}$ with fixed $x_j$ for $j\not=i$. 
This Wilson line can be Taylor expanded in the standard way in terms of (path-ordered) one-dimensional integrals (similarly as 
we have done in Eq.~(\ref{Phiabe2})), or in terms of 
(formal) two dimensional integrals (see, for instance, Ref.~\cite{Freidel:2006qz}):
\begin{IEEEeqnarray}{rCl}
M_i(\vec x) &= 1&-e\int_y G_i(\vec x;\vec y) A_i(\vec y) + e^2\int_{y,z}G_i(\vec x;\vec z)A_i(\vec z)G_i(\vec z;\vec y)A_i(\vec y)+\ldots, \label{M_iofA_i}\\
M_i^{-1}(\vec x) &= 1&+e\int_y G_i(\vec x;\vec y) A_i(\vec y) - e^2\int_{y,z}G_i(\vec x;\vec z)A_i(\vec z)G_i(\vec z;\vec y)A_i(\vec y) \nn\\
&& +  e^2\int_{y,z}G_i(\vec x;\vec z)A_i(\vec z)G_i(\vec x;\vec y)A_i(\vec y)+\ldots\nn\\
\Bigg(&= 1&+e\int_y G_i(\vec x;\vec y) A_i(\vec y) +  e^2\int_{y,z}G_i(\vec x;\vec y)G_i(\vec y;\vec z)A_i(\vec z)A_i(\vec x)+\ldots\Bigg) \,, \label{M_i-1}
\end{IEEEeqnarray}
where
\be
G_1(\vec x;\vec y)\equiv G_1(\vec x-\vec y)=\theta(x_1-y_1)\de(x_2-y_2)\quad \mathrm{and}\quad G_2(\vec x;\vec y)\equiv G_(\vec x-\vec y)=\de(x_1-y_1)\theta(x_2-y_2) \label{G_i}\,.
\ee
$M_i^{ab}=2\tr(T^aM_iT^bM_i^{-1})$ is the Euclidean analogue of Eq.~(\ref{Mab}). 
With these definitions
\bE{rCl}
A_i &=& -{1\over e}\p_iM_iM_i^{-1}  \label{Ai} \\
\Longleftrightarrow D_iM_i &=& 0\,. \label{D_iA_i}
\eE
Note that Eqs.~(\ref{M_iofA_i}-\ref{Ai}) are the Euclidean versions of Eqs.~(\ref{Mexp}-\ref{G}) and (\ref{AofM}), respectively, except for the fact that unlike $G(\bar x;\bar y)$ and $\bar G(x;y)$, $G_i(\vec x;\vec y)$ is not antisymmetric.

Variating Eq.~(\ref{D_iA_i}) one finds (see App.~\ref{sec:MoverA})
\be
\frac{\de M_j(y)}{\de A_i^a(x)} = -\de^{ij} M_i^{ab}(x) G_i(y,x) M_i(y)T_b \,. \label{MjOverAi}
\ee
The functional derivative of $A_i$ acting on the Wilson line in Eq.~(\ref{Treg_middleString}) is ill-defined if both the derivative and the Wilson line are defined at the same point. Therefore, we have to regularize it, taking the coincidence limit only after the functional derivative has been applied:
\bE{rl}
\int_{x,v}&\de_\mu(\vec x,\vec v) \left[\frac{\de}{\de A_i^a(\vec x)}\Phi_{ab}(\vec x,\vec v)\right]\frac{\de}{\de A_i^b(\vec v)}  \nn\\
&:= \lim_{\nu\to\infty} \int_{x,v,X}\de_\mu(\vec x,\vec v)\de_\nu(\vec X)\Phi_{ar}(\vec x,\vec x+\vec X) \left[\frac{\de}{\de A_i^r(\vec x+\vec X)}\Phi_{ab}(\vec x,\vec v)\right]\frac{\de}{\de A_i^b(\vec v)} \,. \label{regDeriv}
\eE
This way of regularizing is analogous to the regularizations used in Eq.~(3.24) of Ref.~\cite{Karabali:1997wk} and in Eqs.~(100-101) of Ref.~\cite{Freidel:2006qz}.

Using Eqs.~(\ref{Phi}) and (\ref{MjOverAi}) in Eq.~(\ref{regDeriv}) one finds
\be
\int_{x,v}\de_\mu(\vec x,\vec v) \left[\frac{\de}{\de A_i^a(\vec x)}\Phi_{ab}(\vec x,\vec v)\right]\frac{\de}{\de A_i^b(\vec v)}  =0 \,,
\ee
such that the regularized kinetic operator Eq.~(\ref{Treg_middleString}) reduces to
\be
\mathcal{T}_{reg}=-{1\over2}\int_{x,v} \delta_\mu(\vec x, \vec v) \Phi_{ab}(\vec x,\vec v) \frac{\delta}{\delta A_i^a(\vec x)}  \frac{\delta}{\delta A_i^b(\vec v)}\,.
\label{Treg}
\ee
This is shown in App.~\ref{sec:herm} in detail.

\medskip

Once we have regularized the kinetic operator we turn to the determination of the vacuum wave functional. Realizing that the vacuum wave functional for the kinetic operator $\T$ alone is the identity, one can write the complete wave functional as
\be
\Psi=e^{-F}\mathds{1}\,.
\ee
Therefore, instead of solving 
\be
\H\Psi=(\T+\V)\Psi=0
\,,
\ee
one can solve (see, for instance, Ref.~\cite{Karabali:1998yq})
\be
\label{HamEq}
\widetilde{\H}\mathds{1}=e^F(\T+\V)e^{-F}\mathds{1}=\left(\T+\V-[\T,F]+\half\left[[\T,F],F\right]\right)
\mathds{1}=0\,,
\ee
since $\T$ contains at most two functional derivatives:
\be
\T=\int_{x}\omega_i^a(\vec x)\frac{\delta}{\delta A_i^a(\vec x)}+\int_{x,y}\Omega_{ij}^{ab}(\vec x,\vec y)\frac{\delta^2}{\delta A_i^a(\vec x) \delta A_j^b(\vec y)} \,,
\ee
where $\omega^a_i(\vec x)=0$ and $\Omega_{ij}^{ab}(\vec x,\vec y)=\de_{ij}\Omega^{ab}(\vec x,\vec y)=-\half\de_{ij}\delta_\mu(\vec x,\vec y)\Phi_{ab}(\vec x,\vec y)$. Using this explicit expression, Eq. (\ref{HamEq}) reads
\be
\V-\int_{x}\omega_i^a(\vec x)\frac{\delta F}{\delta A_i^a(\vec x)}-\int_{x,y}\Omega_{ij}^{ab}(\vec x,\vec y)\frac{\delta^2 F}{\delta A_i^a(\vec x) \delta A_j^b(\vec y)}
+ \int_{x,y}\Omega_{ij}^{ab}(\vec x,\vec y)\frac{\delta F}{\delta A_i^a(\vec x)} \frac{\delta F}{\delta A_j^b(\vec y)} = 0  \label{SE}
\,.
\ee

In order to ensure that we restrict ourselves to gauge invariant states we also have to demand that $\Psi$ satisfies the Gauss law constraint Eq.~(\ref{NonAbGL}):
\be
\label{Gausslaw}
I^a\Psi =\,
(\vec{D}\cdot\vec{E})^a\Psi =
i\left(
\vec \nabla \cdot \frac{\delta }{\delta {\vec A}_a}+ef^{abc}{\vec A}_b \cdot \frac{\delta }{\delta {\vec A}_c}
\right)\Psi=0 
\,.
\ee
Equations (\ref{SE}) and (\ref{Gausslaw}) will be our starting point for the determination of the vacuum wave functional. 

As in Chap.~\ref{chap:Comparison}, in the following we will distinguish between methods (A) and (B), and name their solutions $\Psi_{GL}=e^{-F_{GL}}$ and 
 $\Psi_{GI}=e^{-F_{GI}}$, respectively. The first method consists in directly solving Eqs.~(\ref{SE}) and (\ref{Gausslaw}), and will be addressed in the next section. The second method consists in rewriting Eq.~(\ref{SE}) in terms of the gauge invariant variables $J^a$ defined in Eq.~(\ref{J}). 
It will be addressed in Sec.~\ref{sec:KKN}. In both cases we will Taylor expand $F$ in powers of the coupling constant $e$, and solve the resulting equations iteratively, like in Chap.~\ref{chap:Comparison}. 
In this chapter the main focus will be on the novel aspects resulting from the careful introduction of the regularization. 

\section{Determination of $\Psi_{GL}[{\vec A}]$}
\label{sec:Reg:Hatfield} 
We expand $F_{GL}=F_{GL}^{(0)}+eF_{GL}^{(1)}+e^2F_{GL}^{(2)}+{\cal O}(e^3)$ and 
\be
\Omega^{ab}(\vec x,\vec y)=-{1\over2}\delta_\mu(\vec x,\vec y)\Phi_{ab}(\vec x,\vec y)=-{1\over2}\delta_\mu(\vec x,\vec y)\left(\Phi^{(0)}_{ab}(\vec x,\vec y)+e\Phi^{(1)}_{ab}(\vec x,\vec y)+e^2\Phi^{(2)}_{ab}(\vec x,\vec y)+{\cal O}(e^3)\right) \qquad
\ee
in powers of the coupling constant. Considering the contributions order by order in $e$ yields the following equations:

\medskip

At ${\cal O}(e^0)$ we have
\be
\label{eqe0}
\V|_{{\cal O}(e^0)} -{1\over2}\int_{x,y}\delta_\mu(\vec x,\vec y)\delta_{ab} \left(- \frac{\delta^2 F_{GL}^{(0)} }{\delta A_i^a(\vec x) \delta A_i^b(\vec y)}+\frac{\delta F_{GL}^{(0)}}{\delta A_i^a(\vec x)} \frac{\delta F_{GL}^{(0)}}{\delta A_i^b(\vec y)} \right)= 0\,.
\ee

For this equation we can take the $\mu\to\infty$ limit, reducing it to the standard unregularized free field equation, Eq.~(\ref{Sch_Free}), the solution of which is Eq.~(\ref{FGL0}):
\bea
F_{GL}^{(0)}[{\vec A}] &=& \frac{1}{2}\int_\slashed{k}\frac{1}{|\vec k|} (\vec{k}\times\vec{A}^a(\vec{k})) (\vec{k}\times\vec{A}^a(-\vec{k})) \label{FGL0m} \\
&=& \frac{1}{4\pi}\int_{x,y}\frac{1}{|\vec{x}-\vec{y}|} (\vec{\nabla}\times\vec{A}^a(\vec{x}))(\vec{\nabla}\times\vec{A}^a(\vec{y})) 
\,.\label{FGL0p} 
\eea
 
\medskip

At ${\cal O}(e)$ we have
\bE{l}
\V|_{{\cal O}(e)} +{1\over2}\int_{x,y}\delta_\mu(\vec x,\vec y)\delta_{ab}
\left(
\frac{\delta^2 F_{GL}^{(1)}}{\delta A_i^a(\vec x) \delta A_i^b(\vec y)} 
- 2  \frac{\delta F_{GL}^{(0)}}{\delta A_i^a(\vec x)} \frac{\delta F_{GL}^{(1)}}{\delta A_i^b(\vec y)}
\right) 
\nn\\
\quad -{1\over2} \int_{x,y}\delta_\mu(\vec x,\vec y)\Phi^{(1)}_{ab}(\vec x,\vec y) 
\left(
\frac{\delta F_{GL}^{(0)}}{\delta A_i^a(\vec x)} \frac{\delta F_{GL}^{(0)}}{\delta A_i^b(\vec y)} 
- \frac{\delta^2 F_{GL}^{(0)}}{\delta A_i^a(\vec x) \delta A_i^b(\vec y)}
\right) = 0\,. \label{regF1eq}
\eE
Both terms proportional to $\Phi^{(1)}_{ab}(\vec x,\vec y)$ vanish (the second because of contraction of color indices, for the first see App.~\ref{subsec:append4}). For the remaining terms we can take the limit $\mu \rightarrow \infty$. Therefore, this equation also reduces to the unregularized \SE, Eq.~(\ref{OeF1}). It is solved by Eq.~(\ref{F1H}):
\begin{IEEEeqnarray}{rCl}
\label{FGL1}
F_{GL}^{(1)}[{\vec A}] &=&  i f^{abc} \int_{\slashed{k_1},\slashed{k_2},\slashed{k_3}}\slashed{\delta}\left(\sum_{i=1}^3 \vec{k}_i\right) \Bigg\{ \frac{1}{2(\sum_i^3|\vec{k}_i|)} (\vec{k}_1\times\vec{A}^a(\vec{k}_1)) (\vec{A}^b(\vec{k}_2)\times\vec{A}^c(\vec{k}_3))\nn\\
&& -\frac{1}{(\sum_i^3|\vec{k}_i|)|\vec{k}_1||\vec{k}_3|} (\vec{k}_1\cdot\vec{A}^a(\vec{k}_1)) (\vec{k}_3\times\vec{A}^b(\vec{k}_2)) (\vec{k}_3\times\vec{A}^c(\vec{k}_3)) \Bigg\} 
\,.
\end{IEEEeqnarray}

\medskip

At ${\cal O}(e^2)$ we determine $F_{GL}^{(2)}$. $F^{(2)}_{GL}$ can have contributions with four, two and zero fields: $F^{(2)}_{GL}=F^{(2,4)}_{GL}+F^{(2,2)}_{GL}+F^{(2,0)}_{GL}$. As argued in Chap.~\ref{chap:Comparison} there is no need to compute $F^{(2,0)}_{GL}$, as it only changes the normalization of the state, which we do not fix, or alternatively can be absorbed in a redefinition of the ground-state energy. $F^{(2,4)}_{GL}$ is determined by the following equation:
\bE{l}
 \V|_{{\cal O}(e^2)}   -{1\over2} \int_{x,y}\delta_\mu(\vec x,\vec y)\delta_{ab}\left( \frac{\delta F_{GL}^{(1)}}{\delta A_i^a(\vec x)} \frac{\delta F_{GL}^{(1)}}{\delta A_i^b(\vec y)}  + 2\frac{\delta F_{GL}^{(0)}}{\delta A_i^a(\vec x)} \frac{\delta F_{GL}^{(2,4)}}{\delta A_i^b(\vec y)} \right)  \nn\\
\quad  -{1\over2}\int_{x,y}\delta_\mu(\vec x,\vec y)\left(\Phi^{(2)}_{ab}(\vec x,\vec y) \frac{\delta F_{GL}^{(0)}}{\delta A_i^a(\vec x)} \frac{\delta F^{(0)}}{\delta A_i^b(\vec y)} + 2 \Phi^{(1)}_{ab}(\vec x,\vec y) \frac{\delta F_{GL}^{(0)}}{\delta A_i^a(\vec x)} \frac{\delta F_{GL}^{(1)}}{\delta A_i^b(\vec y)} \right) = 0 \,,
\label{regF24eq}
\eE
The two terms in the second line vanish (see App.~\ref{subsec:append5}). For the leftover we can take the $\mu \rightarrow \infty$ limit. Eq.~(\ref{regF24eq}) then reduces to its unregularized version, Eq.~(\ref{Sch_e2_4}), which is solved by Eq.~(\ref{F24GL}). We quote it here for completeness:
\begin{IEEEeqnarray}{l}
\label{F24GL2}
F_{GL}^{(2,4)}=f^{abc}f^{cde}\int_{\slashed{k_1},\slashed{k_2},\slashed{q_1},\slashed{q_2}}
\slashed{\delta}\left(\sum_i (\vec{k}_i+\vec{q}_i)\right) \frac{1}{|\vec{k}_1|+|\vec{k}_2|+|\vec{q}_1|+|\vec{q}_2|} \Bigg\{  \nn\\
\quad \frac{1}{2(|\vec{k}_1|+|\vec{k}_2|+|\vec{k}_1+\vec{k}_2|)(|\vec{q}_1|+|\vec{q}_2|+|\vec{q}_1+\vec{q}_2|)} \Bigg\{\left(\vec{A}^d(\vec{q}_1)\times\vec{A}^e(\vec{q}_2)\right)  \nn\\
\qquad \times \Bigg[ -\frac{1}{4}|\vec{k}_1+\vec{k}_2|^2 \vec{A}^a(\vec{k}_1)\times\vec{A}^b(\vec{k}_2) \nn\\
\qquad\qquad +\frac{|\vec{k}_1+\vec{k}_2|}{|\vec{k}_2|}(\vec{k}_1+\vec{k}_2)\times\vec{A}^a(\vec{k}_1) (\vec{k}_2\cdot\vec{A}^b(\vec{k}_2)) +\frac{(\vec{k}_1+\vec{k}_2)\cdot\vec{k}_2}{|\vec{k}_1||\vec{k}_2|} (\vec{k}_1\cdot\vec{A}^a(\vec{k}_1)) (\vec{k}_2\times\vec{A}^b(\vec{k}_2)) \nn\\
\qquad\qquad +(\vec{k}_1\times\vec{A}^a(\vec{k}_1)) (\vec{k}_1+\vec{k}_2)\cdot\vec{A}^b(\vec{k}_2) \Bigg] \nn\\
\qquad +  (\vec{k}_1\times\vec{A}^a(\vec{k}_1)) (\vec{q}_1\times\vec{A}^d(\vec{q}_1)) \left(\vec{A}^b(\vec{k}_2)\cdot\vec{A}^e(\vec{q}_2)\right)\nn\\
\qquad + \frac{1}{|\vec{k}_1||\vec{k}_2|}\left[2\vec{k}_2\cdot\vec{A}^e(\vec{q}_2)-\frac{\vec{q}_1\cdot\vec{k}_2}{|\vec{q}_1||\vec{k}_2|}\vec{q}_2\cdot\vec{A}^e(\vec{q}_2)\right] (\vec{k}_1\cdot\vec{A}^a(\vec{k}_1)) (\vec{k}_2\times\vec{A}^b(\vec{k}_2)) (\vec{q}_1\times\vec{A}^d(\vec{q}_1)) \nn\\
\qquad + \frac{1}{|\vec{k}_1|} (\vec{k}_1\cdot\vec{A}^a(\vec{k}_1)) (\vec{k}_1+\vec{k}_2) \times\vec{A}^b(\vec{k}_2) \Bigg[  \frac{1}{|\vec{q}_2|} (\vec{q}_1+\vec{q}_2)\times\vec{A}^d(\vec{q}_1) (\vec{q}_2 \cdot\vec{A}^e(\vec{q}_2)) \nn\\
\qquad\qquad + \frac{2}{|\vec{q}_1+\vec{q}_2|} (\vec{q}_1\times\vec{A}^d(\vec{q}_1)) (\vec{q}_1+\vec{q}_2) \cdot\vec{A}^e(\vec{q}_2) \Bigg] \nn\\
\qquad -\frac{2(\vec{q}_1+\vec{q}_2)\cdot\vec{q}_1}{|\vec{k}_1+\vec{k}_2||\vec{k}_1| |\vec{q}_1||\vec{q}_2|}  (\vec{k}_1\cdot\vec{A}^a(\vec{k}_1)) (\vec{k}_1+\vec{k}_2) \times\vec{A}^b(\vec{k}_2) (\vec{q}_1\times\vec{A}^d(\vec{q}_1)) (\vec{q}_2 \cdot\vec{A}^e(\vec{q}_2)) \nn\\
\qquad +\frac{2\vec{k}_1\times\vec{k}_2}{|\vec{k}_1||\vec{k}_2| |\vec{q}_1+\vec{q}_2| |\vec{q}_2|}  (\vec{k}_1\cdot\vec{A}^a(\vec{k}_1)) (\vec{k}_2 \times\vec{A}^b(\vec{k}_2)) (\vec{q}_2\times\vec{A}^d(\vec{q}_1)) (\vec{q}_2 \times\vec{A}^e(\vec{q}_2)) \nn\\
\qquad +\frac{2}{|\vec{q}_1+\vec{q}_2||\vec{q}_2|}  (\vec{k}_1\times\vec{A}^a(\vec{k}_1)) (\vec{k}_1+\vec{k}_2) \times\vec{A}^b(\vec{k}_2) (\vec{q}_2\times\vec{A}^d(\vec{q}_1)) (\vec{q}_2 \times\vec{A}^e(\vec{q}_2)) \nn\\
\qquad -\frac{1}{|\vec{k}_2||\vec{q}_2|}  (\vec{k}_2\times\vec{A}^a(\vec{k}_1)) (\vec{k}_2 \times\vec{A}^b(\vec{k}_2)) (\vec{q}_2\times\vec{A}^d(\vec{q}_1)) (\vec{q}_2 \times\vec{A}^e(\vec{q}_2))
\Bigg\} \nn\\
\quad +\frac{1}{8} \left(\vec{A}^a(\vec{k}_1)\times\vec{A}^b(\vec{k}_2)\right) \left(\vec{A}^d(\vec{q}_1)\times\vec{A}^e(\vec{q}_2)\right) \nn\\
\quad + \frac{1}{|\vec{k}_1|(|\vec{q}_1|+|\vec{q}_2|+|\vec{q}_1+\vec{q}_2|)}  (\vec{k}_1\cdot\vec{A}^a(\vec{k}_1)) \Bigg\{ \frac{1}{2}  (\vec{k}_1+\vec{k}_2) \times\vec{A}^b(\vec{k}_2)  \left(\vec{A}^d(\vec{q}_1)\times\vec{A}^e(\vec{q}_2)\right) \nn\\
\quad\quad 
-  (\vec{q}_1 \times\vec{A}^d(\vec{q}_1) ) \left(\vec{A}^b(\vec{k}_2)\times\vec{A}^e(\vec{q}_2)\right) \nn\\
\quad \quad-  \frac{1}{|\vec{q}_1+\vec{q}_2||\vec{q}_2|} (\vec{k}_1+\vec{k}_2) \times\vec{A}^b(\vec{k}_2) (\vec{q}_1+\vec{q}_2)\times\vec{A}^d(\vec{q}_1) (\vec{q}_2 \cdot\vec{A}^e(\vec{q}_2)) \nn\\
\quad \quad +  \frac{1}{|\vec{q}_1||\vec{q}_2|} (\vec{q}_2 \times\vec{A}^b(\vec{k}_2)) (\vec{q}_1\cdot\vec{A}^d(\vec{q}_1)) (\vec{q}_2 \times\vec{A}^e(\vec{q}_2)) \nn\\
\quad \quad  -  \frac{1}{|\vec{q}_1+\vec{q}_2||\vec{q}_2|} (\vec{k}_1+\vec{k}_2) \cdot\vec{A}^b(\vec{k}_2) (\vec{q}_2\times\vec{A}^d(\vec{q}_1)) (\vec{q}_2 \times\vec{A}^e(\vec{q}_2)) 
\Bigg\}
\Bigg\}
\,. 
\end{IEEEeqnarray}

So far the regularization of the kinetic term has not produced any modification to the results obtained in Chap.~\ref{chap:Comparison}. This could have been expected. If we have to make an analogy 
of this computation with the standard diagrammatic approach, the computations above would correspond to tree-level-like diagrams, for which one can take the cutoff to infinity. It is only when one has internal loops, where the momentum can run to infinity, when regularization effects become important. In our approach those effects are hidden in $F^{(2,2)}_{GL}$, where we have an effect similar to the contraction of two fields. We compute this term in the next subsection.

\subsection{$F^{(2,2)}_{GL}$}
\label{subsec:F22GL}
$F^{(2,2)}_{GL}$ is determined by the following equation:
\bE{l}
\int_{x,y}\delta_\mu(\vec x,\vec y)\Bigg(\Phi^{(0)}_{ab}(\vec x,\vec y)\frac{\delta^2 F_{GL}^{(2,4)} }{\delta A_i^a(\vec x) \delta A_i^b(\vec y)} + \Phi^{(1)}_{ab}(\vec x,\vec y)\frac{\delta^2 F_{GL}^{(1)} }{\delta A_i^a(\vec x) \delta A_i^b(\vec y)}+ \Phi^{(2)}_{ab}(\vec x,\vec y)\frac{\delta^2 F_{GL}^{(0)} }{\delta A_i^a(\vec x) \delta A_i^b(\vec y)} \nn\\
\qquad - 2\Phi^{(0)}_{ab}(\vec x,\vec y) \frac{\delta F_{GL}^{(0)}}{\delta A_i^a(\vec x)} \frac{\delta F_{GL}^{(2,2)}}{\delta A_i^b(\vec y)} \Bigg) = 0 \,.
\label{regF22eq}
\eE
In order to solve this equation it is convenient to rewrite it in momentum space. Then, the last term of Eq.~(\ref{regF22eq}) reads
\bea
&&- 2 \int_{x,y} \delta_\mu(\vec x,\vec y) \Phi^{(0)}_{ab}(\vec x,\vec y) \frac{\delta F_{GL}^{(0)}}{\delta A_i^a(\vec x)} \frac{\delta F_{GL}^{(2,2)}}{\delta A_i^b(\vec y)}  \nn\\
&=&-2 \int_{\slashed{p}}\delta_\mu(\vec p)\delta^{ab} \frac{1}{|{\vec p}|} (\vec{p}\times\vec{A}^a(\vec{p}))  \left(\vec{p}\times \frac{\delta F_{GL}^{(2,2)}[\vec A]}{\delta \vec{A}^b(\vec{p})}\right) 
\nn\\
&=&-2 \int_{\slashed{p}}\delta_\mu(\vec p)\frac{1}{|{\vec p}|} 
\left\{
{\vec p}^2\left(\vec{A}^a(\vec{p})\cdot \frac{\delta F_{GL}^{(2,2)}[\vec A]}{\delta \vec{A}^a(\vec{p})}\right) 
-
\left({\vec p}\cdot {\vec A}^a({\vec p})\right)
\left(\vec{p}\cdot \frac{\delta F_{GL}^{(2,2)}[\vec A]}{\delta \vec{A}^a(\vec{p})}\right) 
\right\}
\,, \label{edd}
\eea
where $\de_\mu(\vec p)=e^{-\frac{\vec{p}^2}{4\mu^2}}$ is the Fourier transform of $\delta_\mu(\vec x,\vec y)$ and we used $\e_{ij}\e_{kl}=\de_{ik}\de_{jl}-\de_{jk}\de_{il}$.\\
The Gauss law implies that the second term on the right-hand side of the last equality of Eq.~(\ref{edd}) vanishes, so Eq.~(\ref{regF22eq}) can be rewritten as 
\bE{rCl}
\label{F22mom}
2 \int_{\slashed{p}}\delta_\mu(\vec p)|{\vec p}| \left(\vec{A}^a(\vec{p})\cdot \frac{\delta F_{GL}^{(2,2)}[\vec A]}{\delta \vec{A}^a(\vec{p})}\right) 
 &=& \int_{x,y}\int_{\slashed{p},\slashed{q}}e^{-i{\vec p}\cdot{\vec x}}e^{-i{\vec q}\cdot{\vec y}}\delta_\mu(\vec x,\vec y)\Bigg(\de^{ab}\frac{\delta^2 F_{GL}^{(2,4)} }{\delta A_i^a(\vec p) \delta A_i^b(\vec q)} \nn\\
&&\quad + \Phi^{(1)}_{ab}(\vec x,\vec y)\frac{\delta^2 F_{GL}^{(1)} }{\delta A_i^a(\vec p) \delta A_i^b(\vec q)}+ \Phi^{(2)}_{ab}(\vec x,\vec y)\frac{\delta^2 F_{GL}^{(0)} }{\delta A_i^a(\vec p) \delta A_i^b(\vec q)} \Bigg) \,. \nn\\
\eE
Before going on we need to compute the right-hand side of this equation (which again is better handled in momentum space). The first term corresponds to the regularized version of the term that already appeared in Eq.~(\ref{F22GLa}). 
As we can see in Eq.~(\ref{F24GL2}), the explicit expression of $F^{(2,4)}_{GL}[\vec A]$ is very lengthy and complicated. 
This made impossible a direct brute force computation of $\frac{\delta^2 F_{GL}^{(2,4)} }{\delta A_i^a(\vec x) \delta A_i^b(\vec y)}$. The strategy we followed 
instead was to rewrite $F^{(2,4)}_{GL}[\vec A]$ in terms of $J$ and $\theta=\frac{1}{\bar\p}\bar A+ \mathcal{O}(e)$ (see Eq.~(\ref{F24GLGI})), which allows for a cleaner arrangement of the terms, in particular between gauge invariant and gauge dependent terms. Proceeding in the same way and using (see Eq.~(\ref{funderiv}))
\begin{IEEEeqnarray}{rCl}
\int_p\frac{\delta^2}{\delta A^a_i(-\vec{p})\delta A^a_i(\vec{p})} 
&=& 4 \int_p\frac{p}{\bar{p}}\frac{\delta^2}{\delta J^a(-\vec{p})\delta J^a(\vec{p})} +2\int_p \bar{p} \frac{\delta^2}{\delta \theta^a(-\vec{p})\delta J^a(\vec{p})} +\mathcal{O}(e)\,,
\end{IEEEeqnarray}
we obtain
\bea
\label{F22Jtheta}
&& 
\hspace{-0.5cm}\int_{x,y}\int_{\slashed{p},\slashed{q}}e^{-i{\vec p}\cdot{\vec x}}e^{-i{\vec q}\cdot{\vec y}}\delta_\mu(\vec x,\vec y)\de^{ab}\frac{\delta^2 F_{GL}^{(2,4)} }{\delta A_i^a({\vec p}) \delta A_i^b({\vec q})} 
 \\
 && =
 4C_A \int_{\slashed{p},\slashed{k}} e^{-\frac{\vec{p}^2}{4\mu^2}} \Bigg\{  \left(-\frac{1}{32} \frac{1}{\bar{p}}  g^{(3)}(k,p,-k-p) - \frac{1}{64} \frac{p}{\bar{p}}  g^{(4)}(p,k;-p,-k) \right) J^{a}(\vec{k})J^{a}(-\vec{k}) \nn\\
&&\quad + \frac{1}{4}\Bigg(\frac{1}{4}\left(2\frac{p}{\bar{p}}+\frac{k}{\bar{k}+\bar{p}}-\frac{p\bar{k}}{\bar{p}(\bar{k}+\bar{p})}\right) g^{(3)}(p,k,-p-k) \nn\\
 &&\quad  -2 \frac{1}{\bar{p}} \frac{(\bar{k}+\bar{p})^2}{|\vec{k}+\vec{p}|} +2 \frac{1}{\bar{p}} \frac{\bar{k}^2}{|\vec{k}|}  - \frac{\bar{k}-\bar{p}}{\bar{p}(\bar{k}+\bar{p})} \frac{\bar{k}^2}{|\vec{k}|} + \frac{\bar{k}-\bar{p}}{\bar{k}+\bar{p}} \frac{\bar{p}}{|\vec{p}|}\Bigg) J^{a}(\vec{k})\theta^{a}(-\vec{k}) \nn\\
 &&\quad + \left( \frac{p}{\bar{p}} \left( \frac{(\bar{p}+\bar{k})^2}{|\vec{p}+\vec{k}|} -  \frac{\bar{p}^2}{|\vec{p}|} \right) - \frac{p}{\bar{p}}\bar{k} \left(\frac{\bar{p}+\bar{k}}{|\vec{p}+\vec{k}|} - \frac{\bar{p}}{|\vec{p}|} \right) + k \left(\frac{\bar{p}+\bar{k}}{|\vec{p}+\vec{k}|} - \frac{\bar{p}}{|\vec{p}|} \right) \right) \theta^{a}(\vec{k}) \theta^{a}(-\vec{k}) \Bigg\} \nn
\,. 
\eea
This expression has an internal loop for the momentum $\vec p$, the integral of which is regularized by $\delta_{\mu}({\vec p})$. 
If we naively take the limit $\mu \rightarrow \infty$ and do formal manipulations (momentum shifts) of the integrals, we find the result obtained in Eq.~(\ref{F2GLb}):
\be
\label{J2I4}
-N\frac{C_A}{\pi}\int_\slashed{k}\frac{\bar{k}^2}{|\vec{k}|^2}J^{a}(\vec{k})J^{a}(-\vec{k}) =-N\frac{C_A}{\pi}\int_\slashed{k}\frac{1}{|\vec{k}|^2} (\vec{k}\times\vec{A}^a(\vec{k})) (\vec{k}\times\vec{A}^a(-\vec{k})) \,,
\ee
where $N$ has been defined in Eq.~(\ref{N}),
whereas the terms proportional to $J \theta$ and $\theta^2$ vanish.

Yet, this is not the whole story. The internal momentum of the loop is characterized by two scales: $|{\vec p}| \sim \mu$ and $|{\vec p}| \sim |{\vec k}|$, and taking the limit $\mu \rightarrow \infty$ before integration neglects contributions from the $|{\vec p}| \sim \mu$ region. Things change once the regularization is taken into account, as the high energy modes $|{\vec p}| \sim \mu$ are now also included in the computation. 
The loop result of the $J^2$ term is not modified by the introduction of the regularization, since the contribution due to $|{\vec p}| \sim \mu$ is subleading. Therefore, Eq.~(\ref{J2I4}) remains unchanged. 
Things are different, however, for the $J\theta$ and $\theta^2$ term. The $\theta^2$ term can be simplified to the following expression
 \begin{IEEEeqnarray}{rCl}
4 C_A\int_{\slashed{p},\slashed{k}} e^{-\frac{\vec{p}^2}{4\mu^2}}  \left( \left(  \frac{p(\bar{p}+\bar{k})}{|\vec{p}+\vec{k}|} -  \frac{1}{4}|\vec{p}|  \right) + k \left(\frac{\bar{p}+\bar{k}}{|\vec{p}+\vec{k}|} \right)  + \frac{\bar{k}p-k\bar{p}}{|\vec{p}|}\right) \theta^{a}(\vec{k}) \theta^{a}(-\vec{k}) 
\,.
\end{IEEEeqnarray}
The last term vanishes under $\vec p\to-\vec p$ and the first and the third can be combined to yield (note that the integral is dominated by $|{\vec p}| \sim \mu$ and that the $|{\vec p}| \sim |{\vec k}|$ region gives subleading contributions) 
\begin{IEEEeqnarray}{rCl}
\label{theta2}
C_A  \int_{\slashed{p},\slashed{k}} e^{-\frac{\vec{p}^2}{4\mu^2}}  \left(|\vec{p}+\vec{k}| -|\vec{p}|  \right)  \theta^{a}(\vec{k}) \theta^{a}(-\vec{k}) 
=
\int_{ \slashed{k}}\frac{C_A\mu}{8\sqrt{\pi}}  |\vec{k}|^2  \theta^{a}(\vec{k}) \theta^{a}(-\vec{k})+{\cal O}(1/\mu)\,.
\end{IEEEeqnarray}
We can deal with the $J\theta$ term of Eq.~(\ref{F22Jtheta}) in a very similar way (though with lengthier expressions). 
As before, the integral is dominated by the $|{\vec p}| \sim \mu$ region, whereas the $|{\vec p}| \sim |{\vec k}|$ region of momentum gives a subleading contribution\footnote{Actually statements of this sort are not true in general, as finite momentum shifts in the integrals may produce corrections from the $|{\vec p}| \sim |{\vec k}|$ region. Such shifts do not change the leading order contribution, which in our case is of ${\cal O}(\mu)$ but may change the individual ${\cal O}(\mu^0)$ contributions due to the $|{\vec p}| \sim |{\vec k}|$ and $|{\vec p}| \sim \mu$ regions (but in such a way that the total sum remains the same), which is the precision we seek. Therefore, such statements should be understood for a specific routing of momenta.}. Using
\bE{rCl}
{1\over2}\left(  J^{a}(\vec{k})\theta^{a}(-\vec{k}) -  J^{a}(-\vec{k})\theta^{a}(\vec{k}) \right) &=& -\frac{1}{2\bar{k}}\vec{A}^a(\vec{k})\cdot\vec{A}^a(-\vec{k})+2k\;\theta^a(\vec{k})\theta^a(-\vec{k}) +O(e)\,,
\eE
we rewrite the result in terms of $\vec{A}$ and $\theta$, and obtain
\begin{IEEEeqnarray}{rCl}
\label{jtheta}
 -\frac{C_A}{8\sqrt{\pi}} \mu  \int_{\slashed{k}}&    \Bigg(-\vec{A}^a(\vec{k})\cdot\vec{A}^a(-\vec{k})+ |\vec{k}|^2\, \theta^a(\vec{k})\theta^a(-\vec{k}) \Bigg) \,.
\end{IEEEeqnarray}
The bilinear terms in $\theta$ in Eqs.~(\ref{theta2}) and (\ref{jtheta}) cancel each other. Therefore, summing the contributions from Eqs.~(\ref{J2I4}), (\ref{theta2}) and (\ref{jtheta}) we obtain 
\bea
\label{F2I}
&&
\int_{x,y}\int_{\slashed{p},\slashed{q}}e^{-i{\vec p}\cdot{\vec x}}e^{-i{\vec q}\cdot{\vec y}}\delta_\mu(\vec x,\vec y)\de^{ab}\frac{\delta^2 F_{GL}^{(2,4)} }{\delta A_i^a({\vec p}) \delta A_i^b(\vec q)} 
 = 
 \\
 \nn
 &&
-N\frac{C_A}{\pi}\int_\slashed{k}\frac{1}{|\vec{k}|} (\vec{k}\times\vec{A}^a(\vec{k})) (\vec{k}\times\vec{A}^a(-\vec{k})) +\frac{C_A}{8\sqrt{\pi}}\mu \int_{\slashed{k}}  \vec{A}^a(\vec k)\cdot\vec{A}^a(-\vec k) +O\left(\mu^{-1}\right) 
\,.
\eea

We now compute the second term of the right-hand side of Eq.~(\ref{F22mom})
\bE{l}
\label{F2II}
\int_{x,y}\int_{\slashed{p},\slashed{q}}e^{-i{\vec p}\cdot{\vec x}}e^{-i{\vec q}\cdot{\vec y}}\delta_\mu(\vec x,\vec y) \Phi^{(1)}_{ab}(\vec x,\vec y)\frac{\delta^2 F_{GL}^{(1)} }{\delta A_i^a(\vec p) \delta A_i^b(\vec q)}
\nn\\
 = {1\over 2} f^{abd}  \int_{u,v,y}\int_{\slashed{k},\slashed{q},\slashed{p}}\delta_\mu(\vec u,\vec v)   \Bigg\{(G_1(\vec u;\vec y)-G_1(v_1,u_2;\vec y)+G_1(u_1,v_2;\vec y)-G_1(\vec v;\vec y)) A_1^d(\vec y)  \nn\\
\qquad  + (G_2(v_1,u_2;\vec y)-G_2(\vec v;\vec y)+G_2(\vec u;\vec y)-G_2(u_1,v_2;\vec y)) A_2^d(\vec y) \Bigg\} \nn\\
\quad 
\times
 i f^{abc} e^{-i\vec{p}\cdot\vec{u}} e^{-i\vec{q}\cdot\vec{v}} \frac{\slashed{\delta}\left(\vec{k}+\vec{p}+\vec{q}\right)}{|\vec{k}|+|\vec{p}|+|\vec{q}|} \Bigg\{  (\vec{q}-\vec{p})\cdot  \vec{A}^c(\vec{k})   -\frac{|\vec{q}|-|\vec{p}|}{|\vec{k}|} \vec{k}\cdot\vec{A}^c(\vec{k}) \nn\\
\qquad + \frac{1}{|\vec{q}||\vec{k}|}  (\vec{k}\times\vec{q}) (\vec{k}\times\vec{A}^c(\vec{k})) + \frac{1}{|\vec{q}||\vec{p}|} (\vec{p}\times\vec{q}) (\vec{k}\times\vec{A}^c(\vec{k}))  -\frac{1}{|\vec{p}||\vec{k}|}   (\vec{k}\times\vec{p})  (\vec{k}\times\vec{A}^c(\vec{k}))   \Bigg\} 
\nn
\quad\\
 = {C_A\over 2}\int_{\slashed{p},\slashed{q}} \Bigg\{\left(e^{-\frac{(p_1-q_1)^2}{4 \mu ^2}}-e^{-\frac{q_1^2}{4\mu^2}} \right)\left(e^{-\frac{(p_2-q_2)^2}{4 \mu ^2}}+e^{-\frac{q_2^2}{4\mu^2}} \right) \frac{1}{p_1} A_1^c(\vec p)  \nn\\
\qquad  + \left(e^{-\frac{(p_2-q_2)^2}{4 \mu ^2}}-e^{-\frac{q_2^2}{4\mu^2}} \right)\left(e^{-\frac{(p_1-q_1)^2}{4 \mu ^2}}+e^{-\frac{q_1^2}{4\mu^2}} \right) \frac{1}{p_2} A_2^c(\vec p)  \Bigg\}  \nn\\
\quad  
\times
\frac{1}{|\vec{q}|+|\vec{p}|+|\vec{q}-\vec{p}|} \Bigg\{  (\vec{p}-2\vec{q})\cdot  \vec{A}^c(-\vec{p})   -\frac{|\vec{q}-\vec{p}|-|\vec{q}|}{|\vec{p}|} (-\vec{p})\cdot\vec{A}^c(-\vec{p}) \nn\\
\qquad + \Bigg(\frac{ (-\vec{q}\times\vec{p})}{|\vec{p}||\vec{q}-\vec{p}|}  -\frac{(-\vec{p}\times\vec{q})}{|\vec{q}||\vec{p}|}  + \frac{(\vec{q}\times\vec{p})}{|\vec{q}||\vec{q}-\vec{p}|}   \Bigg) (-\vec{p})\times\vec{A}^c(-\vec{p})   \Bigg\} 
\nn\\
\nn\\
 =- {C_A\over 4\sqrt{\pi}}\mu \int_{\slashed{p}}  \vec{A}^c(-p)\cdot\vec{A}^c(p) - {C_A\over 8 \pi} \int_{\slashed{p}}   \frac{1}{|\vec{p}|}  (\vec{p}\times\vec{A}^c(-\vec{p})) (\vec{p}\times\vec{A}^c(\vec{p}))   + O\left(\mu^{-1}\right)
\,.
\eE
The third term of the right-hand side of Eq.~(\ref{F22mom}) reads
\bE{rCl}
\label{F2III}
&\int_{x,y}& \int_{\slashed{p},\slashed{q}}e^{-i{\vec p}\cdot{\vec x}}e^{-i{\vec q}\cdot{\vec y}}\delta_\mu(\vec x,\vec y) \Phi^{(2)}_{ab}(\vec x,\vec y)\frac{\delta^2 F_{GL}^{(0)} }{\delta A_i^a(\vec p) \delta A_i^b(\vec q)}
\nn\\
 &=& 2\frac{1}{4\pi}\int_{u.vx,w} \delta_\mu(\vec u,\vec v) \frac{1}{|\vec{x}-\vec{w}|} \p_{x_i}\delta(\vec x-\vec u) \p_{w_i}\delta(\vec w-\vec v) \delta^{ab} \nn\\
 && \Bigg\{  {1\over 2}f^{adc}f^{dbe}\int_{y,z}\Big( (G_1(\vec u;\vec z)-G_1(v_1,u_2;\vec z))(G_1(\vec z;\vec y)-G_1(v_1,u_2;\vec y))\nn\\
&& \qquad + (G_1(u_1,v_2;\vec z)-G_1(\vec v;\vec z))(G_1(\vec z;\vec y)-G_1(\vec v;\vec y))\Big)A_1^c(\vec z)A_1^e(\vec y)  \nn\\
&& + {1\over 2}f^{adc}f^{dbe}\int_{y,z}\Big( (G_2(v_1,u_2;\vec z)-G_2(\vec v;\vec z))(G_2(\vec z;\vec y)-G_2(\vec v;\vec y))\nn\\
&& \qquad+ (G_2(\vec u;\vec z)-G_2(u_1,v_2;\vec z))(G_2(\vec z;\vec y)-G_2(u_1,v_2;\vec y))\Big)A_2^c(\vec z)A_2^e(\vec y)  \nn\\
&& + {1\over 2}f^{adc}f^{dbe}\int_{y,z} (G_1(\vec u;\vec y)-G_1(v_1,u_2;\vec y))(G_2(v_1,u_2;\vec z)-G_2(\vec v;\vec z))A_1^c(\vec y) A_2^e(\vec z) \nn\\
&& + {1\over 2}f^{ade}f^{dbc}\int_{y,z} (G_2(\vec u;\vec z)-G_2(u_1,v_2;\vec z))(G_1(u_1,v_2;\vec y)-G_1(\vec v;\vec y))A_2^e(\vec z)A_1^c(\vec y) \Bigg\}  \qquad 
\nn
\\
&=& {C_A\over 8\sqrt{\pi }}\mu  \int_{\slashed{p}} \vec{A}^c(\vec p)\cdot\vec{A}^c(-\vec p) + O\left(\mu^{-1}\right)
\,.
\eE

Combining Eqs.~(\ref{F2I}), (\ref{F2II}) and (\ref{F2III}) we obtain 
\be
\int_{\slashed{p}}\delta_\mu(\vec p)|{\vec p}| \left(\vec{A}^a(\vec{p})\cdot \frac{\delta F_{GL}^{(2,2)}[\vec A]}{\delta \vec{A}^a(\vec{p})}\right) 
=
-\left(N+{1\over8}\right) {C_A\over 2 \pi} \int_{\slashed{p}}   \frac{1}{|\vec{p}|}  (\vec{p}\times\vec{A}^a(-\vec{p})) (\vec{p}\times\vec{A}^a(\vec{p})) \,.
\ee
Note that the divergent term has disappeared on the right-hand side so we can take the $\mu \rightarrow \infty$ limit. This equation can be solved using Eqs.~(\ref{density1}) and (\ref{density2}). We obtain
\bE{rCl}
F^{(2,2)}_{GL}[\vec A]  &=&   -\left(N+{1\over8}\right) {C_A\over 4 \pi} \int_{\slashed{p}}   \frac{1}{|\vec{p}|^2}  (\vec{p}\times\vec{A}^a(-\vec{p})) (\vec{p}\times\vec{A}^a(\vec{p})) \label{F22GL} \,.
\eE
This concludes the computation of the wave functional with ${\cal O}(e^2)$ precision. The complete result is summarized in Eqs.~(\ref{FGL0p}), (\ref{FGL1}), (\ref{F24GL2}) and (\ref{F22GL}). Note that the result is different from the one obtained in Sec.~\ref{sec:Comp:Hatfield}. The reason is that the 
prefactor of $F^{(2,2)}_{GL}$ has changed from Eq.~(\ref{FGL22}) to Eq.~(\ref{F22GL}): $N \rightarrow N+1/8$. This highlights the importance of doing the regularization of the theory from the very beginning. The existence of very lengthy and complicated expressions in the intermediate steps impedes in practice the identification of the divergences. 
Therefore, one could easily miss some contributions (and yet get a finite result) if formally manipulating the integrals as if they were finite before regularizing them.

\section{Determination of $\Psi_{GI}[J]$}
\label{sec:KKN}
In Sec.~\ref{sec:KNY} we reformulated the approximation scheme worked out in Ref.~\cite{Karabali:2009rg} to provide with a systematic expansion of the weak coupling limit. This method uses a change of field variables to the gauge invariant variables $J$, 
 which has the great advantage that the Gauss law constraint is trivially satisfied. 
%

\subsection{Regularizing the kinetic term}
\label{sec:RegulatingKNY}
One important consequence of this approach is that, since the vacuum wave functional is gauge invariant, it only depends on $J$. It is also 
possible to obtain an explicit and compact expression for the Hamiltonian in terms of $J$ fields. This was done
in Refs.~\cite{Karabali:1995ps,Karabali:1996je,Karabali:1996iu, Karabali:1997wk,Karabali:1998yq,Karabali:2009rg}, starting with a regularized Hamiltonian.
Interestingly enough, the regularization of the kinetic operator produced a finite extra term in the Hamiltonian. Yet, the expression found in those references will prove to be insufficient for our purposes. Therefore, since the regularization is an important point for us, we will rederive the Hamiltonian in terms of the $J$ fields. 
In several aspects the derivation will be identical to the one carried out in Refs.~\cite{Karabali:1995ps,Karabali:1996je,Karabali:1996iu, Karabali:1997wk,Karabali:1998yq,Karabali:2009rg},  but we will see that we need to consider some extra terms. 
Our starting point is the regularized kinetic operator  $\mathcal{T}_{reg}$ defined in Eq.~(\ref{Treg}). 
We then write the kinetic operator in terms of holomorphic and anti-holomorphic gauge fields\footnote{In Refs.~\cite{Karabali:1995ps,Karabali:1996je,Karabali:1996iu, Karabali:1997wk,Karabali:1998yq,Karabali:2009rg} the second term of Eq.~(\ref{TAbarA}) is not incorporated, but trivially considered to be equal to the first term. Yet, we find it illustrative to show their equality, as it is not evident from the actual computation after the change of variables.}:
\bE{l}
\label{TAbarA}
\T_{\mathrm{reg}}  = -\frac{1}{4}\int_{x,v}\delta_\mu(\vec x, \vec v) \Phi_{ab}(\vec x,\vec v)  \left(\frac{\delta}{\delta \bar{A}^a(\vec x)}\frac{\delta}{\delta A^b(\vec v)} + \frac{\delta}{\delta A^a(\vec x)} \frac{\delta}{\delta \bar{A}^b(\vec v)}\right) \,,
\eE
and transform it to $J$ variables. The functional derivatives of the first term can be rewritten in the following way
\bE{rCl}
&&
\frac{\delta}{\delta \bar{A}^a(\vec x)} \frac{\delta}{\delta A^b(\vec v)}  = \int_{y,z} \left[\frac{\delta J^d(\vec z)}{\delta \bar{A}^a(\vec x)} \frac{\delta}{\delta J^d(\vec z)} +\frac{\delta \bar{A}^d(\vec z)}{\delta \bar{A}^a(\vec x)} \frac{\delta}{\delta \bar{A}^d(\vec z)}  \right]    \left[\frac{\delta J^c(\vec y)}{\delta A^b(\vec v)} \frac{\delta}{\delta J^c(\vec y)} +\frac{\delta \bar{A}^c(\vec y)}{\delta A^b(\vec v)} \frac{\delta}{\delta \bar{A}^c(\vec y)}  \right] 
\nn\\
&&\\
&&\quad= \int_{y,z} \left[-2iM^\dagger_{dh}(\vec z)\left(D_z^{he}\left(\bar{D}^{-1}\right)^{ea}_{zx}\right) \frac{\delta}{\delta J^d(\vec z)} +\delta(\vec x-\vec z) \frac{\delta}{\delta \bar{A}^a(\vec z)}  \right]    \left[2iM^\dagger_{cb}(\vec y)\delta(\vec y-\vec v) \frac{\delta}{\delta J^c(\vec y)}\right]\,.\nn
\eE
using the equalities of Sec.~\ref{sec:KNY}. Accordingly, we find
\bE{rCl}
&&
\hspace{-0.5cm} \Phi_{ab}(\vec x,\vec v) \frac{\delta}{\delta \bar{A}^a(\vec x)} \frac{\delta}{\delta A^b(\vec v)}  \nn\\
&=& 2i  \Phi_{ab}(\vec x,\vec v) \frac{\delta M^\dagger_{cb}(\vec v)}{\delta \bar{A}^a(\vec x)} \frac{\delta}{\delta J^c(\vec v)}     \nn\\
&& +  4\int_{z}  \Phi_{ab}(\vec x,\vec v) \left[\left(\p_z \bar{G}(z-x)\right)M^\dagger_{da}(\vec x) + \frac{ie}{2}\bar{G}(z-x) f^{edf} J^e(\vec z)M^{\dagger}_{fa}(\vec x) \right]  M^{\dagger}_{cb}(\vec v) \frac{\delta^2}{\delta J^d(\vec z) \delta J^c(\vec v)}   
\nn\\
&& +2i   \Phi_{ab}(\vec x,\vec v) M^\dagger_{cb}(\vec v)\frac{\delta^2}{\delta \bar{A}^a(\vec x) \delta J^c(\vec v)}
\,. \label{PhiAbarA}
\eE
The last term is proportional to the Gauss law operator $I^a = i\bar{D}^{ab}\frac{\delta}{\delta \bar{A}^b}= iM^{\dagger-1}_{ad}\bar{\p}\left(M^\dagger_{db}\frac{\delta}{\delta \bar{A}^b}\right)$ (see Sec.~\ref{sec:KNY}), which vanishes on physical wave functionals. For the other two terms we have to take care of the regularization. 
Using Eqs.~(\ref{Mf}) and (\ref{MdaggerOverAbar}) we can rewrite the first term of Eq.~(\ref{PhiAbarA}) in the following way 
\bE{rCl}
2i \Phi_{ab}(\vec x,\vec v) \frac{\delta M^\dagger_{cb}(\vec v)}{\delta \bar{A}^a(\vec x)}
&=& 2ie \Phi_{ab}(\vec x,\vec v) \frac{1}{\pi(v-x)} M^{\dagger-1}_{bd}(\vec v) f^{dch} M^{\dagger-1}_{ah}(\vec x) \label{splitdMdA} \\
&=:& 2ie V_{hd}(\vec x,\vec v) \frac{1}{\pi(v-x)} f^{dch}\,, \label{1stTermV}
\eE
where we defined
\be
V^{dc}(\vec x,\vec v) := M_{da}^\dagger(\vec x)\Phi^{ab}(\vec x,\vec v)M^{\dagger-1}_{bc}(\vec v)\,.
\ee
We now turn to the second term of the regularized kinetic operator, Eq.~(\ref{TAbarA}):
\bE{rCl}
&&\hspace{-0.5cm} \Phi_{ab}(\vec x,\vec v) \frac{\delta}{\delta A^a(\vec x)} \frac{\delta}{\delta \bar{A}^b(\vec v)} \nn\\
&=& \int_{y,z} \Phi_{ab}(\vec x,\vec v)  \left[2iM^\dagger_{ca}(\vec y)\delta(\vec y-\vec x) \frac{\delta}{\delta J^c(\vec y)}\right] \nn\\
&&\quad \left[-2iM^\dagger_{dh}(\vec z)\left(D_z^{he}\left(\bar{D}^{-1}\right)^{eb}_{zv}\right) \frac{\delta}{\delta J^d(\vec z)} +\delta(\vec v-\vec z) \frac{\delta}{\delta \bar{A}^b(\vec z)}  \right]  \\
&=& 2i \Phi_{ab}(\vec x,\vec v) M^\dagger_{ca}(\vec x) \frac{\delta^2}{\delta J^c(\vec x) \delta \bar{A}^b(\vec v)}  \\
&&\quad + 4 \int_z \left[  \left(\p_z \bar{G}(z-v)\right) V^{cd}(\vec x,\vec v) + \frac{ie}{2}\bar{G}(z-v) f^{edf} J^e(\vec z) V^{cf}(\vec x,\vec v) \right] \frac{\delta^2}{\delta J^c(\vec x)\delta J^d(\vec z)}   \nn\\
&&\quad + 4\Phi_{ab}(\vec x,\vec v)  \int_z M^\dagger_{ca}(\vec x)  \frac{\delta}{\delta J^c(\vec x)}    \left[  \left(\p_z \bar{G}(z-v)\right)M^\dagger_{db}(\vec v) \right. \nn\\
&&\qquad\qquad\qquad\qquad\qquad\qquad\qquad\qquad \left. + \frac{ie}{2}\bar{G}(z-v) f^{edf} J^e(\vec z)M^{\dagger}_{fb}(\vec v) \right] \frac{\delta}{\delta J^d(\vec z)}\,.  \nn
\eE
Again, the first term is proportional to the Gauss law operator $I^a$. After renaming $v\leftrightarrow x$ (which can be done under the integral) and using $V^{ba}(v,x)=V^{ab}(x,v)$ the second term is identical to the second term of Eq.~(\ref{PhiAbarA}). The third term, after application of the functional derivative, reduces to
\bE{l}
2ie\Phi_{ab}(\vec x,\vec v) M^\dagger_{ca}(\vec x)  \bar{G}(x-v) f^{cdf} M^{\dagger}_{fb}(\vec v)  \frac{\delta}{\delta J^d(\vec x)} \,.    
\eE
Since $\bar{G}(-x)=-\bar{G}(x)$, this expression is identical to Eq.~(\ref{splitdMdA}). 

Therefore, we find that both subterms of Eq.~(\ref{TAbarA}) are equal. Summing them up and multiplying by $\left(-{1\over4}\right)$ we obtain
the completely regularized kinetic term to all orders in perturbation theory
\bE{rCl}
\label{TregKKNallorders}
\T_\mathrm{reg}
&=& -2  \int_{x,v,z} \delta_{\mu}({\vec x},{\vec v}) \left((\p_z \delta^{df} + \frac{ie}{2} f^{dfa} J^a(\vec z)){\bar G}(z-x)\right)V_{fc}({\vec x},{\vec v})
{\delta \over \delta J^d (\vec z)} {\delta \over \delta J^c (\vec v)}  \nn\\
&&   -ie\int_{x,v} \delta_{\mu}({\vec x},{\vec v})V_{hd}({\vec x},{\vec v})f^{dch}{\bar G}(v-x){\delta \over \delta J^c (\vec v)}\,,
\eE
This is a pure function of $J$, since $V_{dc}(\vec x,\vec v)$ is a gauge invariant object, which makes it possible to rewrite it completely
 in terms of $J$.
The easiest way to proceed is to first consider an infinitesimal path with small ${\vec v}-{\vec x}$. 
By Taylor expansion one finds 
\bE{rCl}
V_{dc}({\vec x},{\vec v}) &=& \de_{dc} - (v-x){e\over2} J_{dc}(\vec x) +{\cal O}(|\vec x -\vec v|^2)\,,
\eE
where we used $J_{dc}=-if^{dce}J^e$. 
By composition of these infinitesimal paths we obtain
\be
V_{dc}({\vec x},{\vec v})=\left( \mathcal{P} e^{{e\over2}\int_{\mathcal{C}}d z J(\vec z)}\right)_{dc} \,.
\ee
Note that the integration is over the holomorphic component only. $V_{dc}(\vec x,\vec v)$ depends on the path, though physical results should not. 
For illustration, we show the explicit expression for small $|\vec x -\vec v|$ for the specific combination of paths that we consider in this chapter:
\bea
\nn
V_{dc}({\vec x},{\vec v})&=&\delta_{dc} +
 {e\over2}\left[
 (x-v) J_{dc}(\vec v)+\frac{(x-v)^2}{2}\partial J_{dc}(\vec v)+\frac{(x-v)(\bar x-\bar v)}{2}\bar \partial J_{dc}(\vec v)
 \right]
\\
&& 
+ {e^2\over4}\frac{(x-v)^2}{2} (J(\vec v)J(\vec v))_{dc}+
{\cal O}(|\vec x -\vec v|^3)
\,.
\label{Ve2}
\eea
The ${\cal O}(e|\vec x -\vec v|)$ and ${\cal O}(e^2|\vec x -\vec v|^2)$ terms are path independent but not the ${\cal O}(e|\vec x -\vec v|^2)$ terms.

\medskip

The kinetic operator $\T_\mathrm{reg}$ admits a Taylor expansion in powers of $e$. We are only interested in keeping the terms that may contribute to the 
wave functional to $\mathcal{O} (e^2)$. We first consider the second term of Eq.~(\ref{TregKKNallorders}). Inserting Eq.~(\ref{Ve2}) in Eq.~(\ref{1stTermV}) we find 
\bE{rCl}
\label{premassterm}
2i \Phi_{ab}(\vec x,\vec v) \frac{\delta M^\dagger_{cb}(\vec v)}{\delta \bar{A}^a(\vec x)}
&=& -\frac{e^2C_A}{\pi} J^c(\vec x) +{\cal O}(e^2|\vec x -\vec v|,e^3|\vec x -\vec v|) \,. \label{1stTermJ}
\eE
Note that regularization is crucial for obtaining a finite contribution, as the leading term from the Wilson line
(proportional to $\delta_{ab}$) vanishes.
Therefore, 
the integration of the regularized delta function times Eq.~(\ref{premassterm}) over $v$ gives
\bE{rCl}
\label{massterm}
-\frac{2}{4}\int_{x,v} \delta_\mu(\vec x, \vec v) 2i \Phi_{ab}(\vec x,\vec v) \frac{\delta M^\dagger_{cb}(\vec v)}{\delta \bar{A}^a(\vec x)} \frac{\delta}{\delta J^c(\vec v)}
&=&  \frac{e^2C_A}{2 \pi}\int_x J^c(\vec x) \frac{\delta}{\delta J^c(\vec x)}+{\cal O}(e^2/\mu,e^3/\mu) \,.
\eE
This contribution to the kinetic operator has been generated by the regularization of the theory, i.e.~it is an effect produced by the high-energy modes. 
It was first obtained in Ref.~\cite{Karabali:1996je}, and it has a nice interpretation in terms of an anomaly-like computation. This term 
has played a major role in the strong coupling analysis carried out in Refs.~\cite{Karabali:1995ps,Karabali:1996je,Karabali:1996iu, Karabali:1997wk,Karabali:1998yq,Karabali:2009rg}, 
where it is argued to be responsible for generating the mass gap. 
Yet, we would like to remark, as is clear from the 
analysis above, that this contribution is obtained from a pure perturbative computation (as anomaly-like effects are anyway), arising from a Taylor 
expansion in powers of $e$. The corrections to this expression are $1/\mu$ suppressed, irrespectively of the power of $e$ (but starting at ${\cal O}(e^2)$). In general we may worry that such $1/\mu$ suppression may be compensated by divergences when applied to the wave functional. This is not the case for this term, as there is a complete factorization between the momentum of the internal loop and the momentum of the fields that will act on the wave functional. Therefore, we will not consider these vanishing contributions explicitly any further (even though they are formally of ${\cal O}(e^2)$). 

\medskip

We now move to the first term of Eq.~(\ref{TregKKNallorders}). The expansion of $V$ around $v=x$ yields
 \bE{rCl}
 \label{OmReg} 
&& -2\int_{z}  \Bigg[\left(\p_z \bar{G}(z-x)\right)\de^{dc} + \frac{ie}{2}\bar{G}(z-x) f^{dce} J^e(\vec z) \\
&&\qquad\qquad +\frac{ie}{2}(v-x) \left(\p_z \bar{G}(z-x)\right)f^{dce}J^e(\vec x) \nn\\
&&\qquad\qquad -  \frac{e^2}{8}f^{dea}f^{ecb} J^b(\vec x)  \Big( (v-x)^2\left(\p_z \bar{G}(z-x)\right) J^a(\vec x) \nn\\
&&\qquad\qquad\qquad\qquad\qquad\qquad\qquad +2(v-x)\bar{G}(z-x) J^a(\vec z) \Big)  \Bigg] \frac{\delta^2}{\delta J^d(\vec z) \delta J^c(\vec v)}\,.\qquad 
\nn
\eE
The third and fourth term are of ${\cal O}(e|\vec x -\vec v|)$ and ${\cal O}(e^2|\vec x -\vec v|^2)$ respectively, but when applied to a functional they can give finite contributions.
We have not included \linebreak ${\cal O}(e|\vec x -\vec v|^2)$ terms in this expansion. In principle they may 
contribute to the wave functional at ${\cal O}(e^2)$. Nevertheless, as we will see in the following, only the ${\cal O}(e|\vec x -\vec v|)$ terms give 
finite contributions at ${\cal O}(e^2)$. Therefore, the ${\cal O}(e|\vec x -\vec v|^2)$ terms would give, at most, ${\cal O}(e^2/\mu)$ corrections to the wave functional. 
In order to maintain the expressions in a manageable way, we will neglect them in the following.

\medskip

After this discussion we can approximate the kinetic operator by an expression suitable to obtain the wave functional with ${\cal O}(e^2)$ accuracy:
\bE{rCl}
\label{TregKKN}
\T_\mathrm{reg}
&=& \frac{e^2C_A}{2\pi}  \int_x J^a (\vec x) {\delta \over \delta J^a (\vec x)}  + {2\over \pi} \int _{x,y} 
 {1\over (y-x)^2} {\delta \over \delta J^a (\vec x)} {\delta \over \delta J^a (\vec y)}  \nn\\
&&\quad  + i e \int_{x,y} f^{abc} {J^c(\vec x) \over \pi (y-x)} {\delta \over \delta J^a (\vec x)} {\delta \over \delta J^b (\vec y)}  \nn\\
&&\quad   +\int_{x,v,y} \delta_\mu(\vec x, \vec v)  \Bigg[ie\, (x-v) \left(\p_y \bar{G}(y-v)\right)f^{abe} J^e(\vec v) \nn\\
&&\quad\qquad\qquad + \frac{e^2}{4}f^{ace}f^{bed} J^c(\vec v) \Big( (x-v)^2\left(\p_y \bar{G}(y-v)\right) J^d(\vec v)  \nn\\
&&\quad\qquad\qquad\qquad\qquad\qquad\qquad\qquad  +2 (x-v)\bar{G}(y-v) J^d(\vec y) \Big)   \Bigg] \frac{\delta^2}{\delta J^a(\vec x)\delta J^b(\vec y)}  \nn\\
 &&\quad -\int_{y,z} \bar{G}(y-z) M^\dagger_{ca}(\vec y) {\delta \over \delta J^c (\vec z)} I^a(\vec y) +{\cal O}(e^3,1/\mu) \\
&=:& \int_x\omega(\vec x)^a {\delta \over \delta J^a (\vec x)} + \int_{x,v,y}\tilde{\Omega}_{ab}^\mathrm{reg}(\vec x,\vec v,\vec y)\frac{\delta^2}{\delta J^a(\vec x) \delta J^b(\vec y)} +{\cal O}(e^3,1/\mu) \label{omega} \\
&=:& \int_x\omega(\vec x)^a {\delta \over \delta J^a (\vec x)} \nn\\
&&+  \int_{x,y,(v)}\left(\Omega_{ab}^{(0)}(\vec x,\vec y) +e\Omega_{ab}^{(1)}(\vec x,\vec y) +e\tilde{\Omega}_{ab}^{(1)}(\vec x,\vec v,\vec y) +e^2\tilde{\Omega}_{ab}^{(2)}(\vec x,\vec v,\vec y) \right) \frac{\delta^2}{\delta J^a(\vec x) \delta J^b(\vec y)} \nn\\
&&+{\cal O}(e^3,1/\mu)  \,, \qquad \label{omega2}
\eE
where we dropped the term proportional to the Gauss law operator in the last two equalities, and we defined $\Omega_{ab}^{(0)}(\vec x,\vec y)$ and $\Omega_{ab}^{(1)}(\vec x,\vec y)$ as the coefficients of the second and the third term of Eq.~(\ref{TregKKN}), respectively, while $\tilde{\Omega}_{ab}^{(1)}(\vec x,\vec v,\vec y)$ is the coefficient of the third line and $\tilde{\Omega}_{ab}^{(2)}(\vec x,\vec v,\vec y)$ is the coefficient of the fourth and fifth line.

Eq.~(\ref{TregKKN}) is different from the expression used in Ref.~\cite{Karabali:2009rg} (given in Eq.~(\ref{HamiltonianNair})). They only coincide when we take the limit $\mu \rightarrow \infty$. 
In which case they agree to any order in perturbation theory. 
Nevertheless, as we will see, this is not enough for our purposes, since we will also have to keep some subleading terms in $1/\mu$.

\subsection{Solving the \SE}

Once we have obtained the regularized kinetic operator we can compute $\Psi_{GI}[J]$. 
After changing to the $J$ variables Eq. (\ref{SE}) reads in our case
\be
\V-\int_{x}
\omega^a(\vec x)\frac{\delta F_{GI}}{\delta J^a(\vec x)}-
\int_{x,v,y}\tilde{\Omega}^\mathrm{reg}_{ab}(\vec x,\vec v,\vec y)\frac{\delta^2 F_{GI}}{\delta J^a(\vec x) \delta J^b(\vec y)}  + \int_{x,v,y}\tilde{\Omega}^\mathrm{reg}_{ab}(\vec x,\vec v,\vec y)\frac{\delta F_{GI}}{\delta J^a(\vec x)} \frac{\delta F_{GI}}{\delta J^b(\vec y)} = 0 \,, \label{SEJ}
\ee
where
\be
\V=\frac{1}{2}\int_x{\bar \partial}J^a(\vec x){\bar \partial}J^a(\vec x)
\,,
\ee
and $\omega^a(\vec x)$ and $\tilde{\Omega}^\mathrm{reg}_{ab}(\vec x,\vec v,\vec y)$ are defined in Eq.~(\ref{omega}).
As before, we expand the exponent of the vacuum wave functional in powers of the coupling constant
\be
F_{GI}=F_{GI}^{(0)}+eF_{GI}^{(1)}+e^2F_{GI}^{(2)}+{\cal O}(e^3)\,,
\ee
and separate the \SE $ $ order by order in the coupling constant.

At ${\cal O}(e^0)$ we have
\bE{l}
\int_{x,y}\Omega^{(0)}_{ab}(\vec x,\vec y) \left(\frac{\delta^2 F_{GI}^{(0)}}{\delta J^a(\vec x) \delta J^b(\vec y)} - \frac{\delta F_{GI}^{(0)}}{\delta J^a(\vec x)} \frac{\delta F_{GI}^{(0)}}{\delta J^b(\vec y)} \right)= {1\over 2} \int_z  \bar{\p} J^a(\vec z) \bar{\p} J^a(\vec z)  \,. 
\eE 
This, as before, is the unregularized lowest order \SE. Its solution is the leading order computed in Sec.~\ref{sec:KNY} (see Eq.~(\ref{rec7})). It corresponds to the weak coupling limit of the leading order of Ref.~\cite{Karabali:2009rg}:
\bE{rCl}
\label{FGI0}
F^{(0)}_{GI} &=&{1\over2} \int_\slashed{k} \frac{\bar{k}^2}{E_k} J^a(\vec{k}) J^a(-\vec{k})   = {1\over2} \int_\slashed{k}\frac{1}{E_k} (\vec{k}\times\vec{A}^a(\vec{k})) (\vec{k}\times\vec{A}^a(-\vec{k}))+ {\mathcal O}(e) \\
  &=& F_{GL}^{(0)}[{\vec A}] + {\mathcal O}(e) \,, \qquad\nn 
\eE
 where $E_k \equiv |\vec k|$.

At ${\cal O}(e)$ we have
\bE{l}
-\int_{x,y}\Omega^{(0)}_{ab}(\vec x,\vec y) \left(\frac{\delta^2 F_{GI}^{(1)}}{\delta J^a(\vec x) \delta J^b(\vec y)} - 2 \frac{\delta F_{GI}^{(0)}}{\delta J^a(\vec x)} \frac{\delta F_{GI}^{(1)}}{\delta J^b(\vec y)} \right)\nn\\
\quad -\int_{x,y} \Omega^{(1)}_{ab}(\vec x,\vec y)\left(\frac{\delta^2 F_{GI}^{(0)}}{\delta J^a(\vec x) \delta J^b(\vec y)} - \frac{\delta F_{GI}^{(0)}}{\delta J^a(\vec x)} \frac{\delta F_{GI}^{(0)}}{\delta J^b(\vec y)} \right) \nn\\
\quad -\int_{x,v,y}\tilde{\Omega}^{(1)}_{ab}(\vec x,\vec v,\vec y) \left(\frac{\delta^2 F_{GI}^{(0)}}{\delta J^a(\vec x) \delta J^b(\vec y)} - \frac{\delta F_{GI}^{(0)}}{\delta J^a(\vec x)} \frac{\delta F_{GI}^{(0)}}{\delta J^b(\vec y)} \right) = 0 \,. \label{regF1GIeq}
\eE 
The first term of the last line vanishes under contraction of the color indices. The second term is of $O(\mu^{-2})$ (see App.~\ref{subsec:append1}). So, as for the leading order, this equation reduces to the unregularized version of Sec.~\ref{sec:KNY}. Thus, its solution is Eq.~(\ref{rec50}), which also corresponds to the ${\cal O}(e)$ weak coupling limit of the solution shown in 
Ref.~\cite{Karabali:2009rg}:
\be
F^{(1)}_{GI}=-\frac{1}{4}\int_{\slashed{k_1},\slashed{k_2},\slashed{k_3}} \frac{f^{a_1 a_2 a_3}}{24} \sd (\vec k_1+\vec k_2+\vec k_3)\  g^{(3)}(\vec k_1,\vec k_2,\vec k_3) J^{a_1}(\vec k_1)J^{a_2}(\vec k_2)J^{a_3}(\vec k_3)\,, \label{FGI1}
\ee
where
\be
g^{(3)}(\vec k_1,\vec k_2,\vec k_3) = \frac{16}{E_{k_1}\! + E_{k_2}\! + E_{k_3}}\left \{ \frac{\bar k_1 \bar k_2 (\bar k_1 - \bar k_2)}{E_{k_1} E_{k_2}} + {cycl.\ perm.} \right \}
\,. 
\ee

At ${\cal O}(e^2)$ we determine $F_{GI}^{(2)}$. As in the previous section, $F^{(2)}_{GI}$ can have contributions with four, two and zero $J$'s:
$F^{(2)}_{GI}=F^{(2,4)}_{GI}+F^{(2,2)}_{GI}+F^{(2,0)}_{GI}$. Again, there is no need to compute $F^{(2,0)}_{GI}$, as it only changes the normalization of the state, which we do not fix, or alternatively can be absorbed in a redefinition of the ground-state energy. $F^{(2,4)}_{GI}$ is determined by the following equation (where $\Omega^{(1)}_{ab}(\vec x,\vec y)$ and $\tilde{\Omega}^{(1)}_{ab}(\vec x,\vec v,\vec y)$ should be understood in a symmetrized way):
\bE{l}
 \int_{x,y}\Omega^{(0)}_{ab}(\vec x,\vec y)\left(\frac{\delta F_{GI}^{(1)}}{\delta J^a(\vec x)} \frac{\delta F_{GI}^{(1)}}{\delta J^b(\vec y)}  + 2\frac{\delta F_{GI}^{(0)}}{\delta J^a(\vec x)} \frac{\delta F_{GI}^{(2,4)}}{\delta J^b(\vec y)} \right)  +2\int_{x,y}\Omega^{(1)}_{ab}(\vec x,\vec y)\frac{\delta F_{GI}^{(0)}}{\delta J^a(\vec x)} \frac{\delta F_{GI}^{(1)}}{\delta J^b(\vec y)}  \nn\\
\quad  +2\int_{x,v,y}\tilde{\Omega}^{(1)}_{ab}(\vec x,\vec v,\vec y)\frac{\delta F_{GI}^{(0)}}{\delta J^a(\vec x)} \frac{\delta F_{GI}^{(1)}}{\delta J^b(\vec y)}  +\int_{x,v,y}\tilde{\Omega}^{(2)}_{ab}(\vec x,\vec v,\vec y)\frac{\delta F_{GI}^{(0)}}{\delta J^a(\vec x)} \frac{\delta F_{GI}^{(0)}}{\delta J^b(\vec y)}  = 0 \,, \qquad \label{regF24GIeq}
\eE 
The last line vanishes for $\mu\to\infty$ (see App.~\ref{subsec:append2}), and again the equation reduces to the unregularized equation with the solution
\bE{rCl}
\label{FGI24}
F^{(2,4)}_{GI}&=-\frac{1}{8}\int_{\slashed{k_1},\slashed{k_2},\slashed{q_1},\slashed{q_2}} \frac{f^{a_1 a_2 c} f^{b_1 b_2 c}}{64} & \sd (\vec k_1+\vec k_2+\vec q_1+\vec q_2)  g^{(4)}(\vec k_1, \vec k_2; \vec q_1, \vec q_2) \nn\\
&&\times J^{a_1}(\vec k_1)J^{a_2}(\vec k_2)J^{b_1}(\vec q_1)J^{b_2}(\vec q_2)\,,
\eE
where
\begin{equation}
\begin{array}{cl}
g^{(4)}(\vec k_1, \vec k_2; \vec q_1, \vec q_2)& =\ \vspace{.2in} \displaystyle \frac{1}{E_{k_1}\! + E_{k_2}\! + E_{q_1}\! + E_{q_2}} \\
\vspace{.2in}
&\!\!\!\!\displaystyle \left \{ g^{(3)}(\vec k_1, \vec k_2, -\vec k_1-\vec k_2)\ \frac{k_1 + k_2}{\bar k_1 +\bar k_2}\ g^{(3)}(\vec q_1, \vec q_2, -\vec q_1-\vec q_2) \right . \\
\vspace{.2in}
&\displaystyle -  \left [ \frac{(2\bar k_1 + \bar k_2)\,\bar k_1}{E_{k_1}} - \frac{(2\bar k_2 + \bar k_1)\,\bar k_2}{ E_{k_2}}\right ]\frac{4}{\bar k_1+\bar k_2}\  g^{(3)}(\vec q_1, \vec q_2, -\vec q_1-\vec q_2) \\
&\displaystyle -  \left .   g^{(3)}(\vec k_1, \vec k_2, -\vec k_1-\vec k_2)\ \frac{4}{\bar q_1+\bar q_2}\left [ \frac{(2\bar q_1 + \bar q_2)\,\bar q_1}{ E_{q_1}} - \frac{(2\bar q_2 + \bar q_1)\,\bar q_2}{ E_{q_2}}\right ] \right\} \,.
\\
\end{array}
\end{equation}
Again, this term corresponds to the weak coupling limit of the the analogous expression in Ref.~\cite{Karabali:2009rg}, and to the expression already found in Chap.~\ref{chap:Comparison}.

\bigskip

So far the regularization of the kinetic term has not produced any modification of the results obtained in Sec.~\ref{sec:KNY}. 
The reason is the same as in the previous section, in the sense that, so far, all computations we did were tree-level-like. ``Loop" effects (sensitive to the hard modes) are hidden in $F^{(2,2)}_{GI}$, where we have a kind of contraction of two fields. We compute this term in the 
next subsection.

\newpage 
 \subsection{$F^{(2,2)}_{GI}$}
\label{subsec:F22GI}
 $F^{(2,2)}_{GI}$ is determined by the following equation
\bE{l}
-{C_A\over2\pi}\int_x J^a(\vec x)\frac{\delta F_{GI}^{(0)}}{\delta J^a(\vec x)} - \int_{x,y}\Omega^{(0)}_{ab}(\vec x,\vec y)\left(\frac{\delta^2 F_{GI}^{(2,4)}}{\delta J^a(\vec x) \delta J^b(\vec y)}  - 2\frac{\delta F_{GI}^{(0)}}{\delta J^a(\vec x)} \frac{\delta F_{GI}^{(2,2)}}{\delta J^b(\vec y)} \right) \nn\\
\quad  -\int_{x,y}\Omega^{(1)}_{ab}(\vec x,\vec y) \frac{\delta^2 F_{GI}^{(1)}}{\delta J^a(\vec x) \delta J^b(\vec y)}  \nn\\
\quad  -\int_{x,v,y}\tilde{\Omega}^{(1)}_{ab}(\vec x,\vec v,\vec y)\frac{\delta^2 F_{GI}^{(1)}}{\delta J^a(\vec x) \delta J^b(\vec y)}  -\int_{x,v,y}\tilde{\Omega}^{(2)}_{ab}(\vec x,\vec v,\vec y)\frac{\delta^2 F_{GI}^{(0)}}{\delta J^a(\vec x) \delta J^b(\vec y)}  = 0 \,.
\label{SEF22}
\eE 
The last term vanishes in the $\mu\to\infty$ limit (see App.~\ref{subsec:append3}), the next-to-last term, however, does not. 
With Eqs.~(\ref{omega2}) and (\ref{FGI1}) we find
\bE{rl}
 &\hspace{-1cm} \int_{x,v,y}\tilde{\Omega}^{(1)}_{ab}(\vec x,\vec v,\vec y)\frac{\delta^2 F_{GI}^{(1)}}{\delta J^a(\vec x) \delta J^b(\vec y)}   \nn\\
= & 3{C_A\over48\mu^2} \int_{\slashed{k},\slashed{p}} \frac{p(\bar{k}+\bar{p})}{\bar{p}} e^{-\frac{(\vec{k}+\vec{p})^2}{4\mu^2}} g^{(3)}(\vec p,\vec k,-\vec k-\vec p)  J^a(-\vec k)J^a(\vec k)\,.
\eE
In order to compute the loop integral over the internal ${\vec p}$ momentum, we again factorize the modes according to the two scales of the problem: $|{\vec p}| \sim \mu$ and $|{\vec p}| \sim |{\vec k}|$.
The integral is dominated by $|{\vec p}| \sim \mu$, while the $|{\vec p}| \sim |{\vec k}|$ region gives subleading contributions.
Overall we obtain (here $\alpha$ is the angular component of $\vec{k}$, such that $\bar{k}={1\over2}|\vec{k}|e^{-i\alpha}$)
\bea
\label{new}
&& \hspace{-1cm}
\int_{x,v,y}\tilde{\Omega}^{(1)}_{ab}(\vec x,\vec v,\vec y)\frac{\delta^2 F_{GI}^{(1)}}{\delta J^a(\vec x) \delta J^b(\vec y)}  
\\
\nn
&=&
{C_A\over16\mu^2(2\pi)^2} \int_{\slashed{k}} \Big(-7 e^{-2 i \alpha } |\vec{k}| \pi  \mu ^2 + \frac{9}{4} e^{-2 i \alpha } |\vec{k}|^2 \pi ^{3/2} \mu \Big) 
J^a(-\vec k)J^a(\vec k) +{\cal O}(1/\mu^{2})
 \\
\nn
 &=&
 -{7\over8}{C_A\over2\pi} \int_{\slashed{k}} \frac{\bar{k}^2}{|\vec{k}|}  J^a(-\vec k)J^a(\vec k) +{\cal O}(1/\mu)\,.
\eea

We now have all the ingredients to determine $f^{(2,2)}_{a_1a_2}(k)$ from Eq.~(\ref{SEF22}), which now reads
\bE{l}
{C_A\over2\pi}\int_x J^a(\vec x)\frac{\delta F_{GI}^{(0)}}{\delta J^a(\vec x)} + \int_{x,y}\Omega^{(0)}_{ab}(\vec x,\vec y)\left(\frac{\delta^2 F_{GI}^{(2,4)}}{\delta J^a(\vec x) \delta J^b(\vec y)}  - 2\frac{\delta F_{GI}^{(0)}}{\delta J^a(\vec x)} \frac{\delta F_{GI}^{(2,2)}}{\delta J^b(\vec y)} \right) \nn\\
\quad  +\int_{x,y}\Omega^{(1)}_{ab}(\vec x,\vec y) \frac{\delta^2 F_{GI}^{(1)}}{\delta J^a(\vec x) \delta J^b(\vec y)}  -{7\over8}{C_A\over2\pi} \int_{\slashed{k}} \frac{\bar{k}^2}{|\vec{k}|}  J^a(-\vec k)J^a(\vec k)   = 0  \qquad
\eE 
\bE{rCl}
\Longleftrightarrow 2\int_{\slashed{k}} |\vec{k}|  f^{(2,2)}_{a_1a_2}(\vec k) J^{a_1}(-\vec k)J^{a_2}(\vec k) &=& -\frac{C_A}{32} \int_{\slashed{k},\slashed{p}}\frac{p}{\bar{p}}  g^{(4)}(\vec k,\vec p,-\vec k,-\vec p)  J^a(-\vec k)J^a(\vec k)  \nn\\
&&  -\frac{C_A}{16} \int_{\slashed{k},\slashed{p}} \frac{1}{\bar{p}} g^{(3)}(\vec k,\vec p,-\vec p-\vec k) J^a(-\vec k)J^a(\vec k)  \nn\\
&&  -\left(1-{7\over8}\right){C_A\over2\pi} \int_{\slashed{k}} \frac{\bar{k}^2}{|\vec{k}|} J^a(-\vec k)J^a(\vec k)\,,   \qquad
\eE 
and it is solved by
\bE{rCl}
 f^{(2,2)}_{a_1a_2}(\vec k) &=&- {C_A\over4\pi} \Bigg( N  +{1\over8}\Bigg) \frac{\bar{k}^2}{|\vec{k}|^2} \de_{a_1a_2} \,,
\eE 
where $N= 0.025999\,(8\pi) $ was defined in Eq.~(\ref{N}). Therefore, $e^2F_{GI}^{(2,2)}$ reads
\bE{rCl}
\label{FGI22}
 e^2F_{GI}^{(2,2)} &=& -\Bigg( N +{1\over8}\Bigg)  {e^2C_A\over4\pi} \int_\slashed{k} \frac{\bar{k}^2}{|\vec{k}|^2}  J^a(-\vec k)J^a(\vec k) \\
 &=& -\Bigg( N +{1\over8}\Bigg)  {e^2C_A\over4\pi} \int_\slashed{k} \frac{1}{|\vec{k}|^2}  (\vec{k}\times\vec{A}^a(-\vec{k})) (\vec{k}\times\vec{A}^a(\vec{k})) +{\cal O}(e^3) \\
  &=& e^2F^{(2,2)}_{GL} +{\cal O}(e^3) \nn\,.
\eE 
This concludes the computation of the wave functional with ${\cal O}(e^2)$ precision in terms of $J$ fields. The complete result is summarized in Eqs.~(\ref{FGI0}), (\ref{FGI1}), (\ref{FGI24}) and (\ref{FGI22}).
This result differs from the expression obtained in Sec.~\ref{sec:KNY}, and from the weak coupling limit of the expression obtained in Ref.~\cite{Karabali:2009rg}. 
The reason is that the prefactor of $F_{GI}^{(2,2)}$ has changed from Eq.~(\ref{F22GI}) to Eq.~(\ref{FGI22}): $N +1 \rightarrow N+1/8$. This is important, as now the new prefactors of Eqs.~(\ref{F22GL})  and
(\ref{FGI22}) agree with each other. This was the missing ingredient to claim complete agreement between both computations, which now we do:
The vacuum wave functional computed with methods (A) and (B) agree with each other with ${\cal O}(e^2)$ precision (when written with the same variables, either $J$ or ${\vec A}$). In other words
\be
F^{(0)}_{GI} + e F^{(1)}_{GI} + e^2 (F^{(2,2)}_{GI}+F^{(2,4)}_{GI}) = F^{(0)}_{GL} + e F^{(1)}_{GL} + e^2 (F^{(2,2)}_{GL}+F^{(2,4)}_{GL})+{\cal O}(e^3)\,.
\ee

Considering the recursion relations given in Chap.~\ref{chap:Comparison}, the above result implies that Eq.~(\ref{rec4}) should be replaced by
\beqar
&& 2\frac{e^2C_A}{2\pi}~ f^{(2)}_{a_1 a_2}(\vec x_1, \vec x_2) + 4 \int_{x,y}  f^{(2)}_{a_1 a}(\vec x_1, \vec x) (\bar{\Omega}^0)_{ab}(\vec x,\vec y) f^{(2)}_{b a_2}(\vec y, \vec x_2) +V_{ab}  \nn
\\
&&+e^2 \left[ 6 \int_{x,y} \!\! f^{(4)}_{a_1 a_2 a b }(\vec x_1, \vec x_2, \vec x,\vec y) (\bar{\Omega}^0)_{ab}(\vec x,\vec y) + 3 \int_{x,y} \!\! f^{(3)}_{a_1 a b  }(\vec x_1, \vec x,\vec y) (\bar{\Omega}^1)_{ab a_2}(\vec x,\vec y, \vec x_2) \right.\nn\\
&&\qquad\qquad \left. + 3 \int_{x,y,v} \!\! f^{(3)}_{a_1 a b  }(\vec x_1, \vec x,\vec y) (\bar{\tilde\Omega}^1)_{ab a_2}(\vec x,\vec y, \vec v, \vec x_2)  \right] +\O\left(e^3,\mu^{-1}\right)
= 0\,, \label{recreg}
\eeqar
with
\bE{rCl}
(\bar{\tilde\Omega}^1)_{ab a_2}(\vec x,\vec y, \vec v, \vec x_2) 
&=& \frac{i}{2} f^{aba_2}  \de(\vec v-\vec x_2) \\
&&  \times\left(\delta_\mu(\vec x, \vec v)  (x-v) \left(\p_y \bar{G}(y-v)\right) -\delta_\mu(\vec y, \vec v)  (y-v) \left(\p_x \bar{G}(x-v)\right)\right)\,, \nn
\eE
while Eq.~(\ref{rec5}) remains valid up to $\O(e^2)$. Note that Eqs.~(\ref{rec4}-\ref{rec5}) were taken to be correct to all orders in Ref.~\cite{Karabali:2009rg}, while here we only consider perturbation theory up to $\O(e^2)$, dropping terms that would modify  Eqs.~(\ref{rec4}-\ref{rec5}) at higher orders.

Finally, let us note that  the ``mass term" Eq.~(\ref{massterm}), which is taken to be responsible for generating the mass gap in a strong coupling analysis, is not a special term from the point of view of weak coupling, as there are more terms in the Hamiltonian Eq.~(\ref{TregKKN}) that produce identical terms to the wave functional (see, for instance, Eq.~(\ref{new})).

\section{Conclusions}
\label{sec:Conclusions:Regularization}

We have obtained the complete expression for the Yang-Mills vacuum wave functional in three dimensions at weak coupling with ${\cal O}(e^2)$ precision. We have used two different methods to solve the functional Schr\"odinger  equation: 
(A) One of them generalizes to ${\cal O}(e^2)$ the method followed by Hatfield at ${\cal O}(e)$~\cite{Hatfield:1984dv}. We have named the 
result obtained $\Psi_{GL}[{\vec A}]$. 
(B) The other uses the weak coupling version of the gauge invariant formulation of the Schr\"odinger equation and the ground-state wave functional followed by Karabali, Nair, and Yelnikov \cite{Karabali:2009rg}. We have named the result obtained $\Psi_{GI}[J]$. 
We addressed this computation in Chap.~\ref{chap:Comparison},
obtaining conflicting results between both methods, because effects associated to the regularization of the Hamiltonian were not studied. In this chapter we have carried out this study in full detail. This has led in both cases to new (but different) contributions emanating from the regularization of the theory. 
The final results for both methods now agree with each other. This is a 
very strong check of the computations and of the regularization procedure used here. We can now claim that we have obtained the complete 
expression of the Yang-Mills vacuum wave functional in three dimensions with ${\cal O}(e^2)$ precision for the first time. In terms of the ${\vec A}$ fields the vacuum wave functional can be found in Eqs.~(\ref{FGL0p}), (\ref{FGL1}), (\ref{F24GL2}) and (\ref{F22GL}),  and in terms of the gauge invariant $J$ variable in Eqs.~(\ref{FGI0}), (\ref{FGI1}), (\ref{FGI24}) and (\ref{FGI22}). Both results are equal to ${\cal O}(e^2)$. To our knowledge this is the first time that a full fledge (including regularization) computation of the wave functional of a gauge theory has been undertaken.

That the result obtained here differs from the one obtained in Chap.~\ref{chap:Comparison} with method (A) should not be so surprising, 
as the regularization of the kinetic operator was not considered there. More surprising is the fact that a new term has been found 
using method (B), the regularization of which had been studied in detail in the past  (see, for instance, the discussions in Refs.~\cite{Karabali:1997wk,Agarwal:2007ns}, in particular in the appendix of the last reference). In those references an intermediate cutoff $\mu'  \ll \mu $ was introduced in the wave functional, damping the modes with energies greater than $\mu'$. This procedure eliminates the extra contribution we found with method (B) in Sec.~\ref{subsec:F22GI}. However, if the same procedure is applied to method (A), it also eliminates the mass term obtained in Sec.~\ref{subsec:F22GL}, producing the two incompatible results of Chap.~\ref{chap:Comparison}. Instead, we advocate doing the whole computation with a single cutoff $\mu$ that regularizes the kinetic operator and the ground-state wave functional (and all excitations) at the same time. It is only after solving the Schr\"odinger equation that we can take the cutoff $\mu$ to infinity compared with any finite momentum of the system. In other words, the momenta of the fields of the wave functional can be large. As one goes to higher orders in perturbation theory, loops appear, whose integrals run up to infinity, and all of these modes have to be taken into account, producing new contributions, as we have seen in Eq.~(\ref{new}). 
In a different language, in order to be able to give meaning to the theory we need to regularize the Hamiltonian. This defines a (regularized) Hilbert space, in which both the Hamiltonian and the states depend on the same regulator. Preserving unitarity requires all states to be considered in the computation. In particular, cutting them off with a second regulator impairs the completeness relation.

 In any case it is clear that regularization of the wave functional in the Schr\"odinger formalism is still in its infancy, and more work is needed to put the formalism on more solid ground. In this respect we would like to mention possible additional checks of our wave functional. One could be a numerical study at short distances, similar to the ones executed in Refs.~\cite{Greensite:2011pj} and \cite{Greensite:2013rva}, but it is unclear whether it is possible to obtain conclusive results in this way, since one might not find a sufficiently large difference between $\Psi_{GI}$ of Chap.~\ref{chap:Comparison} (i.e.~the weak coupling limit of the wave functional proposed in \cite{Karabali:2009rg}) and the fully regularized wave functional given in this chapter. Another test, this one analytical, could be the computation of the static potential in a weak coupling expansion up to ${\cal O}(e^2)$ from the expectation value of the Wilson loop, and subsequent comparison with known results computed in other representations of QFT.

Finally, we cannot avoid making some considerations of the possible significance of the mass-like term (\ref{FGI22}). Its mass prefactor is gauge independent. Following Refs.\cite{Karabali:1995ps,Karabali:1996je,Karabali:1996iu, Karabali:1997wk,Karabali:1998yq} 
one may argue about its relation with the magnetic screening mass. If we do so, we obtain
\be
m=\left(\frac{1}{8} + (8\pi)0.025999\right)\frac{C_A e^2}{2\pi}=0.778426\frac{C_A e^2}{2\pi}=0.247781\frac{C_A }{2}e^2 \label{mag-screen}
\,.
\ee
This value is in the same ballpark as the values obtained from some resummation schemes of perturbation theory at one loop \cite{Alexanian:1995rp,Jackiw:1995nf,Buchmuller:1996pp,Cornwall:1997dc}\footnote{At two loops the result depends on the renormalization scale, see Table I of Ref.~\cite{Bieletzki:2012rd}, but the agreement is still reasonable.}. In particular, it is remarkably close to the value quoted in Ref.~\cite{Cornwall:1997dc}. It is also not far from the mass value proposed in Ref.~\cite{Karabali:1995ps}: $m=\frac{C_A e^2}{2\pi}$, which was obtained from a strong coupling computation at leading order.

\bigskip


\chapter{Towards the Non-perturbative Regime}
\label{chap:Potential}
\section{Introduction}

Solving the \SE $ $ is of course not an end in itself. Once the vacuum wave functional, or a suitable approximation, has been found, it can be used to compute observables, as discussed in Chap.~\ref{chap:Schrödinger}. In particular it is possible to calculate the vacuum expectation value of an operator $\hat{\cal O}$, using Eq.~(\ref{VEV}):
\be
\langle\hat{\cal O}\rangle = \frac{ \int\D\phi\, \Psi_0^*[\phi]{\cal O}\, \Psi_0[\phi]}{ \int\D\phi\, \Psi_0^*[\phi]\Psi_0[\phi]}  \,. 
\ee
Computations at weak coupling can obviously check results obtained with other representations, but it is in the non-perturbative regime, where the Schr\"odinger representation can realize its full potential, since it allows for a straightforward way to go beyond perturbation theory. In this chapter we will use a trial wave functional to illustrate how the Schr\"odinger picture can be used to calculate relevant QCD observables in the regimes beyond weak coupling. 

A very important quantity in QCD which in principle can be calculated from the Yang-Mills vacuum wave functional is the static potential $E_s$ between two static color sources. This object is at the center of the mechanism by which confinement takes place. So, an analytical understanding of the static potential is crucial for a quantitative explanation of this process. For sources in the fundamental representation (e.g.~heavy quarks) it is assumed that the static potential is linear at long distances, as long as there are no dynamical quarks in the theory. This has not been proven analytically, but only confirmed numerically by lattice calculations (see e.g.~Refs.~\cite{Meyer:2006gm,HariDass:2007tx}). If dynamical quarks are present, and also for sources in the adjoint representation, on the other hand, we expect screening of the color charge of the source, meaning that the potential should approach a constant at long distances.

In order to investigate the static potential at long distances (and possibly other non-perturbative observables) a strong coupling expansion for the vacuum wave functional was developed in Ref.~\cite{Karabali:1998yq}. It was based on an interesting fact, easy to see in the formulation in terms of $J$ fields: The  potential term $\V$ of the Yang-Mills Hamiltonian viewed as a functional is an eigenfunction of the kinetic operator $\T$.
Remarkably enough the leading order (LO) term of the vacuum wave functional in this expansion predicted a linear potential at long distances. The proportionality coefficient $\sigma$, called string tension, was also obtained, finding agreement within one or two percent with lattice simulations, which obviously is an outstanding result. 

In Chap.~\ref{chap:Regularization}, however, we found that the kinetic operator has to be modified in order to incorporate the full regularization (see Eq.~(\ref{TregKKNallorders})). This raises the question whether the eigenvalue equation is affected by this change in the operator. In Sec.~\ref{sec:TV} we compute the action of $\T$ on $\V$ in a perturbative expansion in terms of the original gauge fields. While we can show that $\V$ is still an eigenfunction of $\T$, we also find, however, that the eigenvalue depends on the regularization used. This sheds some doubt on the straightforward use of the eigenvalue equation in the determination of the vacuum wave functional.

Independent of this, because of other issues of the strong coupling expansion, and in order to provide with an expression for the ground-state wave functional that interpolates between the weak coupling and the strong coupling regimes, a new expansion scheme was developed in Ref.~\cite{Karabali:2009rg}. The idea of which is to define 
\be
m:=\m
\ee
as a parameter independent from $e$ and to perform an expansion in $e^2/m$ (note that $e^2/m$ is of $\O(1)$, yet the success of the LO result may suggest that this is a good expansion). With this, the Hamiltonian of Eq.~(\ref{HamiltonianNair}) could be split in a way different from the splitting used in Chap.~\ref{chap:Comparison}, including the term with one derivative in $\H^{(0)}$ and only taking the $\O(e)$ term as the perturbative Hamiltonian $\H_I$.  Maintaining $m$ as an independent parameter, the \SE $ $ was solved up to $\O\left(e^2\right)$, yielding a new vacuum wave functional, $\Psi_{KNY}[J]$, which can be considered a result from resummation of perturbation theory.\footnote{The functional $\Psi_{GI}[J]$ of Sec.~\ref{sec:KNY} is the Taylor expansion up to $\O\left(e^2\right)$ (writing $m$ as $\m$) of  $\Psi_{KNY}[J]$.}  Using this wave functional, a partial set of the $\O\left(e^2/m\right)$ corrections to the static potential were computed in Ref.~\cite{Karabali:2009rg}. These corrections were still consistent
with a linear potential, but there are several points of concern for this result. First, it is not complete: not all of the $\O\left(e^2/m\right)$ corrections were computed, since in the expansion scheme used there, it would require an infinite number of diagrams.
Moreover, some of the corrections were found to be ambiguous, since they depend on the factorization scale (even though it was argued that the ambiguity was small). Actually, in Ref.~\cite{Greensite:2007ij}
the string tension was computed numerically using a gauge invariant version of the leading order
of $\Psi_{KNY}[J]$ in terms of the chromomagnetic fields and covariant
derivatives. The authors concluded that the string tension obtained from such a functional
would diverge in the continuum limit. A third point of concern regarding the computation performed in Ref.~\cite{Karabali:2009rg} is that the ground state wave functional was assumed to be real. Whereas this is true for both
the exact result, and the approximate expressions in the weak coupling limit (as we have shown in Sec.~\ref{sec:comp}), the approximate expressions with $m$ as an independent parameter have a non-vanishing imaginary part. Finally, there may be issues with the regularization of $\Psi_{KNY}[J]$. As we have seen in Chap.~\ref{chap:Regularization}, there are problems in the weak coupling limit, and up to now it is unclear how this translates to other regimes.

Clarifying these questions is very important, since if it were possible to show that all corrections to all orders are compatible with a linear potential, this would prove confinement in three dimensions\footnote{Provided that the sum of all contributions is different from zero.}. In order to shed light on them, we rewrite $\Psi_{KNY}[J]$ in terms of the gauge fields, which allows us to compute all of the $\O(e^2/m)$ corrections. In view of the issues mentioned above, in particular the fact that the weak coupling limit of $\Psi_{KNY}[J]$ does not agree completely with the vacuum wave functional of Chap.~\ref{chap:Regularization}, we do not claim that the wave functional obtained in this way is the actual Yang-Mills vacuum wave functional, rather we use it as a trial functional to test the proposal of Ref.~\cite{Karabali:2009rg}. We find, unsurprisingly, that at LO a linear potential is predicted with the same coefficient as obtained in Refs.~\cite{Karabali:1998yq} and \cite{Karabali:2009rg}. The next-to-leading order (NLO), however, includes terms of a cubic potential. We know that when the potential is computed in perturbation theory, it contains terms of all powers, while in the full expression all of these terms should add up to produce the linear potential:
\be
E_s^\mathrm{pert}(r)=c_0\ln r+ c_1 r + c_2 r^2 + c_3 r^3+\ldots \stackrel{r\to\infty}{=} \sigma r +\O(1)\,.
\ee
Hence, higher order terms in the potential are not a problem per se, but they suggest that either this resummation scheme is not sufficient to prove confinement, or that the trial functional does not have the correct long distance limit.

We investigate the strong coupling expansion of Ref.~\cite{Karabali:1998yq} in Sec.~\ref{sec:TV}. In Sec.~\ref{sec:action} we explore the interpolating wave functional proposed in Ref.~\cite{Karabali:2009rg} and develop a method to compute expectation values. As an illustration of the method, we calculate the correlator of the chromomagnetic field and the gluon condensate at LO in Sec.~\ref{sec:condensate}. We then perform the computation of the static potential up to NLO in Sec.~\ref{sec:potential}. We summarize the results of this chapter in Sec.~\ref{sec:ConclusionsPot}.

\section{A strong coupling expansion: The Yang-Mills potential as an eigenfunction of the kinetic operator}
\label{sec:TV}
Considering the Yang-Mills Hamiltonian in the language of the currents $J$ reveals an interesting property: The potential term $\V$ viewed as a functional is an eigenfunction of the kinetic operator. The potential term considered in Ref.~\cite{Karabali:1998yq} is 
\bea
\V &=& \frac{\pi}{m C_A} \int_x \bar{\p} J^a(x) \bar{\p} J^a(x) \label{potJ} 
\eea
and the kinetic operator is 
\bE{rCl}
\T_{KKN} &=& {m C_A\over \pi} \int _{x,y} 
 {1\over (y-x)^2} {\delta \over \delta J^a (x)} {\delta \over \delta J^a (y)} + i m \int_{x,y} f^{abc} {J^c(x) \over \pi (y-x)} {\delta \over \delta J^a (x)} {\delta \over \delta J^b (y)} \nn\\
&&\qquad +m  \int_x J^a (x) {\delta \over \delta J^a (x)}  \,,\qquad \label{HamiltonianNair2}
\eE
where $m=\m$. They are obtained from the Hamiltonian of Eq.~(\ref{HamiltonianNair}) (which is the $\mu\to\infty$ limit of Eq.~(\ref{TregKKNallorders})) by a rescaling of the currents $J^a\to\frac{2\pi}{e C_A}J^a$. The last term of Eq.~(\ref{HamiltonianNair2}) counts the number of $J$ fields in any functional it is applied to. When applied to Eq.~(\ref{potJ}), the first term of the kinetic operator produces an infinite constant and the second term vanishes. Therefore
\be
\T_{KKN}:\V: =2m:\V:\,, \label{NairTV}
\ee
where we subtracted an infinite constant in the definition of the normally ordered potential $:\V:$. Note that this equality is exact to all orders in perturbation theory. It was taken as the starting point of a strong coupling expansion of the vacuum wave functional in Ref.~\cite{Karabali:1998yq}, in order to solve Eq.~(\ref{HamEq}):
\be
\widetilde{\H}\mathds{1}=e^F(\T+\V)e^{-F}\mathds{1}=\left(\T+\V-[\T,F]+\half\left[[\T,F],F\right]\right)
\mathds{1}=0\,. \label{HamEq2}
\ee
For momentum modes $k\ll m$ (in the regime of $e^2\gg J$) the potential can be treated perturbatively, leading to
\be
F=\frac{1}{2m}\V+\O(m^{-2})\,.
\ee
This is the opposite limit of what we considered in Chaps.~\ref{chap:Comparison} and \ref{chap:Regularization}, where we took $e^2\ll J$. The LO vacuum wave functional obtained in this way allowed for the prediction of a static potential with a string tension within one or two percent of the results of lattice computations.

This is an impressive result, but it has been obtained with the kinetic operator in the $\mu\to\infty$ limit, while the momenta $\vec k$ of the potential term were taken to be $|\vec k| \ll \mu$. In Chap.~\ref{chap:Regularization}, however, we found that the $\mu\to\infty$ limit should only be taken at the end of the computation. This raises the question whether this new regularization method changes the property of $\V$ being an eigenfunction of $\T$ or, if not, whether it modifies the eigenvalue. 

In Ref.~\cite{Karabali:1997wk} the computation was done with a differently regularized kinetic operator in terms of $J$ variables (which after removal of the regulator reduces to Eq.~(\ref{HamiltonianNair2})) and a regularized potential, but maintaining the assumption that the momenta in $\V$ are much smaller than $\mu$. In this computation an eigenvalue was found which depends logarithmically on the regulators. While it was argued that the regulator dependence could be fixed in such a way that the eigenvalue was $2m$ (reproducing the result of Eq.~(\ref{NairTV})), it is interesting to see how this relates to the regularization method of Chap.~\ref{chap:Regularization}.



To investigate these questions we turn to the formulation of the Yang-Mills Hamiltonian in terms of the original gauge fields (method (A)). When looking at these expressions, however, one finds that the action of $\T$ on $\V$ is ill-defined -- as long as only unregularized operators are considered. But when regularizing both the kinetic operator (Eq.~(\ref{Treg}))
\be
\mathcal{T}_{reg}=-{1\over2}\int_{u,v} \delta_\mu(\vec u, \vec v) \Phi_{ab}(\vec u,\vec v) \frac{\delta}{\delta A_i^a(\vec u)}  \frac{\delta}{\delta A_i^b(\vec v)}\,, \label{Treg2}
\ee
and the potential term (where a Wilson line is necessary for gauge invariant point splitting, and we use an independent cutoff $\mu'$ for the potential in order to keep the discussion as general as possible)
\be
\mathcal{V}_{reg}={1\over2}\int_{x,y} \delta_{\mu'}(\vec x, \vec y) B^a(\vec x) \Phi_{ab}(\vec x,\vec y)  B^b(\vec y)\,, \label{Vreg}
\ee
we also find that $\mathcal{V}_{reg}$ is an eigenfunction of $\mathcal{T}_{reg}$ at $\O(e^2)$. The eigenvalue, however, is different, and in particular it depends on the regulators. Note that Eq.~(\ref{Vreg}) is different from the regularized potential used in Ref.~\cite{Karabali:1997wk}. The computation goes as follows. We look at the terms order by order in $e$. For this we split the chromomagnetic field as
\be
B^a = {\vec \nabla}\times\vec{A}^a +\frac{e}{2}f^{abc}\vec{A}^b\times\vec{A}^c =: B^{(0)}_a + eB^{(1)}_a
\,,
\ee
and we also expand the Wilson lines in both the potential and kinetic terms up to $\O(e^2)$, using Eqs.~(\ref{Phi}-\ref{M_i-1}).
With this we can write $\T\V$ up to $\O(e^2)$:
\bE{rll}
\T_\mathrm{reg}(\mu) \V_\mathrm{reg}(\mu') =& -\frac{1}{4} \int_{u,v,x,y}&  \delta_{\mu}(\vec u,\vec v)\Phi^{ab}(\vec u,\vec v)  \frac{\delta^2}{\delta A^a_i(\vec u)\delta A^b_i(\vec v)} \delta_{\mu'}(\vec x,\vec y)B^c(\vec x)\Phi^{cd}(\vec x,\vec y)B^d(\vec y) \nn\\
&&\\
=& -\frac{1}{4} \int_{u,v,x,y} & \delta_{\mu}(\vec u,\vec v)\delta_{\mu'}(\vec x,\vec y) \delta^{ab}  \frac{\delta^2}{\delta A^a_i(\vec u)\delta A^b_i(\vec v)} B^{(0)}_c(\vec x)\de^{cd}B^{(0)}_d(\vec y) \nn\\
  & -\frac{e}{4} \int_{u,v,x,y} &\delta_{\mu}(\vec u,\vec v)\delta_{\mu'}(\vec x,\vec y) \Bigg\{2\delta^{ab}  \frac{\delta^2}{\delta A^a_i(\vec u)\delta A^b_i(\vec v)} B^{(0)}_c(\vec x)\de^{cd}B^{(1)}_d(\vec y) \nn\\
  && + \de^{ab} \frac{\delta^2}{\delta A^a_i(\vec u)\delta A^b_i(\vec v)} B^{(0)}_c(\vec x)\Phi^{(1)}_{cd}(\vec x,\vec y)B^{(0)}_d(\vec y) \nn\\ 
  && +  \Phi^{(1)}_{ab}(\vec u,\vec v) \frac{\delta^2}{\delta A^a_i(\vec u)\delta A^b_i(\vec v)} B^{(0)}_c(\vec x)\delta^{cd}B^{(0)}_d(\vec y) \Bigg\} \nn\\
& -\frac{e^2}{4} \int_{u,v,x,y}& \delta_{\mu}(\vec u,\vec v)\delta_{\mu'}(\vec x,\vec y) \Bigg\{\delta^{ab}  \frac{\delta^2}{\delta A^a_i(\vec u)\delta A^b_i(\vec v)} B^{(1)}_c(\vec x)\de^{cd}B^{(1)}_d(\vec y) \nn\\
&& + 2 \delta^{ab}  \frac{\delta^2}{\delta A^a_i(\vec u)\delta A^b_i(\vec v)} B^{(0)}_c(\vec x)\Phi^{(1)}_{cd}(\vec x,\vec y)B^{(1)}_d(\vec y) \nn\\
&& + \delta^{ab}  \frac{\delta^2}{\delta A^a_i(\vec u)\delta A^b_i(\vec v)} B^{(0)}_c(\vec x)\Phi^{(2)}_{cd}(\vec x,\vec y)B^{(0)}_d(\vec y) \nn\\
&& + 2 \Phi^{(1)}_{ab}(\vec u,\vec v) \frac{\delta^2}{\delta A^a_i(\vec u)\delta A^b_i(\vec v)} B^{(0)}_c(\vec x) \de^{cd}B^{(1)}_d(\vec y) \nn\\ 
&& + \Phi^{(1)}_{ab}(\vec u,\vec v) \frac{\delta^2}{\delta A^a_i(\vec u)\delta A^b_i(\vec v)} B^{(0)}_c(\vec x)\Phi^{(1)}_{cd}(\vec x,\vec y)B^{(0)}_d(\vec y) \nn\\ 
&& +  \Phi^{(2)}_{ab}(\vec u,\vec v) \frac{\delta^2}{\delta A^a_i(\vec u)\delta A^b_i(\vec v)} B^{(0)}_c(\vec x)\delta^{cd}B^{(0)}_d(\vec y) \Bigg\} \nn\\
& +\O(e^3)\,. & 
\eE
The term at $\O(e^0)$ is an infinite constant which we call $c$ and absorb in the definition of the normally ordered potential. Also, it can be checked easily, that the terms at $\O(e)$ vanish under color symmetry. The terms at $\O(e^2)$ however, turn out to be proportional to the potential. The computation of these terms is lengthy but straightforward (see App.~\ref{app:TV} for details).
\bE{rCl}
\T_\mathrm{reg}(\mu) \V_\mathrm{reg}(\mu') &=& c\nn\\
&&  -\frac{e^2C_A}{2\pi} \int_{y}  \Bigg(\frac{\mu^2 \mu'^2}{ \left(\mu^2+\mu'^2\right)} \vec{A}^a(\vec y)\cdot\vec{A}^a(\vec y)  \nn\\
&& \qquad  -\frac{\mu^2 \mu'^2}{4 \left(\mu^2+\mu'^2\right)^2}\Big((\p_1A_1^a(\vec y))^2 + (\p_2A_2^a(\vec y))^2 + (\p_1A_2^a(\vec y))^2 + (\p_2A_1^a(\vec y))^2      \Big) \Bigg) \nn\\
&& + {e^2 C_A\over2\pi} \int_{y}\Bigg( 2\frac{\mu ^4 \mu'^2}{ \left(\mu^2+\mu'^2\right)^2} \vec{A}^a(\vec y)\cdot\vec{A}^a(\vec y) -  \frac{\mu^4 \mu'^2}{2 \left(\mu^2+\mu'^2\right)^3}(\vec \nabla\cdot\vec{A}^a(\vec y))^2  \nn\\
&&\qquad + B^{(0)a}(y)B^{(0)a}(y) \left( \frac{\mu ^2 }{2 \mu' \sqrt{\mu^2+\mu'^2}} + \frac{\mu ^2 }{2 \left(\mu ^2+\mu'^2\right)} - \frac{\mu^4 \mu'^2}{4 \left(\mu^2+\mu'^2\right)^3} \right)  \Bigg) \nn\\
&& + \frac{e^2 C_A}{2\pi}  \int_{y}   \Bigg(  B^{(0)a}(\vec y)B^{(0)a}(\vec y) \Big(\sqrt{1+\frac{\mu^2}{\mu'^2}}-1\Big) \nn\\
&&\qquad + \frac{\mu^4 \mu'^2}{ \left(\mu^2+\mu'^2\right)^3} \Big((\mu'^2-\mu^2) \vec{A}^a(\vec y)\cdot\vec{A}^a(\vec y) - \frac{\mu'^2-2\mu^2}{8 \left(\mu^2+\mu'^2\right)}(\vec \nabla\cdot\vec{A}^a(\vec y))^2\Big)   \nn\\
&&\qquad  -{1\over2} \left(\vec \nabla\times \vec{A}^a(\vec y)\right)^2 \frac{\mu^4}{\left(\mu ^2+\mu'^2\right)^2} \Bigg) \nn\\
&& +  {e^2 C_A\over2\pi} \int_{y} \Bigg( 2\frac{\mu^2 \mu'^4}{ \left(\mu^2+\mu'^2\right)^2} \vec{A}^a(\vec y)\cdot\vec{A}^a(\vec y) \nn\\
&&\qquad - \frac{\mu^2 \mu'^4}{2 \left(\mu^2+\mu'^2\right)^3}(\vec \nabla\cdot\vec{A}^a(\vec y))^2 +\frac{\mu^2 \mu'^4}{4 \left(\mu^2+\mu'^2\right)^3}(\vec \nabla\times\vec{A}^a(\vec y))^2  \Bigg) \nn\\
&&  +\frac{e^2 C_A}{2\pi} \int_{y}\Bigg( -\frac{4\mu^4 \mu'^4}{ \left(\mu^2+\mu'^2\right)^3}  \vec{A}^a(\vec y)\cdot\vec{A}^a(\vec y)    \nn\\
&& \qquad   +   \frac{3\mu^4 \mu'^4}{4 \left(\mu^2+\mu'^2\right)^4} \Big((\p_1A_1^a(\vec y))^2 + (\p_2A_2^a(\vec y))^2\Big)    \nn\\
&& \qquad   + \frac{3\mu^4 \mu'^4}{4 \left(\mu^2+\mu'^2\right)^4}   \Big((\p_1A_2^a(\vec y))^2 + (\p_2A_1^a(\vec y))^2\Big)          \nn\\
&& \qquad +   B^{(0)a}(\vec y)  (\vec \nabla\times\vec{A}^a(\vec y)) \Big( {1\over2} \frac{\mu ^2}{\mu ^2+\mu'^2} - {1\over2} \frac{\mu^2 \left(\mu^2+2\mu'^2\right)\sqrt{\mu ^2+\mu'^2}}{ \mu'  \left(\mu ^2+\mu'^2\right)^2}  \Big) \Bigg) \nn\\
&& + {e^2 C_A\over 2\pi} \int_{y} \frac{\mu^2 \mu'^4}{\left(\mu^2+\mu'^2\right)^3} \Bigg(  (\mu^2 -\mu'^2) \vec{A}^a(\vec y)\cdot\vec{A}^a(\vec y) -\frac{\mu^2 -2\mu'^2}{8 \left(\mu^2+\mu'^2\right)}(\vec \nabla\cdot\vec{A}^a(\vec y))^2\Bigg) \nn\\
&&+\O\left(\mu^{-1}\right)+\O\left(e^3\right) \label{TVB}
\eE
Summing up all the terms one finds
\bE{rCl}
\T_\mathrm{reg}(\mu) \V_\mathrm{reg}(\mu')
&=& c + \frac{e^2 C_A}{2\pi} \half \int_{y} B^{(0)a}(\vec y)B^{(0)a}(\vec y) \nn\\
&&\qquad\times \Bigg(-1+\frac{3 \mu'^8}{2\left(\mu ^2+\mu'^2\right)^4}-\frac{3 \mu'^6}{\left(\mu ^2+\mu'^2\right)^3}+\frac{\mu'^4}{2\left(\mu ^2+\mu'^2\right)^2} \nn\\
&& \qquad\qquad +\frac{\mu'^3}{\left(\mu ^2+\mu'^2\right)^{3/2}}-\frac{\mu'}{\sqrt{\mu ^2+\mu'^2}}+2\frac{\sqrt{\mu ^2+\mu'^2}}{\mu'}\Bigg)  \nn\\
&& +\O(e^3)\,.
\eE
Since we used a gauge invariant regularization and $\T_\mathrm{reg}$ and $\V_\mathrm{reg}$ are gauge invariant operators, it was to be expected that $\T_\mathrm{reg} \V_\mathrm{reg}$ would result in a gauge invariant object. That this would be the potential, or any local quantity at all, however, was not obvious.

While we find that $\V_\mathrm{reg}(\mu')$ is also an eigenfunction of $\T_\mathrm{reg}(\mu)$ in this formulation, its eigenvalue depends on the regulators that are used. In particular the dependence is different from the one found in Ref.~\cite{Karabali:1997wk}.  In the case of equal regulators $\mu'=\mu$, the eigenvalue equation is 
\be
\T_\mathrm{reg}(\mu) :\V_\mathrm{reg}(\mu):  = \frac{1}{32} \left(-37+56 \sqrt{2}\right)\m:\V: + \O(e^3) \,,
\ee
In the limit of $\mu\gg\mu'$ it reduces to
\be
\T_\mathrm{reg}(\mu) :\V_\mathrm{reg}(\mu'):  = \left(2{\mu\over\mu'}-1+\O(\mu^{-1})\right)\m:\V: + \O(e^3) \,,
\ee
and it vanishes in the limit of $\mu'\gg\mu$.

The computation in Ref.~\cite{Karabali:1998yq} was done in the limit of $\mu\gg\mu'$, which in our case leads to a divergent eigenvalue. As $\T\V$ is not a physical observable this is no fundamental problem, but the discrepancy of this result with Eq.~(\ref{NairTV}) and, more importantly, its regulator dependence, suggest that the straightforward use of the eigenvalue equality in a strong coupling expansion of Eq.~(\ref{HamEq2}) may be problematic. 

Another problem with this expansion, which was already mentioned in Ref.~\cite{Karabali:1998yq}, is the fact that the contribution of momentum modes $k\gtrsim e^2$ were completely neglected. This is not justified, even if we were only interested in the long distance behavior of the static potential, because the effect of those modes is of the same order as the effect already included in the previous approximation, and could go from changing the value of the coefficient of the linear potential to completely changing the asymptotic behavior of the potential at long distances. 

In order to overcome this problem and to find an expression for the vacuum wave functional that interpolates between the weak and the strong coupling regimes, a new approach was developed in Ref.~\cite{Karabali:2009rg}, which we investigate in the following section.

\section{An interpolating wave functional}
\label{sec:action}

Taking $m=\m$ as an independent parameter, in Ref.~\cite{Karabali:2009rg}, the Hamiltonian of Eq.~(\ref{HamiltonianNair}) 
\bea
\H &=& \V+ {2\over \pi} \int _{w,z} 
 {1\over (z-w)^2} {\d \over {\d J_a (\vw)}} {\d \over {\d
J_a (\vz)}} +m  \int J_a (\vz) {\d \over {\d J_a (\vz)}} 
\nn\\
&& \quad + i e \int_{w,z} f_{abc} {J^c(w) \over \pi (z-w)} {\d \over {\d J_a (\vw)}} {\d \over {\d
J_b (\vz)}} \\
&=&\H^{(0)}+\H_I \nn
\eea
was split in a way different from the splitting used in Chap.~\ref{chap:Comparison}, including the term with one derivative in $\H^{(0)}$ and only taking the $\O(e)$ term as $\H_I$. Using a double expansion of the vacuum wave functional in e and the number of $J$ fields (Eqs.~(\ref{rec2}) and (\ref{rec3})), the authors obtained the recursion relations Eqs.~(\ref{rec4}) and (\ref{rec5}) with $\m$ replaced by $m$. Maintaining $m$ as an independent parameter, the recursion relations were solved up to $\O(e^2)$, yielding the (resummed) vacuum wave functional
\be
\label{WFJ}
\Psi_{KNY}[J] = \exp (-F_{KNY}[J])\,,
\ee 
\bea
-2F_{KNY}[J] &=& \int f^{(2)}_{a_1 a_2}(\vec x_1, \vec x_2)\ J^{a_1}(\vec x_1) J^{a_2}(\vec x_2) ~+~
\frac{e}{2}\ f^{(3)}_{a_1 a_2 a_3}(\vec x_1,\vec  x_2,\vec  x_3)\ J^{a_1}(\vec x_1) J^{a_2}(\vec x_2) J^{a_3}(\vec x_3)
\nn\\
&&\hskip .2in~+~
\frac{e^2}{4}\ f^{(4)}_{a_1 a_2 a_3 a_4}(\vec x_1,\vec  x_2,\vec  x_3,\vec  x_4)\ J^{a_1}(\vec x_1) J^{a_2}(\vec x_2) J^{a_3}(\vec x_3)
J^{a_4}(\vec x_4)~+~\ldots\,, \\
&=& -2\left(F^{(0)}_{KNY}[J]+eF^{(1)}_{KNY}[J]+e^2F^{(2)}_{KNY}[J]+\ldots\right)
\eea
\bea
f^{(2)}_{a_1 a_2}(\vec x_1, \vec x_2) &=& f^{(2)}_{0~a_1 a_2}(\vec x_1, \vec x_2) +
e^2 f^{(2)}_{2~a_1 a_2}(\vec x_1, \vec x_2) +\ldots\,,
\nn 
\eea
\bea
f^{(3)}_{a_1 a_2 a_3}(\vec x_1, \vec x_2, \vec x_3)&=& -\frac{f^{a_1 a_2 a_3}}{24}\int_{\slashed{k_1} \cdots \slashed{k_3}} \exp\left( i \sum_i^3 \vec k_i\cdot \vec x_i\right)  \sd\left(\sum_i^3 \vec k_i\right) g^{(3)}(\vec k_1,\vec k_2,\vec k_3) \nn\\ &&+\O(e^2)
\,,  \\
f^{(4)}_{a_1 a_2 a_3 a_4}(\vec x_1, \vec x_2, \vec x_3, \vec x_4) &=&  \frac{f^{a_1 a_2 c} f^{b_1 b_2 c}}{64} \int_{\slashed{k_1} \cdots \slashed{k_4}} \exp\left( i \sum_i^4 \vec k_i\cdot \vec x_i\right) \sd\left(\sum_i^4 \vec k_i\right) g^{(4)}(\vec k_1, \vec k_2; \vec k_3, \vec k_4) \nn\\ &&+\O(e^2)\,, 
\eea
 with
\be
 f^{(2)}_{0\ a_1 a_2}(\vec k) = -\frac{\bar{k}^2}{m+E_k} \delta_{a_1 a_2} \,,
\ee
\bea
f_{2\ a_1 a_2}^{(2)}(\vec k) &=&  \delta_{a_1 a_2}{1 \over E_k} {C_A\over 2\pi} \left(\int \frac{d^2 p}{32\pi}\ \frac{1}{\bar p}\ g^{(3)}(\vec k,\vec p,-\vec p-\vec k)\ +\ \int \frac{d^2 p}{64\pi}\ \frac{p}{\bar p}\ g^{(4)}(\vec k,\vec p;-\vec k,-\vec p)\right)\nn\\
&=&  \delta_{a_1 a_2}\frac{\bar{k}^2}{m^2}{C_A\over 2\pi}\tilde N +\O(\bar k^2 \vec k^2)\,, \label{f22expanded} \\
\tilde N &=& \left(-\frac{63}{32}+\frac{25}{4}\ln\frac{3}{2}\right) \approx 0.565407\,,
\eea
\beq 
g^{(3)}(\vec k_1,\vec k_2,\vec k_3) = \frac{16}{E_{k_1}\! + E_{k_2}\! + E_{k_3}}\left \{ \frac{\bar k_1 \bar k_2 (\bar k_1 - \bar k_2)}{(m+E_{k_1})(m+ E_{k_2})} + {cycl.\ perm.} \right \}
\,,
\eeq
\begin{equation}
\begin{array}{cl}
g^{(4)}(\vec k_1, \vec k_2; \vec q_1, \vec q_2)& =\ \vspace{.2in} \displaystyle \frac{1}{E_{k_1}\! + E_{k_2}\! + E_{q_1}\! + E_{q_2}} \\
\vspace{.2in}
&\!\!\!\!\displaystyle \left \{ g^{(3)}(\vec k_1, \vec k_2, -\vec k_1-\vec k_2)\ \frac{k_1 + k_2}{\bar k_1 +\bar k_2}\ g^{(3)}(\vec q_1, \vec q_2, -\vec q_1-\vec q_2) \right . \\
\vspace{.2in}
&\displaystyle -  \left [ \frac{(2\bar k_1 + \bar k_2)\,\bar k_1}{m+E_{k_1}} - \frac{(2\bar k_2 + \bar k_1)\,\bar k_2}{m+ E_{k_2}}\right ]\frac{4}{\bar k_1+\bar k_2}\  g^{(3)}(\vec q_1, \vec q_2, -\vec q_1-\vec q_2) \\
&\displaystyle -  \left .   g^{(3)}(\vec k_1, \vec k_2, -\vec k_1-\vec k_2)\ \frac{4}{\bar q_1+\bar q_2}\left [ \frac{(2\bar q_1 + \bar q_2)\,\bar q_1}{ m+E_{q_1}} - \frac{(2\bar q_2 + \bar q_1)\,\bar q_2}{ m+E_{q_2}}\right ] \right\} \,,
\\
\end{array}
\end{equation}
where now in all of the above 
\be 
E_k=\sqrt{m^2+\vec k^2}\,.
\ee
 This vacuum wave functional was claimed to interpolate between the weak and the strong coupling regimes, and to be a good approximation for all scales. In the strong coupling limit it reduces to the wave functional proposed in Ref.~\cite{Karabali:1998yq}. We have seen in \ref{sec:TV} that the latter may be problematic conceptually, but on the other hand, it led to an impressive prediction for the string tension. In the weak coupling limit $\Psi_{KNY}$ yields\footnote{Note that Eqs.~(\ref{rec7}) and (\ref{rec52}-\ref{f22Weak}) are just the Taylor expansions of the above expressions to $\O(e^2)$, after setting $m=\m$ again.}  $\Psi_{GI}$ of Chap.~\ref{chap:Comparison}, which we found to be correct up to $\O(e)$, but slightly different from the true vacuum wave functional at $\O(e^2)$. So, while there are issues with this proposal, we still think it is worthwhile to use it as a trial functional to test it on different observables. 

In order to do so, we again use Eq.~(\ref{JofB}): 
\be
\bar \partial^nJ=-iM^{\dagger}(\bar D^{n-1}B)M^{\dagger-1}\,, 
\ee
and Eq.~(\ref{MdaggerofAbar}):
\bE{rCl}
M^{\dagger}(\vec x) &=&1+e\int_y \bar{G}(x;y) \bar A(\vec y) + e^2\int_{y,z}\bar G(x;z) \bar G(z;y)\bar A(\vec y) \bar A(\vec z)+ O(e^3) 
\,,
\eE
to transform $J$ fields into $\vec A$ fields. We then obtain the trial functional
\bE{rCl}
\Psi_\mathrm{trial}[\vec A] &=& e^{-F_\mathrm{trial}[{\vec A}]} = e^{-F_\mathrm{trial}^{(0)}[{\vec A}]-eF_\mathrm{trial}^{(1)}[{\vec A}]-e^2F_\mathrm{trial}^{(2)}[{\vec A}]+{\cal O}(e^3)} \nn\\
&=& e^{-F_{KNY}^{(0)}[J({\vec A})]-eF_{KNY}^{(1)}[J({\vec A})]-e^2F_{KNY}^{(2)}[J({\vec A})]+{\cal O}(e^3)} \nn\\
&=& \Psi_{KNY}[J(\vec A)]+{\cal O}(e^3)\,.
\eE
Up to ${\cal O}(e)$ we find
\bE{rCl}
F^{(0)}_\mathrm{trial} &=& -\int_\slashed{k}\frac{1}{m+E_k}\mathrm{Tr}\left[B(\vec{k})B(\vec{-k})\right] \nn\\
&& -2 i e \int_{\slashed{k_1},\slashed{k_2},\slashed{k_3}} \frac{\slashed{\delta}\left(\sum_{i=1}^3\vec{k}_i\right) }{(m+E_3)\vec{k}_1^2} \mathrm{Tr}\Bigg[\left[\vec{k}_1\cdot \vec{A}(\vec{k}_1)-i\vec{k}_1\times\vec{A}(\vec{k}_1), B(\vec{k}_2) \right]B(\vec{k}_3)\Bigg] +\O(e^2)\quad\qquad
\end{IEEEeqnarray}
and
\begin{IEEEeqnarray}{rCl}
eF^{(1)}_\mathrm{trial} &=& -2e \int_{\slashed{k_1},\slashed{k_2},\slashed{k_3}}  \slashed{\delta}\left(\sum_{i=1}^3\vec{k}_i\right) \frac{\vec{k}_1\times\vec{k}_2+i\vec{k}_1\cdot\vec{k}_2}{(\sum_i E_i) (m + E_2)(m + E_3)\vec{k}_1^2}  \mathrm{Tr} \left[B(\vec{k}_1) \left[ B(\vec{k}_2),B(\vec{k}_3) \right]\right] \nn\\
&& +\O(e^2)\,.
\end{IEEEeqnarray}
Hence there is a non-trivial imaginary part at ${\cal O}(e)$
\begin{IEEEeqnarray}{rCl}
 \mathrm{Im}\,F^{(0)}_\mathrm{trial}\Big|_{\O(e)} &=& -i  \int_{\slashed{k_1},\slashed{k_2},\slashed{k_3}} \frac{\slashed{\delta}\left(\sum_{i=1}^3 \vec{k}_i\right)}{(m+E_3)\vec{k}_1^2} \left(1+ \frac{\vec{k}_1\cdot\vec{k}_2}{(E_1+E_2+E_3)(m+E_2)}\right)  \nn\\
&& \qquad\qquad\times \left(\vec{k}_1\times\vec{A}^a(\vec{k}_1)\right) \left(\vec{k}_2\times\vec{A}^b(\vec{k}_2)\right) \left(\vec{k}_3\times\vec{A}^c(\vec{k}_3)\right) f^{abc}\,,
\end{IEEEeqnarray}
which only vanishes in the limit of $m\to0$ (as shown in App.~\ref{F1comp}), and the same is true at $\O(e^2)$. The ground-state wave functional is real (see e.g.~Ref.~\cite{Feynman:1981ss}), so it could be argued that the non-vanishing imaginary part is an artifact of the expansion, which should drop out in the complete expression. In any case, in Ref.~\cite{Karabali:2009rg} $\Psi$ was set to be real.

In the method which we use to compute VEVs it is built in from the beginning that only the real part of the wave functional is needed, as long as the operator whose expectation value we want to compute does not contain any functional derivatives. In this case
\be
\langle\hat{\cal O}[\vec A]\rangle = \frac{ \int\D A_1\D A_2 \Psi_\mathrm{trial}^*[\vec A]{\cal O}[\vec A]\, \Psi_\mathrm{trial}[\vec A]}{ \int\D A_1\D A_2 \Psi_\mathrm{trial}^*[\vec A] \Psi_\mathrm{trial}[\vec A]}  = \frac{ \int\D A_1\D A_2 e^{-(F_\mathrm{trial}^\dagger[\vec A]+F_\mathrm{trial}[\vec A])}{\cal O}[\vec A]}{ \int\D A_1\D A_2 e^{-(F_\mathrm{trial}^\dagger[\vec A]+F_\mathrm{trial}[\vec A])}}  \,. 
\ee

The relevant terms are therefore
\bE{rCl}
F^{(0)}_\mathrm{trial}+F^{(0)\dagger}_\mathrm{trial} &=& -2\int_\slashed{k}\frac{1}{m+E_k}\mathrm{Tr}\left[B(\vec{k})B(\vec{-k})\right] \nn\\
&& -4 i e \int\int_{\slashed{k_1},\slashed{k_2},\slashed{k_3},\slashed{k_4}}\slashed{\delta}\left(\sum_{i=1}^3\vec{k}_i\right)  \frac{1}{(m+E_3)\vec{k}_1^2} \mathrm{Tr}\Bigg[\left[\vec{k}_1\cdot \vec{A}(\vec{k}_1), B(\vec{k}_2) \right]B(\vec{k}_3)\Bigg] \nn\\
&& +\frac{e^2}{2} f^{a_1a_2c}f^{b_1b_2c} \int_{\slashed{k_1},\slashed{k_2},\slashed{k_3},\slashed{k_4}}\slashed{\delta}\left(\sum_{i=1}^4\vec{k}_i\right) \Bigg\{\left(\frac{1}{m+E_{1+2}}-\frac{1}{m+E_3}\right)\frac{1}{\vec{k_2}^2\vec{k_4}^2}\nn\\ 
  &&\quad\Bigg((\vec{k}_1\times\vec{A}^{a_1}(\vec{k}_1))(\vec{k_2}\cdot\vec{A}^{a_2}(\vec{k_2})) (\vec{k}_3\times\vec{A}^{b_1}(\vec{k}_3))(\vec{k}_4\cdot\vec{A}^{b_2}(\vec{k}_4)) \nn\\
  &&\qquad - (\vec{k}_1\times\vec{A}^{a_1}(\vec{k}_1))(\vec{k_2}\times\vec{A}^{a_2}(\vec{k_2})) (\vec{k}_3\times\vec{A}^{b_1}(\vec{k}_3))(\vec{k}_4\times\vec{A}^{b_2}(\vec{k}_4)) \Bigg)\nn\\
  &&\quad + \frac{1}{(m+E_2)(\vec{k}_3+\vec{k}_4)^2\vec{k}_3^2}(\vec{k}_1\times\vec{A}^{a_1}(\vec{k}_1)) (\vec{k}_2\times\vec{A}^{a_2}(\vec{k}_2)) \nn\\
  &&\quad \Bigg((\vec{k}_3\cdot\vec{A}^{b_1}(\vec{k}_3)) (\vec{k}_3+\vec{k}_4)\cdot\vec{A}^{b_2}(\vec{k}_4) - (\vec{k}_3\times\vec{A}^{b_1} (\vec{k}_3)) (\vec{k}_3+\vec{k}_4)\times\vec{A}^{b_2}(\vec{k}_4)\Bigg)\Bigg\} \nn\\
&&+\O(e^3)\,,
\end{IEEEeqnarray}
\begin{IEEEeqnarray}{rCl}
F^{(1)}_\mathrm{trial}+F^{(1)\dagger}_\mathrm{trial} &=& -4 \int_{\slashed{k_1},\slashed{k_2},\slashed{k_3}}  \slashed{\delta}\left(\sum_{i=1}^3\vec{k}_i\right) \frac{\vec{k}_1\times\vec{k}_2}{(\sum_i E_i) (m + E_2)(m + E_3)\vec{k}_1^2}  \mathrm{Tr} \left[B(\vec{k}_1) \left[ B(\vec{k}_2),B(\vec{k}_3) \right]\right] \nn\\
&& +4ie \int_{\slashed{k_1},\slashed{k_2},\slashed{k_3},\slashed{k_4}}  \slashed{\delta}\left(\sum_{i=1}^4\vec{k}_i\right) \frac{1}{(E_{1+2} +E_3 + E_4)(m + E_3)\vec{k}_1^2} \nn\\
&&\quad \mathrm{Tr} \Bigg[ \left[ B(\vec{k}_3),B(\vec{k}_4) \right] \nn\\
&&\quad\qquad \Bigg\{ \frac{1}{(m+E_{1+2})\vec{k}_4^2}\Bigg(-\vec{k}_1^2 \left[\vec{k}_4\times \vec{A}(\vec{k}_1),B(\vec{k}_2)\right]  \nn\\
&&\qquad\qquad + ((\vec{k}_3-\vec{k}_2)\cdot\vec{k}_4) \left[\vec{k}_1\times \vec{A}(\vec{k}_1),B(\vec{k}_2)\right]  \nn\\
&&\qquad\qquad + ((\vec{k}_3-\vec{k}_2)\times\vec{k}_4)\left[ \vec{k}_1\cdot \vec{A}(\vec{k}_1),B(\vec{k}_2)\right] \Bigg)    \nn\\
&&\quad\qquad +\frac{1}{(m+E_4)(\vec{k}_1+\vec{k}_2)^2\vec{k}_2^2}  \nn\\
&&\qquad\qquad \Bigg( (\vec{k}_2\cdot\vec{k}_4)((2\vec{k}_1\cdot\vec{k}_2+\vec{k}_2^2)\; \vec{k}_1-\vec{k}_1^2\; \vec{k}_2)\times \left[\vec{A}(\vec{k}_1),B(\vec{k}_2)\right] \nn\\
&&\qquad\qquad\; -(\vec{k}_2\times\vec{k}_4)((2\vec{k}_1\cdot\vec{k}_2+\vec{k}_2^2)\; \vec{k}_1 -\vec{k}_1^2\; \vec{k}_2)\cdot \left[\vec{A}(\vec{k}_1),B(\vec{k}_2)\right]\Bigg) \Bigg\}\Bigg] \nn\\
&&+\O(e^2)\,,
\end{IEEEeqnarray}
and
\begin{IEEEeqnarray}{l}
F^{(2)}_\mathrm{trial}+F^{(2)\dagger}_\mathrm{trial} = \nn\\
\tilde N{C_A\over \pi}\int_\slashed{k}\frac{1}{m^2}\mathrm{Tr}\left[B(\vec{k})B(\vec{-k})\right]  +\O(\vec k^2)\nn\\
-2 \int_{\slashed{k_1},\slashed{k_2},\slashed{k_3},\slashed{k_4}}  \slashed{\delta}\left(\sum_{i=1}^4\vec{k}_i\right) \mathrm{Tr} \left[ \left[ B(\vec{k}_1),B(\vec{k}_2) \right] \left[ B(\vec{k}_3),B(\vec{k}_4) \right] \right] \nn\\
\quad\times\frac{1}{(\sum_i E_i) (E_1 + E_2 + E_{3+4})(E_3 + E_4 + E_{1+2})(m+E_1)(m+E_3)}  \nn\\
\quad\times\Bigg\{ \frac{1}{(m+E_2)(m+E_4)} \frac{\vec{k}_1^2 \vec{k}_3^2-(\vec{k}_1\times\vec{k}_2)(\vec{k}_3\times\vec{k}_4)}{(\vec{k}_1+\vec{k}_2)^2} \nn\\
\qquad + \frac{\vec{k}_2^2}{(m+E_2)} \left(-2\left(2\frac{\vec{k}_3\cdot\vec{k}_4}{\vec{k}_4^2} +1\right) -4\frac{(\vec{k}_1\times\vec{k}_2)(\vec{k}_3\times\vec{k}_4)}{\vec{k}_2^2\vec{k}_4^2}\right) \left(\frac{1}{m+E_{1+2}} - \frac{E_3+E_4+E_{1+2}}{(\vec{k}_3+\vec{k}_4)^2} \right) \nn\\
\qquad+ \frac{(\vec{k}_3 + \vec{k}_4)^2}{(m+E_{3+4})} \left(\left(2\frac{\vec{k}_1\cdot\vec{k}_2}{\vec{k}_2^2} +1\right) \left(2\frac{\vec{k}_3\cdot\vec{k}_4}{\vec{k}_4^2} +1\right) -4\frac{(\vec{k}_1\times\vec{k}_2)(\vec{k}_3\times\vec{k}_4)}{\vec{k}_2^2\vec{k}_4^2}  \right) \nn\\
\qquad\qquad\qquad \times \left(\frac{1}{m+E_{1+2}} - 2\frac{E_3+E_4+E_{1+2}}{(\vec{k}_3+\vec{k}_4)^2} \right) \Bigg\}  \nn\\
+\O(e)\,,
\end{IEEEeqnarray}
where, for brevity, we use
\be
E_{1+2} =\sqrt{m^2+\left(\vec k_1+\vec k_2\right)^2}\,.
\ee 

There is still a residual gauge freedom in $\Psi_\mathrm{trial}$, which we fix by going to the axial gauge $A_1=0$, in order to be able to actually perform the calculation. For convenience, we shall then call $A_2=A$, which should not be confused with the holomorphic component.

In practice, the computation of a VEV in 2+1 dimensions in the Schr\"odinger picture is then similar to a computation in the path-integral formalism with a complicated two dimensional euclidean effective action $S[A]:=(F_\mathrm{trial}^\dagger[\vec A]+F_\mathrm{trial}[\vec A])_{A_1=0, A_2=A}$:
\bea
\langle\hat{\cal O}[\vec A]\rangle  &=& \frac{ \int\D A_1\D A_2 \de(A_1) e^{-(F_\mathrm{trial}^\dagger[\vec A]+F_\mathrm{trial}[\vec A])}{\cal O}[\vec A]}{ \int\D A_1\D A_2 \de(A_1) e^{-(F_\mathrm{trial}^\dagger[\vec A]+F_\mathrm{trial}[\vec A])}}  = \frac{ \int\D A \,e^{-S[A]}{\cal O}[A]}{ \int\D A \,e^{-S[A]}} \\
&=&  \frac{ \int\D A \,e^{-S^{(0)}[A]-eS^{(1)}[A]-e^2S^{(2)}[A]+\O(e^3)}{\cal O}[A]}{ \int\D A \,e^{-S^{(0)}[A]-eS^{(1)}[A]-e^2S^{(2)}[A]+\O(e^3)}}  \\
&=&  \frac{ \int\D A \,e^{-S^{(0)}[A]}{\cal O}[A]\left(1-eS^{(1)}[A]-e^2S^{(2)}[A]+\frac{e^2}{2}\left( S^{(1)}[A]\right)^2 + \O(e^3)\right)}{ \int\D A \,e^{-S^{(0)}[A]-eS^{(1)}[A]-e^2S^{(2)}[A]+\O(e^3)}}  \,,
\eea
where 
\begin{IEEEeqnarray}{rCl}
S^{(0)} &=& \frac{1}{2} \int_{\slashed{k_1},\slashed{k_2}}\slashed{\delta}\left(\vec{k}_1+\vec{k}_2\right) \frac{2\left(k^{(1)}\right)^2}{m+E_{k_1}}\delta^{ab}  A^a(\vec{k}_1)A^b(\vec{k}_2)\\
S^{(1)} &=& 2i \int_{\slashed{k_1},\slashed{k_2},\slashed{k_3}}\slashed{\delta}\left(\sum_{i=1}^3\vec{k}_i\right) \frac{1}{m+E_3}\frac{k^{(1)}_2 k^{(1)}_3}{\vec{k}_1^2} \left(k^{(2)}_1 -\frac{k^{(1)}_1 \vec{k}_1\times \vec{k}_2}{(\sum_{i=1}^3 E_i) (m+E_2)}\right) \nn\\
&&\qquad\qquad \times f^{abc}  A^a(\vec{k}_1) A^b(\vec{k}_2) A^c(\vec{k}_3) \label{S1} \\
S^{(2)} &=& -\tilde N {C_A\over 2\pi}\int_{\slashed{k_1},\slashed{k_2}}\slashed{\delta}\left(\vec{k}_1+\vec{k}_2\right) \frac{\left(k^{(1)}_1\right)^2}{m^2} \delta^{ab}  A^a(\vec{k}_1)A^b(\vec{k}_2)  +\O(\bar k^2 \vec k^2)\nn\\
&&+\int_{\slashed{k_1},\slashed{k_2},\slashed{k_3},\slashed{k_4}} \slashed{\delta}\left(\sum_{i=1}^4\vec{k}_i\right) f^{a_1a_2c} f^{b_1b_2c} A^{a_1}(\vec{k}_1)A^{a_2}(\vec{k}_2)A^{b_1}(\vec{k}_3)A^{b_2}(\vec{k}_4) \nn\\
&&\Bigg\{-\left(\frac{1}{m+E_4}-\frac{1}{m+E_{3+4}}\right) \frac{k^{(1)}_2 k^{(1)}_4}{\vec{k}_1^2\vec{k}_3^2} \left(k_1^{(2)}k_3^{(2)}-k_1^{(1)}k_3^{(1)}\right) \nn\\
&&\qquad -\frac{1}{m+E_2}\frac{k^{(1)}_1 k^{(1)}_2}{(\vec{k}_3+\vec{k}_4)^2 \vec{k}_4^2}\left((k_3^{(2)}+k_4^{(2)})k_4^{(2)}-(k_3^{(1)}+k_4^{(1)})k_4^{(1)}\right) \nn\\
&&+\frac{2}{(E_{1+2} +E_3 + E_4)(m + E_3)} \frac{k_2^{(1)}k_3^{(1)}k_4^{(1)}}{\vec{k}_1^2} \Bigg[\frac{1}{(m+E_4)(\vec{k}_1+\vec{k}_2)^2\vec{k}_2^2}\nn\\
&&\qquad\times \Bigg( (\vec{k}_2\cdot\vec{k}_4)((2\vec{k}_1\cdot\vec{k}_2+\vec{k}_2^2)\; k_1^{(1)}-\vec{k}_1^2\; k_2^{(1)}) + (\vec{k}_2\times\vec{k}_4)((2\vec{k}_1\cdot\vec{k}_2+\vec{k}_2^2)\; k_1^{(2)}- \vec{k}_1^2\; k_2^{(2)})\Bigg)  \nn\\
&&\quad + \frac{1}{(m+E_{1+2})\vec{k}_4^2}\Bigg(-\vec{k}_1^2 k^{(1)}_4 +((\vec{k}_3-\vec{k}_2)\cdot\vec{k}_4) k_1^{(1)}- ((\vec{k}_3-\vec{k}_2)\times\vec{k}_4) k_1^{(2)} \Bigg) \Bigg] \nn\\
&&+\frac{k_1^{(1)}k_2^{(1)}k_3^{(1)}k_4^{(1)}}{(\sum_i E_i) (E_1 + E_2 + E_{3+4})(E_3 + E_4 + E_{1+2})(m+E_1)(m+E_3)} \nn\\
&& \quad \Bigg[\frac{1}{(m+E_2)(m+E_4)} \frac{\vec{k}_1^2 \vec{k}_3^2-(\vec{k}_1\times\vec{k}_2)(\vec{k}_3\times\vec{k}_4)}{(\vec{k}_1+\vec{k}_2)^2} \nn\\
&& \qquad +\frac{\vec{k}_2^2}{(m+E_2)} \left(-2\left(2\frac{\vec{k}_3\cdot\vec{k}_4}{\vec{k}_4^2} +1\right) -4\frac{(\vec{k}_1\times\vec{k}_2)(\vec{k}_3\times\vec{k}_4)}{\vec{k}_2^2\vec{k}_4^2}\right) \nn\\
&&\qquad\qquad\times \left(\frac{1}{m+E_{1+2}} - \frac{E_3+E_4+E_{1+2}}{(\vec{k}_3+\vec{k}_4)^2} \right) \nn\\
&& \qquad +\frac{(\vec{k}_3 + \vec{k}_4)^2}{(m+E_{3+4})} \left(\left(2\frac{\vec{k}_1\cdot\vec{k}_2}{\vec{k}_2^2} +1\right) \left(2\frac{\vec{k}_3\cdot\vec{k}_4}{\vec{k}_4^2} +1\right) -4\frac{(\vec{k}_1\times\vec{k}_2)(\vec{k}_3\times\vec{k}_4)}{\vec{k}_2^2\vec{k}_4^2}  \right) \nn\\
&& \qquad\qquad\times \left(\frac{1}{m+E_{1+2}} - 2\frac{E_3+E_4+E_{1+2}}{(\vec{k}_3+\vec{k}_4)^2} \right) \Bigg] \Bigg\}\,, \label{S2}
\end{IEEEeqnarray}
and $k_j^{(i)}$ indicates component $i$ of vector $\vec k_j$.

In terms of this effective action the LO correlator of the $A$ field is thus
\be
\langle A^a(\vec{x})A^b(\vec{y}) \rangle = \int_\slashed{k} \frac{m+E_{k}}{2\left(k^{(1)}\right)^2} e^{i\vec k\cdot(\vec x -\vec y)}\delta^{ab} \,. \label{Aprop}
\ee


\section{The magnetic field correlator and the gluon condensate at leading order}
\label{sec:condensate}
As a warm up we will calculate the correlator of the chromomagnetic field
\be
\langle B^a(\vec x)\Phi^{ab}(\vec x, \vec y)B^b(\vec y)\rangle
\ee
at leading order, which is a special case of the field strength correlator
\be
D_{\mu\nu\alpha\beta}(x, y):=\langle G_{\mu\nu}^a(x)\Phi^{ab}(x, y)G_{\alpha\beta}^b(y)\rangle \,.
\ee
 This is an interesting object, since it appears in non-perturbative models of QCD (see e.g.~\cite{Dosch1994}), in the gluelump spectrum (\cite{Bali:2003jq}) and in the hybrid static potential (\cite{Brambilla:1999xf}, \cite{Pineda:2011db}). In order to compute the chromomagnetic field correlator in the Schr\"odinger picture, we make use of Eq.~(\ref{Aprop}). At leading order, where $\Phi^{ab}(\vec x, \vec y)=\de^{ab}$ and in the $A_1=0$ gauge, the correlator reads
\bE{rCl}
\langle B^a(\vec x)\Phi^{ab}(\vec x, \vec y)B^b(\vec y)\rangle &=& \p_1^x\p_1^y\de^{ab}\langle A^a(\vec{x})A^b(\vec{y}) \rangle +\O\left(e^2/m\right) \,, \label{BBcor}
\eE
and with Eq.~(\ref{Aprop}) this is computed to 
\bE{rCl}
\langle B^a(\vec x)\Phi^{ab}(\vec x, \vec y)B^b(\vec y)\rangle&=& -\frac{e^{-m |\vec z|} (1+m |\vec z|)}{4\pi |\vec z|^3}(N_c^2-1) +\O\left(e^2/m\right) \,, \label{GCKNY}
\eE
where $\vec z = \vec x-\vec y$.

We can compare this to the leading order result of the field strength correlator $D_{\mu\nu\alpha\beta}(x, y) = D_{\mu\nu\alpha\beta}(z)$, calculated in the operator approach in Ref.~\cite{Eidemuller:1997bb}. From the Lorentz structure of this object it is clear that it can be written as
\bE{rCl}
D_{\mu\nu\alpha\beta}(z) &=& (\eta_{\mu\al}\eta_{\nu\beta}-\eta_{\mu\beta}\eta_{\nu\al})(D_0(z^2)+D_1(z^2)) \nn\\
&&+  (\eta_{\mu\al}z_\nu z_\beta-\eta_{\mu\beta}z_\nu z_\al - \eta_{\nu\al}z_\mu z_\beta +\eta_{\nu\beta}z_\mu z_\al) \frac{\p}{\p z^2} D_1(z^2)\,,
\eE
see Ref.~\cite{Eidemuller:1997bb} and references therein. There the LO in perturbation theory was computed in $D=4-2\e$ dimensions. Taking $\epsilon\to\half$ instead of 0, the 2+1 dimensional result is 
\bE{rCl}
D_{\mu\nu\alpha\beta}(z) &=& (\eta_{\mu\al}\eta_{\nu\beta}-\eta_{\mu\beta}\eta_{\nu\al}) \frac{N_c^2-1}{2 \pi  (z^2)^\frac{3}{2} }  \nn\\
&&+  (\eta_{\mu\al}z_\nu z_\beta-\eta_{\mu\beta}z_\nu z_\al - \eta_{\nu\al}z_\mu z_\beta +\eta_{\nu\beta}z_\mu z_\al) \frac{-3}{4 \pi  (z^2)^\frac{5}{2} }(N_c^2-1)  +\O(e)  \,. \label{GCJamin}
\eE
With $B^a=\half\e_{ij}G_{ij}^a$, and taking $x$ and $y$ at the same time, we can compare this to Eq.~(\ref{GCKNY}):
\bE{rCl}
\langle B^a(\vec x)\Phi^{ab}(\vec x, \vec y)B^b(\vec y)\rangle&=& \frac{1}{4}\e_{ij}\e_{kl}D_{ijkl}(z_0=0,\vec z) =-\frac{1}{4\pi |\vec z|^3}(N_c^2-1) +\O(e) \,.
\eE
This is equal to the $m\to0$ limit of Eq.~(\ref{GCKNY}).

The gluon condensate is the $z\to0$ limit of $D_{\mu\nu}^{\phantom{\mu\nu}\mu\nu}(z)$ and is also relevant for non-perturbative QCD models (\cite{Shifman:1978bx}). In three dimensions it appears in the computation of the three dimensional static potential (see Ref.~\cite{Pineda:2010mb}) as well as in the computation of the thermodynamic pressure of four dimensional QCD, being hence relevant for the expansion rate of the universe (see e.g.~\cite{Miller:2006hr}).  The chromomagnetic part of 
\be
\left\langle G^{\mu\nu,a}(\vec x)G_{\mu\nu}^a(\vec x) \right\rangle = -2 \left\langle \left(\vec E^a(\vec x)\right)^2 - \left(B^a(\vec x)\right)^2 \right\rangle
\ee
can also be computed from Eq.~(\ref{BBcor}) by first taking the limit of $\vec z\to0$, and then doing the integration over $\vec k$, making use of dimensional regularization to eliminate the power divergences:
\be
\left\langle \left(B^a(\vec x)\right)^2 \right\rangle = \half \de^{aa} \int\frac{d^d k}{(2\pi)^d}\left(m+\sqrt{m^2+\vec k^2}\right) \Bigg|_{d\to2} = -\frac{m^3}{12\pi}(N_c^2-1) +\O\left(e^2/m\right)\,.
\ee

The VEV of the squared chromoelectric field can be computed by taking the functional derivative of the wave functional. In this case, one can set $A_1=0$ only after the derivative has been taken:
\bE{rCl}
\left\langle \left(\vec E^a(\vec x)\right)^2 \right\rangle &=& \frac{ \int\D A_1\D A_2 \de(A_1) \Psi_\mathrm{trial}^*[\vec A]\left(-\frac{\de^2}{(\de A_i^a)^2}\right)  \Psi_\mathrm{trial}[\vec A]}{ \int\D A_1\D A_2 \de(A_1)  \Psi_\mathrm{trial}^*[\vec A] \Psi_\mathrm{trial}[\vec A]} \\
 &=& - \int_{\slashed{p},\slashed{q}} e^{i(\vec p+ \vec q)\cdot\vec x} \frac{(\vec p\cdot\vec q)\ p_1q_1}{(m+E_p)(m+E_q)}  \langle A^a(\vec p) A^a(\vec q) \rangle \\
 &=& \frac{m^3}{12\pi}(N_c^2-1) +\O\left(e^2/m\right)\,.
\eE
Summing up both contributions leaves us with
\bE{rCl}
\left\langle G^{\mu\nu,a}(\vec x)G_{\mu\nu}^a(\vec x) \right\rangle &=& -\frac{m^3}{3\pi}(N_c^2-1)  +\O\left(e^2/m\right)
\,.
\eE
This has the same color structure and is numerically in the same ballpark as the finite term of the result of Ref.~\cite{DiRenzo:2006nh}.

\newpage
\section{The static potential at leading and next-to-leading order}
\label{sec:potential}

After this brief illustration we shall now come to the main point of this chapter: The computation of the static potential up to NLO. The usual way to do this is by calculation of the VEV of a rectangular Wilson loop $\langle W_\square \rangle$ with sides of lengths $r$ and $T$. The static potential at long distances $r$ is then given by
\be
E_s(r)=-\lim_{T\to\infty}\frac{1}{T}\ln\langle W_\square \rangle\,.
\ee
If the VEV of the Wilson loop satisfies the famous area law, i.e.
\be
\langle W_\square \rangle\propto\exp(-\sigma \A_\square)\,,
\ee
where $\A_\square=r\times T$ is the area enclosed by the Wilson loop, then the potential is linear in $r$ and thus confining. While a dependence on higher powers in $r$ would still result in confinement, it was shown in Ref.~\cite{Seiler:1978ur} that, in principle, the potential cannot rise faster than linearly in the limit of $r\to\infty$, and lattice calculations (see e.g.~Refs.~\cite{Meyer:2006gm,HariDass:2007tx}) confirm that this is the actual behavior in 2+1 dimensions.

We will follow a different, but equivalent approach here. First it will be convenient to consider one of the two euclidean dimensions as time, so we choose $x_2=it$ purely imaginary. We then add static fermionic sources in the (anti-)fundamental representation of color, which can be thought of as heavy (anti-)quarks, to the gluonic action
\bE{rCl}
S_\mathrm{tot}&=&S_\mathrm{stat}[\psi, \chi, A]+S_\mathrm{gl}[A] \\
&=& \int\left(\psi^\dagger (i\p_2+eA_2)\psi + \chi_c^\dagger (i\p_2-eA_2^T)\chi_c\right) +\left(F_\mathrm{trial}^\dagger[\vec A]+F_\mathrm{trial}[\vec A]\right)\Bigg|_{A_1=0,A_2=A} \label{Stot} 
\\
&=& \int\left(\psi^\dagger (i\p_2+eA)\psi + \chi_c^\dagger (i\p_2-eA^T)\chi_c\right) \nn\\
&& + \frac{1}{2} \int_{\slashed{k_1},\slashed{k_2}}\slashed{\delta}\left(\vec{k}_1+\vec{k}_2\right) \frac{2\left(k^{(1)}\right)^2}{m+E_{k_1}}\delta^{ab}  A^a(\vec{k}_1)A^b(\vec{k}_2) \nn\\
&& \raisebox{-0.2cm}{$+ eS^{(1)}[A] + e^2S^{(2)}[A] \,,$}
\eE
where $\psi$ is the Pauli spinor field that annihilates the fermion and $\chi_c=C\chi^*$ (with $C$ being the charge conjugation matrix) is the Pauli spinor field that annihilates the antifermion. $S^{(1)}[A]$ and $S^{(2)}[A]$ are given in Eqs.~(\ref{S1}) and (\ref{S2}), respectively.

We create a color singlet state $S(x; x'; t)$ depending on time $t$ as well as (anti-)quark positions $x$ and $x'$:
\be
|S(x; x'; t) \rangle 
= \psi^\dagger(x,t)\Phi(x,x')\chi(x',t) |\Psi_0\rangle \,,
\ee
where $\Phi(x, x')$ is the Wilson line in the fundamental representation. We consider the time evolution amplitude $I(T)$ of $S$:
\bE{rCl}
I(T)&=&\langle S(y; y'; 0) | S(x; x'; T)\rangle  \\
&=& \langle S(y; y'; 0) | e^{-i\H T} |S(x; x'; 0)\rangle \\
&=& \sum_n \langle S(y; y'; 0) | \Psi_n\rangle\langle \Psi_n |S(x; x'; 0)\rangle e^{-iE_n T} \\
&\stackrel{T\to\infty}{\longrightarrow}& \langle S(y; y'; 0) | \Psi_0\rangle\langle \Psi_0 |S(x; x'; 0)\rangle e^{-iE_0 T}
\eE
Since the sources are static, we can take $|x-x'|=|y-y'|=:r$ and find the static potential as
\be
E_s(r)=\lim_{T\to\infty}\frac{i}{T}\ln I(T) \,.
\ee
This computation is equivalent to the approach using the VEV of a rectangular Wilson loop, see Ref.~\cite{Lucha:1991vn}.

Making use of the framework of potential non-relativistic QCD (pNRQCD), developed in Ref.~\cite{Pineda:1997bj} (for a review see Ref.~\cite{Brambilla:2004jw}), we will match this computation on an effective theory with
\be
\mathcal{L} = S^\dagger(i\p_2+E_s(r))S\,. \label{EFT}
\ee
We can compute the potential in momentum space order by order in $e^2$. Since we are
interested in very long distances this is equivalent to very low (external) momentum $q\to0$. This will
simplify the computation, yet there are two momentum scales that have to be considered in the
loops: $k \sim q$ (soft) and $k \sim m$ (hard).

\subsection{The leading order}

At tree-level, there is only one diagram:

\begin{figure}[h]
\centering
\includegraphics[width=0.25\textwidth]{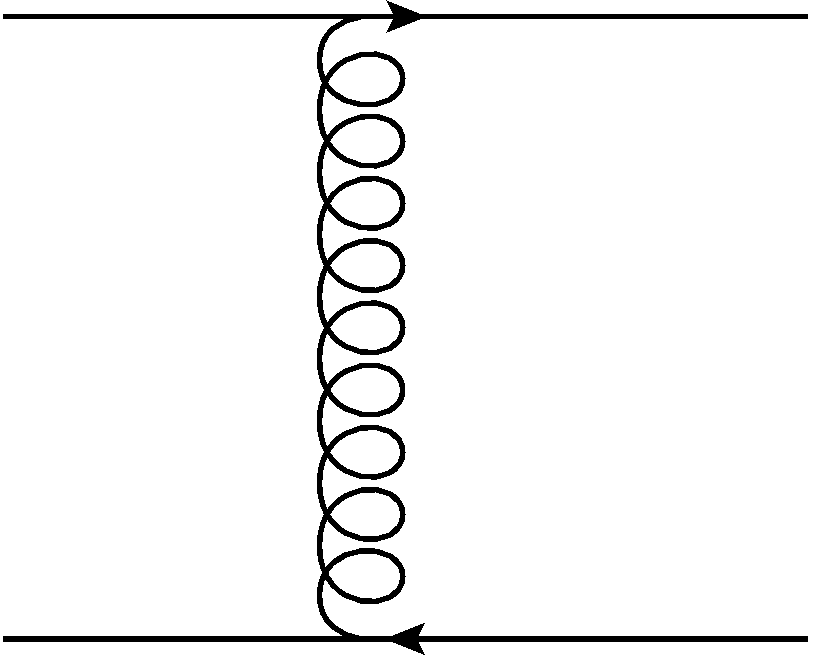}
\caption{Tree-level gluon exchange.}
\end{figure}

In momentum space, at leading order in the exchanged momentum, the result is given by
\be
\tilde E_s^{(0)} = -e^2 C_F \frac{m}{q_1^2} + \O(q^0)\,, \label{tree-level-potential}
\ee
which in position space becomes the sought after linear potential
\be
E_s^{(0)} = \frac{e^2}{2}m C_F r\,,
\ee
with a string tension of
\be
\sigma =\frac{e^4C_AC_F}{4\pi} \,.
\ee
Unsurprisingly, we just reproduced the result of Ref.~\cite{Karabali:1998yq}, using a modified approach.

\newpage
\subsection{The next-to-leading order}
\label{sec:NLO}

At NLO there is a variety of diagrams:

\begin{figure}[!ht]
\begin{tabular}[c]{cccc}
a) \includegraphics[width=0.19\textwidth]{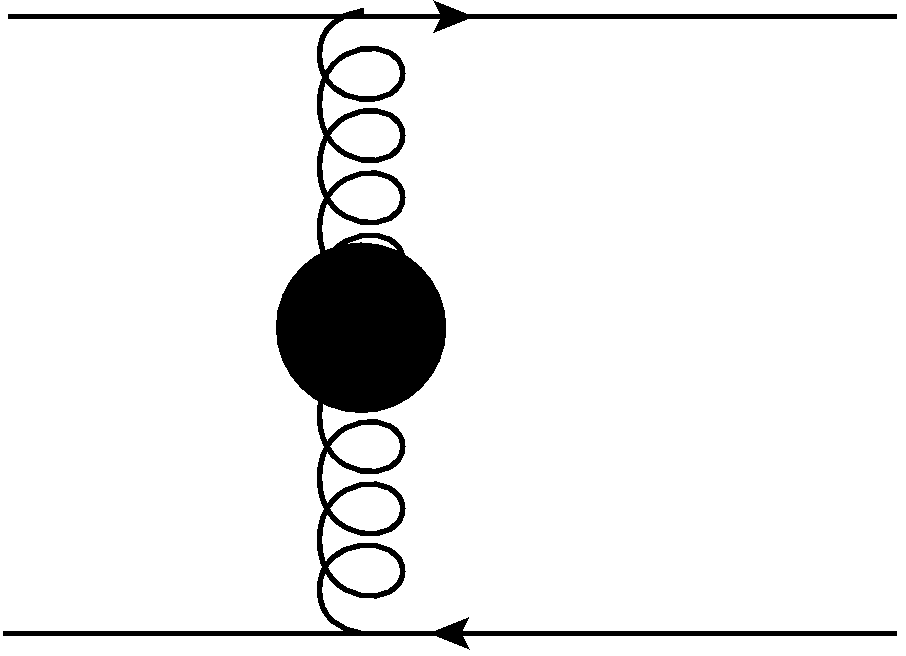} & \multicolumn{2}{c}{ b) \includegraphics[width=0.19\textwidth]{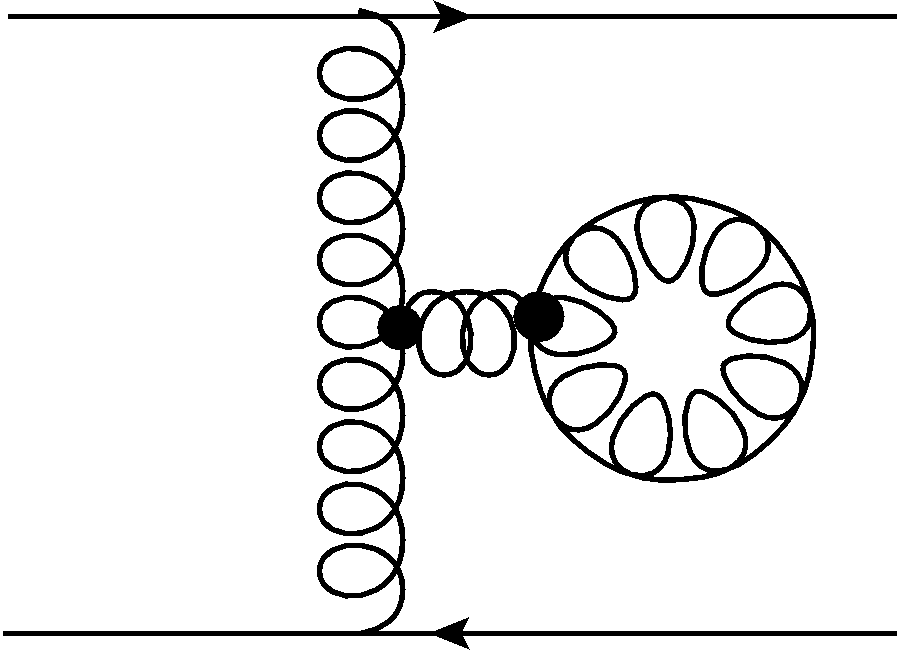}}  & c) \includegraphics[width=0.19\textwidth]{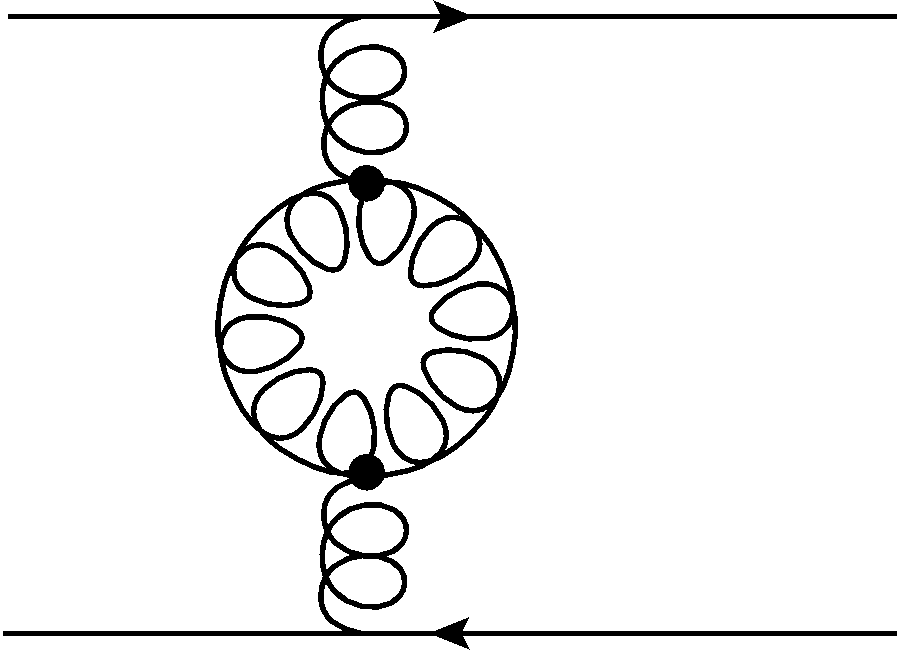}   \\
&&&\\
d) \includegraphics[width=0.19\textwidth]{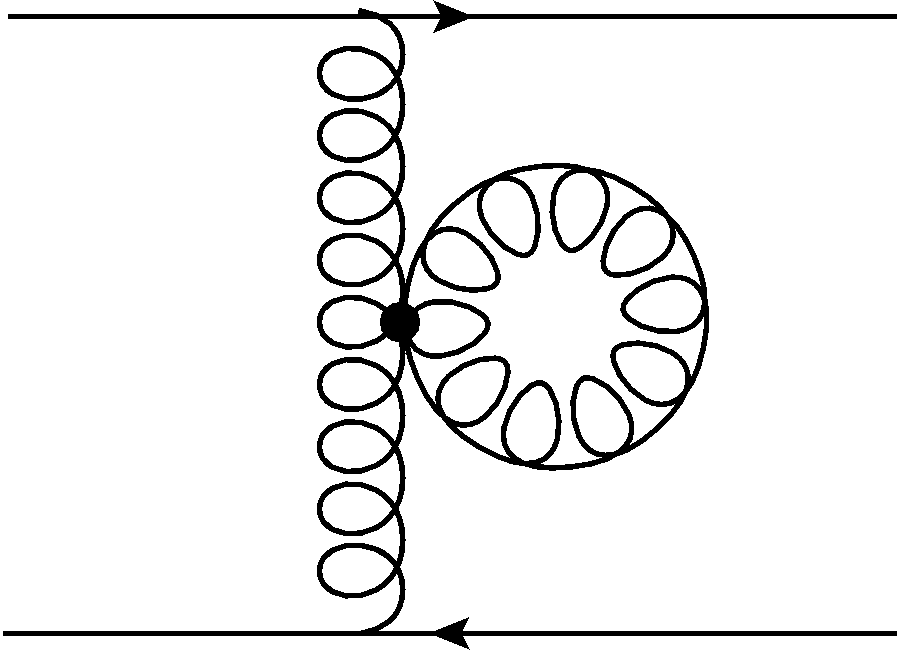} &  \multicolumn{2}{c}{ e) \includegraphics[width=0.19\textwidth]{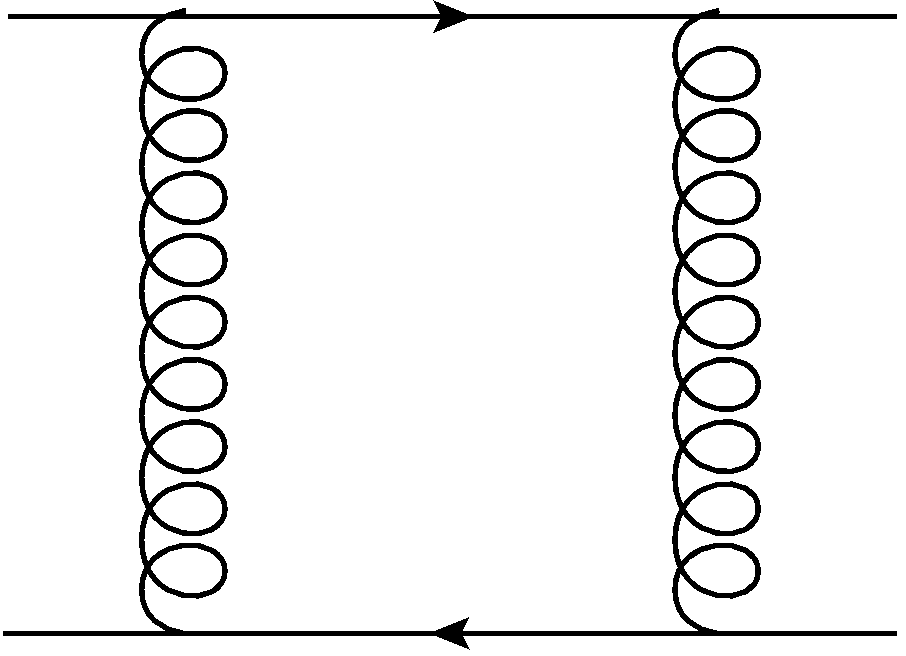} } & f) \includegraphics[width=0.19\textwidth]{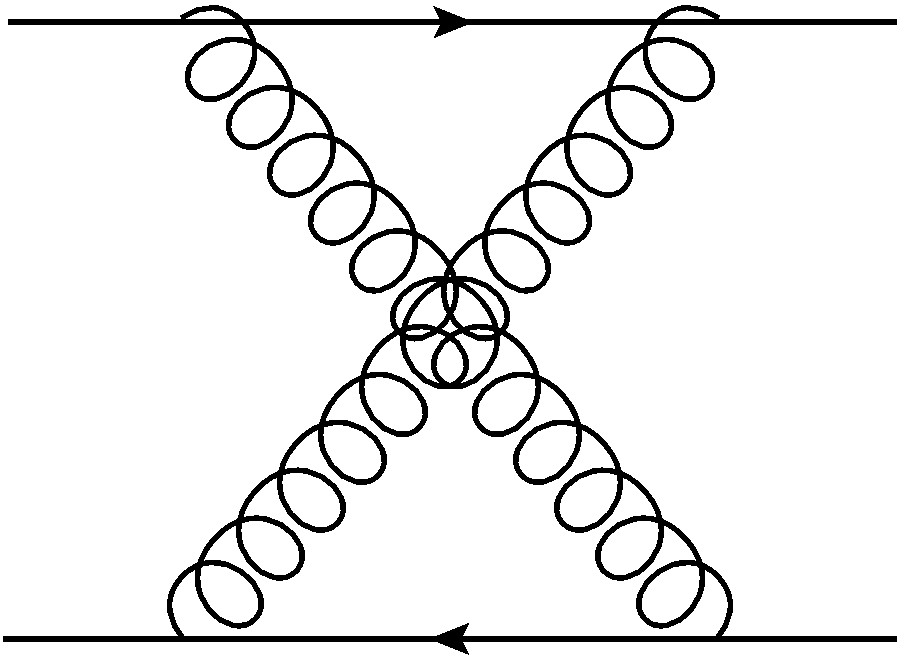}  \\
&&&\\
g) \includegraphics[width=0.19\textwidth]{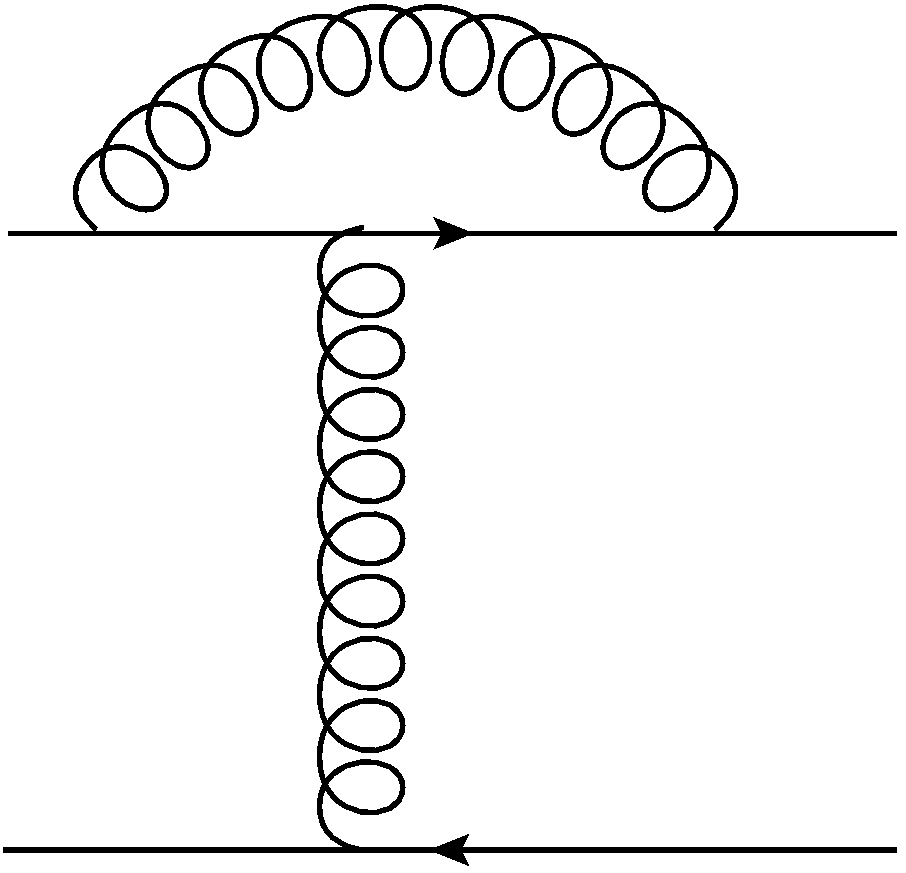} & h) \raisebox{-0.8cm}{\includegraphics[width=0.19\textwidth]{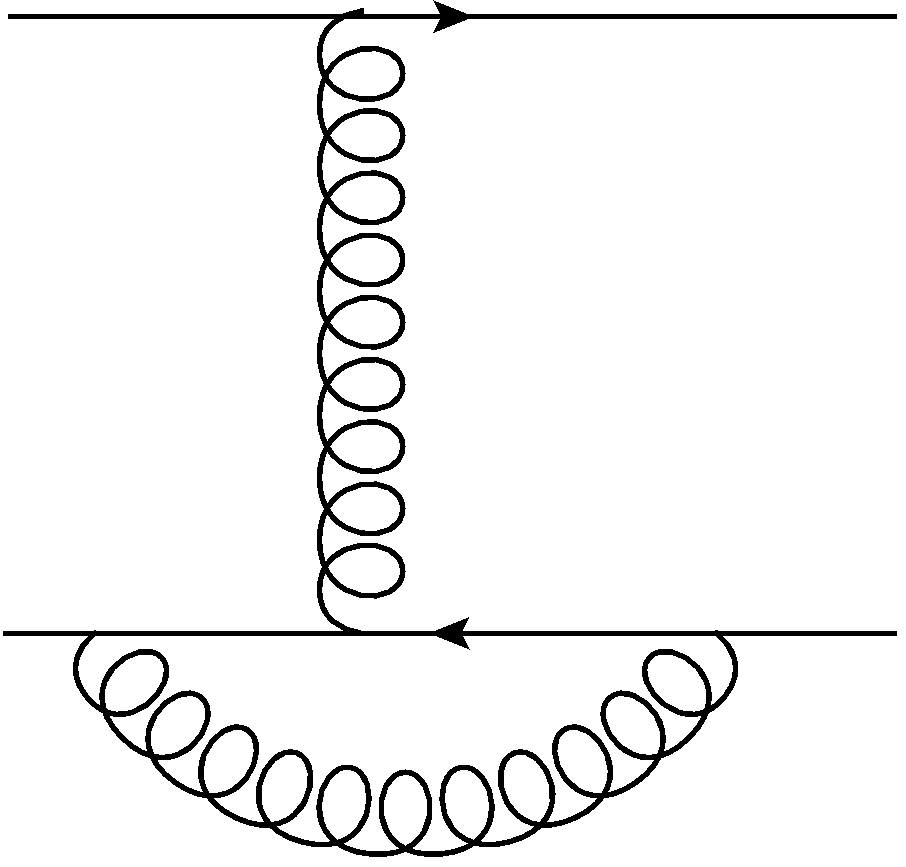}} & i) \includegraphics[width=0.19\textwidth]{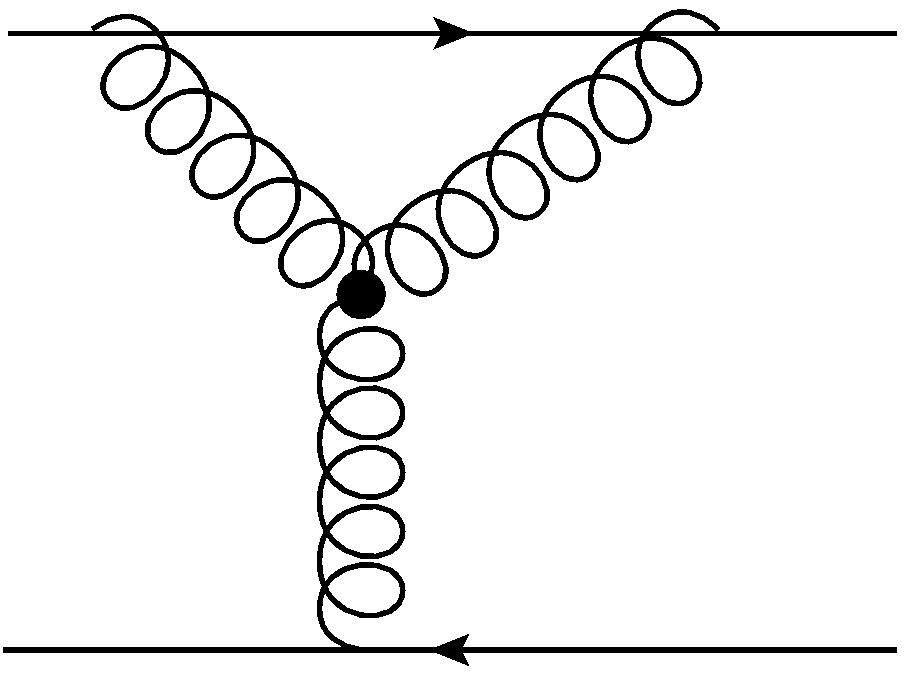}&  j) \includegraphics[width=0.19\textwidth]{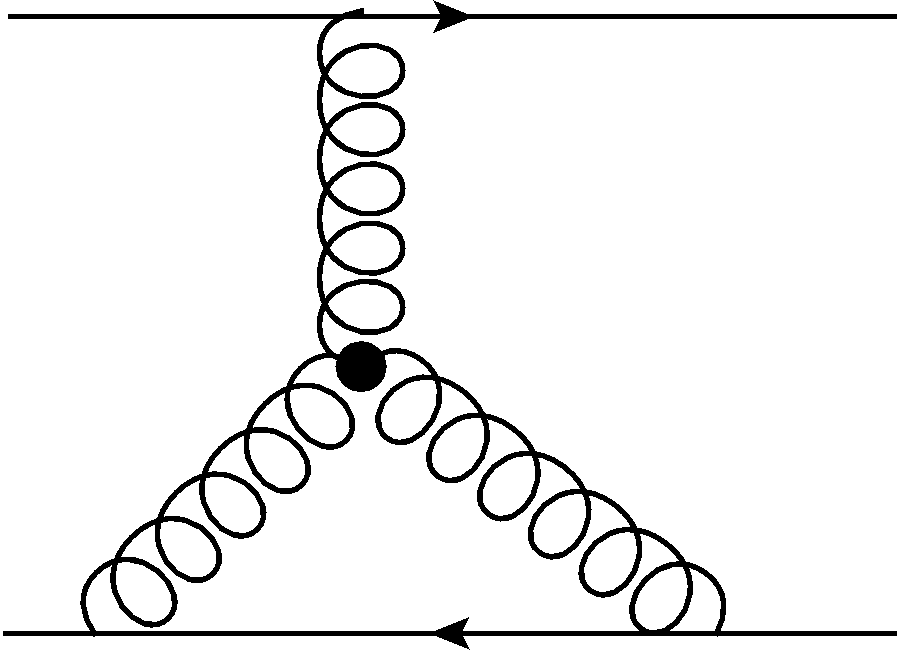} \\
k) \raisebox{-0.8cm}{\includegraphics[width=0.19\textwidth]{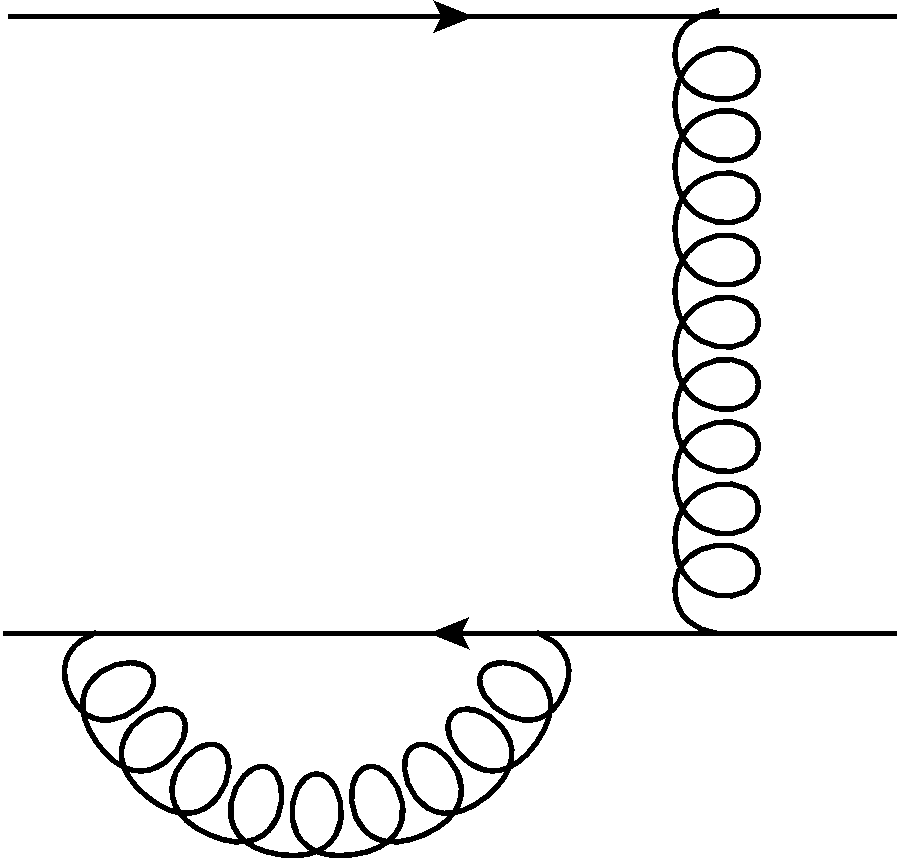}} & l) \raisebox{-0.8cm}{\includegraphics[width=0.19\textwidth]{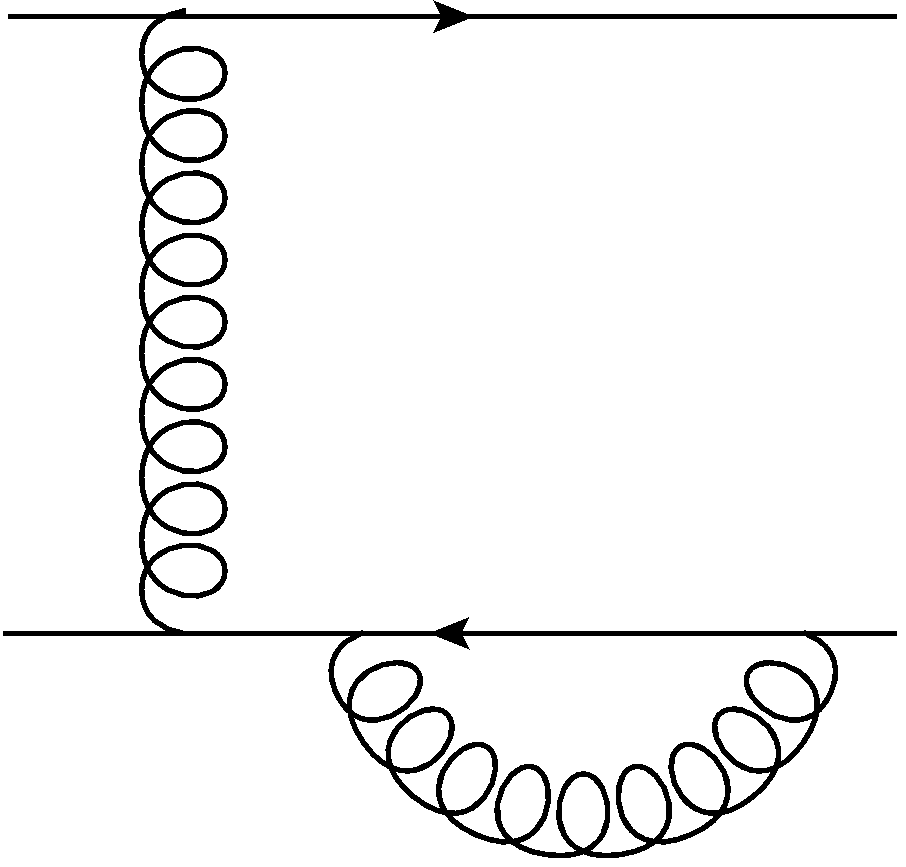}} & m) \includegraphics[width=0.19\textwidth]{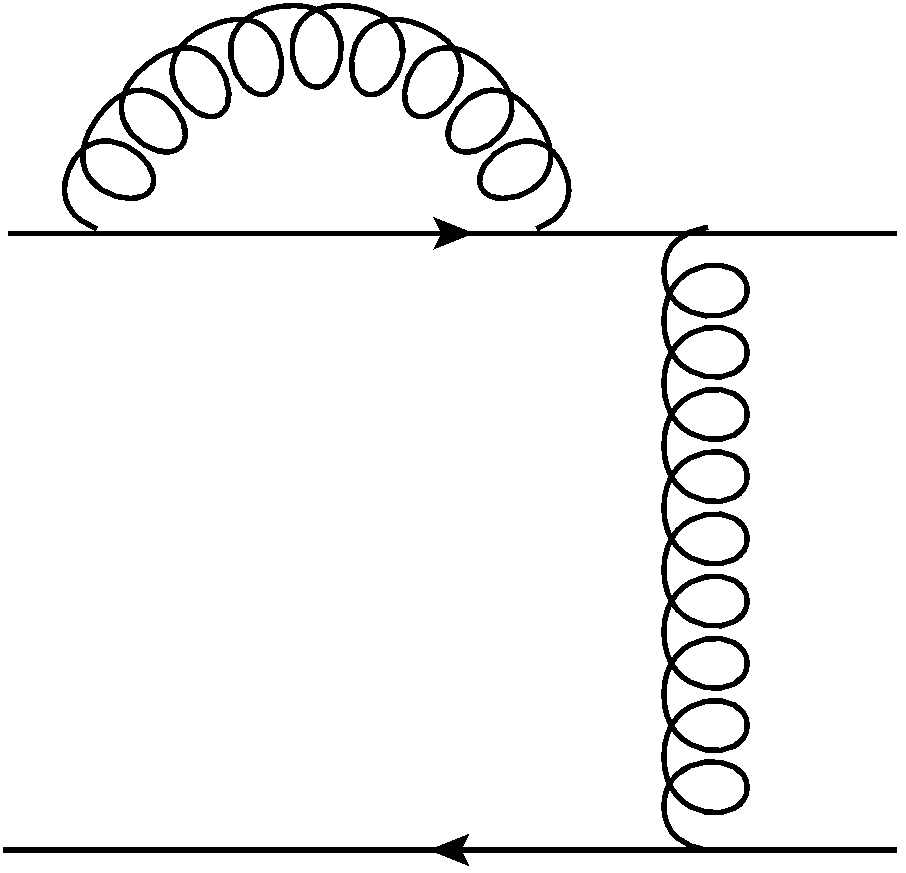} & n) \includegraphics[width=0.19\textwidth]{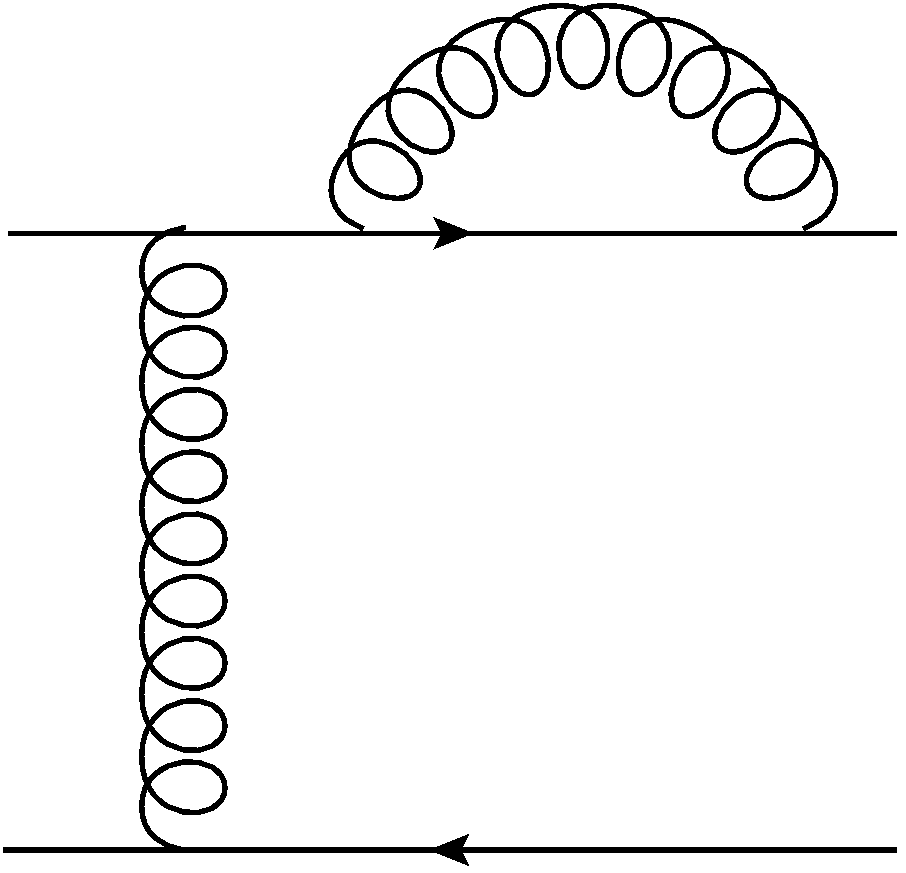}\\
\end{tabular}
\caption{Diagrams a) to n) of the NLO.}
\end{figure}

Except for diagram a), all of these diagrams have a hard ($k \sim m$) and a soft ($k \sim q$) contribution. Once again, in order to keep the chapter uncluttered, we will give the actual computations in the appendix (App.~\ref{app:Diagrams}), and only present the results here.

\bigskip
In diagram a) we insert the $\O(e^2A^2)$ term of the action (the first term of Eq.~(\ref{S2})) as a 2-field vertex. We are only interested in the leading order in the exchanged momentum, which is why we already expanded the term in $k/m$ in Eq.~(\ref{f22expanded}). The result of this diagram is thus
\be
\de \tilde E_s^{(1),a} =  -2\tilde N e^2 C_F \frac{m}{q_1^2} \frac{e^2C_A}{2\pi m} + \O(q^0)\,.
\ee
While this strongly modifies the numeric value of the leading order result (since $2 \tilde N\approx 1.1308$), it maintains the linear dependence of the potential on the separation. This diagram is equivalent to the correction found in Eq.~(34) of Ref.~\cite{Karabali:2009rg}. The other diagrams cannot be related in such a direct fashion to the computation done there.

\medskip
Diagram b) does not contribute, neither in the hard nor in the soft regime, because the 3-field vertex is proportional to $f^{abc}$, while the propagator in the loop is proportional to $\de^{ab}$, hence the diagram vanishes:
\be
\de\tilde E_s^{(1),b} = 0\,.
\ee

\medskip
Diagram c) has to be evaluated in the soft and in the hard regime. In the hard regime we find
\be
\de\tilde E_s^{(1),c} = e^2 C_F \frac{m}{q_1^2} \left[K^{(3)} \frac{4\pi m^2}{q_1^2}+\O(q_1^0)\right] \frac{e^2C_A}{2\pi m}\,,
\ee
with
\be
K^{(3)} = -\half\int_\slashed{k} \frac{E_k-m}{(m+E_k)(m+2 E_k)^2}\,,
\ee
giving a non-linear contribution for the potential.

\medskip
Diagram d) also has to be computed in both regimes. The contribution of the hard regime is
\be
\de\tilde E_s^{(1),d} = e^2 C_F \frac{m}{q_1^2} \left[K^{(4)} \frac{4\pi m^2}{q_1^2}+\O(q_1^0)\right] \frac{e^2C_A}{2\pi m}\,,
\ee
with
\be
K^{(4)} = \half\int_\slashed{k} \left\{ \frac{m^3+3m^2E_k+mE_k^2-E_k^3}{4E_k^3(m+E_k)^2} -2\frac{2m^3+4m^2E_k+mE_k^2-E_k^3}{(m+2 E_k)^2E_k(m+E_k)^2} \right\}\,.
\ee

While $K^{(3)}$ and $K^{(4)}$ are divergent quantities in $d=2$, their divergences cancel once they are added up and we obtain
\be
K^{(3)}+K^{(4)} = -\frac{2+\log(3/8)}{16\pi}\,.
\ee
 So we find that the NLO term in the $e^2/m$ expansion leads to terms of $\O\left(q_1^{-4}\right)$:
\be
\de\tilde E_s^{(1),c} + \de\tilde E_s^{(1),d} = -e^2 C_F \frac{(2+\log(3/8))}{4} \frac{m^3}{q_1^4}  \frac{e^2C_A}{2\pi m} +\O\left(q^{-2}\right) \,. \label{q-4}
\ee
In position space this corresponds to a term cubic in the separation. This is in contradiction to the result of Ref.~\cite{Karabali:2009rg}, where only linear contributions were found at NLO. Moreover it crushes our hope of computing the string tension from first principles, at least with this wave functional.
The only remedy would be more terms of $\O\left(q_1^{-4}\right)$ that cancel Eq.~(\ref{q-4}), coming from the other diagrams. Unfortunately, however, this does not happen (see App.~\ref{app:Diagrams} for details):

\begin{itemize}
\item Both diagrams c) and d) are of $\O\left(q_1^{-2}\right)$ in the soft regime.

\item The soft contribution of diagram e) is the iteration of the potential $E_s(r)$ of Eq.~(\ref{EFT}) and its hard contribution is of $\O\left(q_1^0\right)$. 

\item Diagram f) vanishes in the soft regime, while in the hard regime it is of $\O\left(q_1^0\right)$. 

\item Diagrams g) and h) are of $\O\left(q_1^{-2}\right)$ in both the soft and the hard regime.

\item The same is true for diagrams i) and j).

\item Diagrams k) to n) are all of $\O\left(q_1^{-2}\right)$ in both the soft and the hard regime.
%
\end{itemize}
So we conclude that the $e^2/m$ correction to the static potential of Eq.~(\ref{tree-level-potential}) is of $\O\left(q_1^{-4}\right)$, or in position space, of $\O\left(r^3\right)$:
\bE{rCCl}
\tilde E_s &= -e^2 C_F \frac{m}{q_1^2} &\left[1 + \frac{e^2}{m}\left(C_A \frac{(2+\log(3/8))}{8\pi} \frac{m^2}{q_1^2}  +\O\left(q^0\right) \right) \right]& +\O\left(\left( \frac{e^2}{m}\right)^2\right) \nn\\
\Longleftrightarrow E_s &= \frac{e^2}{2}m C_F r &\left[1 - \frac{e^2}{m}\left( C_A \frac{(2+\log(3/8))}{48\pi} m^2r^2  +\O\left(r^0\right) \right)\right]&  +\O\left(\left( \frac{e^2}{m}\right)^2\right) \,. \label{corrected_potential}
\eE
This is not a fundamental problem, as these higher order terms should combine in such a way that the sum grows at most linearly for long distances. It could also be that they are canceled by higher order terms in the expansion in $e^2/m$, which as mentioned before is of $\O(1)$. The possible conclusions to be drawn from Eq.~(\ref{corrected_potential}) are, however, either that the expansion in $e^2/m$ is not helpful in proving confinement, or that $\Psi_{KNY}[J]$ does not have the correct long distance behavior. This last possibility is supported by our findings of Chap.~\ref{chap:Regularization}. Of course, it is also possible that both explanations are true.

\section{Conclusions}
\label{sec:ConclusionsPot}

In this chapter we have explored the non-perturbative regime and we have illustrated how observables can be computed from the vacuum wave functional. Mathematically the computation is identical to a computation in the path integral formalism, where the action is given by the exponent of the vacuum wave functional. Effectively the problem is reduced to a calculation in two euclidean dimensions, the price to pay, however, is that it has to be done with a very complicated action.

First we looked at the potential term $\V$ of the Yang-Mills Hamiltonian and found that also in a perturbative expansion in terms of the gauge fields, it is an eigenfunction of the kinetic term $\T$. Regularization of both terms is crucial for this property to exhibit itself. It was used in Ref.~\cite{Karabali:1998yq} as the starting point of a strong coupling expansion of the vacuum wave functional. We find, however, that the eigenvalue depends on the regulators, which might make its use in the determination of the vacuum wave functional problematic. This issue should be clarified before relying on a strong coupling expansion along these lines.

In a second step, we then used a trial wave functional obtained from $\Psi_{KNY}[J]$, which was proposed in Ref.~\cite{Karabali:2009rg}, via a transformation of the field variables from gauge invariant currents $J^a$ to gluon fields $\vec A^a$. This functional is claimed to be a good approximation at all scales. It is given as a series with $e^2/m\sim\O(1)$ as expansion parameter, and it enables us to estimate the correlator of the chromomagnetic field, the gluon condensate, and the string tension. At LO the result for the gluon condensate is in reasonable agreement with the result of Ref.~\cite{DiRenzo:2006nh}, and the string tension agrees exactly with the result of \cite{Karabali:2009rg}, which itself is in very good agreement with results from lattice computations. At NLO, however, we run into trouble. While the NLO computation of the static potential in Ref.~\cite{Karabali:2009rg} produced only terms compatible with a linear potential, we do find terms that are cubic in the separation. We presume that this difference is due to one or several of the problems of the calculation done there: First, not all contributions at NLO were computed, second, some of the contributions were ambiguous, and third, the wave functional was assumed to be real, which is true for the complete functional, but not for the individual terms of this approximation.

The failure to produce only linear terms at $\O\left(e^2/m\right)$ in this computation leads us to two possible conclusions. The obvious one seems to be that $e^2/m$ is not a good expansion parameter, given that it is of $\O(1)$. On the one hand this has the effect that, like in ordinary (non-resummed) perturbation theory, higher order terms appear which then should add up to the linear potential. On the other hand, even if only linear terms were found, we could never be sure of the numeric value of the string tension, as higher order terms might give big contributions to it, so the justification for its use could only be given a posteriori, when comparing with results obtained with lattice calculations. Still, given the success of the LO result and the tantalizing outlook to compute the string tension analytically from first principles keeps us from dismissing this approach. 

In light of our conclusions of Chap.~\ref{chap:Regularization}, a second (not necessarily exclusive) explanation for the appearance of the cubic term comes to mind: As we had to use a different regularization for method (B) than the one used in Ref.~\cite{Karabali:2009rg} to obtain the correct vacuum wave functional at weak coupling, this different regularization method will probably also modify the wave functional in the non-perturbative regime. This would imply that in this chapter we did not use a good approximation, and a different functional, obtained from the Hamiltonian (\ref{TregKKNallorders}) might actually lead to a purely linear potential.

In order to determine the vacuum wave functional in the non-perturbative regime, while incorporating the regularization method that we developed in Chap.~\ref{chap:Regularization}, a possibility could be to explore an approach proposed in Ref.~\cite{Zarembo:1998bp}. The idea is to apply the background field method to the Schr\"odinger representation: splitting the fields into hard and soft modes, treating the hard modes perturbatively and integrating them out. This then leads to an effective potential for the soft modes. This approach is particularly appealing, since its validity in 3+1 dimensions is straightforward, given that the coupling constant is indeed small for the hard modes, as long as the factorization scale is set high enough. In light of this, a good choice of variables for this kind of calculation might be the real gauge invariant currents, proposed by Freidel in Ref.~\cite{Freidel:2006qz}, which can be extended to 3+1 dimensions without conceptual problems. We will discuss them in the following chapter.

\chapter{Towards Four Dimensions}
\label{chap:TonV}
\section{Introduction}

As we have seen, considering Yang-Mills theory in three dimensions and at weak coupling is important, since it advances our understanding of the theory. In particular, in Chap.~\ref{chap:Regularization} the simplifications provided by the weak coupling limit and the super-renormalizability of the three dimensional theory enabled us to clarify how Yang-Mills theory in the Schr\"odinger representation should be regularized. This knowledge translates to other dimensions and to the strong coupling regime, as well as to other theories. Furthermore, it is an alternative approach to compute observables in three dimensions in the weak coupling regime, but it may also allow us to compute physically relevant objects in four dimensions like the magnetic screening mass (see Eq.~(\ref{mag-screen})). Possible extensions to the non-perturbative regime have been explored in Chap.~\ref{chap:Potential}.

Nevertheless, it is of course of major importance to devise computationally useful schemes that can be applied to the Schr\"odinger representation in the physical case of four dimensions. A possible way to do this is to use real gauge invariant variables instead of complex ones: Inspired by, and in order to profit from, the computational power of the approach developed by Karabali et al.~in Refs.~\cite{Karabali:1995ps,Karabali:1996je,Karabali:1996iu,Karabali:1997wk,Karabali:1998yq,Karabali:2009rg}, a modified approach was devised in Ref.~\cite{Freidel:2006qz}. The field variables in this case are real, which has the advantage that any wave functional obtained in this way is real and gauge invariant by construction. As we have seen in Chap.~\ref{chap:Comparison}, depending on the computational method, neither of these properties is necessarily evident. More importantly, this approach allows for a generalization to any dimension. 
We shall call it method (C). We will begin in 2+1 dimensions and then see how this formulation can be extended to 3+1 dimensions. For both cases we obtain the Hamiltonian in these variables. It differs from the one proposed in Ref.~\cite{Freidel:2006qz}, because we employ the regularization developed in Chap.~\ref{chap:Regularization}.

\section{Real gauge invariant variables} 
\label{sec:Freidel}

The principal idea of method (C) is that the variable transformation used by Karabali et al.~does not rely on the variables being complex. So instead of finding complex solutions $M$ and $M^\dagger$ for Eq.~(\ref{AofM}) one can also start with Eq.~(\ref{Ai})  (no sum over repeated spatial indices in all of this chapter):
\be
A_i=-{1\over e}\p_iM_iM_i^{-1}  \,,
\ee
which is the Euclidean analogue of Eq.~(\ref{AofM}). It is solved by the Bars variables (see Ref.~\cite{Bars:1978xy}), given in Eq.~(\ref{M_iPathordered}):
\be
M_i(\vec x)=\mathcal{P}e^{-e\int^{\vec x}_{\infty}d z_iA_i(\vec z)}\,, 
\ee
where the integral is a straight spatial contour for fixed $x_j$ for $j\neq i$, explicitly 
\bea
M_1(x) &=& 1+\sum_{n=1}^{\infty}(-e)^n\int_{x_1>t_1>\ldots>t_n}A_1(t_1,x_2)\cdots A_1(t_n,x_2)\, d t_1\cdots d t_n \nn\\
&=& \sum_{n=0}^{\infty}(-e)^n\int_y (G_1 A_1)^n(\vec x,\vec y)\,,
\eea
with
\be
(G_1A_1)^2=\int_z G_1(x,z)A_1(z)G_1(z,y)A_1(y)\quad\mathrm{etc.}
\ee
and analogously for $M_2$. The Green's functions are (see Eq.~(\ref{G_i}))
\be
G_1(\vec x;\vec y)\equiv G_1(\vec x-\vec y)=\theta(x_1-y_1)\de(x_2-y_2)\quad \mathrm{and}\quad G_2(\vec x;\vec y)\equiv G_(\vec x-\vec y)=\de(x_1-y_1)\theta(x_2-y_2) \,.
\ee
Note that they are not antisymmetric under exchange of $\vec x$ and $\vec y$.

Gauge transformations (\ref{GaugeTrafo}) act on the $M_i$'s like on their (anti-)holomorphic counterparts as
\be
M_i\to gM_i\,,
\ee
so one can define gauge invariant variables
\be
H_{ij}=M_i^{-1}M_j
\ee
and currents
\be
J_{ij}={1\over e}(\p_j H_{ij})H_{ij}^{-1}\,.
\ee
Note that $H_{ii}=1$ and $H_{ji}=H_{ij}^{-1}$.

There is a ``reality condition" on the currents' derivatives (analogous to Eq.~(\ref{RealityCon})):
\be
\p_iJ_{ij}=-H_{ij}(\p_jJ_{ji})H_{ji}\,, \label{RealCon}
\ee
which in 2+1 dimensions just means that there is only one physical degree of freedom. We choose to work with
$J_{12}$. It is related to the magnetic field by
\be
B =-M_1(\p_1J_{12})M_1^{-1}\,,
\ee
thus the potential term in terms of these new fields is
\be
\V  = \half \int_{x}   \p_1 J_{12}^a(x) \p_1 J_{12}^a(x) \,. \label{VFreidel}
\ee

It is somewhat more involved to find the kinetic operator.
%
To obtain the regularized kinetic operator we start again from Eq.~(\ref{Treg}) 
\be
\mathcal{T}_{reg}=-{1\over2}\int_{x,v} \delta_\mu(\vec x, \vec v) \Phi_{ab}(\vec x,\vec v) \frac{\delta}{\delta A_i^a(\vec x)}  \frac{\delta}{\delta A_i^b(\vec v)}\,, 
\ee
with the Wilson line 
\bE{rCl}
 \Phi_{ab}(\vec x,\vec v) &=&{1\over 2}\left((M_1(\vec x)M_1^{-1}(v_1,x_2)M_2(v_1,x_2)M_2^{-1}(\vec v))^{ab} \right. \nn\\
 &&\qquad\qquad \left.+(M_2(\vec x)M_2^{-1}(x_1,v_2)M_1(x_1,v_2)M_1^{-1}(\vec v))^{ab}\right)\,, 
\eE
defined in Eq.~(\ref{Phi}), and transform the fields $(A_1,A_2)\to(A_1,J_{12})$:
\bE{rCl}
\frac{\delta^2}{\delta A^a_i(\vec x)\delta A^b_i(\vec v)} &=&\int_{z,w} \Bigg\{ \left[\frac{\de A^c_1(\vec z)}{\de A^a_1(\vec x)} \frac{\de}{\de A^c_1(\vec z)} + \frac{\de J^c_{12}(\vec z)}{\de A^a_1(\vec x)} \frac{\de}{\de J^c_{12}(\vec z)} \right] \nn\\
&&\qquad\qquad\times \left[\frac{\de A^d_1(\vec w)}{\de A^b_1(\vec v)} \frac{\de}{\de A^d_1(\vec w)} + \frac{\de J^d_{12}(\vec w)}{\de A^b_1(\vec v)} \frac{\de}{\de J^d_{12}(\vec w)} \right] \nn\\
&& + \left[\frac{\de A^c_1(\vec z)}{\de A^a_2(\vec x)} \frac{\de}{\de A^c_1(\vec z)} + \frac{\de J^c_{12}(\vec z)}{\de A^a_2(\vec x)} \frac{\de}{\de J^c_{12}(\vec z)} \right] \nn\\
&&\qquad\qquad\times \left[\frac{\de A^d_1(\vec w)}{\de A^b_2(\vec v)} \frac{\de}{\de A^d_1(\vec w)} + \frac{\de J^d_{12}(\vec w)}{\de A^b_2(\vec v)} \frac{\de}{\de J^d_{12}(\vec w)} \right] \Bigg\}\,. 
\eE
Similar to Eqs.~(\ref{JOverA}) and (\ref{JOverAbar}) we find
\bE{rCl}
 \frac{\de J^c_{12}(\vec z)}{\de A^a_1(\vec x)} &=& \left[\D_{12}^{ce}(\vec z)G_1(\vec z,\vec x)\right] M_1^{ae}(\vec x)\,, \label{J12OverA1} \\
 \frac{\de J^c_{12}(\vec z)}{\de A^a_2(\vec x)} &=& - M_1^{ac}(\vec x)\de(\vec x-\vec z)\,, \label{J12OverA2} \\
\D_{12}^{ce} &=& \p_2\de^{ce}-eJ_{12}^{ce} = \p_2\de^{ce}+eJ_{12}^ff^{cef} = (H_{12}\p_2H_{21})^{ce} \,. \label{D12}
\eE
With these three equalities one can show that the Gauss law operator reads
\bE{rCl}
I^a(\vec x) &=& i\vec D^{ab}\cdot\frac{\delta}{\delta \vec A^b(\vec x)}\\
 &=& iD_1^{ab}(\vec x)\left[\frac{\delta}{\delta A^b_1(\vec x)} + \int_y \frac{\delta J_{12}^c(\vec y)}{\delta A^b_1(\vec x)}\frac{\delta}{\delta J^c_{12}(\vec y)}\right]+ i\int_y D_2^{ab}(\vec x)\frac{\delta J_{12}^c(\vec y)}{\delta A^b_2(\vec x)}\frac{\delta}{\delta J^c_{12}(\vec y)} \qquad \\
&=& iD_1^{ab}(\vec x)\frac{\delta}{\delta A^b_1(\vec x)} \,.
\eE
This reduces the kinetic operator to 
\bE{rCl}
\T&=&  -\half\int_{x,v,z}\delta_\mu(\vec x, \vec v) \Phi_{ab}(\vec x,\vec v) \frac{\de^2 J^c_{12}(\vec z)}{\de A^a_1(\vec x) \de A^b_1(\vec v)} \frac{\de}{\de J^c_{12}(\vec z)} +\int_x c^aI^a(\vec x)\nn\\
&& -\half \int_{x,v,z,w} \delta_\mu(\vec x, \vec v) \Phi_{ab}(\vec x,\vec v)\Bigg\{ \frac{\de J^c_{12}(\vec z)}{\de A^a_1(\vec x)} \frac{\de J^d_{12}(\vec w)}{\de A^b_1(\vec v)} + \frac{\de J^c_{12}(\vec z)}{\de A^a_2(\vec x)} \frac{\de J^d_{12}(\vec w)}{\de A^b_2(\vec v)} \Bigg\} \frac{\de}{\de J^c_{12}(\vec z)} \frac{\de}{\de J^d_{12}(\vec w)} \nn\\
&&\\
 &=&  -{e\over 4}\int_{x,v,z} \left[\Lambda_1(\vec x,\vec v)^{eg}+\Lambda_1(\vec v,\vec x)^{ge}\right] \delta_{\mu}(\vec x,\vec v) \nn\\
&&\qquad  \left[ f^{cef}\left[\D_{12}^{fg}(\vec z) G_1(\vec z,\vec v)\right] G_1(\vec z,\vec x)- \left[\D_{12}^{cf}(\vec z)G_1(\vec z,\vec x)\right] G_1(\vec x,\vec v) f^{gfe} \right] \frac{\de}{\de J^c_{12}(\vec z)} \nn\\
&& -{1\over 4}\int_{x,v,z,w} \left[\Lambda_1(\vec x,\vec v)^{ef}+\Lambda_1(\vec v,\vec x)^{fe}\right] \delta_{\mu}(\vec x,\vec v)\nn\\
&& \qquad \left[ \left[\D_{12}^{ce}(\vec z)G_1(\vec z,\vec x)\right] \left[\D_{12}^{df}(\vec w)G_1(\vec w,\vec v)\right]  +  \de^{ce}\de^{fd}\de(\vec x-\vec z)\de(\vec v-\vec w) \right]  \frac{\de}{\de J^c_{12}(\vec z)} \frac{\de}{\de J^d_{12}(\vec w)}\,,\nn\\
\eE
where we dropped the terms proportional to the Gauss law operator in the second equality, since it vanishes on physical wave functionals and, following \cite{Freidel:2006qz}, we defined
\be
\Lambda_1(\vec x,\vec v) := H_{12}(v_1,x_2)H_{21}(\vec v)\,.
\ee
Note that $\p_1^x\Lambda_1(\vec x,\vec v)=0$ and $\Lambda_1(\vec x,\vec x)=\mathds{1}$. With some simplification we find
\bE{rCl}
\T_C &=&-{e\over 4}\int_{x,v,z} \left[\Lambda_1(\vec x,\vec v)^{eg}+\Lambda_1(\vec v,\vec x)^{ge}\right] \delta_{\mu}(\vec x,\vec v)   f^{cef}\left[\D_{12}^{fg}(\vec z) G_1(\vec z,\vec v)\right] G_1(\vec z,\vec x) \frac{\de}{\de J^c_{12}(\vec z)} \nn\\
&&-{1\over 2}\int_{v,z,w} \left[ \left[\D_{12}^{ce}(\vec z)\Lambda_1(\vec z,\vec v)^{ef} G_1^{\mu}(\vec z,\vec v)\right] \left[\D_{12}^{df}(\vec w)G_1(\vec w,\vec v)\right]  \right]  \frac{\de}{\de J^c_{12}(\vec z)} \frac{\de}{\de J^d_{12}(\vec w)}\,, \label{TFreidelNew} \nn\\
&&-{1\over 2}\int_{z,w} \Lambda_1(\vec z,\vec w)^{cd} \delta_{\mu}(\vec z,\vec w)   \frac{\de}{\de J^c_{12}(\vec z)} \frac{\de}{\de J^d_{12}(\vec w)}\,, \label{TFreidelNew} 
\eE
where we introduced the regularized Greens function
\be
G_1^{\mu}(\vec x,\vec y) = \theta_\mu(x_1-y_1)\de_\mu(x_2-y_2) = \half\left(1+\mathrm{Erf}[\mu (x_1-y_1)]\right) \frac{\mu}{\sqrt{\pi}}e^{-\mu^2(x_2-y_2)^2}\,. 
\ee
Note that this kinetic operator differs from the 2+1 dimensional counterpart of the 3+1 dimensional one used in Ref.~\cite{Freidel:2006qz}. We will explain this discrepancy later when we discuss the 3+1 dimensional Hamiltonian.

The main advantage of this approach is that it can be extended to 3+1 dimensions. In order to do so, we generalize everything we did in this section. This provides no problems for the $M_i$ and $J_{ij}$ fields. In particular we have
\be
A_3=-{1\over e}\p_3M_3M_3^{-1}
\ee
and
\be
J_{23}={1\over e}(\p_3 H_{23})H_{23}^{-1}\,,\qquad J_{31}={1\over e}(\p_1 H_{31})H_{31}^{-1}\,.
\ee 
While in theory we also have $J_{13}$ and $J_{32}$, these are not independent degrees of freedom, since they are constraint by the ``reality condition" Eq.~(\ref{RealCon}). The generalizations of Eqs.~(\ref{J12OverA1}-\ref{D12}) are 
\bE{rCl}
 \frac{\de J^c_{ij}(\vec z)}{\de A^a_i(\vec x)} &=& \left[\D_{ij}^{ce}(\vec z)G_i(\vec z,\vec x)\right] M_i^{ae}(\vec x) \,,\\
 \frac{\de J^c_{ji}(\vec z)}{\de A^a_i(\vec x)} &=& - M_j^{ac}(\vec x)\de(\vec x-\vec z) \,, \\
  \frac{\de J^c_{jk}(\vec z)}{\de A^a_i(\vec x)} &=& 0, \quad i\neq j,k\,,  \\
\D_{ij}^{ce} &=& \p_j\de^{ce}-eJ_{ij}^{ce} = \p_j\de^{ce}+eJ_{ij}^ff^{cef} = (H_{ij}\p_jH_{ji})^{ce} \,. 
\eE
The components of the chromomagnetic field in 3+1 dimensions are given by
\be
B_i =-M_{i+1}(\p_{i+1}J_{i+1,i+2})M_{i+1}^{-1}\,,
\ee
where addition in the indices is {\it modulo 3}. The potential operator in 3+1 dimensions in these variables is hence
\be
\V_{3+1}=\half\sum_{i=1}^3\int_x B_i^a(\vec x) B_i^a(\vec x) = \half\sum_{i=1}^3\int_x (\p_iJ_{i,i+1}^a(\vec x))(\p_iJ_{i,i+1}^a(\vec x)) \,.
\ee

In order to obtain the kinetic operator we first compute
\bE{rCl}
&&\sum_{i=1}^3 \frac{\delta^2}{\delta A^a_i(\vec x)\delta A^b_i(\vec v)} \nn\\
 &=&\sum_{i,j,k=1}^3 \int_{z,w}  \frac{\de J^c_{j,j+1}(\vec z)}{\de A^a_i(\vec x)} \frac{\de}{\de J^c_{j,j+1}(\vec z)}\  \frac{\de J^d_{k,k+1}(\vec w)}{\de A^b_i(\vec v)} \frac{\de}{\de J^d_{k,k+1}(\vec w)} \\
&=&\sum_{i=1}^3 \int_{z,w} \left( \left[\D_{i,i+1}^{ce}(\vec z)G_i(\vec z,\vec x)\right] M_i^{ae}(\vec x)  \frac{\de}{\de J^c_{i,i+1}(\vec z)} - M_{i-1}^{ac}(\vec x)\de(\vec x,\vec z) \frac{\de}{\de J^c_{i-1,i}(\vec z)}\right) \nn\\
&&\quad\times \left( \left[\D_{i,i+1}^{df}(\vec w)G_i(\vec w,\vec v)\right] M_i^{bf}(\vec v)  \frac{\de}{\de J^d_{i,i+1}(\vec w)} - M_{i-1}^{bd}(\vec v)\de(\vec v,\vec w) \frac{\de}{\de J^d_{i-1,i}(\vec w)}\right)  \\
&=&\sum_{i=1}^3 \int_{z,w} \Bigg\{ eM_i^{ae}(\vec x) M_i^{bf}(\vec v) \left[\D_{i,i+1}^{ce}(\vec z)G_i(\vec z,\vec x)\right]   \left[ f^{dfc}\de(\vec w,\vec z) G_i(\vec w,\vec v)\right]   \frac{\de}{\de J^d_{i,i+1}(\vec w)}  \nn\\
&& + M_i^{ae}(\vec x)  M_i^{bf}(\vec v) \Bigg(  \left[\D_{i,i+1}^{ce}(\vec z)G_i(\vec z,\vec x)\right]  \left[\D_{i,i+1}^{df}(\vec w)G_i(\vec w,\vec v)\right] \nn\\
&&\qquad\qquad\qquad\qquad + \de^{ce}\de^{df}\de(\vec x,\vec z)\de(\vec v,\vec w) \Bigg)\frac{\de^2}{\de J^c_{i,i+1}(\vec z)\de J^d_{i,i+1}(\vec w) } \nn\\
&& - \Bigg(  M_i^{ae}(\vec x) M_{i-1}^{bd}(\vec v)\de(\vec v,\vec w) \left[\D_{i,i+1}^{ce}(\vec z)G_i(\vec z,\vec x)\right] \nn\\
&& \qquad\quad + M_{i}^{ac}(\vec x) M_{i+1}^{bf}(\vec v)  \de(\vec x,\vec z) \left[\D_{i+1,i-1}^{df}(\vec w)G_{i+1}(\vec w,\vec v)\right] \Bigg) \frac{\de^2}{\de J^c_{i,i+1}(\vec z) \de J^d_{i-1,i}(\vec w)} \Bigg\} \nn\\
\eE

Generalizing the Wilson line of Eq.~(\ref{Phi}) we now move along the edges of a rectangular hexahedron instead of a rectangle, leading to 
\bE{rCl}
 \Phi_{ab}^\mathrm{3d}(\vec x,\vec v) &=&{1\over 6}\Big(M_1(\vec x)H_{12}(v_1,x_2,x_3)H_{23}(v_1,v_2,x_3)M_3^{-1}(\vec v)  \nn\\
 &&\quad + M_2(\vec x)H_{23}(x_1,v_2,x_3)H_{31}(x_1,v_2,v_3)M_1^{-1}(\vec v)   \nn\\
 &&\quad + M_3(\vec x)H_{31}(x_1,x_2,v_3)H_{12}(v_1,x_2,v_3)M_2^{-1}(\vec v)  \nn\\
 &&\quad + M_1(\vec x)H_{13}(v_1,x_2,x_3)H_{32}(v_1,x_2,v_3)M_2^{-1}(\vec v)  \nn\\
 &&\quad + M_2(\vec x)H_{21}(x_1,v_2,x_3)H_{13}(v_1,v_2,x_3)M_3^{-1}(\vec v)   \nn\\
 &&\quad + M_3(\vec x)H_{32}(x_1,x_2,v_3)H_{21}(x_1,v_2,v_3)M_1^{-1}(\vec v)\Big )^{ab}\,, 
\eE
which equally satisfies $ \Phi_{ab}^\mathrm{3d}(\vec x,\vec v) =  \Phi_{ba}^\mathrm{3d}(\vec v,\vec x)$. The Hamiltonian operator in terms of gauge invariant variables in 3+1 dimensions is hence
\bE{rCl}
&&\H_{3+1} \nn\\
&=&\half\sum_{i=1}^3 \int_{x,v,z,w} \de_\mu(\vec x,\vec v) \nn\\
&& \times  \Bigg\{ e\Big(M_i^{-1}(\vec x)  \Phi^\mathrm{3d}(\vec x,\vec v) M_i(\vec v) \Big)^{ef} \left[\D_{i,i+1}^{ce}(\vec z)G_i(\vec z,\vec x)\right]    f^{dfc}\de(\vec w,\vec z) G_i(\vec w,\vec v)   \frac{\de}{\de J^d_{i,i+1}(\vec w)}  \nn\\
&&\quad + \Big(M_i^{-1}(\vec x)  \Phi^\mathrm{3d}(\vec x,\vec v) M_i(\vec v) \Big)^{ef} \nn\\
&&\qquad \times \Bigg(  \left[\D_{i,i+1}^{ce}(\vec z)G_i(\vec z,\vec x)\right]  \left[\D_{i,i+1}^{df}(\vec w)G_i(\vec w,\vec v)\right]  + \de^{ce}\de^{df}\de(\vec x,\vec z)\de(\vec v,\vec w) \Bigg)\frac{\de^2}{\de J^c_{i,i+1}(\vec z)\de J^d_{i,i+1}(\vec w) } \nn\\
&&\quad - \Bigg(  \Big(M_i^{-1}(\vec x)  \Phi^\mathrm{3d}(\vec x,\vec v) M_{i-1}(\vec v) \Big)^{ed}  \de(\vec v,\vec w) \left[\D_{i,i+1}^{ce}(\vec z)G_i(\vec z,\vec x)\right] \nn\\
&& \qquad + \Big(M_i^{-1}(\vec x)  \Phi^\mathrm{3d}(\vec x,\vec v) M_{i+1}(\vec v) \Big)^{ce} \de(\vec x,\vec z) \left[\D_{i+1,i-1}^{de}(\vec w)G_{i+1}(\vec w,\vec v)\right] \Bigg) \frac{\de^2}{\de J^c_{i,i+1}(\vec z) \de J^d_{i-1,i}(\vec w)} \Bigg\} \nn\\
&& + \half\sum_{i=1}^3\int_x (\p_iJ_{i,i+1}^a(\vec x))(\p_iJ_{i,i+1}^a(\vec x)) \,. \label{TFreidel3d}
\eE
While this is a complicated expression, it allows for the translation of method (B), and therefore for analytic computations in the non-perturbative regime, to 3+1 dimensions. Note that it differs from the Hamiltonian proposed in Ref.~\cite{Freidel:2006qz}, where a different regularization was used and the one derivative term was argued to be subleading. In light of the results of Chap.~\ref{chap:Regularization} we argue, however, that all terms should be maintained until the end of a computation and only then should the regulator be removed.

\section{Conclusions}

In this chapter we have considered a modification of method (B), using real gauge invariant currents instead of complex ones (method (C)). As demonstrated in Chap.~\ref{chap:Comparison}, proving that the wave functional $\Psi_{GI}$ obtained with complex variables is actually real, is a tedious exercise, and $\Psi_\mathrm{trial}$ of Chap.~\ref{chap:Potential} even does have a non-trivial imaginary part. Using real currents from the beginning guarantees a vacuum wave functional, which is both real and gauge invariant, thus eliminating this problem right away. The main advantage of method (C), however, is that it allows for the generalization to 3+1 dimensions. While this extension results in a complicated expression for the Hamiltonian it is in principle possible and should be explored. 

It is tempting to directly do computations in 3+1 dimensions where observables have immediate physical relevance, but the Schr\"odinger representation, though promising, is still not fully understood. We have seen in this thesis that conceptual questions, in this case regularization, can be clarified in 2+1 dimensions, and it seems worthwhile to fully understand the Schr\"odinger representation before moving on to more complicated problems. Due to its super-renormalizability and the less complicated Hamiltonian, 2+1 dimensional Yang-Mills theory proves to be an ideal testing ground for different approaches, which can then, hopefully, be translated to other dimensions.

\chapter{Conclusions}
\label{chap:Conclusions}
In this thesis we have investigated the Yang-Mills vacuum wave functional in 2+1 dimensions, focusing mainly on the weak coupling regime. 2+1 dimensional Yang-Mills theory is relevant because it is the lowest dimensional Yang-Mills theory with propagating degrees of freedom. Put in another way, three is the lowest dimension in which the non-abelian nature of the theory has an effect. This allows us to draw information about the four dimensional case from it. 
On the other hand, three dimensional Yang-Mills theory is important in its own right because its euclidean version constitutes the high temperature limit of four dimensional QCD. 
The framework of the Schr\"odinger representation, which we considered in this thesis, is interesting because it allows for analytical computations in the non-perturbative regime. Yet, it is rarely considered in the literature, and even perturbative computations are not well developed. Moreover, regularization and renormalization are also not well understood in this framework. In this thesis we aimed to put both perturbation theory and regularization in the Schr\"odinger picture on more solid ground.

\bigskip

In Chap.~\ref{chap:Comparison} we computed the ground-state wave functional in a perturbative expansion to $\O(e^2)$, using two different methods. First we started from the usual gauge field Hamiltonian and computed the vacuum wave functional directly in perturbation theory, generalizing the method developed in Ref.~\cite{Hatfield:1984dv} (method (A)). We then compared this to the corresponding result obtained from a weak coupling expansion of the wave functional proposed in Ref.~\cite{Karabali:2009rg} (method (B)). Each method has its own advantages and drawbacks: The wave functional obtained with method (A), which we called $\Psi_{GL}$, is explicitly real, but its gauge invariance cannot be guaranteed a priori. The result of method (B), called $\Psi_{GI}$, on the other hand is gauge invariant by construction, but it has a non-trivial imaginary part. Comparing the results of the two approaches in a systematic fashion (as the expressions are too complicated for a straightforward comparison) we were able to show in Chap.~\ref{chap:Comparison} that they agree up to a real, gauge invariant term. This proves on the one hand the gauge invariance of $\Psi_{GL}$, and on the other hand the reality of $\Psi_{GI}$.

\bigskip

Still, as we found a difference between the results of these two methods, and since regularization in the Schr\"odinger representation is not a well-developed subject, we had to reconsider the regularization method used, and we did so in Chap.~\ref{chap:Regularization}. No regularization was used for method (A) in Chap.~\ref{chap:Comparison}, and even though the result was finite, we found in Chap.~\ref{chap:Regularization} that without regularization some contributions were missed. Moreover, we found  that the regularization scheme for method (B), used in Ref.~\cite{Karabali:2009rg}, also had to be modified. We developed a new regularization scheme in Chap.~\ref{chap:Regularization}. Applying it in the same way to both methods we found new contributions for both approaches, such that the new results are identical, as expected. This is a strong check of our computation, and we therefore claim that the wave functional given in Eqs.~(\ref{FGL0p}), (\ref{FGL1}), (\ref{F24GL2}) and (\ref{F22GL}) in terms of the gauge fields $\vec A^a$ (and in Eqs.~(\ref{FGI0}), (\ref{FGI1}), (\ref{FGI24}) and (\ref{FGI22}) in terms of the gauge invariant variables $J^a$) is the correct Yang-Mills vacuum wave functional to ${\cal O}(e^2)$, given here for the first time. This is one of the major results of this thesis. Using it, we were able to give an estimate of the magnetic screening mass.

That the result for method (A) differs from Chap.~\ref{chap:Comparison} to Chap.~\ref{chap:Regularization} is not very surprising, as the regularization of the kinetic operator was not considered in Chap.~\ref{chap:Comparison}. More surprising is the fact that we had to modify the result of method (B), the regularization of which had been studied in detail in the past. In Refs.~\cite{Karabali:1997wk,Agarwal:2007ns} an intermediate cutoff $\mu'  \ll \mu $ was introduced in the wave functional, damping the modes with energies greater than $\mu'$. This procedure eliminates the extra contribution we found with method (B) in Sec.~\ref{subsec:F22GI}. However, if the same procedure is applied to method (A), it also eliminates the mass term obtained in Sec.~\ref{subsec:F22GL}, producing the two incompatible results of Chap.~\ref{chap:Comparison}. Instead, we advocate doing the whole computation with a single cutoff $\mu$ that regularizes the kinetic operator and the ground-state wave functional (and all excitations) at the same time. It is only after solving the Schr\"odinger equation that we can take the cutoff $\mu$ to infinity compared with any finite momentum of the system. In other words, the momenta of the fields of the wave functional can be large. As one goes to higher orders in perturbation theory, loops appear, whose integrals run up to infinity, and all of these modes have to be taken into account, producing new contributions, as we have seen in Eq.~(\ref{new}). 
In a different language, in order to be able to give meaning to the theory we need to regularize the Hamiltonian. This defines a (regularized) Hilbert space, in which both the Hamiltonian and the states depend on the same regulator. Preserving unitarity requires all states to be considered in the computation. In particular, cutting them off with a second regulator impairs the completeness relation.
This regularization procedure is not specific to Yang-Mills theory or to weak coupling or to three dimensions. It should be applied in the same way to any QFT in the Schr\"odinger picture.

\bigskip

In Chap.~\ref{chap:Potential} we investigated the non-perturbative regime, which is where the Schr\"odinger representation can develop its full power. In Ref.~\cite{Karabali:1998yq} it was found that the potential $\V$ of the Yang-Mills Hamiltonian is an eigenfunction of the kinetic operator $\T$. We tested the robustness of this result after regularization in perturbation theory. We found that $\V$ is still an eigenfunction of $\T$, but the eigenvalue is different, and in particular, regulator-dependent. This suggests that a strong coupling expansion along the lines of Ref.~\cite{Karabali:1998yq} may be problematic.

We then considered an interpolating trial functional, which was obtained by transforming the proposed wave functional of Ref.~\cite{Karabali:2009rg} to gauge field variables. This trial functional stems from an expansion in $e^2/m$ and is claimed to be a good approximation at all scales. As a test, we used it to compute the correlator of the chromomagnetic field at leading order, which in the weak coupling limit agrees with the perturbative computation, and to estimate the gluon condensate. We then turned our attention to the static potential. At leading order we found a linear potential, but the next order in the $e^2/m$ expansion leads to contributions cubic in the separation, contrary to what was found in Ref.~\cite{Karabali:2009rg}, where all corrections were compatible with a linear potential. This makes it impossible to compute the string tension analytically from first principles in this fashion. While it is perfectly possible that the cubic terms are due to the fact that the $e^2/m$ expansion is not the appropriate one to only contain linear terms, another explanation might be that the trial functional does not have the correct long distance behavior. 
As was shown in Chap.~\ref{chap:Regularization} (in particular in Eq.~(\ref{new})), the crucial ``mass term" which led to the specific form of the vacuum wave functional of Ref.~\cite{Karabali:2009rg} is not the only term of this sort in the weak coupling regime: more terms in the Hamiltonian Eq.~(\ref{TregKKN}) produce this type of terms in the wave functional. Taking these into account in the non-perturbative regime 
probably leads to a different vacuum wave functional, which might exhibit the desired behavior. In any case the consequences of the different regularization method employed in Chap.~\ref{chap:Regularization} for the approximate resummation scheme analysis carried out in Ref.~\cite{Karabali:2009rg} should be explored.

\bigskip

A third approach, the formulation of Yang-Mills theory in terms of real gauge invariant variables developed in Ref.~\cite{Freidel:2006qz}, was presented in Chap.~\ref{chap:TonV}. It combines the advantages of having a vacuum wave functional which is both manifestly real and gauge invariant by construction with the possibility of a straightforward extension to 3+1 dimensions. Following our result of Chap.~\ref{chap:Regularization} we claim, however, that the correct Hamiltonian in this formulation is not the one proposed in Ref.~\cite{Freidel:2006qz}, but is given by Eqs.~(\ref{TFreidelNew}) and (\ref{TFreidel3d}) for 2+1 and 3+1 dimensions, respectively, since the regulator should only be removed after the determination of the vacuum wave functional. 

\bigskip

The main contribution of this thesis is that we clarify how regularization in the Schr\"odinger picture should be implemented. We have demonstrated that it is worthwhile to investigate theories outside of their physically relevant regime, since the resulting simplifications can help to understand conceptual problems, whose solutions, as in this case, may then be generalized to other regimes.

\appendix
\appendixpage
\noappendicestocpagenum
\addappheadtotoc

\chapter{Comparison of $\Psi_{GL}$ and $\Psi_{GI}$ at $\O(e)$}
\label{F1comp}
In this appendix we will show that $\Psi_{GL}$ and $\Psi_{GI}$ are equal at $\O(e)$.
At this order we obtain in the gauge invariant approach Eq.~(\ref{F1Nair}):
\begin{IEEEeqnarray}{rCl}
F^{(1)}_{GI}[{\vec A}]
&=& i f^{abc} \int_{\slashed{k_1},\slashed{k_2},\slashed{k_3}}\slashed{\delta}\left(\sum_{i=1}^3 \vec{k}_i\right) \Bigg\{ \frac{1}{2|\vec{k}_1|} (\vec{k}_1\times\vec{A}^a(\vec{k}_1)) (\vec{A}^b(\vec{k}_2)\times\vec{A}^c(\vec{k}_3))\nn\\
&&- \frac{1}{|\vec{k}_3|\vec{k}_1^2} \left(  \frac{\vec{k}_1\times\vec{k}_2+i\vec{k}_1\cdot\vec{k}_2}{(|\vec{k}_1|+|\vec{k}_2|+|\vec{k}_3|)|\vec{k}_2|} +i  \right) (\vec{k}_1\times\vec{A}^a(\vec{k}_1))(\vec{k}_2\times\vec{A}^b(\vec{k}_2)) (\vec{k}_3\times\vec{A}^c(\vec{k}_3))\nn\\
&& + \frac{1}{|\vec{k}_3|\vec{k}_1^2} (\vec{k}_1\cdot\vec{A}^a(\vec{k}_1)) (\vec{k}_2\times\vec{A}^b(\vec{k}_2)) (\vec{k}_3\times\vec{A}^c(\vec{k}_3)) \Bigg\}
\,. \label{F1app}
\end{IEEEeqnarray}

The imaginary part
\begin{IEEEeqnarray}{rCl}
\mathrm{Im}[F^{(1)}_{GI}[{\vec A}]] &=&  if^{abc} \int_{\slashed{k_1},\slashed{k_2},\slashed{k_3}}\slashed{\delta}\left(\sum_{i=1}^3 \vec{k}_i\right) \left( \frac{\vec{k}_1\cdot\vec{k}_2}{(\sum_i|\vec{k}_i|)\vec{k}_1^2|\vec{k}_2||\vec{k}_3|} +\frac{1}{|\vec{k}_3|\vec{k}_1^2}  \right)  \nn\\
&&\qquad  (\vec{k}_1\times\vec{A}^a(\vec{k}_1))(\vec{k}_2\times\vec{A}^b(\vec{k}_2)) (\vec{k}_3\times\vec{A}^c(\vec{k}_3)) 
\eE
vanishes identically as we now show. Because of the delta function we can write $\vec k_1$ as $-\vec k_2 -\vec k_3$ under the integral:
\bE{rCl}
&=& if^{abc} \int_{\slashed{k_1},\slashed{k_2},\slashed{k_3}}\slashed{\delta}\left(\sum_{i=1}^3 \vec{k}_i\right) \left( \frac{\vec{k}_2^2}{(\sum_i|\vec{k}_i|) \vec{k}_1^2|\vec{k}_2||\vec{k}_3|}  - \frac{\vec{k}_3\cdot\vec{k}_2}{(\sum_i|\vec{k}_i|)\vec{k}_1^2|\vec{k}_2||\vec{k}_3|} +\frac{1}{|\vec{k}_3|\vec{k}_1^2}  \right) \nn\\
&&\qquad (\vec{k}_1\times\vec{A}^a(\vec{k}_1))(\vec{k}_2\times\vec{A}^b(\vec{k}_2)) (\vec{k}_3\times\vec{A}^c(\vec{k}_3))\,.
\eE
The second term vanishes when interchanging $\vec k_2 \leftrightarrow \vec k_3$, hence
\bE{rCl}
\mathrm{Im}[F^{(1)}_{GI}[{\vec A}]] &=& if^{abc} \int_{\slashed{k_1},\slashed{k_2},\slashed{k_3}}\slashed{\delta}\left(\sum_{i=1}^3 \vec{k}_i\right)  \frac{1}{(\sum_i|\vec{k}_i|) \vec{k}_1^2|\vec{k}_3|}\left(-|\vec{k}_2| +\sum_i|\vec{k}_i|  \right)  \nn\\
&&\qquad (\vec{k}_1\times\vec{A}^a(\vec{k}_1))(\vec{k}_2\times\vec{A}^b(\vec{k}_2)) (\vec{k}_3\times\vec{A}^c(\vec{k}_3))\qquad \\
&=& if^{abc} \int_{\slashed{k_1},\slashed{k_2},\slashed{k_3}}\slashed{\delta}\left(\sum_{i=1}^3 \vec{k}_i\right)  \left(\frac{1}{(\sum_i|\vec{k}_i|)|\vec{k}_1||\vec{k}_3|} +\frac{1}{(\sum_i|\vec{k}_i|) \vec{k}_1^2}  \right)  \nn\\
&&\qquad (\vec{k}_1\times\vec{A}^a(\vec{k}_1))(\vec{k}_2\times\vec{A}^b(\vec{k}_2)) (\vec{k}_3\times\vec{A}^c(\vec{k}_3)) \\
&=& 0\,.
\end{IEEEeqnarray}
The first term vanishes under $\vec k_1 \leftrightarrow \vec k_3$, the second under $\vec k_2 \leftrightarrow \vec k_3$.

We now look at (the real part of) the second line of Eq.~(\ref{F1app}):
\begin{IEEEeqnarray}{rCl}
&& i f^{abc} \int_{\slashed{k_1},\slashed{k_2},\slashed{k_3}}\slashed{\delta}\left(\sum_{i=1}^3 \vec{k}_i\right) \frac{-\vec{k}_1\times\vec{k}_2}{(\sum_i|\vec{k}_i|)\vec{k}_1^2|\vec{k}_2||\vec{k}_3|}(\vec{k}_1\times\vec{A}^a(\vec{k}_1))(\vec{k}_2\times\vec{A}^b(\vec{k}_2)) (\vec{k}_3\times\vec{A}^c(\vec{k}_3)) \qquad \\
&=& -i f^{abc} \int_{\slashed{k_1},\slashed{k_2},\slashed{k_3}}\slashed{\delta}\left(\sum_{i=1}^3 \vec{k}_i\right) \frac{1}{(\sum_i|\vec{k}_i|)|\vec{k}_2||\vec{k}_3|}(\vec{k}_2\cdot\vec{A}^a(\vec{k}_1))(\vec{k}_2\times\vec{A}^b(\vec{k}_2)) (\vec{k}_3\times\vec{A}^c(\vec{k}_3)) \qquad \\
&&\quad + i f^{abc} \int_{\slashed{k_1},\slashed{k_2},\slashed{k_3}}\slashed{\delta}\left(\sum_{i=1}^3 \vec{k}_i\right) \frac{\vec{k}_1\cdot\vec{k}_2}{(\sum_i|\vec{k}_i|)\vec{k}_1^2|\vec{k}_2||\vec{k}_3|}(\vec{k}_1\cdot\vec{A}^a(\vec{k}_1))(\vec{k}_2\times\vec{A}^b(\vec{k}_2)) (\vec{k}_3\times\vec{A}^c(\vec{k}_3))\,, \nn
\eE
where we used $\epsilon_{ij}\epsilon_{kl}=\delta_{ik}\delta_{jl}-\delta_{il}\delta_{jk}$. We again write  $\vec k_1$ as $-\vec k_2 -\vec k_3$ and note that as above the $\vec k_3 \cdot\vec k_2$ term vanishes due to symmetry under $\vec k_2 \leftrightarrow \vec k_3$:
\bE{rCl}
&=& -i f^{abc} \int_{\slashed{k_1},\slashed{k_2},\slashed{k_3}}\slashed{\delta}\left(\sum_{i=1}^3 \vec{k}_i\right) \frac{1}{(\sum_i|\vec{k}_i|)|\vec{k}_2||\vec{k}_3|}(\vec{k}_2\cdot\vec{A}^a(\vec{k}_1))(\vec{k}_2\times\vec{A}^b(\vec{k}_2)) (\vec{k}_3\times\vec{A}^c(\vec{k}_3)) \nn\\
&&\quad - i f^{abc} \int_{\slashed{k_1},\slashed{k_2},\slashed{k_3}}\slashed{\delta}\left(\sum_{i=1}^3 \vec{k}_i\right) \frac{|\vec{k}_2|}{(\sum_i|\vec{k}_i|)\vec{k}_1^2|\vec{k}_3|}(\vec{k}_1\cdot\vec{A}^a(\vec{k}_1))(\vec{k}_2\times\vec{A}^b(\vec{k}_2)) (\vec{k}_3\times\vec{A}^c(\vec{k}_3))\,.\qquad
\end{IEEEeqnarray}
Plugging this back into Eq.~(\ref{F1app}) gives
\begin{IEEEeqnarray}{rCl}
F^{(1)}_{GI}[{\vec A}] &=& i f^{abc} \int_{\slashed{k_1},\slashed{k_2},\slashed{k_3}}\slashed{\delta}\left(\sum_{i=1}^3 \vec{k}_i\right) \Bigg\{ \frac{1}{2|\vec{k}_1|} (\vec{k}_1\times\vec{A}^a(\vec{k}_1)) (\vec{A}^b(\vec{k}_2)\times\vec{A}^c(\vec{k}_3)) \\
&&-\frac{1}{(\sum_i|\vec{k}_i|)|\vec{k}_2||\vec{k}_3|}(\vec{k}_2\cdot\vec{A}^a(\vec{k}_1))(\vec{k}_2\times\vec{A}^b(\vec{k}_2)) (\vec{k}_3\times\vec{A}^c(\vec{k}_3))\nn\\
&& + \left(\frac{1}{(\sum_i|\vec{k}_i|)|\vec{k}_1||\vec{k}_3|}+\frac{1}{(\sum_i|\vec{k}_i|)\vec{k}_1^2}\right) (\vec{k}_1\cdot\vec{A}^a(\vec{k}_1)) (\vec{k}_2\times\vec{A}^b(\vec{k}_2)) (\vec{k}_3\times\vec{A}^c(\vec{k}_3)) \Bigg\}\,.\nn
\end{IEEEeqnarray}
The last term vanishes under $\vec k_2 \leftrightarrow \vec k_3$, and in the next-to-last we replace $(\vec{k}_2\times\vec{A}^b(\vec{k}_2)) \rightarrow -(\vec{k}_1+\vec{k}_3)\times\vec{A}^b(\vec{k}_2)$. Then we have
\begin{IEEEeqnarray}{rCl}
F^{(1)}_{GI}[{\vec A}] &=& i f^{abc} \int_{\slashed{k_1},\slashed{k_2},\slashed{k_3}}\slashed{\delta}\left(\sum_{i=1}^3 \vec{k}_i\right) \Bigg\{ \frac{1}{2|\vec{k}_1|} (\vec{k}_1\times\vec{A}^a(\vec{k}_1)) (\vec{A}^b(\vec{k}_2)\times\vec{A}^c(\vec{k}_3))\nn\\
&&-\frac{1}{(\sum_i|\vec{k}_i|)|\vec{k}_2||\vec{k}_3|}(\vec{k}_2\cdot\vec{A}^a(\vec{k}_1))(\vec{k}_2\times\vec{A}^b(\vec{k}_2)) (\vec{k}_3\times\vec{A}^c(\vec{k}_3))\nn\\
&& -\frac{1}{(\sum_i|\vec{k}_i|)|\vec{k}_1||\vec{k}_3|} (\vec{k}_1\cdot\vec{A}^a(\vec{k}_1)) (\vec{k}_1\times\vec{A}^b(\vec{k}_2)) (\vec{k}_3\times\vec{A}^c(\vec{k}_3))\nn\\
&& -\frac{1}{(\sum_i|\vec{k}_i|)|\vec{k}_1||\vec{k}_3|} (\vec{k}_1\cdot\vec{A}^a(\vec{k}_1)) (\vec{k}_3\times\vec{A}^b(\vec{k}_2)) (\vec{k}_3\times\vec{A}^c(\vec{k}_3)) \Bigg\} \,.
\end{IEEEeqnarray}
Exchanging $\vec k_2 \leftrightarrow \vec k_1$ in the next-to-last line and making use of \be ({\vec k}\cdot {\vec A})({\vec k}\times {\vec B})-({\vec k}\times {\vec A})({\vec k}\cdot {\vec B})={\vec k}^2
({\vec A}\times {\vec B})\ee we find
\begin{IEEEeqnarray}{rCl}
F^{(1)}_{GI}[{\vec A}] &=& i f^{abc} \int_{\slashed{k_1},\slashed{k_2},\slashed{k_3}}\slashed{\delta}\left(\sum_{i=1}^3 \vec{k}_i\right) \Bigg\{ \frac{1}{2|\vec{k}_1|} (\vec{k}_1\times\vec{A}^a(\vec{k}_1)) (\vec{A}^b(\vec{k}_2)\times\vec{A}^c(\vec{k}_3))\nn\\
&&-\frac{1}{(\sum_i|\vec{k}_i|)|\vec{k}_2||\vec{k}_3|}\left(\vec{k}_2^2\; \vec{A}^a(\vec{k}_1)\times\vec{A}^b(\vec{k}_2)\right)(\vec{k}_3\times\vec{A}^c(\vec{k}_3))\nn\\
&& -\frac{1}{(\sum_i|\vec{k}_i|)|\vec{k}_1||\vec{k}_3|} (\vec{k}_1\cdot\vec{A}^a(\vec{k}_1)) (\vec{k}_3\times\vec{A}^b(\vec{k}_2)) (\vec{k}_3\times\vec{A}^c(\vec{k}_3)) \Bigg\} \\
&=& i f^{abc} \int_{\slashed{k_1},\slashed{k_2},\slashed{k_3}}\slashed{\delta}\left(\sum_{i=1}^3 \vec{k}_i\right) \Bigg\{ \frac{1}{2|\vec{k}_1|} (\vec{k}_1\times\vec{A}^a(\vec{k}_1)) (\vec{A}^b(\vec{k}_2)\times\vec{A}^c(\vec{k}_3))\nn\\
&&-\frac{|\vec{k}_2|+|\vec{k}_3|}{2(\sum_i|\vec{k}_i|)|\vec{k}_1|}(\vec{k}_1\times\vec{A}^a(\vec{k}_1)) (\vec{A}^b(\vec{k}_2)\times\vec{A}^c(\vec{k}_3))\nn\\
&& -\frac{1}{(\sum_i|\vec{k}_i|)|\vec{k}_1||\vec{k}_3|} (\vec{k}_1\cdot\vec{A}^a(\vec{k}_1)) (\vec{k}_3\times\vec{A}^b(\vec{k}_2)) (\vec{k}_3\times\vec{A}^c(\vec{k}_3)) \Bigg\} \\
&=& i f^{abc} \int_{\slashed{k_1},\slashed{k_2},\slashed{k_3}}\slashed{\delta}\left(\sum_{i=1}^3 \vec{k}_i\right) \Bigg\{ \frac{1}{2(\sum_i|\vec{k}_i|)} (\vec{k}_1\times\vec{A}^a(\vec{k}_1)) (\vec{A}^b(\vec{k}_2)\times\vec{A}^c(\vec{k}_3))\nn\\
&& -\frac{1}{(\sum_i|\vec{k}_i|)|\vec{k}_1||\vec{k}_3|} (\vec{k}_1\cdot\vec{A}^a(\vec{k}_1)) (\vec{k}_3\times\vec{A}^b(\vec{k}_2)) (\vec{k}_3\times\vec{A}^c(\vec{k}_3)) \Bigg\} \\
&=&F^{(1)}_{GL}[{\vec A}] \,. \label{F1equal}
\end{IEEEeqnarray}
This is $F^{(1)}_{GL}[{\vec A}] $ as found in Eq.~(\ref{F1H}), so $\Psi_{GL}$ and $\Psi_{GI}$ are equal at $\O(e)$.
\chapter{Comparison of $\Psi_{GL}$ and $\Psi_{GI}$ at $\O(e^2)$}
\label{F2comp}

In this appendix we will show that $\Psi_{GL}$ and $\Psi_{GI}$ are equal at $\O(e^2A^4)$. In order to do so we rewrite $F^{(0)}_{GL}[\vec A]$ (Eq.~(\ref{F2})), $F^{(1)}_{GL}[\vec A]$ (Eq.~(\ref{F1H})), and $F^{(2,4)}_{GL}[\vec A]$ (Eq.~(\ref{F24GL})) in terms of $J$ and $\theta$.

We first consider $F^{(1)}_{GL}[J(\vec A),\theta(\vec A)]$ at $\O(e^0)$ in Sec.~\ref{sec:B1}, which allows us to compute $F^{(2,4)}_{GL}[J,\theta]$ in Sec.~\ref{sec:B2} from Eq.~(\ref{F4fromF3}):
\begin{IEEEeqnarray}{l}
F_{GL}^{(2,4)} = -\frac{1}{2}\int_{\slashed{p},\slashed{k_1},\slashed{k_2},\slashed{q_1},\slashed{q_2}}\frac{1}{\sum^2_i(|\vec{k}_i|+|\vec{q}_i|)}\left(\frac{\delta F_{GL}^{(1)}}{\delta A^a_i(\vec{p})}\right)[\vec{k}_1,\vec{k}_2]\left(\frac{\delta F_{GL}^{(1)}}{\delta A^a_i(-\vec{p})}\right)[\vec{q}_1,\vec{q}_2] \nn\\
\qquad -if^{b_1b_2c}\int_{\slashed{p},\slashed{k_1},\slashed{k_2},\slashed{q_1},\slashed{q_2}}\frac{\slashed{\delta}(\vec{q}_1+\vec{q}_2-\vec{p})}{\sum_i(|\vec{k}_i|+|\vec{q}_i|)|\vec{q}_1|}(\vec{q}_1\cdot\vec{A}^{b_1}(\vec{q}_1))\left(\vec{A}^{b_2}(\vec{q}_2)\cdot\frac{\delta F_{GL}^{(1)}}{\delta \vec{A}^c(\vec{p})}[\vec{k}_1,\vec{k}_2]\right) \nn\\
\qquad +\frac{1}{8}f^{a_1a_2c}f^{b_1b_2c}\int_{\slashed{k_1},\slashed{k_2},\slashed{q_1},\slashed{q_2}}\frac{\slashed{\delta}(\sum_i(\vec{k}_i+\vec{q}_i))}{\sum_i(|\vec{k}_i|+|\vec{q}_i|)}(\vec{A}^{a_1}(\vec{k}_1)\times\vec{A}^{a_2}(\vec{k}_2))(\vec{A}^{b_1}(\vec{q}_1)\times\vec{A}^{b_2}(\vec{q}_2))
\,. \label{F4fromF3App}
\end{IEEEeqnarray}
This we split up order by order in $\theta$ and rewrite it in such a way that the prefactor $\displaystyle \frac{1}{|\vec{k}_1|+|\vec{k}_2|+|\vec{q}_1|+|\vec{q}_2|}$ drops out.
These expressions then give us guiding lines on the form in which we need to bring $F^{(1)}_{GL}[J,\theta]$ at $\O(e)$, which we do in Sec.~\ref{sec:B3}, and $F^{(0)}_{GL}[J,\theta]$ at $\O(e^2)$ (in Sec.~\ref{sec:B4}). We will then see, that adding up all the terms cancels the $\theta$ dependent terms, while the $\theta$ independent term is equal to $F^{(2,4)}_{GI}[J]$.

We use Eqs.~(\ref{AinJ}) and (\ref{AbarinJ}):
\bE{rCl}
A^a(\vec{k}) &=& -\frac{i}{2}J^a(\vec{k}) +ik\theta^a(\vec{k})
+\frac{ie}{2}f^{abc}\int_\slashed{q}\theta^b(\vec{k}-\vec{q})J^c(\vec{q}) 
-\frac{ie}{2}f^{abc}\int_\slashed{q}q\,\theta^b(\vec{k}-\vec{q})\theta^c(\vec{q})
\nn\\
&& + \frac{ie^2}{4}f^{bcd}f^{dea}\int_\slashed{q}\int_\slashed{p} \theta^b(\vec{k}-\vec{q}-\vec{p}) J^c(\vec{q}) \theta^e(\vec{p})  \nn\\
&& - \frac{ie^2}{3!}f^{bcd}f^{dea}\int_\slashed{q}\int_\slashed{p} \theta^b(\vec{k}-\vec{q}-\vec{p}) q\theta^c(\vec{q}) \theta^e(\vec{p}) 
\nn\\
&&
+{\cal O}(e^3) 
\,,\\
\bar{A}^a(\vec{k}) &=& i\bar{k}\theta^a(\vec{k}) 
-\frac{ie}{2}f^{abc}\int_\slashed{q}\bar{q}\,\theta^b(\vec{k}-\vec{q})\theta^c(\vec{q}) \nn\\
&& -\frac{e^2}{3!}f^{bcd}f^{dea}\int_{\slashed{q},\slashed{p}}[k\bar{q}-\bar{k}q]\,\theta^b(\vec{k}-\vec{q}-\vec{p})\theta^c(\vec{q})\theta^e(\vec{p})
\nn\\
&&
+{\cal O}(e^3)
\,,
\eE
So the terms that actually appear in $F_{GL}$ are
\bE{rCl}
\vec{k}\times\vec{A}^a(\vec{k}) &=& 2i(k\bar{A}^a(\vec{k})-\bar{k}A^a(\vec{k})) \nn\\
&=& 2i\Bigg(\frac{i}{2}\bar{k}J^a(\vec{k}) -\frac{ie}{2}f^{abc}\int_\slashed{q}\bar{k}\theta^b(\vec{k}-\vec{q})J^c(\vec{q})  \\
&&\qquad - \bar{k} \frac{ie^2}{4} f^{bcd}f^{dea} \int_\slashed{q}\int_\slashed{p} \theta^b(\vec{k}-\vec{q}-\vec{p}) J^c(\vec{q}) \theta^e(\vec{p}) \nn\\
&&\qquad  +\frac{ie}{2}f^{abc}\int_\slashed{q}(k\bar{q}-\bar{k}q)\,\theta^b(\vec{q})\theta^c(\vec{k}-\vec{q}) \nn\\
&&\qquad - \frac{ie^2}{6}f^{bcd}f^{dea}\int_\slashed{q}\int_\slashed{p}(k\bar{q}-\bar{k}q) \theta^b(\vec{k}-\vec{q}-\vec{p}) \theta^c(\vec{q}) \theta^e(\vec{p}) +\O(e^3) \Bigg)\,, \nn\\
\vec{k}\cdot\vec{A}^a(\vec{k}) &=& 2(k\bar{A}^a(\vec{k})+\bar{k}A^a(\vec{k})) \nn\\
&=& 2\Bigg(-\frac{i}{2} \bar{k} J^a(\vec{k}) +2ik\bar{k}\theta^a(\vec{k}) +\frac{ie}{2} f^{abc}\int_\slashed{q}\bar{k}\theta^b(\vec{k}-\vec{q})J^c(\vec{q})  \nn\\
&&\qquad  +\frac{ie}{2}f^{abc}\int_\slashed{q}(k\bar{q}+\bar{k}q)\,\theta^b(\vec{q})\theta^c(\vec{k}-\vec{q}) +\O(e^2)\Bigg)\,,\\
\vec{p}\times\vec{A}^a(\vec{k}) &=& 2i(p\bar{A}^a(\vec{k})-\bar{p}A^a(\vec{k})) \nn\\
&=& 2i\Bigg(\frac{i}{2} \bar{p} J^a(\vec{k}) +i (p\bar{k} - \bar{p} k) \theta^a(\vec{k}) -\frac{ie}{2} f^{abc}\int_\slashed{q}\bar{p}\theta^b(\vec{k}-\vec{q})J^c(\vec{q})   \nn\\
&&\qquad  \ +\frac{ie}{2} f^{abc}\int_\slashed{q}(p\bar{q}-\bar{p}q)\,\theta^b(\vec{q})\theta^c(\vec{k}-\vec{q})  +\O(e^2)\Bigg)\,,\qquad\mathrm{and}\\
\vec{A}^a(\vec{k}_1)\times\vec{A}^b(\vec{k}_2) &=& 2i(A^a(\vec{k}_1)\bar{A}^b(\vec{k}_2)-\bar{A}^a(\vec{k}_1){A}^b(\vec{k}_2)) \nn\\
& = & iJ^a(\vec{k}_1)\bar{k_2}\theta^b(\vec{k}_2)-i\bar{k_1}\theta^a(\vec{k}_1)J^b(\vec{k}_2)  - 2i(k_1\bar{k}_2-\bar{k}_1 k_2)\theta^a(\vec{k}_1)\theta^b(\vec{k}_2) \nn\\
&& + \frac{ie}{2}J^a(\vec{k}_1) f^{bcd}\int_\slashed{q}\bar{q}\,\theta^c(\vec{q})\theta^d(\vec{k}_2-\vec{q}) - \frac{ie}{2} f^{acd}\int_\slashed{q}\bar{q}\,\theta^c(\vec{q})\theta^d(\vec{k}_1-\vec{q}) J^b(\vec{k}_2) \nn\\
&& -ief^{acd}\int_\slashed{q}\theta^c(\vec{k}_1-\vec{q})J^d(\vec{q}) \bar{k}_2\theta^b(\vec{k}_2) \nn\\
&& +ie\bar{k}_1\theta^a(\vec{k}_1)f^{bcd}\int_\slashed{q}\theta^c(\vec{k}_2-\vec{q})J^d(\vec{q})  +\O(e^2)\,.
\eE
If $\vec k_1$ and $\vec k_2$ can be interchanged such that interchanging $a\leftrightarrow b$ gives a minus, this last product becomes:
\bE{rCl}
\int_{k_1,k_2}f^{abc}
\vec{A}^a(\vec{k}_1)\times\vec{A}^b(\vec{k}_2) &= & \int_{k_1,k_2}f^{abc} \Bigg\{ 2i\bar{k_2}J^a(\vec{k}_1) \theta^b(\vec{k}_2)  - 4ik_1\bar{k}_2\theta^a(\vec{k}_1)\theta^b(\vec{k}_2)    \nn\\
&&\qquad + ieJ^a(\vec{k}_1) f^{bde}\int_\slashed{q}\bar{q}\,\theta^d(\vec{q})\theta^e(\vec{k}_2-\vec{q}) \nn\\
&&\qquad -2ie\bar{k}_2 f^{ade}\int_\slashed{q}\theta^d(\vec{k}_1-\vec{q})J^e(\vec{q}) \theta^b(\vec{k}_2) +\O(e^2) \Bigg\} \,.
\eE

\section{$F^{(1)}_{GL}[J(\vec A),\theta(\vec A)]$ at $\O(e^0)$}
\label{sec:B1}

We begin with Eq.~(\ref{F1H}):
\begin{IEEEeqnarray}{rCl}
F_{GL}^{(1)}[{\vec A}] &=&  i f^{abc} \int_{\slashed{k_1},\slashed{k_2},\slashed{k_3}}\slashed{\delta}\left(\sum_{i=1}^3 \vec{k}_i\right) \Bigg\{ \frac{1}{2(\sum_i^3|\vec{k}_i|)} (\vec{k}_1\times\vec{A}^a(\vec{k}_1)) (\vec{A}^b(\vec{k}_2)\times\vec{A}^c(\vec{k}_3))\nn\\
&& -\frac{1}{(\sum_i^3|\vec{k}_i|)|\vec{k}_1||\vec{k}_3|} (\vec{k}_1\cdot\vec{A}^a(\vec{k}_1)) (\vec{k}_3\times\vec{A}^b(\vec{k}_2)) (\vec{k}_3\times\vec{A}^c(\vec{k}_3)) \Bigg\} 
\end{IEEEeqnarray}
and use the above relations to rewrite it in terms of $J$ and $\theta$, finding
\bE{rCl}
F^{(1)}_{GL}[J,\theta] &=&  -f^{abc} \int_{\slashed{k_1},\slashed{k_2},\slashed{k_3}}\slashed{\delta}\left(\sum_{i=1}^3 \vec{k}_i\right)\Bigg\{ \frac{1}{\sum_{i=1}^3 |\vec{k}_i|} \frac{\bar{k}_1\bar{k}_3^2}{|\vec{k}_1||\vec{k}_3|} J^a(\vec{k}_1)J^b(\vec{k}_2)J^c(\vec{k}_3) \nn\\
&&\qquad+\frac{\bar{k}_3^2}{|\vec{k}_3|} J^a(\vec{k}_1)J^c(\vec{k}_3) \theta^b(\vec{k}_2) -2\frac{\bar{k}_1k_2\bar{k}_3}{\sum_{i=1}^3 |\vec{k}_i|} J^a(\vec{k}_1)\theta^b(\vec{k}_2)\theta^c(\vec{k}_3)  \nn\\
&&\qquad +2 \frac{(k_3\bar{k}_2-\bar{k}_3k_2)\bar{k}_3 |\vec{k}_1|}{\sum_{i=1}^3 |\vec{k}_i||\vec{k}_3|}\theta^a(\vec{k}_1)\theta^b(\vec{k}_2)J^c(\vec{k}_3)\Bigg\}  + \O(e)\,.
\eE
Note that the term cubic in $J$ is $F^{(1)}_{GI}[J]$, also note that there is no term cubic in $\theta$.

In the parenthesis of the last term, $k_3$ and $\bar{k}_3$ can be replaced by $-(k_1+k_2)$ and $-(\bar{k}_1+\bar{k}_2)$ respectively, where the $\bar{k}_2k_2$ terms cancel:
\bE{rCl}
F^{(1)}_{GL}[J,\theta] &=&  -f^{abc} \int_{\slashed{k_1},\slashed{k_2},\slashed{k_3}}\slashed{\delta}\left(\sum_{i=1}^3 \vec{k}_i\right)\Bigg\{ \frac{1}{\sum_{i=1}^3 |\vec{k}_i|} \frac{\bar{k}_1\bar{k}_3^2}{|\vec{k}_1||\vec{k}_3|} J^a(\vec{k}_1)J^b(\vec{k}_2)J^c(\vec{k}_3) \nn\\
&&\qquad+\frac{\bar{k}_3^2}{|\vec{k}_3|} J^a(\vec{k}_1)J^c(\vec{k}_3) \theta^b(\vec{k}_2) -2\frac{\bar{k}_1k_2\bar{k}_3}{\sum_{i=1}^3 |\vec{k}_i|} J^a(\vec{k}_1)\theta^b(\vec{k}_2)\theta^c(\vec{k}_3)  \nn\\
&&\qquad +2 \frac{(k_1\bar{k}_2-\bar{k}_1k_2)\bar{k}_3 |\vec{k}_1|}{\sum_{i=1}^3 |\vec{k}_i||\vec{k}_3|}\theta^a(\vec{k}_1)\theta^b(\vec{k}_2)J^c(\vec{k}_3)\Bigg\}  + \O(e)\,.
\eE
Renaming $\vec k_1\leftrightarrow \vec k_3$ in the last term results in
\bE{rCl}
&=&  -f^{abc} \int_{\slashed{k_1},\slashed{k_2},\slashed{k_3}}\slashed{\delta}\left(\sum_{i=1}^3 \vec{k}_i\right)\Bigg\{ \frac{1}{\sum_{i=1}^3 |\vec{k}_i|} \frac{\bar{k}_1\bar{k}_3^2}{|\vec{k}_1||\vec{k}_3|} J^a(\vec{k}_1)J^b(\vec{k}_2)J^c(\vec{k}_3) + \frac{\bar{k}_3^2}{|\vec{k}_3|} J^a(\vec{k}_1)J^c(\vec{k}_3) \theta^b(\vec{k}_2) \Bigg\} \nn\\
&& -2f^{abc}\int_{\slashed{k_1},\slashed{k_2},\slashed{k_3}}\frac{\slashed{\delta}\left(\sum_{i=1}^3 \vec{k}_i\right)}{\sum_{i=1}^3 |\vec{k}_i|}\Bigg\{\bar{k}_1k_2\bar{k}_3 - \frac{|\vec{k}_3|}{|\vec{k}_1|}k_3\bar{k}_2\bar{k}_1 +\frac{|\vec{k}_3|}{|\vec{k}_1|}\bar{k}_3k_2\bar{k}_1\Bigg\}J^a(\vec{k}_1)\theta^b(\vec{k}_2)\theta^c(\vec{k}_3) \nn\\
&&+ \O(e)\,.\qquad 
\eE
Renaming $k_2\leftrightarrow k_3$ in the 2nd term of the 2nd line gives a sum over momenta-moduli, thus canceling the sum in the denominator.
\bE{rCl}
&=&  -f^{abc} \int_{\slashed{k_1},\slashed{k_2},\slashed{k_3}}\slashed{\delta}\left(\sum_{i=1}^3 \vec{k}_i\right)\Bigg\{ \frac{1}{\sum_{i=1}^3 |\vec{k}_i|} \frac{\bar{k}_1\bar{k}_3^2}{|\vec{k}_1||\vec{k}_3|} J^a(\vec{k}_1)J^b(\vec{k}_2)J^c(\vec{k}_3) + \frac{\bar{k}_3^2}{|\vec{k}_3|} J^a(\vec{k}_1)J^c(\vec{k}_3) \theta^b(\vec{k}_2) \Bigg\} \nn\\
&& -2f^{abc}\int_{\slashed{k_1},\slashed{k_2},\slashed{k_3}}\slashed{\delta}\left(\sum_{i=1}^3 \vec{k}_i\right)\frac{\bar{k}_1k_2\bar{k}_3}{|\vec{k}_1|}J^a(\vec{k}_1)\theta^b(\vec{k}_2)\theta^c(\vec{k}_3)  +\O(e) \,.
\eE

\subsection{$\frac{\delta F^{(1)}_{GL}}{\delta A^a_i(\vec{p})}$}

We also transform the functional derivatives:
\bE{rCl}
\frac{\delta F^{(1)}_{GL}}{\delta A^a_i(\vec{p})} &=& \int_q\frac{\delta A^b(\vec{q})}{\delta A^a_i(\vec{p})}\frac{\delta F^{(1)}_{GL}}{\delta A^b(\vec{q})} + \int_q\frac{\delta \bar{A}^b(\vec{q})}{\delta A^a_i(\vec{p})}\frac{\delta F^{(1)}_{GL}}{\delta \bar{A}^b(\vec{q})} \nn\\
&=& \int_{q_1,q_2}\frac{\delta A^b(\vec{q}_1)}{\delta A^a_i(\vec{p})}\frac{\delta J^c(\vec{q}_2)}{\delta A^b(\vec{q}_1)}\frac{\delta F^{(1)}_{GL}}{\delta J^c(\vec{q}_2)} \nn\\
&&+ \int_{q_1,q_2}\frac{\delta \bar{A}^b(\vec{q}_1)}{\delta A^a_i(\vec{p})}\left(\frac{\delta J^c(\vec{q}_2)}{\delta \bar{A}^b(\vec{q}_1)}\frac{\delta F^{(1)}_{GL}}{\delta J^c(\vec{q}_2)}+\delta(\vec{q}_1-\vec{q}_2)\frac{\delta F^{(1)}_{GL}}{\delta \bar{A}^b(\vec{q}_2)}\right) \nn\\
&=&\frac{1}{2}\left(\delta_{1i}+i\delta_{2i}\right)(2i)\frac{\delta F^{(1)}_{GL}}{\delta J^a(\vec{p})}   + \frac{1}{2}\left(\delta_{1i}-i\delta_{2i}\right) \left(-2i\frac{p}{\bar{p}}\frac{\delta F^{(1)}_{GL}}{\delta J^a(\vec{p})}+\frac{\delta F^{(1)}_{GL}}{\delta \bar{A}^a(\vec{p})}\right)  \nn\\
&& +\O(e) \\
&=&\frac{1}{2}\left(\delta_{1i}+i\delta_{2i}\right)(2i)\frac{\delta F^{(1)}_{GL}}{\delta J^a(\vec{p})}   + \frac{1}{2}\left(\delta_{1i}-i\delta_{2i}\right) \left(-2i\frac{p}{\bar{p}}\frac{\delta F^{(1)}_{GL}}{\delta J^a(\vec{p})}+\frac{1}{i\bar p}\frac{\delta F^{(1)}_{GL}}{\delta \theta^a(\vec{p})}\right) \nn\\
&&+\O(e) \,.
\eE

\bE{l}
\frac{\delta F^{(1)}_{GL}}{\delta J^a(\vec{p})} = -f^{aa_1a_2}\int_{\slashed{k_1},\slashed{k_2}}\slashed{\delta}\left(\vec{k}_1+\vec{k}_2+\vec{p}\right) \Bigg\{\frac{g^{(3)}(\vec{k}_1,\vec{k}_2,\vec{p})}{32} J^{a_1}(\vec{k}_1)J^{a_2}(\vec{k}_2) \nn\\
\qquad\qquad\qquad +\left( \frac{\bar{p}^2}{|\vec{p}|}-\frac{\bar{k}_1^2}{|\vec{k}_1|}   \right)J^{a_1}(\vec{k}_1)\theta^{a_2}(\vec{k}_2) +2\frac{\bar{p}k_1\bar{k}_2}{|\vec{p}|}\theta^{a_1}(\vec{k}_1)\theta^{a_2}(\vec{k}_2)\Bigg\} +  \O(e) \,.
\eE

\bE{l}
\frac{\delta F^{(1)}_{GL}}{\delta \bar{A}^a(\vec{p})} = -if^{aa_1a_2} \int \slashed{\delta}\left(\vec{k}_1+\vec{k}_2+\vec{p}\right) \frac{\bar{k}_2^2}{|\vec{k}_2|}\frac{1}{\bar{p}} J^{a_1}(\vec{k}_1)J^{a_2}(\vec{k}_2) \nn\\
\qquad\qquad\qquad +2if^{aa_1a_2}\int\slashed{\delta}\left(\vec{k}_1+\vec{k}_2+\vec{p}\right)\Bigg\{-\frac{\bar{k}_1\bar{k}_2}{|\vec{k}_1|}\frac{p}{\bar{p}}+\frac{\bar{k}_1k_2}{|\vec{k}_1|}\Bigg\} J^{a_1}(\vec{k}_1)\theta^{a_2}(\vec{k}_2) +  \O(e) \,.\qquad
\eE

 Putting them together results in
\bE{rCl}
\frac{\delta F^{(1)}_{GL}}{\delta A^a_i(\vec{p})}&=& -if^{aa_1a_2} \int_{\slashed{k_1},\slashed{k_2}}\slashed{\delta}\left(\vec{k}_1+\vec{k}_2+\vec{p}\right) \nn\\
&&\Bigg\{\Bigg\{ (\delta_{1i}+i\delta_{2i})\frac{g^{(3)}(\vec{k}_1,\vec{k}_2,\vec{p})}{32} \nn\\
&& \qquad+ (\delta_{1i}-i\delta_{2i})\left(-\frac{p}{\bar{p}}\frac{g^{(3)}(\vec{k}_1,\vec{k}_2,\vec{p})}{32}+\frac{\bar{k}_2^2}{2\bar{p}|\vec{k}_2|}\right)\Bigg\}J^{a_1}(\vec{k}_1)J^{a_2}(\vec{k}_2) \nn\\
&&+\Bigg\{(\delta_{1i}+i\delta_{2i})\left( \frac{\bar{p}^2}{|\vec{p}|}-\frac{\bar{k}_1^2}{|\vec{k}_1|}\right)  \nn\\
&&\qquad - (\delta_{1i}-i\delta_{2i}) \left(\frac{1}{4}|\vec{p}|+\frac{\bar{k}_1}{|\vec{k}_1|}\left(-\frac{p}{\bar{p}}(\bar{k}_1+\bar{k}_2)+k_2\right) \right)\Bigg\}J^{a_1}(\vec{k}_1)\theta^{a_2}(\vec{k}_2) \nn\\
&&+\Bigg\{(\delta_{1i}+i\delta_{2i})2\frac{\bar{p}k_1\bar{k}_2}{|\vec{p}|} -(\delta_{1i}-i\delta_{2i}) 2\frac{p k_1\bar{k}_2}{|\vec{p}|}\Bigg\} \theta^{a_1}(\vec{k}_1)\theta^{a_2}(\vec{k}_2) \Bigg\}+\O(e)\,.  
\eE

\section{$F^{(2,4)}_{GL}[J,\theta]$}
\label{sec:B2}

From the above we can compute $F^{(2,4)}_{GL}[J,\theta]$ using Eq.~(\ref{F4fromF3App}), and we do so, order by order in $\theta$.

\subsection{Orders $\theta^0$ and $\theta$:}

\subsubsection{$\left(\frac{\delta F^{(1)}_{GL}}{\delta A}\right)^2$-term}

We shall use that
\bE{l}
\Bigg( (\delta_{1i}+i\delta_{2i})C(\vec{p}) + (\delta_{1i}-i\delta_{2i})D(\vec{p})\Bigg) \Bigg( (\delta_{1i}+i\delta_{2i})C(-\vec{p}) + (\delta_{1i}-i\delta_{2i})D(-\vec{p})\Bigg) \nn\\
= 2C(\vec{p})D(-\vec{p}) + 2D(\vec{p})C(-\vec{p}) \,.
\eE

\bE{l}
\Longrightarrow \int_p\frac{\delta F^{(1)}_{GL}}{\delta A^a_i(\vec{p})}\frac{\delta F^{(1)}_{GL}}{\delta A^a_i(-\vec{p})} \nn\\
\quad = -\frac{1}{16}f^{aa_1a_2}f^{ab_1b_2} \int_{\slashed{k_1},\slashed{k_2},\slashed{q_1},\slashed{q_2},p}\slashed{\delta}\left(\vec{k}_1+\vec{k}_2+\vec{p}\right)\slashed{\delta}\left(\vec{q}_1+\vec{q}_2-\vec{p}\right) J^{a_1}(\vec{k}_1)J^{a_2}(\vec{k}_2)J^{b_1}(\vec{q}_1)J^{b_2}(\vec{q}_2) \nn\\
\qquad  \Bigg\{ g^{(3)}(\vec{k}_1,\vec{k}_2,\vec{p})\left(\frac{\bar{q}_2^2}{2(-\bar{p})|\vec{q}_2|} - \frac{p}{\bar{p}}\frac{g^{(3)}(\vec{q}_1,\vec{q}_2,-\vec{p})}{32}\right)  \nn\\
\qquad\qquad + \left(\frac{\bar{k}_2^2}{2\bar{p}|\vec{k}_2|} - \frac{p}{\bar{p}}\frac{g^{(3)}(\vec{k}_1,\vec{k}_2,\vec{p})}{32}\right)g^{(3)}(\vec{q}_1,\vec{q}_2,-\vec{p})\Bigg\} \nn\\
\qquad -2f^{aa_1a_2}f^{ab_1b_2} \int_{\slashed{k_1},\slashed{k_2},\slashed{q_1},\slashed{q_2},p}\slashed{\delta}\left(\vec{k}_1+\vec{k}_2+\vec{p}\right)\slashed{\delta}\left(\vec{q}_1+\vec{q}_2-\vec{p}\right) \nn\\
\qquad \Bigg\{\left( \frac{\bar{p}^2}{|\vec{p}|}-\frac{\bar{k}_1^2}{|\vec{k}_1|} \right) \left(\frac{\bar{q}_2^2}{2(-\bar{p})|\vec{q}_2|} - \frac{p}{\bar{p}}\frac{g^{(3)}(\vec{q}_1,\vec{q}_2,-\vec{p})}{32}\right)J^{a_1}(\vec{k}_1)\theta^{a_2}(\vec{k}_2)J^{b_1}(\vec{q}_1)J^{b_2}(\vec{q}_2) \nn\\
\qquad\qquad + \left( \frac{\bar{p}^2}{|\vec{p}|}-\frac{\bar{q}_1^2}{|\vec{q}_1|} \right) \left(\frac{\bar{k}_2^2}{2\bar{p}|\vec{k}_2|} - \frac{p}{\bar{p}}\frac{g^{(3)}(\vec{k}_1,\vec{k}_2,\vec{p})}{32}\right)J^{a_1}(\vec{k}_1)J^{a_2}(\vec{k}_2)J^{b_1}(\vec{q}_1)\theta^{b_2}(\vec{q}_2)\Bigg\}\\
\qquad +\frac{1}{16}f^{aa_1a_2}f^{ab_1b_2} \int_{\slashed{k_1},\slashed{k_2},\slashed{q_1},\slashed{q_2},p}\slashed{\delta}\left(\vec{k}_1+\vec{k}_2+\vec{p}\right)\slashed{\delta}\left(\vec{q}_1+\vec{q}_2-\vec{p}\right) \nn\\
\qquad \Bigg\{g^{(3)}(\vec{k}_1,\vec{k}_2,\vec{p}) \left(\frac{1}{4}|\vec{p}|+\frac{\bar{q}_1}{|\vec{q}_1|}\left(-\frac{p}{\bar{p}}(\bar{q}_1+\bar{q}_2)+q_2\right) \right) J^{a_1}(\vec{k}_1)J^{a_2}(\vec{k}_2)J^{b_1}(\vec{q}_1)\theta^{b_2}(\vec{q}_2)\nn\\
\qquad + g^{(3)}(\vec{q}_1,\vec{q}_2,-\vec{p}) \left(\frac{1}{4}|\vec{p}|+\frac{\bar{k}_1}{|\vec{k}_1|}\left(-\frac{p}{\bar{p}}(\bar{k}_1+\bar{k}_2)+k_2\right) \right) J^{a_1}(\vec{k}_1)\theta^{a_2}(\vec{k}_2)J^{b_1}(\vec{q}_1)J^{b_2}(\vec{q}_2)\Bigg\} \nn\\
\qquad +\O(\theta^2)+\O(e)\nn\\
\quad = -\frac{1}{256}f^{aa_1a_2}f^{ab_1b_2} \int_{\slashed{k_1},\slashed{k_2},\slashed{q_1},\slashed{q_2}}\slashed{\delta}\left(\vec{k}_1+\vec{k}_2+\vec{q}_1+\vec{q}_2\right) J^{a_1}(\vec{k}_1)J^{a_2}(\vec{k}_2)J^{b_1}(\vec{q}_1)J^{b_2}(\vec{q}_2)\nn\\
\qquad\quad \left(16\frac{\bar{k}_2^2}{(-\bar{k}_1-\bar{k}_2)|\vec{k}_2|} - \frac{k_1+k_2}{\bar{k}_1+\bar{k}_2}g^{(3)}(\vec{k}_1,\vec{k}_2,-\vec{k}_1-\vec{k}_2)\right)g^{(3)}(\vec{q}_1,\vec{q}_2,-\vec{q}_1-\vec{q}_2)  \nn\\
\qquad -\frac{1}{8}f^{aa_1a_2}f^{ab_1b_2} \int_{\slashed{k_1},\slashed{k_2},\slashed{q_1},\slashed{q_2}}\slashed{\delta}\left(\vec{k}_1+\vec{k}_2+\vec{q}_1+\vec{q}_2\right) J^{a_1}(\vec{k}_1)J^{a_2}(\vec{k}_2)J^{b_1}(\vec{q}_1)\theta^{b_2}(\vec{q}_2)\nn\\
\qquad\qquad  \left( \frac{(\bar{q}_1+\bar{q}_2)^2}{|\vec{q}_1+\vec{q}_2|}-\frac{\bar{q}_1^2}{|\vec{q}_1|} \right) \left(-16\frac{\bar{k}_2^2}{(\bar{k}_1+\bar{k}_2)|\vec{k}_2|} - \frac{k_1+k_2}{\bar{k}_1+\bar{k}_2}g^{(3)}(\vec{k}_1,\vec{k}_2,-\vec{k}_1-\vec{k}_2)\right) \nn\\
\qquad +\frac{1}{8}f^{aa_1a_2}f^{ab_1b_2} \int_{\slashed{k_1},\slashed{k_2},\slashed{q_1},\slashed{q_2}}\slashed{\delta}\left(\vec{k}_1+\vec{k}_2+\vec{q}_1+\vec{q}_2\right) J^{a_1}(\vec{k}_1)J^{a_2}(\vec{k}_2)J^{b_1}(\vec{q}_1)\theta^{b_2}(\vec{q}_2) \nn\\
\qquad\qquad g^{(3)}(\vec{k}_1,\vec{k}_2,-\vec{k}_1-\vec{k}_2) \left(\frac{1}{4}|\vec{q}_1+\vec{q}_2|+\frac{\bar{q}_1}{|\vec{q}_1|}\left(-(q_1+q_2)+q_2\right) \right) +\O(\theta^2)+\O(e) \,.
\eE

\subsubsection{$\left(\vec{A}\cdot\frac{\delta F^{(1)}_{GL}}{\delta \vec{A}}\right)$-term}

\bE{l}
-if^{b_1b_2c}\int_{\slashed{p},\slashed{k_1},\slashed{k_2},\slashed{q_1},\slashed{q_2}}\frac{\slashed{\delta}(\vec{q}_1+\vec{q}_2-\vec{p})}{\sum_i(|\vec{k}_i|+|\vec{q}_i|)|\vec{q}_1|}(\vec{q}_1\cdot\vec{A}^{b_1}(\vec{q}_1))\left(\vec{A}^{b_2}(\vec{q}_2)\cdot\frac{\delta F^{(1)}_{GL}}{\delta \vec{A}^c(\vec{p})}[\vec{k}_1,\vec{k}_2]\right) \nn\\
= -i^2 f^{b_1b_2c}f^{a_1a_2c}\int_{\slashed{p},\slashed{k_1},\slashed{k_2},\slashed{q_1},\slashed{q_2}}\frac{\slashed{\delta}(\vec{q}_1+\vec{q}_2-\vec{p})\slashed{\delta}(\vec{k}_1+\vec{k}_2+\vec{p})}{(|\vec{k}_1|+|\vec{k}_2|+|\vec{q}_1|+|\vec{q}_2|)|\vec{q}_1|} \nn\\
\quad \times (-i\bar{q}_1 J^{b_1}(\vec{q}_1)+i|\vec{q}_1|^2\theta^{b_1}(\vec{q}_1)) A^{b_2}_i(\vec{q}_2) (-i) \nn\\
\quad  \Bigg\{\Bigg\{ (\delta_{1i}+i\delta_{2i})\frac{g^{(3)}(\vec{k}_1,\vec{k}_2,\vec{p})}{32} + (\delta_{1i}-i\delta_{2i})\left(-\frac{p}{\bar{p}}\frac{g^{(3)}(\vec{k}_1,\vec{k}_2,\vec{p})}{32}+\frac{\bar{k}_2^2}{2\bar{p}|\vec{k}_2|}\right)\Bigg\}J^{a_1}(\vec{k}_1)J^{a_2}(\vec{k}_2) \nn\\
\qquad\qquad  +(\delta_{1i}+i\delta_{2i})\left( \frac{\bar{p}^2}{|\vec{p}|}-\frac{\bar{k}_1^2}{|\vec{k}_1|}   \right)J^{a_1}(\vec{k}_1)\theta^{a_2}(\vec{k}_2) +(\delta_{1i}-i\delta_{2i})\O(\theta)\Bigg\}+\O(e)\\
= (-i)^3 f^{b_1b_2c}f^{a_1a_2c}\int_{\slashed{p},\slashed{k_1},\slashed{k_2},\slashed{q_1},\slashed{q_2}}\frac{\slashed{\delta}(\vec{q}_1+\vec{q}_2-\vec{p})\slashed{\delta}(\vec{k}_1+\vec{k}_2+\vec{p})}{(|\vec{k}_1|+|\vec{k}_2|+|\vec{q}_1|+|\vec{q}_2|)|\vec{q}_1|}\left(\bar{q}_1 J^{b_1}(\vec{q}_1)-|\vec{q}_1|^2\theta^{b_1}(\vec{q}_1)\right) \nn\\
\quad \Bigg\{\Bigg\{ 2A^{b_2}(\vec{q}_2)\frac{g^{(3)}(\vec{k}_1,\vec{k}_2,\vec{p})}{32} + 2\bar{A}^{b_2}(\vec{q}_2)\left(-\frac{p}{\bar{p}}\frac{g^{(3)}(\vec{k}_1,\vec{k}_2,\vec{p})}{32}+\frac{\bar{k}_2^2}{2\bar{p}|\vec{k}_2|}\right)\Bigg\}J^{a_1}(\vec{k}_1)J^{a_2}(\vec{k}_2) \nn\\
\qquad\qquad  +2A^{b_2}(\vec{q}_2)\left( \frac{\bar{p}^2}{|\vec{p}|}-\frac{\bar{k}_1^2}{|\vec{k}_1|}   \right)J^{a_1}(\vec{k}_1)\theta^{a_2}(\vec{k}_2)\Bigg\} +\O(\theta^2)+\O(e)\\
= (-i)^4 f^{b_1b_2c}f^{a_1a_2c}\int_{\slashed{p},\slashed{k_1},\slashed{k_2},\slashed{q_1},\slashed{q_2}}\frac{\slashed{\delta}(\vec{q}_1+\vec{q}_2-\vec{p})\slashed{\delta}(\vec{k}_1+\vec{k}_2+\vec{p})}{(|\vec{k}_1|+|\vec{k}_2|+|\vec{q}_1|+|\vec{q}_2|)|\vec{q}_1|}\left(\bar{q}_1 J^{b_1}(\vec{q}_1)-|\vec{q}_1|^2\theta^{b_1}(\vec{q}_1)\right) \nn\\
\quad \Bigg\{\Bigg\{ (J^{b_2}(\vec{q}_2)-2q_2\theta^{b_2}(\vec{q}_2))\frac{g^{(3)}(\vec{k}_1,\vec{k}_2,\vec{p})}{32} - 2\bar{q}_2\theta^{b_2}(\vec{q}_2)\left(-\frac{p}{\bar{p}}\frac{g^{(3)}(\vec{k}_1,\vec{k}_2,\vec{p})}{32}+\frac{\bar{k}_2^2}{2\bar{p}|\vec{k}_2|}\right)\Bigg\} \nn\\
\qquad\qquad \times J^{a_1}(\vec{k}_1)J^{a_2}(\vec{k}_2) \nn\\
\qquad\quad  +J^{b_2}(\vec{q}_2)\left( \frac{\bar{p}^2}{|\vec{p}|}-\frac{\bar{k}_1^2}{|\vec{k}_1|}   \right)J^{a_1}(\vec{k}_1)\theta^{a_2}(\vec{k}_2)\Bigg\} +\O(\theta^2)+\O(e)\\
= \frac{1}{32} f^{a_1a_2c}f^{b_1b_2c}\int_{\slashed{k_1},\slashed{k_2},\slashed{q_1},\slashed{q_2}}\frac{\slashed{\delta}(\vec{k}_1+\vec{k}_2+\vec{q}_1+\vec{q}_2)\bar{k}_1}{(|\vec{k}_1|+|\vec{k}_2|+|\vec{q}_1|+|\vec{q}_2|)|\vec{k}_1|}g^{(3)}(\vec{q}_1,\vec{q}_2,-\vec{q}_1-\vec{q}_2) \nn\\
\qquad\qquad  J^{a_1}(\vec{k}_1)J^{a_2}(\vec{k}_2)J^{b_1}(\vec{q}_1) J^{b_2}(\vec{q}_2)  \nn\\
\quad +\frac{1}{16} f^{a_1a_2c}f^{b_1b_2c}\int_{\slashed{k_1},\slashed{k_2},\slashed{q_1},\slashed{q_2}}\frac{\slashed{\delta}(\vec{k}_1+\vec{k}_2+\vec{q}_1+\vec{q}_2)}{|\vec{k}_1|+|\vec{k}_2|+|\vec{q}_1|+|\vec{q}_2|} J^{a_1}(\vec{k}_1)J^{a_2}(\vec{k}_2)J^{b_1}(\vec{q}_1)\theta^{b_2}(\vec{q}_2)\nn\\
\qquad \Bigg\{\frac{1}{2}|\vec{q}_2|g^{(3)}(\vec{k}_1,\vec{k}_2,-\vec{k}_1-\vec{k}_2)  +\frac{\bar{q}_1}{|\vec{q}_1|}\left(-q_2+\bar{q}_2\frac{k_1+k_2}{\bar{k}_1+\bar{k}_2}\right)g^{(3)}(\vec{k}_1,\vec{k}_2,-\vec{k}_1-\vec{k}_2) \nn\\
\qquad\qquad +16\frac{\bar{q}_1}{|\vec{q}_1|}\frac{\bar{q}_2\bar{k}_2^2}{(\bar{k}_1+\bar{k}_2)|\vec{k}_2|}+16\frac{\bar{k}_1}{|\vec{k}_1|}\left( \frac{(\bar{q}_1+\bar{q}_2)^2}{|\vec{q}_1+\vec{q}_2|}-\frac{\bar{q}_1^2}{|\vec{q}_1|}   \right)\Bigg\} +\O(\theta^2)+\O(e) \,.
\eE

\subsubsection{Both of them together}
The $(A\times A)^2$-term is of $\O(\theta^2)$, so

\bE{l}
F^{(2,4)}_{GL} = -\frac{1}{2}\int_{\slashed{p},\slashed{k_1},\slashed{k_2},\slashed{q_1},\slashed{q_2}}\frac{1}{\sum_i(|\vec{k}_i|+|\vec{q}_i|)}\left(\frac{\delta F^{(1)}_{GL}}{\delta A^a_i(\vec{p})}\right)[\vec{k}_1,\vec{k}_2]\left(\frac{\delta F^{(1)}_{GL}}{\delta A^a_i(-\vec{p})}\right)[\vec{q}_1,\vec{q}_2] \nn\\
\qquad -if^{b_1b_2c}\int_{\slashed{p},\slashed{k_1},\slashed{k_2},\slashed{q_1},\slashed{q_2}}\frac{\slashed{\delta}(\vec{q}_1+\vec{q}_2-\vec{p})}{\sum_i(|\vec{k}_i|+|\vec{q}_i|)|\vec{q}_1|}(\vec{q}_1\cdot\vec{A}^{b_1}(\vec{q}_1))\left(\vec{A}^{b_2}(\vec{q}_2)\cdot\frac{\delta F^{(1)}_{GL}}{\delta \vec{A}^c(\vec{p})}[\vec{k}_1,\vec{k}_2]\right) \nn\\
\qquad +\O(\theta^2)+\O(e)\\
= -\frac{1}{512}f^{aa_1a_2}f^{ab_1b_2} \int_{\slashed{k_1},\slashed{k_2},\slashed{q_1},\slashed{q_2}} \frac{\slashed{\delta}\left(\vec{k}_1+\vec{k}_2+\vec{q}_1+\vec{q}_2\right)}{|\vec{k}_1|+|\vec{k}_2|+|\vec{q}_1|+|\vec{q}_2|} J^{a_1}(\vec{k}_1)J^{a_2}(\vec{k}_2)J^{b_1}(\vec{q}_1)J^{b_2}(\vec{q}_2)\nn\\
\qquad\quad \Bigg\{g^{(3)}(\vec{k}_1,\vec{k}_2,-\vec{k}_1-\vec{k}_2)\frac{k_1+k_2}{\bar{k}_1+\bar{k}_2} g^{(3)}(\vec{q}_1,\vec{q}_2,-\vec{q}_1-\vec{q}_2) \nn\\
\qquad\qquad +16\left(\frac{\bar{k}_2^2}{(\bar{k}_1+\bar{k}_2)|\vec{k}_2|} +\frac{\bar{k}_2}{|\vec{k}_2|}\right)g^{(3)}(\vec{q}_1,\vec{q}_2,-\vec{q}_1-\vec{q}_2) \Bigg\} \nn\\
\quad +\frac{1}{16} f^{a_1a_2c}f^{b_1b_2c}\int_{\slashed{k_1},\slashed{k_2},\slashed{q_1},\slashed{q_2}}\frac{\slashed{\delta}(\vec{k}_1+\vec{k}_2+\vec{q}_1+\vec{q}_2)}{|\vec{k}_1|+|\vec{k}_2|+|\vec{q}_1|+|\vec{q}_2|} J^{a_1}(\vec{k}_1)J^{a_2}(\vec{k}_2)J^{b_1}(\vec{q}_1)\theta^{b_2}(\vec{q}_2)\nn\\
\quad\Bigg\{\left( \frac{(\bar{q}_1+\bar{q}_2)^2}{|\vec{q}_1+\vec{q}_2|}-\frac{\bar{q}_1^2}{|\vec{q}_1|} \right) \left(-16\frac{\bar{k}_2^2}{(\bar{k}_1+\bar{k}_2)|\vec{k}_2|} - \frac{k_1+k_2}{\bar{k}_1+\bar{k}_2}g^{(3)}(\vec{k}_1,\vec{k}_2,-\vec{k}_1-\vec{k}_2)\right) \nn\\
\quad\quad -\frac{1}{4}g^{(3)}(\vec{k}_1,\vec{k}_2,-\vec{k}_1-\vec{k}_2)(|\vec{q}_1+\vec{q}_2|-|\vec{q}_1|)\nn\\
\quad\quad +\frac{1}{2}|\vec{q}_2|g^{(3)}(\vec{k}_1,\vec{k}_2,-\vec{k}_1-\vec{k}_2)+\frac{\bar{q}_1}{|\vec{q}_1|}\left(-q_2+\bar{q}_2\frac{k_1+k_2}{\bar{k}_1+\bar{k}_2}\right)g^{(3)}(\vec{k}_1,\vec{k}_2,-\vec{k}_1-\vec{k}_2) \nn\\
\qquad\quad +16\frac{\bar{q}_1}{|\vec{q}_1|}\frac{\bar{q}_2\bar{k}_2^2}{(\bar{k}_1+\bar{k}_2)|\vec{k}_2|}+16\frac{\bar{k}_1}{|\vec{k}_1|}\left( \frac{(\bar{q}_1+\bar{q}_2)^2}{|\vec{q}_1+\vec{q}_2|}-\frac{\bar{q}_1^2}{|\vec{q}_1|}   \right)\Bigg\} +\O(\theta^2)+\O(e) \,.
\eE

\subsubsection{Only order $\theta^0$}


\bE{l}
F^{(2,4)}_{GL}|_{\O(\theta^0)} = -\frac{1}{512}f^{aa_1a_2}f^{ab_1b_2} \int_{\slashed{k_1},\slashed{k_2},\slashed{q_1},\slashed{q_2}} \slashed{\delta}\left(\vec{k}_1+\vec{k}_2+\vec{q}_1+\vec{q}_2\right) J^{a_1}(\vec{k}_1)J^{a_2}(\vec{k}_2)J^{b_1}(\vec{q}_1)J^{b_2}(\vec{q}_2)\nn\\
\qquad\quad \frac{1}{|\vec{k}_1|+|\vec{k}_2|+|\vec{q}_1|+|\vec{q}_2|} \Bigg\{g^{(3)}(\vec{k}_1,\vec{k}_2,-\vec{k}_1-\vec{k}_2)\frac{k_1+k_2}{\bar{k}_1+\bar{k}_2} g^{(3)}(\vec{q}_1,\vec{q}_2,-\vec{q}_1-\vec{q}_2) \nn\\
\qquad\qquad +\left(\frac{\bar{k}_2(2\bar{k}_2+\bar{k}_1)}{|\vec{k}_2|}-\frac{\bar{k}_1(2\bar{k}_1+\bar{k}_2)}{|\vec{k}_1|}\right)\frac{4}{\bar{k}_1+\bar{k}_2}g^{(3)}(\vec{q}_1,\vec{q}_2,-\vec{q}_1-\vec{q}_2) \nn\\
\qquad\qquad +\left(\frac{\bar{q}_2(2\bar{q}_2+\bar{q}_1)}{|\vec{q}_2|}-\frac{\bar{q}_1(2\bar{q}_1+\bar{q}_2)}{|\vec{q}_1|}\right)\frac{4}{\bar{q}_1+\bar{q}_2}g^{(3)}(\vec{k}_1,\vec{k}_2,-\vec{k}_1-\vec{k}_2) \Bigg\} + \O(e) \label{F24theta0} \\
=F^{(2,4)}_{GI}[J] +O(e)\,, 
\eE
which is what we expect.

\subsubsection{Order $\theta$}

\bE{l}
F^{(2,4)}_{GL}|_{\O(\theta)} = \frac{1}{16} f^{a_1a_2c}f^{b_1b_2c}\int_{\slashed{k_1},\slashed{k_2},\slashed{q_1},\slashed{q_2}}\frac{\slashed{\delta}(\vec{k}_1+\vec{k}_2+\vec{q}_1+\vec{q}_2)}{|\vec{k}_1|+|\vec{k}_2|+|\vec{q}_1|+|\vec{q}_2|} J^{a_1}(\vec{k}_1)J^{a_2}(\vec{k}_2)J^{b_1}(\vec{q}_1)\theta^{b_2}(\vec{q}_2)\nn\\
\qquad\Bigg\{\left( -\frac{1}{4}|\vec{q}_1+\vec{q}_2|+\frac{\bar{q}_1}{|\vec{q}_1|} (\bar{q}_1+\bar{q}_2)\frac{k_1+k_2}{\bar{k}_1+\bar{k}_2}-\frac{\bar{q}_1q_2}{|\vec{q}_1|}-\frac{1}{4}|\vec{q}_1+\vec{q}_2|+\frac{1}{4}|\vec{q}_1|+\frac{1}{2}|\vec{q}_2|\right) \nn\\
\qquad\quad \times g^{(3)}(\vec{k}_1,\vec{k}_2,-\vec{k}_1-\vec{k}_2)\nn\\
\qquad +16\left(\frac{(\bar{q}_1+\bar{q}_2)}{|\vec{q}_1+\vec{q}_2|}\frac{\bar{k}_2^2}{|\vec{k}_2|} + \frac{\bar{q}_1\bar{k}_2^2}{|\vec{q}_1|(\bar{k}_1+\bar{k}_2)|\vec{k}_2|}(\bar{q}_1+\bar{q}_2)+\frac{\bar{k}_1}{|\vec{k}_1|} \frac{(\bar{q}_1+\bar{q}_2)^2}{|\vec{q}_1+\vec{q}_2|}-\frac{\bar{k}_1}{|\vec{k}_1|}\frac{\bar{q}_1^2}{|\vec{q}_1|}   \right)\Bigg\}
\nn\\\qquad + \O(e) \\
=\frac{1}{16} f^{a_1a_2c}f^{b_1b_2c}\int_{\slashed{k_1},\slashed{k_2},\slashed{q_1},\slashed{q_2}}\frac{\slashed{\delta}(\vec{k}_1+\vec{k}_2+\vec{q}_1+\vec{q}_2)}{|\vec{k}_1|+|\vec{k}_2|+|\vec{q}_1|+|\vec{q}_2|} J^{a_1}(\vec{k}_1)J^{a_2}(\vec{k}_2)J^{b_1}(\vec{q}_1)\theta^{b_2}(\vec{q}_2)\nn\\
\qquad\Bigg\{\left( -\frac{1}{2}|\vec{q}_1+\vec{q}_2|+\frac{1}{2}|\vec{q}_1|+\frac{1}{2}|\vec{q}_2| \right) g^{(3)}(\vec{k}_1,\vec{k}_2,-\vec{k}_1-\vec{k}_2)\nn\\
\qquad +16\left(\frac{\bar{k}_2^2}{|\vec{k}_2|}\frac{(-\bar{k}_1-\bar{k}_2)}{|-\vec{k}_1-\vec{k}_2|}   +\frac{(-\bar{k}_1-\bar{k}_2)^2}{|-\vec{k}_1-\vec{k}_2|}\frac{\bar{k}_1}{|\vec{k}_1|} - \frac{\bar{q}_1\bar{k}_2^2}{|\vec{q}_1||\vec{k}_2|}  -\frac{\bar{k}_1}{|\vec{k}_1|}\frac{\bar{q}_1^2}{|\vec{q}_1|}   \right)\Bigg\}
\quad +  \O(e)\\
=\frac{1}{16} f^{a_1a_2c}f^{b_1b_2c}\int_{\slashed{k_1},\slashed{k_2},\slashed{q_1},\slashed{q_2}}\frac{\slashed{\delta}(\vec{k}_1+\vec{k}_2+\vec{q}_1+\vec{q}_2)}{|\vec{k}_1|+|\vec{k}_2|+|\vec{q}_1|+|\vec{q}_2|} J^{a_1}(\vec{k}_1)J^{a_2}(\vec{k}_2)J^{b_1}(\vec{q}_1)\theta^{b_2}(\vec{q}_2)\nn\\
\qquad\Bigg\{\left( -\frac{1}{2}|\vec{q}_1+\vec{q}_2|+\frac{1}{2}|\vec{q}_1| +\frac{1}{2}|\vec{q}_2|\right) g^{(3)}(\vec{k}_1,\vec{k}_2,-\vec{k}_1-\vec{k}_2)\nn\\
\qquad +\frac{1}{2}\left(|\vec{k}_1|+|\vec{k}_2|+|\vec{k}_1+\vec{k}_2|\right)g^{(3)}(\vec{k}_1,\vec{k}_2,-\vec{k}_1-\vec{k}_2)\nn\\
\qquad +16\left(-\frac{\bar{k}_1^2\bar{k}_2}{|\vec{k}_1||\vec{k}_2|} - \frac{\bar{q}_1\bar{k}_2^2}{|\vec{q}_1||\vec{k}_2|}  -\frac{\bar{k}_1}{|\vec{k}_1|}\frac{\bar{q}_1^2}{|\vec{q}_1|}   \right)\Bigg\}
\quad + \O(e)\\
=\frac{1}{16}f^{a_1a_2c}f^{b_1b_2c}\int_{\slashed{k_1},\slashed{k_2},\slashed{q_1},\slashed{q_2}}\frac{\slashed{\delta}(\vec{k}_1+\vec{k}_2+\vec{q}_1+\vec{q}_2)}{|\vec{k}_1|+|\vec{k}_2|+|\vec{q}_1|+|\vec{q}_2|} J^{a_1}(\vec{k}_1)J^{a_2}(\vec{k}_2)J^{b_1}(\vec{q}_1)\theta^{b_2}(\vec{q}_2)\nn\\
\qquad\Bigg\{\frac{1}{2}\left(|\vec{k}_1|+|\vec{k}_2|+|\vec{q}_1|+|\vec{q}_2|\right)g^{(3)}(\vec{k}_1,\vec{k}_2,-\vec{k}_1-\vec{k}_2)\nn\\
\qquad -\frac{1}{2}\left(|\vec{k}_1|+|\vec{k}_2|+|\vec{q}_1|\right)g^{(3)}(\vec{k}_1,\vec{k}_2,\vec{q}_1)\Bigg\}
\quad + \O(e) \\
=\frac{1}{32}f^{a_1a_2c}f^{b_1b_2c}\int_{\slashed{k_1},\slashed{k_2},\slashed{q_1},\slashed{q_2}} \slashed{\delta}\left(\sum_i(\vec{k}_i+\vec{q}_i)\right) J^{a_1}(\vec{k}_1)J^{a_2}(\vec{k}_2)J^{b_1}(\vec{q}_1)\theta^{b_2}(\vec{q}_2) g^{(3)}(\vec{k}_1,\vec{k}_2,-\vec{k}_1-\vec{k}_2)\nn\\
\quad -\frac{1}{32}\frac{1}{3}\left(f^{a_1a_2c}f^{b_1b_2c} +f^{b_1a_1c}f^{a_2b_2c} +f^{a_2b_1c}f^{a_1b_2c}\right)\int_{\slashed{k_1},\slashed{k_2},\slashed{q_1},\slashed{q_2}} \slashed{\delta}\left(\sum_i(\vec{k}_i+\vec{q}_i)\right)\nn\\
\qquad\frac{|\vec{k}_1|+|\vec{k}_2|+|\vec{q}_1|}{|\vec{k}_1|+|\vec{k}_2|+|\vec{q}_1|+|\vec{q}_2|}g^{(3)}(\vec{k}_1,\vec{k}_2,\vec{q}_1) J^{a_1}(\vec{k}_1)J^{a_2}(\vec{k}_2)J^{b_1}(\vec{q}_1)\theta^{b_2}(\vec{q}_2)
\quad + \O(e) \,.
\eE
The last equality is true because $g^{(3)}(\vec{k}_1,\vec{k}_2,\vec{q}_1) = g^{(3)}(\vec{q}_1,\vec{k}_1,\vec{k}_2) = g^{(3)}(\vec{k}_2,\vec{q}_1,\vec{k}_1)$. The sum of the structure constants is the Jacobi-Identity, which vanishes, so
\bE{l}
F^{(2,4)}_{GL}|_{\O(\theta)}=\frac{f^{a_1a_2c}f^{b_1b_2c}}{32}\int_{\slashed{k_1},\slashed{k_2},\slashed{q_1},\slashed{q_2}} \slashed{\delta}\left(\sum_i(\vec{k}_i+\vec{q}_i)\right)g^{(3)}(\vec{k}_1,\vec{k}_2,-\vec{k}_1-\vec{k}_2) \nn\\
\qquad\qquad\qquad J^{a_1}(\vec{k}_1)J^{a_2}(\vec{k}_2)J^{b_1}(\vec{q}_1)\theta^{b_2}(\vec{q}_2) \,. \qquad \label{F24theta1}
\eE

\subsection{Order $\theta^2$:}

\color{black}

\subsubsection{$\left(\frac{\delta F^{(1)}_{GL}}{\delta A}\right)^2$-term}

\bE{l}
\int_p\frac{\delta F^{(1)}_{GL}}{\delta A^a_i(\vec{p})}\frac{\delta F^{(1)}_{GL}}{\delta A^a_i(-\vec{p})} \Bigg|_{\O(\theta^2)} \nn\\
\quad = 4(-i)^2 f^{aa_1a_2}f^{ab_1b_2} \int_{\slashed{k_1},\slashed{k_2},\slashed{q_1},\slashed{q_2},p}\slashed{\delta}\left(\vec{k}_1+\vec{k}_2+\vec{p}\right)\slashed{\delta}\left(\vec{q}_1+\vec{q}_2-\vec{p}\right) J^{a_1}(\vec{k}_1)\theta^{a_2}(\vec{k}_2)J^{b_1}(\vec{q}_1)\theta^{b_2}(\vec{q}_2) \nn\\
\qquad\qquad  \Bigg\{ \left( \frac{\bar{p}^2}{|\vec{p}|}-\frac{\bar{k}_1^2}{|\vec{k}_1|}\right) (-1)\left(\frac{1}{4}|\vec{p}|+\frac{\bar{q}_1}{|\vec{q}_1|}\left(-\frac{p}{\bar{p}}(\bar{q}_1+\bar{q}_2)+q_2\right) \right)\Bigg\}\nn\\
\qquad 4(-i)^2 f^{aa_1a_2}f^{ab_1b_2} \int_{\slashed{k_1},\slashed{k_2},\slashed{q_1},\slashed{q_2},p}\slashed{\delta}\left(\vec{k}_1+\vec{k}_2+\vec{p}\right)\slashed{\delta}\left(\vec{q}_1+\vec{q}_2-\vec{p}\right) J^{a_1}(\vec{k}_1)J^{a_2}(\vec{k}_2)\theta^{b_1}(\vec{q}_1)\theta^{b_2}(\vec{q}_2) \nn\\
\qquad\qquad  \Bigg\{ \frac{g^{(3)}(\vec{k}_1,\vec{k}_2,\vec{p})}{32}  (-2)\frac{(-p) q_1\bar{q}_2}{|\vec{p}|} + 2\frac{(-\bar{p})q_1\bar{q}_2}{|\vec{p}|} \left(-\frac{p}{\bar{p}}\frac{g^{(3)}(\vec{k}_1,\vec{k}_2,\vec{p})}{32}+\frac{\bar{k}_2^2}{2\bar{p}|\vec{k}_2|}\right)\Bigg\} \nn\\
\qquad + \O(e)\qquad \\
= f^{aa_1a_2}f^{ab_1b_2} \int_{\slashed{k_1},\slashed{k_2},\slashed{q_1},\slashed{q_2}}\slashed{\delta}\left(\vec{k}_1+\vec{k}_2+\vec{q}_1+\vec{q}_2\right) J^{a_1}(\vec{k}_1)\theta^{a_2}(\vec{k}_2)J^{b_1}(\vec{q}_1)\theta^{b_2}(\vec{q}_2) \nn\\
\qquad\qquad  \Bigg\{ \left( \frac{(\bar{k}_1+\bar{k}_2)^2}{|\vec{k}_1+\vec{k}_2|}-\frac{\bar{k}_1^2}{|\vec{k}_1|}\right) \left(|\vec{q}_1+\vec{q}_2|-|\vec{q}_1| \right) \Bigg\}\nn\\
\qquad -f^{aa_1a_2}f^{ab_1b_2} \int_{\slashed{k_1},\slashed{k_2},\slashed{q_1},\slashed{q_2},p}\slashed{\delta}\left(\vec{k}_1+\vec{k}_2+\vec{q}_1+\vec{q}_2\right) J^{a_1}(\vec{k}_1)J^{a_2}(\vec{k}_2)\theta^{b_1}(\vec{q}_1)\theta^{b_2}(\vec{q}_2)  \nn\\
\qquad\qquad \Bigg\{g^{(3)}(\vec{k}_1,\vec{k}_2,-\vec{k}_1-\vec{k}_2) \frac{(q_1+q_2) q_1\bar{q}_2}{2|\vec{q}_1+\vec{q}_2|} -\frac{4q_1\bar{q}_2}{|\vec{q}_1+\vec{q}_2|}\frac{\bar{k}_2^2}{|\vec{k}_2|} \Bigg\} + \O(e)\,.
\eE

\subsubsection{$\left(\vec{A}\cdot\frac{\delta F^{(1)}_{GL}}{\delta \vec{A}}\right)$-term}

\bE{l}
-if^{b_1b_2c}\int_{\slashed{p},\slashed{k_1},\slashed{k_2},\slashed{q_1},\slashed{q_2}}\frac{\slashed{\delta}(\vec{q}_1+\vec{q}_2-\vec{p})}{\sum_i(|\vec{k}_i|+|\vec{q}_i|)|\vec{q}_1|}(\vec{q}_1\cdot\vec{A}^{b_1}(\vec{q}_1))\left(\vec{A}^{b_2}(\vec{q}_2)\cdot\frac{\delta F^{(1)}_{GL}}{\delta \vec{A}^c(\vec{p})}[\vec{k}_1,\vec{k}_2]\right)  \nn\\
= -i f^{b_1b_2c}f^{a_1a_2c}\int_{\slashed{p},\slashed{k_1},\slashed{k_2},\slashed{q_1},\slashed{q_2}}\frac{\slashed{\delta}(\vec{q}_1+\vec{q}_2-\vec{p})\slashed{\delta}(\vec{k}_1+\vec{k}_2+\vec{p})}{(|\vec{k}_1|+|\vec{k}_2|+|\vec{q}_1|+|\vec{q}_2|)|\vec{q}_1|} \nn\\
\quad \left(-i\bar{q}_1 J^{b_1}(\vec{q}_1)+i|\vec{q}_1|^2\theta^{b_1}(\vec{q}_1) +i f^{b_1de}\int_\slashed{l}(q_1\bar{l}+\bar{q}_1 l)\theta^d(\vec{l})\theta^e(\vec{l}+\vec{q}_1)\right) (-i) \nn\\
\quad A^{b_2}_i(\vec{q}_2)\Bigg\{\Bigg\{ (\delta_{1i}+i\delta_{2i})\frac{g^{(3)}(\vec{k}_1,\vec{k}_2,\vec{p})}{32} \nn\\
\qquad\qquad\qquad + (\delta_{1i}-i\delta_{2i})\left(-\frac{p}{\bar{p}}\frac{g^{(3)}(\vec{k}_1,\vec{k}_2,\vec{p})}{32}+\frac{\bar{k}_2^2}{2\bar{p}|\vec{k}_2|}\right)\Bigg\}J^{a_1}(\vec{k}_1)J^{a_2}(\vec{k}_2) \nn\\
\qquad\qquad  +\Bigg\{(\delta_{1i}+i\delta_{2i})\left( \frac{\bar{p}^2}{|\vec{p}|}-\frac{\bar{k}_1^2}{|\vec{k}_1|}\right) \nn\\
\qquad\qquad\qquad - (\delta_{1i}-i\delta_{2i}) \left(\frac{1}{4}|\vec{p}|+\frac{\bar{k}_1}{|\vec{k}_1|}\left(-\frac{p}{\bar{p}}(\bar{k}_1+\bar{k}_2)+k_2\right) \right)\Bigg\}J^{a_1}(\vec{k}_1)\theta^{a_2}(\vec{k}_2) \nn\\
\qquad\qquad  +\Bigg\{(\delta_{1i}+i\delta_{2i})2\frac{\bar{p}k_1\bar{k}_2}{|\vec{p}|} -(\delta_{1i}-i\delta_{2i}) 2\frac{p k_1\bar{k}_2}{|\vec{p}|}\Bigg\} \theta^{a_1}(\vec{k}_1)\theta^{a_2}(\vec{k}_2)\Bigg\}\\
= (-i)^3 f^{b_1b_2c}f^{a_1a_2c}\int_{\slashed{p},\slashed{k_1},\slashed{k_2},\slashed{q_1},\slashed{q_2}}\frac{\slashed{\delta}(\vec{q}_1+\vec{q}_2-\vec{p})\slashed{\delta}(\vec{k}_1+\vec{k}_2+\vec{p})}{(|\vec{k}_1|+|\vec{k}_2|+|\vec{q}_1|+|\vec{q}_2|)|\vec{q}_1|} \nn\\
\quad \left(\bar{q}_1 J^{b_1}(\vec{q}_1)-|\vec{q}_1|^2\theta^{b_1}(\vec{q}_1)\right) \nn\\
\quad \Bigg\{\Bigg\{ 2A^{b_2}(\vec{q}_2)\frac{g^{(3)}(\vec{k}_1,\vec{k}_2,\vec{p})}{32} + 2\bar{A}^{b_2}(\vec{q}_2)\left(-\frac{p}{\bar{p}}\frac{g^{(3)}(\vec{k}_1,\vec{k}_2,\vec{p})}{32}+\frac{\bar{k}_2^2}{2\bar{p}|\vec{k}_2|}\right)\Bigg\}J^{a_1}(\vec{k}_1)J^{a_2}(\vec{k}_2) \nn\\
\qquad  +\Bigg\{2A^{b_2}(\vec{q}_2)\left( \frac{\bar{p}^2}{|\vec{p}|}-\frac{\bar{k}_1^2}{|\vec{k}_1|}\right)  - 2\bar{A}^{b_2}(\vec{q}_2) \left(\frac{1}{4}|\vec{p}|+\frac{\bar{k}_1}{|\vec{k}_1|}\left(-\frac{p}{\bar{p}}(\bar{k}_1+\bar{k}_2)+k_2\right) \right)\Bigg\}J^{a_1}(\vec{k}_1)\theta^{a_2}(\vec{k}_2) \nn\\
\qquad  +\Bigg\{2A^{b_2}(\vec{q}_2)2\frac{\bar{p}k_1\bar{k}_2}{|\vec{p}|} -O(\theta)\Bigg\} \theta^{a_1}(\vec{k}_1)\theta^{a_2}(\vec{k}_2)\Bigg\}\nn\\
\quad +\O(e)\\
= (-i)^3 f^{b_1b_2c}f^{a_1a_2c}\int_{\slashed{p},\slashed{k_1},\slashed{k_2},\slashed{q_1},\slashed{q_2}}\frac{\slashed{\delta}(\vec{q}_1+\vec{q}_2-\vec{p})\slashed{\delta}(\vec{k}_1+\vec{k}_2+\vec{p})}{(|\vec{k}_1|+|\vec{k}_2|+|\vec{q}_1|+|\vec{q}_2|)|\vec{q}_1|} \nn\\
\quad \Bigg\{\Bigg\{\left(O(J^2)+O(\theta J)-|\vec{q}_1|^2\theta^{b_1}(\vec{q}_1)2iq_2\theta^{b_2}(\vec{q}_2)\right)\frac{g^{(3)}(\vec{k}_1,\vec{k}_2,\vec{p})}{32}\nn\\ 
\qquad\quad -\left(O(\theta J)+|\vec{q}_1|^2\theta^{b_1}(\vec{q}_1) \; 2i\bar{q}_2\theta^{b_2}(\vec{q}_2)\right) \left(-\frac{p}{\bar{p}}\frac{g^{(3)}(\vec{k}_1,\vec{k}_2,\vec{p})}{32}+\frac{\bar{k}_2^2}{2\bar{p}|\vec{k}_2|}\right)\Bigg\}J^{a_1}(\vec{k}_1)J^{a_2}(\vec{k}_2) \nn\\
\qquad  +\Bigg\{\left(\bar{q}_1 J^{b_1}(\vec{q}_1)-|\vec{q}_1|^2\theta^{b_1}(\vec{q}_1)\right)(-iJ^{b_2}(\vec{q}_2)+2iq_2\theta^{b_2}(\vec{q}_2))\left( \frac{\bar{p}^2}{|\vec{p}|}-\frac{\bar{k}_1^2}{|\vec{k}_1|}\right) \nn\\
\qquad\quad -(\bar{q}_1 J^{b_1}(\vec{q}_1)\; 2i\bar{q}_2\theta^{b_2}(\vec{q}_2) + O(\theta^2)) \left(\frac{1}{4}|\vec{p}|+\frac{\bar{k}_1}{|\vec{k}_1|}\left(-\frac{p}{\bar{p}}(\bar{k}_1+\bar{k}_2)+k_2\right) \right)\Bigg\}J^{a_1}(\vec{k}_1)\theta^{a_2}(\vec{k}_2) \nn\\
\qquad  +\bar{q}_1 J^{b_1}(\vec{q}_1) (-i)J^{b_2}(\vec{q}_2)2\frac{\bar{p}k_1\bar{k}_2}{|\vec{p}|} \theta^{a_1}(\vec{k}_1)\theta^{a_2}(\vec{k}_2)\Bigg\}\nn\\
\quad  +O(J^4)+O(J^3\theta)+O(J\theta^3)+O(\theta^4)+O(e)\,.
\eE
With this we find
\bE{l}
-if^{b_1b_2c}\int_{\slashed{p},\slashed{k_1},\slashed{k_2},\slashed{q_1},\slashed{q_2}}\frac{\slashed{\delta}(\vec{q}_1+\vec{q}_2-\vec{p})}{\sum_i(|\vec{k}_i|+|\vec{q}_i|)|\vec{q}_1|}(\vec{q}_1\cdot\vec{A}^{b_1}(\vec{q}_1))\left(\vec{A}^{b_2}(\vec{q}_2)\cdot\frac{\delta F^{(1)}_{GL}}{\delta \vec{A}^c(\vec{p})}[\vec{k}_1,\vec{k}_2]\right)\Bigg|_{\O(\theta^2)}   \nn\\
= f^{a_1a_2c}f^{b_1b_2c}\int_{\slashed{k_1},\slashed{k_2},\slashed{q_1},\slashed{q_2}}\frac{\slashed{\delta}\left(\sum_i^4\vec{k}_i\right)}{\sum_i^4|\vec{k}_i|} \nn\\
 \Bigg\{\left(|\vec{q}_1|\frac{q_2\bar{q}_1-\bar{q}_2q_1}{\bar{q}_1+\bar{q}_2} \frac{g^{(3)}(\vec{k}_1,\vec{k}_2, -\vec{k}_1-\vec{k}_2)}{16} -\frac{|\vec{q}_1|\bar{q}_2\bar{k}_2^2}{(\bar{k}_1+\bar{k}_2)|\vec{k}_2|}  +2 \frac{\bar{k}_1(\bar{k}_1+\bar{k}_2)q_1\bar{q}_2}{|\vec{k}_1||\vec{k}_1+\vec{k}_2|} \right) \nn\\
\qquad\qquad\times J^{a_1}(\vec{k}_1)J^{a_2}(\vec{k}_2)\theta^{b_1}(\vec{q}_1) \theta^{b_2}(\vec{q}_2) \nn\\
\quad  +\left\{\left(|\vec{q}_2| -2\frac{\bar{q}_1q_2}{|\vec{q}_1|} \right) \left( \frac{(\bar{k}_1+\bar{k}_2)^2}{|\vec{k}_1+\vec{k}_2|} -\frac{\bar{k}_1^2}{|\vec{k}_1|}\right) +2 \frac{\bar{q}_1\bar{q}_2}{|\vec{q}_1|} \left(\frac{1}{4}|\vec{k}_1+\vec{k}_2| -\frac{1}{4}|\vec{k}_1| \right)\right\}  \nn\\
\qquad\qquad\times J^{a_1}(\vec{k}_1)\theta^{a_2}(\vec{k}_2)  J^{b_1}(\vec{q}_1)\theta^{b_2}(\vec{q}_2) \Bigg\}
  +\O(e)\,.
\eE

\subsubsection{$ (A\times A)(A\times A)$-term}

\bE{l}
\frac{1}{8}f^{a_1a_2c}f^{b_1b_2c}\int_{\slashed{p},\slashed{k_1},\slashed{k_2},\slashed{q_1},\slashed{q_2}}\frac{\slashed{\delta}(\sum_i(\vec{k}_i+\vec{q}_i))}{\sum_i(|\vec{k}_i|+|\vec{q}_i|)}(\vec{A}^{a_1}(\vec{k}_1)\times\vec{A}^{a_2}(\vec{k}_2))(\vec{A}^{b_1}(\vec{q}_1)\times\vec{A}^{b_2}(\vec{q}_2))\nn\\
= -\frac{1}{2}f^{a_1a_2c}f^{b_1b_2c}\int_{\slashed{p},\slashed{k_1},\slashed{k_2},\slashed{q_1},\slashed{q_2}}\frac{\slashed{\delta}(\sum_i(\vec{k}_i+\vec{q}_i))}{\sum_i(|\vec{k}_i|+|\vec{q}_i|)} \bar{k_2}\bar{q_2}J^{a_1}(\vec{k}_1)\theta^{a_2}(\vec{k}_2) J^{b_1}(\vec{q}_1)\theta^{b_2}(\vec{q}_2)  \nn\\
\qquad\quad+\O(\theta^3)+\O(e)\,.
\eE

\subsubsection{All three together}

\bE{l}
F^{(2,4)}_{GL}|_{\O(\theta^2)}
= f^{aa_1a_2}f^{ab_1b_2} \int_{\slashed{k_1},\slashed{k_2},\slashed{q_1},\slashed{q_2}}\frac{\slashed{\delta}\left(\vec{k}_1+\vec{k}_2+\vec{q}_1+\vec{q}_2\right) }{\left(|\vec{k}_1|+|\vec{k}_2|+|\vec{q}_1|+|\vec{q}_2|\right)} \nn\\
\qquad  \Bigg\{-\frac{1}{2} \left( \frac{(\bar{k}_1+\bar{k}_2)^2}{|\vec{k}_1+\vec{k}_2|}-\frac{\bar{k}_1^2}{|\vec{k}_1|}\right) \left(|\vec{q}_1+\vec{q}_2|-|\vec{q}_1| \right)J^{a_1}(\vec{k}_1)\theta^{a_2}(\vec{k}_2)J^{b_1}(\vec{q}_1)\theta^{b_2}(\vec{q}_2)\nn\\
\qquad\quad +\frac{1}{2} \Bigg\{g^{(3)}(\vec{k}_1,\vec{k}_2,-\vec{k}_1-\vec{k}_2) \frac{(q_1+q_2) q_1\bar{q}_2}{2|\vec{q}_1+\vec{q}_2|} -\frac{4q_1\bar{q}_2}{|\vec{q}_1+\vec{q}_2|}\frac{\bar{k}_2^2}{|\vec{k}_2|} \Bigg\} 
 J^{a_1}(\vec{k}_1)J^{a_2}(\vec{k}_2)\theta^{b_1}(\vec{q}_1)\theta^{b_2}(\vec{q}_2)\Bigg\}\nn\\
+ f^{a_1a_2c}f^{b_1b_2c}\int_{\slashed{k_1},\slashed{k_2},\slashed{q_1},\slashed{q_2}}\frac{\slashed{\delta}\left(\sum_i^4\vec{k}_i\right)}{\sum_i^4|\vec{k}_i|} \nn\\
\quad \Bigg\{\left(|\vec{q}_1|\frac{q_2\bar{q}_1-\bar{q}_2q_1}{\bar{q}_1+\bar{q}_2} \frac{g^{(3)}(\vec{k}_1,\vec{k}_2, -\vec{k}_1-\vec{k}_2)}{16} -|\vec{q}_1|\bar{q}_2\frac{\bar{k}_2^2}{(\bar{k}_1+\bar{k}_2)|\vec{k}_2|}  +2\bar{k}_1 \frac{(\bar{k}_1+\bar{k}_2)q_1\bar{q}_2}{|\vec{k}_1||\vec{k}_1+\vec{k}_2|} \right)\nn\\
\qquad\qquad \qquad J^{a_1}(\vec{k}_1)J^{a_2}(\vec{k}_2)\theta^{b_1}(\vec{q}_1) \theta^{b_2}(\vec{q}_2) \nn\\
\qquad  +\left\{-\left(-|\vec{q}_2| +2\frac{\bar{q}_1q_2}{|\vec{q}_1|} \right) \left( \frac{(\bar{k}_1+\bar{k}_2)^2}{|\vec{k}_1+\vec{k}_2|} -\frac{\bar{k}_1^2}{|\vec{k}_1|}\right) +2 \frac{\bar{q}_1\bar{q}_2}{|\vec{q}_1|} \left(\frac{1}{4}|\vec{k}_1+\vec{k}_2| -\frac{1}{4}|\vec{k}_1| \right)\right\} \nn\\
\qquad\qquad \qquad J^{a_1}(\vec{k}_1)\theta^{a_2}(\vec{k}_2)  J^{b_1}(\vec{q}_1)\theta^{b_2}(\vec{q}_2) \Bigg\}\nn\\
-\frac{1}{2}f^{a_1a_2c}f^{b_1b_2c}\int_{\slashed{p},\slashed{k_1},\slashed{k_2},\slashed{q_1},\slashed{q_2}}\frac{\slashed{\delta}(\sum_i(\vec{k}_i+\vec{q}_i))}{\sum_i(|\vec{k}_i|+|\vec{q}_i|)} \bar{k_2}\bar{q_2}J^{a_1}(\vec{k}_1)\theta^{a_2}(\vec{k}_2) J^{b_1}(\vec{q}_1)\theta^{b_2}(\vec{q}_2) +\O(e)\nn\\
= f^{a_1a_2c}f^{b_1b_2c} \int_{\slashed{k_1},\slashed{k_2},\slashed{q_1},\slashed{q_2}}\frac{\slashed{\delta}\left(\vec{k}_1+\vec{k}_2+\vec{q}_1+\vec{q}_2\right) }{\left(|\vec{k}_1|+|\vec{k}_2|+|\vec{q}_1|+|\vec{q}_2|\right)} J^{a_1}(\vec{k}_1)\theta^{a_2}(\vec{k}_2)J^{b_1}(\vec{q}_1)\theta^{b_2}(\vec{q}_2) \nn\\
\qquad   \Bigg\{-\frac{1}{2} \left( \frac{(\bar{k}_1+\bar{k}_2)^2}{|\vec{k}_1+\vec{k}_2|}-\frac{\bar{k}_1^2}{|\vec{k}_1|}\right) \left(|\vec{q}_1+\vec{q}_2|-|\vec{q}_1| \right) -\left(-|\vec{q}_2| +2\frac{\bar{q}_1q_2}{|\vec{q}_1|} \right) \left( \frac{(\bar{k}_1+\bar{k}_2)^2}{|\vec{k}_1+\vec{k}_2|} -\frac{\bar{k}_1^2}{|\vec{k}_1|}\right) \nn\\
\qquad\quad +\frac{1}{2} \frac{\bar{q}_1\bar{q}_2}{|\vec{q}_1|} \left(|\vec{k}_1+\vec{k}_2| -|\vec{k}_1| \right) -\frac{\bar{k_2}\bar{q_2}}{2 }\Bigg\}\nn\\
\quad + f^{a_1a_2c}f^{b_1b_2c} \int_{\slashed{k_1},\slashed{k_2},\slashed{q_1},\slashed{q_2}}\frac{\slashed{\delta}\left(\vec{k}_1+\vec{k}_2+\vec{q}_1+\vec{q}_2\right) }{\left(|\vec{k}_1|+|\vec{k}_2|+|\vec{q}_1|+|\vec{q}_2|\right)} J^{a_1}(\vec{k}_1)J^{a_2}(\vec{k}_2)\theta^{b_1}(\vec{q}_1)\theta^{b_2}(\vec{q}_2) \nn\\
\qquad \Bigg\{g^{(3)}(\vec{k}_1,\vec{k}_2,-\vec{k}_1-\vec{k}_2) \frac{(q_1+q_2) q_1\bar{q}_2}{4|\vec{q}_1+\vec{q}_2|}-2 \frac{q_1\bar{q}_2}{|\vec{q}_1+\vec{q}_2|}\frac{\bar{k}_2^2}{|\vec{k}_2|}  \nn\\
\qquad\qquad +|\vec{q}_1|\frac{q_2\bar{q}_1-\bar{q}_2q_1}{\bar{q}_1+\bar{q}_2} \frac{g^{(3)}(\vec{k}_1,\vec{k}_2, -\vec{k}_1-\vec{k}_2)}{16} \nn\\
\qquad\qquad -|\vec{q}_1|\bar{q}_2\frac{\bar{k}_2^2}{(\bar{k}_1+\bar{k}_2)|\vec{k}_2|} +2\bar{k}_1\bar{q}_2 \frac{(\bar{k}_1+\bar{k}_2)q_1}{|\vec{k}_1||\vec{k}_1+\vec{k}_2|}\Bigg\} +\O(e)\\
= f^{a_1a_2c}f^{b_1b_2c} \int_{\slashed{k_1},\slashed{k_2},\slashed{q_1},\slashed{q_2}}\frac{\slashed{\delta}\left(\vec{k}_1+\vec{k}_2+\vec{q}_1+\vec{q}_2\right) }{\left(|\vec{k}_1|+|\vec{k}_2|+|\vec{q}_1|+|\vec{q}_2|\right)} J^{a_1}(\vec{k}_1)\theta^{a_2}(\vec{k}_2)J^{b_1}(\vec{q}_1)\theta^{b_2}(\vec{q}_2) \nn\\
\qquad \Bigg\{-\frac{1}{2} \left( \frac{(\bar{k}_1+\bar{k}_2)^2}{|\vec{k}_1+\vec{k}_2|}-\frac{\bar{k}_1^2}{|\vec{k}_1|}\right) \left(|\vec{q}_1+\vec{q}_2|-|\vec{q}_1| - 2|\vec{q}_2| +4\frac{\bar{q}_1q_2}{|\vec{q}_1|} \right)   \nn\\
\qquad\qquad\qquad +\frac{1}{2} \frac{\bar{k}_1\bar{k}_2}{|\vec{k}_1|} \left(|\vec{q}_1+\vec{q}_2| -|\vec{q}_1| \right) -\frac{\bar{k_2}\bar{q_2}}{2 }\Bigg\}\nn\\
\quad + f^{a_1a_2c}f^{b_1b_2c} \int_{\slashed{k_1},\slashed{k_2},\slashed{q_1},\slashed{q_2}}\frac{\slashed{\delta}\left(\vec{k}_1+\vec{k}_2+\vec{q}_1+\vec{q}_2\right) }{\left(|\vec{k}_1|+|\vec{k}_2|+|\vec{q}_1|+|\vec{q}_2|\right)} J^{a_1}(\vec{k}_1)J^{a_2}(\vec{k}_2)\theta^{b_1}(\vec{q}_1)\theta^{b_2}(\vec{q}_2) \nn\\
\quad \Bigg\{-2  \frac{q_1\bar{q}_2}{|\vec{q}_1+\vec{q}_2|}\frac{\bar{k}_2^2}{|\vec{k}_2|}  -|\vec{q}_1|\bar{q}_2\frac{\bar{k}_2^2}{(\bar{k}_1+\bar{k}_2)|\vec{k}_2|} +2\bar{k}_1\bar{q}_2 \frac{(\bar{k}_1+\bar{k}_2)q_1}{|\vec{k}_1||\vec{k}_1+\vec{k}_2|}\nn\\
\quad\quad +\frac{q_1\bar{q}_2}{\bar{q}_1+\bar{q}_2}\left(|\vec{q}_1+\vec{q}_2|-|\vec{q}_1|-|\vec{q}_2|\right)  \frac{g^{(3)}(\vec{k}_1,\vec{k}_2, -\vec{k}_1-\vec{k}_2)}{16} \Bigg\} +\O(e)\,.
\eE

\subsubsection{Manipulations of $F^{(2,4)}_{GL}$ at order $\theta^2$}
We now manipulate the obtained expression with the objective of getting rid of the \linebreak $\frac{1}{\left(|\vec{k}_1|+|\vec{k}_2|+|\vec{q}_1|+|\vec{q}_2|\right)}$ prefactor.

We replace $q_2\rightarrow -q_1-k_1-k_2$ in the $4\frac{\bar{q}_1q_2}{|\vec{q}_1|}$-term of the second line.

\bE{l}
= f^{a_1a_2c}f^{b_1b_2c} \int_{\slashed{k_1},\slashed{k_2},\slashed{q_1},\slashed{q_2}}\frac{\slashed{\delta}\left(\vec{k}_1+\vec{k}_2+\vec{q}_1+\vec{q}_2\right) }{\left(|\vec{k}_1|+|\vec{k}_2|+|\vec{q}_1|+|\vec{q}_2|\right)} J^{a_1}(\vec{k}_1)\theta^{a_2}(\vec{k}_2)J^{b_1}(\vec{q}_1)\theta^{b_2}(\vec{q}_2) \nn\\
\qquad \Bigg\{-\frac{1}{2} \left( \frac{(\bar{k}_1+\bar{k}_2)^2}{|\vec{k}_1+\vec{k}_2|}-\frac{\bar{k}_1^2}{|\vec{k}_1|}\right) \left(|\vec{q}_1+\vec{q}_2|-2|\vec{q}_1| - 2|\vec{q}_2| -4\frac{\bar{q}_1(k_1+k_2)}{|\vec{q}_1|} \right) \nn\\
\qquad\qquad +\frac{1}{2} \frac{\bar{k}_1\bar{k}_2}{|\vec{k}_1|} \left(|\vec{q}_1+\vec{q}_2| -|\vec{q}_1| \right) -\frac{\bar{k_2}\bar{q_2}}{2 }\Bigg\}\nn\\
\quad + f^{a_1a_2c}f^{b_1b_2c} \int_{\slashed{k_1},\slashed{k_2},\slashed{q_1},\slashed{q_2}}\frac{\slashed{\delta}\left(\vec{k}_1+\vec{k}_2+\vec{q}_1+\vec{q}_2\right) }{\left(|\vec{k}_1|+|\vec{k}_2|+|\vec{q}_1|+|\vec{q}_2|\right)} J^{a_1}(\vec{k}_1)J^{a_2}(\vec{k}_2)\theta^{b_1}(\vec{q}_1)\theta^{b_2}(\vec{q}_2) \nn\\
\quad \Bigg\{\frac{\bar{q}_2}{\bar{k}_1+\bar{k}_2}\Bigg( -2  q_1 \frac{\bar{k}_2^2(\bar{k}_1+\bar{k}_2)}{|\vec{k}_2||\vec{k}_1+\vec{k}_2|}  -|\vec{q}_1|\frac{\bar{k}_2^2}{|\vec{k}_2|} +2q_1 \frac{(\bar{k}_1+\bar{k}_2)^2 \bar{k}_1}{|\vec{k}_1||\vec{k}_1+\vec{k}_2|}\Bigg) \nn\\
\quad\quad +\frac{q_1\bar{q}_2}{\bar{q}_1+\bar{q}_2}\left(|\vec{q}_1+\vec{q}_2|-|\vec{q}_1|-|\vec{q}_2|\right)  \frac{g^{(3)}(\vec{k}_1,\vec{k}_2, -\vec{k}_1-\vec{k}_2)}{16} \Bigg\} +\O(e)
\eE

\bE{l}
= f^{a_1a_2c}f^{b_1b_2c} \int_{\slashed{k_1},\slashed{k_2},\slashed{q_1},\slashed{q_2}}\slashed{\delta}\left(\vec{k}_1+\vec{k}_2+\vec{q}_1+\vec{q}_2\right)  J^{a_1}(\vec{k}_1)\theta^{a_2}(\vec{k}_2)J^{b_1}(\vec{q}_1)\theta^{b_2}(\vec{q}_2) \frac{1}{2}  \frac{(\bar{k}_1+\bar{k}_2)^2}{|\vec{k}_1+\vec{k}_2|}\nn\\
\quad +f^{a_1a_2c}f^{b_1b_2c} \int_{\slashed{k_1},\slashed{k_2},\slashed{q_1},\slashed{q_2}}\frac{\slashed{\delta}\left(\vec{k}_1+\vec{k}_2+\vec{q}_1+\vec{q}_2\right) }{\left(|\vec{k}_1|+|\vec{k}_2|+|\vec{q}_1|+|\vec{q}_2|\right)} J^{a_1}(\vec{k}_1)\theta^{a_2}(\vec{k}_2)J^{b_1}(\vec{q}_1)\theta^{b_2}(\vec{q}_2) \nn\\
\qquad \Bigg\{-\frac{1}{2}  \frac{(\bar{k}_1+\bar{k}_2)^2}{|\vec{k}_1+\vec{k}_2|} \left(|\vec{q}_1+\vec{q}_2| -4\frac{\bar{q}_1(k_1+k_2)}{|\vec{q}_1|} \right)   \nn\\
\qquad\quad+\frac{1}{2}\frac{\bar{k}_1^2}{|\vec{k}_1|} \left(|\vec{q}_1+\vec{q}_2|-2|\vec{q}_1| - 2|\vec{q}_2| -4\frac{\bar{q}_1(k_1+k_2)}{|\vec{q}_1|} \right) +\frac{1}{2} \frac{\bar{k}_1\bar{k}_2}{|\vec{k}_1|} \left(|\vec{q}_1+\vec{q}_2| -|\vec{q}_1| \right) -\frac{\bar{k_2}\bar{q_2}}{2 }\Bigg\}\nn\\
\quad + f^{a_1a_2c}f^{b_1b_2c} \int_{\slashed{k_1},\slashed{k_2},\slashed{q_1},\slashed{q_2}}\frac{\slashed{\delta}\left(\vec{k}_1+\vec{k}_2+\vec{q}_1+\vec{q}_2\right) }{\left(|\vec{k}_1|+|\vec{k}_2|+|\vec{q}_1|+|\vec{q}_2|\right)} J^{a_1}(\vec{k}_1)J^{a_2}(\vec{k}_2)\theta^{b_1}(\vec{q}_1)\theta^{b_2}(\vec{q}_2) \nn\\
\quad \Bigg\{\frac{\bar{q}_2}{\bar{k}_1+\bar{k}_2}\left(|\vec{k}_1|+|\vec{k}_2|+|\vec{k}_1+\vec{k}_2|\right)  \frac{g^{(3)}(\vec{k}_1,\vec{k}_2, -\vec{k}_1-\vec{k}_2)}{16} \nn\\
\qquad +\frac{\bar{q}_2}{\bar{k}_1+\bar{k}_2}\Bigg( -2  q_1 \frac{\bar{k}_1^2\bar{k}_2}{|\vec{k}_1||\vec{k}_2|}  -|\vec{q}_1|\frac{\bar{k}_2^2}{|\vec{k}_2|} \Bigg) \nn\\
\quad\quad +\frac{q_1\bar{q}_2}{\bar{q}_1+\bar{q}_2}\left(|\vec{q}_1+\vec{q}_2|-|\vec{q}_1|-|\vec{q}_2|\right)  \frac{g^{(3)}(\vec{k}_1,\vec{k}_2, -\vec{k}_1-\vec{k}_2)}{16} \Bigg\}  +\O(e)
\eE

\bE{l}
= f^{a_1a_2c}f^{b_1b_2c} \int_{\slashed{k_1},\slashed{k_2},\slashed{q_1},\slashed{q_2}}\slashed{\delta}\left(\vec{k}_1+\vec{k}_2+\vec{q}_1+\vec{q}_2\right)  J^{a_1}(\vec{k}_1)\theta^{a_2}(\vec{k}_2)J^{b_1}(\vec{q}_1)\theta^{b_2}(\vec{q}_2) \frac{1}{2}  \frac{(\bar{k}_1+\bar{k}_2)^2}{|\vec{k}_1+\vec{k}_2|}\nn\\
\quad +f^{a_1a_2c}f^{b_1b_2c} \int_{\slashed{k_1},\slashed{k_2},\slashed{q_1},\slashed{q_2}}\frac{\slashed{\delta}\left(\vec{k}_1+\vec{k}_2+\vec{q}_1+\vec{q}_2\right) }{\left(|\vec{k}_1|+|\vec{k}_2|+|\vec{q}_1|+|\vec{q}_2|\right)} J^{a_1}(\vec{k}_1)\theta^{a_2}(\vec{k}_2)J^{b_1}(\vec{q}_1)\theta^{b_2}(\vec{q}_2) \nn\\
\qquad \Bigg\{-\frac{1}{2} (\bar{k}_1+\bar{k}_2)^2 -\frac{1}{2}  (\bar{k}_1+\bar{k}_2)|\vec{k}_1+\vec{k}_2|\frac{\bar{k}_1}{|\vec{k}_1|}   \nn\\
\qquad\quad +\frac{1}{2}\frac{\bar{k}_1^2}{|\vec{k}_1|} \left(|\vec{q}_1+\vec{q}_2|-2|\vec{q}_1| - 2|\vec{q}_2| -4\frac{\bar{q}_1(k_1+k_2)}{|\vec{q}_1|} \right)\nn\\
\qquad\quad  +\frac{1}{2} \frac{\bar{k}_1\bar{k}_2}{|\vec{k}_1|} \left(|\vec{q}_1+\vec{q}_2| -|\vec{q}_1| \right) -\frac{\bar{k_2}\bar{q_2}}{2 }\Bigg\}\nn\\
\quad - f^{a_1a_2c}f^{b_1b_2c} \int_{\slashed{k_1},\slashed{k_2},\slashed{q_1},\slashed{q_2}} \slashed{\delta}\left(\vec{k}_1+\vec{k}_2+\vec{q}_1+\vec{q}_2\right)  \frac{q_1\bar{q}_2}{\bar{q}_1+\bar{q}_2} \frac{g^{(3)}(\vec{k}_1,\vec{k}_2, -\vec{k}_1-\vec{k}_2)}{16}   \nn\\
\qquad\qquad\qquad  J^{a_1}(\vec{k}_1)J^{a_2}(\vec{k}_2)\theta^{b_1}(\vec{q}_1)\theta^{b_2}(\vec{q}_2) \nn\\
\quad - f^{a_1a_2c}f^{b_1b_2c} \int_{\slashed{k_1},\slashed{k_2},\slashed{q_1},\slashed{q_2}}\frac{\slashed{\delta}\left(\vec{k}_1+\vec{k}_2+\vec{q}_1+\vec{q}_2\right) }{\left(|\vec{k}_1|+|\vec{k}_2|+|\vec{q}_1|+|\vec{q}_2|\right)} J^{a_1}(\vec{k}_1)J^{a_2}(\vec{k}_2)\theta^{b_1}(\vec{q}_1)\theta^{b_2}(\vec{q}_2) \nn\\
\qquad\qquad \frac{\bar{q}_2}{\bar{k}_1+\bar{k}_2}\Bigg( 2  q_1 \frac{\bar{k}_1^2\bar{k}_2}{|\vec{k}_1||\vec{k}_2|}  +|\vec{q}_1|\frac{\bar{k}_2^2}{|\vec{k}_2|} \Bigg)  +\O(e)
\eE

\bE{l}
= f^{a_1a_2c}f^{b_1b_2c} \int_{\slashed{k_1},\slashed{k_2},\slashed{q_1},\slashed{q_2}}\slashed{\delta}\left(\vec{k}_1+\vec{k}_2+\vec{q}_1+\vec{q}_2\right)  J^{a_1}(\vec{k}_1)\theta^{a_2}(\vec{k}_2)J^{b_1}(\vec{q}_1)\theta^{b_2}(\vec{q}_2) \frac{1}{2}  \frac{(\bar{k}_1+\bar{k}_2)^2}{|\vec{k}_1+\vec{k}_2|}\nn\\
\quad +f^{a_1a_2c}f^{b_1b_2c} \int_{\slashed{k_1},\slashed{k_2},\slashed{q_1},\slashed{q_2}}\frac{\slashed{\delta}\left(\vec{k}_1+\vec{k}_2+\vec{q}_1+\vec{q}_2\right) }{\left(|\vec{k}_1|+|\vec{k}_2|+|\vec{q}_1|+|\vec{q}_2|\right)} J^{a_1}(\vec{k}_1)\theta^{a_2}(\vec{k}_2)J^{b_1}(\vec{q}_1)\theta^{b_2}(\vec{q}_2) \nn\\
\qquad \Bigg\{-\frac{1}{2} (\bar{k}_1+\bar{k}_2)^2   +\frac{1}{2}\frac{\bar{k}_1^2}{|\vec{k}_1|} \left(-2|\vec{q}_1| - 2|\vec{q}_2| -4\frac{\bar{q}_1(k_1+k_2)}{|\vec{q}_1|} \right) -\frac{1}{2} \frac{\bar{k}_1\bar{k}_2}{|\vec{k}_1|}|\vec{q}_1|  -\frac{\bar{k_2}\bar{q_2}}{2 }\Bigg\}\nn\\
\quad - f^{a_1a_2c}f^{b_1b_2c} \int_{\slashed{k_1},\slashed{k_2},\slashed{q_1},\slashed{q_2}} \slashed{\delta}\left(\vec{k}_1+\vec{k}_2+\vec{q}_1+\vec{q}_2\right)  \frac{q_1\bar{q}_2}{\bar{q}_1+\bar{q}_2} \frac{g^{(3)}(\vec{k}_1,\vec{k}_2, -\vec{k}_1-\vec{k}_2)}{16}   \nn\\
\qquad\qquad\qquad  J^{a_1}(\vec{k}_1)J^{a_2}(\vec{k}_2)\theta^{b_1}(\vec{q}_1)\theta^{b_2}(\vec{q}_2) \nn\\
\quad - f^{a_1a_2c}f^{b_1b_2c} \int_{\slashed{k_1},\slashed{k_2},\slashed{q_1},\slashed{q_2}}\frac{\slashed{\delta}\left(\vec{k}_1+\vec{k}_2+\vec{q}_1+\vec{q}_2\right) }{\left(|\vec{k}_1|+|\vec{k}_2|+|\vec{q}_1|+|\vec{q}_2|\right)} J^{a_1}(\vec{k}_1)J^{a_2}(\vec{k}_2)\theta^{b_1}(\vec{q}_1)\theta^{b_2}(\vec{q}_2) \nn\\
\qquad\qquad \frac{ 2  q_1 \bar{q}_2}{\bar{k}_1+\bar{k}_2}\Bigg( \frac{\bar{k}_1^2\bar{k}_2}{|\vec{k}_1||\vec{k}_2|}  +2\bar{q}_1 \frac{\bar{k}_2^2}{|\vec{q}_1||\vec{k}_2|} \Bigg)  +\O(e)\,.
\eE
Using the Jacobi-Identity $f^{a_1a_2c}f^{b_1b_2c}={1\over 2}\left(f^{a_1a_2c}f^{b_1b_2c} -f^{a_1b_1c}f^{b_2a_2c} -f^{a_1b_2c}f^{a_2b_1c}\right)$ in the second line, then renaming $b_1 \leftrightarrow a_2$, $q_1 \leftrightarrow k_2$ in the second term; and $b_2 \leftrightarrow a_2$, $q_2 \leftrightarrow k_2$ in the third term leads to
\bE{l}
= f^{a_1a_2c}f^{b_1b_2c} \int_{\slashed{k_1},\slashed{k_2},\slashed{q_1},\slashed{q_2}}\slashed{\delta}\left(\vec{k}_1+\vec{k}_2+\vec{q}_1+\vec{q}_2\right)  J^{a_1}(\vec{k}_1)\theta^{a_2}(\vec{k}_2)J^{b_1}(\vec{q}_1)\theta^{b_2}(\vec{q}_2) \frac{1}{2}  \frac{(\bar{k}_1+\bar{k}_2)^2}{|\vec{k}_1+\vec{k}_2|}\nn\\
\quad +{1\over 2}f^{a_1a_2c}f^{b_1b_2c} \int_{\slashed{k_1},\slashed{k_2},\slashed{q_1},\slashed{q_2}}\frac{\slashed{\delta}\left(\vec{k}_1+\vec{k}_2+\vec{q}_1+\vec{q}_2\right) }{\left(|\vec{k}_1|+|\vec{k}_2|+|\vec{q}_1|+|\vec{q}_2|\right)} J^{a_1}(\vec{k}_1)\theta^{a_2}(\vec{k}_2)J^{b_1}(\vec{q}_1)\theta^{b_2}(\vec{q}_2) \nn\\
\qquad \Bigg\{-\frac{1}{2} (\bar{k}_1+\bar{k}_2)^2   +\frac{1}{2}\frac{\bar{k}_1^2}{|\vec{k}_1|} \left(-2|\vec{q}_1| - 2|\vec{q}_2| -4\frac{\bar{q}_1(k_1+k_2)}{|\vec{q}_1|} \right) -\frac{1}{2} \frac{\bar{k}_1\bar{k}_2}{|\vec{k}_1|}|\vec{q}_1|  -\frac{\bar{k_2}\bar{q_2}}{2 }\Bigg\}\nn\\
\quad -{1\over 2}f^{a_1a_2c}f^{b_2b_1c} \int_{\slashed{k_1},\slashed{k_2},\slashed{q_1},\slashed{q_2}}\frac{\slashed{\delta}\left(\vec{k}_1+\vec{k}_2+\vec{q}_1+\vec{q}_2\right) }{\left(|\vec{k}_1|+|\vec{k}_2|+|\vec{q}_1|+|\vec{q}_2|\right)} J^{a_1}(\vec{k}_1) J^{a_2}(\vec{k}_2)\theta^{b_1}(\vec{q}_1)\theta^{b_2}(\vec{q}_2) \nn\\
\qquad \Bigg\{-\frac{1}{2} (\bar{k}_1+\bar{q}_1)^2   +\frac{1}{2}\frac{\bar{k}_1^2}{|\vec{k}_1|} \left(-2|\vec{k}_2| - 2|\vec{q}_2| -4\frac{\bar{k}_2(k_1+q_1)}{|\vec{k}_2|} \right) -\frac{1}{2} \frac{\bar{k}_1\bar{q}_1}{|\vec{k}_1|}|\vec{k}_2|  -\frac{\bar{q_1}\bar{q_2}}{2 }\Bigg\}\nn\\
\quad -{1\over 2}f^{a_1a_2c}f^{b_2b_1c} \int_{\slashed{k_1},\slashed{k_2},\slashed{q_1},\slashed{q_2}}\frac{\slashed{\delta}\left(\vec{k}_1+\vec{k}_2+\vec{q}_1+\vec{q}_2\right) }{\left(|\vec{k}_1|+|\vec{k}_2|+|\vec{q}_1|+|\vec{q}_2|\right)} J^{a_1}(\vec{k}_1) \theta^{a_2}(\vec{k}_2) J^{b_1}(\vec{q}_1) \theta^{b_2}(\vec{q}_2)\nn\\
\qquad \Bigg\{-\frac{1}{2} (\bar{k}_1+\bar{q}_2)^2   +\frac{1}{2}\frac{\bar{k}_1^2}{|\vec{k}_1|} \left(-2|\vec{q}_1| - 2|\vec{k}_2| -4\frac{\bar{q}_1(k_1+q_2)}{|\vec{q}_1|} \right) -\frac{1}{2} \frac{\bar{k}_1\bar{q}_2}{|\vec{k}_1|}|\vec{q}_1|  -\frac{\bar{q_2}\bar{k_2}}{2 }\Bigg\}\nn\\
\quad - f^{a_1a_2c}f^{b_1b_2c} \int_{\slashed{k_1},\slashed{k_2},\slashed{q_1},\slashed{q_2}} \slashed{\delta}\left(\vec{k}_1+\vec{k}_2+\vec{q}_1+\vec{q}_2\right)  \frac{q_1\bar{q}_2}{\bar{q}_1+\bar{q}_2} \frac{g^{(3)}(\vec{k}_1,\vec{k}_2, -\vec{k}_1-\vec{k}_2)}{16}    \nn\\
\qquad\qquad\qquad J^{a_1}(\vec{k}_1)J^{a_2}(\vec{k}_2)\theta^{b_1}(\vec{q}_1)\theta^{b_2}(\vec{q}_2) \nn\\
\quad -f^{a_1a_2c}f^{b_1b_2c} \int_{\slashed{k_1},\slashed{k_2},\slashed{q_1},\slashed{q_2}}\frac{\slashed{\delta}\left(\vec{k}_1+\vec{k}_2+\vec{q}_1+\vec{q}_2\right) }{\left(|\vec{k}_1|+|\vec{k}_2|+|\vec{q}_1|+|\vec{q}_2|\right)} J^{a_1}(\vec{k}_1)J^{a_2}(\vec{k}_2)\theta^{b_1}(\vec{q}_1)\theta^{b_2}(\vec{q}_2) \nn\\
\qquad\qquad \frac{ 2  q_1 \bar{q}_2}{\bar{k}_1+\bar{k}_2}\Bigg( \frac{\bar{k}_1^2\bar{k}_2}{|\vec{k}_1||\vec{k}_2|}  +2\bar{q}_1 \frac{\bar{k}_2^2}{|\vec{q}_1||\vec{k}_2|} \Bigg) +\O(e)
\eE

\bE{l}
= f^{a_1a_2c}f^{b_1b_2c} \int_{\slashed{k_1},\slashed{k_2},\slashed{q_1},\slashed{q_2}}\slashed{\delta}\left(\vec{k}_1+\vec{k}_2+\vec{q}_1+\vec{q}_2\right)  J^{a_1}(\vec{k}_1)\theta^{a_2}(\vec{k}_2)J^{b_1}(\vec{q}_1)\theta^{b_2}(\vec{q}_2) \frac{1}{2}  \frac{(\bar{k}_1+\bar{k}_2)^2}{|\vec{k}_1+\vec{k}_2|}\nn\\
\quad +{1\over 2}f^{a_1a_2c}f^{b_1b_2c} \int_{\slashed{k_1},\slashed{k_2},\slashed{q_1},\slashed{q_2}}\frac{\slashed{\delta}\left(\vec{k}_1+\vec{k}_2+\vec{q}_1+\vec{q}_2\right) }{\left(|\vec{k}_1|+|\vec{k}_2|+|\vec{q}_1|+|\vec{q}_2|\right)} J^{a_1}(\vec{k}_1)\theta^{a_2}(\vec{k}_2)J^{b_1}(\vec{q}_1)\theta^{b_2}(\vec{q}_2) \nn\\
\qquad \Bigg\{-\frac{1}{2} (\bar{k}_1+\bar{k}_2)^2   +\frac{1}{2}\frac{\bar{k}_1^2}{|\vec{k}_1|} \left(-2|\vec{q}_1| - 2|\vec{q}_2| -4\frac{\bar{q}_1(k_1+k_2)}{|\vec{q}_1|} \right) -\frac{1}{2} \frac{\bar{k}_1\bar{k}_2}{|\vec{k}_1|}|\vec{q}_1|  -\frac{\bar{k_2}\bar{q_2}}{2 }\nn\\
\qquad\quad -\frac{1}{2} (\bar{k}_1+\bar{q}_2)^2   +\frac{1}{2}\frac{\bar{k}_1^2}{|\vec{k}_1|} \left(-2|\vec{q}_1| - 2|\vec{k}_2| -4\frac{\bar{q}_1(k_1+q_2)}{|\vec{q}_1|} \right) -\frac{1}{2} \frac{\bar{k}_1\bar{q}_2}{|\vec{k}_1|}|\vec{q}_1|  -\frac{\bar{q_2}\bar{k_2}}{2} \Bigg\}\nn\\
\quad - f^{a_1a_2c}f^{b_1b_2c} \int_{\slashed{k_1},\slashed{k_2},\slashed{q_1},\slashed{q_2}} \slashed{\delta}\left(\vec{k}_1+\vec{k}_2+\vec{q}_1+\vec{q}_2\right)  \frac{q_1\bar{q}_2}{\bar{q}_1+\bar{q}_2} \frac{g^{(3)}(\vec{k}_1,\vec{k}_2, -\vec{k}_1-\vec{k}_2)}{16}  \nn\\
\qquad\qquad\qquad   J^{a_1}(\vec{k}_1)J^{a_2}(\vec{k}_2)\theta^{b_1}(\vec{q}_1)\theta^{b_2}(\vec{q}_2) \nn\\
\quad +{1\over 2} f^{a_1a_2c}f^{b_1b_2c} \int_{\slashed{k_1},\slashed{k_2},\slashed{q_1},\slashed{q_2}}\frac{\slashed{\delta}\left(\vec{k}_1+\vec{k}_2+\vec{q}_1+\vec{q}_2\right) }{\left(|\vec{k}_1|+|\vec{k}_2|+|\vec{q}_1|+|\vec{q}_2|\right)} J^{a_1}(\vec{k}_1)J^{a_2}(\vec{k}_2)\theta^{b_1}(\vec{q}_1)\theta^{b_2}(\vec{q}_2) \nn\\
\qquad\Bigg\{ - \frac{4  q_1 \bar{q}_2}{\bar{k}_1+\bar{k}_2}\Bigg( \frac{\bar{k}_1^2\bar{k}_2}{|\vec{k}_1||\vec{k}_2|}  +2\bar{q}_1 \frac{\bar{k}_2^2}{|\vec{q}_1||\vec{k}_2|} \Bigg) -\frac{1}{2} (\bar{k}_1+\bar{q}_1)^2 \nn\\
\qquad\quad   +\frac{1}{2}\frac{\bar{k}_1^2}{|\vec{k}_1|} \left(-2|\vec{k}_2| - 2|\vec{q}_2| -4\frac{\bar{k}_2(k_1+q_1)}{|\vec{k}_2|} \right) -\frac{1}{2} \frac{\bar{k}_1\bar{q}_1}{|\vec{k}_1|}|\vec{k}_2|  -\frac{\bar{q_1}\bar{q_2}}{2} \Bigg\}  +\O(e)\,.
\eE
In the last term of the 5th line it was necessary to rename $\vec q_1 \leftrightarrow \vec q_2$. We now take $(\vec k_1+\vec q_2) = -(\vec k_2+\vec q_1)$ in the 4th line. In the last line there are several terms that are either independent of $\vec q$ or $\vec k$. These vanish under $\vec k_1 \leftrightarrow \vec k_2$ and $\vec q_1 \leftrightarrow \vec q_2$, respectively.
\bE{l}
= f^{a_1a_2c}f^{b_1b_2c} \int_{\slashed{k_1},\slashed{k_2},\slashed{q_1},\slashed{q_2}}\slashed{\delta}\left(\vec{k}_1+\vec{k}_2+\vec{q}_1+\vec{q}_2\right)  J^{a_1}(\vec{k}_1)\theta^{a_2}(\vec{k}_2)J^{b_1}(\vec{q}_1)\theta^{b_2}(\vec{q}_2) \frac{1}{2}  \frac{(\bar{k}_1+\bar{k}_2)^2}{|\vec{k}_1+\vec{k}_2|}\nn\\
\quad +{1\over 2}f^{a_1a_2c}f^{b_1b_2c} \int_{\slashed{k_1},\slashed{k_2},\slashed{q_1},\slashed{q_2}}\frac{\slashed{\delta}\left(\vec{k}_1+\vec{k}_2+\vec{q}_1+\vec{q}_2\right) }{\left(|\vec{k}_1|+|\vec{k}_2|+|\vec{q}_1|+|\vec{q}_2|\right)} J^{a_1}(\vec{k}_1)\theta^{a_2}(\vec{k}_2)J^{b_1}(\vec{q}_1)\theta^{b_2}(\vec{q}_2) \nn\\
\qquad \Bigg\{-\frac{1}{2} (\bar{k}_1^2+2\bar{k}_1\bar{k}_2+\bar{k}_2^2)   +\frac{1}{2}\frac{\bar{k}_1^2}{|\vec{k}_1|} \left(-4|\vec{q}_1| - 2|\vec{q}_2|- 2|\vec{k}_2| -4\frac{\bar{q}_1(k_1-q_1)}{|\vec{q}_1|} \right) \nn\\
\qquad\quad -\frac{1}{2}  (\bar{k}_1^2+2\bar{k}_1\bar{q}_2+\bar{q}_2^2)     -\frac{1}{2} \frac{\bar{k}_1(\bar{k}_2+\bar{q}_2)}{|\vec{k}_1|}|\vec{q}_1|  -\bar{q_2}\bar{k_2}\Bigg\}\nn\\
\quad - f^{a_1a_2c}f^{b_1b_2c} \int_{\slashed{k_1},\slashed{k_2},\slashed{q_1},\slashed{q_2}} \slashed{\delta}\left(\vec{k}_1+\vec{k}_2+\vec{q}_1+\vec{q}_2\right)  \frac{q_1\bar{q}_2}{\bar{q}_1+\bar{q}_2} \frac{g^{(3)}(\vec{k}_1,\vec{k}_2, -\vec{k}_1-\vec{k}_2)}{16}   \nn\\
\qquad\qquad\qquad  J^{a_1}(\vec{k}_1)J^{a_2}(\vec{k}_2)\theta^{b_1}(\vec{q}_1)\theta^{b_2}(\vec{q}_2) \nn\\
\quad +{1\over 2} f^{a_1a_2c}f^{b_1b_2c} \int_{\slashed{k_1},\slashed{k_2},\slashed{q_1},\slashed{q_2}}\frac{\slashed{\delta}\left(\vec{k}_1+\vec{k}_2+\vec{q}_1+\vec{q}_2\right) }{\left(|\vec{k}_1|+|\vec{k}_2|+|\vec{q}_1|+|\vec{q}_2|\right)} J^{a_1}(\vec{k}_1)J^{a_2}(\vec{k}_2)\theta^{b_1}(\vec{q}_1)\theta^{b_2}(\vec{q}_2) \nn\\
\qquad\Bigg\{ - \frac{4  q_1 \bar{q}_2}{\bar{k}_1+\bar{k}_2}\Bigg( \frac{\bar{k}_1^2\bar{k}_2}{|\vec{k}_1||\vec{k}_2|}  +2\bar{q}_1 \frac{\bar{k}_2^2}{|\vec{q}_1||\vec{k}_2|} \Bigg) - \bar{k}_1\bar{q}_1   -\frac{\bar{k}_1^2}{|\vec{k}_1|}  |\vec{q}_2| -2\frac{\bar{k}_1^2}{|\vec{k}_1|}\frac{\bar{k}_2q_1}{|\vec{k}_2|} -\frac{1}{2} \frac{\bar{k}_1\bar{q}_1}{|\vec{k}_1|}|\vec{k}_2|  \Bigg\} \nn\\
\qquad +\O(e)\,. 
\eE
In the last line we combine the 1st term with the 5th and the 2nd with the 4th.
\bE{l}
= f^{a_1a_2c}f^{b_1b_2c} \int_{\slashed{k_1},\slashed{k_2},\slashed{q_1},\slashed{q_2}}\slashed{\delta}\left(\vec{k}_1+\vec{k}_2+\vec{q}_1+\vec{q}_2\right)  J^{a_1}(\vec{k}_1)\theta^{a_2}(\vec{k}_2)J^{b_1}(\vec{q}_1)\theta^{b_2}(\vec{q}_2) \frac{1}{2}  \frac{(\bar{k}_1+\bar{k}_2)^2}{|\vec{k}_1+\vec{k}_2|}\nn\\
\quad +{1\over 2}f^{a_1a_2c}f^{b_1b_2c} \int_{\slashed{k_1},\slashed{k_2},\slashed{q_1},\slashed{q_2}}\frac{\slashed{\delta}\left(\vec{k}_1+\vec{k}_2+\vec{q}_1+\vec{q}_2\right) }{\left(|\vec{k}_1|+|\vec{k}_2|+|\vec{q}_1|+|\vec{q}_2|\right)} J^{a_1}(\vec{k}_1)\theta^{a_2}(\vec{k}_2)J^{b_1}(\vec{q}_1)\theta^{b_2}(\vec{q}_2) \nn\\
\qquad \Bigg\{\frac{1}{2}\frac{\bar{k}_1^2}{|\vec{k}_1|} \left(-4|\vec{q}_1| - 2|\vec{q}_2|- 2|\vec{k}_2| +|\vec{q}_1| \right) -\frac{1}{2}\bar{k}_1\bar{q}_1 \frac{|\vec{k}_1|}{|\vec{q}_1|}\nn\\
\qquad\quad -\frac{1}{2}  (2\bar{k}_1^2+2\bar{k}_1\bar{q}_2+2\bar{k}_1\bar{k}_2+2\bar{k}_2\bar{q}_2+\bar{k}_2^2+\bar{q}_2^2)     +\frac{1}{2} \frac{\bar{k}_1^2}{|\vec{k}_1|}|\vec{q}_1| 
+\frac{1}{2} \frac{\bar{k}_1\bar{q}_1}{|\vec{k}_1|}|\vec{q}_1|  \Bigg\}\nn\\
\quad - f^{a_1a_2c}f^{b_1b_2c} \int_{\slashed{k_1},\slashed{k_2},\slashed{q_1},\slashed{q_2}} \slashed{\delta}\left(\vec{k}_1+\vec{k}_2+\vec{q}_1+\vec{q}_2\right)  \frac{q_1\bar{q}_2}{\bar{q}_1+\bar{q}_2} \frac{g^{(3)}(\vec{k}_1,\vec{k}_2, -\vec{k}_1-\vec{k}_2)}{16}   \nn\\
\qquad\qquad\qquad  J^{a_1}(\vec{k}_1)J^{a_2}(\vec{k}_2)\theta^{b_1}(\vec{q}_1)\theta^{b_2}(\vec{q}_2) \nn\\
\quad -{1\over 2} f^{a_1a_2c}f^{b_1b_2c} \int_{\slashed{k_1},\slashed{k_2},\slashed{q_1},\slashed{q_2}} \frac{\slashed{\delta}\left(\vec{k}_1+\vec{k}_2+\vec{q}_1+\vec{q}_2\right) }{\left(|\vec{k}_1|+|\vec{k}_2|+|\vec{q}_1|+|\vec{q}_2|\right)} J^{a_1}(\vec{k}_1)J^{a_2}(\vec{k}_2)\theta^{b_1}(\vec{q}_1)\theta^{b_2}(\vec{q}_2) \nn\\
\qquad\Bigg\{  \bar{k}_1\bar{q}_1   +\frac{\bar{k}_2^2}{|\vec{k}_2|}|\vec{q}_1| \left(1+2\frac{\bar{q}_2}{\bar{k}_1+\bar{k}_2}\right) +2\frac{\bar{k}_1^2\bar{k}_2q_1}{|\vec{k}_1||\vec{k}_2|} \left(1+2\frac{\bar{q}_2}{\bar{k}_1+\bar{k}_2}\right) +\frac{1}{2} \frac{\bar{k}_1\bar{q}_1}{|\vec{k}_1|}|\vec{k}_2|  \Bigg\}  \nn\\
\quad+\O(e)\,.
\eE
The last term of the third line cancels with the last term of the fourth line (after $\vec k\leftrightarrow \vec q$), the next-to-last term in the fourth line goes into the parenthesis of the third line, and the $\frac{1}{2}2\bar{k}_1^2$ term in the fourth line becomes $\frac{1}{2}2\bar{k}_1^2\frac{|\vec{k}_1|}{|\vec{k}_1|}$, so
\bE{l}
= f^{a_1a_2c}f^{b_1b_2c} \int_{\slashed{k_1},\slashed{k_2},\slashed{q_1},\slashed{q_2}}\slashed{\delta}\left(\vec{k}_1+\vec{k}_2+\vec{q}_1+\vec{q}_2\right)  \frac{1}{2} \left( \frac{(\bar{k}_1+\bar{k}_2)^2}{|\vec{k}_1+\vec{k}_2|} -  \frac{\bar{k}_1^2}{|\vec{k}_1|} \right) \nn\\
\qquad\qquad\qquad J^{a_1}(\vec{k}_1)\theta^{a_2}(\vec{k}_2)J^{b_1}(\vec{q}_1)\theta^{b_2}(\vec{q}_2) \nn\\
\quad -{1\over 4}f^{a_1a_2c}f^{b_1b_2c} \int_{\slashed{k_1},\slashed{k_2},\slashed{q_1},\slashed{q_2}}\frac{\slashed{\delta}\left(\vec{k}_1+\vec{k}_2+\vec{q}_1+\vec{q}_2\right) }{\left(|\vec{k}_1|+|\vec{k}_2|+|\vec{q}_1|+|\vec{q}_2|\right)} J^{a_1}(\vec{k}_1)\theta^{a_2}(\vec{k}_2)J^{b_1}(\vec{q}_1)\theta^{b_2}(\vec{q}_2) \nn\\
\qquad\quad  (2\bar{k}_1\bar{q}_2+2\bar{k}_1\bar{k}_2+2\bar{k}_2\bar{q}_2+\bar{k}_2^2+\bar{q}_2^2)     \nn\\
\quad - f^{a_1a_2c}f^{b_1b_2c} \int_{\slashed{k_1},\slashed{k_2},\slashed{q_1},\slashed{q_2}} \slashed{\delta}\left(\vec{k}_1+\vec{k}_2+\vec{q}_1+\vec{q}_2\right)  \frac{q_1\bar{q}_2}{\bar{q}_1+\bar{q}_2} \frac{g^{(3)}(\vec{k}_1,\vec{k}_2, -\vec{k}_1-\vec{k}_2)}{16}   \nn\\
\qquad\qquad\qquad  J^{a_1}(\vec{k}_1)J^{a_2}(\vec{k}_2)\theta^{b_1}(\vec{q}_1)\theta^{b_2}(\vec{q}_2) \nn\\
\quad -{1\over 2} f^{a_1a_2c}f^{b_1b_2c} \int_{\slashed{k_1},\slashed{k_2},\slashed{q_1},\slashed{q_2}} \frac{\slashed{\delta}\left(\vec{k}_1+\vec{k}_2+\vec{q}_1+\vec{q}_2\right) }{\left(|\vec{k}_1|+|\vec{k}_2|+|\vec{q}_1|+|\vec{q}_2|\right)} J^{a_1}(\vec{k}_1)J^{a_2}(\vec{k}_2)\theta^{b_1}(\vec{q}_1)\theta^{b_2}(\vec{q}_2) \nn\\
\qquad\Bigg\{  \bar{k}_1\bar{q}_1   +\frac{\bar{k}_2^2}{|\vec{k}_2|}|\vec{q}_1| \frac{\bar{q}_2-\bar{q}_1}{\bar{k}_1+\bar{k}_2} +2\frac{\bar{k}_1^2\bar{k}_2q_1}{|\vec{k}_1||\vec{k}_2|} \frac{\bar{q}_2-\bar{q}_1}{\bar{k}_1+\bar{k}_2} +\frac{1}{2} \frac{\bar{k}_1\bar{q}_1}{|\vec{k}_1|}|\vec{k}_2|  \Bigg\}  +\O(e)\,.
\eE
We interchange $\vec k\leftrightarrow \vec q$ in the first and the last term of line 3 and write $\bar{k}_1\rightarrow(-\bar{k}_2-\bar{q}_1-\bar{q}_2)$ in the second term. In the last line we interchange $\vec q_1\leftrightarrow \vec q_2$ in the 1st, 2nd, and 4th term, finding
\bE{l}
= f^{a_1a_2c}f^{b_1b_2c} \int_{\slashed{k_1},\slashed{k_2},\slashed{q_1},\slashed{q_2}}\slashed{\delta}\left(\vec{k}_1+\vec{k}_2+\vec{q}_1+\vec{q}_2\right)   \frac{1}{2} \left( \frac{(\bar{k}_1+\bar{k}_2)^2}{|\vec{k}_1+\vec{k}_2|} -  \frac{\bar{k}_1^2}{|\vec{k}_1|} \right) \nn\\
\qquad\qquad\qquad  J^{a_1}(\vec{k}_1)\theta^{a_2}(\vec{k}_2)J^{b_1}(\vec{q}_1)\theta^{b_2}(\vec{q}_2) \nn\\
\quad -{1\over 4}f^{a_1a_2c}f^{b_1b_2c} \int_{\slashed{k_1},\slashed{k_2},\slashed{q_1},\slashed{q_2}}\frac{\slashed{\delta}\left(\vec{k}_1+\vec{k}_2+\vec{q}_1+\vec{q}_2\right) }{\left(|\vec{k}_1|+|\vec{k}_2|+|\vec{q}_1|+|\vec{q}_2|\right)} J^{a_1}(\vec{k}_1)\theta^{a_2}(\vec{k}_2)J^{b_1}(\vec{q}_1)\theta^{b_2}(\vec{q}_2) \nn\\
\qquad\quad  (2\bar{q}_1\bar{k}_2+2(-\bar{k}_2-\bar{q}_1-\bar{q}_2)\bar{k}_2+2\bar{k}_2\bar{q}_2+2\bar{k}_2^2)     \nn\\
\quad - f^{a_1a_2c}f^{b_1b_2c} \int_{\slashed{k_1},\slashed{k_2},\slashed{q_1},\slashed{q_2}} \slashed{\delta}\left(\vec{k}_1+\vec{k}_2+\vec{q}_1+\vec{q}_2\right)  \frac{q_1\bar{q}_2}{\bar{q}_1+\bar{q}_2} \frac{g^{(3)}(\vec{k}_1,\vec{k}_2, -\vec{k}_1-\vec{k}_2)}{16}   \nn\\
\qquad\qquad\qquad  J^{a_1}(\vec{k}_1)J^{a_2}(\vec{k}_2)\theta^{b_1}(\vec{q}_1)\theta^{b_2}(\vec{q}_2) \nn\\
\quad -{1\over 2} f^{a_1a_2c}f^{b_1b_2c} \int_{\slashed{k_1},\slashed{k_2},\slashed{q_1},\slashed{q_2}} \frac{\slashed{\delta}\left(\vec{k}_1+\vec{k}_2+\vec{q}_1+\vec{q}_2\right) }{\left(|\vec{k}_1|+|\vec{k}_2|+|\vec{q}_1|+|\vec{q}_2|\right)} J^{a_1}(\vec{k}_1)J^{a_2}(\vec{k}_2)\theta^{b_1}(\vec{q}_1)\theta^{b_2}(\vec{q}_2) \nn\\
\qquad\Bigg\{ \frac{1}{2} \bar{k}_1(\bar{q}_1-\bar{q}_2) \frac{\bar{k}_1+\bar{k}_2}{\bar{k}_1+\bar{k}_2}  +\frac{\bar{k}_2^2}{|\vec{k}_2|} \frac{1}{2}\left(|\vec{q}_1|+|\vec{q}_2|\right) \frac{\bar{q}_2-\bar{q}_1}{\bar{k}_1+\bar{k}_2} +\frac{\bar{k}_1^2\bar{k}_2(q_1-q_2-k_1-k_2)}{|\vec{k}_1||\vec{k}_2|} \frac{\bar{q}_2-\bar{q}_1}{\bar{k}_1+\bar{k}_2} \nn\\
\qquad\qquad +\frac{1}{4} \frac{\bar{k}_1(\bar{k}_1+\bar{k}_2)}{|\vec{k}_1|}\frac{\bar{q}_1-\bar{q}_2}{\bar{k}_1+\bar{k}_2}|\vec{k}_2|  \Bigg\} +\O(e)\\
= \frac{1}{2} f^{a_1a_2c}f^{b_1b_2c} \int_{\slashed{k_1},\slashed{k_2},\slashed{q_1},\slashed{q_2}} \slashed{\delta}\left(\vec{k}_1+\vec{k}_2+\vec{q}_1+\vec{q}_2\right)   \left( \frac{(\bar{k}_1+\bar{k}_2)^2}{|\vec{k}_1+\vec{k}_2|} -  \frac{\bar{k}_1^2}{|\vec{k}_1|} \right)  \nn\\
\qquad\qquad\qquad  J^{a_1}(\vec{k}_1)\theta^{a_2}(\vec{k}_2)J^{b_1}(\vec{q}_1)\theta^{b_2}(\vec{q}_2) \nn\\
\quad - f^{a_1a_2c}f^{b_1b_2c} \int_{\slashed{k_1},\slashed{k_2},\slashed{q_1},\slashed{q_2}} \slashed{\delta}\left(\vec{k}_1+\vec{k}_2+\vec{q}_1+\vec{q}_2\right)  \frac{q_1\bar{q}_2}{\bar{q}_1+\bar{q}_2} \frac{g^{(3)}(\vec{k}_1,\vec{k}_2, -\vec{k}_1-\vec{k}_2)}{16}   \nn\\
\qquad\qquad\qquad   J^{a_1}(\vec{k}_1)J^{a_2}(\vec{k}_2)\theta^{b_1}(\vec{q}_1)\theta^{b_2}(\vec{q}_2) \nn\\
\quad -{1\over 2} f^{a_1a_2c}f^{b_1b_2c} \int_{\slashed{k_1},\slashed{k_2},\slashed{q_1},\slashed{q_2}} \frac{\slashed{\delta}\left(\vec{k}_1+\vec{k}_2+\vec{q}_1+\vec{q}_2\right) }{\left(|\vec{k}_1|+|\vec{k}_2|+|\vec{q}_1|+|\vec{q}_2|\right)} J^{a_1}(\vec{k}_1)J^{a_2}(\vec{k}_2)\theta^{b_1}(\vec{q}_1)\theta^{b_2}(\vec{q}_2) \nn\\
\qquad\Bigg\{ \frac{1}{2}(\bar{q}_1-\bar{q}_2) \frac{\bar{k}_1\bar{k}_2}{\bar{k}_1+\bar{k}_2}  +\frac{\bar{k}_2^2}{2|\vec{k}_2|} \frac{\bar{q}_2-\bar{q}_1}{\bar{k}_1+\bar{k}_2}\left(|\vec{k}_2|+|\vec{q}_1|+|\vec{q}_2|\right) +\frac{\bar{k}_1^2\bar{k}_2(q_1-q_2)}{|\vec{k}_1||\vec{k}_2|} \frac{\bar{q}_2-\bar{q}_1}{\bar{k}_1+\bar{k}_2} \nn\\
\qquad\qquad -\frac{\bar{k}_1\bar{k}_2|\vec{k}_1|}{4|\vec{k}_2|}\frac{\bar{q}_2-\bar{q}_1}{\bar{k}_1+\bar{k}_2} -\frac{\bar{k}_1^2|\vec{k}_2|}{4|\vec{k}_1|} \frac{\bar{q}_2-\bar{q}_1}{\bar{k}_1+\bar{k}_2} +\frac{1}{4} \frac{\bar{k}_1(\bar{k}_1+\bar{k}_2)}{|\vec{k}_1|}\frac{\bar{q}_1-\bar{q}_2}{\bar{k}_1+\bar{k}_2}|\vec{k}_2|  \Bigg\} +\O(e)\,.
\eE
The first term of the 4th line vanishes under $\vec k_1\leftrightarrow \vec k_2$ and the last term under $\vec q_1\leftrightarrow \vec q_2$. In the last line we interchange $\vec k_1\leftrightarrow \vec k_2$ in the two last terms (note the change from $\bar{q}_1-\bar{q}_2$ to $\bar{q}_2-\bar{q}_1$ in the very last term).
\bE{l}
= \frac{1}{2} f^{a_1a_2c}f^{b_1b_2c} \int_{\slashed{k_1},\slashed{k_2},\slashed{q_1},\slashed{q_2}} \slashed{\delta}\left(\vec{k}_1+\vec{k}_2+\vec{q}_1+\vec{q}_2\right)   \left( \frac{(\bar{k}_1+\bar{k}_2)^2}{|\vec{k}_1+\vec{k}_2|} -  \frac{\bar{k}_1^2}{|\vec{k}_1|} \right)  \nn\\
\qquad\qquad\qquad  J^{a_1}(\vec{k}_1)\theta^{a_2}(\vec{k}_2)J^{b_1}(\vec{q}_1)\theta^{b_2}(\vec{q}_2) \nn\\
\quad - f^{a_1a_2c}f^{b_1b_2c} \int_{\slashed{k_1},\slashed{k_2},\slashed{q_1},\slashed{q_2}} \slashed{\delta}\left(\vec{k}_1+\vec{k}_2+\vec{q}_1+\vec{q}_2\right)  \frac{q_1\bar{q}_2}{\bar{q}_1+\bar{q}_2} \frac{g^{(3)}(\vec{k}_1,\vec{k}_2, -\vec{k}_1-\vec{k}_2)}{16}  \nn\\
\qquad\qquad\qquad  J^{a_1}(\vec{k}_1)J^{a_2}(\vec{k}_2)\theta^{b_1}(\vec{q}_1)\theta^{b_2}(\vec{q}_2) \nn\\
\quad -{1\over 2} f^{a_1a_2c}f^{b_1b_2c} \int_{\slashed{k_1},\slashed{k_2},\slashed{q_1},\slashed{q_2}} \frac{\slashed{\delta}\left(\vec{k}_1+\vec{k}_2+\vec{q}_1+\vec{q}_2\right) }{\left(|\vec{k}_1|+|\vec{k}_2|+|\vec{q}_1|+|\vec{q}_2|\right)} J^{a_1}(\vec{k}_1)J^{a_2}(\vec{k}_2)\theta^{b_1}(\vec{q}_1)\theta^{b_2}(\vec{q}_2) \nn\\
\qquad\Bigg\{ \frac{\bar{k}_2^2}{2|\vec{k}_2|} \frac{\bar{q}_2-\bar{q}_1}{\bar{k}_1+\bar{k}_2}\left(|\vec{k}_2|+|\vec{q}_1|+|\vec{q}_2|\right)  -\frac{\bar{k}_1\bar{k}_2|\vec{k}_1|}{4|\vec{k}_2|}\frac{\bar{q}_2-\bar{q}_1}{\bar{k}_1+\bar{k}_2} +\frac{\bar{k}_2^2|\vec{k}_1|}{4|\vec{k}_2|} \frac{\bar{q}_2-\bar{q}_1}{\bar{k}_1+\bar{k}_2} \nn\\
\qquad\qquad +\frac{1}{4} \frac{\bar{k}_2(\bar{k}_1+\bar{k}_2)}{|\vec{k}_2|}\frac{\bar{q}_2-\bar{q}_1}{\bar{k}_1+\bar{k}_2}|\vec{k}_1|  \Bigg\} \qquad \\
= \frac{1}{2} f^{a_1a_2c}f^{b_1b_2c} \int_{\slashed{k_1},\slashed{k_2},\slashed{q_1},\slashed{q_2}} \slashed{\delta}\left(\vec{k}_1+\vec{k}_2+\vec{q}_1+\vec{q}_2\right)   \left( \frac{(\bar{k}_1+\bar{k}_2)^2}{|\vec{k}_1+\vec{k}_2|} -  \frac{\bar{k}_1^2}{|\vec{k}_1|} \right)  \nn\\
\qquad\qquad\qquad J^{a_1}(\vec{k}_1)\theta^{a_2}(\vec{k}_2)J^{b_1}(\vec{q}_1)\theta^{b_2}(\vec{q}_2) \nn\\
\quad - f^{a_1a_2c}f^{b_1b_2c} \int_{\slashed{k_1},\slashed{k_2},\slashed{q_1},\slashed{q_2}} \slashed{\delta}\left(\vec{k}_1+\vec{k}_2+\vec{q}_1+\vec{q}_2\right)  \frac{q_1\bar{q}_2}{\bar{q}_1+\bar{q}_2} \frac{g^{(3)}(\vec{k}_1,\vec{k}_2, -\vec{k}_1-\vec{k}_2)}{16}   \nn\\
\qquad\qquad\qquad  J^{a_1}(\vec{k}_1)J^{a_2}(\vec{k}_2)\theta^{b_1}(\vec{q}_1)\theta^{b_2}(\vec{q}_2) \nn\\
\quad -{1\over 4} f^{a_1a_2c}f^{b_1b_2c} \int_{\slashed{k_1},\slashed{k_2},\slashed{q_1},\slashed{q_2}} \slashed{\delta}\left(\vec{k}_1+\vec{k}_2+\vec{q}_1+\vec{q}_2\right) \frac{\bar{k}_2^2}{|\vec{k}_2|} \frac{\bar{q}_2-\bar{q}_1}{\bar{k}_1+\bar{k}_2}  J^{a_1}(\vec{k}_1)J^{a_2}(\vec{k}_2)\theta^{b_1}(\vec{q}_1)\theta^{b_2}(\vec{q}_2) \nn\\
\quad +\O(e)\,.\qquad
\eE

So we can simplify $F^{(2,4)}_{GL}|_{\O(\theta^2)}$ to
\bE{l}
F^{(2,4)}_{GL}|_{\O(\theta^2)} \nn\\
= \frac{1}{2} f^{a_1a_2c}f^{b_1b_2c} \int_{\slashed{k_1},\slashed{k_2},\slashed{q_1},\slashed{q_2}} \slashed{\delta}\left(\vec{k}_1+\vec{k}_2+\vec{q}_1+\vec{q}_2\right)   \left( \frac{(\bar{k}_1+\bar{k}_2)^2}{|\vec{k}_1+\vec{k}_2|} -  \frac{\bar{k}_1^2}{|\vec{k}_1|} \right)  \nn\\
\qquad\qquad\qquad J^{a_1}(\vec{k}_1)\theta^{a_2}(\vec{k}_2)J^{b_1}(\vec{q}_1)\theta^{b_2}(\vec{q}_2) \nn\\
\quad - f^{a_1a_2c}f^{b_1b_2c} \int_{\slashed{k_1},\slashed{k_2},\slashed{q_1},\slashed{q_2}} \slashed{\delta}\left(\vec{k}_1+\vec{k}_2+\vec{q}_1+\vec{q}_2\right)    J^{a_1}(\vec{k}_1)J^{a_2}(\vec{k}_2)\theta^{b_1}(\vec{q}_1)\theta^{b_2}(\vec{q}_2) \nn\\
\qquad\qquad \left\{\frac{q_1\bar{q}_2}{\bar{q}_1+\bar{q}_2} \frac{g^{(3)}(\vec{k}_1,\vec{k}_2, -\vec{k}_1-\vec{k}_2)}{16} -   \frac{\bar{q}_2}{\bar{q}_1+\bar{q}_2} \frac{\bar{k}_2^2}{2|\vec{k}_2|}\right\} \label{F24theta2}
\eE

\subsection{Order $\theta^3$:}

\subsubsection{$\left(\frac{\delta F^{(1)}_{GL}}{\delta A}\right)^2$-term}

\bE{l}
-\frac{1}{2}\int_{\slashed{p},\slashed{k_1},\slashed{k_2},\slashed{q_1},\slashed{q_2}}\frac{1}{\sum_i(|\vec{k}_i|+|\vec{q}_i|)}\left(\frac{\delta F^{(1)}_{GL}}{\delta A^a_i(\vec{p})}\right)[\vec{k}_1,\vec{k}_2]\left(\frac{\delta F^{(1)}_{GL}}{\delta A^a_i(-\vec{p})}\right)[\vec{q}_1,\vec{q}_2] \Bigg|_{\O(\theta^3)} \nn\\
\quad = -\frac{-4}{2}f^{a_1a_2c}f^{b_1b_2c}\int_{\slashed{p},\slashed{k_1},\slashed{k_2},\slashed{q_1},\slashed{q_2}}\frac{\slashed{\delta}\left(\vec{k}_1+\vec{k}_2+\vec{p}\right)\slashed{\delta}\left(\vec{q}_1+\vec{q}_2-\vec{p}\right)}{\sum_i(|\vec{k}_i|+|\vec{q}_i|)}J^{a_1}(\vec{k}_1)\theta^{a_2}(\vec{k}_2)\theta^{b_1}(\vec{q}_1)\theta^{b_2}(\vec{q}_2) \nn\\
\qquad \Bigg\{\left( \frac{\bar{p}^2}{|\vec{p}|}-\frac{\bar{k}_1^2}{|\vec{k}_1|}\right)(-2)\frac{-p q_1\bar{q}_2}{|\vec{p}|} - \left(\frac{1}{4}|\vec{p}|+\frac{\bar{k}_1}{|\vec{k}_1|}\left(-\frac{p}{\bar{p}}(\bar{k}_1+\bar{k}_2)+k_2\right) \right) 2\frac{-\bar{p}q_1\bar{q}_2}{|\vec{p}|} \Bigg\} \\
\quad = 4f^{a_1a_2c}f^{b_1b_2c}\int_{\slashed{p},\slashed{k_1},\slashed{k_2},\slashed{q_1},\slashed{q_2}}\frac{\slashed{\delta}\left(\vec{k}_1+\vec{k}_2+\vec{q}_1+\vec{q}_2\right)}{\sum_i(|\vec{k}_i|+|\vec{q}_i|)}J^{a_1}(\vec{k}_1)\theta^{a_2}(\vec{k}_2)\theta^{b_1}(\vec{q}_1)\theta^{b_2}(\vec{q}_2) \nn\\
\qquad \Bigg\{\left( \frac{(\bar{k}_1+\bar{k}_2)^2}{|\vec{k}_1+\vec{k}_2|}-\frac{\bar{k}_1^2}{|\vec{k}_1|}\right)\frac{(q_1+q_2) q_1\bar{q}_2}{|\vec{q}_1+\vec{q}_2|} + \frac{1}{4}\left(|\vec{k}_1+\vec{k}_2|-|\vec{k}_1| \right) \frac{(\bar{q}_1+\bar{q}_2) q_1\bar{q}_2}{|\vec{q}_1+\vec{q}_2|} \Bigg\} \\
\quad = 4f^{a_1a_2c}f^{b_1b_2c}\int_{\slashed{p},\slashed{k_1},\slashed{k_2},\slashed{q_1},\slashed{q_2}}\frac{\slashed{\delta}\left(\sum_i(\vec{k}_i+\vec{q}_i)\right)}{\sum_i(|\vec{k}_i|+|\vec{q}_i|)}J^{a_1}(\vec{k}_1)\theta^{a_2}(\vec{k}_2)\theta^{b_1}(\vec{q}_1)\theta^{b_2}(\vec{q}_2) \nn\\
\qquad \Bigg\{-\frac{1}{4} (\bar{k}_1+\bar{k}_2) q_1\bar{q}_2 + \frac{\bar{k}_1^2}{|\vec{k}_1|}\frac{(k_1+k_2) q_1\bar{q}_2}{|\vec{q}_1+\vec{q}_2|} + \frac{1}{4}\left(|\vec{k}_1+\vec{k}_2|-|\vec{k}_1| \right) \frac{(\bar{q}_1+\bar{q}_2) q_1\bar{q}_2}{|\vec{q}_1+\vec{q}_2|} \Bigg\} \\
\quad = 4f^{a_1a_2c}f^{b_1b_2c}\int_{\slashed{p},\slashed{k_1},\slashed{k_2},\slashed{q_1},\slashed{q_2}}\frac{\slashed{\delta}\left(\sum_i(\vec{k}_i+\vec{q}_i)\right)}{\sum_i(|\vec{k}_i|+|\vec{q}_i|)}J^{a_1}(\vec{k}_1)\theta^{a_2}(\vec{k}_2)\theta^{b_1}(\vec{q}_1)\theta^{b_2}(\vec{q}_2) \nn\\
\qquad \Bigg\{\frac{1}{2}(\bar{q}_1+\bar{q}_2) q_1\bar{q}_2 + \frac{\bar{k}_1^2}{|\vec{k}_1|}\frac{k_2 q_1\bar{q}_2}{|\vec{q}_1+\vec{q}_2|} + \frac{1}{4}|\vec{k}_1| \frac{(-\bar{q}_1-\bar{q}_2+\bar{k}_1) q_1\bar{q}_2}{|\vec{q}_1+\vec{q}_2|} \Bigg\} \,.
\eE

\subsubsection{$\left(\vec{A}\cdot\frac{\delta F^{(1)}_{GL}}{\delta \vec{A}}\right)$-term}

\bE{l}
 -if^{b_1b_2c}\int_{\slashed{p},\slashed{k_1},\slashed{k_2},\slashed{q_1},\slashed{q_2}}\frac{\slashed{\delta}(\vec{q}_1+\vec{q}_2-\vec{p})}{\sum_i(|\vec{k}_i|+|\vec{q}_i|)|\vec{q}_1|}(\vec{q}_1\cdot\vec{A}^{b_1}(\vec{q}_1))\left(\vec{A}^{b_2}(\vec{q}_2)\cdot\frac{\delta F^{(1)}_{GL}}{\delta \vec{A}^c(\vec{p})}[\vec{k}_1,\vec{k}_2]\right) \nn\\
\quad = -i f^{b_1b_2c}f^{a_1a_2c}\int_{\slashed{p},\slashed{k_1},\slashed{k_2},\slashed{q_1},\slashed{q_2}}\frac{\slashed{\delta}(\vec{q}_1+\vec{q}_2-\vec{p})\slashed{\delta}(\vec{k}_1+\vec{k}_2+\vec{p})}{(|\vec{k}_1|+|\vec{k}_2|+|\vec{q}_1|+|\vec{q}_2|)|\vec{q}_1|} \nn\\
\qquad \left(-i\bar{q}_1 J^{b_1}(\vec{q}_1)+i|\vec{q}_1|^2\theta^{b_1}(\vec{q}_1) +\O(\theta^2)\right) \nn\\
\qquad A^{b_2}_i(\vec{q}_2) (-i) \Bigg\{O(J^2) \nn\\
\qquad +\Bigg\{(\delta_{1i}+i\delta_{2i})\left( \frac{\bar{p}^2}{|\vec{p}|}-\frac{\bar{k}_1^2}{|\vec{k}_1|}\right)   \nn\\
\qquad\qquad - (\delta_{1i}-i\delta_{2i}) \left(\frac{1}{4}|\vec{p}|+\frac{\bar{k}_1}{|\vec{k}_1|}\left(-\frac{p}{\bar{p}}(\bar{k}_1+\bar{k}_2)+k_2\right) \right)\Bigg\}J^{a_1}(\vec{k}_1)\theta^{a_2}(\vec{k}_2) \nn\\
\qquad +\Bigg\{(\delta_{1i}+i\delta_{2i})2\frac{\bar{p}k_1\bar{k}_2}{|\vec{p}|} -(\delta_{1i}-i\delta_{2i}) 2\frac{p k_1\bar{k}_2}{|\vec{p}|}\Bigg\} \theta^{a_1}(\vec{k}_1)\theta^{a_2}(\vec{k}_2)\Bigg\} +\O(e)\\
\quad = (-i)^3 f^{b_1b_2c}f^{a_1a_2c}\int_{\slashed{p},\slashed{k_1},\slashed{k_2},\slashed{q_1},\slashed{q_2}}\frac{\slashed{\delta}(\vec{q}_1+\vec{q}_2-\vec{p})\slashed{\delta}(\vec{k}_1+\vec{k}_2+\vec{p})}{(|\vec{k}_1|+|\vec{k}_2|+|\vec{q}_1|+|\vec{q}_2|)|\vec{q}_1|} \nn\\
\qquad \left(\bar{q}_1 J^{b_1}(\vec{q}_1)-|\vec{q}_1|^2\theta^{b_1}(\vec{q}_1) \right) \nn\\
\qquad \Bigg\{\Bigg\{2A^{b_2}(\vec{q}_2) \left( \frac{\bar{p}^2}{|\vec{p}|}-\frac{\bar{k}_1^2}{|\vec{k}_1|}\right)  - 2\bar{A}^{b_2}(\vec{q}_2) \left(\frac{1}{4}|\vec{p}|+\frac{\bar{k}_1}{|\vec{k}_1|}\left(-\frac{p}{\bar{p}}(\bar{k}_1+\bar{k}_2)+k_2\right) \right)\Bigg\}J^{a_1}(\vec{k}_1)\theta^{a_2}(\vec{k}_2) \nn\\
\qquad\qquad +\Bigg\{2A^{b_2}(\vec{q}_2)2\frac{\bar{p}k_1\bar{k}_2}{|\vec{p}|} -2\bar{A}^{b_2}(\vec{q}_2) 2\frac{p k_1\bar{k}_2}{|\vec{p}|}\Bigg\} \theta^{a_1}(\vec{k}_1)\theta^{a_2}(\vec{k}_2)\Bigg\}\nn\\
\qquad +\O(J^4)+\O(J^3\theta)+\O(J^2\theta^2)+\O(e)\\
\quad = (-i)^3 f^{b_1b_2c}f^{a_1a_2c}\int_{\slashed{p},\slashed{k_1},\slashed{k_2},\slashed{q_1},\slashed{q_2}} \frac{\slashed{\delta}\left(\sum_i(\vec{k}_i+\vec{q}_i)\right)}{\sum_i(|\vec{k}_i|+|\vec{q}_i|)} \nn\\
\qquad \left(\bar{q}_1 J^{b_1}(\vec{q}_1)-|\vec{q}_1|^2\theta^{b_1}(\vec{q}_1) \right) \nn\\
\qquad \frac{1}{|\vec{q}_1|} \Bigg\{(\O(J) +2iq_2\theta^{b_2}(\vec{q}_2)) \left( \frac{(\bar{k}_1+\bar{k}_2)^2}{|\vec{k}_1+\vec{k}_2|}-\frac{\bar{k}_1^2}{|\vec{k}_1|}\right)   \nn\\
\qquad\qquad - 2i\bar{q}_2\theta^{b_2}(\vec{q}_2) \frac{1}{4} \left(|\vec{k}_1+\vec{k}_2|-|\vec{k}_1| \right)J^{a_1}(\vec{k}_1)\theta^{a_2}(\vec{k}_2) \nn\\
\qquad\qquad +\Bigg\{(-iJ^{b_2}(\vec{q}_2)+2iq_2\theta^{b_2}(\vec{q}_2)) 2\frac{(-\bar{k}_1-\bar{k}_2)k_1\bar{k}_2}{|\vec{k}_1+\vec{k}_2|}  \nn\\
\qquad\qquad\qquad  -2i\bar{q}_2\theta^{b_2}(\vec{q}_2) 2\frac{(-k_1-k_2) k_1\bar{k}_2}{|\vec{k}_1+\vec{k}_2|}\Bigg\} \theta^{a_1}(\vec{k}_1)\theta^{a_2}(\vec{k}_2)\Bigg\}\nn\\
\qquad +\O(J^4)+\O(J^3\theta)+\O(J^2\theta^2)+\O(e)\,.
\eE
This gives us
\bE{l}
 -if^{b_1b_2c}\int_{\slashed{p},\slashed{k_1},\slashed{k_2},\slashed{q_1},\slashed{q_2}}\frac{\slashed{\delta}(\vec{q}_1+\vec{q}_2-\vec{p})}{\sum_i(|\vec{k}_i|+|\vec{q}_i|)|\vec{q}_1|}(\vec{q}_1\cdot\vec{A}^{b_1}(\vec{q}_1))\left(\vec{A}^{b_2}(\vec{q}_2)\cdot\frac{\delta F^{(1)}_{GL}}{\delta \vec{A}^c(\vec{p})}[\vec{k}_1,\vec{k}_2]\right)  \Bigg|_{\O(\theta^3)} \nn\\
\quad = 2 f^{b_1b_2c}f^{a_1a_2c}\int_{\slashed{p},\slashed{k_1},\slashed{k_2},\slashed{q_1},\slashed{q_2}} \frac{\slashed{\delta}\left(\sum_i(\vec{k}_i+\vec{q}_i)\right)}{\sum_i(|\vec{k}_i|+|\vec{q}_i|)} J^{a_1}(\vec{k}_1)\theta^{a_2}(\vec{k}_2) \theta^{b_1}(\vec{q}_1)\theta^{b_2}(\vec{q}_2) \nn\\
\qquad  \Bigg\{|\vec{q}_1|q_2 \left( \frac{(\bar{k}_1+\bar{k}_2)^2}{|\vec{k}_1+\vec{k}_2|}-\frac{\bar{k}_1^2}{|\vec{k}_1|}\right) -|\vec{q}_1|\bar{q}_2 \frac{1}{4} \left(|\vec{k}_1+\vec{k}_2|-|\vec{k}_1| \right) \nn\\
\qquad\qquad -|\vec{k}_2| \frac{(\bar{q}_1+\bar{q}_2)q_1\bar{q}_2}{|\vec{q}_1+\vec{q}_2|} + 2\frac{\bar{k}_1}{|\vec{k}_1|}k_2  \frac{(\bar{q}_1+\bar{q}_2)q_1\bar{q}_2}{|\vec{q}_1+\vec{q}_2|}  -2\frac{\bar{k}_1}{|\vec{k}_1|} \bar{k}_2  \frac{(q_1+q_2) q_1\bar{q}_2}{|\vec{q}_1+\vec{q}_2|} \Bigg\} +\O(e)\,.\qquad
\eE

\subsubsection{$ (A\times A)(A\times A)$-term}

\bE{l}
\frac{1}{8}f^{a_1a_2c}f^{b_1b_2c}\int_{\slashed{k_1},\slashed{k_2},\slashed{q_1},\slashed{q_2}}\frac{\slashed{\delta}(\sum_i(\vec{k}_i+\vec{q}_i))}{\sum_i(|\vec{k}_i|+|\vec{q}_i|)}(\vec{A}^{a_1}(\vec{k}_1)\times\vec{A}^{a_2}(\vec{k}_2))(\vec{A}^{b_1}(\vec{q}_1)\times\vec{A}^{b_2}(\vec{q}_2)) \Bigg|_{\O(\theta^3)} \nn\\
= 2f^{a_1a_2c}f^{b_1b_2c}\int_{\slashed{k_1},\slashed{k_2},\slashed{q_1},\slashed{q_2}}\frac{\slashed{\delta}(\sum_i(\vec{k}_i+\vec{q}_i))}{\sum_i(|\vec{k}_i|+|\vec{q}_i|)} \bar{k_2}q_1\bar{q}_2 J^{a_1}(\vec{k}_1)\theta^{a_2}(\vec{k}_2) \theta^{b_1}(\vec{q}_1)\theta^{b_2}(\vec{q}_2) +\O(e)\,.
\eE

\subsubsection{All three together}

\bE{rCl}
F^{(2,4)}_{GL} \Bigg|_{\O(\theta^3)} &=& 2 f^{b_1b_2c}f^{a_1a_2c}\int_{\slashed{p},\slashed{k_1},\slashed{k_2},\slashed{q_1},\slashed{q_2}} \frac{\slashed{\delta}\left(\sum_i(\vec{k}_i+\vec{q}_i)\right)}{\sum_i(|\vec{k}_i|+|\vec{q}_i|)} J^{a_1}(\vec{k}_1)\theta^{a_2}(\vec{k}_2) \theta^{b_1}(\vec{q}_1)\theta^{b_2}(\vec{q}_2) \nn\\
&&\qquad  \Bigg\{(\bar{q}_1+\bar{q}_2) q_1\bar{q}_2 + 2\frac{\bar{k}_1^2}{|\vec{k}_1|}\frac{k_2 q_1\bar{q}_2}{|\vec{q}_1+\vec{q}_2|} + \frac{1}{2}|\vec{k}_1| \frac{(-\bar{q}_1-\bar{q}_2+\bar{k}_1) q_1\bar{q}_2}{|\vec{q}_1+\vec{q}_2|} \nn\\
&&\qquad\quad + |\vec{q}_1|q_2 \left( \frac{(\bar{k}_1+\bar{k}_2)^2}{|\vec{k}_1+\vec{k}_2|}-\frac{\bar{k}_1^2}{|\vec{k}_1|}\right) -|\vec{q}_1|\bar{q}_2 \frac{1}{4} \left(|\vec{k}_1+\vec{k}_2|-|\vec{k}_1| \right) \nn\\
&&\qquad\qquad\quad -|\vec{k}_2| \frac{(\bar{q}_1+\bar{q}_2)q_1\bar{q}_2}{|\vec{q}_1+\vec{q}_2|} + 2\frac{\bar{k}_1}{|\vec{k}_1|}k_2  \frac{(\bar{q}_1+\bar{q}_2)q_1\bar{q}_2}{|\vec{q}_1+\vec{q}_2|}  -2\frac{\bar{k}_1}{|\vec{k}_1|} \bar{k}_2  \frac{(q_1+q_2) q_1\bar{q}_2}{|\vec{q}_1+\vec{q}_2|}\nn\\
&&\qquad\qquad + \bar{k_2}q_1\bar{q}_2 \Bigg\}+\O(e) \,.
\eE

Again, we now manipulate this term, in order to get rid of the $\frac{1}{\sum_i(|\vec{k}_i|+|\vec{q}_i|)}$ prefactor:
\bE{l}
=  2 f^{b_1b_2c}f^{a_1a_2c}\int_{\slashed{p},\slashed{k_1},\slashed{k_2},\slashed{q_1},\slashed{q_2}} \frac{\slashed{\delta}\left(\sum_i(\vec{k}_i+\vec{q}_i)\right)}{\sum_i(|\vec{k}_i|+|\vec{q}_i|)} J^{a_1}(\vec{k}_1)\theta^{a_2}(\vec{k}_2) \theta^{b_1}(\vec{q}_1)\theta^{b_2}(\vec{q}_2) \\
\qquad  \Bigg\{-\bar{k}_1 q_1\bar{q}_2 + 2\frac{\bar{k}_1^2}{|\vec{k}_1|}\frac{k_2 q_1\bar{q}_2}{|\vec{q}_1+\vec{q}_2|} + \frac{1}{2}|\vec{k}_1| \frac{\bar{k}_1 q_1\bar{q}_2}{|\vec{q}_1+\vec{q}_2|} \nn\\
\qquad + |\vec{q}_1|q_2 \left( \frac{(\bar{k}_1+\bar{k}_2)^2}{|\vec{k}_1+\vec{k}_2|}-\frac{\bar{k}_1^2}{|\vec{k}_1|}\right) -|\vec{q}_1|\bar{q}_2  \left( \frac{(\bar{k}_1+\bar{k}_2)(k_1+k_2)}{|\vec{k}_1+\vec{k}_2|}-\frac{\bar{k}_1k_1}{|\vec{k}_1|}\right) \nn\\
\qquad -\left(|\vec{k}_2|+ \frac{1}{2}|\vec{k}_1| \right) \frac{(\bar{q}_1+\bar{q}_2)q_1\bar{q}_2}{|\vec{q}_1+\vec{q}_2|} + 2\frac{\bar{k}_1}{|\vec{k}_1|}k_2  \frac{(\bar{q}_1+\bar{q}_2)q_1\bar{q}_2}{|\vec{q}_1+\vec{q}_2|}  -2\frac{\bar{k}_1}{|\vec{k}_1|} \bar{k}_2  \frac{(q_1+q_2) q_1\bar{q}_2}{|\vec{q}_1+\vec{q}_2|} \Bigg\}+\O(e) \nn\\
=  2 f^{b_1b_2c}f^{a_1a_2c}\int_{\slashed{p},\slashed{k_1},\slashed{k_2},\slashed{q_1},\slashed{q_2}} \frac{\slashed{\delta}\left(\sum_i(\vec{k}_i+\vec{q}_i)\right)}{\sum_i(|\vec{k}_i|+|\vec{q}_i|)} J^{a_1}(\vec{k}_1)\theta^{a_2}(\vec{k}_2) \theta^{b_1}(\vec{q}_1)\theta^{b_2}(\vec{q}_2) \nn\\
\qquad  \Bigg\{-\bar{k}_1 q_1\bar{q}_2 + |\vec{q}_1| \frac{q_2(\bar{k}_1+\bar{k}_2)(-\bar{q}_1)}{|\vec{k}_1+\vec{k}_2|} -|\vec{q}_1|  \frac{\bar{q}_2(\bar{k}_1+\bar{k}_2)(-q_1)}{|\vec{k}_1+\vec{k}_2|} \nn\\
\qquad +  \bar{k}_1\frac{|\vec{q}_1|}{|\vec{k}_1|}(k_1\bar{q}_2 - \bar{k}_1q_2) -\left(|\vec{k}_2|+ \frac{1}{2}|\vec{k}_1| \right) \frac{(\bar{q}_1+\bar{q}_2)q_1\bar{q}_2}{|\vec{q}_1+\vec{q}_2|} \nn\\
\qquad + 2\frac{\bar{k}_1q_1\bar{q}_2}{|\vec{k}_1||\vec{q}_1+\vec{q}_2|} \left(\bar{k}_1 k_2+ \bar{k}_1 k_1 - k_2  (\bar{k}_1+\bar{k}_2)  + \bar{k}_2 (k_1+k_2) \right) \Bigg\} +\O(e) \\
 =  2 f^{b_1b_2c}f^{a_1a_2c}\int_{\slashed{p},\slashed{k_1},\slashed{k_2},\slashed{q_1},\slashed{q_2}} \frac{\slashed{\delta}\left(\sum_i(\vec{k}_i+\vec{q}_i)\right)}{\sum_i(|\vec{k}_i|+|\vec{q}_i|)} J^{a_1}(\vec{k}_1)\theta^{a_2}(\vec{k}_2) \theta^{b_1}(\vec{q}_1)\theta^{b_2}(\vec{q}_2) \nn\\
\qquad  \Bigg\{-\bar{k}_1 q_1\bar{q}_2  +(|\vec{q}_1|+|\vec{q}_2|)  \frac{\bar{q}_2q_1}{|\vec{k}_1+\vec{k}_2|} (\bar{k}_1+\bar{k}_2)\nn\\
\qquad +  \bar{k}_1\frac{|\vec{q}_1|}{|\vec{k}_1|}(k_1\bar{q}_2 - \bar{k}_1q_2) -\left(|\vec{k}_2|+ \frac{1}{2}|\vec{k}_1| \right) \frac{(\bar{q}_1+\bar{q}_2)q_1\bar{q}_2}{|\vec{q}_1+\vec{q}_2|} \nn\\
\qquad + 2\frac{\bar{k}_1q_1\bar{q}_2}{|\vec{k}_1||\vec{q}_1+\vec{q}_2|} \left(\bar{k}_1 k_1 + \bar{k}_2 k_1 \right) \Bigg\} +\O(e) \\
 =  2 f^{b_1b_2c}f^{a_1a_2c}\int_{\slashed{p},\slashed{k_1},\slashed{k_2},\slashed{q_1},\slashed{q_2}} \frac{\slashed{\delta}\left(\sum_i(\vec{k}_i+\vec{q}_i)\right)}{\sum_i(|\vec{k}_i|+|\vec{q}_i|)} J^{a_1}(\vec{k}_1)\theta^{a_2}(\vec{k}_2) \theta^{b_1}(\vec{q}_1)\theta^{b_2}(\vec{q}_2) \nn\\
\qquad  \Bigg\{-\bar{k}_1 q_1\bar{q}_2  +(|\vec{q}_1|+|\vec{q}_2|)  \frac{\bar{q}_2q_1}{|\vec{k}_1+\vec{k}_2|} (\bar{k}_1+\bar{k}_2) +  \bar{k}_1\frac{|\vec{q}_1|}{|\vec{k}_1|}(k_1\bar{q}_2 - \bar{k}_1q_2)\nn\\
\qquad +\left(|\vec{k}_2|+ \frac{1}{2}|\vec{k}_1| \right) \frac{(\bar{k}_1+\bar{k}_2)q_1\bar{q}_2}{|\vec{q}_1+\vec{q}_2|}   + \frac{1}{2}\frac{|\vec{k}_1|q_1\bar{q}_2}{|\vec{q}_1+\vec{q}_2|} \left(\bar{k}_1 + \bar{k}_2 \right) \Bigg\} +\O(e) \\
 =  2 f^{b_1b_2c}f^{a_1a_2c}\int_{\slashed{p},\slashed{k_1},\slashed{k_2},\slashed{q_1},\slashed{q_2}} \frac{\slashed{\delta}\left(\sum_i(\vec{k}_i+\vec{q}_i)\right)}{\sum_i(|\vec{k}_i|+|\vec{q}_i|)} J^{a_1}(\vec{k}_1)\theta^{a_2}(\vec{k}_2) \theta^{b_1}(\vec{q}_1)\theta^{b_2}(\vec{q}_2) \nn\\
\qquad  \Bigg\{-\bar{k}_1 q_1\bar{q}_2  +(|\vec{q}_1|+|\vec{q}_2|+|\vec{k}_2|+|\vec{k}_1|)  \frac{\bar{q}_2q_1}{|\vec{k}_1+\vec{k}_2|} (\bar{k}_1+\bar{k}_2) +  \bar{k}_1\frac{|\vec{q}_1|}{|\vec{k}_1|}(k_1\bar{q}_2 - \bar{k}_1q_2)\Bigg\}  \nn\\
\qquad+\O(e)\qquad \\
=  2 f^{b_1b_2c}f^{a_1a_2c}\int_{\slashed{p},\slashed{k_1},\slashed{k_2},\slashed{q_1},\slashed{q_2}} \slashed{\delta}\left(\sum_i(\vec{k}_i+\vec{q}_i)\right) \frac{\bar{q}_2q_1}{|\vec{k}_1+\vec{k}_2|} (\bar{k}_1+\bar{k}_2) J^{a_1}(\vec{k}_1)\theta^{a_2}(\vec{k}_2) \theta^{b_1}(\vec{q}_1)\theta^{b_2}(\vec{q}_2) \nn\\
\quad + 2 f^{b_1b_2c}f^{a_1a_2c}\int_{\slashed{p},\slashed{k_1},\slashed{k_2},\slashed{q_1},\slashed{q_2}} \frac{\slashed{\delta}\left(\sum_i(\vec{k}_i+\vec{q}_i)\right)}{\sum_i(|\vec{k}_i|+|\vec{q}_i|)} J^{a_1}(\vec{k}_1)\theta^{a_2}(\vec{k}_2) \theta^{b_1}(\vec{q}_1)\theta^{b_2}(\vec{q}_2) \nn\\
\qquad  \Bigg\{-\bar{k}_1 q_1\bar{q}_2  -  \bar{k}_1\frac{|\vec{q}_1|}{|\vec{k}_1|}((q_1+k_2)\bar{q}_2 - (\bar{q}_1+\bar{k}_2)q_2) \Bigg\}+\O(e) \\
 =  2 f^{b_1b_2c}f^{a_1a_2c}\int_{\slashed{p},\slashed{k_1},\slashed{k_2},\slashed{q_1},\slashed{q_2}} \slashed{\delta}\left(\sum_i(\vec{k}_i+\vec{q}_i)\right) \frac{\bar{q}_2q_1}{|\vec{k}_1+\vec{k}_2|} (\bar{k}_1+\bar{k}_2) J^{a_1}(\vec{k}_1)\theta^{a_2}(\vec{k}_2) \theta^{b_1}(\vec{q}_1)\theta^{b_2}(\vec{q}_2) \nn\\
\quad + 2 f^{b_1b_2c}f^{a_1a_2c}\int_{\slashed{p},\slashed{k_1},\slashed{k_2},\slashed{q_1},\slashed{q_2}} \frac{\slashed{\delta}\left(\sum_i(\vec{k}_i+\vec{q}_i)\right)}{\sum_i(|\vec{k}_i|+|\vec{q}_i|)} J^{a_1}(\vec{k}_1)\theta^{a_2}(\vec{k}_2) \theta^{b_1}(\vec{q}_1)\theta^{b_2}(\vec{q}_2) \nn\\
\qquad  \Bigg\{-\bar{k}_1 q_1\bar{q}_2\left(1+\frac{|\vec{q}_1|}{|\vec{k}_1|}+\frac{|\vec{q}_2|}{|\vec{k}_1|}\right)  -  \bar{k}_1\frac{|\vec{q}_1|}{|\vec{k}_1|}(k_2\bar{q}_2 - \bar{k}_2q_2)  \Bigg\} +\O(e) \,.
\eE
We add and subtract the missing term.
\bE{l}
=  2 f^{b_1b_2c}f^{a_1a_2c}\int_{\slashed{p},\slashed{k_1},\slashed{k_2},\slashed{q_1},\slashed{q_2}} \slashed{\delta}\left(\sum_i(\vec{k}_i+\vec{q}_i)\right) q_1\bar{q}_2 \left(\frac{\bar{k}_1+\bar{k}_2}{|\vec{k}_1+\vec{k}_2|} - \frac{\bar{k}_1}{|\vec{k}_1|} \right)  \nn\\
\qquad\qquad\qquad  J^{a_1}(\vec{k}_1)\theta^{a_2}(\vec{k}_2) \theta^{b_1}(\vec{q}_1)\theta^{b_2}(\vec{q}_2) \nn\\
\quad + 2 f^{b_1b_2c}f^{a_1a_2c}\int_{\slashed{p},\slashed{k_1},\slashed{k_2},\slashed{q_1},\slashed{q_2}} \frac{\slashed{\delta}\left(\sum_i(\vec{k}_i+\vec{q}_i)\right)}{\sum_i(|\vec{k}_i|+|\vec{q}_i|)} J^{a_1}(\vec{k}_1)\theta^{a_2}(\vec{k}_2) \theta^{b_1}(\vec{q}_1)\theta^{b_2}(\vec{q}_2) \nn\\
\qquad  \Bigg\{\bar{k}_1 q_1\bar{q}_2 \frac{|\vec{k}_2|}{|\vec{k}_1|}  -  \bar{k}_1\frac{|\vec{q}_1|}{|\vec{k}_1|}(k_2\bar{q}_2 - \bar{k}_2q_2)  \Bigg\}+\O(e) \\
=  2 f^{b_1b_2c}f^{a_1a_2c}\int_{\slashed{p},\slashed{k_1},\slashed{k_2},\slashed{q_1},\slashed{q_2}} \slashed{\delta}\left(\sum_i(\vec{k}_i+\vec{q}_i)\right) q_1\bar{q}_2 \left(\frac{\bar{k}_1+\bar{k}_2}{|\vec{k}_1+\vec{k}_2|} - \frac{\bar{k}_1}{|\vec{k}_1|} \right)  \nn\\
\qquad\qquad\qquad  J^{a_1}(\vec{k}_1)\theta^{a_2}(\vec{k}_2) \theta^{b_1}(\vec{q}_1)\theta^{b_2}(\vec{q}_2) \nn\\
\quad + 2 f^{b_1b_2c}f^{a_1a_2c}\int_{\slashed{p},\slashed{k_1},\slashed{k_2},\slashed{q_1},\slashed{q_2}} \frac{\slashed{\delta}\left(\sum_i(\vec{k}_i+\vec{q}_i)\right)}{\sum_i(|\vec{k}_i|+|\vec{q}_i|)} J^{a_1}(\vec{k}_1)\theta^{a_2}(\vec{k}_2) \theta^{b_1}(\vec{q}_1)\theta^{b_2}(\vec{q}_2) \nn\\
\qquad  \Bigg\{ \frac{\bar{k}_1}{|\vec{k}_1|} (q_1\bar{q}_2|\vec{k}_2| -  |\vec{q}_1| k_2\bar{q}_2 +|\vec{q}_1| \bar{k}_2q_2)  \Bigg\}+\O(e) \,.
\eE
Using the Jacobi-Identity $f^{a_1a_2c}f^{b_1b_2c}= -f^{a_1b_1c}f^{b_2a_2c} -f^{a_1b_2c}f^{a_2b_1c}$ in the second term then renaming $b_1 \leftrightarrow a_2$, $\vec q_1 \leftrightarrow \vec k_2$ in the (new) second term; and $b_2 \leftrightarrow a_2$, $\vec q_2 \leftrightarrow \vec k_2$ in the (new) third term, we obtain
\bE{l}
 =  2 f^{b_1b_2c}f^{a_1a_2c}\int_{\slashed{p},\slashed{k_1},\slashed{k_2},\slashed{q_1},\slashed{q_2}} \slashed{\delta}\left(\sum_i(\vec{k}_i+\vec{q}_i)\right) q_1\bar{q}_2 \left(\frac{\bar{k}_1+\bar{k}_2}{|\vec{k}_1+\vec{k}_2|} - \frac{\bar{k}_1}{|\vec{k}_1|} \right)  \nn\\
\qquad\qquad\qquad J^{a_1}(\vec{k}_1)\theta^{a_2}(\vec{k}_2) \theta^{b_1}(\vec{q}_1)\theta^{b_2}(\vec{q}_2) \nn\\
\quad + 2 f^{b_1b_2c}f^{a_1a_2c}\int_{\slashed{p},\slashed{k_1},\slashed{k_2},\slashed{q_1},\slashed{q_2}} \frac{\slashed{\delta}\left(\sum_i(\vec{k}_i+\vec{q}_i)\right)}{\sum_i(|\vec{k}_i|+|\vec{q}_i|)} J^{a_1}(\vec{k}_1)\theta^{a_2}(\vec{k}_2) \theta^{b_1}(\vec{q}_1)\theta^{b_2}(\vec{q}_2) \nn\\
\qquad  \Bigg\{ \frac{\bar{k}_1}{|\vec{k}_1|} (q_1\bar{q}_2|\vec{k}_2| -  |\vec{k}_2| q_1\bar{q}_2 -  |\vec{q}_1| q_2\bar{k}_2 +|\vec{q}_1| \bar{k}_2q_2)  \Bigg\}+\O(e) \,.
\eE
The last term vanishes, so we find
\bE{l}
F^{(2,4)}_{GL}|_{\O(\theta^3)} =  2 f^{b_1b_2c}f^{a_1a_2c}\int_{\slashed{p},\slashed{k_1},\slashed{k_2},\slashed{q_1},\slashed{q_2}} \slashed{\delta}\left(\sum_i(\vec{k}_i+\vec{q}_i)\right) q_1\bar{q}_2 \left(\frac{\bar{k}_1+\bar{k}_2}{|\vec{k}_1+\vec{k}_2|} - \frac{\bar{k}_1}{|\vec{k}_1|} \right) \nn\\ \qquad\qquad\qquad J^{a_1}(\vec{k}_1)\theta^{a_2}(\vec{k}_2) \theta^{b_1}(\vec{q}_1)\theta^{b_2}(\vec{q}_2)  +\O(e) \label{F24theta3}
\eE

\subsection{Order $\theta^4$:}

\subsubsection{$\left(\frac{\delta F^{(1)}_{GL}}{\delta A}\right)^2$-term}

\bE{l}
-\frac{1}{2}\int_{\slashed{p},\slashed{k_1},\slashed{k_2},\slashed{q_1},\slashed{q_2}}\frac{1}{\sum_i(|\vec{k}_i|+|\vec{q}_i|)}\left(\frac{\delta F^{(1)}_{GL}}{\delta A^a_i(\vec{p})}\right)[\vec{k}_1,\vec{k}_2]\left(\frac{\delta F^{(1)}_{GL}}{\delta A^a_i(-\vec{p})}\right)[\vec{q}_1,\vec{q}_2]  \Bigg|_{\O(\theta^4)}\nn\\
\quad = 2f^{a_1a_2c}f^{b_1b_2c}\int_{\slashed{p},\slashed{k_1},\slashed{k_2},\slashed{q_1},\slashed{q_2}}\frac{\slashed{\delta}\left(\sum_i(\vec{k}_i+\vec{q}_i)\right)}{\sum_i(|\vec{k}_i|+|\vec{q}_i|)} k_1\bar{k}_2 q_1\bar{q}_2 \;\theta^{a_1}(\vec{k}_1)\theta^{a_2}(\vec{k}_2)\theta^{b_1}(\vec{q}_1)\theta^{b_2}(\vec{q}_2) +\O(e)\,.\qquad
\eE

\subsubsection{$\left(\vec{A}\cdot\frac{\delta F^{(1)}_{GL}}{\delta \vec{A}}\right)$-term}

\bE{l}
 -if^{b_1b_2c}\int_{\slashed{p},\slashed{k_1},\slashed{k_2},\slashed{q_1},\slashed{q_2}}\frac{\slashed{\delta}(\vec{q}_1+\vec{q}_2-\vec{p})}{\sum_i(|\vec{k}_i|+|\vec{q}_i|)|\vec{q}_1|}(\vec{q}_1\cdot\vec{A}^{b_1}(\vec{q}_1))\left(\vec{A}^{b_2}(\vec{q}_2)\cdot\frac{\delta F^{(1)}_{GL}}{\delta \vec{A}^c(\vec{p})}[\vec{k}_1,\vec{k}_2]\right) \nn\\
\quad = (-i)^3 f^{b_1b_2c}f^{a_1a_2c}\int_{\slashed{p},\slashed{k_1},\slashed{k_2},\slashed{q_1},\slashed{q_2}}\frac{\slashed{\delta}(\vec{q}_1+\vec{q}_2-\vec{p})\slashed{\delta}(\vec{k}_1+\vec{k}_2+\vec{p})}{(|\vec{k}_1|+|\vec{k}_2|+|\vec{q}_1|+|\vec{q}_2|)} \nn\\
\qquad \left(-|\vec{q}_1|\theta^{b_1}(\vec{q}_1) \right) \Bigg\{2A^{b_2}(\vec{q}_2)2\frac{\bar{p}k_1\bar{k}_2}{|\vec{p}|} -2\bar{A}^{b_2}(\vec{q}_2) 2\frac{p k_1\bar{k}_2}{|\vec{p}|}\Bigg\} \theta^{a_1}(\vec{k}_1)\theta^{a_2}(\vec{k}_2) +\O(e)\\
\quad = (-i)^3 f^{b_1b_2c}f^{a_1a_2c}\int_{\slashed{k_1},\slashed{k_2},\slashed{q_1},\slashed{q_2}} \frac{\slashed{\delta}\left(\sum_i(\vec{k}_i+\vec{q}_i)\right)}{\sum_i(|\vec{k}_i|+|\vec{q}_i|)} \nn\\
\qquad \left(-|\vec{q}_1|\theta^{b_1}(\vec{q}_1) \right) \Bigg\{(\O(J)+2iq_2\theta^{b_2}(\vec{q}_2)) 2\frac{(-\bar{k}_1-\bar{k}_2)k_1\bar{k}_2}{|\vec{k}_1+\vec{k}_2|} \nn\\
\qquad\qquad\qquad -2i\bar{q}_2\theta^{b_2}(\vec{q}_2) 2\frac{(-k_1-k_2) k_1\bar{k}_2}{|\vec{k}_1+\vec{k}_2|}\Bigg\} \theta^{a_1}(\vec{k}_1)\theta^{a_2}(\vec{k}_2)\nn\\
\qquad +\O(J^4)+\ldots+\O(J\theta^3)+\O(e)\,.
\eE
Thus
\bE{l}
 -if^{b_1b_2c}\int_{\slashed{p},\slashed{k_1},\slashed{k_2},\slashed{q_1},\slashed{q_2}}\frac{\slashed{\delta}(\vec{q}_1+\vec{q}_2-\vec{p})}{\sum_i(|\vec{k}_i|+|\vec{q}_i|)|\vec{q}_1|}(\vec{q}_1\cdot\vec{A}^{b_1}(\vec{q}_1))\left(\vec{A}^{b_2}(\vec{q}_2)\cdot\frac{\delta F^{(1)}_{GL}}{\delta \vec{A}^c(\vec{p})}[\vec{k}_1,\vec{k}_2]\right)  \Bigg|_{\O(\theta^4)} \nn\\
\quad = -4 f^{b_1b_2c}f^{a_1a_2c}\int_{\slashed{k_1},\slashed{k_2},\slashed{q_1},\slashed{q_2}} \frac{\slashed{\delta}\left(\sum_i(\vec{k}_i+\vec{q}_i)\right)}{\sum_i(|\vec{k}_i|+|\vec{q}_i|)} k_1\bar{k}_2q_1\bar{q}_2\frac{ |\vec{q}_1|+ |\vec{q}_2|}{|\vec{k}_1+\vec{k}_2|} \theta^{a_1}(\vec{k}_1)\theta^{a_2}(\vec{k}_2)\theta^{b_1}(\vec{q}_1)  \theta^{b_2}(\vec{q}_2)  \nn\\
\qquad\qquad +\O(e)\,.\nn
\eE

\subsubsection{$ (A\times A)(A\times A)$-term}

\bE{l}
\frac{1}{8}f^{a_1a_2c}f^{b_1b_2c}\int_{\slashed{k_1},\slashed{k_2},\slashed{q_1},\slashed{q_2}}\frac{\slashed{\delta}(\sum_i(\vec{k}_i+\vec{q}_i))}{\sum_i(|\vec{k}_i|+|\vec{q}_i|)}(\vec{A}^{a_1}(\vec{k}_1)\times\vec{A}^{a_2}(\vec{k}_2))(\vec{A}^{b_1}(\vec{q}_1)\times\vec{A}^{b_2}(\vec{q}_2))\nn\\
= -2f^{a_1a_2c}f^{b_1b_2c}\int_{\slashed{k_1},\slashed{k_2},\slashed{q_1},\slashed{q_2}}\frac{\slashed{\delta}(\sum_i(\vec{k}_i+\vec{q}_i))}{\sum_i(|\vec{k}_i|+|\vec{q}_i|)} k_1\bar{k_2}q_1\bar{q}_2 \theta^{a_1}(\vec{k}_1)\theta^{a_2}(\vec{k}_2) \theta^{b_1}(\vec{q}_1)\theta^{b_2}(\vec{q}_2) +\O(e)\,.\qquad
\eE

\subsubsection{All three together}

\bE{l}
F^{(2,4)}_{GL}|_{\O(\theta^4)} =  f^{b_1b_2c}f^{a_1a_2c}\int_{\slashed{k_1},\slashed{k_2},\slashed{q_1},\slashed{q_2}} \frac{\slashed{\delta}\left(\sum_i(\vec{k}_i+\vec{q}_i)\right)}{\sum_i(|\vec{k}_i|+|\vec{q}_i|)} k_1\bar{k}_2q_1\bar{q}_2\left(2 -4\frac{ |\vec{q}_1|+ |\vec{q}_2|}{|\vec{k}_1+\vec{k}_2|} -2\right) \nn\\
\qquad\qquad\qquad  \theta^{a_1}(\vec{k}_1)\theta^{a_2}(\vec{k}_2)\theta^{b_1}(\vec{q}_1)  \theta^{b_2}(\vec{q}_2) +\O(e^2)\\
=  -2 f^{a_1a_2c}f^{b_1b_2c}\int_{\slashed{k_1},\slashed{k_2},\slashed{q_1},\slashed{q_2}} \slashed{\delta}\left(\sum_i(\vec{k}_i+\vec{q}_i)\right) \frac{ k_1\bar{k}_2q_1\bar{q}_2}{|\vec{k}_1+\vec{k}_2|} \theta^{a_1}(\vec{k}_1)\theta^{a_2}(\vec{k}_2)\theta^{b_1}(\vec{q}_1)  \theta^{b_2}(\vec{q}_2) +\O(e) \,.\qquad \label{F24theta4}
\eE

\subsection{All orders}
Adding up Eqs.~(\ref{F24theta0}), (\ref{F24theta1}), (\ref{F24theta2}), (\ref{F24theta3}) and  (\ref{F24theta4}) we find
\begin{IEEEeqnarray}{rCl}
\label{F24GLApp}
F^{(2,4)}_{GL} &=& -\frac{1}{512}f^{a_1a_2c}f^{b_1b_2c} \int_{\slashed{k_1},\slashed{k_2},\slashed{q_1},\slashed{q_2}}
\slashed{\delta}\left(\sum^2_i(\vec{k}_i+\vec{q}_i)\right) g^{(4)}(\vec k_1,\vec k_2;\vec q_1,\vec q_2)  \nn\\
&&\qquad\qquad\qquad  J^{a_1}(\vec{k}_1)J^{a_2}(\vec{k}_2)J^{b_1}(\vec{q}_1)J^{b_2}(\vec{q}_2) \nn\\
&& +\frac{1}{32}f^{a_1a_2c}f^{b_1b_2c}\int_{\slashed{k_1},\slashed{k_2},\slashed{q_1},\slashed{q_2}} 
\slashed{\delta}\left(\sum^2_i(\vec{k}_i+\vec{q}_i)\right) g^{(3)}(\vec{k}_1,\vec{k}_2,-\vec{k}_1-\vec{k}_2)  \nn\\
&&\qquad\qquad\qquad  J^{a_1}(\vec{k}_1)J^{a_2}(\vec{k}_2)J^{b_1}(\vec{q}_1)\theta^{b_2}(\vec{q}_2)\nn\\
&& + \frac{1}{2} f^{a_1a_2c}f^{b_1b_2c} \int_{\slashed{k_1},\slashed{k_2},\slashed{q_1},\slashed{q_2}} 
\slashed{\delta}\left(\sum^2_i(\vec{k}_i+\vec{q}_i)\right)   \left( \frac{(\bar{k}_1+\bar{k}_2)^2}{|\vec{k}_1+\vec{k}_2|} -  \frac{\bar{k}_1^2}{|\vec{k}_1|} \right)  \nn\\
&&\qquad\qquad\qquad  J^{a_1}(\vec{k}_1)\theta^{a_2}(\vec{k}_2)J^{b_1}(\vec{q}_1)\theta^{b_2}(\vec{q}_2) \nn\\
&& - f^{a_1a_2c}f^{b_1b_2c} \int_{\slashed{k_1},\slashed{k_2},\slashed{q_1},\slashed{q_2}} \slashed{\delta}
\left(\sum^2_i(\vec{k}_i+\vec{q}_i)\right)    J^{a_1}(\vec{k}_1)J^{a_2}(\vec{k}_2)\theta^{b_1}(\vec{q}_1)\theta^{b_2}(\vec{q}_2) \nn\\
&& \qquad\qquad \times \left(\frac{q_1\bar{q}_2}{\bar{q}_1+\bar{q}_2} \frac{g^{(3)}(\vec{k}_1,\vec{k}_2, -\vec{k}_1-\vec{k}_2)}{16} -   \frac{\bar{q}_2}{\bar{q}_1+\bar{q}_2} \frac{\bar{k}_2^2}{2|\vec{k}_2|}\right) \nn\\
&& + 2 f^{a_1a_2c}f^{b_1b_2c}\int_{\slashed{p},\slashed{k_1},\slashed{k_2},\slashed{q_1},\slashed{q_2}} 
\slashed{\delta}\left(\sum_i(\vec{k}_i+\vec{q}_i)\right) q_1\bar{q}_2 \left(\frac{\bar{k}_1+\bar{k}_2}{|\vec{k}_1+\vec{k}_2|} - \frac{\bar{k}_1}{|\vec{k}_1|} \right)   \nn\\
&&\qquad\qquad\qquad J^{a_1}(\vec{k}_1)\theta^{a_2}(\vec{k}_2) \theta^{b_1}(\vec{q}_1)\theta^{b_2}(\vec{q}_2) 
\nn\\
&&  -2 f^{a_1a_2c}f^{b_1b_2c}\int_{\slashed{k_1},\slashed{k_2},\slashed{q_1},\slashed{q_2}} \slashed{\delta}\left(\sum_i(\vec{k}_i+\vec{q}_i)\right) \frac{ k_1\bar{k}_2q_1\bar{q}_2}{|\vec{k}_1+\vec{k}_2|} \theta^{a_1}(\vec{k}_1)\theta^{a_2}(\vec{k}_2)\theta^{b_1}(\vec{q}_1)  \theta^{b_2}(\vec{q}_2)   \nn\\
&&  +O(e)\,.
\end{IEEEeqnarray}

\section{$F^{(0)}_{GL}[J,\theta]$ at $\O(e^2)$}
\label{sec:B3}

We now compute $F^{(0)}_{GL}[J,\theta]$ at $\O(e^2)$, split it by powers in $\theta$ and bring it into a form similar to Eq.~(\ref{F24GLApp}):

\bE{rCl}
F^{(0)}_{GL} &=& \frac{1}{2}\int_\slashed{k}\frac{1}{|\vec{k}|} (\vec{k}\times\vec{A}^a(\vec{k}))(\vec{k}\times\vec{A}^a(-\vec{k})) \\
&=& -2\int_\slashed{k}\frac{1}{|\vec{k}|} \Bigg(\frac{i}{2}\bar{k}J^a(\vec{k}) -\frac{ie}{2}f^{abc}\int_\slashed{q}\bar{k}\theta^b(\vec{k}-\vec{q})J^c(\vec{q}) \nn\\
&&\qquad\qquad\quad - \bar{k} \frac{ie^2}{4} f^{bcd}f^{dea} \int_\slashed{q}\int_\slashed{p} \theta^b(\vec{k}-\vec{q}-\vec{p}) J^c(\vec{q}) \theta^e(\vec{p})   \nn\\
&&\qquad\qquad\quad +\frac{ie}{2}f^{abc}\int_\slashed{q}(k\bar{q}-\bar{k}q)\,\theta^b(\vec{q})\theta^c(\vec{k}-\vec{q}) \nn\\
&&\qquad\qquad\quad - \frac{ie^2}{6}f^{bcd}f^{dea}\int_\slashed{q}\int_\slashed{p}(k\bar{q}-\bar{k}q) \theta^b(\vec{k}-\vec{q}-\vec{p}) \theta^c(\vec{q}) \theta^e(\vec{p}) \Bigg)  \nn\\ 
&&\qquad \times \Bigg(\frac{i}{2}\bar{k}J^a(-\vec{k}) -\frac{ie}{2}f^{ade}\int_\slashed{p}\bar{k}\theta^d(-\vec{k}-\vec{p})J^e(\vec{p})  \nn\\
&&\qquad\qquad - \bar{k} \frac{ie^2}{4} f^{bcd}f^{dea} \int_\slashed{q}\int_\slashed{p} \theta^b(-\vec{k}-\vec{q}-\vec{p}) J^c(\vec{q}) \theta^e(\vec{p})   \nn\\
&&\qquad\qquad +\frac{ie}{2}f^{abc}\int_\slashed{q}(k\bar{q}-\bar{k}q)\,\theta^b(\vec{q})\theta^c(-\vec{k}-\vec{q}) \nn\\
&&\qquad\qquad - \frac{ie^2}{6}f^{bcd}f^{dea}\int_\slashed{q}\int_\slashed{p}(k\bar{q}-\bar{k}q) \theta^b(-\vec{k}-\vec{q}-\vec{p}) \theta^c(\vec{q}) \theta^e(\vec{p}) \Bigg)  \nn\\
&&\qquad +\O(e^3)\,.
\eE

At $\O(e^2)$ there are no terms with no or with exactly one $\theta$.

\subsection{Order $\theta^2$}
\bE{rCl}
F^{(0)}_{GL}|_{\O(e^2\theta^2)} &=&
\frac{1}{2}f^{abc}f^{ade}\int_{\slashed{k},\slashed{q},\slashed{p}}\frac{\bar{k}^2}{|\vec{k}|}\theta^b(\vec{k}-\vec{q})J^c(\vec{q})
\theta^d(-\vec{k}-\vec{p})J^e(\vec{p}) \\
&& -  \frac{1}{4} f^{bcd}f^{dea} \int_{\slashed{k},\slashed{q},\slashed{p}}\frac{\bar{k}^2}{|\vec{k}|}\Big(J^a(\vec{k}) \theta^b(-\vec{k}-\vec{q}-\vec{p}) + J^a(-\vec{k}) \theta^b(\vec{k}-\vec{q}-\vec{p})\Big) J^c(\vec{q}) \theta^e(\vec{p}) \nn\\
&=& 
 \frac{1}{2}f^{a_1a_2c}f^{b_1b_2e}\int_{\slashed{k_1},\slashed{k_2},\slashed{q_1},\slashed{q_2}} \slashed{\delta}\left(\vec{k}_1+\vec{k}_2+\vec{q}_1+\vec{q}_2\right) \left( \frac{(\bar{k}_1+\bar{k}_2)^2}{|\vec{k}_1+\vec{k}_2|} - \frac{\bar{k}_1^2}{|\vec{k}_1|} \right) \nn\\
&&\qquad\qquad J^{a_1}(\vec{k}_1) \theta^{a_2}(\vec{k}_2)J^{b_1}(\vec{q}_1)\theta^{b_2}(\vec{q}_2)  \,.\label{F0theta2}
\eE

\subsection{Order $\theta^3$}
\bE{rCl} 
F^{(0)}_{GL}|_{\O(e^2\theta^3)} &=&
 -2 \int_\slashed{k}\frac{1}{|\vec{k}|} \Bigg\{\frac{i}{2}\bar{k}J^a(\vec{k}) \Bigg( - \frac{i}{6}f^{bcd}f^{dea}\int_\slashed{q}\int_\slashed{p}(k\bar{q}-\bar{k}q) \theta^b(-\vec{k}-\vec{q}-\vec{p}) \theta^c(\vec{q}) \theta^e(\vec{p}) \Bigg)  \nn\\
&&\qquad -\frac{i}{2}f^{abc}\int_\slashed{q}\bar{k}\theta^b(\vec{k}-\vec{q})J^c(\vec{q}) \Bigg( \frac{i}{2}f^{abc}\int_\slashed{q}(k\bar{q}-\bar{k}q)\,\theta^b(\vec{q})\theta^c(-\vec{k}-\vec{q})     \Bigg) \nn\\
&&\qquad  +\frac{i}{2}f^{abc}\int_\slashed{q}(k\bar{q}-\bar{k}q)\,\theta^b(\vec{q})\theta^c(\vec{k}-\vec{q}) \Bigg(  -\frac{i}{2}f^{ade}\int_\slashed{p}\bar{k}\theta^d(-\vec{k}-\vec{p})J^e(\vec{p})  \Bigg) \nn\\
&&\qquad - \frac{i}{6}f^{bcd}f^{dea}\int_\slashed{q}\int_\slashed{p}(k\bar{q}-\bar{k}q) \theta^b(\vec{k}-\vec{q}-\vec{p}) \theta^c(\vec{q}) \theta^e(\vec{p}) \Bigg(\frac{i}{2}\bar{k}J^a(-\vec{k}) \Bigg) \Bigg\} \\
&=&  -\frac{1}{2}\int_{\slashed{k},\slashed{q},\slashed{p}} \Bigg\{\frac{1}{3}f^{bcd}f^{dea}\frac{1}{|\vec{k}|}\bar{k} (k\bar{q}-\bar{k}q) J^a(\vec{k}) \theta^b(-\vec{k}-\vec{q}-\vec{p}) \theta^c(\vec{q}) \theta^e(\vec{p})   \nn\\
&&\qquad +f^{abc}f^{ade}\frac{1}{|\vec{k}|}\bar{k} (k\bar{p}-\bar{k}p)\,\theta^b(\vec{k}-\vec{q})J^c(\vec{q})\theta^d(\vec{p})\theta^e(-\vec{k}-\vec{p})    \nn\\
&&\qquad  +f^{abc}f^{ade}\frac{1}{|\vec{k}|}(k\bar{q}-\bar{k}q)\bar{k} \,\theta^b(\vec{q})\theta^c(\vec{k}-\vec{q}) \theta^d(-\vec{k}-\vec{p})J^e(\vec{p})  \nn\\
&&\qquad + \frac{1}{3}f^{bcd}f^{dea}\frac{1}{|\vec{k}|}(k\bar{q}-\bar{k}q) \bar{k} \, \theta^b(\vec{k}-\vec{q}-\vec{p}) \theta^c(\vec{q}) \theta^e(\vec{p}) J^a(-\vec{k})  \Bigg\} \\
&=&  f^{a_1a_2c}f^{b_1b_2c}\int_{\slashed{k}_1,\slashed{k}_2,\slashed{q}_1,\slashed{q}_2}\slashed{\delta}\left(\sum_i(\vec{k}_i+\vec{q}_i)\right)  J^{a_1}(\vec{k}_1)\theta^{a_2}(\vec{k}_2) \theta^{b_1}(\vec{q}_1) \theta^{b_2}(\vec{q}_2)    \nn\\
&&\qquad \Bigg\{ \frac{1}{3} \frac{1}{|\vec{k}_1|} \bar{k}_1(k_1\bar{q}_2-\bar{k}_1q_2)  +\frac{1}{|\vec{q}_1+\vec{q}_2|}(\bar{q}_1+\bar{q}_2) (q_2\bar{q}_1-\bar{q}_2q_1)\, \Bigg\} \,.\label{F0theta3}
\eE

\subsection{Order $\theta^4$}

\bE{rCl}
F^{(0)}_{GL}|_{\O(e^2\theta^3)} &=& \frac{1}{2}f^{abc}f^{ade}\int_{\slashed{k},\slashed{p},\slashed{q}} \frac{1}{|\vec{k}|} (k\bar{q}-\bar{k}q)  (k\bar{p}-\bar{k}p)\,\theta^b(\vec{q})\theta^c(\vec{k}-\vec{q}) \theta^d(\vec{p})\theta^e(-\vec{k}-\vec{p})   \\ 
&=& \frac{1}{2}f^{abc}f^{ade}\int_{\slashed{k_1},\slashed{k_2},\slashed{q_1},\slashed{q_2}} \slashed{\delta}\left(\sum_i(\vec{k}_i+\vec{q}_i)\right) \theta^b(\vec{k}_1)\theta^c(\vec{k}_2) \theta^d(\vec{q}_1)\theta^e(\vec{q}_2)   \nn\\ 
&&  \frac{1}{|\vec{k}_1+\vec{k}_2|} (k_2\bar{k_1}-\bar{k_2}k_1)  ((k_2+k_1)\bar{q_1}-(\bar{k_2}+\bar{k_1})q_1) \\
&=&  -2f^{a_1a_2c}f^{b_1b_2c}\int_{\slashed{k_1},\slashed{k_2},\slashed{q_1},\slashed{q_2}} \slashed{\delta}\left(\sum_i(\vec{k}_i+\vec{q}_i)\right)   \frac{\bar{k_2}k_1\bar{q_2}q_1}{|\vec{k}_1+\vec{k}_2|}  \theta^{a_1}(\vec{k}_1)\theta^{a_2}(\vec{k}_2)\theta^{b_1}(\vec{q}_1)  \theta^{b_2}(\vec{q}_2)   \,.\nn\\ \label{F0theta4}
\eE

\subsection{All orders}

Summing the results of the previous subsections, we find

\begin{IEEEeqnarray}{rCl}
F^{(0)}_{GL}|_{\O(e^2)} &=&  \frac{1}{2}f^{a_1a_2c}f^{b_1b_2e}\int_{\slashed{k_1},\slashed{k_2},\slashed{q_1},\slashed{q_2}} 
\slashed{\delta}\left(\sum^2_i(\vec{k}_i+\vec{q}_i)\right) \left( \frac{(\bar{k}_1+\bar{k}_2)^2}{|\vec{k}_1+\vec{k}_2|} - \frac{\bar{k}_1^2}{|\vec{k}_1|} \right) \label{F0GLApp} \\
&&\qquad\qquad J^{a_1}(\vec{k}_1) \theta^{a_2}(\vec{k}_2)J^{b_1}(\vec{q}_1)\theta^{b_2}(\vec{q}_2) 
\nn\\
&&+ f^{a_1a_2c}f^{b_1b_2c}\int_{\slashed{k}_1,\slashed{k}_2,\slashed{q}_1,\slashed{q}_2}\slashed{\delta}\left(\sum_i(\vec{k}_i+\vec{q}_i)\right)  J^{a_1}(\vec{k}_1)\theta^{a_2}(\vec{k}_2) \theta^{b_1}(\vec{q}_1) \theta^{b_2}(\vec{q}_2)    \nn\\
&&\qquad\qquad \times \Bigg( \frac{1}{3} \frac{1}{|\vec{k}_1|} \bar{k}_1(k_1\bar{q}_2-\bar{k}_1q_2)  +\frac{1}{|\vec{q}_1+\vec{q}_2|}(\bar{q}_1+\bar{q}_2) (q_2\bar{q}_1-\bar{q}_2q_1)\, \Bigg)
 \nn\\
&& -2f^{a_1a_2c}f^{b_1b_2c}\int_{\slashed{k_1},\slashed{k_2},\slashed{q_1},\slashed{q_2}} 
\slashed{\delta}\left(\sum_i(\vec{k}_i+\vec{q}_i)\right)    \frac{\bar{k_2}k_1\bar{q_2}q_1}{|\vec{k}_1+\vec{k}_2|} \theta^{a_1}(\vec{k}_1)\theta^{a_2}(\vec{k}_2)\theta^{b_1}(\vec{q}_1)  \theta^{b_2}(\vec{q}_2)  \,.  \nn
\end{IEEEeqnarray}

\section{$F^{(1)}_{GL}[J,\theta]$ at $\O(e)$}
\label{sec:B4}
 
Finally, we compute $F^{(1)}_{GL}[J,\theta]$ at $\O(e)$, split it by powers in $\theta$ and bring it in a form which makes it possible to see the cancellation with the corresponding terms of $F^{(2,4)}_{GL}|  + F^{(0)}_{GL}|_{\O(e^2)}$.

\bE{l}
F^{(1)}_{GL} \nn\\
=  if^{abc} \int_{\slashed{k_1},\slashed{k_2},\slashed{k_3}}\slashed{\delta}\left(\sum_{i=1}^3 \vec{k}_i\right) \Bigg\{ \frac{(\vec{k}_1\times\vec{A}^a(\vec{k}_1)) (\vec{A}^b(\vec{k}_2)\times\vec{A}^c(\vec{k}_3))} {2(|\vec{k}_1|+|\vec{k}_2|+|\vec{k}_3|)} \nn\\
\qquad\qquad - \frac{(\vec{k}_1\cdot\vec{A}^a(\vec{k}_1)) (\vec{k}_3\times\vec{A}^b(\vec{k}_2)) (\vec{k}_3\times\vec{A}^c(\vec{k}_3))}{|\vec{k}_1||\vec{k}_3|(|\vec{k}_1|+|\vec{k}_2|+|\vec{k}_3|)}  \Bigg\} \\
=  -2if^{abc} \int_{\slashed{k_1},\slashed{k_2},\slashed{k_3}}\slashed{\delta}\left(\sum_{i=1}^3 \vec{k}_i\right) \frac{1}{\sum_{i=1}^3 |\vec{k}_i|} \Bigg\{ \left(k_1\bar{A}^a(\vec{k}_1)-\bar{k_1}A^a(\vec{k}_1)\right)  \nn\\
\qquad\qquad \times \left(A^b(\vec{k}_2)\bar{A}^c(\vec{k}_3)-\bar{A}^b(\vec{k}_2){A}^c(\vec{k}_3)\right)  \nn\\
\quad - \frac{4}{|\vec{k}_1||\vec{k}_3|} \left(k_1\bar{A}^a(\vec{k}_1)+\bar{k_1}A^a(\vec{k}_1)\right) \left(k_3\bar{A}^a(\vec{k}_2)-\bar{k_3}A^a(\vec{k}_2)\right) \left(k_3\bar{A}^a(\vec{k}_3)-\bar{k_3}A^a(\vec{k}_3)\right)  \Bigg\} \\
= -2if^{abc} \int_{\slashed{k_1},\slashed{k_2},\slashed{k_3}}\slashed{\delta}\left(\sum_{i=1}^3 \vec{k}_i\right) \frac{1}{\sum_{i=1}^3 |\vec{k}_i|} \Bigg\{ \Bigg(\frac{i}{2}\bar{k_1}J^a(\vec{k_1}) -\frac{ie}{2}\bar{k_1}f^{aa_1a_2}\int_\slashed{q}\theta^{a_1}(\vec{k}_1-\vec{q})J^{a_2}(\vec{q}) \nn\\
\qquad\qquad\qquad +\frac{ie}{2} f^{ac_1c_2}\int_\slashed{q}(k_1\bar{q}-\bar{k}_1q)\,\theta^{c_1}(\vec{q})\theta^{c_2}(\vec{k}_1-\vec{q})\Bigg) \nn\\
\qquad\quad \Bigg(\bar{k_3}J^b(\vec{k}_2) \theta^c(\vec{k}_3)   - 2k_2\bar{k}_3\theta^b(\vec{k}_2)\theta^c(\vec{k}_3) \nn\\
\qquad\qquad +  \frac{e}{2} J^b(\vec{k}_2) f^{cde}\int_\slashed{q}\bar{q}\,\theta^d(\vec{q})\theta^e(\vec{k}_3-\vec{q})  -e\bar{k}_3 f^{bde}\int_\slashed{q}\theta^d(\vec{k}_2-\vec{q})J^e(\vec{q}) \theta^c(\vec{k}_3) \nn\\
\qquad\qquad + e \theta^b(\vec{k}_2)f^{cde}\int_\slashed{q}(\bar{k_2}q-k_2\bar{q})\,\theta^d(\vec{q})\theta^e(\vec{k}_3-\vec{q})\Bigg)  \nn\\
\quad - \frac{4}{|\vec{k}_1||\vec{k}_3|} \Bigg(2ik_1\bar{k_1}\theta^a(\vec{k}_1)-\frac{i}{2}\bar{k_1}J^a(\vec{k}_1)  \nn\\
\qquad\qquad +\frac{ie}{2}f^{ade}\int_\slashed{q}\bar{k_1}\theta^d(\vec{k_1}-\vec{q})J^e(\vec{q})+ \frac{ie}{2}f^{ade}\int_\slashed{q}(k_1\bar{q}+\bar{k}_1q)\,\theta^d(\vec{q})\theta^e(\vec{k}_1-\vec{q}) \Bigg) \nn\\ 
\qquad\quad\Bigg(i(k_3\bar{k_2}-\bar{k_3}k_2)\theta^b(\vec{k}_2)+\frac{i}{2}\bar{k_3}J^b(\vec{k}_2)   \nn\\
\qquad\qquad  -\frac{i}{2}f^{bde}\int_\slashed{q}\bar{k_3}\theta^d(\vec{k_2}-\vec{q})J^e(\vec{q})  +\frac{ie}{2} f^{bb_1b_2}\int_\slashed{q}(k_3\bar{q}-\bar{k}_3q)\,\theta^{b_1}(\vec{q})\theta^{b_2}(\vec{k}_2-\vec{q})\Bigg) \nn\\
\qquad\quad\Bigg(\frac{i}{2}\bar{k_3}J^c(\vec{k}_3) -\frac{i}{2}f^{cde}\int_\slashed{q}\bar{k_3}\theta^d(\vec{k_3}-\vec{q})J^e(\vec{q}) +\frac{ie}{2} f^{cc_1c_2}\int_\slashed{q}(k_3\bar{q}-\bar{k}_3q)\,\theta^{c_1}(\vec{q})\theta^{c_2}(\vec{k}_3-\vec{q})\Bigg)\Bigg\} \,.\nn\\ \label{F1GLref}
\eE

At $\O(e)$, there are no terms without $\theta$-dependence.

\subsection{Order $\theta$}

We need this term to cancel Eq.~(\ref{F24theta1}):
\bE{l}
F^{(2,4)}_{GL}|_{\O(\theta)}=\frac{f^{a_1a_2c}f^{b_1b_2c}}{32}\int_{\slashed{k_1},\slashed{k_2},\slashed{q_1},\slashed{q_2}} \slashed{\delta}\left(\sum_i(\vec{k}_i+\vec{q}_i)\right)g^{(3)}(\vec{k}_1,\vec{k}_2,-\vec{k}_1-\vec{k}_2) \nn\\
\qquad\qquad\qquad J^{a_1}(\vec{k}_1)J^{a_2}(\vec{k}_2)J^{b_1}(\vec{q}_1)\theta^{b_2}(\vec{q}_2)\,. \qquad 
\eE

We extract the $\O(e\theta)$ portion from Eq.~(\ref{F1GLref}) and bring it in the above form:

\bE{rCl}
F^{(1)}_{GL}|_{\O(e\theta)} &=& -\int_{\slashed{k_1},\slashed{k_2},\slashed{k_3}}\slashed{\delta}\left(\sum_{i=1}^3 \vec{k}_i\right)  f^{abc} J^a(\vec{k}_1)J^c(\vec{k}_3)  
\Bigg\{\frac{\bar{k}_3^2}{|\vec{k}_3|} \theta^b(\vec{k}_2)  \nn\\
&&\qquad\quad 
-\left(\frac{\bar{k}_1\bar{k}_3^2}{|\vec{k}_1||\vec{k}_2||\vec{k}_3|}  - \frac{\bar{k}_1^2 +\bar{k}_1\bar{k}_3 +\bar{k}_3^2}{|\vec{k}_2|\sum_{i=1}^3 |\vec{k}_i|} \left(\frac{\bar{k}_1}{|\vec{k}_1|}-\frac{\bar{k}_3}{|\vec{k}_3|} \right) \right) f^{bde}\int_\slashed{q} \theta^d(\vec{k_2}-\vec{q})J^e(\vec{q}) \Bigg\} \nn\\
&=&  f^{a_1a_2c}f^{b_1b_2c}\int_{\slashed{k_1},\slashed{k_2},\slashed{q_1},\slashed{q_2}}\slashed{\delta}(\vec{k}_1+\vec{k}_2+\vec{q}_1+\vec{q}_2)  J^{a_1}(\vec{k}_1)J^{a_2}(\vec{k}_2) \theta^{b_2}(\vec{q}_2)J^{b_1}(\vec{q}_1) \nn\\
&&\Bigg\{\frac{\bar{k}_1\bar{k}_2^2}{|\vec{k}_1||\vec{q}_1+\vec{q}_2||\vec{k}_2|}  - \frac{\bar{k}_1^2 +\bar{k}_1\bar{k}_2 +\bar{k}_2^2}{|\vec{q}_1+\vec{q}_2|(|\vec{k}_1|+|\vec{q}_1+\vec{q}_2|+|\vec{k}_2|)} \left(\frac{\bar{k}_1}{|\vec{k}_1|}-\frac{\bar{k}_2}{|\vec{k}_2|}\right) \Bigg\}\,,
\eE
where in the second equality we renamed: $q\rightarrow q_1$, $k_2\rightarrow q_1+q_2\;$; $k_3\rightarrow k_2$.
\bE{rCl}
 &=&  f^{a_1a_2c}f^{b_1b_2c}\int_{\slashed{k_1},\slashed{k_2},\slashed{q_1},\slashed{q_2}}\slashed{\delta}\left(\sum(\vec{k}_i+\vec{q}_i)\right)  J^{a_1}(\vec{k}_1)J^{a_2}(\vec{k}_2) J^{b_1}(\vec{q}_1) \theta^{b_2}(\vec{q}_2) \frac{1}{|\vec{k}_1|+|\vec{k}_2|+|\vec{k}_1+\vec{k}_2|}\nn\\
&& \Bigg\{\frac{\bar{k}_1\bar{k}_2^2}{|\vec{k}_1+\vec{k}_2||\vec{k}_2|} +\frac{\bar{k}_1\bar{k}_2^2}{|\vec{k}_1||\vec{k}_1+\vec{k}_2|} +\frac{\bar{k}_1\bar{k}_2^2}{|\vec{k}_1||\vec{k}_2|}  - \frac{\bar{k}_1^3 +\bar{k}_1^2\bar{k}_2 +\bar{k}_1\bar{k}_2^2}{|\vec{k}_1+\vec{k}_2||\vec{k}_1|} + \frac{\bar{k}_1^2\bar{k}_2 +\bar{k}_1\bar{k}_2^2 +\bar{k}_2^3}{|\vec{k}_1+\vec{k}_2||\vec{k}_2|} \Bigg\} \\
&=&  f^{a_1a_2c}f^{b_1b_2c}\int_{\slashed{k_1},\slashed{k_2},\slashed{q_1},\slashed{q_2}}\slashed{\delta}\left(\sum(\vec{k}_i+\vec{q}_i)\right)  J^{a_1}(\vec{k}_1)J^{a_2}(\vec{k}_2) J^{b_1}(\vec{q}_1) \theta^{b_2}(\vec{q}_2) \nn\\
&&\quad\frac{1}{|\vec{k}_1|+|\vec{k}_2|+|\vec{k}_1+\vec{k}_2|} \Bigg\{\frac{\bar{k}_1\bar{k}_2^2}{|\vec{k}_1||\vec{k}_2|}  - \frac{\bar{k}_1^3 +\bar{k}_1^2\bar{k}_2}{|\vec{k}_1+\vec{k}_2||\vec{k}_1|} + \frac{\bar{k}_1^2\bar{k}_2 +2\bar{k}_1\bar{k}_2^2 +\bar{k}_2^3}{|\vec{k}_1+\vec{k}_2||\vec{k}_2|} \Bigg\}  \\
&=&  f^{a_1a_2c}f^{b_1b_2c}\int_{\slashed{k_1},\slashed{k_2},\slashed{q_1},\slashed{q_2}}\slashed{\delta}\left(\sum(\vec{k}_i+\vec{q}_i)\right)  J^{a_1}(\vec{k}_1)J^{a_2}(\vec{k}_2) J^{b_1}(\vec{q}_1) \theta^{b_2}(\vec{q}_2) \nn\\
&&\quad\frac{1}{|\vec{k}_1|+|\vec{k}_2|+|\vec{k}_1+\vec{k}_2|} \Bigg\{\frac{\bar{k}_1\bar{k}_2^2}{|\vec{k}_1||\vec{k}_2|}  + \frac{(-\bar{k}_1 -\bar{k}_2)\bar{k}_1^2}{|\vec{k}_1+\vec{k}_2||\vec{k}_1|} + \frac{\bar{k}_2(-\bar{k}_1-\bar{k}_2)^2}{|\vec{k}_1+\vec{k}_2||\vec{k}_2|} \Bigg\} \\
&=& -f^{a_1a_2c}f^{b_1b_2c}\int_{\slashed{k_1},\slashed{k_2},\slashed{q_1},\slashed{q_2}}\slashed{\delta}\left(\sum(\vec{k}_i+\vec{q}_i)\right) \frac{g^{(3)}(\vec{k}_1,\vec{k}_2,-\vec{k}_1-\vec{k}_2)}{32} J^{a_1}(\vec{k}_1)J^{a_2}(\vec{k}_2) J^{b_1}(\vec{q}_1) \theta^{b_2}(\vec{q}_2) \label{F1theta1}\nn\\
&=& -F^{(2,4)}_{GL}|_{\O(\theta)} \,.\label{F1=F24theta1}
\eE

\subsection{Order $\theta^2$}
We want this term to cancel Eqs.~(\ref{F24theta2}) and (\ref{F0theta2}).
\bE{l}
F^{(2,4)}_{GL}|_{\O(\theta^2)}  + F^{(0)}_{GL}|_{\O(e^2\theta^2)}\nn\\
= \frac{1}{2} f^{a_1a_2c}f^{b_1b_2c} \int_{\slashed{k_1},\slashed{k_2},\slashed{q_1},\slashed{q_2}} \slashed{\delta}\left(\sum(\vec{k}_i+\vec{q}_i)\right)   \left( \frac{(\bar{k}_1+\bar{k}_2)^2}{|\vec{k}_1+\vec{k}_2|} -  \frac{\bar{k}_1^2}{|\vec{k}_1|} \right) J^{a_1}(\vec{k}_1)\theta^{a_2}(\vec{k}_2)J^{b_1}(\vec{q}_1)\theta^{b_2}(\vec{q}_2) \nn\\
\quad - f^{a_1a_2c}f^{b_1b_2c} \int_{\slashed{k_1},\slashed{k_2},\slashed{q_1},\slashed{q_2}} \slashed{\delta}\left(\sum(\vec{k}_i+\vec{q}_i)\right)    J^{a_1}(\vec{k}_1)J^{a_2}(\vec{k}_2)\theta^{b_1}(\vec{q}_1)\theta^{b_2}(\vec{q}_2) \nn\\
\qquad\qquad \left\{\frac{q_1\bar{q}_2}{\bar{q}_1+\bar{q}_2} \frac{g^{(3)}(\vec{k}_1,\vec{k}_2, -\vec{k}_1-\vec{k}_2)}{16} -   \frac{\bar{q}_2}{\bar{q}_1+\bar{q}_2} \frac{\bar{k}_2^2}{2|\vec{k}_2|}\right\} \nn\\
 +\frac{1}{2}e^2f^{a_1a_2c}f^{b_1b_2e}\int_{\slashed{k_1},\slashed{k_2},\slashed{q_1},\slashed{q_2}} \slashed{\delta}\left(\sum(\vec{k}_i+\vec{q}_i)\right)  \left( \frac{(\bar{k}_1+\bar{k}_2)^2}{|\vec{k}_1+\vec{k}_2|} - \frac{\bar{k}_1^2}{|\vec{k}_1|} \right) \nn\\
\qquad\qquad J^{a_1}(\vec{k}_1) \theta^{a_2}(\vec{k}_2)J^{b_1}(\vec{q}_1)\theta^{b_2}(\vec{q}_2)  \\
= f^{a_1a_2c}f^{b_1b_2c} \int_{\slashed{k_1},\slashed{k_2},\slashed{q_1},\slashed{q_2}} \slashed{\delta}\left(\sum(\vec{k}_i+\vec{q}_i)\right)   \left( \frac{(\bar{k}_1+\bar{k}_2)^2}{|\vec{k}_1+\vec{k}_2|} -  \frac{\bar{k}_1^2}{|\vec{k}_1|} \right) J^{a_1}(\vec{k}_1)\theta^{a_2}(\vec{k}_2)J^{b_1}(\vec{q}_1)\theta^{b_2}(\vec{q}_2) \nn\\
\quad - f^{a_1a_2c}f^{b_1b_2c} \int_{\slashed{k_1},\slashed{k_2},\slashed{q_1},\slashed{q_2}} \slashed{\delta}\left(\sum(\vec{k}_i+\vec{q}_i)\right)    J^{a_1}(\vec{k}_1)J^{a_2}(\vec{k}_2)\theta^{b_1}(\vec{q}_1)\theta^{b_2}(\vec{q}_2) \nn\\
\qquad\qquad \left\{\frac{q_1\bar{q}_2}{\bar{q}_1+\bar{q}_2} \frac{g^{(3)}(\vec{k}_1,\vec{k}_2, -\vec{k}_1-\vec{k}_2)}{16} -   \frac{\bar{q}_2}{\bar{q}_1+\bar{q}_2} \frac{\bar{k}_2^2}{2|\vec{k}_2|}\right\} \,. \label{F24+1}
\eE

We extract the $\O(e\theta^2)$ portion from Eq.~(\ref{F1GLref}) and bring it in the above form:

\bE{l}
F^{(1)}_{GL}|_{\O(e\theta^2)} \nn\\
=  \frac{1}{2} f^{abc}  \int_{\slashed{k_1},\slashed{k_2},\slashed{k_3},\slashed{q}} \frac{\slashed{\delta}\left(\sum_{i=1}^3 \vec{k}_i\right)}{\sum_{i=1}^3 |\vec{k}_i|} \bar{k_1}\bar{q} J^a(\vec{k_1}) J^b(\vec{k}_2) f^{cde}\theta^d(\vec{q})  \theta^e(\vec{k}_3-\vec{q}) \nn\\
\quad - f^{abc}  \int_{\slashed{k_1},\slashed{k_2},\slashed{k_3},\slashed{q}} \frac{\slashed{\delta}\left(\sum_{i=1}^3 \vec{k}_i\right)}{\sum_{i=1}^3 |\vec{k}_i|} \bar{k_1}\bar{k}_3 J^a(\vec{k_1})  f^{bde}\theta^d(\vec{k}_2-\vec{q}) J^e(\vec{q})  \theta^c(\vec{k}_3) \nn\\
\quad -f^{abc}f^{aa_1a_2} \int_{\slashed{k_1},\slashed{k_2},\slashed{k_3},\slashed{q}}\slashed{\delta}\left(\sum_{i=1}^3 \vec{k}_i+\vec{q}\right) \frac{1}{|\vec{k}_1+\vec{q}|+|\vec{k}_2|+|\vec{k}_3|} (\bar{k_1}+\bar{q})\bar{k}_3\theta^{a_1}(\vec{k}_1) J^{a_2}(\vec{q}) J^b(\vec{k}_2)\theta^c(\vec{k}_3)\nn\\
\quad +8 f^{abc}  \int_{\slashed{k_1},\slashed{k_2},\slashed{k_3},\slashed{q}} \frac{\slashed{\delta}\left(\sum_{i=1}^3 \vec{k}_i\right)}{\sum_{i=1}^3 |\vec{k}_i||\vec{k}_1||\vec{k}_3|} \nn\\
\quad \Bigg\{\frac{2}{4}|\vec{k}_1|^2\theta^a(\vec{k}_1)\frac{-\bar{k}_3^2}{4} \left(J^b(\vec{k}_2)f^{cde}\theta^d(\vec{k}_3-\vec{q})+J^c(\vec{k}_3)f^{bde}\theta^d(\vec{k}_2-\vec{q})\right) J^e(\vec{q}) \nn\\
\qquad -\frac{1}{2}\bar{k}_1J^a(\vec{k}_1)(k_3\bar{k}_2-\bar{k}_3k_2)\theta^b(\vec{k}_2)\frac{-\bar{k}_3}{2} f^{cde}\theta^d(\vec{k}_3-\vec{q})J^e(\vec{q}) \nn\\
\qquad -\frac{1}{2}\bar{k}_1J^a(\vec{k}_1)\frac{\bar{k}_3}{4}(k_3\bar{q}-\bar{k}_3q) \left(J^b(\vec{k}_2)f^{cc_1c_2}\theta^{c_1}(\vec{q})\theta^{c_2}(\vec{k}_3-\vec{q}) +J^c(\vec{k}_3)f^{bb_1b_2}\theta^{b_1}(\vec{q})\theta^{b_2}(\vec{k}_2-\vec{q})\right) \nn\\
\qquad +\frac{1}{2}\bar{k}_1f^{ade}\theta^d(\vec{k}_1-\vec{q})J^e(\vec{q})(k_3\bar{k}_2-\bar{k}_3k_2)\theta^b(\vec{k}_2)\frac{\bar{k}_3}{2}J^c(\vec{k}_3) \nn\\
\qquad + \frac{1}{2}f^{ade}(k_1\bar{q}+\bar{k}_1q)\,\theta^d(\vec{q})\theta^e(\vec{k}_1-\vec{q}) \frac{\bar{k}_3^2}{4}J^b(\vec{k}_2)J^c(\vec{k}_3)
\Bigg\} \,.
\eE
In the last five lines we shift $k_3 \rightarrow k_3+q$, except for the two terms that are $\theta(k_2-q)$, there $k_2 \rightarrow k_2+q$, and the very last line, there $k_1 \rightarrow k_1+q$.
\bE{l}
=  - f^{a_1a_2c}f^{b_1b_2c}  \int_{\slashed{k_1},\slashed{k_2},\slashed{k_3},\slashed{q}}  \slashed{\delta}\left(\sum_{i=1}^2 (\vec{k}_i+\vec{q}_i)\right) \frac{\bar{k_1}\bar{k}_2}{|\vec{k}_1|+|\vec{k}_2|+|\vec{q}_1+\vec{q}_2|} J^{a_1}(\vec{k}_1) \theta^{a_2}(\vec{k}_2) J^{b_1}(\vec{q}_1)\theta^{b_2}(\vec{q}_2) \nn\\
\quad +f^{a_1a_2c}f^{b_1b_2c} \int_{\slashed{k_1},\slashed{k_2},\slashed{q_1},\slashed{q_2}} \slashed{\delta}\left(\sum_{i=1}^2 (\vec{k}_i+\vec{q}_i)\right) \frac{(\bar{k_1}+\bar{k}_2)\bar{q}_2}{|\vec{k}_1+\vec{k}_2|+|\vec{q}_1|+|\vec{q}_2|}  J^{a_1}(\vec{k}_1) \theta^{a_2}(\vec{k}_2) J^{b_1}(\vec{q}_1)\theta^{b_2}(\vec{q}_2)\nn\\
\quad + f^{abc}  \int_{\slashed{k_1},\slashed{k_2},\slashed{k_3},\slashed{q}} \frac{\slashed{\delta}\left(\vec{k}_1+\vec{k}_2+\vec{k}_3+\vec{q}\right)}{(|\vec{k}_1|+|\vec{k}_2|+|\vec{k}_3+\vec{q}|)|\vec{k}_1||\vec{k}_3+\vec{q}|} \nn\\
\quad \times \Bigg\{-|\vec{k}_1|^2\theta^a(\vec{k}_1) (\bar{k}_3+\bar{q})^2 J^b(\vec{k}_2)f^{cde}\theta^d(\vec{k}_3) J^e(\vec{q}) \nn\\
\qquad +2\bar{k}_1J^a(\vec{k}_1)((k_3+q)\bar{k}_2-(\bar{k}_3+\bar{q})k_2)\theta^b(\vec{k}_2)(\bar{k}_3+\bar{q}) f^{cde}\theta^d(\vec{k}_3)J^e(\vec{q}) \nn\\
\qquad -\bar{k}_1J^a(\vec{k}_1)(\bar{k}_3+\bar{q})(k_3\bar{q}-\bar{k}_3q) J^b(\vec{k}_2)f^{cc_1c_2}\theta^{c_1}(\vec{q})\theta^{c_2}(\vec{k}_3) \nn\\
\qquad +2(\bar{k}_3+\bar{q})f^{ade}\theta^d(\vec{k}_3)J^e(\vec{q})(k_1\bar{k}_2-\bar{k}_1k_2)\theta^b(\vec{k}_2)\bar{k}_1J^c(\vec{k}_1)
\Bigg\}\nn\\
\quad - f^{abc}  \int_{\slashed{k_1},\slashed{k_2},\slashed{k_3},\slashed{q}} \frac{\slashed{\delta}\left(\vec{k}_1+\vec{k}_2+\vec{q}+\vec{k}_3\right)}{(|\vec{k}_1|+|\vec{k}_2+\vec{q}|+|\vec{k}_3|)|\vec{k}_1||\vec{k}_3|} \Bigg\{ |\vec{k}_1|^2\theta^a(\vec{k}_1)\bar{k}_3^2  J^c(\vec{k}_3)f^{bde}\theta^d(\vec{k}_2) J^e(\vec{q}) \nn\\
\qquad +\bar{k}_1J^a(\vec{k}_1)\bar{k}_3(k_3\bar{q}-\bar{k}_3q) J^c(\vec{k}_3)f^{bb_1b_2}\theta^{b_1}(\vec{q})\theta^{b_2}(\vec{k}_2) \Bigg\}\nn\\
\quad + f^{abc}  \int_{\slashed{k_1},\slashed{k_2},\slashed{k_3},\slashed{q}} \frac{\slashed{\delta}\left(\vec{k}_1+\vec{q}+\vec{k}_2+\vec{k}_3\right)}{(|\vec{k}_1+\vec{q}|+|\vec{k}_2|+|\vec{k}_3|)|\vec{k}_1+\vec{q}||\vec{k}_3|} f^{ade}(k_1\bar{q}+\bar{k}_1q+2q\bar{q})\bar{k}_3^2\nn\\
 \qquad\qquad\qquad \theta^d(\vec{q})\theta^e(\vec{k}_1) J^b(\vec{k}_2)J^c(\vec{k}_3) \,. 
\eE
Interchanging $\vec k_1\leftrightarrow \vec q$ and $d \leftrightarrow e$ in the last line eliminates the $k_1\bar{q}+\bar{k}_1q$ part:
\bE{l}
= f^{a_1a_2c}f^{b_1b_2c} \int_{\slashed{k_1},\slashed{k_2},\slashed{q_1},\slashed{q_2}} \slashed{\delta}\left(\sum_{i=1}^2 (\vec{k}_i+\vec{q}_i)\right) \frac{-\bar{k_1}\bar{k}_2 +(\bar{q_1}+\bar{q}_2)\bar{k}_2}{|\vec{q}_1+\vec{q}_2|+|\vec{k}_1|+|\vec{k}_2|}  J^{a_1}(\vec{k}_1) \theta^{a_2}(\vec{k}_2) J^{b_1}(\vec{q}_1)\theta^{b_2}(\vec{q}_2)\nn\\
\quad +f^{a_1a_2c}f^{b_1b_2c} \int_{\slashed{k_1},\slashed{k_2},\slashed{q_1},\slashed{q_2}}  \frac{\slashed{\delta}\left(\sum_{i=1}^2 (\vec{k}_i+\vec{q}_i)\right)}{|\vec{k}_1|+|\vec{k}_2|+|\vec{q}_1+\vec{q}_2|} 
\Bigg(\frac{1}{2}\bar{k_1}\bar{q}_1 -\frac{\bar{k}_1(\bar{q}_1+\bar{q}_2)(q_2\bar{q}_1-\bar{q}_2q_1)}{|\vec{k}_1||\vec{q}_1+\vec{q}_2|} \nn\\ 
\qquad\qquad\qquad  + \frac{\bar{k}_1\bar{k}_2(k_2\bar{q}_1-\bar{k}_2q_1)}{|\vec{k}_1||\vec{k}_2|} +\frac{1}{2}\frac{\bar{k}_2^2|\vec{q}_1|^2}{|\vec{k}_2||\vec{q}_1+\vec{q}_2|}\Bigg)  J^{a_1}(\vec{k_1})  J^{a_2}(\vec{k}_2) \theta^{b_1}(\vec{q}_1) \theta^{b_2}(\vec{q}_2) \nn\\
\quad + f^{a_1a_2c}f^{b_1b_2c}  \int_{\slashed{k_1},\slashed{k_2},\slashed{k_3},\slashed{q}}  \frac{\slashed{\delta}\left(\sum_{i=1}^2 (\vec{k}_i+\vec{q}_i)\right)}{|\vec{k}_1|+|\vec{k}_2|+|\vec{q}_1+\vec{q}_2|}  J^{a_1}(\vec{k}_1) \theta^{a_2}(\vec{k}_2) J^{b_1}(\vec{q}_1)\theta^{b_2}(\vec{q}_2) \Bigg\{-\frac{|\vec{k}_2|^2(\bar{q}_1+\bar{q}_2)^2}{|\vec{k}_2||\vec{q}_1+\vec{q}_2|} \nn\\
\qquad -\frac{2\bar{k}_1((q_1+q_2)\bar{k}_2-(\bar{q}_1+\bar{q}_2)k_2)(\bar{q}_1+\bar{q}_2)}{|\vec{k}_1||\vec{q}_1+\vec{q}_2|} +\frac{2(\bar{q}_1+\bar{q}_2)(k_1\bar{k}_2-\bar{k}_1k_2)\bar{k}_1}{|\vec{k}_1||\vec{q}_1+\vec{q}_2|} +\frac{|\vec{k}_2|^2 \bar{k}_1^2}{|\vec{k}_2||\vec{k}_1|} \Bigg\} \nn\\
\eE
%

\bE{l}
= f^{a_1a_2c}f^{b_1b_2c} \int_{\slashed{k_1},\slashed{k_2},\slashed{q_1},\slashed{q_2}}  \frac{\slashed{\delta}\left(\sum_{i=1}^2 (\vec{k}_i+\vec{q}_i)\right)}{|\vec{k}_1|+|\vec{k}_2|+|\vec{k}_1+\vec{k}_2|} \Bigg\{ J^{a_1}(\vec{k_1})  J^{a_2}(\vec{k}_2) \theta^{b_1}(\vec{q}_1) \theta^{b_2}(\vec{q}_2) \nn\\
\quad \Bigg(\frac{1}{2}\bar{k_1}\bar{q}_1\frac{(\bar{q}_1+\bar{q}_2)}{(\bar{q}_1+\bar{q}_2)} +\frac{2\bar{q}_2q_1}{(\bar{q}_1+\bar{q}_2)} \frac{\bar{k}_1(\bar{k}_1+\bar{k}_2)^2}{|\vec{k}_1||\vec{k}_1+\vec{k}_2|} +\frac{(\bar{q}_1+\bar{q}_2)}{(\bar{q}_1+\bar{q}_2)} \frac{\bar{k}_1\bar{k}_2(k_2\bar{q}_1-\bar{k}_2q_1)}{|\vec{k}_1||\vec{k}_2|} \nn\\
 \qquad\qquad\qquad + \frac{2\bar{k}_2^2\bar{q}_1(-q_2-k_1-k_2)}{|\vec{k}_2||\vec{k}_1+\vec{k}_2|}\Bigg)  \nn\\
\quad\quad +  J^{a_1}(\vec{k}_1) \theta^{a_2}(\vec{k}_2) J^{b_1}(\vec{q}_1)\theta^{b_2}(\vec{q}_2)\nn\\
\quad \Bigg((-\bar{k}_1+\bar{q}_1+\bar{q}_2)\bar{k}_2  -\frac{4\bar{k}_1(k_1\bar{k}_2-\bar{k}_1k_2)(\bar{k}_1+\bar{k}_2)}{|\vec{k}_1||\vec{k}_1+\vec{k}_2|}  +|\vec{k}_2| \left( \frac{ \bar{k}_1^2}{|\vec{k}_1|} - \frac{(\bar{k}_1+\bar{k}_2)^2}{|\vec{k}_1+\vec{k}_2|} \right) \Bigg) \Bigg\} \\
 = f^{a_1a_2c}f^{b_1b_2c} \int_{\slashed{k_1},\slashed{k_2},\slashed{q_1},\slashed{q_2}}  \frac{\slashed{\delta}\left(\sum_{i=1}^2 (\vec{k}_i+\vec{q}_i)\right)}{|\vec{k}_1|+|\vec{k}_2|+|\vec{k}_1+\vec{k}_2|} \Bigg\{ J^{a_1}(\vec{k_1})  J^{a_2}(\vec{k}_2) \theta^{b_1}(\vec{q}_1) \theta^{b_2}(\vec{q}_2) \nn\\
\qquad \Bigg(\frac{1}{2}\frac{\bar{k_1}\bar{q}_1^2}{(\bar{q}_1+\bar{q}_2)} +\frac{2\bar{q}_2q_1}{(\bar{q}_1+\bar{q}_2)}\frac{\bar{k}_1(\bar{k}_1+\bar{k}_2)^2}{|\vec{k}_1||\vec{k}_1+\vec{k}_2|} + \frac{\bar{q}_1^2}{(\bar{q}_1+\bar{q}_2)} \frac{\bar{k}_1|\vec{k}_2|}{4|\vec{k}_1|}  - \frac{|\vec{q}_1|^2}{4(\bar{q}_1+\bar{q}_2)} \frac{\bar{k}_1\bar{k}_2^2}{|\vec{k}_1||\vec{k}_2|} \nn\\
\qquad \qquad  - \frac{\bar{q}_2 q_1}{(\bar{q}_1+\bar{q}_2)} \frac{\bar{k}_1\bar{k}_2^2}{|\vec{k}_1||\vec{k}_2|} + \frac{(\bar{q}_1+\bar{q}_2)}{(\bar{q}_1+\bar{q}_2)}\frac{2\bar{k}_2^2\bar{q}_2q_1}{|\vec{k}_2||\vec{k}_1+\vec{k}_2|} +\frac{(\bar{q}_1+\bar{q}_2)}{(\bar{q}_1+\bar{q}_2)}\frac{2\bar{k}_2^2\bar{q}_1(q_1+q_2)}{|\vec{k}_2||\vec{k}_1+\vec{k}_2|}\Bigg)  \nn\\
\quad\quad + \Bigg(\bar{k}_1^2 -(\bar{k_1}+\bar{k}_2)^2 -\frac{4\bar{k}_1(k_1\bar{k}_2-\bar{k}_1(k_2+k_1-k_1))(\bar{k}_1+\bar{k}_2)}{|\vec{k}_1||\vec{k}_1+\vec{k}_2|}  +|\vec{k}_2| \left( \frac{ \bar{k}_1^2}{|\vec{k}_1|} - \frac{(\bar{k}_1+\bar{k}_2)^2}{|\vec{k}_1+\vec{k}_2|} \right) \Bigg) \nn\\
\qquad\qquad J^{a_1}(\vec{k}_1) \theta^{a_2}(\vec{k}_2) J^{b_1}(\vec{q}_1)\theta^{b_2}(\vec{q}_2) \Bigg\}\,.
\eE
The terms $\propto \frac{\bar{q}_1\bar{q}_2}{\bar{q}_1+\bar{q}_2}$ that would have appeared in the second line vanish under symmetry.

\bE{l}
= f^{a_1a_2c}f^{b_1b_2c} \int_{\slashed{k_1},\slashed{k_2},\slashed{q_1},\slashed{q_2}}  \frac{\slashed{\delta}\left(\sum_{i=1}^2 (\vec{k}_i+\vec{q}_i)\right)}{|\vec{k}_1|+|\vec{k}_2|+|\vec{k}_1+\vec{k}_2|} \Bigg\{ J^{a_1}(\vec{k_1})  J^{a_2}(\vec{k}_2) \theta^{b_1}(\vec{q}_1) \theta^{b_2}(\vec{q}_2)\nn\\
\qquad \Bigg(\frac{2\bar{q}_2q_1}{(\bar{q}_1+\bar{q}_2)}\frac{\bar{k}_1(\bar{k}_1+\bar{k}_2)^2}{|\vec{k}_1||\vec{k}_1+\vec{k}_2|} + \frac{2\bar{q}_2q_1}{(\bar{q}_1+\bar{q}_2)}\frac{\bar{k}_2^2(-\bar{k}_1-\bar{k}_2)}{|\vec{k}_2||\vec{k}_1+\vec{k}_2|} + \frac{\bar{q}_2 q_1}{(\bar{q}_1+\bar{q}_2)} \frac{\bar{k}_1^2\bar{k}_2}{|\vec{k}_1||\vec{k}_2|} \nn\\
\qquad \qquad +\frac{1}{2}\frac{\bar{k_1}\bar{q}_1^2}{(\bar{q}_1+\bar{q}_2)}+ \frac{\bar{q}_1^2}{(\bar{q}_1+\bar{q}_2)} \frac{\bar{k}_1|\vec{k}_2|}{4|\vec{k}_1|}  - \frac{\bar{q}_1q_1}{(\bar{q}_1+\bar{q}_2)} \frac{\bar{k}_1\bar{k}_2^2}{|\vec{k}_1||\vec{k}_2|}   +\frac{1}{(\bar{q}_1+\bar{q}_2)} \frac{\bar{k}_2^2\bar{q}_1|\vec{k}_1+\vec{k}_2|}{2|\vec{k}_2|}\Bigg)  \nn\\
\quad + \Bigg(\bar{k}_1^2 -(\bar{k_1}+\bar{k}_2)^2  -\frac{|\vec{k}_1|(\bar{k}_1+\bar{k}_2)^2}{|\vec{k}_1+\vec{k}_2|} +\frac{\bar{k}_1^2 |\vec{k}_1+\vec{k}_2|}{|\vec{k}_1|}  +|\vec{k}_2| \left( \frac{ \bar{k}_1^2}{|\vec{k}_1|} - \frac{(\bar{k}_1+\bar{k}_2)^2}{|\vec{k}_1+\vec{k}_2|} \right) \Bigg) \nn\\
\qquad\qquad J^{a_1}(\vec{k}_1) \theta^{a_2}(\vec{k}_2) J^{b_1}(\vec{q}_1)\theta^{b_2}(\vec{q}_2) \Bigg\} \\
 = f^{a_1a_2c}f^{b_1b_2c} \int_{\slashed{k_1},\slashed{k_2},\slashed{q_1},\slashed{q_2}}  \frac{\slashed{\delta}\left(\sum_{i=1}^2 (\vec{k}_i+\vec{q}_i)\right)}{|\vec{k}_1|+|\vec{k}_2|+|\vec{k}_1+\vec{k}_2|} \Bigg\{ J^{a_1}(\vec{k_1})  J^{a_2}(\vec{k}_2) \theta^{b_1}(\vec{q}_1) \theta^{b_2}(\vec{q}_2)\nn\\
\qquad \Bigg(\frac{2\bar{q}_2q_1}{(\bar{q}_1+\bar{q}_2)}\frac{\bar{k}_1(\bar{k}_1+\bar{k}_2)^2}{|\vec{k}_1||\vec{k}_1+\vec{k}_2|} + \frac{2\bar{q}_2q_1}{(\bar{q}_1+\bar{q}_2)}\frac{\bar{k}_2^2(-\bar{k}_1-\bar{k}_2)}{|\vec{k}_2||\vec{k}_1+\vec{k}_2|} + \frac{2\bar{q}_2 q_1}{(\bar{q}_1+\bar{q}_2)} \frac{\bar{k}_1^2\bar{k}_2}{|\vec{k}_1||\vec{k}_2|}  \nn\\
\qquad \qquad - \frac{\bar{q}_2 q_1}{(\bar{q}_1+\bar{q}_2)} \frac{\bar{k}_1^2\bar{k}_2}{|\vec{k}_1||\vec{k}_2|} +\frac{1}{2}\frac{\bar{k_1}\bar{q}_1^2}{(\bar{q}_1+\bar{q}_2)}+ \frac{\bar{q}_1^2}{(\bar{q}_1+\bar{q}_2)} \frac{\bar{k}_1|\vec{k}_2|}{4|\vec{k}_1|} \nn\\
\qquad \qquad  - \frac{\bar{q}_1(-q_2-k_1-k_2)}{(\bar{q}_1+\bar{q}_2)} \frac{\bar{k}_1\bar{k}_2^2}{|\vec{k}_1||\vec{k}_2|}   -|\vec{k}_1+\vec{k}_2|\frac{\bar{q}_2}{(\bar{q}_1+\bar{q}_2)} \frac{\bar{k}_2^2}{2|\vec{k}_2|}\Bigg)  \nn\\
\quad\quad + \Bigg(\bar{k}_1^2  +\frac{\bar{k}_1^2 |\vec{k}_1+\vec{k}_2|}{|\vec{k}_1|}  +|\vec{k}_2|  \frac{ \bar{k}_1^2}{|\vec{k}_1|} -(|\vec{k}_1|+|\vec{k}_2|+|\vec{k}_1+\vec{k}_2|) \frac{(\bar{k}_1+\bar{k}_2)^2}{|\vec{k}_1+\vec{k}_2|}  \Bigg) \nn\\
\qquad\qquad   J^{a_1}(\vec{k}_1) \theta^{a_2}(\vec{k}_2) J^{b_1}(\vec{q}_1)\theta^{b_2}(\vec{q}_2) \Bigg\} \\
= f^{a_1a_2c}f^{b_1b_2c} \int_{\slashed{k_1},\slashed{k_2},\slashed{q_1},\slashed{q_2}}  \slashed{\delta}\left(\sum_{i=1}^2 (\vec{k}_i+\vec{q}_i)\right) \frac{\bar{q}_2q_1}{\bar{q}_1+\bar{q}_2} \frac{g^{(3)}(\vec{k}_1,\vec{k}_2, -\vec{k}_1-\vec{k}_2)}{16}  \nn\\
\qquad \qquad\qquad J^{a_1}(\vec{k_1})  J^{a_2}(\vec{k}_2) \theta^{b_1}(\vec{q}_1) \theta^{b_2}(\vec{q}_2)\nn\\
\quad +f^{a_1a_2c}f^{b_1b_2c} \int_{\slashed{k_1},\slashed{k_2},\slashed{q_1},\slashed{q_2}}  \frac{\slashed{\delta}\left(\sum_{i=1}^2 (\vec{k}_i+\vec{q}_i)\right)}{|\vec{k}_1|+|\vec{k}_2|+|\vec{k}_1+\vec{k}_2|} \Bigg\{ J^{a_1}(\vec{k_1})  J^{a_2}(\vec{k}_2) \theta^{b_1}(\vec{q}_1) \theta^{b_2}(\vec{q}_2)\nn\\
\qquad\Bigg( - \frac{\bar{q}_2 q_1}{(\bar{q}_1+\bar{q}_2)} \frac{\bar{k}_1^2\bar{k}_2}{|\vec{k}_1||\vec{k}_2|}+\frac{1}{2}\frac{\bar{k_1}\bar{q}_1^2}{(\bar{q}_1+\bar{q}_2)}+ \frac{\bar{q}_1^2}{(\bar{q}_1+\bar{q}_2)} \frac{\bar{k}_1|\vec{k}_2|}{4|\vec{k}_1|}    \nn\\
\qquad \qquad + \frac{\bar{q}_1q_2}{(\bar{q}_1+\bar{q}_2)} \frac{\bar{k}_1\bar{k}_2^2}{|\vec{k}_1||\vec{k}_2|} + \frac{\bar{q}_1}{(\bar{q}_1+\bar{q}_2)} \frac{|\vec{k}_1|\bar{k}_2^2}{4|\vec{k}_2|} + \frac{\bar{q}_1}{(\bar{q}_1+\bar{q}_2)} \frac{\bar{k}_1\bar{k}_2|\vec{k}_2|}{4|\vec{k}_1|} -|\vec{k}_1+\vec{k}_2|\frac{\bar{q}_2}{(\bar{q}_1+\bar{q}_2)} \frac{\bar{k}_2^2}{2|\vec{k}_2|}\Bigg)  \nn\\
\quad -f^{a_1a_2c}f^{b_1b_2c} \int_{\slashed{k_1},\slashed{k_2},\slashed{q_1},\slashed{q_2}} \slashed{\delta}\left(\sum_{i=1}^2 (\vec{k}_i+\vec{q}_i)\right) \left( \frac{(\bar{k}_1+\bar{k}_2)^2}{|\vec{k}_1+\vec{k}_2|} -\frac{ \bar{k}_1^2}{|\vec{k}_1|} \right) J^{a_1}(\vec{k}_1) \theta^{a_2}(\vec{k}_2) J^{b_1}(\vec{q}_1)\theta^{b_2}(\vec{q}_2) \,.\nn\\
\eE
The last line is already correct to cancel the first line of Eq.~(\ref{F24+1}), while the first line cancels the first term of the last line of Eq.~(\ref{F24+1}).

\bE{l}
= f^{a_1a_2c}f^{b_1b_2c} \int_{\slashed{k_1},\slashed{k_2},\slashed{q_1},\slashed{q_2}}  \slashed{\delta}\left(\sum_{i=1}^2 (\vec{k}_i+\vec{q}_i)\right) \frac{\bar{q}_2q_1}{\bar{q}_1+\bar{q}_2} \frac{g^{(3)}(\vec{k}_1,\vec{k}_2, -\vec{k}_1-\vec{k}_2)}{16}   \\
\qquad \qquad\qquad  J^{a_1}(\vec{k_1})  J^{a_2}(\vec{k}_2) \theta^{b_1}(\vec{q}_1) \theta^{b_2}(\vec{q}_2)\nn\\
\quad +f^{a_1a_2c}f^{b_1b_2c} \int_{\slashed{k_1},\slashed{k_2},\slashed{q_1},\slashed{q_2}}  \frac{\slashed{\delta}\left(\sum_{i=1}^2 (\vec{k}_i+\vec{q}_i)\right)}{|\vec{k}_1|+|\vec{k}_2|+|\vec{k}_1+\vec{k}_2|} \Bigg\{ J^{a_1}(\vec{k_1})  J^{a_2}(\vec{k}_2) \theta^{b_1}(\vec{q}_1) \theta^{b_2}(\vec{q}_2) \nn\\
\quad\Bigg(\frac{1}{2}\frac{\bar{k_1}\bar{q}_1^2}{(\bar{q}_1+\bar{q}_2)}+ \frac{\bar{q}_1^2}{(\bar{q}_1+\bar{q}_2)} \frac{\bar{k}_1|\vec{k}_2|}{4|\vec{k}_1|}    + \frac{\bar{q}_1}{(\bar{q}_1+\bar{q}_2)} \frac{\bar{k}_1\bar{k}_2|\vec{k}_2|}{4|\vec{k}_1|} -\left({1 \over 2} |\vec{k}_1|+|\vec{k}_1+\vec{k}_2|\right) \frac{\bar{q}_2}{(\bar{q}_1+\bar{q}_2)} \frac{\bar{k}_2^2}{2|\vec{k}_2|}\Bigg)  \nn\\
\quad -f^{a_1a_2c}f^{b_1b_2c} \int_{\slashed{k_1},\slashed{k_2},\slashed{q_1},\slashed{q_2}} \slashed{\delta}\left(\sum_{i=1}^2 (\vec{k}_i+\vec{q}_i)\right) \left( \frac{(\bar{k}_1+\bar{k}_2)^2}{|\vec{k}_1+\vec{k}_2|} -\frac{ \bar{k}_1^2}{|\vec{k}_1|} \right) J^{a_1}(\vec{k}_1) \theta^{a_2}(\vec{k}_2) J^{b_1}(\vec{q}_1)\theta^{b_2}(\vec{q}_2) \nn\\
%
 = f^{a_1a_2c}f^{b_1b_2c} \int_{\slashed{k_1},\slashed{k_2},\slashed{q_1},\slashed{q_2}}  \slashed{\delta}\left(\sum_{i=1}^2 (\vec{k}_i+\vec{q}_i)\right) \frac{\bar{q}_2q_1}{\bar{q}_1+\bar{q}_2} \frac{g^{(3)}(\vec{k}_1,\vec{k}_2, -\vec{k}_1-\vec{k}_2)}{16}   \nn\\
\qquad \qquad\qquad J^{a_1}(\vec{k_1})  J^{a_2}(\vec{k}_2) \theta^{b_1}(\vec{q}_1) \theta^{b_2}(\vec{q}_2)\nn\\
\quad +f^{a_1a_2c}f^{b_1b_2c} \int_{\slashed{k_1},\slashed{k_2},\slashed{q_1},\slashed{q_2}}  \frac{\slashed{\delta}\left(\sum_{i=1}^2 (\vec{k}_i+\vec{q}_i)\right)}{|\vec{k}_1|+|\vec{k}_2|+|\vec{k}_1+\vec{k}_2|} \Bigg\{ J^{a_1}(\vec{k_1})  J^{a_2}(\vec{k}_2) \theta^{b_1}(\vec{q}_1) \theta^{b_2}(\vec{q}_2)\nn\\
\quad\Bigg(\frac{1}{2}\frac{\bar{k_1}\bar{q}_1^2}{(\bar{q}_1+\bar{q}_2)} - \frac{\bar{q}_1}{(\bar{q}_1+\bar{q}_2)} \frac{\bar{k}_1^2|\vec{k}_2|}{4|\vec{k}_1|}   - \frac{\bar{q}_1\bar{q}_2}{(\bar{q}_1+\bar{q}_2)} \frac{\bar{k}_1|\vec{k}_2|}{4|\vec{k}_1|}    -\left({1 \over 2} |\vec{k}_1|+|\vec{k}_1+\vec{k}_2|\right) \frac{\bar{q}_2}{(\bar{q}_1+\bar{q}_2)} \frac{\bar{k}_2^2}{2|\vec{k}_2|}\Bigg)  \nn\\
\quad -f^{a_1a_2c}f^{b_1b_2c} \int_{\slashed{k_1},\slashed{k_2},\slashed{q_1},\slashed{q_2}} \slashed{\delta}\left(\sum_{i=1}^2 (\vec{k}_i+\vec{q}_i)\right) \left( \frac{(\bar{k}_1+\bar{k}_2)^2}{|\vec{k}_1+\vec{k}_2|} -\frac{ \bar{k}_1^2}{|\vec{k}_1|} \right) J^{a_1}(\vec{k}_1) \theta^{a_2}(\vec{k}_2) J^{b_1}(\vec{q}_1)\theta^{b_2}(\vec{q}_2) \,. \nn\\
\eE
The $\frac{\bar{q}_1\bar{q}_2}{(\bar{q}_1+\bar{q}_2)}$- term vanishes under symmetry.
\bE{l}
= f^{a_1a_2c}f^{b_1b_2c} \int_{\slashed{k_1},\slashed{k_2},\slashed{q_1},\slashed{q_2}}  \slashed{\delta}\left(\sum_{i=1}^2 (\vec{k}_i+\vec{q}_i)\right) \frac{\bar{q}_2q_1}{\bar{q}_1+\bar{q}_2} \frac{g^{(3)}(\vec{k}_1,\vec{k}_2, -\vec{k}_1-\vec{k}_2)}{16}   \nn\\
\qquad \qquad\qquad  J^{a_1}(\vec{k_1})  J^{a_2}(\vec{k}_2) \theta^{b_1}(\vec{q}_1) \theta^{b_2}(\vec{q}_2)\nn\\
\quad +f^{a_1a_2c}f^{b_1b_2c} \int_{\slashed{k_1},\slashed{k_2},\slashed{q_1},\slashed{q_2}}  \frac{\slashed{\delta}\left(\sum_{i=1}^2 (\vec{k}_i+\vec{q}_i)\right)}{|\vec{k}_1|+|\vec{k}_2|+|\vec{k}_1+\vec{k}_2|} \Bigg\{ J^{a_1}(\vec{k_1})  J^{a_2}(\vec{k}_2) \theta^{b_1}(\vec{q}_1) \theta^{b_2}(\vec{q}_2)\nn\\
\qquad\Bigg(\frac{1}{2}\frac{\bar{k_1}\bar{q}_1^2}{(\bar{q}_1+\bar{q}_2)}  -\left(|\vec{k}_1|+|\vec{k}_1+\vec{k}_2|\right) \frac{\bar{q}_2}{(\bar{q}_1+\bar{q}_2)} \frac{\bar{k}_2^2}{2|\vec{k}_2|}\Bigg)  \nn\\
\quad -f^{a_1a_2c}f^{b_1b_2c} \int_{\slashed{k_1},\slashed{k_2},\slashed{q_1},\slashed{q_2}} \slashed{\delta}\left(\sum_{i=1}^2 (\vec{k}_i+\vec{q}_i)\right) \left( \frac{(\bar{k}_1+\bar{k}_2)^2}{|\vec{k}_1+\vec{k}_2|} -\frac{ \bar{k}_1^2}{|\vec{k}_1|} \right)   \nn\\
\qquad \qquad\qquad J^{a_1}(\vec{k}_1) \theta^{a_2}(\vec{k}_2) J^{b_1}(\vec{q}_1)\theta^{b_2}(\vec{q}_2)  \qquad\\
= f^{a_1a_2c}f^{b_1b_2c} \int_{\slashed{k_1},\slashed{k_2},\slashed{q_1},\slashed{q_2}}  \slashed{\delta}\left(\sum_{i=1}^2 (\vec{k}_i+\vec{q}_i)\right) \frac{\bar{q}_2q_1}{\bar{q}_1+\bar{q}_2} \frac{g^{(3)}(\vec{k}_1,\vec{k}_2, -\vec{k}_1-\vec{k}_2)}{16}   \nn\\
\qquad \qquad\qquad J^{a_1}(\vec{k_1})  J^{a_2}(\vec{k}_2) \theta^{b_1}(\vec{q}_1) \theta^{b_2}(\vec{q}_2)\nn\\
\quad +f^{a_1a_2c}f^{b_1b_2c} \int_{\slashed{k_1},\slashed{k_2},\slashed{q_1},\slashed{q_2}}  \frac{\slashed{\delta}\left(\sum_{i=1}^2 (\vec{k}_i+\vec{q}_i)\right)}{|\vec{k}_1|+|\vec{k}_2|+|\vec{k}_1+\vec{k}_2|} \Bigg\{ J^{a_1}(\vec{k_1})  J^{a_2}(\vec{k}_2) \theta^{b_1}(\vec{q}_1) \theta^{b_2}(\vec{q}_2)\nn\\
\qquad\Bigg(\frac{1}{2}\frac{\bar{k_1}\bar{q}_1(-\bar{k}_1-\bar{k}_2-\bar{q}_2)}{(\bar{q}_1+\bar{q}_2)}  -\left(|\vec{k}_1|+|\vec{k}_1+\vec{k}_2|\right) \frac{\bar{q}_2}{(\bar{q}_1+\bar{q}_2)} \frac{\bar{k}_2^2}{2|\vec{k}_2|}\Bigg)  \nn\\
\quad -f^{a_1a_2c}f^{b_1b_2c} \int_{\slashed{k_1},\slashed{k_2},\slashed{q_1},\slashed{q_2}} \slashed{\delta}\left(\sum_{i=1}^2 (\vec{k}_i+\vec{q}_i)\right) \left( \frac{(\bar{k}_1+\bar{k}_2)^2}{|\vec{k}_1+\vec{k}_2|} -\frac{ \bar{k}_1^2}{|\vec{k}_1|} \right)   \nn\\
\qquad \qquad\qquad J^{a_1}(\vec{k}_1) \theta^{a_2}(\vec{k}_2) J^{b_1}(\vec{q}_1)\theta^{b_2}(\vec{q}_2) \,.
\eE
The $\frac{\bar{q}_1\bar{q}_2}{(\bar{q}_1+\bar{q}_2)}$- term and the $\frac{\bar{k}_1\bar{k}_2}{(\bar{q}_1+\bar{q}_2)}$- term vanish under symmetry, so
\bE{l}
F^{(1)}_{GL}|_{\O(e\theta^2)} = f^{a_1a_2c}f^{b_1b_2c} \int_{\slashed{k_1},\slashed{k_2},\slashed{q_1},\slashed{q_2}}  \slashed{\delta}\left(\sum_{i=1}^2 (\vec{k}_i+\vec{q}_i)\right)   J^{a_1}(\vec{k_1})  J^{a_2}(\vec{k}_2) \theta^{b_1}(\vec{q}_1) \theta^{b_2}(\vec{q}_2)\nn\\
\qquad\qquad  \Bigg(\frac{\bar{q}_2q_1}{\bar{q}_1+\bar{q}_2} \frac{g^{(3)}(\vec{k}_1,\vec{k}_2, -\vec{k}_1-\vec{k}_2)}{16} - \frac{\bar{q}_2}{(\bar{q}_1+\bar{q}_2)} \frac{\bar{k}_2^2}{2|\vec{k}_2|}\Bigg)  \nn\\
\quad -f^{a_1a_2c}f^{b_1b_2c} \int_{\slashed{k_1},\slashed{k_2},\slashed{q_1},\slashed{q_2}} \slashed{\delta}\left(\sum_{i=1}^2 (\vec{k}_i+\vec{q}_i)\right) \left( \frac{(\bar{k}_1+\bar{k}_2)^2}{|\vec{k}_1+\vec{k}_2|} -\frac{ \bar{k}_1^2}{|\vec{k}_1|} \right)   \nn\\
\qquad \qquad\qquad J^{a_1}(\vec{k}_1) \theta^{a_2}(\vec{k}_2) J^{b_1}(\vec{q}_1)\theta^{b_2}(\vec{q}_2)   \label{F1theta2}\qquad\\
=-F^{(2,4)}_{GL}|_{\O(\theta^2)}  - F^{(0)}_{GL}|_{\O(e^2\theta^2)} \,. \label{F1=F24theta2}
\eE

\subsection{Order $\theta^3$}

We want this term to cancel Eqs.~(\ref{F24theta3}) and (\ref{F0theta3}).

\bE{l}
F^{(2,4)}_{GL}|_{\O(\theta^3)}  + F^{(0)}_{GL}|_{\O(e^2\theta^3)} \nn\\
=  2 f^{b_1b_2c}f^{a_1a_2c}\int_{\slashed{p},\slashed{k_1},\slashed{k_2},\slashed{q_1},\slashed{q_2}} \slashed{\delta}\left(\sum_i(\vec{k}_i+\vec{q}_i)\right) q_1\bar{q}_2 \left(\frac{\bar{k}_1+\bar{k}_2}{|\vec{k}_1+\vec{k}_2|} - \frac{\bar{k}_1}{|\vec{k}_1|} \right) \nn\\ \qquad\qquad\qquad J^{a_1}(\vec{k}_1)\theta^{a_2}(\vec{k}_2) \theta^{b_1}(\vec{q}_1)\theta^{b_2}(\vec{q}_2)  \nn\\
\quad +   f^{a_1a_2c}f^{b_1b_2c}\int_{\slashed{k}_1,\slashed{k}_2,\slashed{q}_1,\slashed{q}_2}\slashed{\delta}\left(\sum_i(\vec{k}_i+\vec{q}_i)\right)  J^{a_1}(\vec{k}_1)\theta^{a_2}(\vec{k}_2) \theta^{b_1}(\vec{q}_1) \theta^{b_2}(\vec{q}_2)    \nn\\
\qquad \Bigg\{ \frac{1}{3} \frac{1}{|\vec{k}_1|} \bar{k}_1(k_1\bar{q}_2-\bar{k}_1q_2)  +\frac{1}{|\vec{q}_1+\vec{q}_2|}(\bar{q}_1+\bar{q}_2) (q_2\bar{q}_1-\bar{q}_2q_1)\, \Bigg\}\,. \label{F24+F0theta3}
\eE

We extract the $\O(e\theta^3)$ portion from Eq.~(\ref{F1GLref}) and bring it in a similar form:

\bE{l}
= f^{abc} \int_{\slashed{k_1},\slashed{k_2},\slashed{k_3},\slashed{q}}\slashed{\delta}\left(\sum_{i=1}(\vec{k}_i+\vec{q}_i)\right)  \Bigg\{ \Bigg\{ \frac{1}{|\vec{k}_1|+|\vec{k}_2|+|\vec{k}_3+\vec{q}|}\bar{k_1}(\bar{k_2}q-k_2\bar{q})   \nn\\
\qquad \qquad\qquad J^a(\vec{k_1}) \theta^b(\vec{k}_2)f^{cde}\theta^d(\vec{q})\theta^e(\vec{k}_3) \nn\\
\qquad\qquad +2\frac{1}{|\vec{k}_1+\vec{q}|+|\vec{k}_2|+|\vec{k}_3|}(\bar{k_1}+\bar{q})k_2\bar{k}_3f^{aa_1a_2}\theta^{a_1}(\vec{k}_1)J^{a_2}(\vec{q})\theta^b(\vec{k}_2)\theta^c(\vec{k}_3)  \nn\\
\qquad\qquad + f^{ac_1c_2}\frac{1}{|\vec{k}_1+\vec{q}|+|\vec{k}_2|+|\vec{k}_3|}(k_1\bar{q}-\bar{k}_1q)\bar{k_3}\,\theta^{c_1}(\vec{q})\theta^{c_2}(\vec{k}_1)J^b(\vec{k}_2) \theta^c(\vec{k}_3) \Bigg\} \nn\\
 +   \Bigg\{ \bar{k_3}J^c(\vec{k}_3) \Bigg(\frac{1}{|\vec{k}_1||\vec{k}_3|}\frac{1}{|\vec{k}_1|+|\vec{k}_2+\vec{q}|+|\vec{k}_3|}|\vec{k}_1|^2 (k_3\bar{q}-\bar{k}_3q)\theta^a(\vec{k}_1) f^{bb_1b_2}\,\theta^{b_1}(\vec{q})\theta^{b_2}(\vec{k}_2) \nn\\
\qquad\quad +\frac{1}{|\vec{k}_1+\vec{q}||\vec{k}_3|}\frac{1}{|\vec{k}_1+\vec{q}|+|\vec{k}_2|+|\vec{k}_3|} f^{ade}(2k_1\bar{q}+2\bar{k}_1q+|\vec{q}|^2)(k_3\bar{k_2}-\bar{k_3}k_2)   \nn\\
\qquad \qquad\qquad \theta^d(\vec{q})\theta^e(\vec{k}_1) \theta^b(\vec{k}_2)\Bigg)  \nn\\
\quad-\frac{2}{|\vec{k}_1||\vec{k}_3+\vec{q}|}\frac{|\vec{k}_1|^2((k_3+q)\bar{k_2}-(\bar{k_3}+\bar{q})k_2)}{|\vec{k}_1|+|\vec{k}_2|+|\vec{k}_3+\vec{q}|} (\bar{k_3}+\bar{q})\theta^a(\vec{k}_1) \theta^b(\vec{k}_2) f^{cde}\theta^d(\vec{k_3})J^e(\vec{q}) \nn\\
\quad +\frac{1}{|\vec{k}_1||\vec{k}_3+\vec{q}|}\frac{1}{|\vec{k}_1|+|\vec{k}_2|+|\vec{k}_3+\vec{q}|} \Big(|\vec{k}_1|^2 (\bar{k_3}+\bar{q}) \theta^a(\vec{k}_1)J^b(\vec{k}_2) \nn\\
\qquad\qquad  -2\bar{k_1} ((k_3+q)\bar{k_2}-(\bar{k_3}+\bar{q})k_2) J^a(\vec{k}_1)\theta^b(\vec{k}_2) \Big)  f^{cc_1c_2}(k_3\bar{q}-\bar{k}_3q)\,\theta^{c_1}(\vec{q})\theta^{c_2}(\vec{k}_3)\Bigg\} \Bigg\} \nn\\
 + O(J^3)+\ldots+O(\theta^4)\,.
\eE

\bE{l}
=  \int_{\slashed{k_1},\slashed{k_2},\slashed{k_3},\slashed{q}}\slashed{\delta}\left(\sum_{i=1}(\vec{k}_i+\vec{q}_i)\right)  \Bigg\{ \frac{1}{|\vec{k}_1|+|\vec{k}_2|+|\vec{q}_1+\vec{q}_2|}\bar{k_1}(\bar{k_2}q_1-k_2\bar{q}_1)   \nn\\
\qquad -2\frac{1}{|\vec{k}_1+\vec{k}_2|+|\vec{q}_1|+|\vec{q}_2|}(\bar{k_1}+\bar{k_2})q_1\bar{q}_2 +\frac{1}{|\vec{q}_1+\vec{q}_2|+|\vec{k}_1|+|\vec{k}_2|}(q_2\bar{q}_1-\bar{q}_2q_1)\bar{k_2}  \nn\\
\qquad +   \frac{1}{|\vec{k}_2||\vec{k}_1|}\frac{1}{|\vec{k}_2|+|\vec{q}_1+\vec{q}_2|+|\vec{k}_1|}|\vec{k}_2|^2 (k_1\bar{q}_1-\bar{k}_1q_1)\bar{k_1}  \nn\\
\qquad -\frac{1}{|\vec{q}_1+\vec{q}_2||\vec{k}_1|}\frac{1}{|\vec{q}_1+\vec{q}_2|+|\vec{k}_2|+|\vec{k}_1|}(2q_2\bar{q}_1+2\bar{q}_2q_1+|\vec{q}_1|^2)(k_1\bar{k_2}-\bar{k_1}k_2) \bar{k_1} \nn\\
\qquad +\frac{2}{|\vec{q}_1||\vec{k}_1+\vec{k}_2|}\frac{1}{|\vec{q}_1|+|\vec{q}_2|+|\vec{k}_1+\vec{k}_2|} |\vec{q}_1|^2((k_2+k_1)\bar{q_2}-(\bar{k_2}+\bar{k_1})q_2)(\bar{k_1}+\bar{k_2}) \nn\\
\qquad -\frac{1}{|\vec{k}_2||\vec{q}_1+\vec{q}_2|}\frac{1}{|\vec{k}_2|+|\vec{k}_1|+|\vec{q}_1+\vec{q}_2|} |\vec{k}_2|^2 (\bar{q_2}+\bar{q_1})(q_2\bar{q_1}-\bar{q_2}q_1) \nn\\
\qquad -2\frac{1}{|\vec{k}_1||\vec{q}_1+\vec{q}_2|}\frac{1}{|\vec{k}_1|+|\vec{k}_2|+|\vec{q}_1+\vec{q}_2|} \bar{k_1} ((q_1+q_2)\bar{k_2}-(\bar{q_1}+\bar{q_2})k_2) (q_2\bar{q_1}-\bar{q_2}q_1)\Bigg\}  \nn\\
\quad f^{a_1a_2c}f^{b_1b_2c} J^{a_1}(\vec{k}_1)\theta^{a_2}(\vec{k}_2) \theta^{b_1}(\vec{q}_1)\theta^{b_2}(\vec{q}_2) \,.
\eE
The first two terms in line 4 cancel under $\vec q_1\leftrightarrow \vec q_2$, use the delta function in the 5th and the last line.
\bE{l}
=  \int_{\slashed{k_1},\slashed{k_2},\slashed{k_3},\slashed{q}}\slashed{\delta}\left(\sum_{i=1}(\vec{k}_i+\vec{q}_i)\right)   \Bigg\{ \frac{1}{|\vec{k}_1|+|\vec{k}_2|+|\vec{q}_1+\vec{q}_2|}\bar{k_1}(\bar{k_2}q_1-k_2\bar{q}_1)  \nn\\
\qquad  -2\frac{1}{|\vec{k}_1+\vec{k}_2|+|\vec{q}_1|+|\vec{q}_2|}(\bar{k_1}+\bar{k_2})q_1\bar{q}_2  +\frac{1}{|\vec{q}_1+\vec{q}_2|+|\vec{k}_1|+|\vec{k}_2|}(q_2\bar{q}_1-\bar{q}_2q_1)\bar{k_2}  \nn\\
\qquad +   \frac{1}{|\vec{k}_2||\vec{k}_1|}\frac{1}{|\vec{k}_2|+|\vec{q}_1+\vec{q}_2|+|\vec{k}_1|}|\vec{k}_2|^2 (k_1\bar{q}_1-\bar{k}_1q_1)\bar{k_1}  \nn\\
\qquad -\frac{1}{|\vec{q}_1+\vec{q}_2||\vec{k}_1|}\frac{1}{|\vec{q}_1+\vec{q}_2|+|\vec{k}_2|+|\vec{k}_1|}|\vec{q}_1|^2(k_1\bar{k_2}-\bar{k_1}k_2) \bar{k_1} \nn\\
\qquad -\frac{2}{|\vec{q}_1||\vec{k}_1+\vec{k}_2|}\frac{1}{|\vec{q}_1|+|\vec{q}_2|+|\vec{k}_1+\vec{k}_2|} |\vec{q}_1|^2(q_1\bar{q_2}-\bar{q_1}q_2)(\bar{k_1}+\bar{k_2}) \nn\\
\qquad -\frac{1}{|\vec{k}_2||\vec{q}_1+\vec{q}_2|}\frac{1}{|\vec{k}_2|+|\vec{k}_1|+|\vec{q}_1+\vec{q}_2|} |\vec{k}_2|^2 (\bar{q_2}+\bar{q_1})(q_2\bar{q_1}-\bar{q_2}q_1) \nn\\
\qquad +2\frac{1}{|\vec{k}_1||\vec{q}_1+\vec{q}_2|}\frac{1}{|\vec{k}_1|+|\vec{k}_2|+|\vec{q}_1+\vec{q}_2|} \bar{k_1} (k_1\bar{k_2}-\bar{k_1}k_2) (q_2\bar{q_1}-\bar{q_2}q_1) \Bigg\} \nn\\
\quad f^{a_1a_2c}f^{b_1b_2c} J^{a_1}(\vec{k}_1)\theta^{a_2}(\vec{k}_2) \theta^{b_1}(\vec{q}_1)\theta^{b_2}(\vec{q}_2)
\eE

\bE{l}
=  \int_{\slashed{k_1},\slashed{k_2},\slashed{k_3},\slashed{q}}\slashed{\delta}\left(\sum_{i=1}(\vec{k}_i+\vec{q}_i)\right)   f^{a_1a_2c}f^{b_1b_2c} J^{a_1}(\vec{k}_1)\theta^{a_2}(\vec{k}_2) \theta^{b_1}(\vec{q}_1)\theta^{b_2}(\vec{q}_2)\\
\quad \Bigg\{\frac{1}{|\vec{k}_1|+|\vec{k}_2|+|\vec{q}_1+\vec{q}_2|} \Bigg\{ \bar{k_1}(\bar{k_2}q_1-k_2\bar{q}_1)  +(q_2\bar{q}_1-\bar{q}_2q_1)\bar{k_2} \nn\\
\qquad +   \frac{1}{|\vec{k}_2||\vec{k}_1|}|\vec{k}_2|^2 (k_1\bar{q}_1-\bar{k}_1q_1)\bar{k_1}  -\frac{1}{|\vec{q}_1+\vec{q}_2||\vec{k}_1|}|\vec{q}_1|^2(k_1\bar{k_2}-\bar{k_1}k_2) \bar{k_1} \nn\\
\qquad -\frac{1}{|\vec{k}_2||\vec{q}_1+\vec{q}_2|} |\vec{k}_2|^2 (\bar{q_2}+\bar{q_1})(q_2\bar{q_1}-\bar{q_2}q_1)  +2\frac{1}{|\vec{k}_1||\vec{q}_1+\vec{q}_2|} \bar{k_1} (k_1\bar{k_2}-\bar{k_1}k_2) (q_2\bar{q_1}-\bar{q_2}q_1)\Bigg\}  \nn\\
\quad -2\frac{1}{|\vec{k}_1+\vec{k}_2|+|\vec{q}_1|+|\vec{q}_2|}\Bigg\{ (\bar{k_1}+\bar{k_2})q_1\bar{q}_2 + \frac{1}{|\vec{q}_1||\vec{k}_1+\vec{k}_2|} |\vec{q}_1|^2(q_1\bar{q_2}-\bar{q_1}q_2)(\bar{k_1}+\bar{k_2})\Bigg\}\Bigg\} \nn\\
=  \int_{\slashed{k_1},\slashed{k_2},\slashed{k_3},\slashed{q}}\slashed{\delta}\left(\sum_{i=1}(\vec{k}_i+\vec{q}_i)\right)   f^{a_1a_2c}f^{b_1b_2c} J^{a_1}(\vec{k}_1)\theta^{a_2}(\vec{k}_2) \theta^{b_1}(\vec{q}_1)\theta^{b_2}(\vec{q}_2) \nn\\
\quad \Bigg\{\frac{1}{|\vec{k}_1|+|\vec{k}_2|+|\vec{q}_1+\vec{q}_2|} \Bigg\{ \bar{k_1}(\bar{k_2}q_1-k_2\bar{q}_1)    +(q_2\bar{q}_1-\bar{q}_2q_1)\bar{k_2} \nn\\
\qquad +   \frac{1}{|\vec{k}_2||\vec{k}_1|}|\vec{k}_2|^2 (k_1\bar{q}_1-\bar{k}_1q_1)\bar{k_1}  -\frac{1}{|\vec{q}_1+\vec{q}_2||\vec{k}_1|}|\vec{q}_1|^2(k_1\bar{k_2}-\bar{k_1}k_2) \bar{k_1} \nn\\
\qquad -\frac{1}{|\vec{k}_2||\vec{q}_1+\vec{q}_2|} |\vec{k}_2|^2 (\bar{q_2}+\bar{q_1})(q_2\bar{q_1}-\bar{q_2}q_1)  +2\frac{1}{|\vec{k}_1||\vec{q}_1+\vec{q}_2|} \bar{k_1} (k_1\bar{k_2}-\bar{k_1}k_2) (q_2\bar{q_1}-\bar{q_2}q_1)\Bigg\}  \nn\\
\quad -2\frac{(\bar{k_1}+\bar{k_2})}{|\vec{k}_1+\vec{k}_2|}q_1\bar{q_2}\Bigg\} \\
=  \int_{\slashed{k_1},\slashed{k_2},\slashed{k_3},\slashed{q}}\slashed{\delta}\left(\sum_{i=1}(\vec{k}_i+\vec{q}_i)\right)   f^{a_1a_2c}f^{b_1b_2c} J^{a_1}(\vec{k}_1)\theta^{a_2}(\vec{k}_2) \theta^{b_1}(\vec{q}_1)\theta^{b_2}(\vec{q}_2) \nn\\
\quad \Bigg\{\frac{1}{|\vec{k}_1|+|\vec{k}_2|+|\vec{q}_1+\vec{q}_2|} \Bigg\{ \bar{k_1}(\bar{k_2}q_1+k_2\bar{q}_2)      -2\bar{q}_2q_1\bar{k_2} \nn\\
\qquad - \frac{|\vec{k}_2|}{|\vec{k}_1|} (k_1\bar{q}_2-\bar{k}_1q_2)\bar{k_1} -4\frac{q_1\bar{q_1}}{|\vec{q}_1+\vec{q}_2||\vec{k}_1|}(k_1\bar{k_2}-\bar{k_1}k_2) \bar{k_1} \nn\\
\qquad -2|\vec{k}_2|q_1\bar{q_2}\frac{\bar{k_2}+\bar{k_1}}{|\vec{k}_1+\vec{k}_2|}  -4\frac{q_1\bar{q_2}}{|\vec{k}_1||\vec{q}_1+\vec{q}_2|} \bar{k_1} (k_1\bar{k_2}-\bar{k_1}k_2)  \Bigg\}  -2\frac{(\bar{k_1}+\bar{k_2})}{|\vec{k}_1+\vec{k}_2|}q_1\bar{q_2}\Bigg\} 
\eE

\bE{l}
=  \int_{\slashed{k_1},\slashed{k_2},\slashed{k_3},\slashed{q}}\slashed{\delta}\left(\sum_{i=1}(\vec{k}_i+\vec{q}_i)\right)   f^{a_1a_2c}f^{b_1b_2c} J^{a_1}(\vec{k}_1)\theta^{a_2}(\vec{k}_2) \theta^{b_1}(\vec{q}_1)\theta^{b_2}(\vec{q}_2)\Bigg\{\nn\\
\quad \frac{1}{|\vec{k}_1|+|\vec{k}_2|+|\vec{q}_1+\vec{q}_2|} \Bigg\{ \bar{k_1}((-\bar{k_1}-\bar{q_1}-\bar{q_2})q_1+(-k_1-q_1-q_2)\bar{q}_2)   -2\bar{q}_2q_1\bar{k_2} \nn\\
\qquad - \frac{|\vec{k}_2|}{|\vec{k}_1|} (k_1\bar{q}_2-\bar{k}_1q_2)\bar{k_1}  \nn\\
\qquad -2|\vec{k}_2|q_1\bar{q_2}\frac{\bar{k_2}+\bar{k_1}}{|\vec{k}_1+\vec{k}_2|}  -\frac{q_1(\bar{q_1}+\bar{q_2})}{|\vec{q}_1+\vec{q}_2|} |\vec{k}_1|\bar{k_2}  +4\frac{q_1(\bar{q_1}+\bar{q_2})}{|\vec{k}_1||\vec{q}_1+\vec{q}_2|} \bar{k_1}^2k_2)  \Bigg\}   -2\frac{(\bar{k_1}+\bar{k_2})}{|\vec{k}_1+\vec{k}_2|}q_1\bar{q_2}\Bigg\} \\
=  \int_{\slashed{k_1},\slashed{k_2},\slashed{k_3},\slashed{q}}\slashed{\delta}\left(\sum_{i=1}(\vec{k}_i+\vec{q}_i)\right)   f^{a_1a_2c}f^{b_1b_2c} J^{a_1}(\vec{k}_1)\theta^{a_2}(\vec{k}_2) \theta^{b_1}(\vec{q}_1)\theta^{b_2}(\vec{q}_2)\Bigg\{\nn\\
\quad \frac{1}{|\vec{k}_1|+|\vec{k}_2|+|\vec{q}_1+\vec{q}_2|} \Bigg\{ -\bar{k_1}(\bar{k_1}q_1+k_1\bar{q}_2)  - \frac{|\vec{k}_2|}{|\vec{k}_1|} (k_1\bar{q}_2-\bar{k}_1q_2)\bar{k_1}  \nn\\
\quad -(2|\vec{k}_1+\vec{k}_2|+2|\vec{k}_2|+|\vec{k}_1|)q_1\bar{q_2}\frac{\bar{k_2}+\bar{k_1}}{|\vec{k}_1+\vec{k}_2|}  -\frac{q_1(\bar{q_1}+\bar{q_2})}{|\vec{q}_1+\vec{q}_2||\vec{k}_1|} 4k_1\bar{k_1}(-\bar{k_1}-\bar{q_1})   \nn\\
\quad  +4\frac{q_1(\bar{q_1}+\bar{q_2})}{|\vec{k}_1||\vec{q}_1+\vec{q}_2|} \bar{k_1}^2(-k_1-q_1-q_2))  \Bigg\} -2\frac{(\bar{k_1}+\bar{k_2})}{|\vec{k}_1+\vec{k}_2|}q_1\bar{q_2}\Bigg\} \\
%
=  \int_{\slashed{k_1},\slashed{k_2},\slashed{k_3},\slashed{q}}\slashed{\delta}\left(\sum_{i=1}(\vec{k}_i+\vec{q}_i)\right)   f^{a_1a_2c}f^{b_1b_2c} J^{a_1}(\vec{k}_1)\theta^{a_2}(\vec{k}_2) \theta^{b_1}(\vec{q}_1)\theta^{b_2}(\vec{q}_2)\nn\\
\quad \Bigg\{\frac{1}{|\vec{k}_1|+|\vec{k}_2|+|\vec{q}_1+\vec{q}_2|} \Bigg\{  - \left(1+\frac{|\vec{k}_2|}{|\vec{k}_1|}\right) (k_1\bar{q}_2-\bar{k}_1q_2)\bar{k_1}  -\frac{q_1|\vec{q}_1+\vec{q}_2|}{|\vec{k}_1|} \bar{k_1}^2 \nn\\
\qquad -(2|\vec{k}_1+\vec{k}_2|+2|\vec{k}_2|+|\vec{k}_1|)q_1\bar{q_2}\frac{\bar{k_2}+\bar{k_1}}{|\vec{k}_1+\vec{k}_2|}  +\frac{q_1\bar{q_2}}{|\vec{q}_1+\vec{q}_2|} |\vec{k}_1|(-\bar{q_2}-\bar{k_1}-\bar{k_2})  +\frac{q_1\bar{q_1}}{|\vec{q}_1+\vec{q}_2|} |\vec{k}_1|\bar{q_1}  \Bigg\}  \nn\\
\quad -2\frac{(\bar{k_1}+\bar{k_2})}{|\vec{k}_1+\vec{k}_2|}q_1\bar{q_2}\Bigg\} \\
=  \int_{\slashed{k_1},\slashed{k_2},\slashed{k_3},\slashed{q}}\slashed{\delta}\left(\sum_{i=1}(\vec{k}_i+\vec{q}_i)\right)   f^{a_1a_2c}f^{b_1b_2c} J^{a_1}(\vec{k}_1)\theta^{a_2}(\vec{k}_2) \theta^{b_1}(\vec{q}_1)\theta^{b_2}(\vec{q}_2)\nn\\
\quad \Bigg\{\frac{1}{|\vec{k}_1|+|\vec{k}_2|+|\vec{q}_1+\vec{q}_2|} \Bigg\{ - \left(1+\frac{|\vec{k}_2|}{|\vec{k}_1|}\right) (k_1\bar{q}_2-\bar{k}_1q_2)\bar{k_1}  +\frac{q_2|\vec{q}_1+\vec{q}_2|}{|\vec{k}_1|} \bar{k_1}^2 \nn\\
\qquad -(2|\vec{k}_1+\vec{k}_2|+2|\vec{k}_2|+2|\vec{k}_1|)q_1\bar{q_2}\frac{\bar{k_2}+\bar{k_1}}{|\vec{k}_1+\vec{k}_2|}  -\frac{q_1\bar{q_2}}{|\vec{q}_1+\vec{q}_2|} |\vec{k}_1|\bar{q_2}  +\frac{q_1\bar{q_1}}{|\vec{q}_1+\vec{q}_2|} |\vec{k}_1|\bar{q_1}  \Bigg\}  \nn\\
\quad -2\frac{(\bar{k_1}+\bar{k_2})}{|\vec{k}_1+\vec{k}_2|}q_1\bar{q_2}\Bigg\} \,.
\eE
Again, we add and subtract what is missing.
\bE{l}
=  \int_{\slashed{k_1},\slashed{k_2},\slashed{k_3},\slashed{q}}\slashed{\delta}\left(\sum_{i=1}(\vec{k}_i+\vec{q}_i)\right)   f^{a_1a_2c}f^{b_1b_2c} J^{a_1}(\vec{k}_1)\theta^{a_2}(\vec{k}_2) \theta^{b_1}(\vec{q}_1)\theta^{b_2}(\vec{q}_2)\nn\\
\quad\Bigg\{  - \frac{\bar{k_1}}{|\vec{k}_1|} (k_1\bar{q}_2-\bar{k}_1q_2)  -2q_1\bar{q_2}\frac{\bar{k_2}+\bar{k_1}}{|\vec{k}_1+\vec{k}_2|}\nn\\
\quad +\frac{1}{|\vec{k}_1|+|\vec{k}_2|+|\vec{q}_1+\vec{q}_2|} \Bigg\{ +\frac{\bar{q_2}|\vec{q}_1+\vec{q}_2|}{|\vec{k}_1|} k_1\bar{k_1}  +\frac{\bar{q_1}^2(q_2+q_1)}{|\vec{q}_1+\vec{q}_2|} |\vec{k}_1|  \Bigg\}  \nn\\
\quad -2\frac{(\bar{k_1}+\bar{k_2})}{|\vec{k}_1+\vec{k}_2|}q_1\bar{q_2}\Bigg\}\\
=  \int_{\slashed{k_1},\slashed{k_2},\slashed{k_3},\slashed{q}}\slashed{\delta}\left(\sum_{i=1}(\vec{k}_i+\vec{q}_i)\right)   f^{a_1a_2c}f^{b_1b_2c} J^{a_1}(\vec{k}_1)\theta^{a_2}(\vec{k}_2) \theta^{b_1}(\vec{q}_1)\theta^{b_2}(\vec{q}_2)\nn\\
\quad \Bigg\{\frac{1}{4}\frac{|\vec{k}_1||\vec{q}_1+\vec{q}_2|}{|\vec{k}_1|+|\vec{k}_2|+|\vec{q}_1+\vec{q}_2|}  \frac{\bar{q_2}^2 +\bar{q_1}\bar{q_2}+\bar{q_1}^2}{(\bar{q_1}+\bar{q_2})}  \nn\\
\quad  - \frac{\bar{k_1}}{|\vec{k}_1|} (k_1\bar{q}_2-\bar{k}_1q_2) -4\frac{(\bar{k_1}+\bar{k_2})}{|\vec{k}_1+\vec{k}_2|}q_1\bar{q_2}\Bigg\}  + O(J^3)+\ldots+O(\theta^4) \,.
\eE
The second line vanishes under $\vec q_1\leftrightarrow \vec q_2$, so
\bE{rCl}
F^{(1)}_{GL}|_\O(e\theta^3) &=&  \int_{\slashed{k_1},\slashed{k_2},\slashed{k_3},\slashed{q}}\slashed{\delta}\left(\sum_{i=1}(\vec{k}_i+\vec{q}_i)\right)   f^{a_1a_2c}f^{b_1b_2c} J^{a_1}(\vec{k}_1)\theta^{a_2}(\vec{k}_2) \theta^{b_1}(\vec{q}_1)\theta^{b_2}(\vec{q}_2)\nn\\
&&\quad \Bigg\{ - \frac{\bar{k_1}}{|\vec{k}_1|} (k_1\bar{q}_2-\bar{k}_1q_2) -4\frac{(\bar{k_1}+\bar{k_2})}{|\vec{k}_1+\vec{k}_2|}q_1\bar{q_2}\Bigg\}  \,.\label{F1theta3}
\eE
Adding this to Eq.~(\ref{F24+F0theta3}) we obtain
\bE{l}
F^{(2,4)}_{GL}|_{\O(\theta^3)}  + F^{(0)}_{GL}|_{\O(e^2\theta^3)} + F^{(1)}_{GL}|_\O(e\theta^3) \nn\\
=    f^{a_1a_2c}f^{b_1b_2c}\int_{\slashed{k}_1,\slashed{k}_2,\slashed{q}_1,\slashed{q}_2}\slashed{\delta}\left(\sum_i(\vec{k}_i+\vec{q}_i)\right)  J^{a_1}(\vec{k}_1)\theta^{a_2}(\vec{k}_2) \theta^{b_1}(\vec{q}_1) \theta^{b_2}(\vec{q}_2)    \nn\\
\qquad  \Bigg\{- \frac{2}{3} \frac{\bar{k}_1}{|\vec{k}_1|} (k_1\bar{q}_2-\bar{k}_1q_2)  -2\frac{\bar{k}_1}{|\vec{k}_1|} q_1\bar q_2 \Bigg\} \\
=  -2 f^{b_1b_2c}f^{a_1a_2c}\int_{\slashed{k_1},\slashed{k_2},\slashed{q_1},\slashed{q_2}} \slashed{\delta}\left(\sum_i(\vec{k}_i+\vec{q}_i)\right) J^{a_1}(\vec{k}_1)\theta^{a_2}(\vec{k}_2) \theta^{b_1}(\vec{q}_1)\theta^{b_2}(\vec{q}_2)\nn\\
\qquad\qquad \frac{\bar{k}_1}{|\vec{k}_1|} \left(q_1\bar{q}_2 - \frac{1}{3} ((k_2+q_1)\bar{q}_2-(\bar{k}_2+\bar{q}_1)q_2) \right)  \\
=  -\frac{2}{3} f^{b_1b_2c}f^{a_1a_2c}\int_{\slashed{k_1},\slashed{k_2},\slashed{q_1},\slashed{q_2}} \slashed{\delta}\left(\sum_i(\vec{k}_i+\vec{q}_i)\right)  J^{a_1}(\vec{k}_1)\theta^{a_2}(\vec{k}_2) \theta^{b_1}(\vec{q}_1)\theta^{b_2}(\vec{q}_2)\nn\\
\qquad\qquad\frac{\bar{k}_1}{|\vec{k}_1|} \left(q_1\bar{q}_2 -  k_2\bar{q}_2+\bar{k}_2q_2 \right) \,.
\eE
Using the Jacobi-Identity $f^{a_1a_2c}f^{b_1b_2c}= -f^{a_1b_1c}f^{b_2a_2c} -f^{a_1b_2c}f^{a_2b_1c}$ in the second term then renaming $b_1 \leftrightarrow a_2$, $\vec q_1 \leftrightarrow \vec k_2$ in the (new) second term; and $b_2 \leftrightarrow a_2$, $\vec q_2 \leftrightarrow \vec k_2$ in the (new) third term this reduces to
\bE{l}
=  -\frac{2}{3} f^{b_1b_2c}f^{a_1a_2c}\int_{\slashed{k_1},\slashed{k_2},\slashed{q_1},\slashed{q_2}} \slashed{\delta}\left(\sum_i(\vec{k}_i+\vec{q}_i)\right)  J^{a_1}(\vec{k}_1)\theta^{a_2}(\vec{k}_2) \theta^{b_1}(\vec{q}_1)\theta^{b_2}(\vec{q}_2)\nn\\
\qquad\qquad \frac{\bar{k}_1}{|\vec{k}_1|} \left(q_1\bar{q}_2 -  q_1\bar{q}_2 -  q_2\bar{k}_2 +\bar{k}_2q_2 \right) \\
=0 \,.\label{F1=F24theta3}
\eE

\subsection{Order $\theta^4$}

We want this term to cancel Eqs.~(\ref{F24theta4}) and (\ref{F0theta4}).

\bE{l}
F^{(2,4)}_{GL}|_{\O(\theta^4)}  + F^{(0)}_{GL}|_{\O(e^2\theta^4)} \nn\\
=  -2 f^{a_1a_2c}f^{b_1b_2c}\int_{\slashed{k_1},\slashed{k_2},\slashed{q_1},\slashed{q_2}} \slashed{\delta}\left(\sum_i(\vec{k}_i+\vec{q}_i)\right) \frac{ k_1\bar{k}_2q_1\bar{q}_2}{|\vec{k}_1+\vec{k}_2|} \theta^{a_1}(\vec{k}_1)\theta^{a_2}(\vec{k}_2)\theta^{b_1}(\vec{q}_1)  \theta^{b_2}(\vec{q}_2) \nn\\
 -2f^{a_1a_2c}f^{b_1b_2c}\int_{\slashed{k_1},\slashed{k_2},\slashed{q_1},\slashed{q_2}} \slashed{\delta}\left(\sum_i(\vec{k}_i+\vec{q}_i)\right)   \frac{\bar{k_2}k_1\bar{q_2}q_1}{|\vec{k}_1+\vec{k}_2|}  \theta^{a_1}(\vec{k}_1)\theta^{a_2}(\vec{k}_2)\theta^{b_1}(\vec{q}_1)  \theta^{b_2}(\vec{q}_2)  
\nn\\
 = -4f^{a_1a_2c}f^{b_1b_2c}\int_{\slashed{k_1},\slashed{k_2},\slashed{q_1},\slashed{q_2}} \slashed{\delta}\left(\sum_i(\vec{k}_i+\vec{q}_i)\right)   \frac{\bar{k_2}k_1\bar{q_2}q_1}{|\vec{k}_1+\vec{k}_2|}  \theta^{a_1}(\vec{k}_1)\theta^{a_2}(\vec{k}_2)\theta^{b_1}(\vec{q}_1)  \theta^{b_2}(\vec{q}_2)  \,.
\eE

We extract the $\O(e\theta^4)$ portion from Eq.~(\ref{F1GLref}) and bring it in this form:

\bE{l}
F^{(1)}_{GL}|_{\O(e\theta^4)} \nn\\
= -2if^{abc} \int_{\slashed{k_1},\slashed{k_2},\slashed{k_3},\slashed{q}}\slashed{\delta}\left(\sum_{i=1}^3 \vec{k}_i\right) \frac{1}{\sum_{i=1}^3 |\vec{k}_i|}   \nn\\
\quad \Bigg\{  \frac{-2i}{2} f^{ac_1c_2}(k_1\bar{q}-\bar{k}_1q)\,\theta^{c_1}(\vec{q})\theta^{c_2}(\vec{k}_1-\vec{q}) k_2\bar{k}_3\theta^b(\vec{k}_2)\theta^c(\vec{k}_3)  \nn\\
\quad - \frac{4}{|\vec{k}_1||\vec{k}_3|} k_1\bar{k_1}\theta^a(\vec{k}_1) (k_3\bar{k_2}-\bar{k_3}k_2)\theta^b(\vec{k}_2) \frac{2i^3}{2} f^{cc_1c_2} (k_3\bar{q}-\bar{k}_3q)\,\theta^{c_1}(\vec{q})\theta^{c_2}(\vec{k}_3-\vec{q}) \Bigg\} \\
= -2f^{a_1a_2c}f^{b_1b_2c}\int_{\slashed{k_1},\slashed{k_2},\slashed{q_1},\slashed{q_2}} \slashed{\delta}\left(\sum_{i=1}^2(\vec{k}_i+\vec{q}_i)\right)  \theta^{a_1}(\vec{k}_1)\theta^{a_2}(\vec{k}_2)\theta^{b_1}(\vec{q}_1)  \theta^{b_2}(\vec{q}_2)   \frac{1}{|\vec{q}_1+\vec{q}_2|+|\vec{k}_1|+|\vec{k}_2|}   \nn\\
\qquad \Bigg\{(q_2\bar{q_1}-\bar{q_2}q_1)k_1\bar{k_2} - \frac{|\vec{k}_1|}{|\vec{q}_1+\vec{q}_2|}  ((q_2+q_1)\bar{k_2}-(\bar{q_2}+\bar{q_1})k_2)(q_2\bar{q_1}-\bar{q_2}q_1)  \Bigg\} \\
= 4f^{a_1a_2c}f^{b_1b_2c}\int_{\slashed{k_1},\slashed{k_2},\slashed{q_1},\slashed{q_2}} \slashed{\delta}\left(\sum_i(\vec{k}_i+\vec{q}_i)\right)  \theta^{a_1}(\vec{k}_1)\theta^{a_2}(\vec{k}_2)\theta^{b_1}(\vec{q}_1)  \theta^{b_2}(\vec{q}_2)  \frac{k_1\bar{k_2}q_1\bar{q_2}}{|\vec{q}_1+\vec{q}_2|}   \label{F1theta4}\\
=-F^{(2,4)}_{GL}|_{\O(\theta^4)}  - F^{(0)}_{GL}|_{\O(e^2\theta^4)} \,. \label{F1=F24theta4}
\eE

\subsection{All orders}

Adding up Eqs.~(\ref{F1theta1}), (\ref{F1theta2}), (\ref{F1theta3}) and  (\ref{F1theta4}) we find

\begin{IEEEeqnarray}{rCl}
F^{(1)}_{GL}|_{\O(e)} &=&-f^{a_1a_2c}f^{b_1b_2c}\int_{\slashed{k_1},\slashed{k_2},\slashed{q_1},\slashed{q_2}}\slashed{\delta}\left(\sum(\vec{k}_i+\vec{q}_i)\right) 
\frac{g^{(3)}(\vec{k}_1,\vec{k}_2,-\vec{k}_1-\vec{k}_2)}{32}   \nn\\
&&\qquad\qquad J^{a_1}(\vec{k}_1)J^{a_2}(\vec{k}_2) J^{b_1}(\vec{q}_1) \theta^{b_2}(\vec{q}_2) \nn\\
&&
+f^{a_1a_2c}f^{b_1b_2c} \int_{\slashed{k_1},\slashed{k_2},\slashed{q_1},\slashed{q_2}}  \slashed{\delta}\left(\sum_{i=1}^2 (\vec{k}_i+\vec
{q}_i)\right)   J^{a_1}(\vec{k_1})  J^{a_2}(\vec{k}_2) \theta^{b_1}(\vec{q}_1) \theta^{b_2}(\vec{q}_2)\nn\\
&&
\qquad\qquad  \times\Bigg(\frac{\bar{q}_2q_1}{\bar{q}_1+\bar{q}_2} \frac{g^{(3)}(\vec{k}_1,\vec{k}_2, -\vec{k}_1-\vec{k}_2)}{16} - \frac{\bar{q}_2}{(\bar{q}_1+\bar{q}_2)} \frac{\bar{k}_2^2}{2|\vec{k}_2|}\Bigg)  \nn\\
&&
 -f^{a_1a_2c}f^{b_1b_2c} \int_{\slashed{k_1},\slashed{k_2},\slashed{q_1},\slashed{q_2}} 
\slashed{\delta}\left(\sum_{i=1}^2 (\vec{k}_i+\vec{q}_i)\right) 
\left( \frac{(\bar{k}_1+\bar{k}_2)^2}{|\vec{k}_1+\vec{k}_2|} -\frac{ \bar{k}_1^2}{|\vec{k}_1|} \right)   \nn\\
&& \qquad\qquad J^{a_1}(\vec{k}_1) \theta^{a_2}(\vec{k}_2) J^{b_1}(\vec{q}_1)\theta^{b_2}(\vec{q}_2) \nn\\
&& 
-f^{a_1a_2c}f^{b_1b_2c}\int_{\slashed{k_1},\slashed{k_2},\slashed{k_3},\slashed{q}}\slashed{\delta}\left(\sum_{i=1}(\vec{k}_i+\vec{q}_i)\right)    J^{a_1}(\vec{k}_1)\theta^{a_2}(\vec{k}_2) \theta^{b_1}(\vec{q}_1)\theta^{b_2}(\vec{q}_2)\nn\\
&&
\qquad\qquad \times\Bigg( \frac{\bar{k_1}}{|\vec{k}_1|} (k_1\bar{q}_2-\bar{k}_1q_2) +4\frac{(\bar{k_1}+\bar{k_2})}{|\vec{k}_1+\vec{k}_2|}q_1\bar{q_2}\Bigg) \nn \\
&&
+4f^{a_1a_2c}f^{b_1b_2c}\int_{\slashed{k_1},\slashed{k_2},\slashed{q_1},\slashed{q_2}} \slashed{\delta}\left(\sum^2_i(\vec{k}_i+\vec{q}_i)\right)  \frac{k_1\bar{k_2}q_1\bar{q_2}}{|\vec{q}_1+\vec{q}_2|} \nn\\
&&\qquad\qquad \theta^{a_1}(\vec{k}_1)\theta^{a_2}(\vec{k}_2)\theta^{b_1}(\vec{q}_1)  \theta^{b_2}(\vec{q}_2)   . \label{F1GLApp}
\end{IEEEeqnarray}

\section{Conclusion}

From Eqs.~(\ref{F24theta0}), (\ref{F1=F24theta1}), (\ref{F1=F24theta2}), (\ref{F1=F24theta3}), and  (\ref{F1=F24theta4}) (or equivalently: from summing up Eqs.~(\ref{F24GLApp}), (\ref{F0GLApp}) and (\ref{F1GLApp})) we find
\be
F^{(0)}_{GL}[J,\theta]|_{\O(e^2)}  + F^{(1)}_{GL}[J,\theta]|_{\O(e)} +F^{(2,4)}_{GL}[J,\theta]|_{\O(e^0)} = F^{(2,4)}_{GL}[J] \,.
\ee

This, together with Eqs.~(\ref{F0GI3}) and (\ref{F1equal}) means that
\[
  F^{(0)}_{GL}+eF^{(1)}_{GL}+e^2F^{(2,4)}_{GL} = F^{(0)}_{GI}+eF^{(1)}_{GI}+e^2F^{(2,4)}_{GI} +\O(e^3)\,.
\]

\chapter{Hermiticity of the regularized Hamiltonian}
\label{app:herm}
\section{Functional derivative of $M_i$}
\label{sec:MoverA}

Starting from the adjoint version of Eq.~(\ref{D_iA_i}) (No sum over repeated spatial indices in this appendix) 
\bea
D_i^{ab}(\vec y)M_i^{bc}(\vec y) = (\p_i^y\de^{ab}-ef^{abd}A_i^d(\vec y))M_i^{bc}(\vec y)  =0
\eea
we can compute the functional derivative of this object with respect to $A_j$:
\bE{rCl}
\frac{\de}{\de A_j^e (\vec x)}D_i^{ab}(\vec y)M_i^{bc}(\vec y) &=& -ef^{abe}\de_{ij}\de(\vec y-\vec x)M_i^{bc}(\vec y)  + D_i^{ab}(\vec y) \frac{\de M_i^{bc}(\vec y)}{\de A_j^e (\vec x)}=0 \\
\Longleftrightarrow  \frac{\de M_i^{bc}(\vec y)}{\de A_j^e (\vec x)} &=& e\int_z \left(D_i^{-1}\right)^{ba}_{yz} f^{afe}\de_{ij}\de(\vec z-\vec x)M_i^{fc}(\vec z) \\
&=& e\de_{ij}[M_i(\vec y)G_i(\vec{y}-\vec{x})M_i^{-1}(\vec x)]^{ba} f^{afe} M_i^{fc}(\vec x) \\
&=& e\de_{ij}M_i^{bg}(\vec y)G_i(\vec{y}-\vec{x}) f^{gch} M_i^{eh}(\vec x) \,. \label{M_iOverA_j}
\eE
In the fundamental representation the derivative of $M_j$ is given by 
\bE{rCl}
 \frac{\de M_j(\vec y)}{\de A_i^a (\vec x)} &=& ie \de_{ij} M_j(\vec y)T^d G_i(\vec y, \vec x)M_i^{ed}(\vec x) \,. \label{M_iOverA_j:app}
\eE
This can easily be checked by plugging it into the definition of $M_i^{ab}$ (see Eq.~(\ref{Mab})).

\section{Functional derivative of the string}
\label{sec:herm}
We use Eq.~(\ref{M_iOverA_j:app}) to compute
\be
\half\sum_i\int_{x,v}\de_\mu(\vec x,\vec v) \left[\frac{\de}{\de A_i^a(\vec x)}\Phi_{ab}(\vec x,\vec v)\right]\frac{\de}{\de A_i^b(\vec v)} 
\ee
This is actually an ill-defined quantity, so we have to regularize it. We do this by moving the derivative an infinitesimal step $\vec X$ away from the point $\vec x$ and introduce a new regularized delta function and a second string. We then take the limit $\nu \rightarrow \infty$ for finite $\mu$. 
\bE{ll}
\half&\lim_{\nu\to\infty}\sum_i\int_{x,v,X}\de_\mu(\vec x,\vec v)\de_\nu(\vec X)\Phi_{ar}(\vec x,\vec x+\vec X)\left[\frac{\de}{\de A_i^r(\vec x+\vec X)}\Phi_{ab}(\vec x,\vec v)\right]\frac{\de}{\de A_i^b(\vec v)} \nn\\
&={1\over4}\lim_{\nu\to\infty} \sum_i\int_{x,v}\de_\mu(\vec x,\vec v) \de_\nu(\vec X)\Phi_{ar}(\vec x,\vec x+\vec X)  \nn\\
&\qquad \times \frac{\de}{\de A_i^r(\vec x+\vec X)} \Bigg[M_1(\vec x)M_1^{-1}(v_1,x_2)M_2(v_1,x_2)M_2^{-1}(\vec v) \nn\\
&\qquad \qquad +M_2(\vec x)M_2^{-1}(x_1,v_2)M_1(x_1,v_2)M_1^{-1}(\vec v)\Bigg]^{ab}\frac{\de}{\de A_i^b(\vec v)} \qquad\\
&={e\over4}\lim_{\nu\to\infty} \sum_i\int_{x,v} \de_\mu(\vec x,\vec v) \de_\nu(\vec X)\Phi_{ar}(\vec x,\vec x+\vec X) \nn\\
&\qquad\times\Bigg[\de_{i1}M_1^{ag}(\vec x)G_1(-\vec{X}) f^{gch} M_1^{rh}(\vec x+\vec X) [M_1^{-1}(v_1,x_2)M_2(v_1,x_2)M_2^{-1}(\vec v) ]^{cb} \nn\\
&\qquad \qquad -[M_1(\vec x)M_1^{-1}(v_1,x_2)]^{ac} \de_{i1}M_1^{cg}(v_1,x_2)G_1((v_1,x_2)-\vec{x}-\vec X) f^{gdh} M_1^{rh}(\vec x +\vec X) \nn\\
&\qquad \qquad\qquad \times [M_1^{-1}(v_1,x_2)M_2(v_1,x_2)M_2^{-1}(\vec v)]^{db} \nn\\
&\qquad \qquad +[M_1(\vec x)M_1^{-1}(v_1,x_2)]^{ac} \nn\\
&\qquad \qquad\qquad \times \de_{i2}M_2^{cg}(v_1,x_2)G_2((v_1,x_2)-\vec{x}-\vec X) f^{gdh} M_2^{rh}(\vec x+\vec X) [M_2^{-1}(\vec v)]^{db} \nn\\
&\qquad \qquad -[M_1(\vec x)M_1^{-1}(v_1,x_2)M_2(v_1,x_2)M_2^{-1}(\vec v)]^{ac} \nn\\
&\qquad \qquad\qquad \times \de_{i2}M_2^{cg}(\vec v)G_2(\vec v-\vec{x}-\vec X) f^{gdh} M_2^{rh}(\vec x+\vec X) [M_2^{-1}(\vec v)]^{db} \nn\\ 
&\qquad \qquad +\de_{i2}M_2^{ag}(\vec x)G_2(-\vec X) f^{gch} M_2^{rh}(\vec x+\vec X) [M_2^{-1}(x_1,v_2)M_1(x_1,v_2)M_1^{-1}(\vec v)]^{cb} \nn\\
&\qquad \qquad -[M_2(\vec x)M_2^{-1}(x_1,v_2)]^{ac} \de_{i2}M_2^{cg}(x_1,v_2)G_2((x_1,v_2)-\vec{x}-\vec X) f^{gdh} M_2^{rh}(\vec x+\vec X) \nn\\
&\qquad \qquad\qquad \times [M_2^{-1}(x_1,v_2)M_1(x_1,v_2)M_1^{-1}(\vec v)]^{db} \nn\\
&\qquad \qquad +[M_2(\vec x)M_2^{-1}(x_1,v_2)]^{ac}  \nn\\
&\qquad \qquad\qquad \times \de_{i1}M_1^{cg}(x_1,v_2)G_1((x_1,v_2)-\vec{x}-\vec X) f^{gdh} M_1^{rh}(\vec x+\vec X) [M_1^{-1}(\vec v)]^{db} \nn\\
&\qquad \qquad -[M_2(\vec x)M_2^{-1}(x_1,v_2)M_1(x_1,v_2)M_1^{-1}(\vec v)]^{ac}\nn\\
&\qquad \qquad\qquad \times  \de_{i1}M_1^{cg}(\vec v)G_1(\vec v-\vec{x}-\vec X) f^{gdh} M_1^{rh}(\vec x+\vec X) [M_1^{-1}(\vec v)]^{db} \Bigg] \frac{\de}{\de A_i^b(\vec v)} 
\nn
\\
&
=:\lim_{\nu\to\infty}\sum_{i=1}^8T_i\,.
\eE
In the third, fourth, seventh and and eighth term we can take the limit of $\nu\to\infty$ without problems. With $\Phi_{ar}(\vec x,\vec x) = \de^{ar}$ and after integrating the delta functions inside the Green's functions we find for these terms:
\bE{ll}
\lim_{\nu\to\infty}(&T_3+T_4+T_7+T_8) \nn\\
&={e\over4}\int_{x,v_2}\frac{\mu}{\sqrt{\pi}}\de_\mu(x_2-v_2)  \Bigg[[M_1(\vec x)M_1^{-1}(\vec x)]^{ac} M_2^{cg}(\vec x)\th(0) f^{gdh} M_2^{ah}(\vec x) [M_2^{-1}(x_1,v_2)]^{db} \nn\\
&\qquad \qquad -[M_1(\vec x)M_1^{-1}(\vec x)M_2(\vec x)M_2^{-1}(\vec v)]^{ac} \nn\\
&\qquad \qquad\qquad \times  M_2^{cg}(\vec v)\th(v_2-x_2) f^{gdh} M_2^{ah}(\vec x) [M_2^{-1}(x_1,v_2)]^{db} \Bigg] \frac{\de}{\de A_2^b(x_1,v_2)} \nn\\ 
&\quad + {e\over4}\int_{x,v_1} \de_\mu(x_1-v_1)\frac{\mu}{\sqrt{\pi}}\Bigg[[M_2(\vec x)M_2^{-1}(\vec x)]^{ac} M_1^{cg}(\vec x)\th(0) f^{gdh} M_1^{ah}(\vec x) [M_1^{-1}(v_1,x_2)]^{db} \nn\\
&\qquad \qquad -[M_2(\vec x)M_2^{-1}(\vec x)M_1(\vec x)M_1^{-1}(\vec v)]^{ac}\nn\\
&\qquad \qquad\qquad \times M_1^{cg}(\vec v)\th(v_1-x_1) f^{gdh} M_1^{ah}(\vec x) [M_1^{-1}(v_1,x_2)]^{db} \Bigg] \frac{\de}{\de A_1^b(v_1,x_2)} \\
&=0
\eE
All of these terms vanish under color contraction. We are thus left with
\bE{ll}
\half&\lim_{\nu\to\infty}\sum_i\int_{x,v,X}\de_\mu(\vec x,\vec v)\de_\nu(\vec X)\Phi_{ar}(\vec x,\vec x+\vec X)\left[\frac{\de}{\de A_i^r(\vec x+\vec X)}\Phi_{ab}(\vec x,\vec v)\right]\frac{\de}{\de A_i^b(\vec v)} \nn\\
&=\lim_{\nu\to\infty}(T_1+T_2+T_5+T_6) \nn\\
&={e\over4}\lim_{\nu\to\infty}\int_{X,x,v}\Phi_{ar}(\vec x,\vec x+\vec X)\de_\mu(\vec x,\vec v)\de_\nu(\vec X) \Big[\th(-X_1)-\th(v_1-x_1-X_1)\Big]\de(-X_2) \nn\\
&\qquad\qquad M_1^{ag}(\vec x)f^{gch} M_1^{rh}(\vec x+\vec X) [M_1^{-1}(v_1,x_2)M_2(v_1,x_2)M_2^{-1}(\vec v) ]^{cb} \frac{\de}{\de A_1^b(\vec v)} \nn\\
&\quad +{e\over4}\lim_{\nu\to\infty}\int_{X,x,v}\Phi_{ar}(\vec x,\vec x+\vec X)\de_\mu(\vec x,\vec v)\de_\nu(\vec X) \de(-X_1)\Big[\th(-X_2)-\th(v_2-x_2-X_2)\Big]  \nn\\
&\qquad\qquad M_2^{ag}(\vec x) f^{gch} M_2^{rh}(\vec x +\vec X) [M_2^{-1}(x_1,v_2)M_1(x_1,v_2)M_1^{-1}(\vec v)]^{cb}  \frac{\de}{\de A_2^b(\vec v)} \\
&={e\over4}\lim_{\nu\to\infty}\int_{X_1,x,v}\Phi_{ar}(\vec x,\vec x+\vec X)|_{X_2=0} \de_\mu(\vec x,\vec v)\de_\nu(X_1)\frac{\nu}{\sqrt{\pi}} \Big[\th(-X_1)-\th(v_1-x_1-X_1)\Big] \nn\\
&\qquad\qquad M_1^{ag}(\vec x)f^{gch} M_1^{rh}(x_1+X_1,x_2) [M_1^{-1}(v_1,x_2)M_2(v_1,x_2)M_2^{-1}(\vec v) ]^{cb} \frac{\de}{\de A_1^b(\vec v)} \nn\\
&\quad +{e\over4}\lim_{\nu\to\infty}\int_{X_2,x,v}\Phi_{ar}(\vec x,\vec x+\vec X)|_{X_1=0} \de_\mu(\vec x,\vec v)\de_\nu(X_2) \frac{\nu}{\sqrt{\pi}} \Big[\th(-X_2)-\th(v_2-x_2-X_2)\Big]  \nn\\
&\qquad\qquad M_2^{ag}(\vec x) f^{gch} M_2^{rh}(x_1,x_2+X_2) [M_2^{-1}(x_1,v_2)M_1(x_1,v_2)M_1^{-1}(\vec v)]^{cb}  \frac{\de}{\de A_2^b(\vec v)} \,.
\eE
With Eq.~(\ref{Phi}):
\be
\Phi^{ab}(u,v)= {1\over 2} (M_1(u)M_1^{-1}(v_1,u_2)M_2(v_1,u_2)M_2^{-1}(v)+M_2(u)M_2^{-1}(u_1,v_2)M_1(u_1,v_2)M_1^{-1}(v))^{ab} 
\ee
this is
\bE{ll}
&={e\over4}\lim_{\nu\to\infty}\int_{X_1,x,v}(M_1(\vec x)M_1^{-1}(x_1+X_1,x_2))^{ar}  \de_\mu(\vec x,\vec v)\de_\nu(X_1)\frac{\nu}{\sqrt{\pi}} \Big[\th(-X_1)-\th(v_1-x_1-X_1)\Big] \nn\\
&\qquad\qquad M_1^{ag}(\vec x)f^{gch} M_1^{rh}(x_1+X_1,x_2) [M_1^{-1}(v_1,x_2)M_2(v_1,x_2)M_2^{-1}(\vec v) ]^{cb} \frac{\de}{\de A_1^b(\vec v)} \nn\\
&\quad +{e\over4}\lim_{\nu\to\infty}\int_{X_2,x,v} (M_2(\vec x)M_2^{-1}(x_1,x_2+X_2))^{ar}\de_\mu(\vec x,\vec v)\de_\nu(X_2) \frac{\nu}{\sqrt{\pi}} \Big[\th(-X_2)-\th(v_2-x_2-X_2)\Big]  \nn\\
&\qquad\qquad M_2^{ag}(\vec x) f^{gch} M_2^{rh}(x_1,x_2+X_2) [M_2^{-1}(x_1,v_2)M_1(x_1,v_2)M_1^{-1}(\vec v)]^{cb}  \frac{\de}{\de A_2^b(\vec v)} \\
&={e\over4}\lim_{\nu\to\infty}\int_{X_1,x,v}\de_\mu(\vec x,\vec v)\de_\nu(X_1)\frac{\nu}{\sqrt{\pi}} \Big[\th(-X_1)-\th(v_1-x_1-X_1)\Big] \nn\\
&\qquad\qquad \de^{gh}f^{gch} [M_1^{-1}(v_1,x_2)M_2(v_1,x_2)M_2^{-1}(\vec v) ]^{cb} \frac{\de}{\de A_1^b(\vec v)} \nn\\
&\quad +{e\over4}\lim_{\nu\to\infty}\int_{X_2,x,v} \de_\mu(\vec x,\vec v)\de_\nu(X_2) \frac{\nu}{\sqrt{\pi}} \Big[\th(-X_2)-\th(v_2-x_2-X_2)\Big]  \nn\\
&\qquad\qquad \de^{gh}f^{gch} [M_2^{-1}(x_1,v_2)M_1(x_1,v_2)M_1^{-1}(\vec v)]^{cb}  \frac{\de}{\de A_2^b(\vec v)} \\
&=0
\eE
Again, these terms vanish under color contraction. Hence we conclude that
\be
\mathcal{T}_{reg}=-{1\over2}\int_{x,v} \delta_\mu(\vec x, \vec v)  \frac{\delta}{\delta A_i^a(\vec x)}  \Phi_{ab}(\vec x,\vec v) \frac{\delta}{\delta A_i^b(\vec v)} =-{1\over2}\int_{x,v} \delta_\mu(\vec x, \vec v) \Phi_{ab}(\vec x,\vec v) \frac{\delta}{\delta A_i^a(\vec x)}  \frac{\delta}{\delta A_i^b(\vec v)}\,.
\label{Tregeq}
\ee
to all orders in perturbation theory. This confirms that Eq. (\ref{Treg}) is Hermitian.
Finally, as a check, we have also performed the above computation, using the explicit form of the string, to ${\cal O}(e^2)$. 
\chapter{Computation of the vanishing terms of the regularized Hamiltonians}

\section{$\O(e)$ correction to the gauge field Hamiltonian, Eq.~(\ref{regF1eq})}
\label{subsec:append4}

\bE{l}
-{1\over2}\int_{u,v}\delta_\mu(\vec u,\vec v) \Phi_{ab}^{(1)}(\vec u,\vec v) \frac{\delta F_{GL}^{(0)}}{\delta  A_i^a(\vec u)} \frac{\delta F_{GL}^{(0)}}{\delta A_i^b(\vec v)} \nn\\
= - {1\over 8\pi^2} \int_{u,v,y,w}\p_{u_i}\p_{v_i}  \Big( \delta_\mu(\vec u,\vec v) \Phi_{ab}^{(1)}(\vec u,\vec v)  \Big)  \frac{1}{|\vec{u}-\vec{y}|}\frac{1}{|\vec{v}-\vec{w}|}  (\vec{\nabla}\times\vec{A}^a(\vec{y}))(\vec{\nabla}\times\vec{A}^b(\vec{w}))  \\
= - {e\over 4\pi^2}f^{abc} \int_{U,v,x,y,w}   \delta_\mu(\vec U) \Bigg(  A_1^c(\vec y)  (G_1(\vec U+\vec v-\vec y)-G_1(v_1-y_1,U_2+v_2-y_2)  \nn\\
\qquad \qquad +G_1(U_1+v_1-y_1,v_2-y_2)-G_1(\vec v-\vec y))  \nn\\
\qquad\quad  +  A_2^c(\vec y) (G_2(v_1-y_1,U_2+v_2-y_2)-G_2(\vec v-\vec y)  \nn\\
\qquad \qquad +G_2(\vec U+\vec v-\vec y)-G_2(U_1+v_1-y_1,v_2-y_2))  \Bigg) \nn\\
\quad \frac{\mu^2 -\mu^4|\vec{U}|^2}{|\vec{U}+\vec{v}-\vec{x}||\vec{v}-\vec{w}|}  (\vec{\nabla}\times\vec{A}^a(\vec{x}))(\vec{\nabla}\times\vec{A}^b(\vec{w})) \nn\\
\quad -{e\over 4\pi^2}f^{abc} \int_{U,v,x,w}   \delta_\mu(\vec U)  \frac{\mu^2}{|\vec{U}+\vec{v}-\vec{x}||\vec{v}-\vec{w}|}  (\vec{\nabla}\times\vec{A}^a(\vec{x}))(\vec{\nabla}\times\vec{A}^b(\vec{w}))\nn\\
 \quad \Bigg\{ U_1 (A_1^c(v_1,U_2+v_2)+A_1^c(\vec v)) + U_2 (A_2^c(\vec v)+A_2^c(U_1+v_1,v_2)) \Bigg\} \nn\\
\quad +{e\over 4\pi^2}f^{abc} \int_{U,v,x,y,w}   \delta_\mu(\vec U)  \frac{\mu^2}{|\vec{U}+\vec{v}-\vec{x}||\vec{v}-\vec{w}|}  (\vec{\nabla}\times\vec{A}^a(\vec{x}))(\vec{\nabla}\times\vec{A}^b(\vec{w}))\nn\\ \quad \Bigg\{U_1 (G_2(v_1,U_2+v_1;\vec y)-G_2(\vec v;\vec y)) \p_1A_2^c(\vec y)  \nn\\
\qquad\qquad + U_2 (G_1(U_1+v_1,v_2;\vec y)-G_1(\vec v;\vec y)) \p_2 A_1^c(\vec y)  \Bigg\} 
\eE
Except for $\delta_\mu(\vec U)$, we Taylor expand this expression in powers of $\vec U$. The first integral up to 4th order, the other two up to 2nd order.
\bE{l}
= - {e\over 4\pi^2}f^{abc} \int_{v,x,w} \Bigg(\frac{1}{|\vec{v}-\vec{x}|^2} (\vec{v}-\vec{x})\cdot\vec{A}^c(\vec{v}) +\frac{1}{2}\nabla\cdot\vec{A}^c(\vec{v})   \Bigg) \frac{1}{|\vec{v}-\vec{x}||\vec{v}-\vec{w}|}   \nn\\
\qquad\qquad \times (\vec{\nabla}\times\vec{A}^a(\vec{x}))(\vec{\nabla}\times\vec{A}^b(\vec{w})) \nn\\
\quad -{e\over 4\pi^2}f^{abc} \int_{v,x,w}  \frac{1}{|\vec{v}-\vec{x}||\vec{v}-\vec{w}|}  (\vec{\nabla}\times\vec{A}^a(\vec{x}))(\vec{\nabla}\times\vec{A}^b(\vec{w}))  \Bigg\{ -\frac{1}{|\vec{v}-\vec{x}|^2} (\vec{v}-\vec{x})\cdot\vec{A}^c(\vec{v}) \Bigg\} \nn\\
\quad +{e\over 4\pi^2}f^{abc} \int_{v,x,y,w}    \frac{\mu^2}{|\vec{v}-\vec{x}||\vec{v}-\vec{w}|}  (\vec{\nabla}\times\vec{A}^a(\vec{x}))(\vec{\nabla}\times\vec{A}^b(\vec{w})) \Bigg\{ 0 \Bigg\} +\mathcal{O}(\mu^{-2})
\eE
\bE{l}
 = - {e\over 8\pi^2}f^{abc} \int_{v,x,w}  \frac{1}{|\vec{v}-\vec{x}||\vec{v}-\vec{w}|}  (\vec{\nabla}\times\vec{A}^a(\vec{x}))(\vec{\nabla}\times\vec{A}^b(\vec{w})) (\nabla\cdot\vec{A}^c(\vec{v})) + \mathcal{O}(\mu^{-2}) \\
=\mathcal{O}(\mu^{-2})
\eE
The $\mathcal{O}(\mu^0)$ term vanishes under combined interchange of $\{\vec x\leftrightarrow \vec w,a\leftrightarrow b\}$.

\section{$\O(e^2A^4)$ corrections to the gauge field Hamiltonian, Eq.~(\ref{regF24eq})}
\label{subsec:append5}
Vanishing of the first term:
\bE{l}
-{1\over2}\int_{u,v}\delta_\mu(\vec u,\vec v) \Phi_{ab}^{(2)}(\vec u,\vec v) \frac{\delta F_{GL}^{(0)}}{\delta  A_i^a(\vec u)} \frac{\delta F_{GL}^{(0)}}{\delta A_i^b(\vec v)} \nn\\
= - {1\over 8\pi^2} \int_{u,v,r,w}\p_{u_i}\p_{v_i}  \Big( \delta_\mu(\vec u,\vec v) \Phi_{ab}^{(2)}(\vec u,\vec v)  \Big)  \frac{1}{|\vec{u}-\vec{r}|}\frac{1}{|\vec{v}-\vec{w}|}  (\vec{\nabla}\times\vec{A}^a(\vec{r}))(\vec{\nabla}\times\vec{A}^b(\vec{w}))  \\
= - {1\over 4\pi^2}  f^{adc}f^{dbe} \int_{u,v,r,w,y,z} \delta_\mu(\vec u,\vec v)  \frac{\mu^2-\mu^4(\vec{u}-\vec{v})^2}{|\vec{u}-\vec{r}||\vec{v}-\vec{w}|}  (\vec{\nabla}\times\vec{A}^a(\vec{r}))(\vec{\nabla}\times\vec{A}^b(\vec{w})) \nn\\
\qquad\Bigg\{ \Big( (G_1(\vec u;\vec z)-G_1(v_1,u_2;\vec z))(G_1(\vec z;\vec y)-G_1(v_1,u_2;\vec y))\nn\\
\qquad \qquad + (G_1(u_1,v_2;\vec z)-G_1(\vec v;\vec z))(G_1(\vec z;\vec y)-G_1(\vec v;\vec y))\Big)A_1^c(\vec z)A_1^e(\vec y)  \nn\\
 \qquad+  \Big( (G_2(v_1,u_2;\vec z)-G_2(\vec v;\vec z))(G_2(\vec z;\vec y)-G_2(\vec v;\vec y))\nn\\
 \qquad\qquad+ (G_2(\vec u;\vec z)-G_2(u_1,v_2;\vec z))(G_2(\vec z;\vec y)-G_2(u_1,v_2;\vec y))\Big)A_2^c(\vec z)A_2^e(\vec y)  \nn\\
\qquad +   (G_1(\vec u;\vec y)-G_1(v_1,u_2;\vec y))(G_2(v_1,u_2;\vec z)-G_2(\vec v;\vec z))A_1^c(\vec y) A_2^e(\vec z) \nn\\
 \qquad+  (G_2(\vec u;\vec z)-G_2(u_1,v_2;\vec z))(G_1(u_1,v_2;\vec y)-G_1(\vec v;\vec y))A_2^c(\vec z)A_1^e(\vec y) \Bigg\} \nn\\
 + {1\over 4\pi^2} f^{adc}f^{dbe} \int_{u,v,r,w,y,z}  \frac{\mu^2}{|\vec{u}-\vec{r}| |\vec{v}-\vec{w}|}  (\vec{\nabla}\times\vec{A}^a(\vec{r}))(\vec{\nabla}\times\vec{A}^b(\vec{w}))  \delta_\mu(\vec u,\vec v) \nn\\
\qquad \Bigg\{(u_1-v_1)\Bigg( (G_2(v_1,u_2;\vec z)-G_2(\vec v;\vec z))(G_2(\vec z;\vec y)-G_2(\vec v;\vec y))   \nn\\
\qquad\qquad\qquad\times (\p_1A_2^c(\vec z)A_2^e(\vec y) +A_2^c(\vec z)\p_1A_2^e(\vec y)  ) \nn\\
\qquad +  (G_1(\vec u;\vec y)-G_1(v_1,u_2;\vec y)) (G_2(v_1,u_2;\vec z)-G_2(\vec v;\vec z)) A_1^c(\vec y)  \p_1 A_2^e(\vec z)  \Bigg) \nn\\
\qquad +(u_2-v_2) \Bigg(  (G_1(u_1,v_2;\vec z)-G_1(\vec v;\vec z))(G_1(\vec z;\vec y)-G_1(\vec v;\vec y))   \nn\\
\qquad\qquad\qquad\times(\p_2A_1^c(\vec z)A_1^e(\vec y) + A_1^c(\vec z) \p_2A_1^e(\vec y) ) \nn\\
 \qquad+  (G_2(\vec u;\vec z)-G_2(u_1,v_2;\vec z))(G_1(u_1,v_2;\vec y)-G_1(\vec v;\vec y)) A_2^c(\vec z)\p_2A_1^e(\vec y) \Bigg) \Bigg\} \nn\\ 
- {1\over 16\pi^2} f^{adc}f^{dbe} \int_{u,v,r,w,z}  \frac{1}{|\vec{u}-\vec{r}| |\vec{v}-\vec{w}|}  (\vec{\nabla}\times\vec{A}^a(\vec{r}))(\vec{\nabla}\times\vec{A}^b(\vec{w}))  \delta_\mu(\vec u,\vec v) \nn\\
\qquad \Bigg\{  (G_2(v_1,u_2;\vec z)-G_2(\vec v;\vec z)) A_1^c(\vec u)  \p_1A_2^e(\vec z)  -(G_2(\vec u;\vec z)-G_2(u_1,v_2;\vec z))  \p_1 A_2^c(\vec z)A_1^e(\vec v) \nn\\
\qquad + (G_1(u_1,v_2;\vec z)-G_1(\vec v;\vec z))A_2^c(\vec u)\p_2A_1^e(\vec z)  - (G_1(\vec u;\vec z)-G_1(v_1,u_2;\vec z))  \p_2A_1^c(\vec z) A_2^e(\vec v) \Bigg\} \nn\\
- {1\over 4\pi^2} f^{adc}f^{dbe} \int_{u,v,r,w,z}  \frac{\mu^2}{|\vec{u}-\vec{r}||\vec{v}-\vec{w}|}  (\vec{\nabla}\times\vec{A}^a(\vec{r}))(\vec{\nabla}\times\vec{A}^b(\vec{w}))  \delta_\mu(\vec u,\vec v) \nn\\
\qquad \Bigg\{ (u_1-v_1)\Bigg(  (G_1(\vec u;\vec z)-G_1(v_1,u_2;\vec z))A_1^e(v_1,u_2) A_1^c(\vec z)  \nn\\
\qquad\qquad + (G_1(u_1,v_2;\vec z)-G_1(\vec v;\vec z))A_1^e(\vec v) A_1^c(\vec z)  \nn\\
\qquad  +(G_2(v_1,u_2;\vec z)-G_2(\vec v;\vec z))  A_1^c(v_1,u_2) A_2^e(\vec z) + (G_2(\vec u;\vec z)-G_2(u_1,v_2;\vec z))A_2^c(\vec z)A_1^e(\vec v) \Bigg) \nn\\
\qquad +(u_2-v_2) \Bigg( (G_2(\vec u;\vec z)-G_2(u_1,v_2;\vec z)) A_2^e(u_1,v_2)A_2^c(\vec z)   \nn\\
\qquad\qquad + (G_2(v_1,u_2;\vec z)-G_2(\vec v;\vec z))A_2^e(\vec v) A_2^c(\vec z)  \nn\\
\qquad +   (G_1(u_1,v_2;\vec z)-G_1(\vec v;\vec z)) A_2^c(u_1,v_2) A_1^e(\vec z) +  (G_1(\vec u;\vec z)-G_1(v_1,u_2;\vec z))A_2^e(\vec v) A_1^c(\vec z)  \Bigg) \Bigg\} \nn\\
+ {1\over 8\pi^2} f^{adc}f^{dbe} \int_{u,v,r,w}  \frac{1}{|\vec{u}-\vec{r}|}\frac{1}{|\vec{v}-\vec{w}|}  (\vec{\nabla}\times\vec{A}^a(\vec{r}))(\vec{\nabla}\times\vec{A}^b(\vec{w}))  \delta_\mu(\vec u,\vec v) \nn\\
\qquad \Bigg\{ \Big( A_1^c(\vec u)A_1^e(v_1,u_2) +A_2^c(\vec u) A_2^e(u_1,v_2)\Big)  \Bigg\}  
\eE
\bE{l}
= - {1\over 4\pi^2}  f^{adc}f^{dbe} \int_{r,w,y,z} \frac{ -2\delta(\vec y-\vec z)}{4|\vec{z}-\vec{r}||\vec{z}-\vec{w}|}  (\vec{\nabla}\times\vec{A}^a(\vec{r}))(\vec{\nabla}\times\vec{A}^b(\vec{w}))  \Bigg\{A_1^c(\vec z)A_1^e(\vec y) +A_2^c(\vec z)A_2^e(\vec y)   \Bigg\}  \nn\\
\quad - {1\over 4\pi^2} f^{adc}f^{dbe} \int_{r,w,z}  \frac{2}{2|\vec{z}-\vec{r}||\vec{z}-\vec{w}|}  (\vec{\nabla}\times\vec{A}^a(\vec{r}))(\vec{\nabla}\times\vec{A}^b(\vec{w}))    \Bigg\{ A_1^e(\vec z) A_1^c(\vec z) +A_2^e(\vec z) A_2^c(\vec z)   \Bigg\} \nn\\
\quad + {1\over 8\pi^2} f^{adc}f^{dbe} \int_{u,r,w}  \frac{1}{|\vec{u}-\vec{r}|}\frac{1}{|\vec{u}-\vec{w}|}  (\vec{\nabla}\times\vec{A}^a(\vec{r}))(\vec{\nabla}\times\vec{A}^b(\vec{w}))    \Bigg\{ A_1^c(\vec u)A_1^e(\vec u) +A_2^c(\vec u) A_2^e(\vec u)  \Bigg\} \nn\\
\quad +\mathcal{O}(\mu^{-2})  \nn\\
=0+\mathcal{O}(\mu^{-2}) 
\eE
This vanishes for $\mu\to\infty$.\\

Vanishing of the second term:

\bE{l}
 -\int_{u,v} \delta_\mu(\vec u,\vec v)  \Phi_{ab}^{(1)}(u,v)  \frac{\delta F_{GL}^{(0)}}{\delta A_i^a(\vec u)} \frac{\delta F_{GL}^{(1)}}{\delta A_i^b(\vec v)}  \nn\\ 
 =  - {i\over2} f^{a_1a_2b} f^{abc}\int_{\slashed{k_1},\slashed{k_2},\slashed{q},\slashed{p}} \int_{u,v,y} \delta_\mu(\vec u,\vec v)  \\
\qquad \times \Bigg\{ (G_1(\vec u;\vec y)-G_1(v_1,u_2;\vec y)+G_1(u_1,v_2;\vec y)-G_1(\vec v;\vec y)) A_1^c(\vec y)  \nn\\
\qquad\qquad + (G_2(v_1,u_2;\vec y)-G_2(\vec v;\vec y)+G_2(\vec u;\vec y)-G_2(u_1,v_2;\vec y)) A_2^c(\vec y)  \Bigg\}\nn\\
 \qquad \times  e^{-i\vec{q}\cdot\vec{v}} e^{-i\vec{p}\cdot\vec{u}} \frac{1}{|\vec{p}|}(\vec{p}\times\vec{A}^a(-\vec{p}))  \frac{\slashed{\delta}\left(\vec{k}_1+\vec{k}_2+\vec{q}\right)}{|\vec{k}_1|+|\vec{k}_2|+|\vec{q}|} \Bigg\{ \frac{1}{2}  \vec{p}\cdot\vec{q}  (\vec{A}^{a_1}(\vec{k}_1)\times\vec{A}^{a_2}(\vec{k}_2))\nn\\
\qquad\qquad + (\vec{k}_1\times\vec{A}^{a_1}(\vec{k}_1)) \vec{p}\cdot\vec{A}^{a_2} (\vec{k}_2) -\frac{\vec{p}\times\vec{q}}{|\vec{q}||\vec{k}_2|}   (\vec{k}_2\times\vec{A}^{a_1}(\vec{k}_1)) (\vec{k}_2\times\vec{A}^{a_2}(\vec{k}_2)) \nn\\
\qquad\qquad +\frac{\vec{p}\cdot\vec{k}_2}{|\vec{k}_1||\vec{k}_2|}  (\vec{k}_1\cdot\vec{A}^{a_1}(\vec{k}_1))  (\vec{k}_2\times\vec{A}^{a_2}(\vec{k}_2))  -\frac{\vec{p}\cdot\vec{q}}{|\vec{k}_1||\vec{q}|} (\vec{k}_1\cdot\vec{A}^{a_1}(\vec{k}_1)) (\vec{q}\times\vec{A}^{a_2}(\vec{k}_2))  \Bigg\}   \nn\\
 = {1\over2} f^{a_1a_2b} f^{abc}\int_{\slashed{k_1},\slashed{k_2},\slashed{p},\slashed{r}} \sd(\vec p-\vec k_1-\vec k_2-\vec r) \nn\\
\qquad \times  \Bigg\{ \left(e^{-\frac{p_1^2}{4 \mu ^2}}-e^{-\frac{(k_{1,1}+k_{2,1})^2}{4 \mu ^2}}\right) \left(e^{-\frac{p_2^2}{4 \mu ^2}}+e^{-\frac{(k_{1,2}+k_{2,2})^2}{4 \mu ^2}}\right)\frac{1}{r_1} A_1^c(r)  \nn\\
\qquad\qquad + \left(e^{-\frac{p_1^2}{4 \mu ^2}}+e^{-\frac{(k_{1,1}+k_{2,1})^2}{4 \mu ^2}}\right) \left(e^{-\frac{p_2^2}{4 \mu ^2}}-e^{-\frac{(k_{1,2}+k_{2,2})^2}{4 \mu ^2}}\right)\frac{1}{r_2}  A_2^c(r)  \Bigg\}\nn\\
 \qquad \times   \frac{1}{|\vec{p}|}(\vec{p}\times\vec{A}^a(-\vec{p}))  \frac{1}{|\vec{k}_1|+|\vec{k}_2|+|\vec{k}_1+\vec{k}_2|} \Bigg\{- \frac{1}{2}  \vec{p}\cdot(\vec{k}_1+\vec{k}_2)  (\vec{A}^{a_1}(\vec{k}_1)\times\vec{A}^{a_2}(\vec{k}_2))\nn\\
\qquad\qquad + (\vec{k}_1\times\vec{A}^{a_1}(\vec{k}_1)) \vec{p}\cdot\vec{A}^{a_2} (\vec{k}_2) +\frac{\vec{p}\times(\vec{k}_1+\vec{k}_2)}{|\vec{k}_1+\vec{k}_2||\vec{k}_2|}   (\vec{k}_2\times\vec{A}^{a_1}(\vec{k}_1)) (\vec{k}_2\times\vec{A}^{a_2}(\vec{k}_2)) \nn\\
\qquad\qquad +\frac{\vec{p}\cdot\vec{k}_2}{|\vec{k}_1||\vec{k}_2|}  (\vec{k}_1\cdot\vec{A}^{a_1}(\vec{k}_1))  (\vec{k}_2\times\vec{A}^{a_2}(\vec{k}_2)) \nn\\
\qquad\qquad  -\frac{\vec{p}\cdot(\vec{k}_1+\vec{k}_2)}{|\vec{k}_1||\vec{k}_1+\vec{k}_2|} (\vec{k}_1\cdot\vec{A}^{a_1}(\vec{k}_1)) (\vec{k}_1+\vec{k}_2)\times\vec{A}^{a_2}(\vec{k}_2)  \Bigg\}   
 \eE
 
There is no loop momentum, so we can take $\mu\to\infty$. In this limit the expression vanishes.

\section{$\O(e)$ correction to the gauge invariant Hamiltonian, Eq.~(\ref{regF1GIeq})}
\label{subsec:append1}
Vanishing of the second term:
\bE{rl}
&\int_{x,v,y}\tilde{\Omega}^{(1)}_{ab}(\vec x,\vec v,\vec y)\frac{\delta F_{GI}^{(0)}}{\delta J^a(\vec x)} \frac{\delta F_{GI}^{(0)}}{\delta J^b(\vec y)}  \nn\\
\propto & \int_{x,v,y,w,z,r,s} \delta_\mu(\vec x, \vec v) (x-v) \left(\p_y \bar{G}(y-v)\right)f^{abe}J^e(\vec v) \nn\\
&\qquad\qquad \times  \frac{1}{|\vec{w}-\vec{z}|}\bar{\p}_w\de(\vec w-\vec x)J^a(\vec z) \frac{1}{|\vec{r}-\vec{s}|}\bar{\p}_r\de(\vec r-\vec y)J^b(\vec s) \qquad \\
\propto & \int_{x,v,y,w,z,r,s} \delta_\mu(\vec x, \vec v) \mu^2 (x-v)^2 \left(\p_y \de(\vec y-\vec v)\right)f^{abe}J^e(\vec v)  \nn\\
&\qquad\qquad \times \frac{1}{|\vec{w}-\vec{z}|}\de(\vec w-\vec x)J^a(\vec z) \frac{1}{|\vec{r}-\vec{s}|}\de(\vec r-\vec y)J^b(\vec s) \qquad \\
\propto & \int_{x,v,z,s} \delta_\mu(\vec x, \vec v) \mu^2 (x-v)^2 f^{abe}J^e(\vec v) \frac{1}{|\vec{x}-\vec{z}|} J^a(\vec z) \p_v\frac{1}{|\vec{v}-\vec{s}|}J^b(\vec s) \qquad
\eE 
Expanding $ \frac{1}{|\vec{x}-\vec{z}|}$ around $\vec{x}=\vec{v}$, we obtain
\bE{rl}
\propto & \int_{x,v,z,s} \delta_\mu(\vec x, \vec v) \mu^2 (x-v)^2 f^{abe}J^e(\vec v) J^a(\vec z) J^b(\vec s)  \nn\\
&\qquad\qquad \times \left( \frac{1}{|\vec{v}-\vec{z}|} \p_v\frac{1}{|\vec{v}-\vec{s}|} + (\vec{x}-\vec{v})\cdot\nabla_v \frac{1}{|\vec{v}-\vec{z}|} \p_v\frac{1}{|\vec{v}-\vec{s}|} + \ldots\right) \nn\\
=& O(\mu^{-2})
\eE 
Integration over $\vec x$ vanishes for the first two orders (note that $(x-v)^2$ is only the holomorphic component), while the next order is already $O(\mu^{-2})$.

\section{$\O(e^2J^4)$ corrections to the gauge invariant Hamiltonian, Eq.~(\ref{regF24GIeq})}
\label{subsec:append2}
Vanishing of the first term:
\bE{rl}
&\int_{x,v,y}\tilde{\Omega}^{(1)}_{ab}(\vec x,\vec v,\vec y)\frac{\delta F_{GI}^{(0)}}{\delta J^a(\vec x)} \frac{\delta F_{GI}^{(1)}}{\delta J^b(\vec y)}  \nn\\
\propto & f^{abe} \int_{x,v,y}\int_{\slashed{k_1},\slashed{k_2},\slashed{l},\slashed{p},\slashed{q}} \delta_\mu(\vec x, \vec v) (x-v) \frac{q}{\bar{q}} e^{i\vec{q}\cdot(\vec{y}-\vec{v})} J^e(\vec l) e^{i\vec{l}\cdot\vec{v}} \left(e^{-i\vec{p}\cdot\vec{x}} + e^{i\vec{p}\cdot\vec{x}} \right) \frac{\bar{p}^2}{|\vec p|} J^a(\vec{p}) \nn\\
 & e^{i(\vec{k}_1+\vec{k}_2)\cdot\vec{y}} f^{a_1 a_2 b}  g^{(3)}(\vec k_1,\vec k_2,-\vec k_1-\vec k_2) J^{a_1}(\vec k_1)J^{a_2}(\vec k_2) \\
\propto & \frac{1}{\mu^2} f^{abe} \int_{\slashed{k_1},\slashed{k_2},\slashed{l},}e^{-\frac{(\vec{k}_1+\vec{k}_2+\vec{l})^2}{4\mu^2}}  \frac{k_1+k_2}{\bar{k}_1+\bar{k}_2}  J^e(\vec l)  \frac{(\bar{k}_1+\bar{k}_2+\bar{l})^3}{|\vec{k}_1+\vec{k}_2+\vec{l}|} J^a(\vec{k}_1+\vec{k}_2+\vec{l}) \nn\\
 & f^{a_1 a_2 b}  g^{(3)}(\vec k_1,\vec k_2,-\vec k_1-\vec k_2) J^{a_1}(\vec k_1)J^{a_2}(\vec k_2) 
\eE 
Again, there is no loop momentum, so we can take $\mu\to\infty$. In this limit the expression vanishes.

Vanishing of the second term:
\bE{rl}
&\int_{x,v,y}\tilde{\Omega}^{(2)}_{ab}(\vec x,\vec v,\vec y)\frac{\delta F_{GI}^{(0)}}{\delta J^a(\vec x)} \frac{\delta F_{GI}^{(0)}}{\delta J^b(\vec y)} \nn\\
\propto & \int_{x,v,y}\int_{\slashed{k},\slashed{l},\slashed{p},\slashed{q},\slashed{r}} \delta_\mu(\vec x, \vec v) f^{bec}f^{ead} \frac{1}{\bar{q}} e^{i\vec{q}\cdot(\vec{y}-\vec{v})} \Big( (x-v)^2 q e^{i\vec{l}\cdot\vec{v}} -2i(x-v)e^{i\vec{l}\cdot\vec{y}} \Big)J^c(\vec l) J^d(\vec r) e^{i\vec{r}\cdot\vec{v}} \nn\\
&  \left(e^{-i\vec{k}\cdot\vec{x}} + e^{i\vec{k}\cdot\vec{x}} \right) \left(e^{-i\vec{p}\cdot\vec{y}} + e^{i\vec{p}\cdot\vec{y}}\right) \frac{\bar{k}^2}{|\vec k|} J^a(\vec{k})  \frac{\bar{p}^2}{|\vec p|} J^b(\vec{p}) \qquad \\
\propto & \frac{1}{\mu^2} f^{bec}f^{ead}\int_{\slashed{k},\slashed{p},\slashed{r}} e^{-\frac{\vec{k}^2}{4\mu^2}}  \frac{\bar{k}}{\bar{p}} J^d(r)  \frac{\bar{k}^2}{|\vec k|} J^a(\vec{k})  \frac{\bar{p}^2}{|\vec p|} J^b(\vec{p})  \nn\\
&  \Big(\frac{p \bar{k}}{\mu^2} ( J^c(\vec p-\vec r+\vec k) + J^c(\vec p-\vec r-\vec k) + J^c(-\vec p-\vec r+\vec k) + J^c(-\vec p-\vec r-\vec k)) \nn\\
& -2 ( J^c(\vec p-\vec r+\vec k) - J^c(\vec p-\vec r-\vec k) + J^c(-\vec p-\vec r+\vec k) - J^c(-\vec p-\vec r-\vec k)) \Big)  
\eE 
This also vanishes for $\mu\to\infty$.

\section[$\O(e^2J^2)$ corrections to the gauge invariant Hamiltonian, 2$^{nd}$ term of Eq.~(\ref{SEF22})]{$\O(e^2J^2)$ corrections to the gauge invariant Hamiltonian, 2\boldmath $^{nd}$ \unboldmath term of Eq.~(\ref{SEF22})}
\label{subsec:append3}
We look at the different parts of $\tilde{\Omega}^{(2)}_{ab}(\vec x,\vec v,\vec y)\frac{\delta^2 F^{(0)}}{\delta J^a(\vec v) \delta J^b(\vec y)}$ separately:
\subsubsection{The $\p_y \bar{G}(y-x)$ term}
 \bE{rl}
& \int_{x,v,y}\delta_\mu(\vec x, \vec v) \left(\p_y \bar{G}(y-v)\right) (x-v)^2 (J(\vec v)J(\vec v))_{ab} \frac{\delta^2 F_{GI}^{(0)}}{\delta J^a(\vec y) \delta J^b(\vec x)}  \nn\\
\propto &  f^{cae} f^{deb} \int_{x,v,y} \delta_\mu(\vec x, \vec v)  \left(\p_y \bar{G}(y-v)\right) (x-v)^2 J^c(\vec v)J^d(\vec v) \frac{\delta^2 F_{GI}^{(0)}}{\delta J^a(\vec y) \delta J^b(\vec x)}  \\
\propto& f^{cbe} f^{deb} \int_{x,z,v}\delta_\mu(\vec x, \vec v) (x-v)^2  \nn\\
&\qquad \times \int_{\slashed{p},\slashed{k}_1,\slashed{k}_2,\slashed{q}} \frac{p}{\bar{p}} e^{i\vec{p}\cdot(\vec{z}-\vec{v})}   J^c(\vec k_1)J^d(\vec k_2) e^{i(\vec{k}_1+\vec{k}_2)\cdot\vec{v}}  \frac{\bar{q}^2}{2|\vec{q}|} \left[e^{-i\vec{q}\cdot\vec{z}} e^{i\vec{q}\cdot\vec{x}} + e^{i\vec{q}\cdot\vec{z}} e^{-i\vec{q}\cdot\vec{x}} \right] \quad\qquad \\
\propto& {1\over\mu^4} \int_{\slashed{k},\slashed{q}} J^a(\vec k)J^a(-\vec k)  \frac{q\bar{q}^3}{|\vec{q}|} e^{-\frac{\vec{q}^2}{4\mu^2}} 
\eE
This vanishes under integration of the angular component of $\vec{q}$.

\subsubsection{The $\bar{G}(y-x)$ term}
\bE{rl}
&\int_{x,v,y}\delta_\mu(\vec x, \vec v) \bar{G}(y-v) f^{ebf} J^e(\vec y) (x-v) f^{cfa}J^c(\vec v) \frac{\delta^2 F_{GI}^{(0)}}{\delta J^a(\vec y) \delta J^b(\vec x)}   \nn\\
\propto &  \int_{\slashed{k},\slashed{q}} J^a(\vec k)J^a(-\vec k)   e^{-\frac{\vec{q}^2}{4\mu^2}} \frac{2\bar{k}\bar{q}^3}{\mu^2|\vec{q}|(\bar{q}^2-\bar{k}^2)}   \\
\propto &  {1\over\mu} \int_{\slashed{k}} \bar{k}^2 J^a(\vec k)J^a(-\vec k)  
\eE
This vanishes for $\mu\to\infty$.

\chapter{Computation of $\T\V$ at $\O(e^2)$ in terms of gauge fields}
\label{app:TV}
We give the details of the computation of $\T\V$ at $\O(e^2)$ in terms of the gauge fields, Eq.~(\ref{TVB}):
\bE{l}
\hspace{-12cm} \T_\mathrm{reg}(\mu) \V_\mathrm{reg}(\mu')\Bigg|_{\O(e^2)} \nn
\eE
\bE{rCl}
&=& -\frac{e^2}{4} \int_{u,v,x,y} \delta_{\mu}(\vec u-\vec v)\delta_{\mu'}(\vec x-\vec y) \Bigg\{\delta^{ab}  \frac{\delta^2}{\delta A^a_i(\vec u)\delta A^b_i(\vec v)} B^{(1)}_c(\vec x)\de^{cd}B^{(1)}_d(\vec y) \nn\\
&&\qquad\qquad + 2 \delta^{ab}  \frac{\delta^2}{\delta A^a_i(\vec u)\delta A^b_i(\vec v)} B^{(0)}_c(\vec x)\Phi^{(1)}_{cd}(\vec x,\vec y)B^{(1)}_d(\vec y) \nn\\
&&\qquad\qquad + \delta^{ab}  \frac{\delta^2}{\delta A^a_i(\vec u)\delta A^b_i(\vec v)} B^{(0)}_c(\vec x)\Phi^{(2)}_{cd}(\vec x,\vec y)B^{(0)}_d(\vec y) \nn\\
&&\qquad\qquad + 2 \Phi^{(1)}_{ab}(\vec u,\vec v) \frac{\delta^2}{\delta A^a_i(\vec u)\delta A^b_i(\vec v)} B^{(0)}_c(\vec x) \de^{cd}B^{(1)}_d(\vec y) \nn\\ 
&&\qquad\qquad + \Phi^{(1)}_{ab}(\vec u,\vec v) \frac{\delta^2}{\delta A^a_i(\vec u)\delta A^b_i(\vec v)} B^{(0)}_c(\vec x)\Phi^{(1)}_{cd}(\vec x,\vec y)B^{(0)}_d(\vec y) \nn\\ 
&&\qquad\qquad +  \Phi^{(2)}_{ab}(\vec u,\vec v) \frac{\delta^2}{\delta A^a_i(\vec u)\delta A^b_i(\vec v)} B^{(0)}_c(\vec x)\delta^{cd}B^{(0)}_d(\vec y) \Bigg\} \nn\\
&=&   -\frac{e^2C_A}{2\pi} \int_{y}  \Bigg(\frac{\mu^2 \mu'^2}{ \left(\mu^2+\mu'^2\right)} \vec{A}^a(\vec y)\cdot\vec{A}^a(\vec y)  \nn\\
&& \qquad  -\frac{\mu^2 \mu'^2}{4 \left(\mu^2+\mu'^2\right)^2}\Big((\p_1A_1^a(\vec y))^2 + (\p_2A_2^a(\vec y))^2 + (\p_1A_2^a(\vec y))^2 + (\p_2A_1^a(\vec y))^2      \Big) \Bigg) \nn\\
&& + {e^2 C_A\over2\pi} \int_{y}\Bigg( 2\frac{\mu ^4 \mu'^2}{ \left(\mu^2+\mu'^2\right)^2} \vec{A}^a(\vec y)\cdot\vec{A}^a(\vec y) -  \frac{\mu^4 \mu'^2}{2 \left(\mu^2+\mu'^2\right)^3}(\vec \nabla\cdot\vec{A}^a(\vec y))^2  \nn\\
&&\qquad + B^{(0)a}(y)B^{(0)a}(y) \left( \frac{\mu ^2 }{2 \mu' \sqrt{\mu^2+\mu'^2}} + \frac{\mu ^2 }{2 \left(\mu ^2+\mu'^2\right)} - \frac{\mu^4 \mu'^2}{4 \left(\mu^2+\mu'^2\right)^3} \right)  \Bigg) \nn\\
&& + \frac{e^2 C_A}{2\pi}  \int_{y}   \Bigg(  B^{(0)a}(\vec y)B^{(0)a}(\vec y) \Big(\sqrt{1+\frac{\mu^2}{\mu'^2}}-1\Big) \nn\\
&&\qquad + \frac{\mu^4 \mu'^2}{ \left(\mu^2+\mu'^2\right)^3} \Big((\mu'^2-\mu^2) \vec{A}^a(\vec y)\cdot\vec{A}^a(\vec y) - \frac{\mu'^2-2\mu^2}{8 \left(\mu^2+\mu'^2\right)}(\vec \nabla\cdot\vec{A}^a(\vec y))^2\Big)   \nn\\
&&\qquad  -{1\over2} \left(\vec \nabla\times \vec{A}^a(\vec y)\right)^2 \frac{\mu^4}{\left(\mu ^2+\mu'^2\right)^2} \Bigg) \nn\\
&& +  {e^2 C_A\over2\pi} \int_{y} \Bigg( 2\frac{\mu^2 \mu'^4}{ \left(\mu^2+\mu'^2\right)^2} \vec{A}^a(\vec y)\cdot\vec{A}^a(\vec y) \nn\\
&&\qquad - \frac{\mu^2 \mu'^4}{2 \left(\mu^2+\mu'^2\right)^3}(\vec \nabla\cdot\vec{A}^a(\vec y))^2 +\frac{\mu^2 \mu'^4}{4 \left(\mu^2+\mu'^2\right)^3}(\vec \nabla\times\vec{A}^a(\vec y))^2  \Bigg) \nn\\
&&  +\frac{e^2 C_A}{2\pi} \int_{y}\Bigg( -\frac{4\mu^4 \mu'^4}{ \left(\mu^2+\mu'^2\right)^3}  \vec{A}^a(\vec y)\cdot\vec{A}^a(\vec y)    \nn\\
&& \qquad   +   \frac{3\mu^4 \mu'^4}{4 \left(\mu^2+\mu'^2\right)^4} \Big((\p_1A_1^a(\vec y))^2 + (\p_2A_2^a(\vec y))^2\Big)    \nn\\
&& \qquad   + \frac{3\mu^4 \mu'^4}{4 \left(\mu^2+\mu'^2\right)^4}   \Big((\p_1A_2^a(\vec y))^2 + (\p_2A_1^a(\vec y))^2\Big)          \nn\\
&& \qquad +   B^{(0)a}(\vec y)  (\vec \nabla\times\vec{A}^a(\vec y)) \Big( {1\over2} \frac{\mu ^2}{\mu ^2+\mu'^2} - {1\over2} \frac{\mu^2 \left(\mu^2+2\mu'^2\right)\sqrt{\mu ^2+\mu'^2}}{ \mu'  \left(\mu ^2+\mu'^2\right)^2}  \Big) \Bigg) \nn\\
&& + {e^2 C_A\over 2\pi} \int_{y} \frac{\mu^2 \mu'^4}{\left(\mu^2+\mu'^2\right)^3} \Bigg(  (\mu^2 -\mu'^2) \vec{A}^a(\vec y)\cdot\vec{A}^a(\vec y) -\frac{\mu^2 -2\mu'^2}{8 \left(\mu^2+\mu'^2\right)}(\vec \nabla\cdot\vec{A}^a(\vec y))^2\Bigg) \nn\\
&&+\O\left(\mu^{-1}\right)
\eE

\section{Computation of the first term}

\bE{l}
-\frac{e^2}{4} \int_{u,v,x,y} \delta_{\mu}(\vec u-\vec v)\delta_{\mu'}(\vec x-\vec y) \delta^{ab}  \frac{\delta^2}{\delta A^a_i(\vec u)\delta A^b_i(\vec v)}  B^{(1)c}(\vec x)\delta^{cd}B^{(1)d}(\vec y) \\
= -\frac{e^2}{4} \int_{u,v,x,y} \delta_{\mu}(\vec u-\vec v)\delta_{\mu'}(\vec x-\vec y)  \frac{\delta^2}{\delta A^a_i(\vec u)\delta A^a_i(\vec v)}  \frac{e^2}{4}f^{dec}\vec{A}^d(\vec x)\times\vec{A}^e(\vec x) f^{fgc}\vec{A}^f(\vec y)\times\vec{A}^g(\vec y) \qquad
\eE
If both derivatives act on two fields at the same point ($\vec x$ or $\vec y$) this will vanish due to color contraction, so we can only have the derivatives acting on fields at different points. There are eight terms of this type and they can all be combined in one (due to color symmetry and the symmetry of $\delta_\mu$):  
\bE{l}
= -\frac{e^2C_A}{2} \int_{u,v,x,y} \delta_{\mu}(\vec u-\vec v)\delta_{\mu'}(\vec x-\vec y)  \delta(\vec u-\vec x)\delta(\vec y-\vec v)  \vec{A}^a(\vec x)\cdot\vec{A}^a(\vec y) \,.
\eE
We can expand the first field around $\vec x=\vec y$ up to second order, as higher orders will vanish in the limit of $\mu,\mu'\to\infty$.
\bE{l}
= -\frac{e^2C_A}{2} \int_{y,X} \delta_{\mu}(\vec X)\delta_{\mu'}(\vec X)  \nn\\
\qquad \Big(1+X_1\p_{X_1}+X_2\p_{X_2}+\frac{X_1^2}{2}\p_{X_1}^2+X_1X_2\p_{X_1}\p_{X_2}+\frac{X_2^2}{2}\p_{X_2}^2\Big) \vec{A}^a(\vec X+\vec y)\cdot\vec{A}^a(\vec y) \Bigg|_{\vec X=0} \qquad \\
= -\frac{e^2C_A}{2} \int_{y}  \Bigg(\frac{\mu^2 \mu'^2}{\pi  \left(\mu^2+\mu'^2\right)}+\frac{\mu^2 \mu'^2}{4 \pi  \left(\mu^2+\mu'^2\right)^2}\Big(\p_{X_1}^2+\p_{X_2}^2\Big)\Bigg) \vec{A}^a(\vec X+\vec y)\cdot\vec{A}^a(\vec y) \Bigg|_{\vec X=0} \nn\\
\quad +\O\left(\mu^{-1}\right)\,. 
\eE

\section{Computation of the second term}

\bE{l}
-\frac{2}{4} \int_{u,v,x,y} \delta_{\mu}(\vec u-\vec v)\delta_{\mu'}(\vec x-\vec y) \delta^{ab}  \frac{\delta^2}{\delta A^a_i(\vec u)\delta A^b_i(\vec v)} B^{(0)c}(\vec x)e\Phi^{(1)}_{cd}(\vec x,\vec y)eB^{(1)d}(\vec y) \\
=  -\frac{1}{2} \int_{u,v,x,y,z} \delta_{\mu}(\vec u-\vec v)\delta_{\mu'}(\vec x-\vec y)  \frac{\delta^2}{\delta A^a_i(\vec u)\delta A^a_i(\vec v)} \\
\qquad\Bigg\{B^{(0)c}(\vec x) {e\over 2} f^{cde} \Bigg((G_1(\vec x;\vec z)-G_1(y_1,x_2;\vec z)+G_1(x_1,y_2;\vec z)-G_1(\vec y;\vec z)) A_1^e(z)  \nn\\
\qquad\quad  + (G_2(y_1,x_2;\vec z)-G_2(\vec y;\vec z)+G_2(\vec x;\vec z)-G_2(x_1,y_2;\vec z)) A_2^e(\vec z)  \Bigg)  {e\over 2} f^{fgd}\vec{A}^f(\vec y)\times\vec{A}^g(\vec y) \Bigg\}\,. \nn
\eE
Again, if both derivatives act on $B^{(1)}$ this vanishes, as does one derivative acing on $B^{(0)}$ and one acting on $\Phi^{cd}$, due to $f^{ada}=0$. As interchange of $\vec u$ and $\vec v$ is possible in $\delta_\mu(\vec u-\vec v)$ we let $\frac{\delta}{\delta A^a_i(\vec v)}$ act on $B^{(1)}$ and $\frac{\delta}{\delta A^a_i(\vec u)}$ act on $B^{(0)}\Phi^{cd}$, and multiply by a factor 2:
\bE{l}
=  - \frac{2^2e^2}{2^3} \int_{u,v,x,y,z} \delta_{\mu}(\vec u-\vec v)\delta_{\mu'}(\vec x-\vec y)  \nn\\
\qquad\Bigg\{\p_{x_n}\delta(x-u)\e_{ni}  f^{ade} \Bigg((G_1(\vec x;\vec z)-G_1(y_1,x_2;\vec z)+G_1(x_1,y_2;\vec z)-G_1(\vec y;\vec z)) A_1^e(\vec z)  \nn\\
\qquad\qquad  + (G_2(y_1,x_2;\vec z)-G_2(\vec y;\vec z)+G_2(\vec x;\vec z)-G_2(x_1,y_2;\vec z)) A_2^e(\vec z)  \Bigg)  f^{fad}A_k^f(\vec y)\e_{ki}\delta(\vec y-\vec v) \nn\\
\qquad\quad +B^{(0)c}(\vec x)  f^{cda} \Bigg((G_1(\vec x;\vec z)-G_1(y_1,x_2;\vec z)+G_1(x_1,y_2;\vec z)-G_1(\vec y;\vec z)) \delta_{i1}\delta(\vec z-\vec u))  \nn\\
\qquad\qquad  + (G_2(y_1,x_2;\vec z)-G_2(\vec y;\vec z)+G_2(\vec x;\vec z)-G_2(x_1,y_2;\vec z))  \delta_{i2}\delta(\vec z-\vec u))  \Bigg)  \nn\\
\qquad\qquad\qquad\times  f^{fad}A_k^f(\vec y)\e_{ki}\delta(\vec y-\vec v) \Bigg\}
\eE
This becomes (where we defined $\vec X=\vec x-\vec v$, $\vec Z=\vec z-\vec v$, and $\vec U=\vec u-\vec v$, $\vec V=\vec v-\vec x$)
\bE{l}
=   e^2 C_A \int_{v,X,Z} \Big(X_1 A_1^a(\vec v) +X_2 A_2^a(\vec v) \Big)  \delta_{\mu}(\vec X)\delta_{\mu'}(\vec X)  \mu^2 \nn\\
\qquad   \Bigg((\theta(X_1-Z_1)\delta(X_2-Z_2) -\theta(-Z_1)\delta(X_2-Z_2) +\theta(X_1-Z_1)\delta(Z_2) -\theta(-Z_1)\delta(Z_2))   \nn\\
\qquad\qquad \Big(1+Z_1\p_{Z_1}+Z_2\p_{Z_2}+\frac{Z_1^2}{2}\p_{Z_1}^2+Z_1Z_2\p_{Z_1}\p_{Z_2}+\frac{Z_2^2}{2}\p_{Z_2}^2\Big) A_1^a(\vec Z+\vec v)\Bigg|_{\vec Z=0} \nn\\
\qquad\quad  + (\delta(Z_1)\theta(X_2-Z_2) -\delta(Z_1)\theta(-Z_2) +\delta(X_1-Z_1)\theta(X_2-Z_2) -\delta(X_1-Z_1)\theta(-Z_2))  \nn\\
\qquad\qquad \Big(1+Z_1\p_{Z_1}+Z_2\p_{Z_2}+\frac{Z_1^2}{2}\p_{Z_1}^2+Z_1Z_2\p_{Z_1}\p_{Z_2}+\frac{Z_2^2}{2}\p_{Z_2}^2\Big) A_2^a(\vec Z+\vec v)\Bigg|_{\vec Z=0} \Bigg)  \nn\\
\quad -\frac{e^2 C_A}{2} \int_{x,U,V} B^{(0)a}(\vec x)  \delta_{\mu}(\vec U) \delta_{\mu'}(-\vec V)  \nn\\ 
\qquad  \Bigg((\theta(-V_1-U_1)\delta(-V_2-U_2) -\theta(-U_1) \delta(-V_2-U_2) +\theta(-V_1-U_1)\delta(-U_2)  \nn\\
\qquad\qquad -\theta(-U_1)\delta(U_2) ) \times  \Big(1+V_1\p_{V_1}+V_2\p_{V_2}\Big) A_2^a(\vec V+\vec x)\Bigg|_{\vec V=0}   \nn\\
\qquad\qquad  - (\delta(-U_1)\theta(-V_2-U_2)- \delta(-U_1)\theta(-U_2) +\delta(-V_1-U_1)\theta(-V_2-U_2)  \nn\\
\qquad\qquad -\delta(-V_1-U_1)\theta(-U_2)) \times  \Big(1+V_1\p_{V_1}+V_2\p_{V_2}\Big) A_1^a(\vec V+\vec x)\Bigg|_{\vec V=0}   \Bigg) 
\eE
While the expansion in $\vec Z$ cannot be justified a priori, as only $\vec X$ can be considered a small variable, it turns out a posteriori that it is correct: In the $A_i A_i$ terms, because the $\delta_\mu(\vec X)$ turns into a $\delta_\mu(\vec Z)$, after integration over $X$, and in the $A_1A_2$ terms because it turns out that in an expansion to $n=\infty$ only the $Z_1Z_2$ terms survive the $\mu,\mu'\to\infty$ limits.
\bE{l}
=   e^2 C_A \int_{v} \Bigg\{  A_1^a(\vec v) \Big(\frac{\mu ^4 \mu'^2}{\pi  \left(\mu^2+\mu'^2\right)^2} +\frac{\mu^4 \mu'^2}{4 \pi  \left(\mu^2+\mu'^2\right)^3}\p_1^2+\frac{\mu^4 \mu'^2}{8 \pi  \left(\mu^2+\mu'^2\right)^3}\p_2^2\Big) A_1^a(\vec v) \nn\\
\qquad\qquad + A_2^a(\vec v) \Big(\frac{\mu^4 \mu'^2}{8 \pi  \left(\mu^2+\mu'^2\right)^3}\p_1\p_2\Big) A_1^a(\vec v) \nn\\
\qquad\qquad  + A_1^a(\vec v) \Big(\frac{\mu^4 \mu'^2}{8 \pi  \left(\mu^2+\mu'^2\right)^3}\p_1\p_2\Big) A_2^a(\vec v) \nn\\
\qquad\qquad + A_2^a(\vec v) \Big(\frac{\mu ^4 \mu'^2}{\pi  \left(\mu^2+\mu'^2\right)^2} + \frac{\mu^4 \mu'^2}{8 \pi  \left(\mu^2+\mu'^2\right)^3}\p_1^2 + \frac{\mu^4 \mu'^2}{4 \pi  \left(\mu^2+\mu'^2\right)^3}\p_2^2 \Big) A_2^a(\vec v) \Bigg\}  \nn\\
\quad -\frac{e^2 C_A}{2} \int_{x} B^{(0)a}(\vec x)  \Bigg\{ \Big(-\frac{\mu ^2 \left(\mu^2+\mu' \left(\mu'+\sqrt{\mu^2+\mu'^2}\right)\right)}{2 \pi  \mu' \left(\mu ^2+\mu'^2\right)^{3/2}}\p_1\Big) A_2^a(\vec x)   \nn\\
\qquad\qquad  -   \Big(-\frac{\mu ^2 \left(\mu^2+\mu' \left(\mu'+\sqrt{\mu^2+\mu'^2}\right)\right)}{2 \pi  \mu' \left(\mu ^2+\mu'^2\right)^{3/2}}\p_2\Big) A_1^a(\vec x)     \Bigg\} +\O\left(\mu^{-1}\right)\,. 
\eE

\section{Computation of the third term}
\bE{rll}
=& -\frac{e^2}{4} \int_{u,v,x,y}& \delta_{\mu}(\vec u-\vec v)\delta_{\mu'}(\vec x-\vec y) \delta^{ab}  \frac{\delta^2}{\delta A^a_i(\vec u)\delta A^b_i(\vec v)} B^{(0)}_c(\vec x)\Phi^{(2)}_{cd}(\vec x,\vec y)B^{(0)}_d(\vec y) \\
=&  -{e^2\over4}  \int_{u,v,x,y}& \delta_{\mu}(\vec u-\vec v)\delta_{\mu'}(\vec x-\vec y) \Bigg\{ \left(\frac{\delta^2}{\delta A^a_i(\vec u)\delta A^a_i(\vec v)}\Phi^{(2)}_{cd}(\vec x,\vec y)\right)  B^{(0)}_c(\vec x)B^{(0)}_d(\vec y) \nn\\
 && + \left( \frac{\delta^2   B^{(0)}_c(\vec x)B^{(0)}_d(\vec y)}{\delta A^a_i(\vec u)\delta A^a_i(\vec v)} \right) \Phi^{(2)}_{cd}(\vec x,\vec y) \nn\\
 && + 2\left(\frac{\delta   B^{(0)}_c(\vec x)B^{(0)}_d(\vec y) }{\delta A^a_i(\vec u)}\right) \left(  \frac{\delta}{\delta A^a_i(\vec v)}\Phi^{(2)}_{cd}(\vec x,\vec y)\right) \Bigg\} \qquad\quad
\eE
We consider these three subterms individually.

\subsection{First subterm}

Noting that $\frac{\delta^2}{\delta A^a_i(\vec u)\delta A^b_i(\vec v)} A_1A_2=0$ and making use of the fact that one can rename $\vec u\leftrightarrow \vec v$ under the integral, we find
\bE{l}
\label{eq51}
\quad -{e^2\over4}  \int_{u,v,x,y} \delta_{\mu}(\vec u-\vec v)\delta_{\mu'}(\vec x-\vec y) \Bigg\{ \left(\frac{\delta^2}{\delta A^a_i(\vec u)\delta A^a_i(\vec v)}\Phi^{(2)}_{cd}(\vec x,\vec y)\right)  B^{(0)}_c(\vec x)B^{(0)}_d(\vec y) \nn\\
= \frac{e^2 C_A}{4}  \int_{u,v,x,y} \delta_{\mu}(\vec u-\vec v)\delta_{\mu'}(\vec x-\vec y)  B^{(0)}_c(\vec x)B^{(0)}_c(\vec y) \nn\\
\quad\Bigg\{  (G_1(\vec x;\vec u)-G_1(y_1,x_2;\vec u))(G_1(\vec u;\vec v)-G_1(y_1,x_2;\vec v))\nn\\
\qquad\qquad + (G_1(x_1,y_2;\vec u)-G_1(\vec y;\vec u))(G_1(\vec u;\vec v)-G_1(\vec y;\vec v))  \nn\\
\qquad +  (G_2(y_1,x_2;\vec u)-G_2(\vec y;\vec u))(G_2(\vec u;\vec v)-G_2(\vec y;\vec v))\nn\\
\qquad\qquad + (G_2(\vec x;\vec u)-G_2(x_1,y_2;\vec u))(G_2(\vec u;\vec v)-G_2(x_1,y_2;\vec v)) \Bigg\}
\eE
We can also rename $\vec x\leftrightarrow \vec y$:
\bE{l}
= \frac{e^2 C_A}{4}  \int_{u,v,x,y} \delta_{\mu}(\vec u-\vec v)\delta_{\mu'}(\vec x-\vec y)  B^{(0)}_c(\vec x)B^{(0)}_c(\vec y) \nn\\
\quad\Bigg\{  (G_1(\vec x;\vec u)-G_1(y_1,x_2;\vec u))(G_1(\vec x;\vec v)-G_1(y_1,x_2;\vec v))\nn\\
\qquad + (G_2(y_1,x_2;\vec u)-G_2(\vec y;\vec u))(G_2(y_1,x_2;\vec v)-G_2(\vec y;\vec v))\Bigg\}  \\
= \frac{e^2 C_A}{4}  \int_{y}  B^{(0)}_c(\vec y)B^{(0)}_c(\vec y)  \Bigg\{ \frac{-\mu' +\sqrt{\mu^2+\mu'^2}}{\pi  \mu'} + \frac{-\mu' +\sqrt{\mu^2+\mu'^2}}{\pi  \mu'} \Bigg\} \nn\\ 
\qquad +\O\left(\mu'^{-3}\right) 
\eE

\subsection{Second subterm}

\bE{l}
  -{e^2\over4}  \int_{u,v,x,y} \delta_{\mu}(\vec u-\vec v)\delta_{\mu'}(\vec x-\vec y) \left( \frac{\delta^2   B^{(0)}_c(\vec x)B^{(0)}_d(\vec y)}{\delta A^a_i(u)\delta A^a_i(v)} \right) \Phi^{(2)}_{cd}(\vec x,\vec y) \\
=  \frac{e^2C_A}{4} \int_{u,v,y,z} \left(\p_i^u\p_i^v\delta_{\mu}(\vec u-\vec v)\right) \delta_{\mu'}(\vec u-\vec v)  \nn\\
\quad \Big( (G_1(\vec u;\vec z)-G_1(v_1,u_2;\vec z))(G_1(\vec z;\vec y)-G_1(v_1,u_2;\vec y))\nn\\
 \qquad\qquad + (G_1(u_1,v_2;\vec z)-G_1(\vec v;\vec z))(G_1(\vec z;\vec y)-G_1(\vec v;\vec y))\Big)A_1^c(\vec z)A_1^c(\vec y)  \nn\\
 \quad + \Big( (G_2(v_1,u_2;\vec z)-G_2(\vec v;\vec z))(G_2(\vec z;\vec y)-G_2(\vec v;\vec y))\nn\\
\qquad\qquad + (G_2(\vec u;\vec z)-G_2(u_1,v_2;\vec z))(G_2(\vec z;\vec y)-G_2(u_1,v_2;\vec y))\Big)A_2^c(\vec z)A_2^c(\vec y)  \nn\\
\quad + (G_1(\vec u;\vec y)-G_1(v_1,u_2;\vec y))(G_2(v_1,u_2;\vec z)-G_2(\vec v;\vec z))A_1^c(\vec y) A_2^c(\vec z) \nn\\
\quad + (G_2(\vec u;\vec z)-G_2(u_1,v_2;\vec z))(G_1(u_1,v_2;\vec y)-G_1(\vec v;\vec y))A_2^c(\vec z)A_1^c(\vec y) 
\label{eq73} \\
= \frac{e^2C_A}{4}  \int_{y,U,V,Z} (4\mu^2-4\mu^4(U_1^2+U_2^2)) \delta_{\mu}(\vec U) \delta_{\mu'}(\vec U)  \nn\\
\quad \Bigg\{\Big( \delta(U_2+V_2-Z_2) (\theta(U_1+V_1-Z_1)-\theta(V_1-Z_1))(\theta(Z_1)\delta(Z_2) -\theta(V_1)\delta(U_2+V_2))\nn\\
 \qquad\qquad + \delta(V_2-Z_2) (\theta(U_1+V_1-Z_1)-\theta(V_1-Z_1))(\theta(Z_1)\delta(Z_2) -\theta(V_1)\delta(V_2))\Big) \nn\\
\qquad A_1^c(\vec y) \Big(1+Z_1\p_1+Z_2\p_2+\frac{1}{2}Z_1^2\p^2_1+ Z_1Z_2\p_1\p_2+\frac{1}{2}Z_2^2\p^2_2\Big)A_1^c(\vec y)  \nn\\
 \quad + \Big( \delta(V_1-Z_1) (\theta(U_2+V_2-Z_2)-\theta(V_2-Z_2))(\delta(Z_1)\theta(Z_2) -\delta(V_1)\theta(V_2))\nn\\
\qquad\qquad + \delta(U_1+V_1-Z_1) (\theta(U_2+V_2-Z_2)-\theta(V_2-Z_2)) (\delta(Z_1)\theta(Z_2) -\delta(U1+V_1)\theta(V_2))\Big) \nn\\
\qquad A_2^c(\vec y) \Big(1+Z_1\p_1+Z_2\p_2+\frac{1}{2}Z_1^2\p^2_1+ Z_1Z_2\p_1\p_2+\frac{1}{2}Z_2^2\p^2_2\Big)A_2^c(\vec y)  \nn\\
\quad + \delta(U_2+V_2)\delta(V_1-Z_1)(\theta(U_1+V_1)-\theta(V_1))(\theta(U_2+V_2-Z_2)-\theta(V_2-Z_2)) \nn\\
\qquad A_1^c(\vec y) \Big(1+Z_1\p_1+Z_2\p_2+\frac{1}{2}Z_1^2\p^2_1+ Z_1Z_2\p_1\p_2+\frac{1}{2}Z_2^2\p^2_2\Big)A_2^c(\vec y)  \nn\\
\quad + \delta(U_1+V_1-Z_1) \delta(V_2)(\theta(U_2+V_2-Z_2)-\theta(V_2-Z_2))(\theta(U_1+V_1)-\theta(V_1)) \nn\\
\qquad A_1^c(\vec y) \Big(1+Z_1\p_1+Z_2\p_2+\frac{1}{2}Z_1^2\p^2_1+ Z_1Z_2\p_1\p_2+\frac{1}{2}Z_2^2\p^2_2\Big)A_2^c(\vec y)  \Bigg\} +\O\left(\mu^{-1}\right) \,,
\eE
where $\vec Z=\vec z-\vec y$, $\vec U=\vec u-\vec v$, $\vec V=\vec v-\vec y$ and again the justification for the expansion in $\vec Z$ is a posteriori, as higher order vanish in the limits of $\mu,\mu'\to\infty$. This results in
\bE{l}
= \frac{e^2C_A}{4} \int_{y} \Bigg\{ A_1^c(\vec y) \Big(\frac{2\mu^4 \mu'^2}{\pi  \left(\mu^2+\mu'^2\right)^2} -4\frac{\mu^6 \mu'^2}{\pi  \left(\mu^2+\mu'^2\right)^3}  \nn\\
\qquad\qquad\qquad\qquad +\Big(\frac{\mu^4 \mu'^2}{4 \pi  \left(\mu^2+\mu'^2\right)^3} -4\frac{3 \mu^6 \mu'^2}{16 \pi  \left(\mu^2+\mu'^2\right)^4}\Big) \p^2_1\Big)A_1^c(\vec y)  \nn\\
\qquad +  A_2^c(\vec y)  \Big(\frac{2\mu^4 \mu'^2}{\pi  \left(\mu^2+\mu'^2\right)^2} -4 \frac{\mu^6 \mu'^2}{\pi  \left(\mu^2+\mu'^2\right)^3}  +\Big( \frac{\mu^4 \mu'^2}{4 \pi  \left(\mu^2+\mu'^2\right)^3} -4\frac{3 \mu^6 \mu'^2}{16 \pi  \left(\mu^2+\mu'^2\right)^4}\Big) \p^2_2\Big)A_2^c(\vec y)  \nn\\
\qquad + A_1^c(\vec y)\Big( \frac{\mu^4 \mu'^2}{4 \pi  \left(\mu^2+\mu'^2\right)^3} -4\frac{3 \mu^6 \mu'^2}{16 \pi  \left(\mu^2+\mu'^2\right)^4}\Big) \p_1\p_2 A_2^c(\vec y)  \nn\\
\qquad + A_1^c(\vec y) \Big(\frac{\mu^4 \mu'^2}{4 \pi  \left(\mu^2+\mu'^2\right)^3} -4\frac{3 \mu^6 \mu'^2}{16 \pi  \left(\mu^2+\mu'^2\right)^4} \Big) \p_1\p_2 A_2^c(\vec y) \Bigg\} +\O\left(\mu^{-1}\right)
\eE

\subsection{Third subterm}
\bE{l}
\quad  -{2e^2\over4}  \int_{u,v,x,y} \delta_{\mu}(\vec u-\vec v)\delta_{\mu'}(\vec x-\vec y)  \left(\frac{\delta   B^{(0)}_c(\vec x)B^{(0)}_d(\vec y) }{\delta A^a_i(\vec u)}\right) \left(  \frac{\delta}{\delta A^a_i(\vec v)}\Phi^{(2)}_{cd}(\vec x,\vec y)\right) \\
=  -\frac{e^2C_A}{4}\int_{u,v,x,y,z}\delta_{\mu}(\vec u-\vec v)\delta_{\mu'}(\vec x-\vec y) \Bigg\{ \Big( (G_1(\vec x;\vec v)-G_1(y_1,x_2;\vec v))(G_1(\vec v;\vec z)-G_1(y_1,x_2;\vec z))\nn\\
\qquad\qquad + (G_1(x_1,y_2;\vec v)-G_1(\vec y;\vec v))(G_1(\vec v;\vec z)-G_1(\vec y;\vec z))\Big)A_1^a(\vec z) \left(\vec \nabla^x\times \vec{A}^a(\vec x)\right) (-\p_2^y\delta(\vec y-\vec u) ) \nn\\
\quad + \Big( (G_1(\vec x;\vec z)-G_1(y_1,x_2;\vec z))(G_1(\vec z;\vec v)-G_1(y_1,x_2;\vec v))\nn\\
\qquad\qquad + (G_1(x_1,y_2;\vec z)-G_1(\vec y;\vec z))(G_1(\vec z;\vec v)-G_1(\vec y;\vec v))\Big)A_1^a(\vec z) (-\p_2^x\delta(\vec x-\vec u)) \left(\vec  \nabla^y\times \vec{A}^a(\vec y) \right) \nn\\
\quad +\Big( (G_2(y_1,x_2;\vec v)-G_2(\vec y;\vec v))(G_2(\vec v;\vec z)-G_2(\vec y;\vec z))\nn\\
\qquad\qquad + (G_2(\vec x;\vec v)-G_2(x_1,y_2;\vec v))(G_2(\vec v;\vec z)-G_2(x_1,y_2;\vec z))\Big) A_2^a(\vec z) \left(\vec \nabla^x\times \vec{A}^a(\vec x)\right) \p_1^y\delta(\vec y-\vec u) \nn\\
\quad +\Big( (G_2(y_1,x_2;\vec z)-G_2(\vec y;\vec z))(G_2(\vec z;\vec v)-G_2(\vec y;\vec v))\nn\\
\qquad\qquad + (G_2(\vec x;\vec z)-G_2(x_1,y_2;\vec z))(G_2(\vec z;\vec v)-G_2(x_1,y_2;\vec v))\Big)A_2^a(\vec z) \p_1^x\delta(\vec x-\vec u)\left(\vec  \nabla^y\times \vec{A}^a(\vec y) \right) \nn\\
\quad +  (G_1(\vec x;\vec v)-G_1(y_1,x_2;\vec v))(G_2(y_1,x_2;\vec z)-G_2(\vec y;\vec z)) A_2^a(\vec z) \left(\vec \nabla^x\times \vec{A}^a(\vec x)\right) (-\p_2^y\delta(\vec y-\vec u)) \nn\\
\quad +  (G_1(\vec x;\vec z)-G_1(y_1,x_2;\vec z))(G_2(y_1,x_2;\vec v)-G_2(\vec y;\vec v))A_1^a(\vec z) \p_1^x\delta(\vec x-\vec u) \left( \vec \nabla^y\times \vec{A}^a(\vec y) \right) \nn\\
\quad +  (G_2(\vec x;\vec z)-G_2(x_1,y_2;\vec z))(G_1(x_1,y_2;\vec v)-G_1(\vec y;\vec v))A_2^a(\vec z) (-\p_2^x\delta(\vec x-\vec u))\left( \vec \nabla^y\times \vec{A}^a(\vec y) \right) \nn\\
\quad +(G_2(\vec x;\vec v)-G_2(x_1,y_2;\vec v))(G_1(x_1,y_2;\vec z)-G_1(\vec y;\vec z)) A_1^a(\vec z) \left(\vec \nabla^x\times \vec{A}^a(\vec x)\right) \p_1^y\delta(\vec y-\vec u) \Bigg\}\,.
\eE
After partial integration and renaming $\vec y \leftrightarrow \vec x$ in some terms and defining $\vec Z=\vec z-\vec x$, $\vec Y=\vec y-\vec v$, $\vec V=\vec v-\vec z$, we can write this as
\bE{l}
=  -\frac{e^2C_A}{4}\int_{x,V,Y,Z}2\mu^2\delta_{\mu}(\vec Y) \delta_{\mu'}(\vec Y+\vec V+\vec Z)  \left(\vec \nabla\times \vec{A}^a(\vec x)\right) \nn\\
\quad \Bigg\{Y_2 \Big( (G_1(-\vec Z-\vec V)-G_1(Y_1,-Z_2-V_2))(G_1(\vec V)-G_1(Y_1+V_1,-Z_2))\nn\\
\qquad\qquad + (G_1(-Z_1-V_1,Y_2)-G_1(\vec Y))(G_1(\vec V)-G_1(\vec Y+\vec V)) \nn\\
\qquad\qquad + (G_1(\vec Y+\vec V)-G_1(-Z_1,Y_2+V_2))(G_1(-\vec V)-G_1(-Z_1-V_1,Y_2) )\nn\\
\qquad\qquad + (G_1(Y_1+V_1,-Z_2)-G_1(-\vec Z))(G_1(-\vec V)-G_1(-\vec Z-\vec V))\Big)A_1^a(\vec Z+\vec x)   \nn\\
\quad -Y_1 \Big( (G_2(Y_1,-Z_2-V_2)-G_2(\vec Y))(G_2(\vec V)-G_2(\vec Y+\vec V))\nn\\
\qquad\qquad + (G_2(-\vec Z-\vec V)-G_2(-Z_1-V_1,Y_2))(G_2(\vec V)-G_2(-Z_1,Y_2+V_2))  \nn\\
\qquad\qquad + (G_2(-Z_1,Y_2+V_2)-G_2(-\vec Z))(G_2(-\vec V)-G_2(-\vec Z-\vec V))\nn\\
\qquad\qquad + (G_2(\vec Y+\vec V)-G_2(Y_1+V_1,-Z_2))(G_2(-\vec V)-G_2(Y_1,-Z_2-V_2) )\Big)A_2^a(\vec Z+\vec x)  \nn\\
\quad +2Y_2 \Big( (G_1(-\vec Z-\vec V)-G_1(Y_1,-Z_2-V_2))(G_2(Y_1+V_1,-Z_2)-G_2(\vec Y+\vec V))  \Big) A_2^a(\vec Z+\vec x)   \nn\\
\quad -2Y_1 \Big( (G_1(\vec Y+\vec V)-G_1(-Z_1,Y_2+V_2))(G_2(-Z_1-V_1,Y_2)-G_2(-\vec Z-\vec V)) \Big) A_1^a(\vec Z+\vec x)  \Bigg\} \nn\\
\quad +\O\left(\mu^{-1}\right)
\eE
After expansion this can be integrated to
\bE{l}
=  -\frac{e^2c_A}{4}\int_{x,Y,Z}\left(\vec \nabla\times \vec{A}^a(\vec x)\right) 2\mu^2\delta_{\mu}(Y) \delta_{\mu'}(Z_1)\delta_{\mu'}(Y_2) \nn\\
\qquad 2Y_2 \Big[\theta(-Z_1+Y_1)-\theta(Y_1)\Big] (-\theta(Y_2-Z_2))  (1+Z_1\p_1+Z_2\p_2)  A_2^a(\vec x)   \nn\\
\quad +\frac{e^2c_A}{4}\int_{x,Y,Z}\left(\vec \nabla\times \vec{A}^a(\vec x)\right) 2\mu^2\delta_{\mu}(Y) \delta_{\mu'}(Y_1)\delta_{\mu'}(Z_2)  \nn\\
\qquad 2Y_1 \theta(Y_1-Z_1) \Big[\theta(Y_2)-\theta(-Z_2+Y_2)\Big]  (1+Z_1\p_1+Z_2\p_2)  A_1^a(\vec x)  +\O\left(\mu^{-1}\right) \\
= - \frac{e^2 c_A}{4\pi}  \int_{y}    \left(\vec \nabla\times \vec{A}^a(\vec y)\right)^2 \frac{\mu^4}{\left(\mu ^2+\mu'^2\right)^2} +\O\left(\mu^{-1}\right)\,.
\eE

\section{Computation of the fourth term}
\bE{l}
  -\frac{2e^2}{4} \int_{u,v,x,y} \delta_{\mu}(\vec u-\vec v)\delta_{\mu'}(\vec x-\vec y) \Phi^{(1)}_{ab}(\vec u,\vec v) \frac{\delta^2}{\delta A^a_i(\vec u)\delta A^b_i(\vec v)} B^{(0)}_c(\vec x) \de^{cd}B^{(1)}_d(\vec y) \\
= \frac{e^2 C_A}{2} \int_{u,v,x,y,z} \delta_{\mu}(\vec u-\vec v)\delta_{\mu'}(\vec x-\vec y)  \nn\\
\qquad \times \Bigg( (G_1(\vec u;\vec z)-G_1(v_1,u_2;\vec z)+G_1(u_1,v_2;\vec z)-G_1(\vec v;\vec z)) A_1^a(\vec z)  \nn\\
\qquad\qquad  + (G_2(v_1,u_2;\vec z)-G_2(\vec v;\vec z)+G_2(\vec u;\vec z)-G_2(u_1,v_2;\vec z)) A_2^a(\vec z) \Bigg)   \nn\\
\qquad \times \p_{x_n}\delta(\vec x-\vec u) A_n^a(\vec y)\delta(\vec y-\vec v) \quad\qquad\\
= e^2 C_A \int_{v,U,Z} (U_1 A_1^a(\vec v)+U_2 A_2^a(\vec v))  \delta_{\mu}(\vec U) \mu'^2 \delta_{\mu'}(\vec U)  \nn\\
\quad \Bigg( (G_1(\vec U-\vec Z)-G_1(-Z_1,U_2-Z_2)+G_1(U_1-Z_1,-Z_2)-G_1(-\vec Z)) \nn\\ 
\qquad\quad \Big(1+Z_1\p_1+Z_2\p_2+\frac{1}{2}Z_1^2\p^2_1+ Z_1Z_2\p_1\p_2+\frac{1}{2}Z_2^2\p^2_2\Big) A_1^a(\vec Z+\vec v) \Bigg|_{\vec Z=0} \nn\\
\qquad  + (G_2(-Z_1,U_2-Z_2)-G_2(-\vec Z)+G_2(\vec U-\vec Z)-G_2(U_1-Z_1,-Z_2))\nn\\ \qquad\quad \Big(1+Z_1\p_1+Z_2\p_2+\frac{1}{2}Z_1^2\p^2_1+ Z_1Z_2\p_1\p_2+\frac{1}{2}Z_2^2\p^2_2\Big) A_2^a(\vec Z+\vec v) \Bigg|_{\vec Z=0} \Bigg) \nn\\
\quad +\O\left(\mu^{-1}\right)   \\
= e^2 C_A \int_{v} \Bigg\{  A_1^a(\vec v) \Big(\frac{\mu^2 \mu'^4}{\pi  \left(\mu^2+\mu'^2\right)^2} +\frac{\mu^2 \mu'^4}{4\pi  \left(\mu^2+\mu'^2\right)^3} \p_1^2+\frac{\mu^2 \mu'^4}{8\pi  \left(\mu^2+\mu'^2\right)^3} \p_2^2 \Big) A_1^a(\vec v) \nn\\
 \qquad\quad+A_2^a(\vec v) \Big(\frac{\mu^2 \mu'^4}{8 \pi  \left(\mu^2+\mu'^2\right)^3}\p_1\p_2\Big) A_1^a(\vec v) + A_1^a(\vec v) \Big(\frac{\mu^2 \mu'^4}{8 \pi  \left(\mu^2+\mu'^2\right)^3}\p_1\p_2\Big) A_2^a(\vec v)\nn\\
 \qquad\quad+ A_2^a(\vec v) \Big(\frac{\mu^2 \mu'^4}{\pi  \left(\mu^2+\mu'^2\right)^2} +\frac{\mu^2 \mu'^4}{8\pi  \left(\mu^2+\mu'^2\right)^3} \p_1^2 +\frac{\mu^2 \mu'^4}{4\pi  \left(\mu^2+\mu'^2\right)^3} \p_2^2 \Big) A_2^a(\vec v)  \Bigg\}    \nn\\
\quad +\O\left(\mu^{-1}\right)
\eE

\section{Computation of the fifth term}

\bE{rl}
&  -\frac{e^2}{4} \int_{u,v,x,y} \delta_{\mu}(\vec u-\vec v)\delta_{\mu'}(\vec x-\vec y)  \Phi^{(1)}_{ab}(\vec u,\vec v) \frac{\delta^2}{\delta A^a_i(\vec u)\delta A^b_i(\vec v)} B^{(0)}_c(\vec x)\Phi^{(1)}_{cd}(\vec x,\vec y)B^{(0)}_d(\vec y) \nn\\
=& -\frac{e^2}{4} \int_{u,v,x,y} \delta_{\mu}(\vec u-\vec v)\delta_{\mu'}(\vec x-\vec y)  \Phi^{(1)}_{ab}(\vec u,\vec v)  \nn\\
& \qquad\qquad\times \Bigg\{  \frac{\delta^2( B^{(0)}_c(\vec x)B^{(0)}_d(\vec y))}{\delta A^a_i(\vec u)\delta A^b_i(\vec v)} \Phi^{(1)}_{cd}(\vec x,\vec y) + 4 \frac{\delta  B^{(0)}_c(\vec x)}{\delta A^a_i(\vec u)} \frac{\delta \Phi^{(1)}_{cd}(\vec x,\vec y)}{\delta A^b_i(\vec v)} B^{(0)}_d(\vec y) \Bigg\} 
\eE
We consider the two subterms individually.

\subsection{First subterm}

\bE{rl}
 -\frac{e^2}{4} & \int_{u,v,x,y} \delta_{\mu}(\vec u-\vec v)\delta_{\mu'}(\vec x-\vec y)  \Phi^{(1)}_{ab}(\vec u,\vec v)   \frac{\delta^2( B^{(0)}_c(\vec x)B^{(0)}_d(\vec y))}{\delta A^a_i(\vec u)\delta A^b_i(\vec v)} \Phi^{(1)}_{cd}(\vec x,\vec y) \nn\\
= -\frac{e^2 C_A}{8} & \int_{x,y,z,w} \delta_{\mu}(\vec x-\vec y) \delta_{\mu'}(\vec x-\vec y) \nn\\
&\Bigg\{ (G_m(\vec x;\vec z)-G_m(\vec y;\vec z)) A_m^a(\vec z)   - (G_1(y_1,x_2;\vec z)-G_1(x_1,y_2;\vec z)) A_1^a(\vec z)  \nn\\
&\qquad  + (G_2(y_1,x_2;\vec z)-G_2(x_1,y_2;\vec z)) A_2^a(z\vec ) \Bigg\}  \nn\\
&\times \Bigg\{  4 \mu'^2 (x_i-y_i) \Bigg(  \p_i^x G_n(\vec x;\vec w) A_n^a(\vec w)  \nn\\
&\qquad   - ( \delta_{i2}\p_2^x G_1(y_1,x_2;\vec w)-\delta_{i1}\p_1^xG_1(x_1,y_2;\vec w)) A_1^a(\vec w)  \nn\\
&\qquad  +  (\delta_{i2}\p_2^x G_2(y_1,x_2;\vec w)-\delta_{i1}\p_1^x G_2(x_1,y_2;\vec w)) A_2^a(\vec w) \Bigg)    \nn\\  
& \quad + (4\mu'^2 - 4\mu'^4 (x_i-y_i)^2) \Bigg( (G_n(\vec x;\vec w)-G_n(\vec y;\vec w)) A_n^a(\vec w)   \nn\\
&\qquad - (G_1(y_1,x_2;\vec w)-G_1(x_1,y_2;\vec w)) A_1^a(\vec w)  \nn\\
&\qquad   + (G_2(y_1,x_2;\vec w)-G_2(x_1,y_2;\vec w)) A_2^a(\vec w) \Bigg)    \Bigg\} 
\eE
\bE{rl}
= -\frac{e^2 C_A}{8} & \int_{x,y,z,w} \delta_{\mu}(\vec x-\vec y) \delta_{\mu'}(\vec x-\vec y) \nn\\
&\Bigg\{ - (G_1(\vec y;\vec z)-G_1(\vec x;\vec z)+G_1(y_1,x_2;\vec z)-G_1(x_1,y_2;\vec z)) A_1^a(\vec z)  \nn\\
&\qquad  -  (G_2(\vec y;\vec z) - G_2(\vec x;\vec z) - G_2(y_1,x_2;\vec z) + G_2(x_1,y_2;\vec z)) A_2^a(\vec z) \Bigg\}  \nn\\
& \Bigg\{  4 \mu'^2 \Bigg(  - (x_2-y_2) [G_1(y_1,x_2;\vec w)-G_1(\vec x;\vec w)] \p_2 A_1^a(\vec w)  \nn\\
&\qquad  +(x_1-y_1) [\delta(x_1,y_2;\vec w) +\delta(\vec x;\vec w)] A_1^a(\vec w)  \nn\\
&\qquad  + (x_2-y_2) [\delta(y_1,x_2;\vec w)+\delta(\vec x;\vec w)]A_2^a(\vec w)  \nn\\
&\qquad  - (x_1-y_1) [G_2(x_1,y_2;\vec w)-G_2(\vec x;\vec w)] \p_1 A_2^a(\vec w) \Bigg)    \nn\\  
& \; - (4\mu'^2 - 4\mu'^4 (x_i-y_i)^2)  \nn\\
&\; \times \Bigg((G_1(\vec y;\vec w)-G_1(\vec x;\vec w)+G_1(y_1,x_2;\vec w)-G_1(x_1,y_2;\vec w)) A_1^a(\vec w)  \nn\\
&\qquad  + (G_2(\vec y;\vec w) - G_2(\vec x;\vec w) - G_2(y_1,x_2;\vec w) + G_2(x_1,y_2;\vec w)) A_2^a(\vec w) \Bigg)    \Bigg\} 
\eE
\bE{rl}
= &- \frac{e^2 C_A}{8} (-4\mu'^2) \int_{x,y,z} \delta_{\mu}(\vec x-\vec y) \delta_{\mu'}(\vec x-\vec y) \nn\\
&\Bigg\{ (G_1(\vec y;\vec z)-G_1(\vec x;\vec z)+G_1(y_1,x_2;\vec z)-G_1(x_1,y_2;\vec z)) (x_1-y_1) [A_1^a(x_1,y_2) +A_1^a(\vec x)]  A_1^a(\vec z)  \nn\\
&\quad  + (G_1(\vec y;\vec z)-G_1(\vec x;\vec z)+G_1(y_1,x_2;\vec z)-G_1(x_1,y_2;\vec z)) (x_2-y_2) [A_2^a(y_1,x_2)+A_2^a(\vec x)] A_1^a(\vec z)    \nn\\
&\quad  +  (G_2(\vec y;\vec z) - G_2(\vec x;\vec z) - G_2(y_1,x_2;\vec z) + G_2(x_1,y_2;\vec z)) (x_2-y_2) [A_2^a(y_1,x_2)+A_2^a(\vec x)] A_2^a(\vec z) \nn\\
&\quad  +  (G_2(\vec y;\vec z) - G_2(\vec x;\vec z) - G_2(y_1,x_2;\vec z) + G_2(x_1,y_2;\vec z)) (x_1-y_1) [A_1^a(x_1,y_2) +A_1^a(\vec x)]  A_2^a(\vec z)   \Bigg\}  \nn\\
  &- \frac{e^2 C_A}{8}  \int_{x,y,z,w} \delta_{\mu}(\vec x-\vec y) \delta_{\mu'}(\vec x-\vec y) \nn\\
&\Bigg\{ (G_1(\vec y;\vec z)-G_1(\vec x;\vec z)+G_1(y_1,x_2;\vec z)-G_1(x_1,y_2;\vec z)) (4\mu'^2 - 4\mu'^4 (x_i-y_i)^2)  \nn\\
& \qquad   (G_1(\vec y;\vec w)-G_1(\vec x;\vec w)+G_1(y_1,x_2;\vec w)-G_1(x_1,y_2;\vec w))  A_1^a(\vec z) A_1^a(\vec w) \nn\\
&\quad  + (G_1(\vec y;\vec z)-G_1(\vec x;\vec z)+G_1(y_1,x_2;\vec z)-G_1(x_1,y_2;\vec z))  (4\mu'^2 - 4\mu'^4 (x_i-y_i)^2)   \nn\\
&\qquad(G_2(\vec y;\vec w) - G_2(\vec x;\vec w) - G_2(y_1,x_2;\vec w) + G_2(x_1,y_2;\vec w)) A_1^a(\vec z)A_2^a(\vec w) \nn\\
&\quad  +  (G_2(\vec y;\vec z) - G_2(\vec x;\vec z) - G_2(y_1,x_2;\vec z) + G_2(x_1,y_2;\vec z)) (4\mu'^2 - 4\mu'^4 (x_i-y_i)^2)  \nn\\
&\qquad (G_2(\vec y;\vec w) - G_2(\vec x;\vec w) - G_2(y_1,x_2;\vec w) + G_2(x_1,y_2;\vec w)) A_2^a(\vec z)A_2^a(\vec w)  \nn\\
&\quad  +  (G_2(\vec y;\vec z) - G_2(\vec x;\vec z) - G_2(y_1,x_2;\vec z) + G_2(x_1,y_2;\vec z))  (4\mu'^2 - 4\mu'^4 (x_i-y_i)^2 ) \nn\\
& \qquad  (G_1(\vec y;\vec w)-G_1(\vec x;\vec w)+G_1(y_1,x_2;\vec w)-G_1(x_1,y_2;\vec w)) A_2^a(\vec z) A_1^a(\vec w)\Bigg\}  \nn\\
 &-\frac{e^2 C_A}{8} \int_{x,y,z,w} \delta_{\mu}(\vec x-\vec y) \delta_{\mu'}(\vec x-\vec y) \nn\\
&\Bigg\{  (G_1(\vec y;\vec z)-G_1(\vec x;\vec z)+G_1(y_1,x_2;\vec z)-G_1(x_1,y_2;\vec z)) A_1^a(\vec z)  \nn\\
&\qquad  +  (G_2(\vec y;\vec z) - G_2(\vec x;\vec z) - G_2(y_1,x_2;\vec z) + G_2(x_1,y_2;\vec z)) A_2^a(\vec z) \Bigg\}  \nn\\
& \Bigg\{  4 \mu'^2 \Bigg(   (x_2-y_2) [G_1(y_1,x_2;\vec w)-G_1(\vec x;\vec w)] \p_2 A_1^a(\vec w) \nn\\
&\qquad\qquad + (x_1-y_1) [G_2(x_1,y_2;\vec w)-G_2(\vec x;\vec w)] \p_1 A_2^a(\vec w) \Bigg)    \Bigg\} 
\eE
In the first integral we define $\vec X=\vec x-\vec y$, $\vec Y=\vec y-\vec z$, then shift $\vec X\to \vec X-\vec Y$. In the second and third integral we define $\vec Z=\vec z-\vec w$, $\vec X=\vec x-\vec y$, $\vec Y=\vec y-\vec w$: 
\bE{rl}
= &- \frac{e^2 C_A}{8} (-4\mu'^2) \int_{X,Y,z} \delta_{\mu}(\vec X-\vec Y) \delta_{\mu'}(\vec X-\vec Y) \nn\\
&\Bigg\{ (G_1(\vec Y)-G_1(\vec X)+G_1(Y_1,X_2)-G_1(X_1,Y_2)) 
(X_1-Y_1) A_1^a(\vec z) \nn\\
&\qquad  \left[2+2X_1\p_1+(X_2+Y_2)\p_2+{2\over2}X_1^2\p_1^2+{1\over2}(X_2^2+Y_2^2)\p_2^2 +X_1X_2\p_1\p_2 + X_1Y_2\p_1\p_2\right]  A_1^a(\vec z)   \nn\\
&\quad  + (G_1(\vec Y)-G_1(\vec X)+G_1(Y_1,X_2)-G_1(X_1,Y_2)) (X_2-Y_2)A_1^a(\vec z)  \nn\\
&\qquad \left[2+(Y_1+X_1)\p_1+2X_2\p_2+{1\over2}(Y_1^2+X_1^2)\p_1^2 +{2\over2}X_2^2\p_2^2 +Y_1X_2\p_1\p_2 +X_1X_2\p_1\p_2 \right] A_2^a(\vec z)    \nn\\
&\quad  + (G_2(\vec Y)-G_2(\vec X)-G_2(Y_1,X_2)+G_2(X_1,Y_2)) (X_2-Y_2) A_2^a(\vec z) \nn\\
&\qquad \left[2+(Y_1+X_1)\p_1+2X_2\p_2+{1\over2}(Y_1^2+X_1^2)\p_1^2 +{2\over2}X_2^2\p_2^2 +Y_1X_2\p_1\p_2 +X_1X_2\p_1\p_2 \right] A_2^a(\vec z)    \nn\\
&\quad  + (G_2(\vec Y)-G_2(\vec X)-G_2(Y_1,X_2)+G_2(X_1,Y_2)) (X_1-Y_1) A_2^a(\vec z) \nn\\
&\qquad  \left[2+2X_1\p_1+(X_2+Y_2)\p_2+{2\over2}X_1^2\p_1^2+{1\over2}(X_2^2+Y_2^2)\p_2^2 +X_1X_2\p_1\p_2 + X_1Y_2\p_1\p_2\right]  A_1^a(\vec z)    \Bigg\}  \nn\\
 &- \frac{e^2 C_A}{8}  \int_{w,X,Y,Z} \delta_{\mu}(\vec X) \delta_{\mu'}(\vec X) \nn\\
&\Bigg\{  (G_1(\vec Y-\vec Z)-G_1(\vec X+\vec Y-\vec Z) )\nn\\
&\qquad\qquad +G_1(Y_1-Z_1,X_2+Y_2-Z_2)-G_1(X_1+Y_1-Z_1,Y_2-Z_2))\nn\\
&\qquad  (4\mu'^2 - 4\mu'^4 (X_1^2+X_2^2))  (G_1(\vec Y)-G_1(\vec X+\vec Y)+G_1(Y_1,X_2+Y_2)-G_1(X_1+Y_1,Y_2)) \nn\\
&\qquad  \Big(1+Z_1\p_1+Z_2\p_2+\frac{1}{2}Z_1^2\p^2_1+ Z_1Z_2\p_1\p_2+\frac{1}{2}Z_2^2\p^2_2\Big)  A_1^a(\vec Z+\vec w) A_1^a(\vec w) \Bigg|_{\vec Z=0} \nn\\
&\quad + (G_1(\vec Y-\vec Z)-G_1(\vec X+\vec Y-\vec Z)\nn\\
&\qquad\qquad +G_1(Y_1-Z_1,X_2+Y_2-Z_2)-G_1(X_1+Y_1-Z_1,Y_2-Z_2))\nn\\
&\qquad  (4\mu'^2 - 4\mu'^4 (X_1^2+X_2^2))  (G_2(\vec Y) - G_2(\vec X+\vec Y) - G_2(Y_1,X_2+Y_2) + G_2(X_1+Y_1,Y_2)) \nn\\
&\qquad  \Big(1+Z_1\p_1+Z_2\p_2+\frac{1}{2}Z_1^2\p^2_1+ Z_1Z_2\p_1\p_2+\frac{1}{2}Z_2^2\p^2_2\Big) A_1^a(\vec Z+\vec w)A_2^a(\vec w)  \Bigg|_{\vec Z=0} \nn\\
&\quad + (G_2(\vec Y-\vec Z)-G_2(\vec X+\vec Y-\vec Z)\nn\\
&\qquad\qquad -G_2(Y_1-Z_1,X_2+Y_2-Z_2)+G_2(X_1+Y_1-Z_1,Y_2-Z_2))\nn\\
&\qquad  (4\mu'^2 - 4\mu'^4 (X_1^2+X_2^2))(G_2(\vec Y) - G_2(\vec X+\vec Y) - G_2(Y_1,X_2+Y_2) + G_2(X_1+Y_1,Y_2))  \nn\\
&\qquad  \Big(1+Z_1\p_1+Z_2\p_2+\frac{1}{2}Z_1^2\p^2_1+ Z_1Z_2\p_1\p_2+\frac{1}{2}Z_2^2\p^2_2\Big) A_2^a(\vec Z+\vec w)A_2^a(\vec w)  \Bigg|_{\vec Z=0} \nn\\
&\quad  +  (G_2(\vec Y-\vec Z) - G_2(\vec X+\vec Y-\vec Z) \nn\\
&\qquad\qquad - G_2(Y_1-Z_1,X_2+Y_2-Z_2) + G_2(X_1+Y_1-Z_1,Y_2-Z_2)) \nn\\
& \qquad  (4\mu'^2 - 4\mu'^4 (X_1^2+X_2^2)) (G_1(\vec Y)-G_1(\vec X+\vec Y)+G_1(Y_1,X_2+Y_2)-G_1(X_1+Y_1,Y_2))  \nn\\
&\qquad  \Big(1+Z_1\p_1+Z_2\p_2+\frac{1}{2}Z_1^2\p^2_1+ Z_1Z_2\p_1\p_2+\frac{1}{2}Z_2^2\p^2_2\Big) A_2^a(\vec Z+\vec w)A_1^a(\vec w)  \Bigg|_{\vec Z=0} \Bigg\}  \nn\\
 &-\frac{e^2 C_A}{8} \int_{w,X,Y,Z} \delta_{\mu}(\vec X) \delta_{\mu'}(\vec X) \nn\\
&\Bigg\{  (G_1(\vec Y-\vec Z)-G_1(\vec X+\vec Y-\vec Z)+G_1(Y_1-Z_1,X_2+Y_2-Z_2) \nn\\
&\qquad\qquad -G_1(X_1+Y_1-Z_1,Y_2-Z_2)) \times  \Big(1+Z_1\p_1+Z_2\p_2\Big) A_1^a(\vec Z+\vec w)  \nn\\
&\quad  +  (G_2(\vec Y-\vec Z) - G_2(\vec X+\vec Y-\vec Z) - G_2(Y_1-Z_1,X_2+Y_2-Z_2) \nn\\
&\qquad\qquad + G_2(X_1+Y_1-Z_1,Y_2-Z_2)) \times \Big(1+Z_1\p_1+Z_2\p_2\Big) A_2^a(\vec Z+\vec w) \Bigg\}   \nn\\
&\times \Bigg\{  4 \mu'^2 \Bigg(   X_2 [G_1(Y_1,X_2+Y_2)-G_1(\vec X+\vec Y)] \p_2 A_1^a(\vec w) \nn\\
&\qquad\qquad + X_1 [G_2(X_1+Y_1,Y_2)-G_2(\vec X+\vec Y)] \p_1 A_2^a(\vec w) \Bigg)    \Bigg\} \nn\\
 &  +\O\left(\mu^{-1}\right)
\eE
\bE{rl}
= & \frac{e^2 C_A}{2} \int_{z}\Bigg\{  A_1^a(\vec z)  \left[\mu'^22\frac{- \mu^2 \mu'^2}{\pi  \left(\mu^2+\mu'^2\right)^2} + \mu'^2{2\over2}\frac{- \mu^2 \mu'^2}{2 \pi  \left(\mu^2+\mu'^2\right)^3}\p_1^2 +\mu'^2{1\over2}\frac{- \mu^2 \mu'^2}{2 \pi  \left(\mu^2+\mu'^2\right)^3}\p_2^2 \right]  A_1^a(\vec z) 
  \nn\\
&\qquad  + A_1^a(\vec z) \left[0 \right] A_2^a(\vec z)    \nn\\
&\qquad  + A_2^a(\vec z)  \left[\mu'^2 2\frac{- \mu^2 \mu'^2}{\pi  \left(\mu^2+\mu'^2\right)^2} +\mu'^2 {1\over2}\frac{- \mu^2 \mu'^2}{2 \pi  \left(\mu^2+\mu'^2\right)^3}\p_1^2 +\mu'^2{2\over2}\frac{- \mu^2 \mu'^2}{2 \pi  \left(\mu^2+\mu'^2\right)^3}\p_2^2 \right]  A_2^a(\vec z)   \nn\\
&\qquad  +  A_2^a(\vec z)  \left[0 \right]  A_1^a(\vec z)    \Bigg\}  \nn\\
 &- \frac{e^2 C_A}{8}  \int_{w} \Bigg\{  A_1^a(\vec w)  \Big(\frac{8 \mu^2 (\mu -\mu') \mu'^4 (\mu +\mu')}{\pi  \left(\mu^2+\mu'^2\right)^3} \nn\\
&\qquad\qquad  + \frac{\mu^2 \mu'^4 \left(\mu^2-2 \mu'^2\right)}{\pi  \left(\mu^2+\mu'^2\right)^4} \p^2_1 + \frac{\mu^2 \mu'^4 \left(\mu^2-2 \mu'^2\right)}{\pi  \left(\mu^2+\mu'^2\right)^4} \p^2_2\Big)  A_1^a(\vec w)    \nn\\
&\quad +  \Big(0\Big) A_1^a(\vec w)A_2^a(\vec w)   \nn\\
&\quad + A_2^a(\vec w)  \Big(\frac{8 \mu^2 (\mu -\mu') \mu'^4 (\mu +\mu')}{\pi  \left(\mu^2+\mu'^2\right)^3} + \frac{\mu^2 \mu'^4 \left(\mu^2-2 \mu'^2\right)}{\pi  \left(\mu^2+\mu'^2\right)^4} \p^2_1 + \frac{\mu^2 \mu'^4 \left(\mu^2-2 \mu'^2\right)}{\pi  \left(\mu^2+\mu'^2\right)^4} \p^2_2\Big)  A_2^a(\vec w)  \nn\\
&\quad +  \Big(0\Big) A_2^a(\vec w)A_1^a(\vec w)  \Bigg\}  \nn\\
 &-\frac{e^2 C_A}{8} \int_{w} \Bigg\{-\frac{ \mu^2 \mu'^4}{\pi  \left(\mu ^2+\mu'^2\right)^3} (\p_2 A_1^a(\vec w)) (\p_2 A_1^a(\vec w)) - \frac{\mu ^2 \mu'^4}{\pi  \left(\mu ^2+\mu'^2\right)^3} (\p_1 A_2^a(\vec w))(\p_1 A_2^a(\vec w)) \Bigg\} \nn\\
&    +\O\left(\mu^{-1}\right)
\eE

\subsection{Second subterm}

\bE{rl}
 -\frac{4e^2}{4} & \int_{u,v,x,y} \delta_{\mu}(\vec u-\vec v)\delta_{\mu'}(\vec x-\vec y)  \Phi^{(1)}_{ab}(\vec u,\vec v) \frac{\delta  B^{(0)}_c(\vec x)}{\delta A^a_i(\vec u)} \frac{\delta \Phi^{(1)}_{cd}(\vec x,\vec y)}{\delta A^b_i(\vec v)} B^{(0)}_d(\vec y) \nn\\
= - {e^2 C_A\over 4} & \int_{u,v,y,z} 2\mu^2  \delta_{\mu}(\vec u-\vec v)\delta_{\mu'}(\vec x-\vec y) \nn\\
& \qquad \times \Bigg\{(G_1(\vec u;\vec z)-G_1(v_1,u_2;\vec z)+G_1(u_1,v_2;\vec z)-G_1(\vec v;\vec z)) A_1^d(\vec z)  \nn\\
& \qquad\qquad  + (G_2(v_1,u_2;\vec z)-G_2(\vec v;\vec z)+G_2(\vec u;\vec z)-G_2(u_1,v_2;\vec z)) A_2^d(\vec z) \Bigg\} \nn\\
& \qquad \times \Bigg\{ -(u_2-v_2) (G_1(\vec u;\vec v)-G_1(y_1,u_2;\vec v)+G_1(u_1,y_2;\vec v)-G_1(\vec y;\vec v))  \nn\\
& \qquad\qquad + (u_1-v_1) (G_2(y_1,u_2;\vec v)-G_2(\vec y;\vec v)+G_2(\vec u;\vec v)-G_2(u_1,y_2;\vec v)) \Bigg\}  B^{(0)}_d(\vec y)  \nn\\
+ {e^2 C_A\over 4} & \int_{u,v,y,z}   \delta_{\mu}(\vec u-\vec v)\delta_{\mu'}(\vec x-\vec y) \e_{ki}\p_{u_k} \nn\\
& \qquad \times \Bigg\{(G_1(\vec u;\vec z)-G_1(v_1,u_2;\vec z)+G_1(u_1,v_2;\vec z)-G_1(\vec v;\vec z)) A_1^d(\vec z)  \nn\\
& \qquad\qquad  + (G_2(v_1,u_2;\vec z)-G_2(\vec v;\vec z)+G_2(\vec u;\vec z)-G_2(u_1,v_2;\vec z)) A_2^d(\vec z) \Bigg\} \nn\\
& \qquad \times \Bigg\{ (G_1(\vec u;\vec v)-G_1(y_1,u_2;\vec v)+G_1(u_1,y_2;\vec v)-G_1(\vec y;\vec v)) \delta_{1i}  \nn\\
& \qquad\qquad + (G_2(y_1,u_2;\vec v)-G_2(\vec y;\vec v)+G_2(\vec u;\vec v)-G_2(u_1,y_2;\vec v)) \delta_{2i} \Bigg\}  B^{(0)}_d(\vec y) \\
= -{e^2 C_A\over 4} & \int_{u,v,y,z} 2\mu^2 \delta_{\mu}(\vec u-\vec v) \delta_{\mu'}(\vec u-\vec y) \nn\\
& \qquad \times \Bigg\{(G_1(\vec u;\vec z)-G_1(v_1,u_2;\vec z)+G_1(u_1,v_2;\vec z)-G_1(\vec v;\vec z)) A_1^d(\vec z)  \nn\\
& \qquad\qquad  + (G_2(v_1,u_2;\vec z)-G_2(\vec v;\vec z)+G_2(\vec u;\vec z)-G_2(u_1,v_2;\vec z)) A_2^d(\vec z) \Bigg\} \nn\\
& \qquad \Bigg\{ -(u_2-v_2) (G_1(\vec u;\vec v)-G_1(y_1,u_2;\vec v)+G_1(u_1,y_2;\vec v)-G_1(\vec y;\vec v))  \nn\\
& \qquad\qquad + (u_1-v_1) (G_2(y_1,u_2;\vec v)-G_2(\vec y;\vec v)+G_2(\vec u;\vec v)-G_2(u_1,y_2;\vec v)) \Bigg\}   B^{(0)}_d(\vec y)  \nn\\
-{e^2 C_A\over 4} & \int_{u,v,y,z}  \delta_{\mu}(\vec u-\vec v)\delta_{\mu'}(\vec u-\vec y) (G_1(\vec u;\vec v)-G_1(y_1,u_2;\vec v)+G_1(u_1,y_2;\vec v)-G_1(\vec y;\vec v))  \nn\\
& \qquad (G_1(\vec u;\vec z)-G_1(v_1,u_2;\vec z)) \p_2A_1^d(\vec z)   B^{(0)}_d(\vec y)   \nn\\
-{e^2 C_A\over 4} & \int_{u,v,y}  \delta_{\mu}(\vec u-\vec v)\delta_{\mu'}(\vec u-\vec y) (G_1(\vec u;\vec v)-G_1(y_1,u_2;\vec v)+G_1(u_1,y_2;\vec v)-G_1(\vec y;\vec v))  \nn\\
& \qquad  (A_2^d(v_1,u_2)+A_2^d(\vec u))   B^{(0)}_d(\vec y)  \nn\\
+{e^2 C_A\over 4} & \int_{u,v,y}  \delta_{\mu}(\vec u-\vec v)\delta_{\mu'}(\vec u-\vec y) (G_2(y_1,u_2;\vec v)-G_2(\vec y;\vec v)+G_2(\vec u;\vec v)-G_2(u_1,y_2;\vec v))  \nn\\
&\qquad (A_1^d(\vec u)+A_1^d(u_1,v_2))   B^{(0)}_d(\vec y)\nn\\
+{e^2 C_A\over 4} & \int_{u,v,y,z}  \delta_{\mu}(\vec u-\vec v)\delta_{\mu'}(\vec u-\vec y) (G_2(y_1,u_2;\vec v)-G_2(\vec y;\vec v)+G_2(\vec u;\vec v)-G_2(u_1,y_2;\vec v))  \nn\\
&\qquad  (G_2(\vec u;\vec z)-G_2(u_1,v_2;\vec z)) \p_1 A_2^d(\vec z)   B^{(0)}_d(\vec y)
\eE
We define $\vec Z=\vec z-\vec y$, $\vec U=\vec u-\vec y$, $\vec V=\vec v-\vec u$ and expand the fields:
\bE{rl}
= -&{e^2 C_A\over 4} \int_{y,U,V,Z}   B^{(0)}_d(\vec y)  \delta_{\mu}(\vec V)\delta_{\mu'}(\vec U) \nn\\
& \times (G_1(-\vec V)-G_1(-U_1-V_1,-V_2)+G_1(-V_1,-U_2-V_2)-G_1(-\vec U-\vec V))    \nn\\
&  \times \Bigg\{(G_1(\vec U-\vec Z)-G_1(V_1+U_1-Z_1,U_2-Z_2)) \p_2 A_1^d(\vec y) \nn\\
& +2\mu^2 V_2 \Big(G_1(\vec U-\vec Z)-G_1(V_1+U_1-Z_1,U_2-Z_2) \nn\\
& \qquad\; +G_1(U_1-Z_1,V_2+U_2-Z_2)-G_1(\vec V+\vec U-\vec Z)\Big)   \Big(1+Z_1\p_1+Z_2\p_2\Big)A_1^d(\vec y)  \nn\\
& +2\mu^2 V_2 \Big(G_2(V_1+U_1-Z_1,U_2-Z_2)-G_2(\vec V+\vec U-\vec Z) \nn\\
& \qquad\;  +G_2(\vec U-\vec Z)-G_2(U_1-Z_1,V_2+U_2-Z_2)\Big)  \Big(1+Z_1\p_1+Z_2\p_2\Big)A_2^d(\vec y) \Bigg\}    \nn\\ 
+&{e^2 C_A\over 4} \int_{y,U,V,Z}    B^{(0)}_d(\vec y) \delta_{\mu}(\vec V)\delta_{\mu'}(U)\nn\\
&\times (G_2(-U_1-V_1,-V_2)-G_2(-\vec U-\vec V)+G_2(-\vec V)-G_2(-V_1,-U_2-V_2)) \nn\\
& \times \Bigg\{ (G_2(\vec U-\vec Z)-G_2(U_1-Z_1,V_2+U_2-Z_2)) \p_1 A_2^d(\vec y) \nn\\
& +2\mu^2 V_1 \Big(G_1(\vec U-\vec Z)-G_1(V_1+U_1-Z_1,U_2-Z_2) \nn\\
& \qquad\; +G_1(U_1-Z_1,V_2+U_2-Z_2)-G_1(\vec V+\vec U-\vec Z)\Big)  \Big(1+Z_1\p_1+Z_2\p_2\Big)A_1^d(\vec y)  \nn\\
& +2\mu^2 V_1 \Big(G_2(V_1+U_1-Z_1,U_2-Z_2)-G_2(\vec V+\vec U-\vec Z) \nn\\
& \qquad\;  +G_2(\vec U-\vec Z)-G_2(U_1-Z_1,V_2+U_2-Z_2)\Big)  \Big(1+Z_1\p_1+Z_2\p_2\Big)A_2^d(\vec y) \Bigg\} \nn\\
- &{e^2 C_A\over 4}  \int_{y,U,V}  \delta_{\mu}(-\vec V)\delta_{\mu'}(\vec U)  B^{(0)}_d(\vec y) (2+2U_1\p_1+2U_2\p_2+V_1\p_1)A_2^d(\vec y)  \nn\\
& \quad     (G_1(-\vec V)-G_1(-U_1-V_1,-V_2)+G_1(-V_1,-U_2-V_2)-G_1(-\vec U-\vec V))    \nn\\
+&{e^2 C_A\over 4}  \int_{y,U,V}  \delta_{\mu}(-\vec V)\delta_{\mu'}(\vec U) B^{(0)}_d(\vec y) (2+2U_1\p_1+2U_2\p_2+V_2\p_2)A_1^d(\vec y)   \nn\\
&\quad   (G_2(-U_1-V_1,-V_2)-G_2(-\vec U-\vec V)+G_2(-\vec V)-G_2(-V_1,-U_2-V_2)) \nn\\
+&\O\left(\mu^{-1}\right) \,.
\eE
\bE{rl}
= -{e^2 C_A\over 4}  \int_{y} & B^{(0)}_d(\vec y)    \Bigg\{\frac{\mu ^2}{2 \pi  \mu ^2+2 \pi  \mu'^2} \p_2 A_1^d(\vec y)  -\frac{\mu ^2 \mu'  \left(-\mu' +\sqrt{\mu ^2+\mu'^2}\right)}{2 \pi  \left(\mu ^2+\mu'^2\right)^2} \p_2 A_1^d(\vec y)  \nn\\
& -\frac{\mu ^2 \left(2 \mu ^2+\mu'  \left(\mu' -\sqrt{\mu ^2+\mu'^2}\right)\right)}{2 \pi  \left(\mu ^2+\mu'^2\right)^2} \p_1 A_2^d(\vec y) \Bigg\}    \nn\\ 
+ {e^2 C_A\over 4}  \int_{y} &   B^{(0)}_d(\vec y)  \Bigg\{ \frac{\mu ^2}{2 \pi  \mu ^2+2 \pi  \mu'^2} \p_1 A_2^d(\vec y)  -\frac{\mu^2 \left(2 \mu^2+\mu'  \left(\mu' -\sqrt{\mu^2+\mu'^2}\right)\right)}{2 \pi  \left(\mu^2+\mu'^2\right)^2} \p_2 A_1^d(\vec y)  \nn\\
& -\frac{\mu ^2 \mu'  \left(-\mu' +\sqrt{\mu ^2+\mu'^2}\right)}{2 \pi  \left(\mu ^2+\mu'^2\right)^2} \p_1 A_2^d(\vec y) \Bigg\} \nn\\
 - {e^2 C_A\over 4}  \int_{y} &  B^{(0)}_d(\vec y)  \Big(2 \frac{\mu^2 \left(\mu^2+\mu'  \left(\mu' +\sqrt{\mu^2+\mu'^2}\right)\right)}{2 \pi \mu'  \left(\mu ^2+\mu'^2\right)^{3/2}}\p_1 -\frac{\mu^2}{2 \pi  \mu^2+2 \pi  \mu'^2}\p_1\Big)A_2^d(\vec y) \nn\\
 + {e^2 C_A\over 4}  \int_{y} & B^{(0)}_d(\vec y)  \Big(2 \frac{\mu^2 \left(\mu^2+\mu'  \left(\mu' +\sqrt{\mu^2+\mu'^2}\right)\right)}{2 \pi \mu'  \left(\mu ^2+\mu'^2\right)^{3/2}}\p_2 -\frac{\mu^2}{2 \pi  \mu^2+2 \pi  \mu'^2}\p_2\Big)A_1^d(\vec y) \nn\\
+\O\left(\mu^{-1}\right) &\,.
\eE

\section{Computation of the sixth term}

\bE{rCl}
 &&-\frac{e^2}{4} \int_{u,v,x,y} \delta_{\mu}(\vec u-\vec v)\delta_{\mu'}(\vec x-\vec y)  \Phi^{(2)}_{ab}(\vec u,\vec v) \frac{\delta^2}{\delta A^a_i(\vec u)\delta A^b_i(\vec v)} B^{(0)}_c(\vec x)\delta^{cd}B^{(0)}_d(\vec y)  \\
 &=& {e^2C_A\over 4} \int_{u,v,y,z} \delta_\mu(\vec u-\vec v) (4\mu'^2-4\mu'^4(\vec{u}-\vec{v})^2) \delta_{\mu'}(\vec u-\vec v) \nn\\
&& \Bigg\{\Big( (G_1(\vec u;\vec z)-G_1(v_1,u_2;\vec z))(G_1(\vec z;\vec y)-G_1(v_1,u_2;\vec y))\nn\\
&& \qquad\qquad + (G_1(u_1,v_2;\vec z)-G_1(\vec v;\vec z))(G_1(\vec z;\vec y)-G_1(\vec v;\vec y))\Big)A_1^c(\vec z)A_1^c(\vec y)  \nn\\
&& \quad + \Big( (G_2(v_1,u_2;\vec z)-G_2(\vec v;\vec z))(G_2(\vec z;\vec y)-G_2(\vec v;\vec y))\nn\\
&& \qquad\qquad + (G_2(\vec u;\vec z)-G_2(u_1,v_2;\vec z))(G_2(\vec z;\vec y)-G_2(u_1,v_2;\vec y))\Big)A_2^c(\vec z)A_2^c(\vec y)  \nn\\
&& \quad + (G_1(\vec u;\vec y)-G_1(v_1,u_2;\vec y))(G_2(v_1,u_2;\vec z)-G_2(\vec v;\vec z))A_1^c(\vec y) A_2^c(\vec z) \nn\\
&& \quad + (G_2(\vec u;\vec z)-G_2(u_1,v_2;\vec z))(G_1(u_1,v_2;\vec y)-G_1(\vec v;\vec y))A_2^c(\vec z)A_1^c(\vec y)\Bigg\}
\end{IEEEeqnarray}
Define $\vec U=\vec u-\vec v$, $\vec Z=\vec z-\vec y$, $\vec V=\vec v-\vec z$
\begin{IEEEeqnarray}{rCl}
 &=& {e^2C_A\over 4} \int_{y,U,V,Z} \delta_\mu(\vec U) (4\mu'^2-4\mu'^4 \vec{U}^2) \delta_{\mu'}(\vec U) \nn\\
&& \Bigg\{\Big( (G_1(\vec U+\vec V)-G_1(V_1,U_2+V_2))(G_1(\vec Z)-G_1(V_1+Z_1,U_2+V_2+Z_2))\nn\\
&& \qquad + (G_1(U_1+V_1,V_2)-G_1(\vec V))(G_1(\vec Z)-G_1(\vec V+\vec Z))\Big)A_1^c(\vec Z+\vec y)A_1^c(\vec y)  \nn\\
&&  + \Big( (G_2(V_1,U_2+V_2)-G_2(\vec V))(G_2(\vec Z)-G_2(\vec V+\vec Z))\nn\\
&& \qquad + (G_2(\vec U+\vec V)-G_2(U_1+V_1,V_2)) \nn\\
&&\qquad\qquad\times (G_2(\vec Z)-G_2(U_1+V_1+Z_1,V_2+Z_2))\Big)A_2^c(\vec Z+\vec y)A_2^c(\vec y)  \nn\\
&&  + (G_1(\vec U+\vec V+\vec Z)-G_1(V_1+Z_1,U_2+V_2+Z_2)) \nn\\
&&\qquad\qquad\times (G_2(V_1,U_2+V_2)-G_2(\vec V))A_1^c(\vec y) A_2^c(\vec Z+\vec y) \nn\\
&&  + (G_2(\vec U+\vec V)-G_2(U_1+V_1,V_2)) \nn\\
&&\qquad\qquad\times (G_1(U_1+V_1+Z_1,V_2+Z_2)-G_1(\vec V+\vec Z))A_2^c(\vec Z+\vec y)A_1^c(\vec y)\Bigg\} \\
 &=& {e^2 C_A\over 4} \int_{y}  \Bigg\{2 A_1^c(\vec y)  \Big( \frac{2\mu^4 \mu'^4 -\mu^2 \mu'^6}{\pi  \left(\mu^2+\mu'^2\right)^3} -\frac{\mu^4 \mu'^4 }{\pi  \left(\mu^2+\mu'^2\right)^3}  \nn\\
&&\qquad\qquad +\Big(\frac{2\mu^4 \mu'^4 -\mu^2 \mu'^6}{4 \pi  \left(\mu^2+\mu'^2\right)^4}-\frac{3 \mu^4 \mu'^4}{8 \pi  \left(\mu^2+\mu'^2\right)^4}\Big)\p_1^2\Big)A_1^c(\vec y) \nn\\
&& \quad + 2 A_2^c(\vec y)  \Big( \frac{2\mu^4 \mu'^4 -\mu^2 \mu'^6}{\pi  \left(\mu^2+\mu'^2\right)^3}-\frac{\mu^4 \mu'^4}{\pi  \left(\mu^2+\mu'^2\right)^3}  \nn\\
&&\qquad\qquad +\Big(\frac{2\mu^4 \mu'^4 -\mu^2 \mu'^6}{4 \pi  \left(\mu^2+\mu'^2\right)^4}-\frac{3 \mu^4 \mu'^4}{8 \pi  \left(\mu^2+\mu'^2\right)^4}\Big)\p_2^2\Big)A_2^c(\vec y)  \nn\\
&& \quad + A_1^c(\vec y) \Big(-\frac{\mu^2 \mu'^6}{4 \pi  \left(\mu^2+\mu'^2\right)^4}-\frac{\mu^2 \mu'^6}{4 \pi  \left(\mu^2+\mu'^2\right)^4}+\frac{\mu^4 \mu'^4}{4 \pi  \left(\mu^2+\mu'^2\right)^4}\Big) \p_1\p_2 A_2^c(\vec y) \nn\\
&& \quad + A_1^c(\vec y) \Big(-\frac{\mu^2 \mu'^6}{4 \pi  \left(\mu^2+\mu'^2\right)^4}-\frac{\mu^2 \mu'^6}{4 \pi  \left(\mu^2+\mu'^2\right)^4}+\frac{\mu^4 \mu'^4}{4 \pi  \left(\mu^2+\mu'^2\right)^4}\Big) \p_1\p_2 A_2^c(\vec y)  \Bigg\}  \nn\\
&& +\O\left(\mu^{-1}\right) \,.
\end{IEEEeqnarray}

\chapter{Diagrams of the static potential at next-to-leading order}
\label{app:Diagrams}
The effective action used to compute the static potential is Eq.~(\ref{Stot}) 
\bea
S_\mathrm{eff}&=& \int\left(\psi^\dagger (i\p_2+eA_2)\psi + \chi_c^\dagger (i\p_2-eA_2^T)\chi_c\right) +\left(F_\mathrm{trial}^\dagger[\vec A]+F_\mathrm{trial}[\vec A]\right)\Bigg|_{A_1=0,A_2=A}\\
&=& \int\left(\psi^\dagger (i\p_2+eA)\psi + \chi_c^\dagger (i\p_2-eA^T)\chi_c\right) \nn\\&& + \frac{1}{2} \int_{\slashed{k_1},\slashed{k_2}}\slashed{\delta}\left(\vec{k}_1+\vec{k}_2\right) \frac{2\left(k^{(1)}\right)^2}{m+E_{k_1}}\delta^{ab}  A^a(\vec{k}_1)A^b(\vec{k}_2) + eS^{(1)}[A] + e^2S^{(2)}[A] \,,
\eea
where $S^{(1)}[A]$ and $S^{(2)}[A]$ are 
given in Eqs.~(\ref{S1}) and (\ref{S2}), respectively.

With these rules we can compute the diagrams of Sec.~\ref{sec:NLO} and match them onto the effective theory, Eq.~(\ref{EFT}):
\be
\mathcal{L} = S^\dagger(i\p_2+E_s(r))S\,. 
\ee

\pagebreak
\section{Diagram c)}
\begin{figure}[!h]
\centering
\includegraphics[width=0.25\textwidth]{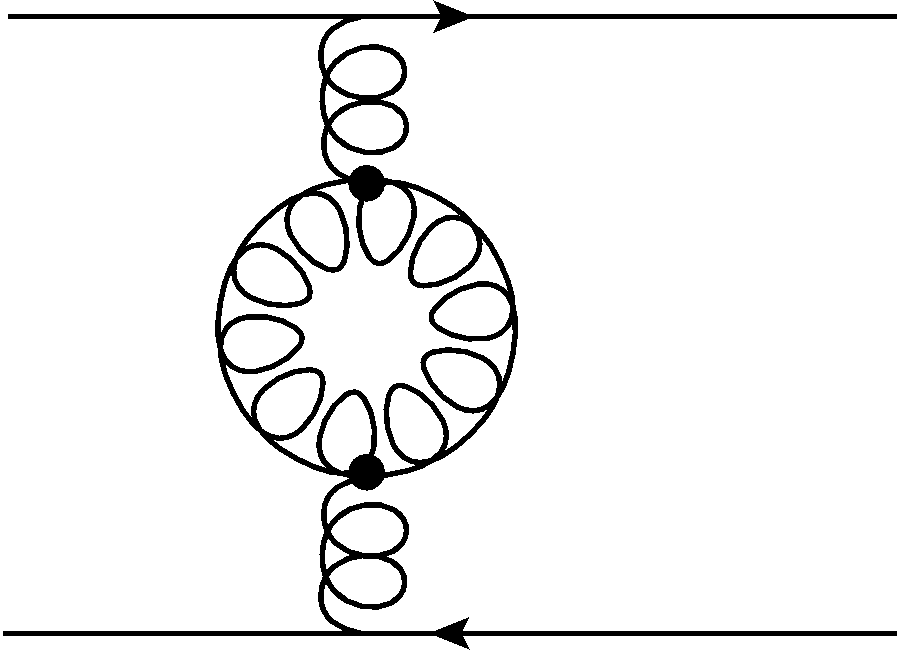} 
\end{figure} 
We write this as an effective bilinear term $e^2C_A A^{a}(\vec{q}) A^{a}(-\vec{q}) K^{(3)}(\vec q)$ and find 
\bE{rCl}
K^{(3)}(\vec q) &=& \frac{1}{2} \int_\slashed{k} \frac{m+\sqrt{m^2+\vec{k}^2}}{2\left(k^{(1)}\right)^2} \frac{m+\sqrt{m^2+(\vec{q}+\vec{k})^2}}{2\left(q^{(1)}+k^{(1)}\right)^2} \nn\\
 &&\times \Big(s^{(3)}(\vec{k},\vec{q},-\vec{k}-\vec{q})
+ s^{(3)}(-\vec{k}-\vec{q},\vec{k},\vec{q})  + s^{(3)}(\vec{q},-\vec{k}-\vec{q},\vec{k}) 
\nn\\
 &&\qquad\qquad -s^{(3)}(\vec{q},\vec{k},-\vec{k}-\vec{q})
- s^{(3)}(-\vec{k}-\vec{q},\vec{q},\vec{k})  - s^{(3)}(\vec{k},-\vec{k}-\vec{q},\vec{q}) \Big)  
\nn\\
 &&\times \Big(s^{(3)}(\vec{k},\vec{q},-\vec{k}-\vec{q})
+ s^{(3)}(-\vec{k}-\vec{q},\vec{k},\vec{q})  + s^{(3)}(\vec{q},-\vec{k}-\vec{q},\vec{k}) \Big) \,, \label{K3app}
\eE
where
\be
s^{(3)}(\vec{k}_1,\vec{k}_2,\vec{k}_3) = 2i \frac{1}{m+E_3}\frac{k^{(1)}_2 k^{(1)}_3}{\vec{k}_1^2} \left(k^{(2)}_1-\frac{k^{(1)}_1 \vec{k}_1\times \vec{k}_2}{(\sum_{i=1}^3 E_i) (m+E_2)}\right)\,. \label{S3app}
\ee
As we first take the limit $T\to\infty$, we can take $\vec q=(q,0)$.

In the hard regime ($k\sim m\gg q$) expansion to the leading order in $q$ gives:
\be
K^{(3)}_\mathrm{hard}(\vec q) = -\frac{1}{2}\int_\slashed{k} \frac{E_k-m}{(m+E_k)(m+2 E_k)^2} +\O(q^2)\,.
\ee
The correction to the potential is thus:
\be
\de\tilde E_s^{(1),c} = e^2C_F\frac{m}{q^2}K^{(3)}\frac{4\pi m^2}{q^2} \frac{e^2C_A}{2\pi m}
\ee

In the soft regime ($k\sim q\ll m$) the combination of $s^{(3)}$'s in the first parenthesis of Eq.~(\ref{K3app}) cancels the first term of Eq.~(\ref{S3app}), making this parenthesis $\O(m^{-3})$ and $K^{(3)}_\mathrm{soft}(\vec q)$ is therefore of $\O(m^{-2})$.
The potential thus has to be
\be
\propto e^4\frac{m^2}{q^4}\frac{q^2}{m^2}=\O(q^{-2}) \,,
\ee
and therefore may contribute to the linear term of the potential, but it cannot compensate the cubic terms.

\section{Diagram d)}
\begin{figure}[!h]
\centering
\includegraphics[width=0.25\textwidth]{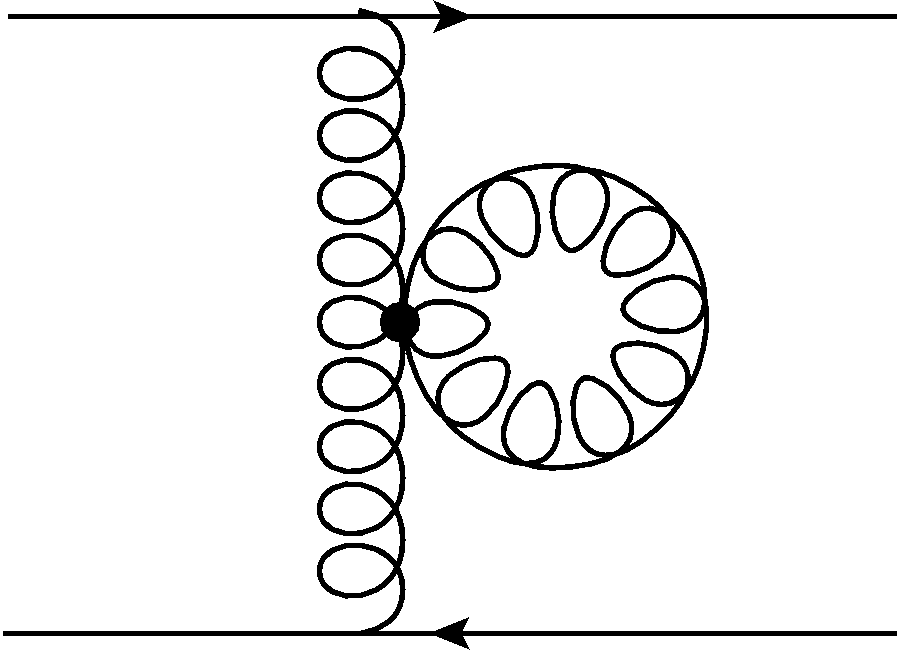} 
\end{figure} 

Analogously, we also write this diagram as an effective two field vertex \linebreak $e^2C_A A^{a}(\vec{q}) A^{a}(-\vec{q}) K^{(4)}(\vec q)$, finding 

\bE{rCl}
K^{(4)}(\vec q)&=& \int_\slashed{k} \frac{m+\sqrt{m^2+\vec{k}^2}}{2\left(k^{(1)}\right)^2} \Big(s^{(4)}(\vec{k},\vec{q};-\vec{k},\vec{q})  - s^{(4)}(\vec{k},\vec{q};-\vec{q},-\vec{k}) \nn\\
 &&\qquad\qquad - s^{(4)}(\vec{q},\vec{k};-\vec{k},-\vec{q}) + s^{(4)}(\vec{q},\vec{k};-\vec{q},\vec{k}) \Big) \,, \label{4fieldvp} 
\eE
with
\bE{l}
\hspace{-12cm} s^{(4)}(\vec{k}_1,\vec{k}_2;\vec{k}_3,\vec{k}_4)  \nn
\eE
\begin{IEEEeqnarray}{rCl}
 &=& \Bigg\{-\left(\frac{1}{m+E_4}-\frac{1}{m+E_{3+4}}\right) \frac{k^{(1)}_2 k^{(1)}_4}{\vec{k}_1^2\vec{k}_3^2} \left(k_1^{(2)}k_3^{(2)}-k_1^{(1)}k_3^{(1)}\right) \nn\\
&&\qquad -\frac{1}{m+E_2}\frac{k^{(1)}_1 k^{(1)}_2}{(\vec{k}_3+\vec{k}_4)^2 \vec{k}_4^2}\left((k_3^{(2)}+k_4^{(2)})k_4^{(2)}-(k_3^{(1)}+k_4^{(1)})k_4^{(1)}\right) \nn\\
&&+\frac{2}{(E_{1+2} +E_3 + E_4)(m + E_3)} \frac{k_2^{(1)}k_3^{(1)}k_4^{(1)}}{\vec{k}_1^2} \Bigg[\frac{1}{(m+E_4)(\vec{k}_1+\vec{k}_2)^2\vec{k}_2^2}\nn\\
&&\qquad\times \Bigg( (\vec{k}_2\cdot\vec{k}_4)((2\vec{k}_1\cdot\vec{k}_2+\vec{k}_2^2)\; k_1^{(1)}-\vec{k}_1^2\; k_2^{(1)}) \nn\\
&&\qquad\;+ (\vec{k}_2\times\vec{k}_4)((2\vec{k}_1\cdot\vec{k}_2+\vec{k}_2^2)\; k_1^{(2)}- \vec{k}_1^2\; k_2^{(2)})\Bigg)  \nn\\
&&\quad + \frac{1}{(m+E_{1+2})\vec{k}_4^2}\Bigg(-\vec{k}_1^2 k^{(1)}_4 +((\vec{k}_3-\vec{k}_2)\cdot\vec{k}_4) k_1^{(1)}- ((\vec{k}_3-\vec{k}_2)\times\vec{k}_4) k_1^{(2)} \Bigg) \Bigg] \nn\\
&&+\frac{k_1^{(1)}k_2^{(1)}k_3^{(1)}k_4^{(1)}}{(\sum_i E_i) (E_1 + E_2 + E_{3+4})(E_3 + E_4 + E_{1+2})(m+E_1)(m+E_3)} \nn\\
&& \quad \Bigg[\frac{1}{(m+E_2)(m+E_4)} \frac{\vec{k}_1^2 \vec{k}_3^2-(\vec{k}_1\times\vec{k}_2)(\vec{k}_3\times\vec{k}_4)}{(\vec{k}_1+\vec{k}_2)^2} \nn\\
&& \qquad +\frac{\vec{k}_2^2}{(m+E_2)} \left(-2\left(2\frac{\vec{k}_3\cdot\vec{k}_4}{\vec{k}_4^2} +1\right) -4\frac{(\vec{k}_1\times\vec{k}_2)(\vec{k}_3\times\vec{k}_4)}{\vec{k}_2^2\vec{k}_4^2}\right) \nn\\
&&\qquad\qquad\times \left(\frac{1}{m+E_{1+2}} - \frac{E_3+E_4+E_{1+2}}{(\vec{k}_3+\vec{k}_4)^2} \right) \nn\\
&& \qquad +\frac{(\vec{k}_3 + \vec{k}_4)^2}{(m+E_{3+4})} \left(\left(2\frac{\vec{k}_1\cdot\vec{k}_2}{\vec{k}_2^2} +1\right) \left(2\frac{\vec{k}_3\cdot\vec{k}_4}{\vec{k}_4^2} +1\right) -4\frac{(\vec{k}_1\times\vec{k}_2)(\vec{k}_3\times\vec{k}_4)}{\vec{k}_2^2\vec{k}_4^2}  \right) \nn\\
&& \qquad\qquad\times \left(\frac{1}{m+E_{1+2}} - 2\frac{E_3+E_4+E_{1+2}}{(\vec{k}_3+\vec{k}_4)^2} \right) \Bigg] \Bigg\}\,, \label{s4}
\end{IEEEeqnarray}

In the hard regime we expand to the leading order in $q$ and obtain
\bE{rCl}
K^{(4)}(\vec q)&=& \int_\slashed{k} \Bigg(\frac{E_k}{4k^2m} -\Bigg\{ \frac{E_k}{4m k^2} +\frac{ k^2 (m-E_k)+3 m^2 (m+E_k)}{E_k (2 k^2+3 m (m+E_k))^2} \Bigg\} \nn\\
&&\qquad\qquad -0 +  \frac{k^2 m +2m^3 -k^2 E_k+2m^2 E_k}{8 E_k^3 (m+E_k)^2}  \Bigg)   +\O(q^2)\,.
\eE
The correction to the potential coming from this diagram in the hard regime is thus:
\be
\de\tilde E_s^{(1),d} = e^2C_F\frac{m}{q^2}K^{(4)}\frac{4\pi m^2}{q^2} \frac{e^2C_A}{2\pi m}
\ee

In the soft regime the first and second line of (\ref{s4}) vanish and cancel in Eq.~(\ref{4fieldvp}), respectively. Hence $K^{(4)}_\mathrm{soft}(\vec q)$ is of $\O(m^{-2})$ and the correction to the potential is 
\be
\propto e^4\frac{m^2}{q^4}\frac{q^2}{m^2}=\O(q^{-2})\,.
\ee

\section{Diagram e)}
\begin{figure}[!h]
\centering
\includegraphics[width=0.25\textwidth]{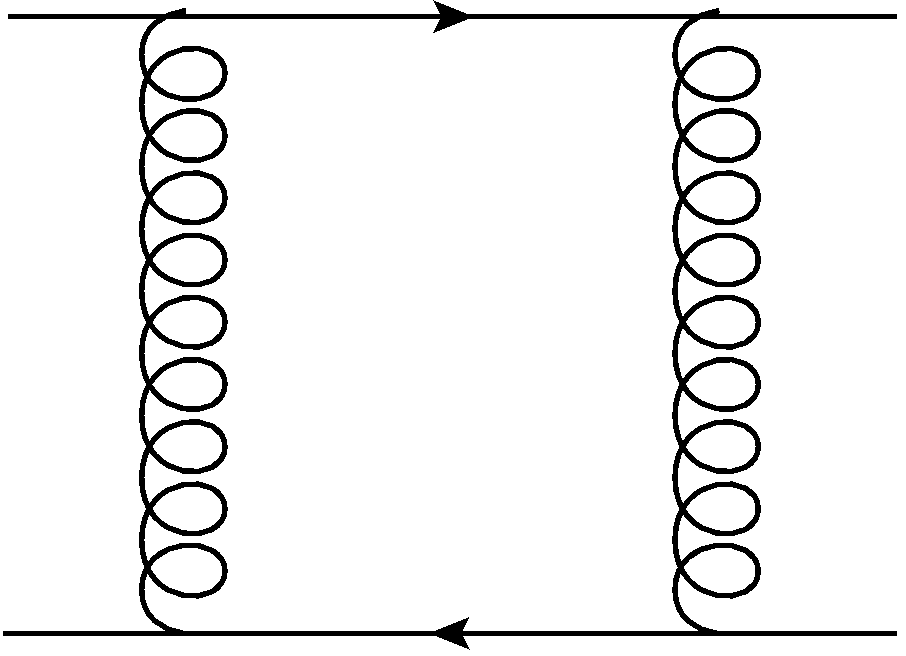} 
\end{figure} 
In the soft regime, this is the iteration of the potential.

In the hard it is 
\bE{rCl}
&\propto& \int_\slashed{k} \frac{m+\sqrt{m^2+\vec{k}^2}}{2\left(k^{(1)}\right)^2} \frac{1}{k^{(2)}+i\e}  \frac{m+\sqrt{m^2+(\vec{k}-\vec{q})^2}}{2\left(k^{(1)}-q^{(1)}\right)^2} \frac{1}{-k^{(2)}+i\e} \\
&\propto& \int_\slashed{k} \left( \frac{m+\sqrt{m^2+\vec{k}^2}}{2\left(k^{(1)}\right)^2}\right)^2 \frac{1}{k^{(2)}+i\e} \frac{1}{-k^{(2)}+i\e} + \O(q^2)\,,
\eE
so it neither contributes to the linear potential, nor to the cubic term.

\section{Diagram f)}
\begin{figure}[!h]
\centering
\includegraphics[width=0.25\textwidth]{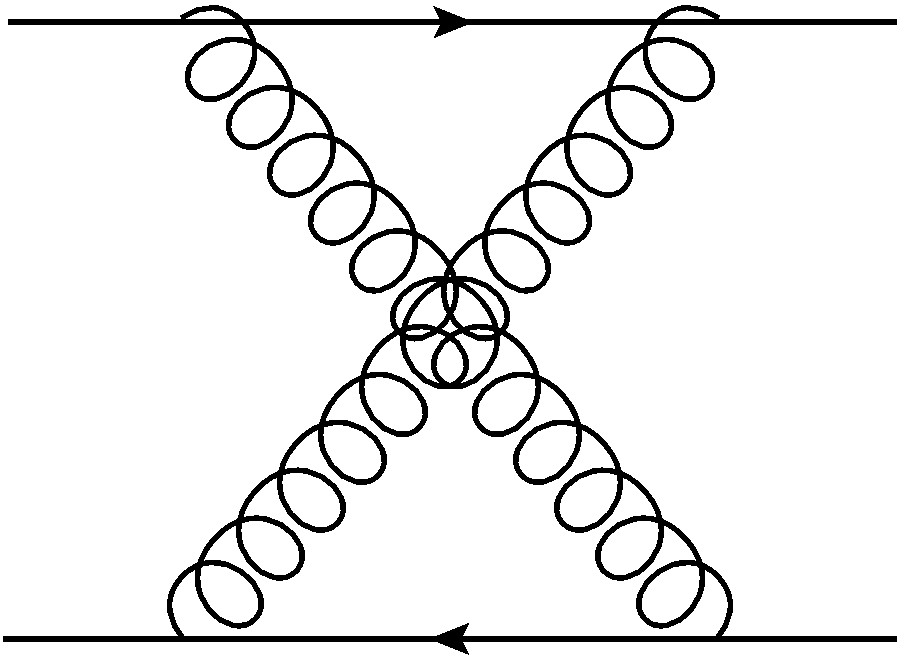} 
\end{figure} 
In the hard regime, this is beyond our accuracy, by the same reasoning used in the previous diagram.

In the soft regime it is
\bE{rCl}
&\propto&  \int_\slashed{k} \frac{m+\sqrt{m^2+\vec{k}^2}}{2\left(k^{(1)}\right)^2} \frac{1}{-k^{(2)}+i\e}  \frac{m+\sqrt{m^2+(\vec{k}+\vec{q})^2}}{2\left(k^{(1)}+q^{(1)}\right)^2} \frac{1}{-k^{(2)}+i\e}\,,
\eE
because we take the limit $q^{(2)}\to 0$ (due to $T\to\infty$). The integral over $k^{(2)}$ is the residue:
\bE{rCl}
&\propto& \int_\slashed{k^{(1)}} \mathrm{Res} \frac{m+\sqrt{m^2+\vec{k}^2}}{2\left(k^{(1)}\right)^2} \frac{m+\sqrt{m^2+(\vec{k}+\vec{q})^2}}{2\left(k^{(1)}+q^{(1)}\right)^2} \left( \frac{1}{-k^{(2)}+i\e} \right)^2\Bigg|_{k^{(2)}=i\e} \nn\\
&=&0 \,.
\eE

\section{Diagrams g) and h)}
\begin{figure}[!h]
\centering
\includegraphics[width=0.25\textwidth]{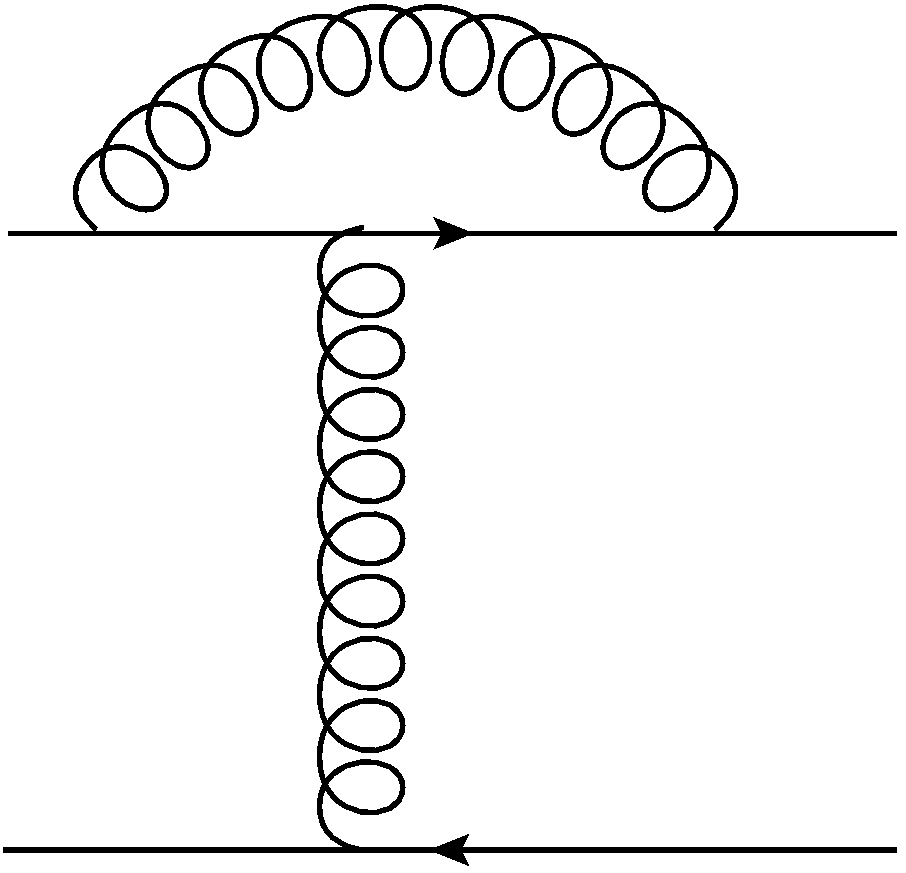} 
\end{figure} 
This diagram is in both regimes
\bE{rCl}
&\propto& \int_\slashed{k} \frac{m+\sqrt{m^2+\vec{k}^2}}{2\left(k^{(1)}\right)^2} \frac{1}{-k^{(2)}+i\e}  \frac{1}{-k^{(2)}+q^{(2)}+i\e}  \frac{m+\sqrt{m^2+\vec{q}^2}}{2\left(q^{(1)}\right)^2}\nn\\
&=& \O\left(q^{-2}\right)\,,
\eE
since $q^{(2)}\to 0$. So it cannot compensate the $q^{-4}$ terms. 

The same holds true for the inverted diagram (diagram h)).

\newpage
\section{Diagrams i) and j)}
\begin{figure}[!h]
\centering
\includegraphics[width=0.25\textwidth]{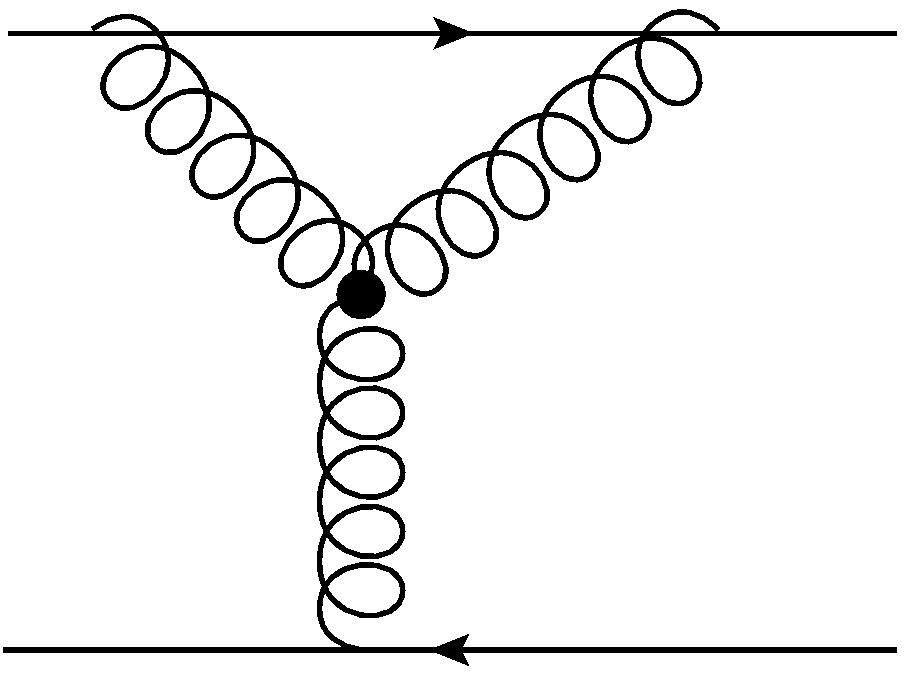} 
\end{figure} 
This diagram is 
\bE{rCl}
&\propto&f^{abc}\tr[[T^a,T^b]T^c] \frac{m+\sqrt{m^2+\vec{q}^2}}{2\left(q^{(1)}\right)^2} \nn\\
 &&\int_\slashed{k} \frac{m+\sqrt{m^2+\vec{k}^2}}{2\left(k^{(1)}\right)^2} \frac{1}{-k^{(2)}+i\e}    \frac{m+\sqrt{m^2+(\vec{k}+\vec{q})^2}}{2\left(q^{(1)}+k^{(1)}\right)^2}  \nn\\
 &&\times \Big(s^{(3)}(\vec{k},\vec{q},-\vec{k}-\vec{q})
+ s^{(3)}(-\vec{k}-\vec{q},\vec{k},\vec{q})  + s^{(3)}(\vec{q},-\vec{k}-\vec{q},\vec{k}) 
\nn\\
 &&-s^{(3)}(\vec{q},\vec{k},-\vec{k}-\vec{q})
- s^{(3)}(-\vec{k}-\vec{q},\vec{q},\vec{k})  - s^{(3)}(\vec{k},-\vec{k}-\vec{q},\vec{q}) \Big)  
\eE

In the hard regime we expand to the leading order in $q$:
\begin{IEEEeqnarray}{rCl}
&\propto& e^4C_A (N^2-1) \frac{m}{q^2} \int_\slashed{k} \frac{1}{(k^{(1)})^2 \left(m+2 \sqrt{m^2+\vec{k}^2}\right)} =\O(q^{-2})
\end{IEEEeqnarray}

In the soft regime, again the leading order terms of the $s^{(3)}$ terms cancel, thus the diagram has to be
\be
\propto e^4\frac{m}{q^2}\frac{1}{m}=\O(q^{-2})\,,
\ee 
and therefore may contribute to the linear term of the potential, but it cannot compensate the cubic terms.

The same holds true for the inverted diagram (diagram j)).

\newpage
\section{Diagrams k), l), m), and n)}
\begin{figure}[!h]
\centering
\includegraphics[width=0.25\textwidth]{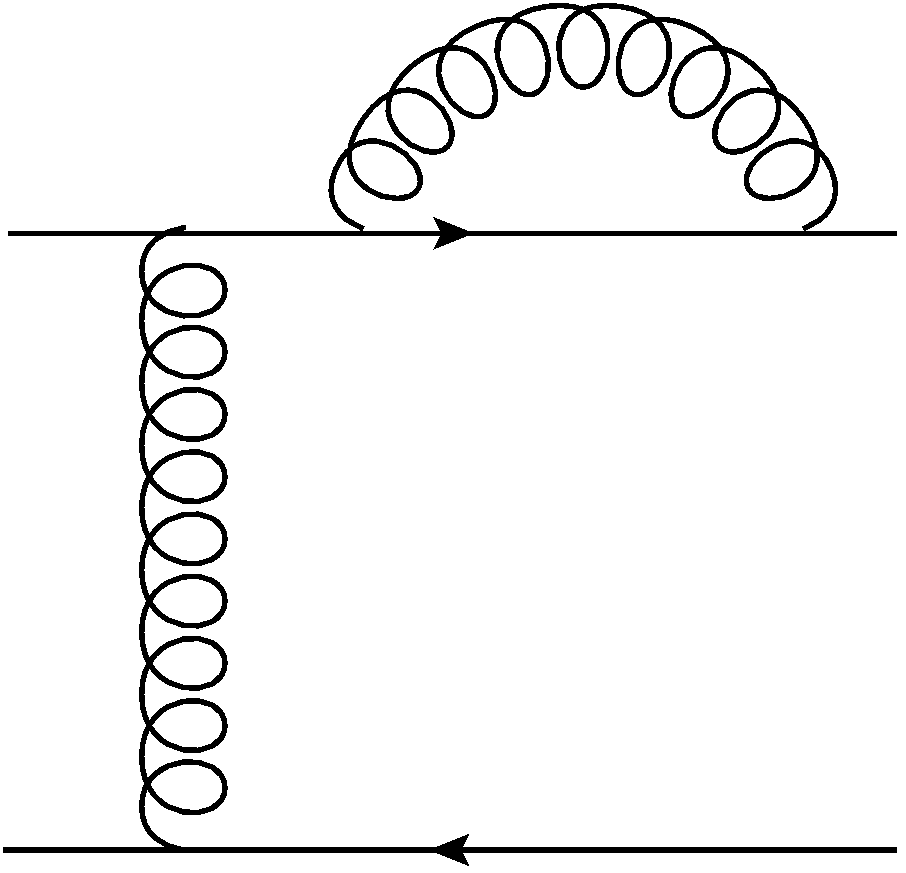}
\end{figure} 

The loop only modifies the overall coefficient, not the momentum transfer, so this diagram and all its permutations are of $\O(q^{-2})$.

\backmatter

\end{document}